\documentclass[twocolumn]{aastex63}
\usepackage{apjfonts}

\usepackage[caption=false]{subfig}
\usepackage{amsmath}
\usepackage[normalem]{ulem} % either use this (simple) or
\usepackage{algpseudocode}
\usepackage{booktabs}
\usepackage{listings}
\usepackage[toc,page]{appendix}

\renewcommand{\thefootnote}{\fnsymbol{footnote}} 
\usepackage{subfiles}

\newcommand{\code}[1]{{\color{blue} \texttt{#1}}}
\newcommand{\pack}[1]{{\color{magenta} \texttt{#1}}}
\newcommand{\lprime}{$L$}

\begin{document}

\shorttitle{The L- and M-band Imaging Atlas}
\shortauthors{Isbell et al.}

\title{The Subarcsecond Mid-Infrared View of Local Active Galactic Nuclei. IV. The L- and M-band Imaging Atlas
\footnote{This atlas makes use of European Southern Observatory (ESO) observing programs 65.P-0519, 67.B-0332, 70.B-0393, 71.B-0379, 71.B-0404, 072.B-0397, 074.B-0166, 085.B-0639, and 290.B-5133. } }

%AAS style
\author[0000-0002-1272-6322]{Jacob W. Isbell } \affiliation{Max-Planck-Insitut f\"ur Astronomie, K\"onigstuhl 17, 69117 Heidelberg, Germany}
\author[0000-0003-1014-043X]{Leonard Burtscher} \affiliation{Sterrewacht Leiden, Niels Bohrweg 2, 2333 CA Leiden, The Netherlands}
\author[0000-0003-0220-2063]{Daniel Asmus} \affiliation{School of Physics \& Astronomy, University of Southampton, Southampton SO17 1BJ, UK} \affiliation{Gymnasium Schwarzenbek, Buschkoppel 7, 21493 Schwarzenbek, Germany}
\author[0000-0003-4291-2078]{J\"org-Uwe Pott} \affiliation{Max-Planck-Insitut f\"ur Astronomie, K\"onigstuhl 17, 69117 Heidelberg, Germany} 
\author{Paul Couzy} \affiliation{Sterrewacht Leiden, Niels Bohrweg 2, 2333 CA Leiden, The Netherlands}
\author[0000-0001-5146-8330]{Marko Stalevski} \affiliation{Astronomical Observatory, Volgina 7, 11060 Belgrade, Serbia} \affiliation{Sterrenkundig Observatorium, Universiteit Gent, Krijgslaan 281-S9, Gent, 9000, Belgium}
\author{Violeta G\'amez Rosas} \affiliation{Sterrewacht Leiden, Niels Bohrweg 2, 2333 CA Leiden, The Netherlands}
\author{Klaus Meisenheimer} \affiliation{Max-Planck-Insitut f\"ur Astronomie, K\"onigstuhl 17, 69117 Heidelberg, Germany}
\correspondingauthor{J. W. Isbell}
\email{isbell@mpia.de}
 
\received{18 September 2020}
\accepted{14 January 2021}
\submitjournal{ApJ}
 

\begin{abstract}
We present the largest currently existing subarcsecond 3-5 \micron~atlas of 119 local ($z < 0.3$) active galactic nuclei (AGN). This atlas includes AGN of 5 subtypes: 22 are Seyfert 1; 5 are intermediate Seyferts; 46 are Seyfert 2; 26 are LINERs; and 20 are composites/starbursts. Each AGN was observed with VLT ISAAC in the $L$- and/or $M$-bands between 2000 and 2013. We detect at 3$\sigma$ confidence 92 sources in the $L$-band and 83 sources in the $M$-band. We separate the flux into unresolved nuclear flux and resolved flux through two-Gaussian fitting. We report the nuclear flux, extended flux, apparent size, and position angle of each source, giving $3\sigma$ upper-limits for sources which are undetected. Using \textit{WISE} \textit{W1}- and \textit{W2}-band photometry we derive relations predicting the nuclear $L$ and $M$ fluxes for Sy1 and Sy2 AGN based on their \textit{W1-W2} color and {\it WISE} fluxes. Lastly, we compare the measured mid-infrared colors to those predicted by dusty torus models SKIRTOR, CLUMPY, CAT3D, and CAT3D-WIND, finding best agreement with the latter. We find that models including polar winds best reproduce the 3-5\micron~colors, indicating that winds are an important component of dusty torus models. We find that several AGN are bluer than models predict. We discuss several explanations for this and find that it is most plausibly stellar light contamination within the ISAAC $L$-band nuclear fluxes.
\end{abstract}

\section{Introduction}
Understanding the dust in the vicinity of central supermassive black holes is instrumental to understanding how active galactic nuclei (AGN) are fed and powered. Large, obscuring dusty structures are held responsible for both funneling material  toward the central engine, and for distinguishing between Type 1 and Type 2 AGN. In the Unified Model of AGN \citep{antonucci1993, urry1995,netzer2015}, a central obscuring torus of dust is oriented such that the broad-line region of the AGN is directly visible (Type 1) or such that its observation is blocked by the torus (Type 2). The dust making up the sublimation ring (the dust closest to the AGN) is $\sim 1500$K and is best observed in the near-infrared (NIR).  The extended dust of the torus, on the other hand, is most readily observable in the thermal infrared (3-25 \micron). 
Interferometric observations of AGN in the $H$-band \citep{weigelt2004}, in the $K$-band \citep{wittkowski1998,  kishimoto2011, gravitycollaboration2020}, and in the $N$-band \citep[e.g.,][]{lopez-gonzaga2014, tristram2014, leftley2019} conclusively show hot ($\gtrsim 100$K) dust in the vicinity of AGN (0.1pc - 100 pc) and provide strong evidence that the torus is clumpy. Clumpy media have moreover been argued in theory as necessary to prevent the destruction of dust grains by the surrounding hot gas \citep{krolik1988}. Following \citet{nenkova2002}, a clumpy formalism has been used in many radiative transfer models of tori \citep[e.g.,][]{nenkova2008,schartmann2008,honig2010,stalevski2016}, reproducing the spectral energy distributions (SEDs) and spectral features of the $N$-band particularly well. The 3-5\micron~bump \citep[see e.g., ][]{edelson1986, kishimoto2011,mor2012,honig2013}, however, has remained difficult to properly model. Recent modeling suggests that this feature can be explained by the inclusion of a wind-driven outflow originating at the sublimation ring and propagating orthogonal to the disk \citep[the disk+wind model;][]{honig2017}. 

 A large body of work using spectral energy distribution fits to local AGN \citep[e.g.,][]{ramosalmeida2009, alonso-herrero2011, lira2013, garcia-gonzalez2016, garcia-bernete2019, martinez-paredes2020} suggests that $L$ and $M$ observations at high sensitivity and angular resolution are required to study the physical properties of the 3-5 \micron~radiation bump. In fact, \citet{lira2013} emphasize that spectral information at 5\micron~is necessary to properly constrain their SED fits. This mid-infrared (MIR) bump is expected to originate from dust radiating at intermediate spatial scales: outside of the accretion disk and the hot dust sublimation zone, but still inside of any extended polar dust emission further out. In the near future, those spatial scales will be directly resolved in detail in the $L$- and $M$-bands with the new instrument Very Large Telescope Interferometer (VLTI) MATISSE, which allows for simultaneous $L$-, $M$-, and $N$-band interferometric observations, but which also requires accurate estimates of nuclear target fluxes \citep{lopez2014}.

A primary goal of this paper is to anticipate such future interferometric investigations of dusty AGN in the thermal infrared. We build on the SubArcSecond MidInfraRed Atlas of Local AGN \citep[SASMIRALA;][hereafter A14]{asmus2014}, which presented an $N$- and $Q$-band imaging atlas of nearby AGN at subarcsecond resolution. In this work, we extend this atlas to the $L$- and $M$- bands for 119 nearby ($z < 0.3$) AGN, at a threefold increase in angular resolution compared to the $N$-band. We derive new spatial flux information at the seeing limit of the excellent Cerro Paranal site, and systematically explore how these fluxes relate to those measured in space with the \textit{WISE} \textit{W1} and \textit{W2} bands. We then investigate how our measurements compare to the expectations derived from existing clumpy torus models.

The examination herein of $L$- and $M$-band fluxes in local AGN in a statistically relevant sample fulfills two goals:
(\textit{i}) direct observational evidence of the fact that \textit{LM}-flux in excess of the classical hot torus radiation is a typical feature of nearby AGN; and furthermore, (\textit{ii}) the presentation of an atlas and systematic characterization of the spatially resolved radiation properties to aide the sample selection for future, detailed interferometric imaging of that excess radiation to further understand its origin.

This paper is structured as follows: in \S \ref{sec:selection} we present the sample, discussing its selection and observation. In \S \ref{sec:reduction} we discuss the data reduction and present the measured fluxes. In \S \ref{sec:fluxcat} we present and describe the $L$ and $M$ flux catalogs. In \S \ref{sec:models}  we compare the MIR colors of our sample to those predicted from various clumpy torus models. In \S \ref{sec:wise} we compare the $L$ and $M$ fluxes to those measured with \textit{WISE} bands \textit{W1} and \textit{W2} respectively.  We summarize and conclude the paper in \S \ref{sec:conc}. Additionally, we present the observing dates and conditions of each source in Appendix A, we further explain the flux calibration procedure in Appendix B, and we present cutouts of all 119 sources in Appendix C.

\section{Sample Selection and Observations}
\label{sec:selection}
 The program from which the majority of sources were observed (ESO ID 290.B-5133(A); PI: Asmus) was a survey of AGN designed to complement the subarcsecond $N-$ and $Q-$band AGN sample of A14 with 3 and 5\micron~images. Out of the original sample of 253 objects, 59 were observed in June and July 2013 with the ISAAC \citep[Infrared  Spectrometer and Array Camera;][]{moorwood1999} instrument on the Very Large Telescope (VLT) before it was decommissioned. Two stars were observed as flux calibrators during each night of the observing program: HD 106965 and HD 130163.  

To supplement this sample, we searched the ESO archive for all $L$- and/or $M$-band ISAAC observations of local ($z<0.3$) active galaxies contained in A14. We focus on 8m class telescopes in the Southern Hemisphere because we need to resolve as much of the central region as possible to properly separate the AGN itself from its host galaxy. We did not include archival observations taken with VLT NaCo because they include only 9 AGN which are not part of this sample, providing a small statistical gain for a large data reduction overhead. The ISAAC archival programs contain 60 of the individual targets and were proposed with a variety of goals, but they each contain nearby, optically-classified active galaxies observed in at least one of the $L$- and $M$-bands. We include 20 AGN observed from the archival programs which were not in the A14 sample. As these are archival data, the selection of calibration stars and the frequency of their observation is inconsistent. Whenever possible, we gather the calibration sources taken with the same instrumental setup on the same night as the AGN. Several sources were repeated in different programs.  

Each AGN was observed with the L$^{\prime}$-filter ($\lambda_{c} = 3.78$\micron; hereafter referred to as the $L$-band), with the narrow M$_{\rm nb}$-filter ($\lambda_c = 4.66$\micron; hereafter referred to as the $M$-band), or with both. We list the ISAAC programs included in this work in Table \ref{tab:eso_progs}, with the principle investigator and number of targets observed in each filter.

\begin{table}[]
    \centering
    \begin{tabular}{l|lcl}
    Prog. ID & PI & $N_{\rm Obs., L}$&$N_{\rm Obs., M}$  \\ \hline \hline \\
       65.P-0519 & Krabbe & 15 & 38\\
       67.B-0332 & Marco & 20&12\\ 
       70.B-0393 & Lira & 40 & 38\\
       71.B-0379 & Lira & 30&36\\ 
       71.B-0404 & Brooks & 3&4\\ 
       072.B-0397 &Galliano & 2&2 \\ 
       074.B-0166 &Galliano & 14 & 0\\ 
       085.B-0639 &Asmus & 14 & 9\\
       290.B-5133 & Asmus & 69 & 69 \\
    \end{tabular}
    \caption{ESO ISAAC observing programs entering into this analysis. The sum of the $N_{\rm obs, X}$ columns can be larger than 119 because several sources were observed in multiple epochs.}
    \label{tab:eso_progs}
\end{table}

\begin{figure}[t]
\centering
\includegraphics[width=0.4\textwidth]{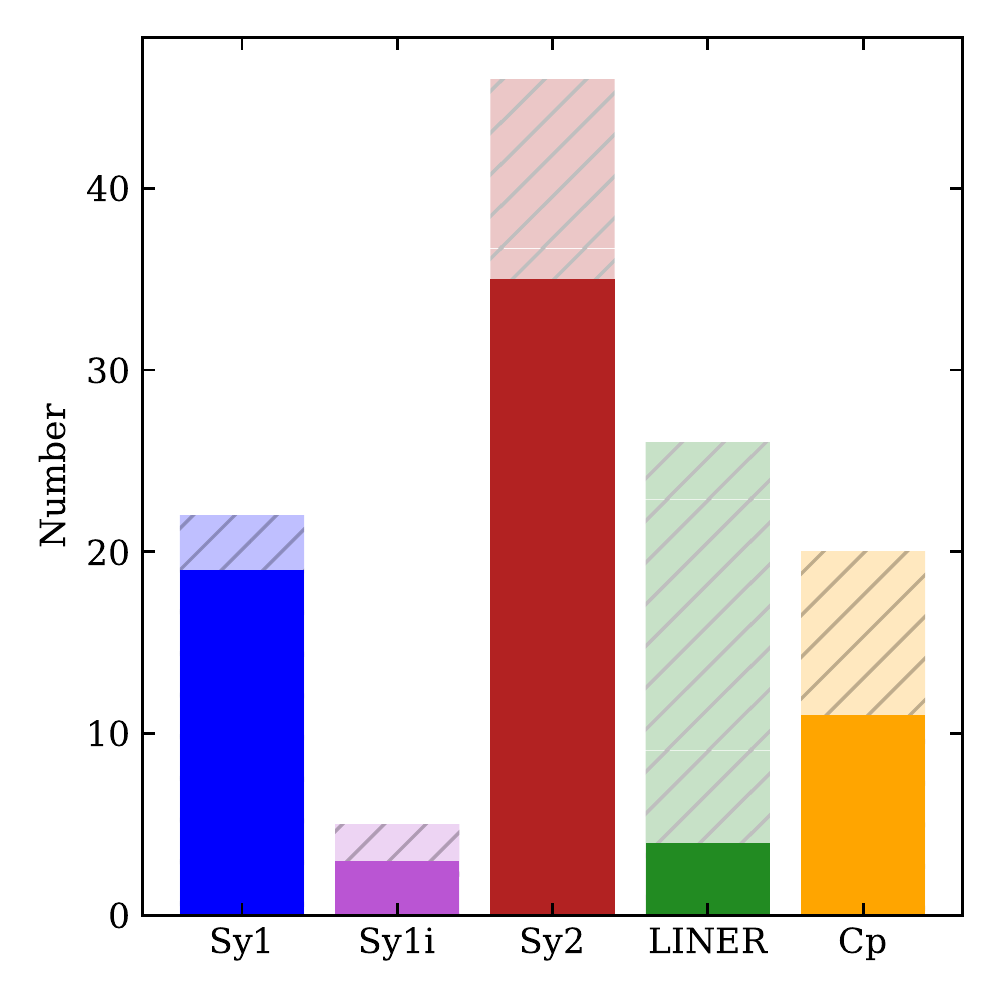}
\caption{Histogram of the AGN optical classifications in this sample. Darkened regions indicate the number of sources detected with SNR$\geq$2 with both the \lprime~and $M$ filters. }
\label{fig:hist}
\end{figure}

The final sample includes 119 AGN of four broad classes, which we group based on their optical classifications as in A14: 
\begin{itemize}
    \item Seyfert 1 (Sy1): contains 1, 1.2, 1.5, 1n
    \item Intermediate Seyferts (Int. Sy): contains 1.8, 1.9, 1.5/2
    \item Seyfert 2 (Sy2): contains 1.8/2, 1.9/2, 2 
    \item Low-ionization nuclear emission region galaxies (LINERs): contains L, L:, L/H, S3
    \item Composites/Starbursts (Cp): contains Cp, Cp:, H.
\end{itemize}
 Optical classifications for each of the AGN come from \citet{asmus2014} when available and are listed with the individual sources otherwise. As A14 compiled all optical classifications from the literature, there are multiple classifications for some objects (e.g., Sy 1.5/2). In Fig. \ref{fig:hist} we show the distribution of AGN classes in the sample.

The final sample is then as follows: 21 are Seyfert 1; 5 are Intermediate Seyferts; 46 are Seyfert 2; 29 are LINERs; and 16 are so-called `Cp' or Composites/Starbursts. Throughout the paper we color-code these types consistently as blue, purple, red, orange, and green, respectively. While there are 119 total AGN, not all of them were observed in both bands; instead there are 95 $L$-band observations and 107 $M$-band observations. The final sample thus includes 87 AGN with measurements in both bands. There are 20 AGN included in this work which were not part of the original A14 $N$- and $Q$-band sample. Their optical classifications (from the literature) are given in Table \ref{tab:nuc}.

\begin{figure*}[ht]
    \centering
    \includegraphics[width=1\textwidth]{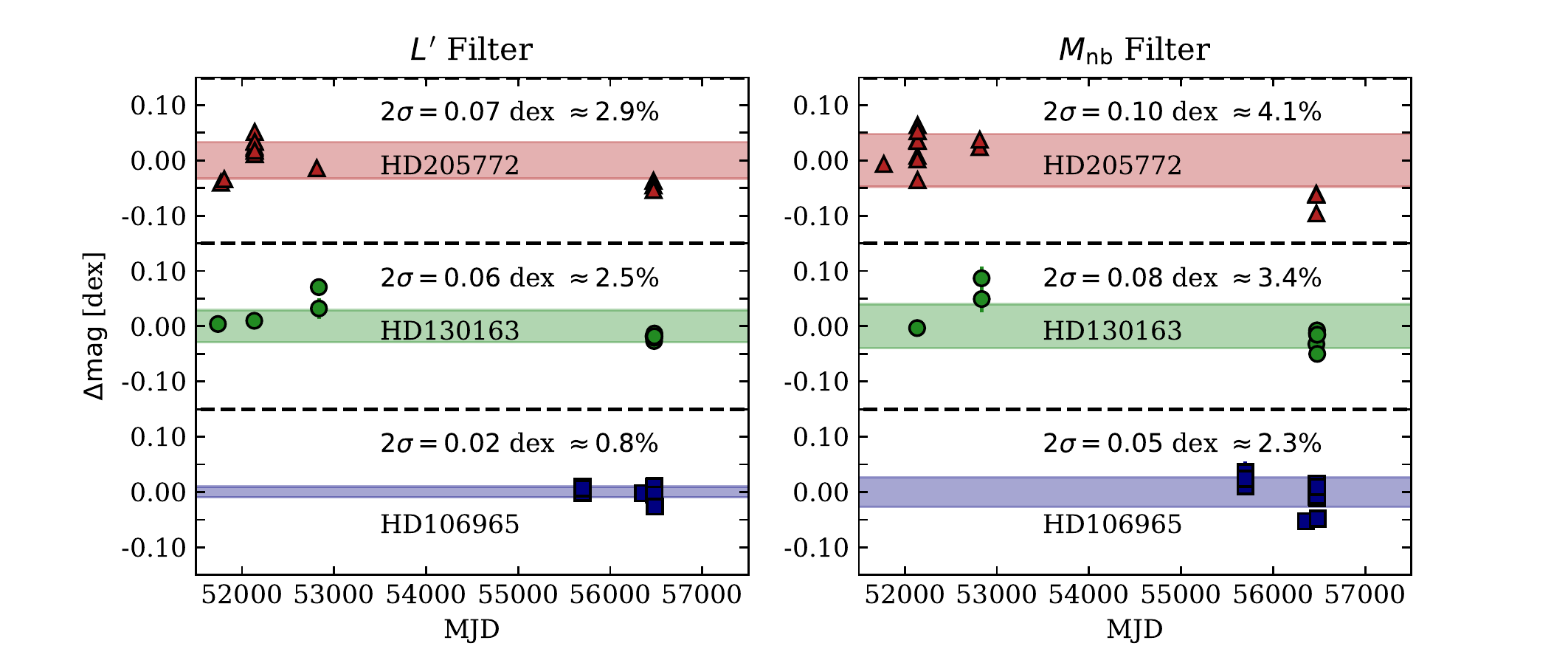}
    \caption{Stability of the 3 most often observed calibrators.}
    \label{fig:cal_stab}
\end{figure*}

\section{Data Reduction} \label{sec:reduction}

The data were reduced using the custom Python tool, VISIR and ISAAC Pipeline Environment\footnote{https://github.com/danielasmus/vipe} (VIPE; Asmus et al., in prep.). This pipeline applies the following algorithm:
\begin{enumerate}
    \item Combine the data at each individual chop (and nod) position
    \item Combine the individual exposures of each nodding cycle into pairs
    \item Combine the nodding pairs into a single exposure, taking the jitter offsets into account
    \item Find the brightest source beam in the total combined image 
    \item Extract the positive and negative beams from the different chop/nod positions by attempting to fit every beam in every nodding pair. 
    If this is not possible, e.g., because the source is too faint, extract and combine beams at calculated chop/nod positions instead.
\end{enumerate}
In comparison to the default pipeline, this method does a better job removing the sky background in especially the M-band, resulting in several novel $\geq 2\sigma$ NIR detections (e.g., of NGC 5278). In the exemplary case of the faintest $L$-band source we detected, 3C321, the signal-to-noise ratio of the detection increases from ${\rm SNR}_{\rm DRS}=1.01$ with the default pipeline to ${\rm SNR}_{\rm VIPE}= 6.71$ with VIPE.

\subsection{Two-Gaussian Fitting}
Our primary interest is the unresolved, nuclear flux capturing the emission of the central engine. While for the nearest AGN, we may detect extended thermal dust emission, for more distant AGN it is likely the nuclear emission also contains significant contribution of stellar light from the host. We aim to separate these two disparate components by
fitting two elliptical Gaussians to each reduced image: one to represent the unresolved emission, and one to represent the extended emission. This method is quite commonly used in MIR interferometric data \citep[e.g.,][]{burtscher2013} to disentangle extended emission and the central engine.

As we were primarily interested in the unresolved component of each AGN, we needed to have an estimate of the point spread function (PSF) of each observation. For this, we did an initial round of fitting only the calibrators with single elliptical Gaussians. We found that the PSF size can vary by up to $\sim 10\%$ within an individual night. We can then set limits on the double-Gaussian fit; one component is set to have the major and minor axes ($\pm 10\%$) of the calibrator measured closest in time, while the second, larger component is required to have axes at least $10\%$ larger than the central component. This accomplishes two things: 1) it effectively ignores the small amount of non-Gaussian central flux in the PSF, and 2) it wholly separates the extended and unresolved components, reducing the number of fit degeneracies and prevents the extended component from mistakenly fitting any PSF residuals. To reduce the number of total parameters, we assumed that the Gaussians are concentric.  

Both the fitted Gaussians' parameters and the error estimates are obtained through Markov-Chain Monte Carlo likelihood maximization. We sample the parameter space using the package \pack{emcee} \citep{foreman-mackey2013}. The log-probability function to be maximized is given by the typical formulation
\begin{equation}
p(\vec{\theta}, c | \vec{x},y,\sigma) \propto p(\vec{\theta}) p(y | \vec{x}, \sigma, \vec{\theta}, c).
\end{equation}
with measurements $y$ at positions $x$, parameters $\vec{\theta}$ and error estimates $\sigma$ scaled by some constant $c$.
For maximum likelihood estimation, the log likelihood function for an arbitrary model $f(x,\vec{\theta})$  can be represented as
\begin{equation}
\ln  p(y | \vec{x},\sigma,\vec{\theta} , c) = -\frac{1}{2} \sum_{n} \Big [ \frac{(y_n - f(x_n,\vec{\theta}))^2}{s_n^2 } + \ln(2 \pi s_n^2) \Big ], 
\end{equation}
where $s_n^2= \sigma_n^2 + c^2f(x_n,\vec{\theta})^2 $, and $c$ represents the underestimation of the variance by some fractional amount. We estimate the best-fit value as the median of each marginalized posterior distribution and the $1\sigma$ errors from the values at $16^{\rm th}$ and $84^{\rm th}$ percentiles. 

Finally, we define the nuclear Gaussian flux ($F_{\rm nuc.gauss}$) as the integrated flux of the PSF-sized, so-called ``unresolved'' component, and the extended Gaussian flux ($F_{\rm ext.gauss}$) as the integrated flux of the second, larger component. In the remainder of this paper, AGN ``nuclear flux'' refers to $F_{\rm nuc.gauss}$, emphasizing that for sources closer than the median distance of $45.6$ Mpc, at the average fitted calibrator size of $\approx 425$ mas, this area covers the central $\leq 100$ pc region of the AGN.

\begin{figure*}[ht]
\centering
\subfloat{\includegraphics[width=0.24\hsize]{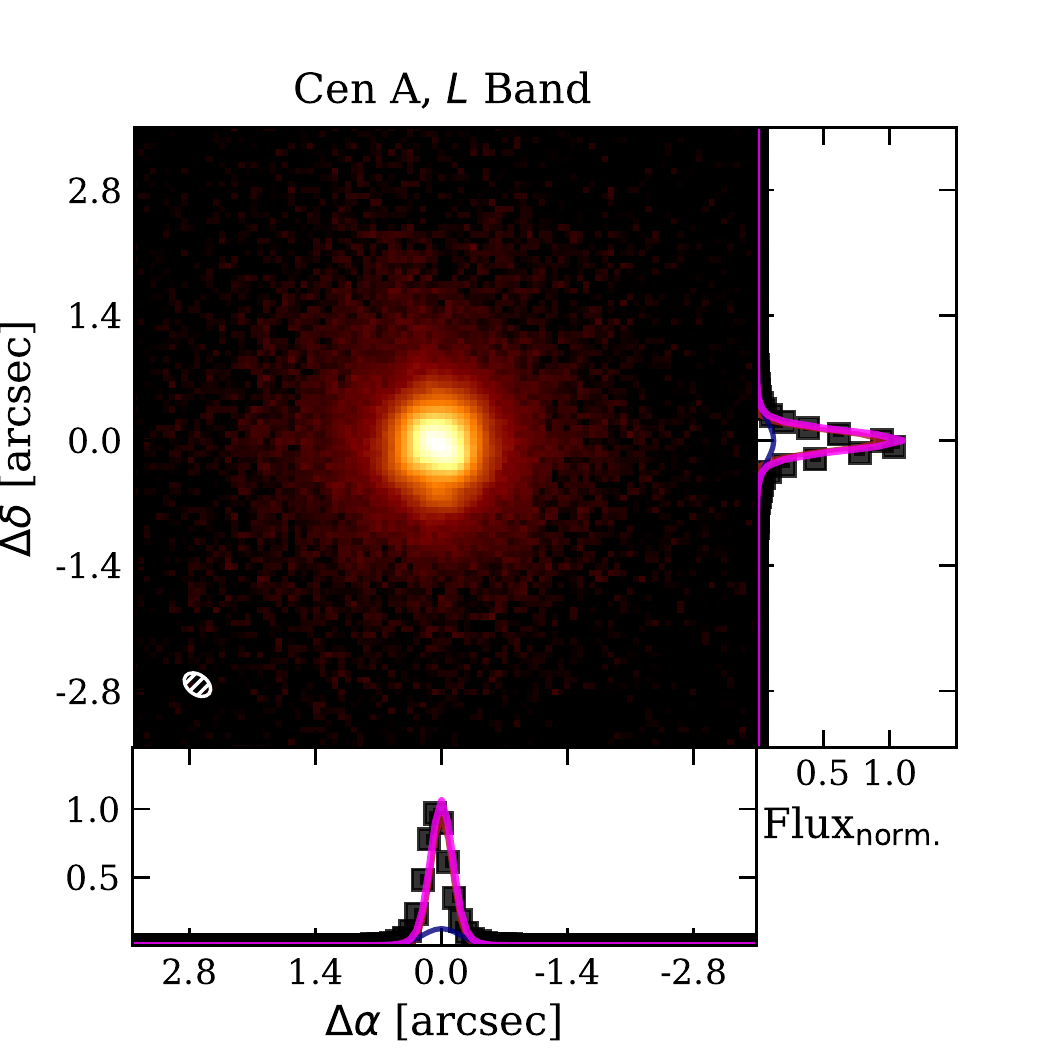}}
\subfloat{\includegraphics[width=0.24\hsize]{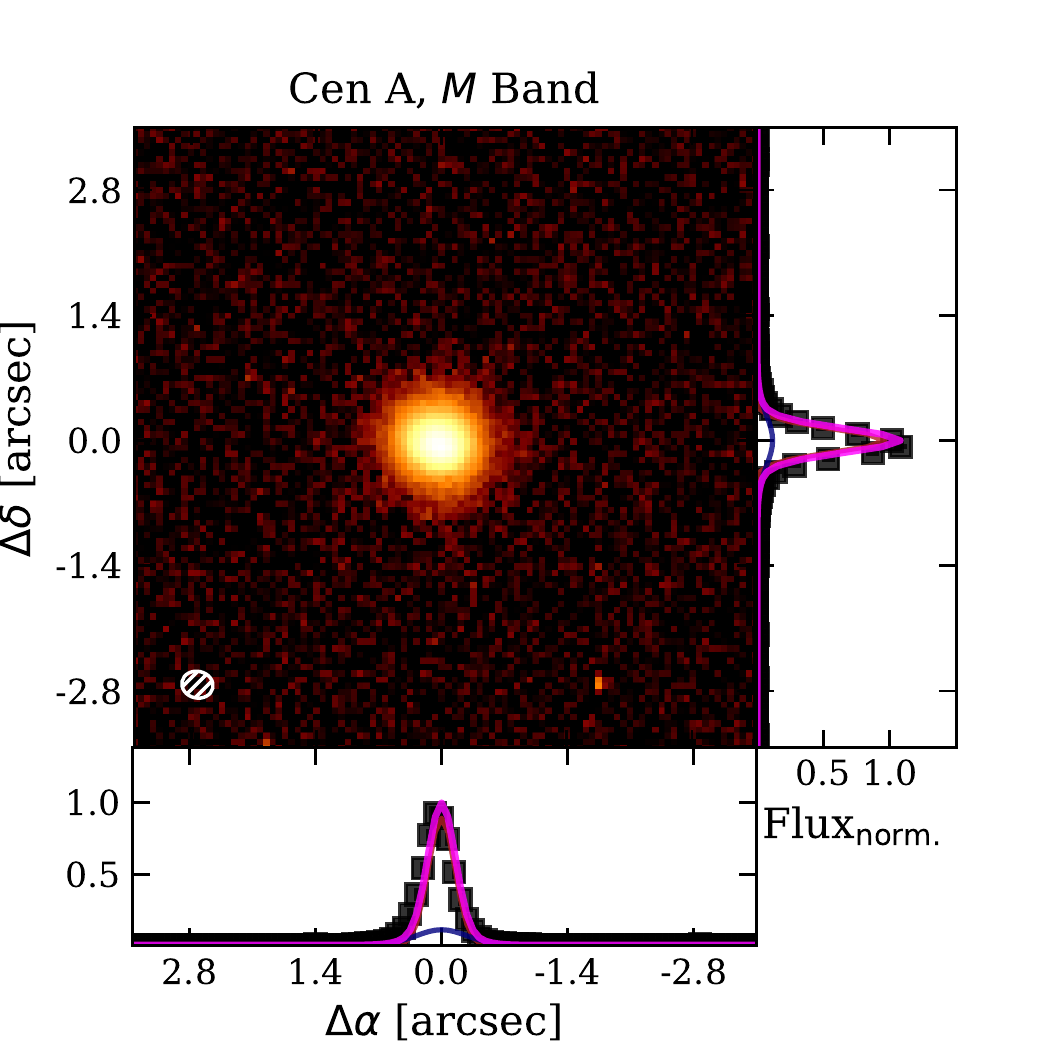} }
\subfloat{\includegraphics[width=0.24\hsize]{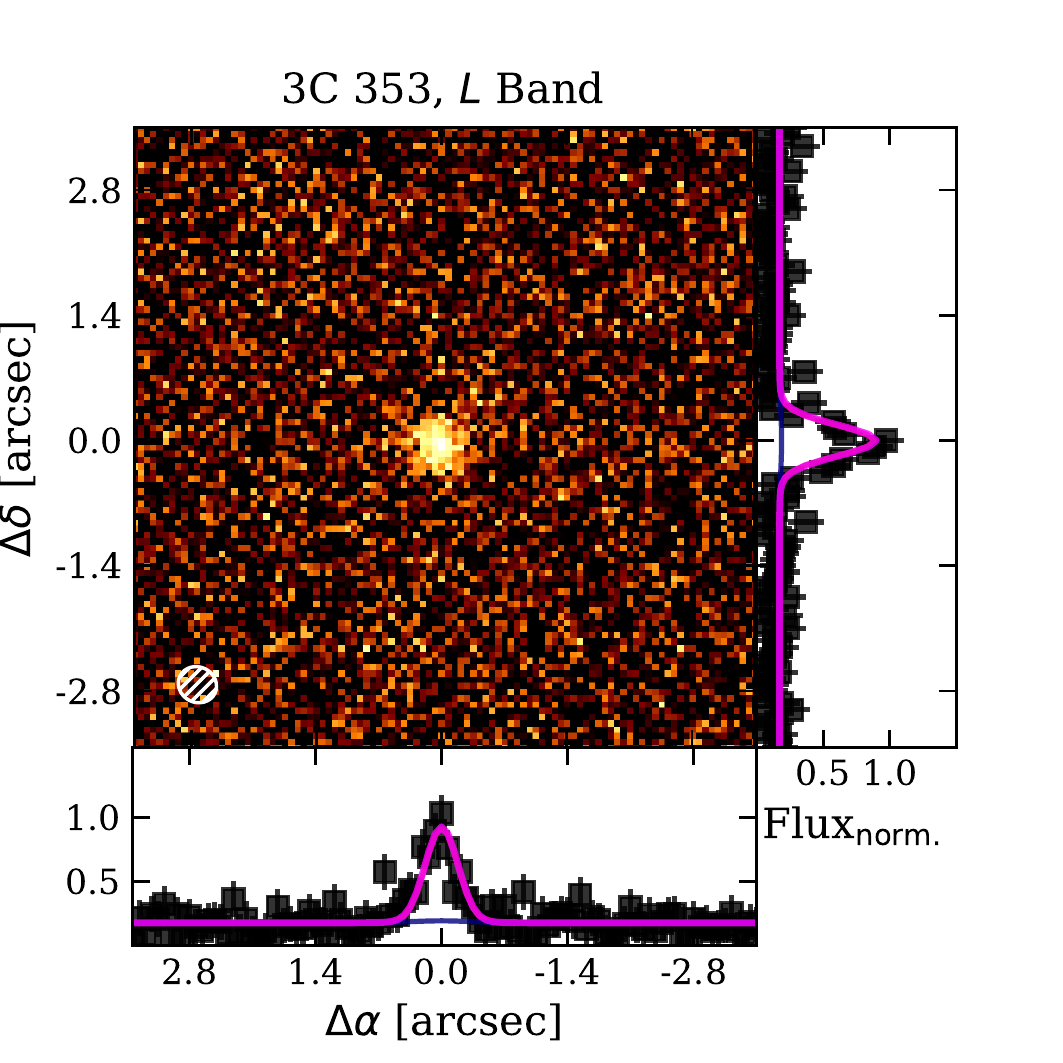}}
\subfloat{\includegraphics[width=0.24\hsize]{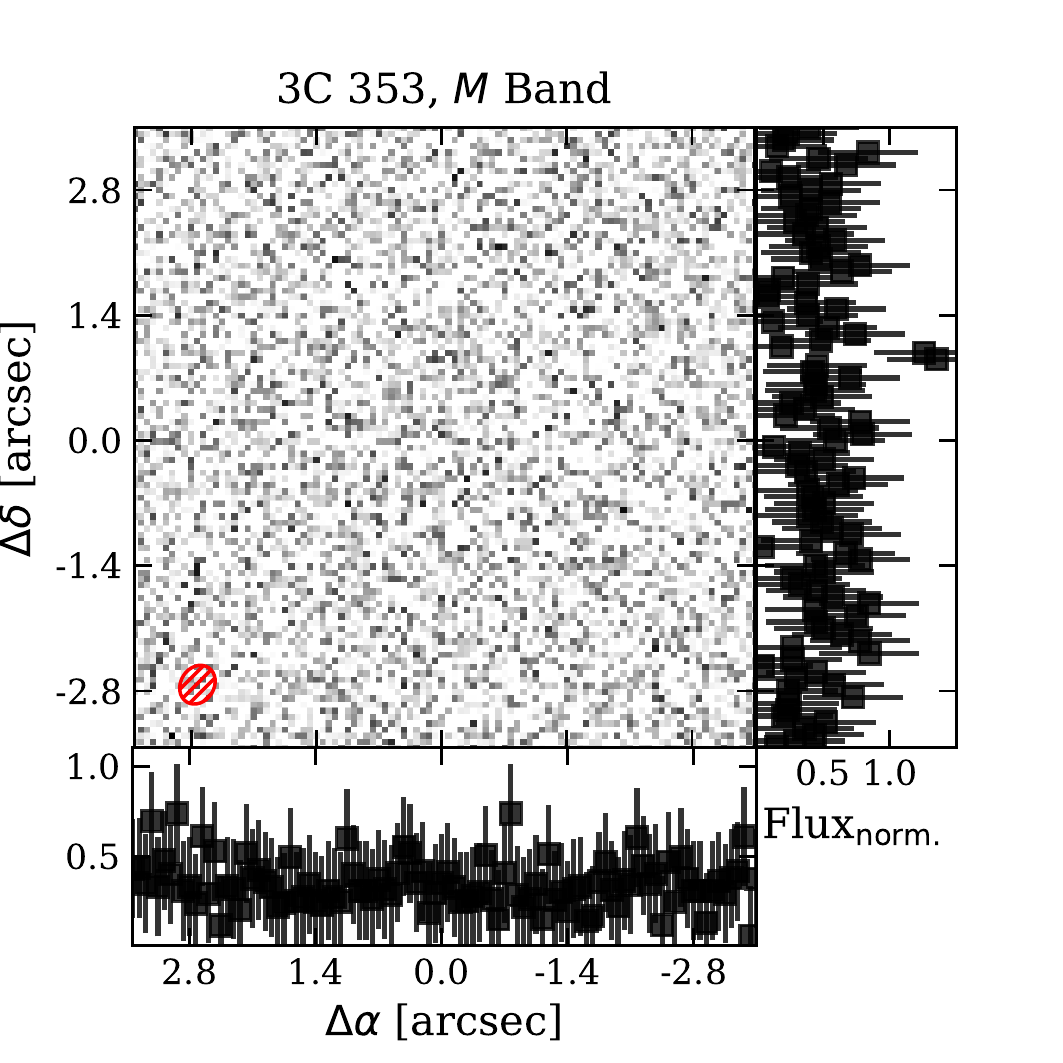}}\\
\subfloat{\includegraphics[width=0.24\hsize]{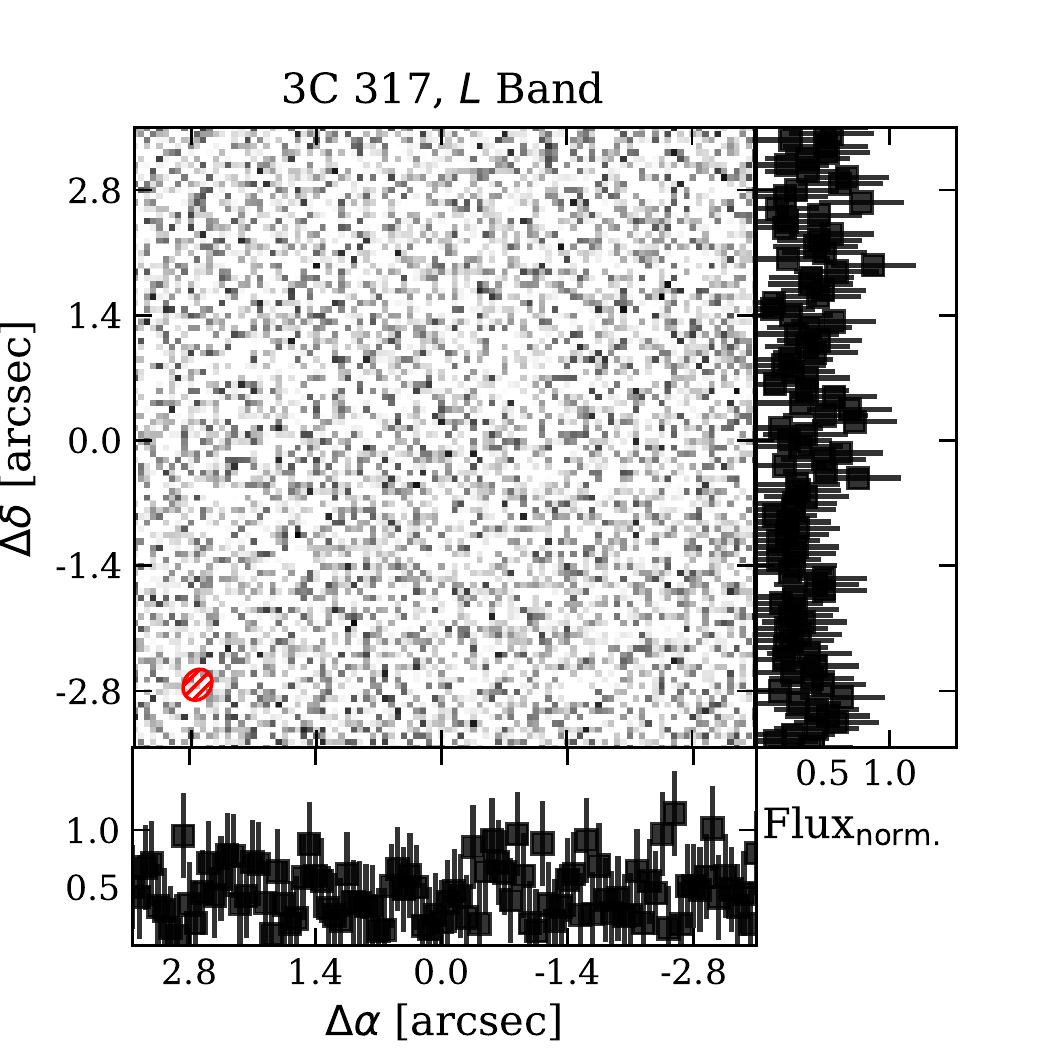}}
\subfloat{\includegraphics[width=0.24\hsize]{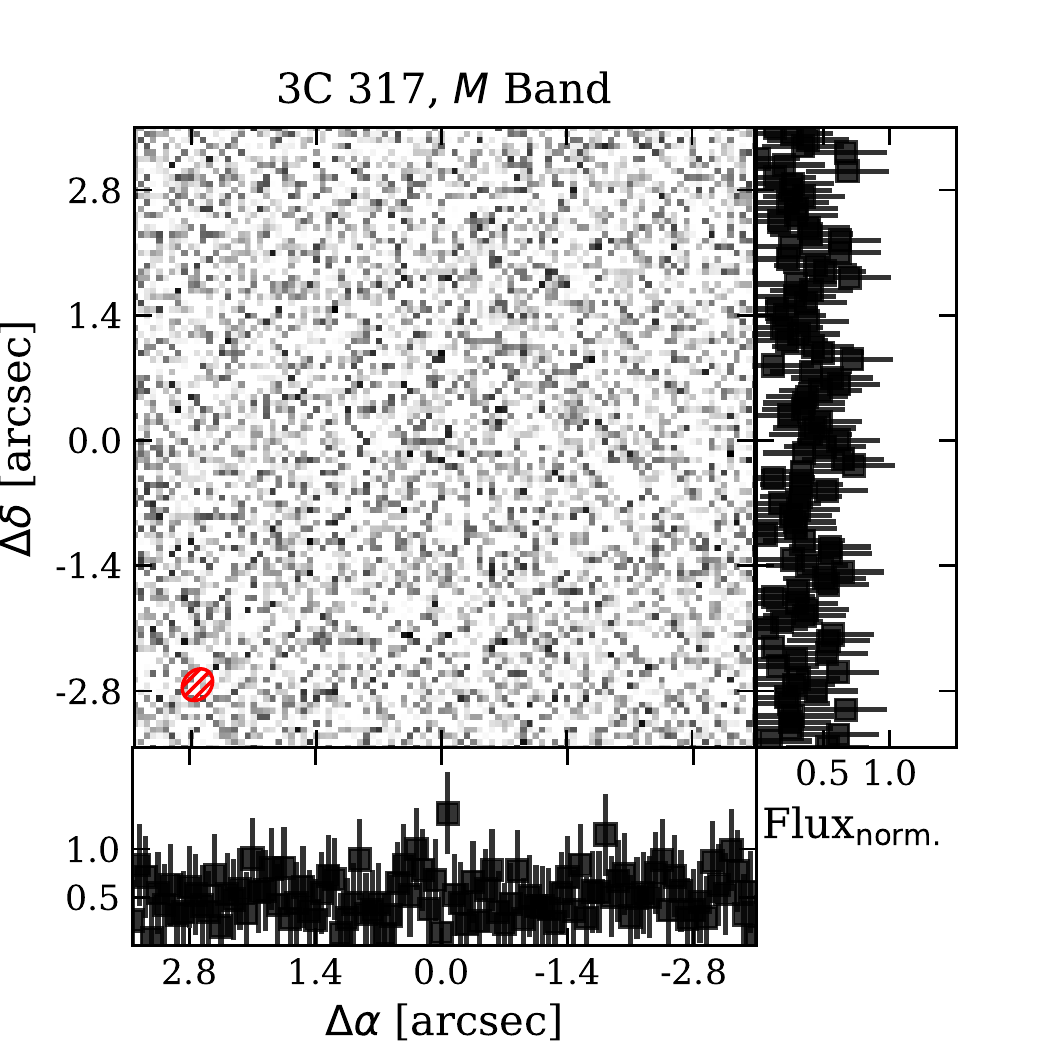}}
\subfloat{\includegraphics[width=0.25\hsize]{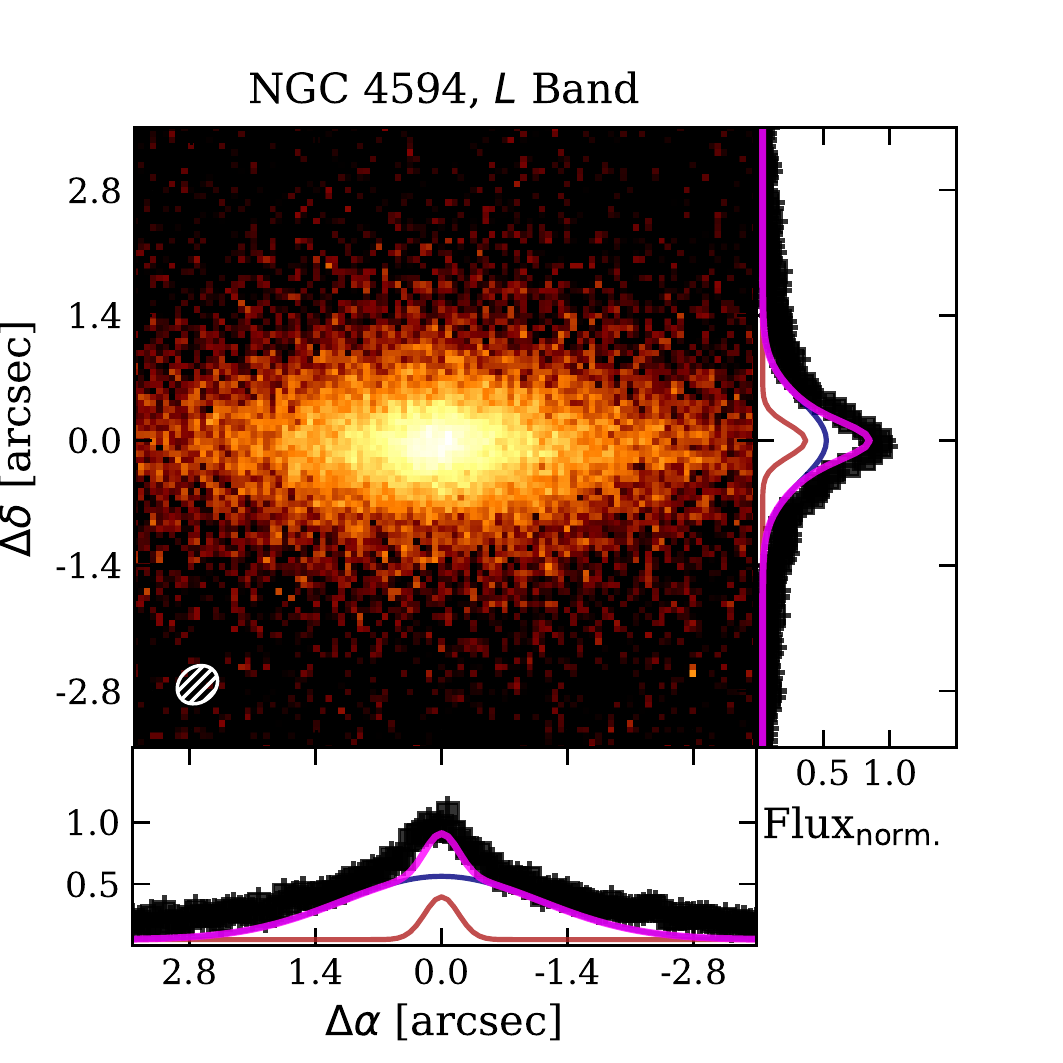}}
\subfloat{\includegraphics[width=0.25\hsize]{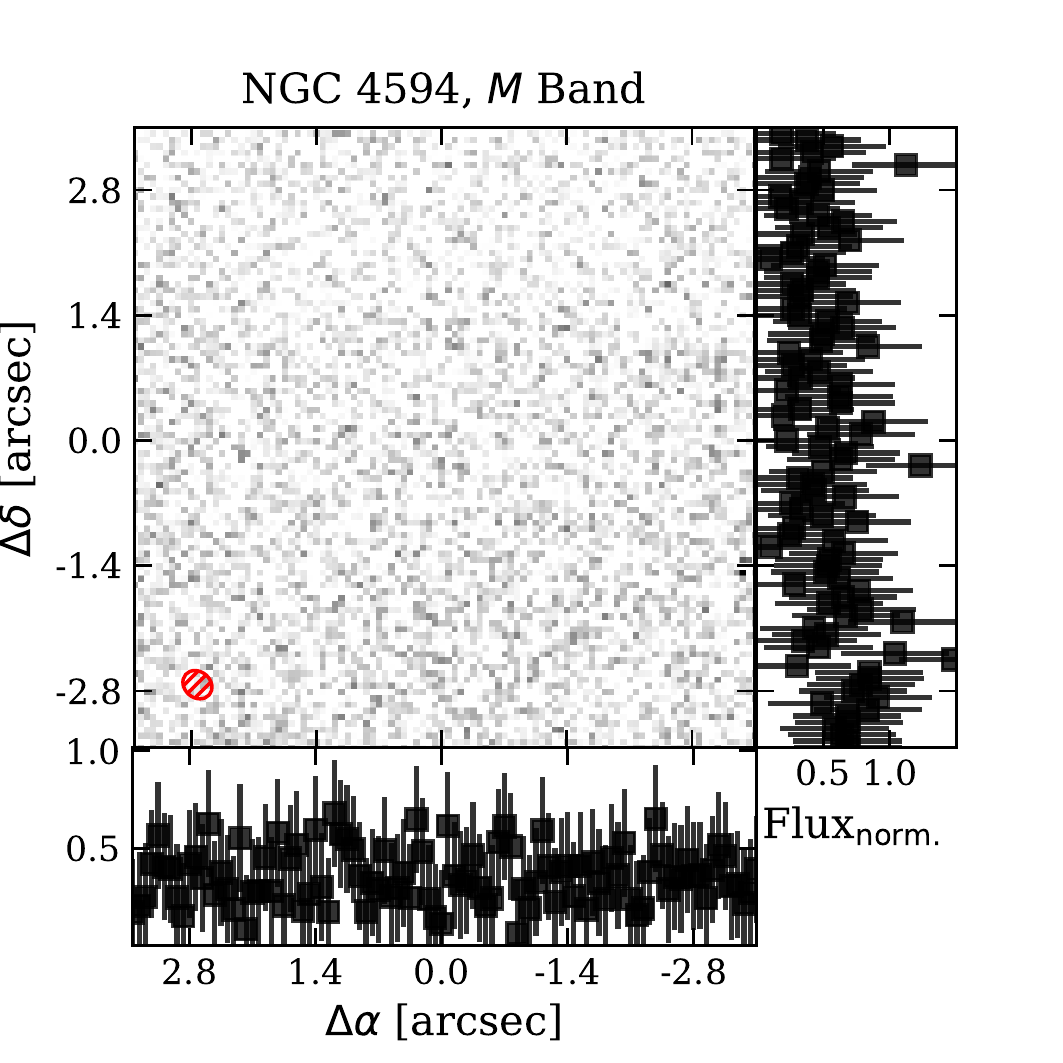}} \\

\caption{Cutouts for 4 representative sources: Cen A (top left) has clear detections in both bands; 3C353 (top right) is detected in only the $L$-band; 3C317 (bottom left) is detected in neither band; and NGC 4594 (bottom right) shows extended emission. Panels in greyscale (with titles in {\it italic}) are classified as non-detections, while those in color have SNR$_{\rm gauss}\geq 3$ in either Gaussian component and are classified as detections. With each cutout we present 1-D slices across the center of the image in both the x- and y-directions. Data in these slices are shown in black, and the profiles of the fitted elliptical Gaussians are plotted in red (nuclear), blue (extended), and magenta (sum). All images are presented with log-scaling. The ellipse in the bottom left of each cutout represents the fitted FWHM of the PSF calibrator. The rest of the cutouts can be found in Appendix C. }
\label{fig:cutouts_one}
\end{figure*}

\subsection{Flux Calibration}
We flux-calibrate each AGN flux measurement ($F_{\rm nuc.gauss}$, and $F_{\rm ext.gauss}$) with the equivalent measurement of the calibration star observed which
\begin{enumerate}
    \item was observed closest in time to the target, to minimize changes in atmospheric transmission and seeing
    \item has {\it either} both $L$- and $M$-band flux in \citet{vanderbliek1996} {or} has spectral type $\in \{O,B,A,F\}$.
\end{enumerate}
The spectral type selection is explained in detail in Appendix A, but in short we choose stars which have an effective temperature high enough such that the NIR color $L - M \approx 0$. This means that even when the catalog is missing a measurement in one of the two bands, the other can be reliably estimated.

In the majority of cases, calibrators were observed within 6hr of the target but there are several nights in which no calibration source was observed. 
For these nights we estimate the long-term stability of the transfer function, by examining the flux stability of a few calibrators over many nights
Three calibrators were observed often between 2000 and 2013: HD 130163, HD 106965, and HD 205772. These three calibrators allow us to examine the stability of the measured flux over time. In Fig. \ref{fig:cal_stab}, we show the flux variations of these sources in the $L$- and M-bands. From this, we see that in the $L$-band the $2\sigma$ flux variation is less than 3\% for all sources, and is as small as 0.8\% for HD 106965. We also find that the M-band $2\sigma$ flux variations are slightly larger, but all smaller than 5\%. 

For the AGN which were calibrated with these ``primary'' stars, we add the 2$\sigma$ flux variation directly to the flux uncertainty. For those calibrated with other stars, which were often observed only once, we cannot derive a similar 2$\sigma$ value. Therefore, we add 3\% and 5\% for the $L$- and M-bands, respectively; values which are slightly larger than the mean standard deviations of the 3 ``primary'' calibrators. Finally, for those sources which have no calibration star observed in the same night, we use whichever of the three often-observed calibrators was observed closest in time and add the $3\sigma$ uncertainty to the flux error estimate. The $3\sigma$ uncertainties are roughly 4\% and 6\% in the $L$- and M-bands, respectively. The presented flux errors combine the fitting uncertainties from both the AGN and the calibration star as well as the flux variations of the calibrators.

When sources were observed on more than one epoch, we report only the ``best'' measurement. We typically select the epoch with smallest seeing. The sample is quite heterogeneous, however, so a simple definition of ``best'' is not satisfactory. We therefore report each source with its observation date, seeing, and any VIPE reduction flags in Appendix A. We opt not to use the mean of the measurements, as sources were often re-observed only because of poor weather or instrumental errors. 
There are, however,  16 total (13 in $L$, 9 in $M$ with some overlap) sources which were observed multiple times under ``good'' conditions (i.e., no instrumental errors, no clouds, seeing $<0.7''$). For these we find the measured fluxes to be quite stable: mean variations of $10.23\%$ are found in the $L$ band total fluxes; and mean variations of $15.63 \%$ are found in the $M$-band total fluxes. These values are comparable to the statistical errors derived from the calibrators and the fitting errors. In none of these sources do we see signs of significant brightening nor dimming over a 10 year span.

Additionally, we fit several targets with special conditions to extract double nuclei (e.g., Arp 220) or to locate the AGN at the center of a much brighter stellar disk (e.g., NGC 7552). We list all of these exceptional sources and the approach used to measure each in Appendix A.

\subsubsection{Extended Emission}
\label{sec:ext}
The ISAAC PSF is slightly non-Gaussian, so not all of the flux is recovered by a Gaussian fit, not even for the spatially unresolved calibrators. We find that on average $87 \pm 6\%$ of the flux measured in a 1'' circular aperture is recovered in the $L$-band by such a Gaussian fit. Comparable point-source fitting in the $M$-band gives a similar value, $88 \pm 5 \%$.
This post-fit PSF residual does not ultimately affect nuclear flux calibration, however, because both the unresolved AGN flux component and the calibrators experience this in the same way. 

However, the small difference between the true PSF and the Gaussian PSF approximation can be ignored when flux calibrating the extended source, since the PSF fine-structure is lost when convolving with a larger source. .  We therefore calibrate only the Gaussian ``unresolved'' fluxes with the fitted fluxes of the calibrators, while we calibrate the extended fluxes with the Jy/cts conversion derived from circular aperture (1'') measurements of the calibrators.

\begin{figure}
\centering
\includegraphics[width=0.5\textwidth]{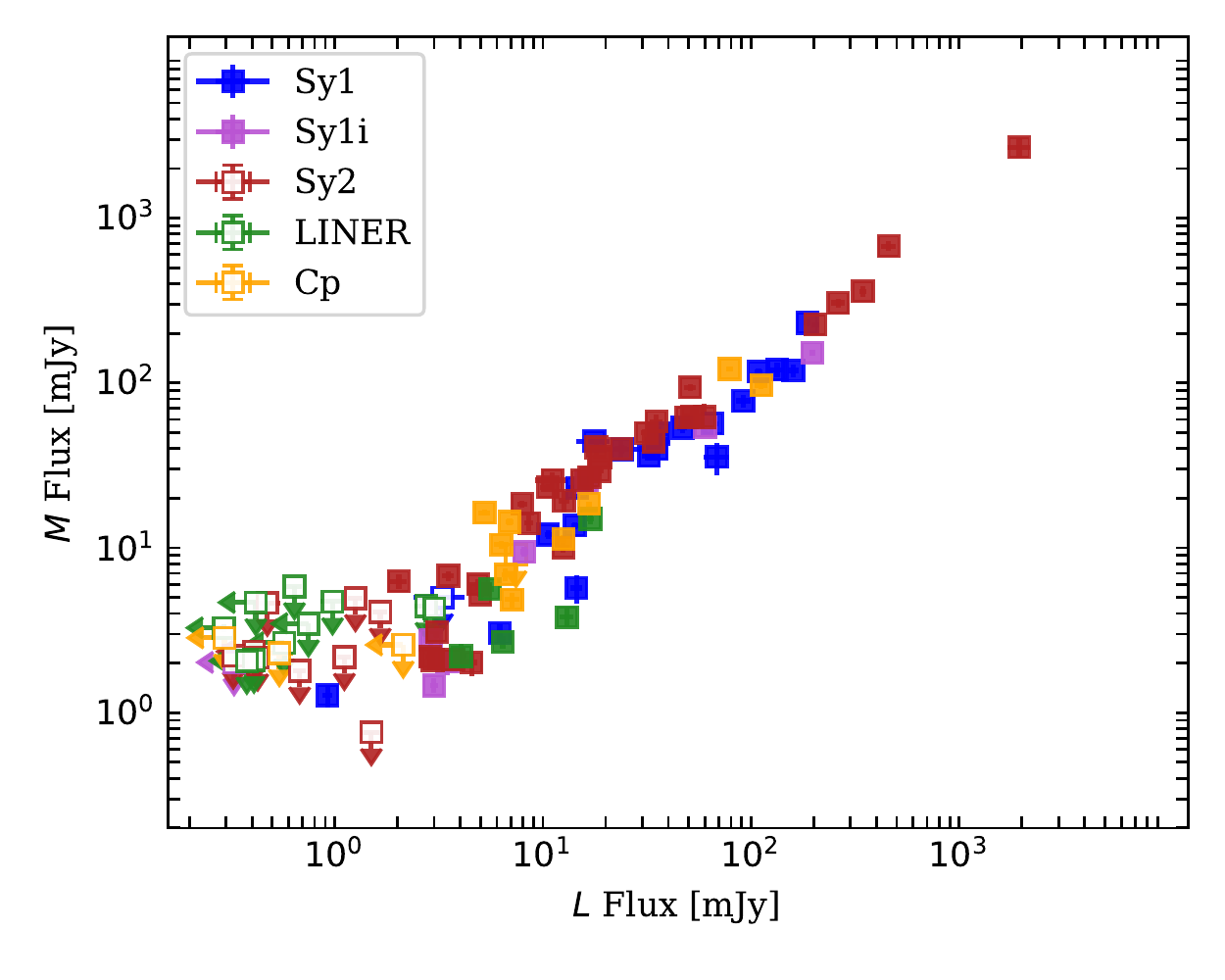}
\caption{\lprime-flux vs $M$-flux for all sources, with non-detections given as upper limits. The colors are the same as in Fig \ref{fig:hist}. Sources with measured fluxes exhibiting SNR$_{\rm gauss} \geq 3$ in both bands are filled. }
\label{fig:fluxflux}
\end{figure}

\section{The Flux Catalogs} \label{sec:fluxcat}

We define a detection for at least one of the fitted flux components ($F_{\rm nuc.gauss}$ or $F_{\rm ext.gauss}$) as having a calibrated SNR $\geq 3$. For non-detections we report only an upper limit of $3\sigma_{\rm gauss}$. We report fluxes with SNR $\geq 2$ only when the accompanying flux component has SNR $\geq 3$; this most often affects the extended flux components. 

\subsection{L- and M-band Nuclear Flux Table}
We present the measured nuclear $L$- and $M$-band fluxes and flux upper-limits of 119 active galaxies in Table \ref{tab:nuc}. We include the $N$-band parent sample fluxes from \cite{asmus2014} for reference. Our reported fluxes are the nuclear, unresolved component of our two-Gaussian fit. We present these fluxes separately, because the unresolved flux is most relevant for study of the AGN and interferometric follow-up observations (such as with VLTI/MATISSE).

We detect 92/95 sources in the $L$-band and 83/119 in the $M$-band. For all AGN with both $L$- and $M$-band measurements, there exist no cases where the AGN was detected in $M$ but not in $L$. We show cutouts and their fitted Gaussians for a representative sample of sources in Fig. \ref{fig:cutouts_one}, and present all of the cutouts in Appendix C. Fig. \ref{fig:fluxflux} shows the measured $L$-band flux versus the $M$-band flux for all sources and/or their upper limits. Here we do not see any significant difference between the different AGN classes, but rather a tight linear relation between the two bands. We detect 19/20 Sy1, 4/5 Int. Sy, 36/46 Sy2, 5/29 LINERs, and 11/16 Cp in both bands.

\startlongtable
\begin{deluxetable*}{lrrc|rr|r}
\tablehead{ \colhead{Target Name}  &\colhead{RA (J2000)} &\colhead{Dec. (J2000)}& \colhead{AGN}& \colhead{$L$ Flux$_{\rm nuc}$}  &\colhead{$M$ Flux$_{\rm nuc}$ } & \colhead{$N$ Flux} \\
\colhead{} &\colhead{[hh:mm:ss]}  & \colhead{[dd:mm:ss]} & \colhead{Type} &  \colhead{[mJy]} & \colhead{[mJy]}  & \colhead{[mJy]}  }
\label{tab:nuc}
\tablecaption{$L$- and $M$-band AGN Nuclear Flux Catalog}
\tabletypesize{\small}
\startdata
3C 273 & 12:29:06.70 & +02:03:09.00 &1 & $91.67 \pm 7.34$&$77.81 \pm 7.09$&$289.5 \pm 51.1$\\
3C 317 & 15:16:44.50 & +07:01:18.00 &2$^{1}$ &$\leq 0.41$&$\leq 2.35$&$\leq 2.7$\\
3C 321 & 15:31:43.50 & +24:04:19.00 &2 & $0.47 \pm 0.07$&$\leq 4.63$&$\leq 50.30$\\
3C 327 & 16:02:27.40 & +01:57:56.00 &1 & $0.92 \pm 0.05$&$1.27 \pm 0.20$&$70.8 \pm 21.3$\\
3C 353 & 17:20:28.20 & -00:58:47.00 &2$^{1}$ &$1.26 \pm 0.10$&$\leq 4.94$&$8.0 \pm 2.3$\\
3C 403 & 19:52:15.80 & +02:30:24.00 &2 & $5.00 \pm 0.19$&$5.21 \pm 0.43$&$94.2 \pm 12.8$\\
3C 424 & 20:48:12.00 & +07:01:17.00 &2 & $\leq 0.43$&$\leq 2.16$&$\leq 1.6$\\
Arp 220E & 15:34:57.07 & +23:30:14.74 &L$^{2}$ &$\leq 0.29$&$3.27 \pm 0.18$&-\\
Arp 220W & 15:34:57.07 & +23:30:14.74 &L$^{2}$ &$13.02 \pm 1.03$&$3.80 \pm 0.48$&-\\
Cen A & 13:25:27.60 & -43:01:08.99 &2 & $261.69 \pm 19.58$&$305.82 \pm 18.28$&$1524.2 \pm 152.4$\\
CGCG381-051 & 23:48:43.90 & +02:14:42.58 &H & $\leq 0.29$&$\leq 2.85$&-\\
Circinus & 14:13:09.90 & -65:20:20.99 &2 & $458.16 \pm 39.18$&$676.41 \pm 44.58$&$8326.9 \pm 1049.2$\\
ESO 103-35 & 18:38:20.30 & -65:25:38.99 &2 & $18.06 \pm 1.97$&$40.67 \pm 4.73$&$\leq 258.70$\\
ESO 138-1 & 16:51:20.10 & -59:14:05.00 &2 & $50.90 \pm 3.45$&$94.10 \pm 4.80$&$759.9 \pm 35.7$\\
ESO 141-55 & 19:21:14.10 & -58:40:13.00 &1.2 & $46.85 \pm 2.53$&$53.51 \pm 4.11$&$148.7 \pm 35.9$\\
ESO 286-19 & 20:58:26.80 & -42:39:00.01 &Cp: & $5.23 \pm 0.33$&$16.33 \pm 0.61$&$411.9 \pm 41.2$\\
ESO 323-32 & 12:53:20.30 & -41:38:08.00 &1.9 & $3.47 \pm 0.12$&$2.04 \pm 0.35$&$56.7 \pm 7.4$\\
ESO 323-77 & 13:06:26.10 & -40:24:53.00 &1.2 & $159.61 \pm 9.27$&$119.07 \pm 10.12$&$346.7 \pm 79.4$\\
ESO 506-27 & 12:38:54.60 & -27:18:28.00 &2 & $10.56 \pm 0.87$&$23.32 \pm 0.89$&$179.5 \pm 17.9$\\
ESO 511-30 & 14:19:22.40 & -26:38:41.00 &1 & $14.23 \pm 0.57$&$13.65 \pm 0.66$&$52.2 \pm 5.2$\\
Fairall 49 & 18:36:58.30 & -59:24:09.00 &2 & $52.25 \pm 5.24$&$62.60 \pm 7.74$&$\leq 213.20$\\
Fairall 51 & 18:44:54.00 & -62:21:53.00 &1.5 & $36.03 \pm 4.05$&$49.31 \pm 5.57$&$391.4 \pm 39.1$\\
IC 3639 & 12:40:52.90 & -36:45:21.00 &2 & $15.39 \pm 0.66$&$25.73 \pm 1.39$&$386.1 \pm 38.6$\\
IC 4329A & 13:49:19.30 & -30:18:34.00 &1.2 & $186.10 \pm 7.07$&$231.37 \pm 9.78$&$1157.7 \pm 99.3$\\
IC 4518W & 14:57:41.20 & -43:07:56.00 &2 & $18.81 \pm 1.41$&$35.44 \pm 1.28$&$199.4 \pm 31.7$\\
IC 5063 & 20:52:02.30 & -57:04:08.00 &2 &$35.06 \pm 2.40$&$58.31 \pm 5.83$&$820.6 \pm 57.8$\\
IC 5179 & 22:16:09.04 & -36:50:36.48 &H$^{2}$ &-&$\leq 2.19$&-\\
IRAS 13349+2438 & 13:37:18.70 & +24:23:03.00 &1n & $133.40 \pm 11.12$&$120.27 \pm 12.80$&$476.4 \pm 63.3$\\
IRASF00198-7926 & 00:21:54.21 & -79:10:09.55 &2 & $11.10 \pm 1.93$&$25.74 \pm 2.99$&-\\
LEDA 170194 & 12:39:06.28 & -16:10:47.09 &2 & $3.09 \pm 0.22$&$3.09 \pm 0.25$&$42.4 \pm 8.2$\\
M87 & 12:30:49.40 & +12:23:28.00 &L & $6.41 \pm 0.24$&$2.69 \pm 0.24$&$20.8 \pm 2.9$\\
MCG+2-4-25 & 01:20:02.66 & +14:21:42.21 &H$^{2}$ &-&$7.34 \pm 1.92$&-\\
MCG-0-29-23 & 11:21:12.44 & -02:59:04.40 &2$^{4}$ &$4.87 \pm 0.17$&$6.00 \pm 0.48$&-\\
MCG-3-34-64 & 13:22:24.50 & -16:43:42.00 &2$^{1}$ &$23.62 \pm 1.67$&$39.81 \pm 5.47$&$530.6 \pm 65.4$\\
MCG-6-30-15 & 13:35:53.70 & -34:17:44.00 &1.5 & $65.24 \pm 4.07$&$56.78 \pm 6.75$&$340.8 \pm 62.9$\\
Mrk 331 & 23:51:26.73 & +20:35:12.51 &Cp$^{1}$ &-&$\leq 2.41$&-\\
Mrk 463 & 13:56:02.97 & +18:22:16.84 &1.5$^{1}$ &$68.53 \pm 9.56$&$35.56 \pm 7.95$&-\\
Mrk 509 & 20:44:09.70 & -10:43:25.00 &1.5 & $108.46 \pm 4.81$&$117.10 \pm 5.40$&$256.4 \pm 29.0$\\
Mrk 841 & 15:04:01.20 & +10:26:16.00 &1.5 & $14.58 \pm 1.48$&$23.03 \pm 1.46$&$163.3 \pm 47.6$\\
Mrk 897 & 21:07:45.80 & +03:52:40.00 &Cp & $7.42 \pm 0.47$&$\leq 9.10$&$8.2 \pm 2.7$\\
NGC 63 & 00:17:45.51 & +11:27:02.73 &H$^{3}$ &-&$0.71 \pm 0.10$&-\\
NGC 253 & 00:47:33.10 & -25:17:18.00 &Cp: & $78.76 \pm 2.68$&$121.74 \pm 4.86$&$\leq 1038.5$\\
NGC 424 & 01:11:27.60 & -38:05:00.00 &2 & $203.46 \pm 7.91$&$225.63 \pm 8.14$&$736.2 \pm 222.9$\\
NGC 520 & 01:24:34.65 & +03:47:32.23 &Cp$^{1}$ &-&$\leq 2.32$&-\\
NGC 660 & 01:43:02.25 & +13:38:48.11 &L$^{1}$ &-&$\leq 2.09$&-\\
NGC 986 & 02:33:34.13 & -39:02:42.14 &H$^{5}$ &-&$4.72 \pm 0.53$&-\\
NGC 1008 & 02:37:57.11 & +02:04:12.59 &L/H$^{1}$ &$\leq 0.05$&$\leq 2.94$&-\\
NGC 1068 & 02:42:40.70 & -00:00:48.00 &2$^{1}$ &$1940.88 \pm 241.52$&$2687.53 \pm 374.86$&$\leq 3567.20$\\
NGC 1097 & 02:46:19.00 & -30:16:30.00 &L & $5.55 \pm 0.24$&$5.63 \pm 0.51$&$16.8 \pm 2.3$\\
NGC 1125 & 02:51:40.48 & -16:39:05.55 &2$^{1}$ &$3.51 \pm 0.23$&$6.76 \pm 0.46$&-\\
NGC 1368 & 03:33:39.77 & -05:05:24.29 &2$^{1}$ &$0.67 \pm 0.06$&-&-\\
NGC 1365 & 03:33:36.40 & -36:08:25.00 &1.8 & $196.43 \pm 6.59$&$152.32 \pm 6.74$&$360.7 \pm 36.1$\\
NGC 1386 & 03:36:46.20 & -35:59:57.01 &2 & $18.65 \pm 0.98$&$29.31 \pm 2.29$&$299.3 \pm 62.3$\\
NGC 1511 & 03:59:32.73 & -67:38:00.00 &H$^{2}$ &-&$2.44 \pm 0.17$&-\\
NGC 1566 & 04:20:00.40 & -54:56:16.00 &1.5 & $10.55 \pm 0.38$&-&$\leq 30.00$\\
NGC 1614 & 04:33:59.90 & -08:34:44.00 &Cp: & -&$7.13 \pm 0.64$&$\leq 345.6$\\
NGC 1667 & 04:48:37.10 & -06:19:12.00 &2 & $1.48 \pm 0.08$&-&$\leq 2.60$\\
NGC 1808 & 05:07:42.30 & -37:30:47.01 &Cp: & $6.30 \pm 0.51$&$10.46 \pm 0.56$&$328.8 \pm 34.2$\\
NGC 3125 & 10:06:33.34 & -29:56:06.31 &L/H$^{2}$ &$\leq 0.75$&$\leq 3.47$&-\\
NGC 3281 & 10:31:52.10 & -34:51:13.00 &2 & -&$116.64 \pm 5.26$&$486.4 \pm 50.5$\\
NGC 3660 & 11:23:32.30 & -08:39:31.00 &1.8$^{1}$ &$2.99 \pm 0.13$&$1.46 \pm 0.15$&$\leq 15.30$\\
NGC 4038/9 & 12:01:54.88 & -18:53:05.95 &1$^{1}$ &-&$2.43 \pm 0.26$&-\\
NGC 4074 & 12:04:29.70 & +20:18:58.00 &2 & $12.53 \pm 0.84$&$10.04 \pm 0.72$&$68.3 \pm 19.2$\\
NGC 4235 & 12:17:09.90 & +07:11:30.00 &1.2 & $14.49 \pm 0.97$&$5.71 \pm 1.15$&$36.0 \pm 6.5$\\
NGC 4253 & 12:18:26.80 & +29:48:45.39 &1n$^{1}$ &$23.99 \pm 6.38$&$39.53 \pm 2.28$&-\\
NGC 4261 & 12:19:23.20 & +05:49:31.00 &L & $0.64 \pm 0.06$&$\leq 5.84$&$13.2 \pm 3.0$\\
NGC 4278 & 12:20:06.80 & +29:16:51.00 &L & $\leq 0.57$&$\leq 2.66$&$2.5 \pm 0.6$\\
NGC 4303 & 12:21:54.90 & +04:28:25.00 &2 & -&$\leq 1.89$&$6.1 \pm 0.8$\\
NGC 4303 & 12:21:54.90 & +04:28:25.00 &2 & $1.65 \pm 0.13$&$\leq 4.07$&$6.1 \pm 0.8$\\
NGC 4374 & 12:25:03.70 & +12:53:13.00 &2$^{1}$ &$0.32 \pm 0.04$&$\leq 2.21$&$\leq 8.0$\\
NGC 4388 & 12:25:46.70 & +12:39:44.00 &2 & $34.11 \pm 1.23$&$44.11 \pm 2.10$&$187.8 \pm 32.8$\\
NGC 4418 & 12:26:54.60 & -00:52:39.00 &2 & $2.03 \pm 0.20$&$6.26 \pm 0.61$&$1426.8 \pm 167.5$\\
NGC 4438 & 12:27:45.60 & +13:00:32.00 &L/H & $4.04 \pm 0.26$&$2.20 \pm 0.43$&$10.3 \pm 2.8$\\
NGC 4457 & 12:28:59.00 & +03:34:14.00 &L & $3.01 \pm 0.22$&$\leq 4.30$&$6.2 \pm 1.7$\\
NGC 4472 & 12:29:46.80 & +08:00:02.00 &2/L & $\leq 0.41$&$\leq 2.09$&$\leq 8.6$\\
NGC 4501 & 12:31:59.20 & +14:25:13.00 &2 & $1.50 \pm 0.16$&$\leq 0.31$&$3.7 \pm 0.5$\\
NGC 4507 & 12:35:36.60 & -39:54:33.01 &2 & $59.67 \pm 5.96$&$62.07 \pm 12.71$&$622.8 \pm 64.3$\\
NGC 4579 & 12:37:43.50 & +11:49:05.00 &L & $16.91 \pm 0.66$&$14.99 \pm 1.02$&$74.6 \pm 4.5$\\
NGC 4593 & 12:39:39.40 & -05:20:39.00 &1 & $32.31 \pm 2.40$&$36.15 \pm 2.54$&$227.4 \pm 37.6$\\
NGC 4594 & 12:39:59.40 & -11:37:23.00 &L & $2.74 \pm 0.32$&$\leq 4.42$&$4.4 \pm 1.4$\\
NGC 4746 & 12:51:55.40 & +12:04:59.00 &L/H & $\leq 0.42$&$\leq 4.66$&$\leq 10.9$\\
NGC 4785 & 12:53:27.30 & -48:44:57.01 &2 & $1.11 \pm 0.08$&$\leq 2.17$&$\leq 17.4$\\
NGC 4941 & 13:04:13.10 & -05:33:06.00 &2 & $4.54 \pm 0.29$&$2.01 \pm 0.34$&$76.1 \pm 8.7$\\
NGC 4945 & 13:05:27.50 & -49:28:06.00 &Cp & $\leq 0.86$&$\leq 2.58$&$22.1 \pm 6.9$\\
NGC 5135 & 13:25:44.10 & -29:50:01.00 &2 & $12.63 \pm 0.70$&$19.31 \pm 2.75$&$132.0 \pm 25.8$\\
NGC 5252 & 13:38:16.00 & +04:32:33.00 &1.9 & $16.17 \pm 1.20$&$25.19 \pm 1.31$&$68.7 \pm 6.9$\\
NGC 5363 & 13:56:07.20 & +05:15:17.00 &L & $0.98 \pm 0.12$&$\leq 4.72$&$\leq 1.30$\\
NGC 5427 & 14:03:26.10 & -06:01:51.00 &2 & $2.95 \pm 0.10$&$2.09 \pm 0.26$&$\leq 20.0$\\
NGC 5506 & 14:13:14.90 & -03:12:27.00 &2 & $343.79 \pm 12.95$&$359.44 \pm 26.41$&$870.8 \pm 65.6$\\
NGC 5548 & 14:17:59.50 & +25:08:12.00 &1.5 & $3.29 \pm 0.89$&$\leq 4.99$&$\leq 77.50$\\
NGC 5643 & 14:32:40.70 & -44:10:27.99 &2 & $8.56 \pm 0.45$&$14.11 \pm 1.55$&$254.1 \pm 68.7$\\
NGC 5728 & 14:42:23.90 & -17:15:11.00 &1.9$^{1}$ &$2.89 \pm 0.22$&$2.76 \pm 0.32$&$49.1 \pm 7.1$\\
NGC 5813 & 15:01:11.20 & +01:42:07.00 &L: & $\leq 0.38$&$\leq 2.06$&$\leq 6.7$\\
NGC 5953 & 15:34:32.40 & +15:11:38.00 &Cp & $6.01 \pm 0.63$&$\leq 2.29$&$\leq 29.5$\\
NGC 5995 & 15:48:25.00 & -13:45:28.00 &1.9 & $60.09 \pm 3.02$&$54.06 \pm 4.99$&$332.4 \pm 46.8$\\
NGC 6000 & 15:49:49.69 & -29:23:14.20 &H$^{1}$ &-&$2.72 \pm 0.46$&-\\
NGC 6221 & 16:52:46.33 & -59:13:00.99 &Cp & $12.53 \pm 0.47$&$11.30 \pm 0.62$&$103.8 \pm 21.1$\\
NGC 6240N & 16:52:58.92 & +02:24:04.78 &Cp & $16.65 \pm 0.56$&$18.47 \pm 0.68$&$\leq 7.40$\\
NGC 6300 & 17:16:59.50 & -62:49:14.00 &2 & $31.34 \pm 1.73$&$49.80 \pm 2.54$&$553.6 \pm 162.1$\\
NGC 6810 & 19:43:34.20 & -58:39:20.00 &Cp & $7.10 \pm 0.56$&$4.85 \pm 0.51$&$44.4 \pm 13.2$\\
NGC 6814 & 19:42:40.60 & -10:19:25.00 &1.5 & $10.57 \pm 0.40$&$12.04 \pm 0.69$&$95.6 \pm 23.5$\\
NGC 6860 & 20:08:46.90 & -61:06:01.00 &1.5 & $35.24 \pm 1.50$&$39.72 \pm 2.47$&$206.1 \pm 24.1$\\
NGC 6890 & 20:18:18.10 & -44:48:24.00 &1.9$^{1}$ &$8.12 \pm 0.35$&$9.45 \pm 0.66$&$116.6 \pm 25.6$\\
NGC 7130 & 21:48:19.50 & -34:57:04.00 &Cp & $6.91 \pm 0.30$&$14.39 \pm 0.75$&$104.5 \pm 21.1$\\
NGC 7172 & 22:02:01.90 & -31:52:11.00 &2 & $48.96 \pm 1.92$&$61.82 \pm 3.30$&$185.0 \pm 18.6$\\
NGC 7213 & 22:09:18.16 & -47:07:59.05 &L/H$^{1}$ &-&$\leq 3.39$&-\\
NGC 7314 & 22:35:46.20 & -26:03:02.00 &2$^{1}$ &$7.94 \pm 0.43$&$18.47 \pm 0.83$&$61.5 \pm 11.8$\\
NGC 7479 & 23:04:56.70 & +12:19:22.00 &2 & $12.84 \pm 0.58$&-&$695.1 \pm 95.1$\\
NGC 7496 & 23:09:47.30 & -43:25:41.00 &Cp & $6.59 \pm 0.26$&$6.89 \pm 0.64$&$169.5 \pm 10.7$\\
NGC 7552 & 23:16:10.80 & -42:35:04.99 &L/H & -&$\leq 2.13$&$\leq 63.3$\\
NGC 7582 & 23:18:23.50 & -42:22:14.00 &Cp & $196.65 \pm 7.49$&$91.64 \pm 5.07$&$443.2 \pm 79.3$\\
NGC 7590 & 23:18:54.80 & -42:14:21.00 &2: & $0.68 \pm 0.05$&$\leq 1.79$&$\leq 10.6$\\
PG 2130+099 & 21:32:27.80 & +10:08:19.00 &1.5 & $17.64 \pm 3.15$&$44.28 \pm 1.44$&$187.8 \pm 20.9$\\
PKS 1417-19 & 14:19:49.70 & -19:28:25.00 &1.5 & $6.20 \pm 0.22$&$3.04 \pm 0.40$&$\leq 10.10$\\
PKS 1814-63 & 18:19:35.00 & -63:45:48.00 &2 & $3.71 \pm 0.12$&$2.09 \pm 0.19$&$27.7 \pm 5.3$\\
PKS 1932-46 & 19:35:56.60 & -46:20:41.00 &1.9 & $\leq 0.33$&$\leq 2.02$&$\leq 2.0$\\
Superantennae S & 19:31:21.47 & -72:39:21.21 &2 & $16.72 \pm 0.81$&$26.91 \pm 1.52$&$221.5 \pm 62.8$\\
UGC 2369 S & 02:54:01.91 & +14:58:17.54 &Cp$^{2}$ &-&$1.18 \pm 0.11$&-\\
Z 41-20 & 12:00:57.90 & +06:48:23.00 &2 & $2.87 \pm 0.13$&$2.18 \pm 0.34$&$29.3 \pm 2.9$\\
\enddata
\tablecomments{Here Flux$_{\rm nuc}$ indicates the flux from the fitted unresolved component. N-band fluxes and most AGN classifications come from \citet{asmus2014} except where otherwise noted. $^{1}$ \citet{veron-cetty2010}; $^{2}$ \citet{yuan2010}; $^{3}$ \citet{ho1997}; $^{4}$ \citet{hernan-caballero2011}; $^{5}$ \citet{hameed1999}  }
\end{deluxetable*}

\begin{figure*}[ht]
\centering
\includegraphics[width=1\textwidth]{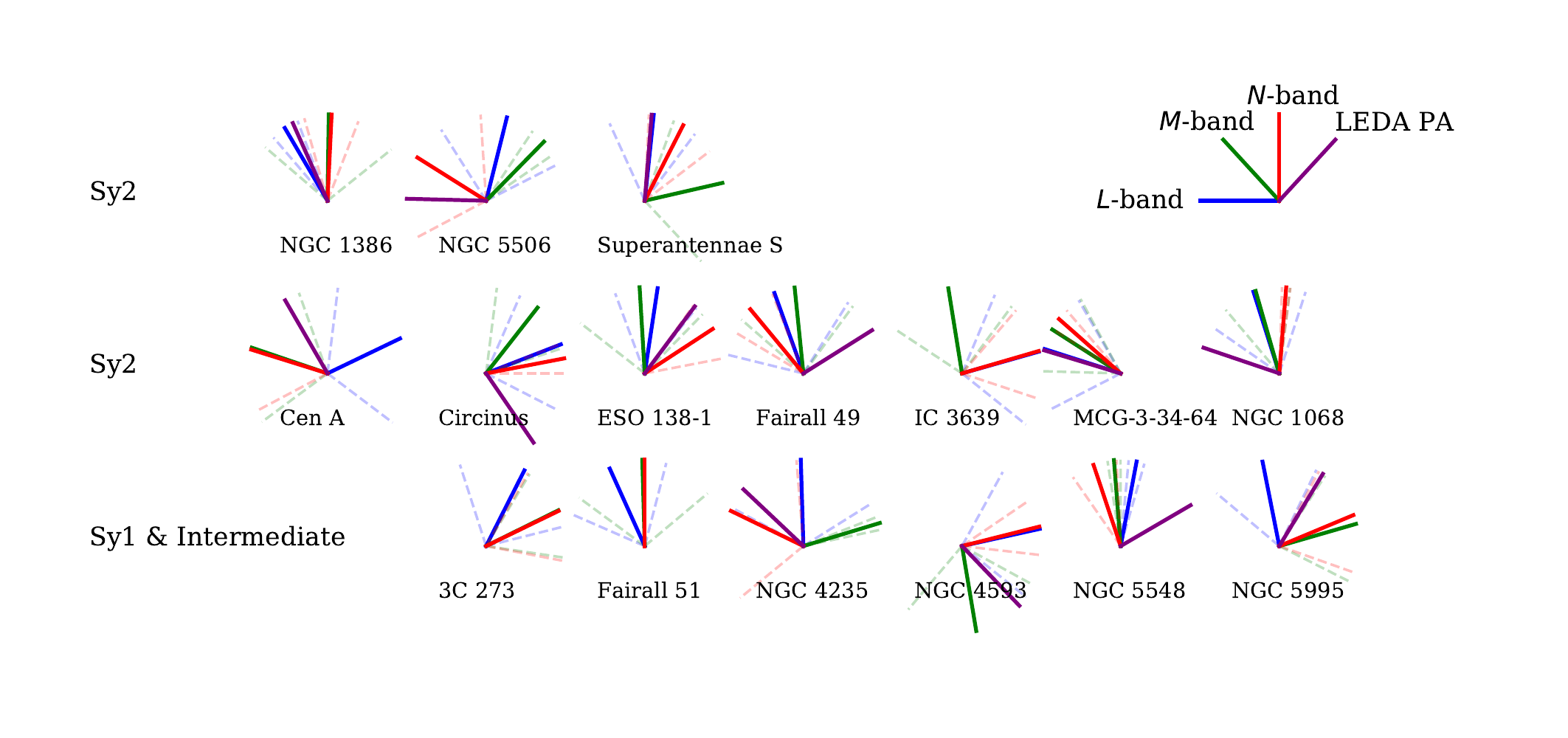}
\caption{The measured infrared position angles (PAs) of sources with 1) SNR$\geq$ 2 extended emission in the $L$- and $M$-bands, and 2) a reported PA in \citet{asmus2016}. We include the optical position angle from HyperLEDA \citep{makarov2014} in purple for reference. The $L,~M$, and $N$ position angles are shown in blue, green, and red, respectively.  $1 \sigma$ errors are given as faint dashed lines in the same color as the reported angle. }
\label{fig:pas}
\end{figure*}

\subsection{L- and M-band Extended Flux Table}

We present the measured \textit{extended} $L$- and $M$-band fluxes of the active galaxies in Table \ref{tab:ext}. In the table we also display the physical extent of the emission (from the fitted Gaussian axes), the minor/major axis ratio, and position angle (PA) of the source.  We also include  $N$-band extended emission PAs from \cite{asmus2016} for reference. In the final column we give the unresolved flux to total flux ratio, $f_{\rm nuc}$.

In total we find significant resolved emission accompanying 73 and 42 AGN in the $L'$ and $M$ bands, respectively. In the $L'$-band, we find that 15 Sy1, 3 Syi, 32 Sy2, 14 LINERs, and 9 Cps exhibit extended emission. In the $M$-band 9 Sy1, 1 Syi, 14 Sy2, 6 LINERs, and 11 Cps had extended emission.  Of the detected AGN, Sy2 were the most likely to show extended emission, with 32/36 resolved.

\subsubsection{A Note on Position Angles}
We note that the errors on the fitted PAs are quite large, and we caution the reader that even bright AGN showed large changes in fitted PA despite having consistent fitted flux and fitted FWHM values across epochs. NGC 1068, for example, exhibited PAs of $-28\pm 38$ deg and $19 \pm 9.1$ deg in the 2001 and 2004 epochs, respectively. The PSF PA varied by a similar amount. The unstable PSF of ISAAC leads us to suggest that the reader uses these PAs cautiously. We show $L$, $M$, and $N$ PAs of the AGN with significant extended emission and PAs reported in \citet{asmus2016} in Fig. \ref{fig:pas}. We include therein the optical PAs from HyperLEDA \citep{makarov2014} for comparison. In a few objects (3C 273, MCG-6-30-15, Circinus, Fairall 49, MGC-3-34-64, NGC 1068, and NGC 1386) there is relatively good agreement between the \lprime, $M$, and $N$ PAs. In only NGC 1386 and Superantennae S do the HyperLEDA angles match the \lprime~angle. The variation per source, however, is so great that we cannot draw any definite conclusions about the PA relations. AO-aided observations with ERIS or with ELT-METIS will be necessary to make progress in this line of inquiry.

\startlongtable
\begin{longrotatetable}
\begin{deluxetable*}{l|rrrr|rrrr|rl}
\tablehead{ \colhead{Target Name}  & \colhead{$L$ Flux$_{\rm ext}$} & \colhead{$\Theta_{L,\rm ext}$} & \colhead{PA$_{L,\rm ext}$}&\colhead{$r_{L}$} &\colhead{M Flux$_{\rm ext}$} & \colhead{$\Theta_{M,\rm ext}$} & \colhead{PA$_{M,\rm ext}$}&\colhead{$r_{M}$} & \colhead{PA$_{N}$} & \colhead{$f_{\rm nuc}$}\\
\colhead{} &  \colhead{[mJy]} & \colhead{[pc]}& \colhead{[deg]} &\colhead{$\equiv \theta/\Theta$} &\colhead{[mJy]} & \colhead{[pc]} & \colhead{[deg]}&\colhead{$\equiv \theta/\Theta$} & \colhead{[deg]} & \colhead{$L$ ($M$) } }
\label{tab:ext}
\tablecaption{$L$-band and $M$-band AGN Extended Flux Catalog}
\tabletypesize{\footnotesize}
\startdata
3C 273 &$69.0 \pm 28.9$&$3160.2 \pm 419.9$&$28.8 \pm 48.2$&$0.9$&$73.6 \pm 22.4$&$3160.2 \pm 292.4$&$65.5 \pm 32.2$&$0.9$&$114.0 \pm 34.0$&0.57 (0.51) \\
3C 317 &$\leq 0.4$&-&-&-&$\leq 2.1$&-&-&-&-&- (-) \\
3C 321 &$\leq 0.7$&-&-&-&$\leq 3.9$&-&-&-&$89.0 \pm 0.0$&- (-) \\
3C 327 &$\leq 0.4$&-&-&-&$\leq 2.8$&-&-&-&$97.0 \pm 0.0$&0.68 (-) \\
3C 353 &$\leq 0.7$&-&-&-&$\leq 4.7$&-&-&-&-&0.66 (-) \\
3C 403 &$\leq 0.7$&-&-&-&$\leq 4.0$&-&-&-&$60.0 \pm 0.0$&0.88 (0.59) \\
3C 424 &$\leq 0.4$&-&-&-&$\leq 2.3$&-&-&-&-&- (-) \\
Arp 220E &$\leq 1.5$&-&-&-&$\leq 2.4$&-&-&-&-&- (0.74) \\
Arp 220W &$2.2 \pm 0.3$&$7.4 \pm 0.2$&$83.8 \pm 1.6$&$0.6$&$\leq 2.4$&-&-&-&-&0.85 (0.77) \\
Cen A &$168.3 \pm 80.0$&$11.6 \pm 1.6$&$66.2 \pm 58.6$&$0.9$&$182.1 \pm 55.1$&$11.6 \pm 1.1$&$5.0 \pm 66.2$&$0.9$&$74.0 \pm 42.0$&0.61 (0.55) \\
CGCG381-051 &$\leq 0.3$&-&-&-&$\leq 2.3$&-&-&-&-&- (-) \\
Circinus &$\leq 1.7$&-&-&-&$591.1 \pm 207.2$&$26.1 \pm 3.2$&$40.4 \pm 32.6$&$0.9$&$100.0 \pm 10.0$&1.00 (0.53) \\
ESO 103-35 &$22.9 \pm 7.2$&$350.3 \pm 29.2$&$-37.1 \pm 42.1$&$0.9$&$\leq 5.3$&-&-&-&$109.0 \pm 12.0$&0.44 (0.91) \\
ESO 138-1 &$51.2 \pm 16.2$&$180.6 \pm 19.6$&$9.3 \pm 31.2$&$0.9$&$39.3 \pm 17.8$&$180.6 \pm 24.9$&$-3.7 \pm 50.6$&$0.9$&$121.0 \pm 21.0$&0.50 (0.70) \\
ESO 141-55 &$37.2 \pm 9.1$&$1062.2 \pm 92.8$&$91.3 \pm 62.1$&$0.9$&$\leq 5.9$&-&-&-&$120.0 \pm 47.0$&0.56 (0.93) \\
ESO 286-19 &$\leq 0.6$&-&-&-&$\leq 4.3$&-&-&-&-&0.90 (0.80) \\
ESO 323-32 &$3.3 \pm 0.8$&$539.9 \pm 54.6$&$-6.8 \pm 9.0$&$0.7$&$\leq 4.1$&-&-&-&$76.0 \pm 33.0$&0.38 (-) \\
ESO 323-77 &$\leq 0.8$&-&-&-&$114.5 \pm 39.3$&$221.1 \pm 24.9$&$85.9 \pm 21.2$&$0.9$&$95.0 \pm 27.0$&1.00 (0.51) \\
ESO 506-27 &$9.0 \pm 4.0$&$658.8 \pm 83.2$&$-1.3 \pm 71.9$&$0.9$&$\leq 4.6$&-&-&-&$93.0 \pm 8.0$&0.54 (0.84) \\
ESO 511-30 &$\leq 0.6$&-&-&-&$\leq 4.4$&-&-&-&$82.0 \pm 3.0$&0.96 (0.77) \\
Fairall 49 &$75.8 \pm 21.5$&$587.7 \pm 46.2$&$-21.4 \pm 55.9$&$0.9$&$87.7 \pm 27.7$&$587.7 \pm 47.0$&$-6.3 \pm 45.1$&$0.9$&$42.0 \pm 19.0$&0.41 (0.41) \\
Fairall 51 &$53.5 \pm 15.9$&$399.6 \pm 33.1$&$-26.2 \pm 42.0$&$0.9$&$\leq 7.4$&-&-&-&$180.0 \pm 0.0$&0.40 (0.90) \\
IC 3639 &$11.0 \pm 2.6$&$253.0 \pm 24.9$&$75.4 \pm 50.9$&$0.9$&$9.0 \pm 4.3$&$253.0 \pm 31.2$&$-9.9 \pm 48.8$&$0.9$&$105.0 \pm 32.0$&0.58 (0.74) \\
IC 4329A &$60.4 \pm 28.6$&$573.3 \pm 92.0$&$4.0 \pm 66.6$&$0.8$&$\leq 9.1$&-&-&-&$65.0 \pm 9.0$&0.76 (0.98) \\
IC 4518W &$10.9 \pm 4.5$&$440.9 \pm 54.9$&$78.4 \pm 47.5$&$0.9$&$\leq 4.6$&-&-&-&-&0.63 (0.89) \\
IC 5063 &$42.4 \pm 5.4$&$203.3 \pm 30.9$&$14.3 \pm 47.0$&$0.9$&$\leq 55.9$&-&-&$-$&$108.0 \pm 5.0$&0.45 (-) \\
IC 5179 &-&-&-&-&$\leq 2.0$&-&-&-&-&- (-) \\
IRAS 13349+2438 &$\leq 0.8$&-&-&-&$110.7 \pm 45.1$&$2641.6 \pm 275.1$&$59.6 \pm 30.7$&$0.9$&$16.0 \pm 19.0$&1.00 (0.52) \\
IRASF00198-7926 &$22.9 \pm 5.8$&$4.8 \pm 0.4$&$24.0 \pm 9.6$&$0.8$&$25.1 \pm 11.5$&$4.8 \pm 0.5$&$15.0 \pm 32.1$&$0.9$&-&0.33 (0.50) \\
LEDA 170194 &$1.6 \pm 0.8$&$642.1 \pm 89.5$&$2.5 \pm 33.2$&$0.8$&$\leq 2.7$&-&-&-&$44.0 \pm 18.0$&0.66 (0.54) \\
M87 &$3.2 \pm 1.4$&$61.3 \pm 8.4$&$72.1 \pm 47.7$&$0.9$&$\leq 4.5$&-&-&-&-&0.56 (-) \\
MCG+2-4-25 &-&-&-&-&$\leq 2.0$&-&-&-&-&- (0.80) \\
MCG-0-29-23 &$13.5 \pm 1.4$&$10.5 \pm 0.6$&$133.6 \pm 3.7$&$0.6$&$\leq 2.0$&-&-&-&-&0.27 (0.77) \\
MCG-3-34-64 &$33.9 \pm 8.9$&$442.9 \pm 33.4$&$106.4 \pm 41.6$&$0.9$&$56.0 \pm 22.3$&$442.9 \pm 45.0$&$120.5 \pm 29.0$&$0.9$&$51.0 \pm 8.0$&0.41 (0.43) \\
MCG-6-30-15 &$38.0 \pm 18.4$&$115.4 \pm 18.1$&$100.0 \pm 44.3$&$0.9$&$53.8 \pm 25.9$&$115.4 \pm 16.2$&$83.4 \pm 13.1$&$0.8$&$106.0 \pm 13.0$&0.63 (0.52) \\
Mrk 509 &$\leq 0.6$&-&-&-&$\leq 4.3$&-&-&-&$107.0 \pm 24.0$&0.99 (0.97) \\
Mrk 841 &$9.9 \pm 4.4$&$598.7 \pm 71.2$&$11.0 \pm 53.8$&$0.9$&$\leq 3.8$&-&-&-&$155.0 \pm 31.0$&0.60 (0.87) \\
Mrk 897 &$3.5 \pm 1.6$&$677.7 \pm 94.8$&$-7.3 \pm 52.1$&$0.9$&-&-&-&-&-&0.68 (-) \\
Mrk 331 &-&-&-&-&$\leq 2.1$&-&-&-&-&- (-) \\
Mrk 463 &$101.4 \pm 43.8$&$5.5 \pm 0.7$&$39.0 \pm 27.7$&$0.9$&$126.4 \pm 27.9$&$5.5 \pm 0.4$&$47.5 \pm 12.5$&$0.9$&-&0.41 (0.21) \\
NGC 6300 &$\leq 0.7$&-&-&-&$\leq 6.4$&-&-&-&$112.0 \pm 4.0$&- (-) \\
NGC 253 &$82.9 \pm 10.7$&$42.8 \pm 3.7$&$130.4 \pm 3.0$&$0.4$&$88.9 \pm 27.8$&$42.8 \pm 4.6$&$134.0 \pm 5.0$&$0.4$&-&0.49 (0.58) \\
NGC 424 &$\leq 0.6$&-&-&-&$81.0 \pm 33.4$&$439.4 \pm 72.7$&$-0.2 \pm 58.0$&$0.8$&$80.0 \pm 0.0$&1.00 (0.74) \\
NGC 520 &-&-&-&-&$\leq 1.9$&-&-&-&-&- (-) \\
NGC 660 &-&-&-&-&$\leq 1.8$&-&-&-&-&- (-) \\
NGC 986 &-&-&-&-&$\leq 4.0$&-&-&-&-&- (-) \\
NGC 1008 &$\leq 0.2$&-&-&-&$\leq 3.2$&-&-&-&-&- (-) \\
NGC 1068 &$1818.3 \pm 874.9$&$83.1 \pm 9.4$&$-19.0 \pm 38.4$&$0.9$&$4204.3 \pm 1488.6$&$83.1 \pm 8.0$&$-17.4 \pm 25.2$&$0.9$&$175.0 \pm 3.0$&0.53 (0.40) \\
NGC 1097 &$8.3 \pm 1.3$&$133.2 \pm 9.1$&$-0.4 \pm 68.5$&$0.9$&$\leq 1.9$&-&-&-&-&0.40 (0.75) \\
NGC 1125 &$6.0 \pm 1.3$&$5.3 \pm 0.4$&$130.5 \pm 10.0$&$0.8$&$\leq 1.6$&-&-&-&-&0.37 (0.82) \\
NGC 1368 &$4.4 \pm 0.3$&$8.1 \pm 0.3$&$104.9 \pm 8.7$&$0.9$&-&-&-&-&-&0.13 (-) \\
NGC 1365 &$48.8 \pm 24.0$&$173.4 \pm 40.9$&$2.7 \pm 58.1$&$0.7$&$\leq 1.7$&-&-&-&$82.0 \pm 25.0$&0.80 (0.99) \\
NGC 1386 &$20.9 \pm 5.0$&$114.2 \pm 8.9$&$-32.6 \pm 10.4$&$0.8$&$18.9 \pm 8.3$&$114.2 \pm 14.1$&$0.9 \pm 52.7$&$0.9$&$177.0 \pm 20.0$&0.47 (0.61) \\
NGC 1511 &-&-&-&-&$\leq 1.5$&-&-&-&-&- (0.63) \\
NGC 1566 &$11.1 \pm 1.8$&$118.2 \pm 9.3$&$1.5 \pm 32.8$&$0.9$&-&-&-&-&$157.0 \pm 37.0$&0.49 (-) \\
NGC 1614 &-&-&-&-&$15.6 \pm 2.1$&-&$72.5 \pm 9.6$&$0.8$&-&- (0.31) \\
NGC 1667 &$2.0 \pm 0.4$&$597.4 \pm 51.5$&$11.4 \pm 3.7$&$0.5$&-&-&-&-&$54.0 \pm 0.0$&0.42 (-) \\
NGC 1808 &$15.7 \pm 2.2$&$72.6 \pm 3.6$&$29.8 \pm 5.2$&$0.7$&$9.0 \pm 2.1$&$72.6 \pm 7.4$&$42.6 \pm 13.9$&$0.8$&-&0.29 (0.54) \\
NGC 3125 &$\leq 0.6$&-&-&-&$\leq 3.1$&-&-&-&-&- (-) \\
NGC 3281 &-&-&-&-&$\leq 5.0$&-&-&-&$176.0 \pm 16.0$&- (0.96) \\
NGC 3660 &$1.8 \pm 0.3$&$324.6 \pm 23.7$&$-8.8 \pm 56.8$&$0.9$&$\leq 1.4$&-&-&-&-&0.49 (-) \\
NGC 4038/9 &-&-&-&-&$\leq 1.9$&-&-&-&-&- (-) \\
NGC 4074 &$\leq 0.7$&-&-&-&$\leq 4.4$&-&-&-&$189.0 \pm 0.0$&0.95 (0.71) \\
NGC 4235 &$11.9 \pm 4.0$&$177.3 \pm 18.8$&$-1.9 \pm 61.9$&$0.9$&$8.7 \pm 2.0$&$177.3 \pm 5.8$&$74.5 \pm 4.5$&$0.5$&$66.0 \pm 61.0$&0.55 (0.39) \\
NGC 4253 &$33.1 \pm 14.4$&$6.5 \pm 0.7$&$-18.0 \pm 29.9$&$0.9$&$\leq 4.6$&-&-&-&-&0.41 (0.90) \\
NGC 4261 &$\leq 0.6$&-&-&-&$\leq 5.2$&-&-&-&-&- (-) \\
NGC 4278 &$\leq 0.5$&-&-&-&$\leq 2.5$&-&-&-&-&- (-) \\
NGC 4303 &-&-&-&-&$\leq 2.1$&-&-&-&-&- (-) \\
NGC 4303 &$5.9 \pm 0.7$&$119.5 \pm 5.4$&$-10.4 \pm 26.3$&$0.9$&$\leq 4.1$&-&-&-&-&0.22 (-) \\
NGC 4374 &$\leq 0.4$&-&-&-&$\leq 2.0$&-&-&-&-&- (-) \\
NGC 4388 &$\leq 0.4$&-&-&-&$\leq 2.6$&-&-&-&$28.0 \pm 34.0$&0.99 (0.95) \\
NGC 4418 &$1.7 \pm 0.8$&$207.9 \pm 27.8$&$99.9 \pm 9.2$&$0.7$&$\leq 4.5$&-&-&-&-&0.55 (0.60) \\
NGC 4438 &$12.6 \pm 1.4$&$133.5 \pm 6.0$&$-23.2 \pm 3.0$&$0.6$&$\leq 5.1$&-&-&-&-&0.24 (-) \\
NGC 4457 &$15.5 \pm 1.2$&$139.6 \pm 4.8$&$112.6 \pm 6.7$&$0.8$&$\leq 3.7$&-&-&-&-&0.16 (-) \\
NGC 4472 &$\leq 0.4$&-&-&-&$\leq 1.9$&-&-&-&-&- (-) \\
NGC 4501 &$13.7 \pm 1.2$&$144.1 \pm 4.9$&$43.1 \pm 5.1$&$0.8$&$\leq 2.4$&-&-&-&$153.0 \pm 0.0$&0.10 (-) \\
NGC 4507 &$51.5 \pm 22.9$&$259.3 \pm 35.5$&$78.1 \pm 45.6$&$0.9$&$\leq 6.4$&-&-&-&$111.0 \pm 22.0$&0.54 (0.92) \\
NGC 4579 &$17.1 \pm 4.3$&$82.5 \pm 9.4$&$85.8 \pm 48.6$&$0.9$&$7.2 \pm 3.3$&$82.5 \pm 13.6$&$87.3 \pm 42.7$&$0.8$&-&0.50 (0.67) \\
NGC 4593 &$22.8 \pm 11.1$&$155.6 \pm 22.4$&$78.5 \pm 47.6$&$0.9$&$25.6 \pm 7.0$&$155.6 \pm 14.9$&$-10.3 \pm 53.0$&$0.9$&$103.0 \pm 19.0$&0.58 (0.58) \\
NGC 4594 &$38.9 \pm 2.0$&$120.0 \pm 3.0$&$90.8 \pm 0.9$&$0.4$&$\leq 3.8$&-&-&-&-&0.07 (-) \\
NGC 4746 &$\leq 0.4$&-&-&-&-&-&-&-&-&- (-) \\
NGC 4785 &$4.0 \pm 0.5$&$436.7 \pm 23.5$&$71.9 \pm 6.7$&$0.8$&$\leq 2.0$&-&-&-&-&0.22 (-) \\
NGC 4941 &$4.2 \pm 1.4$&$115.7 \pm 13.3$&$17.2 \pm 32.7$&$0.9$&$\leq 3.4$&-&-&-&$116.0 \pm 11.0$&0.52 (-) \\
NGC 4945 &$54.9 \pm 3.5$&$42.6 \pm 0.9$&$133.4 \pm 1.0$&$0.4$&$\leq 2.3$&-&-&-&-&0.04 (-) \\
NGC 5135 &$8.0 \pm 3.3$&$202.1 \pm 28.0$&$80.4 \pm 46.0$&$0.9$&$\leq 3.4$&-&-&-&$66.0 \pm 48.0$&0.61 (0.85) \\
NGC 5252 &$13.3 \pm 4.8$&$344.6 \pm 38.6$&$69.0 \pm 50.8$&$0.9$&$\leq 4.8$&-&-&-&$144.0 \pm 24.0$&0.55 (0.85) \\
NGC 5363 &$8.4 \pm 0.7$&$149.6 \pm 5.9$&$46.1 \pm 6.0$&$0.8$&$\leq 4.2$&-&-&-&-&0.10 (-) \\
NGC 5427 &$\leq 0.3$&-&-&-&$\leq 2.1$&-&-&-&-&0.90 (-) \\
NGC 5506 &$123.8 \pm 49.6$&$165.1 \pm 30.0$&$15.2 \pm 49.8$&$0.8$&$152.4 \pm 47.4$&$165.1 \pm 8.0$&$46.8 \pm 10.8$&$0.7$&$60.0 \pm 56.0$&0.73 (0.70) \\
NGC 5548 &$59.0 \pm 3.8$&$410.8 \pm 12.3$&$11.6 \pm 5.8$&$0.9$&$\leq 5.0$&-&-&-&$200.0 \pm 17.0$&0.05 (-) \\
NGC 5643 &$10.9 \pm 2.3$&$91.2 \pm 8.4$&$100.3 \pm 49.2$&$0.9$&$\leq 3.6$&-&-&-&$52.0 \pm 17.0$&0.44 (0.82) \\
NGC 5728 &$3.0 \pm 1.1$&$358.3 \pm 41.7$&$97.7 \pm 29.8$&$0.9$&$\leq 4.5$&-&-&-&$98.0 \pm 25.0$&0.49 (-) \\
NGC 5813 &$\leq 0.3$&-&-&-&$\leq 1.9$&-&-&-&-&- (-) \\
NGC 5953 &$\leq 0.3$&-&-&-&$\leq 2.0$&-&-&-&-&0.96 (-) \\
NGC 5995 &$44.7 \pm 13.4$&$613.3 \pm 69.7$&$-12.2 \pm 40.1$&$0.9$&$53.7 \pm 19.0$&$613.3 \pm 69.4$&$75.1 \pm 39.8$&$0.9$&$111.0 \pm 39.0$&0.57 (0.50) \\
NGC 6000 &-&-&-&-&$\leq 2.1$&-&-&-&-&- (-) \\
NGC 6221 &$12.8 \pm 2.9$&$103.5 \pm 9.0$&$-12.5 \pm 7.5$&$0.6$&$\leq 3.8$&-&-&-&-&0.49 (0.75) \\
NGC 6240N &$12.4 \pm 1.5$&$1510.1 \pm 52.4$&$-12.3 \pm 3.3$&$0.5$&$\leq 1.8$&-&-&-&-&0.57 (0.91) \\
NGC 63 &-&-&-&-&$1.2 \pm 0.5$&-&$56.4 \pm 12.4$&$0.6$&-&- (0.98) \\
NGC 6810 &$28.4 \pm 2.5$&$199.4 \pm 7.1$&$10.7 \pm 2.4$&$0.6$&$10.0 \pm 2.0$&$199.4 \pm 12.3$&$9.3 \pm 3.0$&$0.5$&-&0.20 (0.33) \\
NGC 6814 &$5.4 \pm 1.5$&$108.3 \pm 13.0$&$84.2 \pm 59.7$&$0.9$&$\leq 3.6$&-&-&-&$97.0 \pm 10.0$&0.66 (0.78) \\
NGC 6860 &$15.0 \pm 5.1$&$448.9 \pm 56.5$&$86.1 \pm 60.4$&$0.9$&$\leq 4.9$&-&-&-&$53.0 \pm 49.0$&0.70 (0.91) \\
NGC 6890 &$6.9 \pm 1.5$&$164.6 \pm 15.8$&$-16.3 \pm 39.3$&$0.9$&$\leq 1.3$&-&-&-&$102.0 \pm 0.0$&0.54 (0.88) \\
NGC 7130 &$7.7 \pm 1.2$&$406.2 \pm 28.3$&$9.7 \pm 49.4$&$0.9$&$\leq 2.8$&-&-&-&-&0.47 (0.89) \\
NGC 7172 &$31.5 \pm 7.6$&$215.5 \pm 23.0$&$16.7 \pm 50.5$&$0.9$&$\leq 3.4$&-&-&-&$86.0 \pm 10.0$&0.61 (0.95) \\
NGC 7213 &-&-&-&-&$\leq 3.1$&-&-&-&-&- (-) \\
NGC 7314 &$6.8 \pm 1.5$&$73.4 \pm 6.7$&$91.4 \pm 25.7$&$0.9$&$\leq 2.5$&-&-&-&$87.0 \pm 43.0$&0.54 (0.89) \\
NGC 7479 &$10.0 \pm 2.5$&$168.9 \pm 17.5$&$-3.4 \pm 37.0$&$0.9$&-&-&-&-&$82.0 \pm 0.0$&0.56 (-) \\
NGC 7496 &$4.7 \pm 1.2$&$133.4 \pm 13.9$&$16.6 \pm 48.1$&$0.9$&$\leq 2.3$&-&-&-&-&0.58 (0.77) \\
NGC 7552 &-&-&-&-&$\leq 2.3$&-&-&-&-&- (-) \\
NGC 7582 &$60.2 \pm 26.9$&$246.0 \pm 58.0$&$19.1 \pm 42.0$&$0.7$&$83.2 \pm 22.0$&$246.0 \pm 24.2$&$100.7 \pm 56.8$&$0.9$&-&0.77 (0.52) \\
NGC 7590 &$4.0 \pm 0.3$&$226.6 \pm 8.2$&$-29.0 \pm 3.7$&$0.7$&$\leq 1.7$&-&-&-&-&0.14 (-) \\
PG 2130+099 &$20.7 \pm 9.5$&$1525.4 \pm 180.2$&$67.8 \pm 31.8$&$0.9$&$\leq 4.5$&-&-&-&$158.0 \pm 27.0$&0.45 (0.92) \\
PKS 1417-19 &$\leq 0.7$&-&-&-&$\leq 4.2$&-&-&-&$98.0 \pm 0.0$&0.91 (-) \\
PKS 1814-63 &$\leq 0.4$&-&-&-&$\leq 2.1$&-&-&-&$25.0 \pm 0.0$&0.90 (-) \\
PKS 1932-46 &$\leq 0.3$&-&-&-&$\leq 2.1$&-&-&-&-&- (-) \\
Superantennae S &$9.4 \pm 2.8$&$1553.8 \pm 189.1$&$6.5 \pm 32.8$&$0.8$&$17.4 \pm 4.9$&$1553.8 \pm 132.7$&$78.1 \pm 56.5$&$0.9$&$151.0 \pm 26.0$&0.64 (0.61) \\
UGC 2369 S &-&-&-&-&$\leq 1.9$&-&-&-&-&- (-) \\
Z 41-20 &$\leq 0.7$&-&-&-&$\leq 4.1$&-&-&-&$172.0 \pm 36.0$&0.76 (-) \\
\enddata
\tablecomments{Here Flux$_{\rm ext}$ indicates the flux from the fitted resolved emission, and we only include $\Theta$ and PA for those sources with $>2\sigma$ detections. N-band PAs come from \cite{asmus2016}. Here $r_{X} \equiv \theta_{\rm FWHM,X} / \Theta_{\rm FWHM,X}$. }
\end{deluxetable*}
\end{longrotatetable}

\begin{figure*}[ht]
\centering
\includegraphics[width=0.9\textwidth]{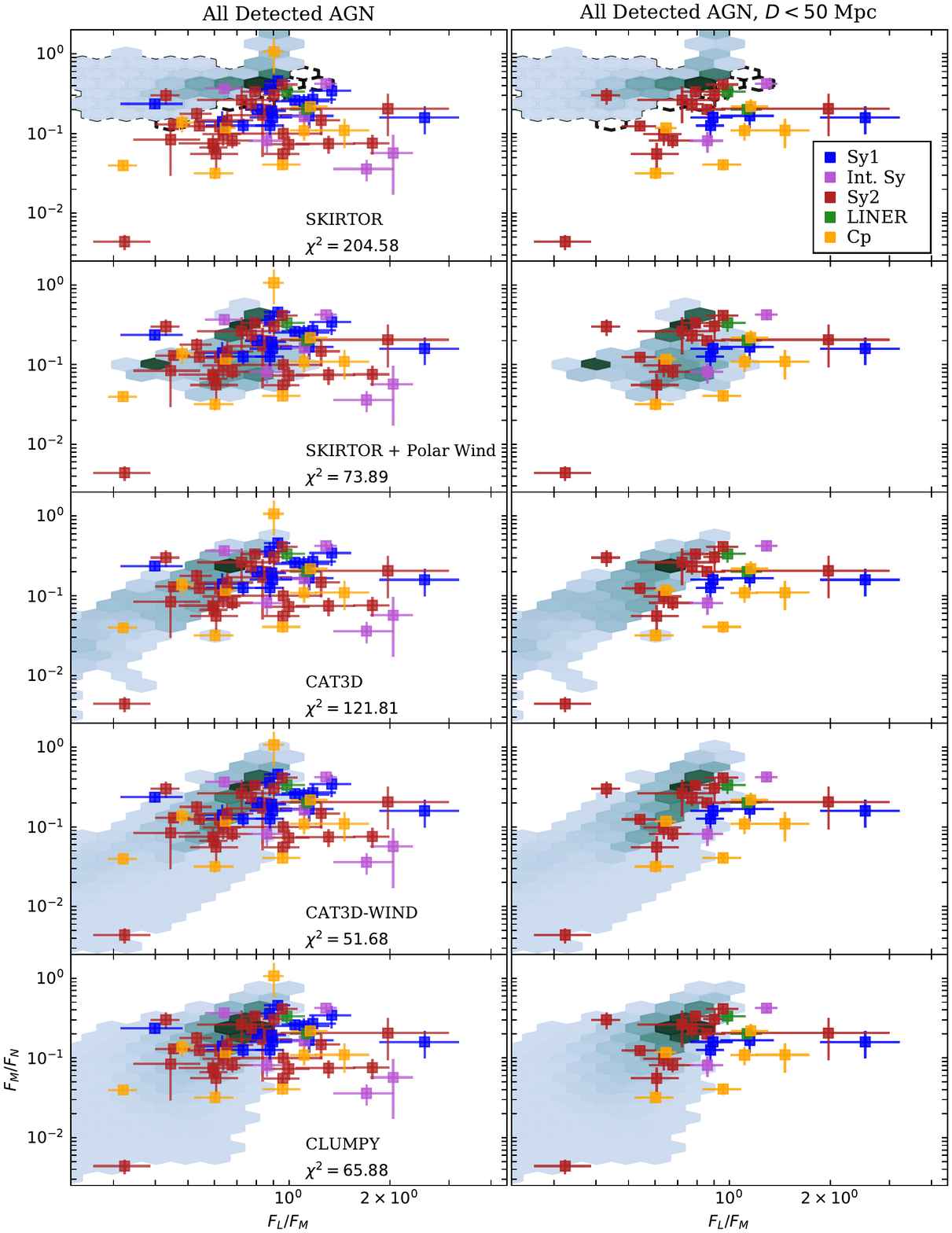}
\caption{Comparisons of $L/M$ and $M/N$ flux ratios to those predicted by torus models SKIRTOR \citep[][]{stalevski2016}, SKIRTOR+Polar Wind, CLUMPY \citep[][]{nenkova2008b}, CAT3D \citep[][]{honig2010}, and CAT3D-WIND \citep[][]{honig2017}. 
Only includes sources with SNR$_{gauss}>3$ in both the $L$- and M-bands. Colors are the same as in Fig. \ref{fig:hist}. In the left panels we show all detected sources in our sample. In the right panels we show only AGN closer than 50 Mpc in our sample. The $\chi^2$ value is measured from each data point to the model SED which provides the best fit. For SKIRTOR we show the effects of using the \citet{kishimoto2008} accretion disk spectrum (described in \S\ref{sec:ad}) as a dashed contour.}  
\label{fig:models}
\end{figure*}

\section{Comparison to Dust Emission Models}

\label{sec:models}
We compare our detected AGN MIR colors to those predicted by various dusty torus models. The SKIRTOR models \citep{stalevski2016} are heterogeneous, consisting of high-density clumps and low-density interclump media. The CAT3D models \citep{honig2010} and the CLUMPY models \citep[][]{nenkova2008b} resemble classical clumpy tori made up of spherical clouds, while CAT3D-WIND \citep{honig2017} includes an additional hollow cone in the polar region, representing a dusty wind driven by radiation pressure. Numerous previous works have performed SED fitting to test various torus models \citep[e.g.,][]{ramosalmeida2009, alonso-herrero2011, lira2013}, and they have emphasized the importance of the 5\micron~flux. We therefore focus on the NIR, utilizing the new $L$- and $M$-band measurements from this work, presenting a spectral slope plot similar to that in e.g., \citet{honig2017}.

In addition, we include in comparison a new, unpublished model library consisting of a flared disk and a polar wind\footnote{For further information on the SKIRTOR+wind model, please contact M. Stalevski.}. We discuss the model only briefly here, and only in comparison to SKIRTOR \citep{stalevski2016}. The disk component is defined by its angular width and optical depth and is similar to that in SKIRTOR, but with dust distributed smoothly. The polar wind takes the shape of a hollow cone parametrized by half-opening angle and radial extent. The dust composition is the same as in SKIRTOR (53 \% silicates and 47 \% graphite), and with the same power-law grain size distribution \citep{mathis1977}, but with larger grains (between 0.1 and 1 \micron), as suggested by flat extinction curves and silicate feature profiles \citep{gaskell2004,shaoz2017,xie2017}. For the hollow cone, we calculated an additional set of models with only graphite grains: if dust in the wind is driven away from the sublimation zone, it might be expected to be silicate-poor. This is a preliminary model library with limited parameter space: the current number of SEDs is an order of magnitude smaller than its precursor, SKIRTOR. However, it is well-motivated by the model of the resolved mid-IR images of Circinus AGN \citep{stalevksi2017,stalevski2019}. Even though this model library is to be developed further and tested against larger datasets, it is useful for our purpose here, since it represents an extension of the SKIRTOR models, thus allowing us to isolate the effect of polar winds on the flux ratios. While at this stage any results obtained with this model are to be taken with caution, it is indicative that, as it will be shown below, even though it is much more limited in terms of parameters and number of SEDs, it provides a significantly better match to the observed color space than its precursor. Also, unlike any other included model library, the color space covered by this model does not extend significantly beyond the observed colors. 

For each model SED produced by these setups, we extract the $L$-, $M$-, and $N$-band fluxes. We then compare the flux ratios $F_{L} / F_{M}$ and $F_{M}/ F_{N}$ to those from the \lprime- and $M$-band fluxes from this sample and the $N$-band fluxes from \cite{asmus2014}, shown in Fig \ref{fig:models}. We compare the models using the $\chi^2$ value, measured from each galaxy to the SED in each model that it matches most closely. We primarily consider Seyfert types 1, intermediate, and 2 in this analysis, as torus orientation aims to explain their differences in AGN Unification. We include the other AGN (e.g., LINERs and Cp) as they may have dusty tori but their mid-infrared emission might not be dominated by them. 

 Our observations favor torus+wind models such at CAT3D-WIND and the new SKIRTOR+wind. Specifically, CAT3D-WIND has the lowest $\chi^2$ value, $\chi_{\rm CAT3D-WIND}^2 = 51.68$, which is a very sharp improvement from the base CAT3D $\chi_{\rm CAT3D}^2 = 121.81$. Similarly, the agreement with SKIRTOR is greatly improved with the inclusion of the polar wind: $\chi_{\rm SKIRTOR}^2 = 204.58 \rightarrow \chi_{\rm SKIRTOR+wind}^2 = 73.89$.  CLUMPY exhibits a similar agreement, $\chi_{\rm CLUMPY}^2 = 65.88$, notably {\it without} the inclusion of winds. The $\chi^2$ values serve here primarily as a means for simple comparison, as the various torus models have vastly different parameter spaces and numbers of SEDs. Our results nonetheless agree with \citet{gonzales-martin2019} who found that CAT3D-WIND best matched their Swift/BAT-selected sample, with CLUMPY performing second-best.  Moreover, similar to \citet{gonzales-martin2019}, we find that the AGN colors are confined to a much smaller region of the parameter space than the torus SEDs. With our simple color-color comparisons, however, we cannot distinguish whether these regions have unphysical parameters or whether the discrepancy is driven by selection effects.

We find 13 of our Seyfert sample are ``bluer'' than predicted by any model $F_{L}/F_{M}$. We note that none of the torus models produces SEDs wherein $F_{L}/F_{M}\gtrsim 1$. We focus on color-color comparisons rather than SED fitting as they utilize the photometric data we have measured in this work, but we note that full SED fitting would be required to fully test the validity of the various torus models. Our comparisons are thus only qualitative, but they do give hints as to a preferred ``torus'' feature: the polar wind.

 There are two explanations for this apparent discrepancy which are discussed in the following subsections: stellar contamination to the nuclear flux and incomplete representation of the 3-5\micron~bump in the torus models.

On average, we find that Sy1 are bluer than Sy2 in the NIR ($\log(F_{\rm L,Sy1} / F_{\rm M,Sy1}) \equiv \log(f_{\rm LM, Sy1}) = -0.05\pm 0.18 $ vs. $\log(f_{\rm LM, Sy2}) = -0.16 \pm 0.18$ and that they are also slightly bluer in the MIR ($\log(F_{\rm M,Sy1} / F_{\rm N,Sy1}) \equiv \log(f_{\rm MN, Sy1}) = -0.70 \pm 0.17 $ vs. $\log(f_{\rm MN, Sy2}) = -0.91 \pm 0.37$. In fact, Sy1 primarily populate the blue limit of the the torus SED models in $F_{L} / F_{M}$. The Sy2 sample shows much more diversity in both flux ratios, and there are several outliers which are much bluer than models predict. Only two LINERs were significantly detected in both \lprime~and $M$, so we cannot draw any conclusions about this population. The extremely mixed Cp subsample shows a large amount of scatter, but tends to be extremely ``blue'' compared to the Seyferts and to the models ($\log(f_{\rm LM, Cp}) = 0.05 \pm 0.24$)). 
It does not seem as if this diagram can be used to reliably separate different AGN types.

\subsection{Stellar Contamination}

\begin{figure*}[ht]
    \centering
    \includegraphics[width=0.9\textwidth]{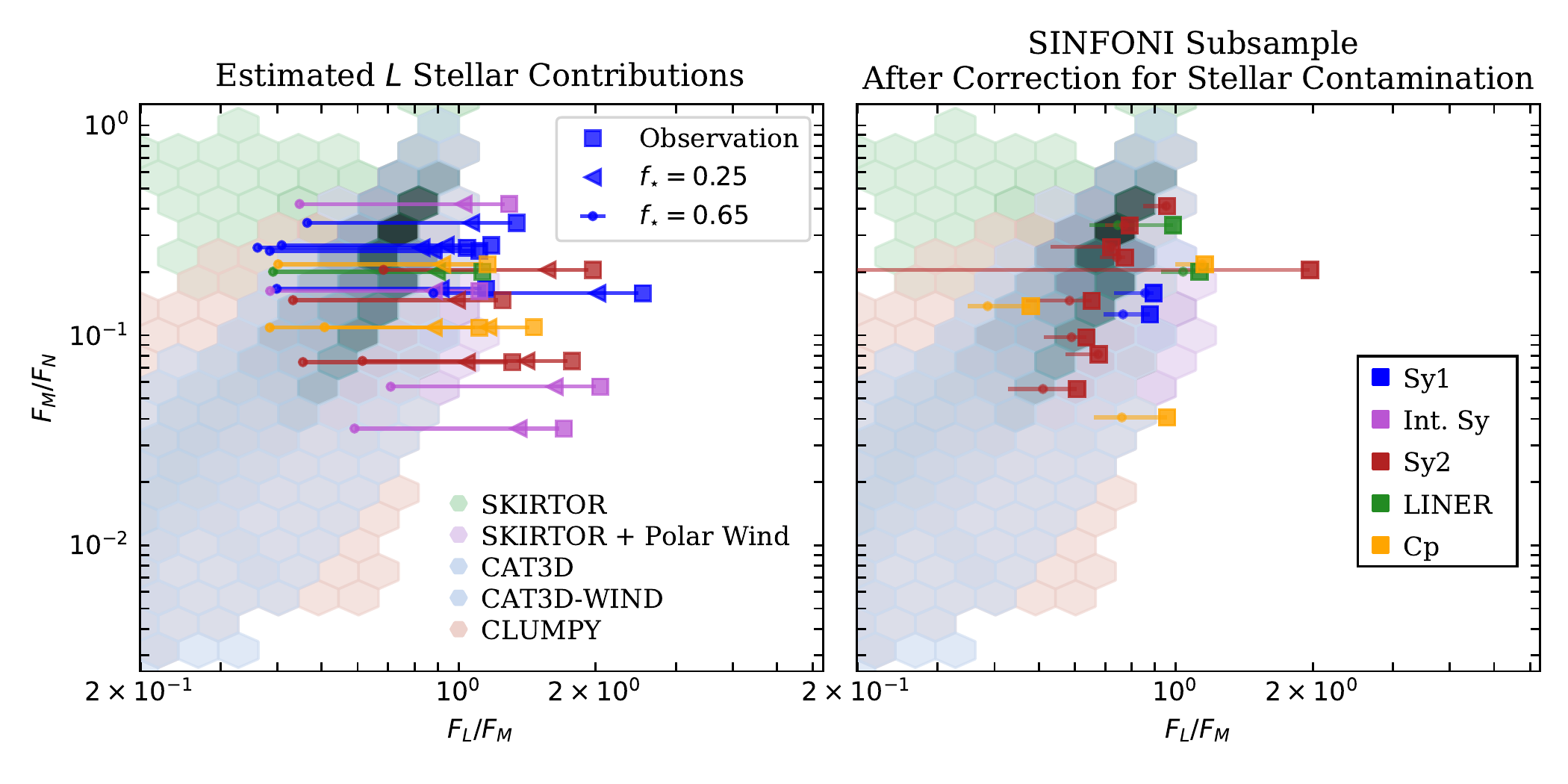}
    \caption{Comparisons of \lprime/M and M/N flux ratios to those predicted by torus models SKIRTOR \citep[green and purple; ][]{stalevski2016}, CAT3D \citep[grey;][]{honig2010}, CLUMPY \citep[orange;][]{nenkova2008b}, and CAT3D-WIND \citep[blue; ][]{honig2017}. Only includes sources with SNR$_{gauss}>2$ in both the $L$- and M-bands. Colors same as in Fig. \ref{fig:hist}. In the left panel we show the sources which are bluer than models allow with corrections made for various $f_{\star,L}$. In the right panel we show the 21 AGN which had SINFONI fluxes in \citet{burtscher2015} which we used to estimate \lprime-band stellar contribution using Eq. 3. }
    \label{fig:stellar_models}
\end{figure*}

The physical scales at which we extract the ``nuclear'' flux in this sample vary greatly, from $\sim 10$pc to more that $1.5$kpc. For essentially all sources at $D_{\rm AGN} \geq 50$Mpc, corresponding to an unresolved area $\theta_{\rm FWHM} \gtrsim 100$pc, we expect a significant contribution from the ``red'' tail of the stellar emission to our $L$-band fluxes. This stellar contribution would serve to make our AGN artificially bluer than the torus models. We therefore perform a simple distance cut, $D \leq 50$ Mpc, in order to focus on the polar dust region. We show this in Fig. \ref{fig:models}. We find that after this cut, only two Seyfert galaxies are strictly $\geq 1\sigma$ bluer than the models. They are NGC 4235 (Sy1) and NGC 1365 (Sy1.8). This simple test then indicates qualitatively that increased stellar emission on larger scales could bias the AGN with $D > 50$Mpc to bluer colors.

More quantitatively, in \cite{alonso-herrero2001} the stellar contribution to the \lprime-band within a $3''$ diameter aperture was estimated to be up to 25$\%$ for a sample of 12 Sy2 AGN (except for NGC\,5252 which had 65$\%$ stellar contribution). In the $M$-band it was estimated to be $<10\%$. This indicates a blueward flux-ratio shift of $\log (F_{\rm L} / F_{\rm M}) \lesssim 0.1$.  For their Sy1  sample (5 AGN) they assumed that $100\%$ of the L-band flux was non-stellar in origin. We show the effects of stellar contamination in Fig. \ref{fig:stellar_models} by plotting the expected flux ratios of the AGN in our sample assuming they have $0 \%$, $25 \%$, and $65\%$ stellar $L$-flux.

With a 25\% $L$-band correction, 7 of the AGN in the full sample are still bluer than the models predict. If we allow for up to $65\%$ of the \lprime-flux to come from a stellar origin \citep[i.e., the maximum percentage found in][]{alonso-herrero2001}, the shift increases to $\log (F_{\rm L} / F_{\rm M}) < 0.5$, which would make all the outliers in Fig. \ref{fig:models} consistent with the models. 

We do not include an estimate of the $M$-band stellar flux because $i$) \citet{alonso-herrero2001} found it to be strictly $< 10\%$; {\it ii}) \citet{assef2010} show that the AGN flux is steeply rising between the $L$- and $M$-bands while the stellar contribution steeply drops, indicating the $M$-band stellar contribution is $\lesssim 5\%$; and {\it iii}) subtracting an $M$-band contribution would actually serve to make AGN in Figs. \ref{fig:models}-\ref{fig:stellar_models} agree even less with the models by shifting them toward bluer colors. 

\subsubsection{SINFONI-Estimated Stellar Fluxes}
For a subset of 21 AGN we can estimate the stellar contamination using $K$-band fluxes and AGN fractions derived by \citet{burtscher2015}. Using SINFONI, these fluxes were measured in the central 1'' diameter aperture of each galaxy. Their near-IR AGN fractions ($f_{\rm AGN, K}$) were derived from spectral fitting in this central region and checked using the stellar CO equivalent width. We then used the average color of old stellar populations in stellar bulges, $K-L = 0.09$ mag, as measured from the empirical elliptical galaxy template of \citet{assef2010}, to calculate the expected $L$-band stellar magnitude as
\begin{equation}
\label{eq:sinfoni}
L_{\rm mag, \star} = K_{\rm mag, total} + {\rm AGN}_{\rm mag, correction} - 0.09,    
\end{equation}
where ${\rm AGN}_{\rm mag,correction}= -\log_{10}((100-f_{\rm AGN, K})/100)$ is a correction for the measured AGN fraction of the galaxy. We converted these magnitudes into ISAAC \lprime~fluxes. 

\begin{figure}
    \centering
    \includegraphics[width=0.5\textwidth]{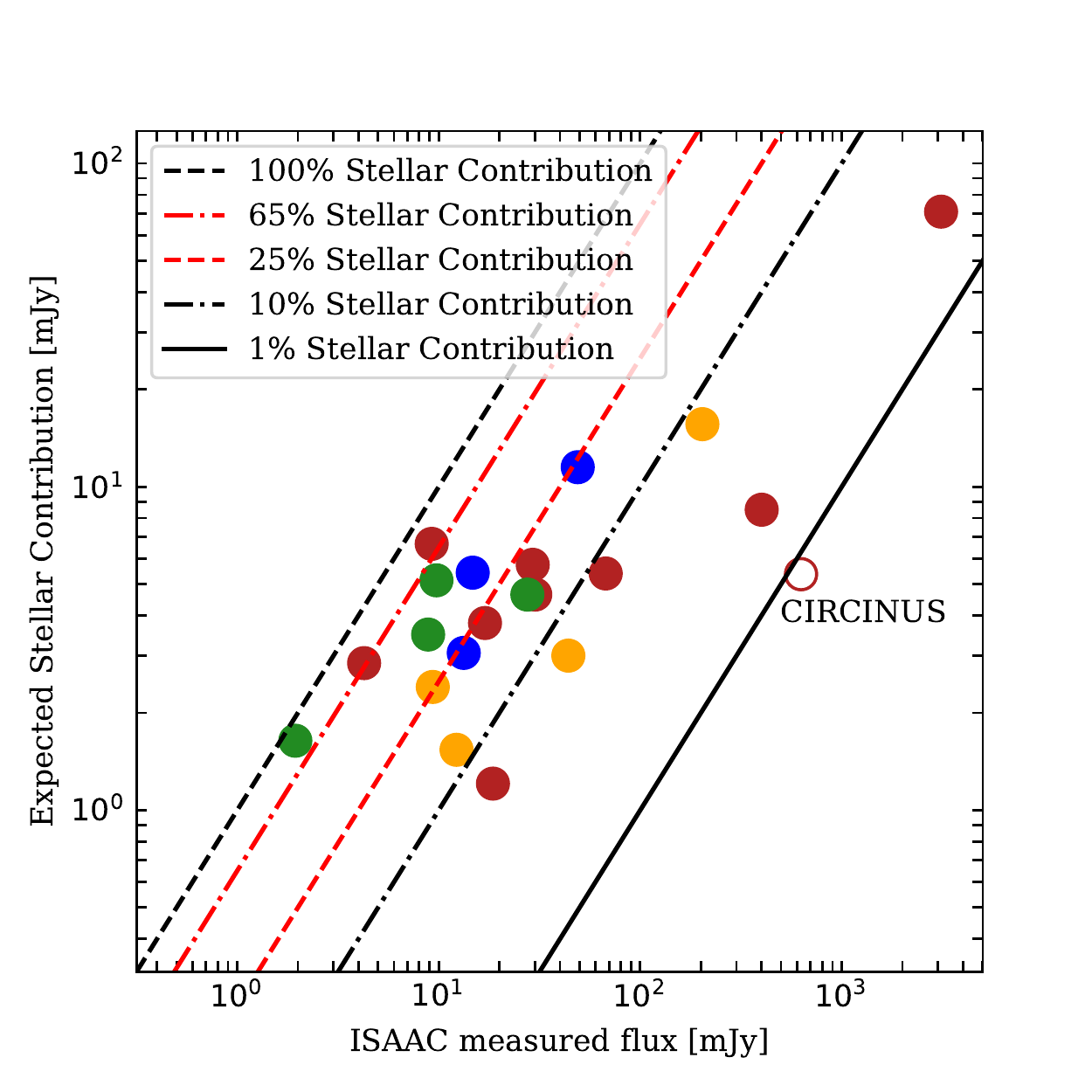}
    \caption{Comparison of SINFONI-estimated $L$-band stellar contributions and our measured 1'' aperture fluxes from ISAAC. Overplotted are the lines showing what we expect if 100\%, 65\%, 25\%, 10\% and 1\% of the flux is stellar in origin. Circinus is marked separately because it was measured with an aperture of 0.5'' in both ISAAC and SINFONI observations. AGN are color coded as in Fig. \ref{fig:hist}.} 
    \label{fig:sinfoni_isaac}
\end{figure}

We measured the ISAAC \lprime~fluxes of these 21 AGN using a central, 1'' diameter aperture. We list the 1'' aperture fluxes, the Gaussian-fitted nuclear fluxes, the estimated stellar fluxes, and the measured AGN fractions in Table \ref{tab:sinfoni_subsample}. We show a comparison in Fig. \ref{fig:sinfoni_isaac} to illustrate the AGN fractions. Note that the typical unresolved nuclear flux in our sample has a FWHM of 0.75'', so the stellar fractions could be slightly less in the unresolved nuclear fluxes on nights with average or better seeing. Note also that Circinus was measured in the SINFONI sample with a $0.5''$ diameter aperture, which we match here; it is marked in Table \ref{tab:sinfoni_subsample} and Fig. \ref{fig:sinfoni_isaac}.

We find that a rather large fraction of the AGN (13/21) should have $> 10\%$  of their ISAAC \lprime~nuclear flux come from stars rather than the unresolved nucleus. Two of these sources, NGC 4303 and NGC 4261, are shown to have $\approx 70\%$ stellar flux in the L-band. This is in rough agreement with the $K$-band $f_{\rm AGN, K} < 10\%$ estimates found by \citet{burtscher2015}. All but two of the AGN in this subsample agree with the findings of \citet{alonso-herrero2001}, i.e., stellar contributions $\leq 65\%$.

We finally compare these AGN once more with the torus models in Fig. \ref{fig:stellar_models}(right panel), but now with their $K$-band estimated stellar flux subtracted. Following this stellar correction, all of the AGN in this subsample are compatible with the torus SED models of CAT3D-WIND, CLUMPY, and SKIRTOR+wind. The stellar corrections required for CLUMPY are larger than for the torus+wind models as shown in Fig. \ref{fig:stellar_models}. To correct all of the AGN nuclear fluxes in this sample, we would need similar SINFONI $K$-band data and stellar CO equivalent width measurements. For the nearest AGN, however, we expect this correction to be on the order of $1 \%$, as exhibited by Circinus and NGC 1068.

\begin{table}[] \small \centering \begin{tabular}{lllll} Target & $L$ Flux$_{\rm 1as}$ &$L$ Flux$_{\rm nuc.gauss}$ & Est. $L$ Flux$_{\star}$ & $f_{\rm AGN}$ \\ 
 & [mJy] & [mJy] & [mJy] & \\ \hline \hline \\
Circinus & $628.0 \pm 6.5$ & $459.3 \pm 42.5$ & $5.4$ & 0.99 \\
IC 5063 & $57.0 \pm 0.1$ & $57.0 \pm 2.2$ & $1.4$ & 0.98 \\
NGC 1068 & $3111.7 \pm 6.2$ & $2012.8 \pm 238.8$ & $70.9$ & 0.98 \\
NGC 1097 & $8.9 \pm 0.1$ & $5.6 \pm 0.2$ & $3.5$ & 0.61 \\
NGC 1386 & $30.1 \pm 0.2$ & $18.8 \pm 1.0$ & $4.6$ & 0.85 \\
NGC 1566 & $14.7 \pm 0.1$ & $10.6 \pm 0.4$ & $5.4$ & 0.63 \\
NGC 4261 & $1.9 \pm 0.1$ & $0.6 \pm 0.1$ & $1.6$ & 0.16 \\
NGC 4303 & $4.3 \pm 0.1$ & $1.6 \pm 0.1$ & $2.8$ & 0.33 \\
NGC 4303 & $4.3 \pm 0.1$ & $3.8 \pm 0.2$ & $2.2$ & 0.50 \\
NGC 4388 & $29.3 \pm 0.1$ & $33.9 \pm 1.3$ & $5.7$ & 0.80 \\
NGC 4501 & $9.2 \pm 0.3$ & $1.5 \pm 0.2$ & $5.1$ & 0.44 \\
NGC 4579 & $27.5 \pm 0.2$ & $16.9 \pm 0.6$ & $6.7$ & 0.76 \\
NGC 4593 & $49.0 \pm 0.2$ & $32.3 \pm 2.5$ & $4.6$ & 0.91 \\
NGC 5135 & $18.6 \pm 0.2$ & $12.6 \pm 0.7$ & $11.5$ & 0.38 \\
NGC 5506 & $400.6 \pm 0.3$ & $343.8 \pm 12.5$ & $1.2$ & 1.00 \\
NGC 5643 & $17.0 \pm 0.1$ & $8.5 \pm 0.5$ & $8.5$ & 0.50 \\
NGC 6814 & $13.3 \pm 0.1$ & $10.6 \pm 0.4$ & $3.0$ & 0.77 \\
NGC 7130 & $12.3 \pm 0.1$ & $6.9 \pm 0.3$ & $3.1$ & 0.75 \\
NGC 7172 & $67.4 \pm 0.2$ & $49.1 \pm 2.0$ & $1.5$ & 0.98 \\
NGC 7496 & $9.3 \pm 0.1$ & $6.6 \pm 0.3$ & $5.4$ & 0.42 \\
NGC 7582 & $150.8 \pm 0.6$ & $86.1 \pm 5.3$ & $2.4$ & 0.98 \\
 \end{tabular} \caption{Measured 1'' diameter aperture flux, measured nuclear Gaussian flux, estimated stellar flux using SINFONI $K$-band measurements and Eq. \ref{eq:sinfoni}, and measured AGN fraction for 21 cross-matched AGN. } \label{tab:sinfoni_subsample} \end{table}

\subsection{Accretion Disk Spectra}
\label{sec:ad}
The assumed spectrum of the accretion disk can play a large role in the final model SED. Traditionally a broken power-law has been assumed, i.e.,
\begin{eqnarray}
\lambda F_{\lambda} \propto 
\begin{cases}
    \lambda^{1.2} & 0.001 \leq \lambda \leq 0.01\micron \\
    \lambda^{0} & 0.01 < \lambda \leq 0.1\micron \\
    \lambda^{-0.5} & 0.1 < \lambda \leq 5\micron \\
    \lambda^{-3} & 5 < \lambda \leq 50\micron
\end{cases}
\label{eq:brokenPLaw}
\end{eqnarray}
\citep[see e.g.,][and references therein]{sanders1989, kishimoto2008}. The spectrum produced by Eq. \ref{eq:brokenPLaw} was used in the models CLUMPY \citep{nenkova2008b} and SKIRTOR \citep{stalevski2016}. More recently, however, a bluer accretion SED has been adopted which is strongly supported by QSO observations \citep[e.g.,][]{zheng1997,manske1998,vandenberk2001,scott2004, kishimoto2008}. This SED is relatively blue as it has a shallower power-law falloff $\nu F_{\nu} \propto \nu^{4/3}$ between 0.3 and 3$\micron$:
\begin{eqnarray}
\lambda F_{\lambda} \propto 
\begin{cases}
    \lambda^{1} &  \lambda < 0.03 \\
    \lambda^{0} & 0.03 < \lambda \leq .3 \\
    \lambda^{-4/3} & 0.3 < \lambda \leq 3 \\
    \lambda^{-3} & 3 < \lambda, 
\end{cases}
\label{eq:modernSED}
\end{eqnarray}
and it was used in both CAT3D and CAT3D-WIND \citep[ respectively]{honig2010, honig2017}.

In order to test the role the model accretion disk spectrum plays in the $L$- and $M$-bands, we replaced Eq. \ref{eq:brokenPLaw} with Eq. \ref{eq:modernSED} in the SKIRTOR models using the method outlined in \citet{yang2020}. The results of this replacement are shown in Fig.\ref{fig:models}. The overall effect of this replacement is to shift the models' flux ratios $F_{L}/F_{M}$ ``blueward'', i.e., to the right in our plots. After this replacement, the SKIRTOR models with the bluer accretion disk are a better match to the data $\chi_{\rm SKIRTOR}^2=204.58 \rightarrow \chi_{\rm Bluer~AD}^2 = 174.17$, but this shift is not enough to explain observations with $F_{L}/F_{M} > 1$. We show the effects of changing the accretion disk spectrum in Fig. \ref{fig:models} as a dashed contour in the SKIRTOR panels. The CLUMPY library provides two sets of models: one with only dust emission from the torus, and another which includes the accretion disk. However, the accretion disk emission is not traced in the radiative transfer calculations. Rather, it is up to the user to decide which one to use, e.g. based on the probability to have a clear line of sight to the central source for the given model. While this is not fully consistent treatment, we tested both versions and found the difference to be very small ($\Delta \chi_{\rm CLUMPY}^2 < 1$). We conclude that including the accretion disk flux in CLUMPY models has an insignificant effect on the model predictions.
In both cases, only a radically different accretion disk spectrum could explain these data, so stellar contamination is still favored.

\subsection{Polar Elongation and the 3-5 Micron Bump}
It is also possible that the MIR flux is underestimated in modern torus models. It has been well documented that many QSOs and Sy1 galaxies exhibit what is called the 3-5 \micron~bump \citep[e.g.,][]{edelson1986, kishimoto2011, mor2012, honig2013}. This feature was thought to be caused by the presence of inter-clump dust, but recent work by \citet{honig2017} claims that the addition of wind orthogonal to the accretion disk can explain it. In CAT3D-WIND, \citet{honig2017} this feature is apparently caused by a large amount of dust in a ``puffed-up'' region in the vicinity of the sublimation ring or by winds which remove clouds from the central few pc and cause the SED to be split into the hot dust emission from the center and the cold dust emission from a region elongated in the polar direction. A similar conclusion is valid for the models based on SKIRTOR with polar winds: the additional dust of low optical depth at the base of the wind is responsible for additional 3-5 \micron~emission and "bluer" colors.

It appears that within our sample, the differences between observations and the models are sufficiently explained by stellar contamination in the \lprime-band. The AGN in our sample agree best with CAT3D-WIND\citep{honig2017}, indicating that the inclusion of the polar wind describes AGN at this resolution quite well. Moreover, the SKIRTOR \citep{stalevski2016} models fit the data much better after the inclusion of a polar wind. On the other hand, we find that CLUMPY \citep{nenkova2008b} provides good fits to the data without polar winds. It will be interesting to further investigate the nuclear regions of NGC 1365 and NGC 4235 with MATISSE, testing whether these ``too blue'' Seyferts are sufficiently explained by stellar contamination.

\section{Estimating VLT Fluxes from \textit{WISE}}

\label{sec:wise}
The Wide-field Infrared Survey Explorer (\textit{WISE}) space telescope \citep{wright2010} has observed a large number of AGN in the MIR at high sensitivity but limited spatial resolution. Moreover, the ALLWISE AGN catalog contains 1.4 million AGN selected using 3 \textit{WISE} bands \citep{secrest2015}. This catalog and others like it present a rich source from which to draw AGN for further study in the MIR. The median point-spread-function in the {\it W1}-band has a FWHM of 6'', making it nearly impossible to spatially distinguish thermal torus emission from star formation in the foreground or even the central few hundred parsecs of even nearby AGN. The large spatial areas probed by \textit{WISE} in each AGN will bias the flux to be comparatively larger than seen from the 8-10m class telescopes used in MIR interferometry. In order to select AGN from a much larger parent sample for MATISSE followup from \textit{WISE}, we study the relationships between \textit{WISE} colors, \textit{WISE} fluxes, and the fraction of nuclear flux observable at the VLT(I). 

\begin{figure}[ht]
\centering
\includegraphics[width=0.5\textwidth]{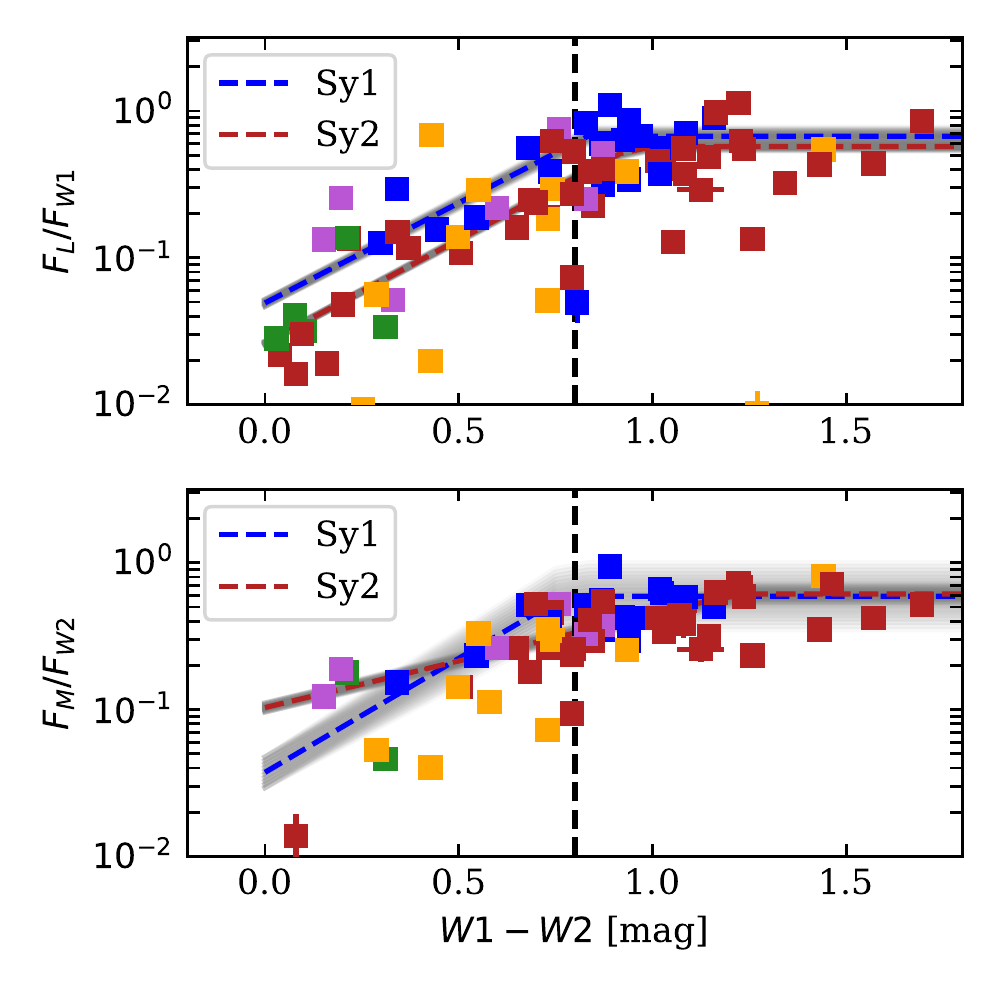}
\caption{ \textit{Top)} {\it L}/{\it W1} flux ratio vs. {\it W1-W2}. \textit{Bottom)} \textit{M}/{\it W2} flux ratio vs. {\it W1-W2}. For each panel only detected (SNR$_{gauss}>2$ in each band) sources are included. Colors same as Fig 1. The black line at \textit{W1-W2}$=0.8$ indicates the AGN selection cutoff of \cite{stern2012}, which is also the point after which $F_{\rm L} / F_{\rm W1} \approx 1$ in our Sy1 sample. Blue and red dashed lines are fitted piece-wise functions (Eq. 5-8) for the Sy1 and Sy2 samples, respectively. Grey shaded regions are the $1\sigma$ fit errors. 
}
\label{fig:cc}
\end{figure}

We explore the relationships between the measured ISAAC fluxes and the existing \textit{WISE} fluxes for each source in order to determine a scaling between the space-based MIR measurements and nuclear fluxes measured at the VLTI. By providing an estimate for the fluxes observed at the VLT, we can select AGN for follow-up observations directly from the \textit{WISE} catalog. This is not straightforward because the \textit{WISE} telescope has a lower resolution than the VLT and because the filters \textit{W1} and \textit{W2} are both wider than and offset from either $L^{\prime}$ or $M_{\rm nb}$. The resolution in the \textit{W1} band is $6.1''$ and is $6.4''$ in the \textit{W2} band \citep{wright2010}, while these ISAAC observations are seeing limited at $\lesssim 1''$. This means that in general we expect $F_{\rm ISAAC}/F_{\rm WISE} \leq 1$, as there will be a contribution from non-nuclear flux, especially at shorter wavelengths wherein the stellar contribution is higher. In Fig. \ref{fig:cc} we show the ratios $F_{\rm L} / F_{\rm W1}$ and $F_{\rm M} / F_{\rm W2}$ for each detected (SNR$_{gauss}>2$) AGN in our sample. Note that $F_{\rm L} / F_{\rm W1}$ and $F_{\rm M} / F_{\rm W2}$ are occasionally $> 1$. This is due to the offset in the filters: \lprime~is redder than \textit{W1}, and it therefore measures a larger relative flux from the nuclear dust emission in the same AGN. Nevertheless, for both the Sy1 and the Sy2 subsamples we see a linearly increasing trend in $F_{\rm ISAAC} / F_{\rm WISE}$ for blue \textit{W1-W2} colors until a ``saturation'' point at \textit{W1-W2}$\approx 0.8$ mag. This point occurs at roughly the same \textit{W1-W2} color in both subsamples in both bands. To further quantify this, we fit a linear piece-wise function to each subsample of the form
\begin{equation}
    f(W1-W2) = 
    \begin{cases}
    ap+b,&~\text{if}~(W1-W2) \geq p \\
    a (W1-W2) + b,&~\text{else}%\text{if}~(W1-W2) < p \\
    \end{cases}
    \label{eq:piecewise}
\end{equation}
where $p$ is the fitted point at which $F_{\rm ISAAC} / F_{\rm WISE} \approx 1$. The fitted lines are shown in Fig. \ref{fig:cc} with $1\sigma$ errors shaded in gray. Fitting was done using \code{curve\_fit} from \pack{scipy.optimize}, employing the Levenberg-Marquardt algorithm. Errors are estimated from the diagonal of the covariance matrix, with the implicit assumption that they are uncorrelated. We find that for the Sy1 sample, the \lprime~flux matches the \textit{W1} flux for $\textit{W1}-\textit{W2} \geq 1.05 \pm 0.03$ mag. For the Sy2 sample this cutoff is $\textit{W1}-\textit{W2} \geq 0.90 \pm 0.01$ mag, much redder than for the Sy1 sample. For Sy1 and Sy2 galaxies we can estimate the \lprime~nuclear flux from the \textit{W1} flux as
\begin{align}
\label{eq:lband_est} 
&\log_{10} F_{{\rm L}, {\rm Sy1}} = \\
&\begin{cases} 
  \log_{10} F_{W1} - 0.05 \pm 0.002,~\text{if}~(W1-W2) \geq 1.05\pm 0.03 \nonumber \\
  \log_{10} F_{W1} + (1.12 \pm 0.03)(W1-W2) - (1.23\pm 0.02),~\text{else} \nonumber \\ 
\end{cases} \nonumber \\
&\log_{10} F_{{\rm L}, {\rm Sy2}} = \\
&\begin{cases} 
  \log_{10} F_{W1} - 0.23 \pm 0.004,~\text{if}~(W1-W2) \geq 0.90 \pm 0.01 \nonumber \\
  \log_{10} F_{W1} + (1.51 \pm 0.03)(W1-W2) - (1.61 \pm 0.01),~\text{else}. \nonumber \\ 
\end{cases}
\end{align}

The WISE \textit{W1-W2} color at which the ISAAC and \textit{WISE} fluxes match is similar to \textit{W1-W2} color used to select AGN in WISE. \cite{stern2012} define an AGN selection color cut of W1-W2$\geq 0.8$ mag. We plot the \citet{stern2012} criterion in Fig. \ref{fig:cc} as a black dashed line for comparison. This cut was motivated by the desire to find high-redshift AGN ($z < 3$), so it is unsurprising that local AGN may exhibit much bluer \textit{W1-W2} colors. Indeed, \citet{mateos2012} show that \textit{W1-W2} is redder for high-redshift AGN.

In the $M$-band we find similar trends, albeit with significantly more scatter. The Sy1 sample shows a match between M and \textit{W2} when $\textit{W1}-\textit{W2} \geq 0.78 \pm 0.02$ mag; while the Sy2 sample does the same at the much redder $\textit{W1}-\textit{W2} \geq 1.19 \pm 0.02$ mag.  For Sy1 and Sy2 galaxies we estimate the M-band nuclear flux from the \textit{W2} flux as

\begin{align}
\label{eq:mband_est}
&\log_{10} F_{\rm M, Sy1} = \\
&\begin{cases} 
  \log_{10} F_{W2} - 0.23 \pm 0.03,~\text{if}~(W1-W2) \geq 0.78\pm 0.02 \nonumber \\
  \log_{10} F_{W2} + (1.48 \pm 0.15)(W1-W2) - (1.38\pm 0.10),~\text{else} \nonumber \\ 
\end{cases} \nonumber \\
&\log_{10} F_{\rm M, Sy2} = \\
&\begin{cases} 
  \log_{10} F_{W2} - 0.16 \pm 0.01,~\text{if}~(W1-W2) \geq 1.19\pm 0.02 \nonumber \\
  \log_{10} F_{W2} + (0.73 \pm 0.03)(W1-W2) - (1.03 \pm 0.03),~\text{else}. \nonumber \\ 
\end{cases}
\end{align}

We find that in both bands, Circinus is ``underluminous'' relative to the general trend. It is likely here that we are over-resolving emission that \textit{WISE} cannot spatially distinguish, as it is at a distance of only 4 Mpc. Another of the outliers, NGC 4355 (a.k.a. NGC 4418), is a compact obscured nucleus (CON) known to exhibit an SED unusual for Sy2 galaxies \citep[e.g.,][]{costagliola2011, ohyama2019}. We also note that NGC 5953 (the extremely ``blue'' galaxy above) is underluminous for either the Sy1 or Sy2 trends, but roughly agrees with other Cp AGN in our sample.

In summary, for Sy1 with WISE colors \textit{W1-W2}$ \geq 1.05$ mag the \textit{WISE} \textit{W1} flux is a good proxy for the \lprime~flux at the VLT. For Sy2 with \textit{WISE} colors \textit{W1-W2}$ \geq 0.90$ mag the \textit{WISE W1} flux is a good proxy for the \lprime~flux at the VLT. We present functions which use the \textit{W1-W2} color and the \textit{W1} (or \textit{W2}) flux to reliably estimate the \lprime-band (or M-band) fluxes observable at the VLT with instruments such as VLTI MATISSE, ERIS, and CRIRES.

\subsection{Potential VLTI/MATISSE Targets}
The short atmospheric coherence times in the $L$,  $M,$ and $N$ bands  severely limit the sensitivity of self-tracked interferometric observations in these bands with VLTI/MATISSE. This necessitates bright targets for detailed study. The recommended limits for $L$-band target flux are 75 mJy when using the UTs and 1 Jy when using the ATs. Using these limits, we can identify 13 potential MATISSE targets in our sample, 2 of which are observable with the ATs. Most of these (10/13) have been previously observed either with MIDI or already with MATISSE as part of commissioning and/or guaranteed time observations. Fortunately, we can use the above formulae to estimate $L$-band fluxes from various {\it WISE} catalogs and identify potential targets. Here a potential target means $L$-band fluxes greater than those listed above and declination $\delta < 20$ deg. 

Using the mid-IR photometrically-selected {\it WISE} AGN Candidates Catalogs\citep{assef2018}, we apply Eq. \ref{eq:lband_est} to esimate the $L$-flux. We use the \citet{assef2018} R90 catalog ($\sim 4$ million AGN), which has 90\% reliability, to identify 57 potential targets, 2 of which can be observed with the ATs.  

In addition to using photometrically selected catalogs, we cross-matched the \citet{veron-cetty2010} AGN catalog with the ALLWISE catalog to create a sample of optically classified AGN that had been detected using {\it WISE}. Using then the {\it W1} and {\it W2} fluxes, we computed the $L$-band fluxes using Eq. \ref{eq:lband_est}. In this way, we identify 44 AGN suitable for follow-up, 4 of which can be observed with the ATs. Many of the AGN in the \citet{veron-cetty2010} catalog are not included in color-selected {\it WISE} AGN catalogs \citep[e.g.,][]{assef2018, secrest2015} because the color cuts employed focus on high-redshift sources \citep[e.g.,][]{mateos2012}. Thus this second approach is helpful if one wishes to focus on nearby AGN.

\section{Summary and Conclusions} \label{sec:conc}
In this work we present a MIR flux catalog of 119 AGN, extending the work done by \citet{asmus2014} from the $N$- and $Q$-bands to the \lprime - and $M$-bands. This is the largest existing subarcsecond catalog of AGN in these bands. Each AGN was observed using VLT ISAAC in at least one of the \lprime-~and $M$-bands between 2000 and 2013. We include local ($z < 0.3$) AGN of 5 optical classifications: 21 are Seyfert 1; 5 are Intermediate Seyferts; 46 are Seyfert 2; 29 are LINERs; and 16 are so-called `Cp' or Composites. We report two tables: one with nuclear fluxes in the $L$- and $M$-bands, and one with resolved emission fluxes, sizes, and PAs in both bands. The nuclear and resolved emission were separated by fitting one Gaussian to the PSF and one to the flux on scales larger than the PSF. We detect 98 sources in the $L$-band and 81 sources in the $M$-band. We found resolved $L$-band emission in 73 of the AGN. 

We compared the flux ratios $F_{\rm L} / F_{\rm M}$ and $F_{\rm M} / F_{\rm N}$ to those predicted by several suites of AGN torus models:  CAT3D \citep{honig2010}, CAT3D-WIND \citep[][]{honig2017}, CLUMPY \citep{nenkova2008b}, SKIRTOR \citep{stalevski2016}, and SKIRTOR models with polar winds. We find that the inclusion of a polar wind component significantly improves the agreement between our measurements and model predictions. CAT3D-WIND provides the best overall match to our measurements.
The SKIRTOR \citep{stalevski2016} models fit the data much better after the inclusion of a polar wind. Notably, we find that CLUMPY \citep{nenkova2008b} provides good fits to the data without polar winds.
No set of models, however, produces values of $F_{\rm L} / F_{\rm M} \gtrsim 1$, which we measure in 10 Seyfert galaxies. We discuss two possible explanations of this: stellar contamination in the relatively large physical scales probed; and underestimation of $L$-band accretion disk flux in torus models. 

We favor the stellar contamination hypothesis, as several of the AGN in this sample (e.g., NGC5252) were shown in \citet{alonso-herrero2001} to have up to 65\% of their $L$-band emission come from stellar sources rather than the AGN. We also use $K$-band flux measurements of 21 cross-matched AGN from \citet{burtscher2015} to estimate the $L$-flux contribution from stars using the \citet{assef2010} $K-L$ colors for old stellar populations. After removing this estimated stellar contribution, all of the AGN become consistent with CAT3D-WIND, CLUMPY, and SKIRTOR+wind. We find that the stellar corrections necessary to match CLUMPY are $\approx 10\%$ larger than for the torus+wind models. 

We measured the effect of changing the accretion disk spectrum in the SKIRTOR models to a QSO-motivated and relatively bluer spectrum. We find that while it does improve the agreement between these models and the observations, it does not fully reproduce the observed $F_{L}/F_{M}$. Nonetheless, follow-up observations of these NIR ``blue'' AGN may provide insight into the mechanisms driving the 3-5\micron~bump and the formation of these dusty tori. 

We lastly derived relations between the reported WISE \textit{W1} and \textit{W2} fluxes and the $L$ and $M$ fluxes observable at the VLT. These relations (Eq. 7-10) can be used to estimate the NIR fluxes for sources not included in this survey, using only the WISE \textit{W1} and \textit{W2} bands. This is especially useful for potential VLTI/MATISSE targets, allowing one to determine their observability and the possibility of resolving nuclear extended dusty structures.

This MIR AGN atlas holds a significant portion of local AGN of all optical classifications. It represents a statistically relevant sample suitable for AGN unification studies and interferometric follow-up. The need for such interferometric follow-up with VLTI/MATISSE is evinced by the fact that even our high-resolution 8m telescope data has a significant fraction of sources with unresolved MIR dust emission.  

%\acknowledgements
\section*{Acknowledgements}
L.B. acknowledges discussions with Dieter Lutz. 
D.A. acknowledges funding through the European Union’s Horizon 2020 and Innovation program 
under the Marie Sklodowska-Curie grant agreement no. 793499 (DUSTDEVILS).
M.S. acknowledges support by the Ministry of Education, Science and Technological Development of the Republic of Serbia through the contract no. 451-03-68/2020-14/200002 and the Science Fund of the Republic of Serbia, PROMIS 6060916, BOWIE. 
J.I. and L.B. acknowledge {\it Astronomers for Planet Earth} and other efforts to raise awareness of the impact of astronomy on global CO$_2$ emissions. J.I. and L.B. pledge to not travel via airplane to promote this paper.
This research has made use of the services of the ESO Science Archive Facility.

\facilities{VLT:Melipal (ISAAC), WISE}

\software{astropy \citep{astropycollaboration2018}, matplotlib (Hunter 2007), Scipy (Jones et al. 2001), emcee \citep{foreman-mackey2013}, and VIPE (https://github.com/danielasmus/vipe)}.
%This paper made extensive use of the python packages scipy, matplotlib, astropy\footnote{\cite{astropycollaboration2018} }, and emcee.

%------------------------------------------------------------------------
%-------------------------Bibliography-----------------------------
%------------------------------------------------------------------------

\bibliographystyle{aasjournal}
\bibliography{mybib}

\clearpage
\appendix

\section{Selected Epochs and Observing Conditions}

\startlongtable
\begin{deluxetable*}{llllrrrrl}
\tablehead{ \colhead{Target Name}  & \colhead{Obs. Date} & Obs. Program & \colhead{Filter} & \colhead{Exp. Time} &\colhead{PSF FWHM} & \colhead{Nuc. Flux} & \colhead{Ext. Flux} & \colhead{Flag} \\
\colhead{} & \colhead{} & \colhead{} & \colhead{}& \colhead{sec} & \colhead{"}& \colhead{[mJy]} & \colhead{[mJy]} & \colhead{}  }
\tablecaption{Observational Data for all Sources and Epochs}
\tabletypesize{\footnotesize}
\startdata
{\bf 3C 273} & 2013-07-07T00:46:53 & 290.B-5113(A) & $L^{\prime}$& 118 & 0.46 & $91.67 \pm 7.34$ & $69.71 \pm 30.55$ & 0\\
{\bf 3C 273} & 2013-07-07T00:51:00 & 290.B-5113(A) & $M_{\rm nb}$& 118 & 0.42 & $77.81 \pm 7.09$ & $74.14 \pm 22.28$ & 0\\
{\bf 3C 317} & 2013-07-03T02:16:39 & 290.B-5113(A) & $L^{\prime}$& 472 & 0.41 & $\leq 0.41$ & $\leq 0.33$ & 1\\
{\bf 3C 317} & 2013-07-03T02:29:07 & 290.B-5113(A) & $M_{\rm nb}$& 944 & 0.42 & $\leq 2.35$ & $\leq 2.05$ & 1\\
3C 321 & 2013-06-27T03:06:14 & 290.B-5113(A) & $L^{\prime}$& 118 & 0.47 & $0.37 \pm 0.06$ & $\leq 0.69$ & 2 \\
3C 321 & 2013-06-27T03:10:22 & 290.B-5113(A) & $M_{\rm nb}$& 236 & 0.36 & $\leq 4.95$ & $\leq 4.30$ & 3 \\
{\bf 3C 321} & 2013-07-04T02:02:53 & 290.B-5113(A) & $L^{\prime}$& 118 & 0.32 & $0.47 \pm 0.07$ & $\leq 0.66$ & 1\\
{\bf 3C 321} & 2013-07-04T02:07:02 & 290.B-5113(A) & $M_{\rm nb}$& 236 & 0.30 & $\leq 4.63$ & $\leq 3.83$ & 3\\
{\bf 3C 327} & 2013-07-04T02:32:55 & 290.B-5113(A) & $L^{\prime}$& 236 & 0.34 & $0.92 \pm 0.05$ & $\leq 0.42$ & 0\\
{\bf 3C 327} & 2013-07-04T02:39:44 & 290.B-5113(A) & $M_{\rm nb}$& 472 & 0.37 & $1.27 \pm 0.20$ & $\leq 2.69$ & 1\\
{\bf 3C 353} & 2013-07-02T05:08:55 & 290.B-5113(A) & $L^{\prime}$& 118 & 0.54 & $1.26 \pm 0.10$ & $\leq 0.67$ & 0\\
{\bf 3C 353} & 2013-07-02T05:13:02 & 290.B-5113(A) & $M_{\rm nb}$& 236 & 0.52 & $\leq 4.94$ & $\leq 4.54$ & 3\\
{\bf 3C 403} & 2013-07-02T06:03:52 & 290.B-5113(A) & $L^{\prime}$& 118 & 0.54 & $5.00 \pm 0.19$ & $\leq 0.69$ & 0\\
{\bf 3C 403} & 2013-07-02T06:07:58 & 290.B-5113(A) & $M_{\rm nb}$& 236 & 0.52 & $5.21 \pm 0.43$ & $\leq 3.72$ & 0\\
3C 424 & 2013-06-27T08:52:04 & 290.B-5113(A) & $L^{\prime}$& 472 & 0.68 & $0.17 \pm 0.04$ & $\leq 0.32$ & 3 \\
3C 424 & 2013-06-27T09:04:25 & 290.B-5113(A) & $M_{\rm nb}$& 944 & 0.66 & $\leq 5.48$ & $\leq 5.27$ & 3 \\
{\bf 3C 424} & 2013-07-03T05:41:24 & 290.B-5113(A) & $L^{\prime}$& 472 & 0.41 & $\leq 0.43$ & $\leq 0.34$ & 1\\
{\bf 3C 424} & 2013-07-03T05:53:51 & 290.B-5113(A) & $M_{\rm nb}$& 944 & 0.42 & $\leq 2.16$ & $\leq 1.88$ & 3\\
{\bf Arp 220} & 2000-07-13T01:26:21 & 65.P-0519(A) & $L^{\prime}$& 720 & 0.46 & $\leq 0.29$ & $12.41 \pm 0.49$ & 0\\
{\bf Arp 220} & 2000-07-13T01:49:23 & 65.P-0519(A) & $M_{\rm nb}$& 2340 & 0.56 & $\leq 1.16$ & $6.13 \pm 0.50$ & 0\\
Arp 220 & 2001-06-11T03:33:02 & 67.B-0332(A) & $L^{\prime}$& 90 & 0.64 & $\leq 3.91$ & $9.96 \pm 1.21$ & 2 \\
Arp 220 & 2001-06-11T03:38:27 & 67.B-0332(A) & $L^{\prime}$& 600 & 0.64 & $3.95 \pm 0.44$ & $4.28 \pm 0.92$ & 0 \\
Arp 220 & 2001-06-11T03:54:05 & 67.B-0332(A) & $M_{\rm nb}$& 120 & 0.56 & $15.24 \pm 1.69$ & $\leq 26.05$ & 3 \\
Arp 220 & 2001-06-12T03:07:40 & 67.B-0332(A) & $L^{\prime}$& 570 & 0.64 & $3.75 \pm 0.30$ & $\leq 1.55$ & 0 \\
Arp 220 & 2001-08-11T23:16:52 & 67.B-0332(A) & $L^{\prime}$& 600 & 0.45 & $\leq 0.52$ & $7.06 \pm 0.41$ & 3 \\
Arp 220 & 2001-08-11T23:36:03 & 67.B-0332(A) & $L^{\prime}$& 600 & 0.45 & $\leq 0.36$ & $12.43 \pm 0.52$ & 0 \\
Arp 220 & 2001-08-11T23:51:35 & 67.B-0332(A) & $M_{\rm nb}$& 480 & 0.45 & $\leq 2.99$ & $\leq 2.55$ & 3 \\
Arp 220 & 2001-08-12T00:04:32 & 67.B-0332(A) & $M_{\rm nb}$& 480 & 0.45 & $2.40 \pm 0.33$ & $\leq 3.05$ & 0 \\
{\bf CGCG381-051} & 2003-06-20T10:04:01 & 71.B-0379(A) & $L^{\prime}$& 600 & 0.61 & $\leq 0.29$ & $\leq 0.25$ & 1\\
Cen A & 2000-07-13T00:50:45 & 65.P-0519(A) & $M_{\rm nb}$& 720 & 0.56 & $224.41 \pm 13.50$ & $182.84 \pm 51.80$ & 0 \\
{\bf Cen A} & 2010-04-09T05:05:29 & 085.B-0639(A) & $L^{\prime}$& 118 & 0.36 & $261.69 \pm 19.58$ & $171.32 \pm 74.21$ & 0\\
Cen A & 2013-07-03T02:06:00 & 290.B-5113(A) & $L^{\prime}$& 118 & 0.41 & $263.97 \pm 16.50$ & $\leq 0.82$ & 0 \\
{\bf Cen A} & 2013-07-03T02:10:07 & 290.B-5113(A) & $M_{\rm nb}$& 118 & 0.42 & $305.82 \pm 18.28$ & $214.63 \pm 82.72$ & 0\\
{\bf Circinus} & 2000-06-22T01:01:03 & 65.P-0519(A) & $M_{\rm nb}$& 1200 & 0.50 & $676.41 \pm 44.58$ & $594.97 \pm 211.14$ & 0\\
{\bf Circinus} & 2001-06-09T01:23:22 & 67.B-0332(A) & $L^{\prime}$& 600 & 0.63 & $458.16 \pm 39.18$ & $345.07 \pm 158.43$ & 0\\
Circinus & 2001-06-09T01:38:47 & 67.B-0332(A) & $M_{\rm nb}$& 480 & 0.56 & $811.51 \pm 109.62$ & $\leq 4.49$ & 0 \\
{\bf ESO 103-35} & 2013-06-26T07:59:58 & 290.B-5113(A) & $L^{\prime}$& 118 & 0.78 & $18.06 \pm 1.97$ & $22.66 \pm 6.85$ & 0\\
{\bf ESO 103-35} & 2013-06-26T08:04:05 & 290.B-5113(A) & $M_{\rm nb}$& 236 & 0.72 & $40.67 \pm 4.73$ & $\leq 4.06$ & 0\\
{\bf ESO 138-1} & 2013-06-27T03:21:47 & 290.B-5113(A) & $L^{\prime}$& 118 & 0.47 & $50.90 \pm 3.45$ & $51.66 \pm 16.89$ & 0\\
ESO 138-1 & 2013-06-27T03:25:53 & 290.B-5113(A) & $M_{\rm nb}$& 118 & 0.36 & $30.43 \pm 6.20$ & $82.40 \pm 21.42$ & 0 \\
ESO 138-1 & 2013-07-03T04:38:50 & 290.B-5113(A) & $L^{\prime}$& 118 & 0.41 & $64.62 \pm 3.23$ & $42.92 \pm 12.28$ & 0 \\
{\bf ESO 138-1} & 2013-07-03T04:43:01 & 290.B-5113(A) & $M_{\rm nb}$& 118 & 0.42 & $94.10 \pm 4.80$ & $39.54 \pm 16.81$ & 0\\
{\bf ESO 141-55} & 2013-06-26T08:25:56 & 290.B-5113(A) & $L^{\prime}$& 118 & 0.78 & $46.85 \pm 2.52$ & $37.00 \pm 9.02$ & 0\\
{\bf ESO 141-55} & 2013-06-26T08:30:05 & 290.B-5113(A) & $M_{\rm nb}$& 236 & 0.72 & $53.51 \pm 4.11$ & $29.64 \pm 13.94$ & 0\\
{\bf ESO 286-19} & 2013-06-26T09:46:44 & 290.B-5113(A) & $L^{\prime}$& 118 & 0.67 & $5.23 \pm 0.33$ & $\leq 0.57$ & 0\\
{\bf ESO 286-19} & 2013-06-26T09:50:55 & 290.B-5113(A) & $M_{\rm nb}$& 236 & 0.65 & $16.33 \pm 0.61$ & $\leq 4.03$ & 0\\
{\bf ESO 323-32} & 2013-07-03T01:01:33 & 290.B-5113(A) & $L^{\prime}$& 118 & 0.53 & $3.47 \pm 0.12$ & $\leq 0.64$ & 0\\
{\bf ESO 323-32} & 2013-07-03T01:05:43 & 290.B-5113(A) & $M_{\rm nb}$& 236 & 0.43 & $2.04 \pm 0.35$ & $\leq 3.82$ & 3\\
{\bf ESO 323-77} & 2013-07-04T01:15:28 & 290.B-5113(A) & $L^{\prime}$& 118 & 0.32 & $159.61 \pm 9.27$ & $\leq 0.65$ & 0\\
{\bf ESO 323-77} & 2013-07-04T01:19:36 & 290.B-5113(A) & $M_{\rm nb}$& 118 & 0.30 & $119.07 \pm 10.12$ & $114.69 \pm 39.73$ & 0\\
ESO 506-27 & 2013-06-27T02:41:19 & 290.B-5113(A) & $L^{\prime}$& 118 & 0.47 & $7.89 \pm 1.44$ & $\leq 0.62$ & 0 \\
ESO 506-27 & 2013-06-27T02:45:27 & 290.B-5113(A) & $M_{\rm nb}$& 236 & 0.36 & $\leq 4.13$ & $14.73 \pm 2.79$ & 0 \\
{\bf ESO 506-27} & 2013-07-06T01:47:44 & 290.B-5113(A) & $L^{\prime}$& 118 & 0.55 & $10.56 \pm 0.87$ & $8.95 \pm 3.78$ & 0\\
{\bf ESO 506-27} & 2013-07-06T01:51:48 & 290.B-5113(A) & $M_{\rm nb}$& 236 & 0.64 & $23.32 \pm 0.89$ & $\leq 4.32$ & 0\\
{\bf ESO 511-30} & 2013-07-06T02:10:20 & 290.B-5113(A) & $L^{\prime}$& 118 & 0.73 & $14.23 \pm 0.57$ & $\leq 0.59$ & 0\\
{\bf ESO 511-30} & 2013-07-06T02:14:28 & 290.B-5113(A) & $M_{\rm nb}$& 236 & 0.74 & $13.65 \pm 0.66$ & $\leq 4.07$ & 0\\
{\bf Fairall 9} & 2003-08-17T06:01:03 & 71.B-0379(A) & $M_{\rm nb}$& 1200 & 0.53 & $25.74 \pm 2.99$ & $25.17 \pm 11.49$ & 0\\
{\bf Fairall 49} & 2013-06-26T07:48:34 & 290.B-5113(A) & $L^{\prime}$& 118 & 0.78 & $52.25 \pm 5.24$ & $75.45 \pm 21.23$ & 0\\
{\bf Fairall 49} & 2013-06-26T07:52:45 & 290.B-5113(A) & $M_{\rm nb}$& 118 & 0.72 & $62.60 \pm 7.74$ & $87.04 \pm 26.79$ & 0\\
{\bf Fairall 51} & 2013-06-26T08:14:21 & 290.B-5113(A) & $L^{\prime}$& 118 & 0.78 & $36.03 \pm 4.05$ & $53.01 \pm 15.93$ & 0\\
{\bf Fairall 51} & 2013-06-26T08:18:27 & 290.B-5113(A) & $M_{\rm nb}$& 118 & 0.72 & $49.31 \pm 5.57$ & $38.76 \pm 16.41$ & 0\\
IC 3639 & 2003-07-05T01:00:12 & 70.B-0393(B) & $L^{\prime}$& 590 & 0.46 & $0.06 \pm 0.01$ & $107.99 \pm 15.03$ & 3 \\
IC 3639 & 2003-07-05T01:17:26 & 70.B-0393(B) & $M_{\rm nb}$& 1180 & 0.57 & $0.61 \pm 0.07$ & $\leq 145.29$ & 3 \\
{\bf IC 3639} & 2003-07-15T00:48:05 & 70.B-0393(B) & $L^{\prime}$& 600 & 0.47 & $15.38 \pm 0.66$ & $10.88 \pm 2.61$ & 0\\
{\bf IC 3639} & 2003-07-15T01:08:01 & 70.B-0393(B) & $M_{\rm nb}$& 1200 & 0.55 & $25.73 \pm 1.39$ & $9.14 \pm 4.30$ & 0\\
{\bf IC 4329A} & 2013-07-06T02:00:20 & 290.B-5113(A) & $L^{\prime}$& 118 & 0.73 & $186.10 \pm 7.07$ & $60.91 \pm 26.08$ & 0\\
{\bf IC 4329A} & 2013-07-06T02:04:28 & 290.B-5113(A) & $M_{\rm nb}$& 118 & 0.74 & $231.37 \pm 9.78$ & $\leq 5.83$ & 0\\
{\bf IC 4518W} & 2013-07-06T02:53:30 & 290.B-5113(A) & $L^{\prime}$& 118 & 0.73 & $18.81 \pm 1.41$ & $10.98 \pm 4.63$ & 1\\
{\bf IC 4518W} & 2013-07-06T02:57:40 & 290.B-5113(A) & $M_{\rm nb}$& 236 & 0.74 & $35.44 \pm 1.28$ & $\leq 4.34$ & 1\\
IC 5063 & 2001-06-24T10:33:18 & 67.B-0332(A) & $L^{\prime}$& 60 & 0.64 & $119.87 \pm 28.70$ & $231.57 \pm 72.47$ & 1 \\
IC 5063 & 2001-06-24T10:39:23 & 67.B-0332(A) & $L^{\prime}$& 600 & 0.64 & $259.46 \pm 20.78$ & $\leq 1.41$ & 0 \\
IC 5063 & 2001-07-19T04:23:41 & 67.B-0332(A) & $L^{\prime}$& 600 & 0.83 & $58.92 \pm 2.42$ & $18.26 \pm 8.55$ & 0 \\
{\bf IC 5063} & 2001-07-19T04:40:43 & 67.B-0332(A) & $L^{\prime}$& 600 & 0.83 & $57.60 \pm 2.18$ & $22.68 \pm 8.54$ & 1\\
IC 5063 & 2001-07-19T04:56:37 & 67.B-0332(A) & $M_{\rm nb}$& 480 & 0.53 & $49.81 \pm 6.29$ & $52.15 \pm 20.36$ & 1 \\
IC 5063 & 2001-08-17T04:32:31 & 67.B-0332(A) & $L^{\prime}$& 600 & 0.41 & $31.93 \pm 2.50$ & $30.11 \pm 12.11$ & 0 \\
IC 5063 & 2001-08-17T04:48:50 & 67.B-0332(A) & $M_{\rm nb}$& 480 & 0.40 & $55.95 \pm 6.96$ & $\leq 1.99$ & 0 \\
IC 5063 & 2003-07-14T05:36:14 & 71.B-0379(A) & $L^{\prime}$& 600 & 0.50 & $33.13 \pm 2.25$ & $31.70 \pm 10.57$ & 0 \\
{\bf IC 5063} & 2003-07-14T05:56:11 & 71.B-0379(A) & $M_{\rm nb}$& 1200 & 0.50 & $53.10 \pm 6.33$ & $\leq 1.24$ & 0\\
IC 5063 & 2004-10-31T00:06:54 & 074.B-0166(A) & $L^{\prime}$& 1534 & 0.45 & $16.59 \pm 1.92$ & $39.34 \pm 6.41$ & 0 \\
{\bf IC 5179} & 2000-08-13T05:14:52 & 65.P-0519(A) & $M_{\rm nb}$& 720 & 0.63 & $\leq 2.19$ & $\leq 1.89$ & 3\\
{\bf IRAS 13349+2438} & 2013-07-02T01:20:04 & 290.B-5113(A) & $L^{\prime}$& 118 & 0.54 & $133.40 \pm 11.12$ & $\leq 0.66$ & 0\\
{\bf IRAS 13349+2438} & 2013-07-02T01:24:12 & 290.B-5113(A) & $M_{\rm nb}$& 236 & 0.52 & $120.27 \pm 12.79$ & $110.77 \pm 44.54$ & 0\\
{\bf IRASF00198-7926} & 2003-08-17T05:40:25 & 71.B-0379(A) & $L^{\prime}$& 600 & 0.50 & $11.10 \pm 1.93$ & $22.90 \pm 5.89$ & 0\\
{\bf LEDA 170194} & 2013-07-07T00:25:15 & 290.B-5113(A) & $L^{\prime}$& 236 & 0.46 & $3.09 \pm 0.22$ & $\leq 0.44$ & 0\\
{\bf LEDA 170194} & 2013-07-07T00:32:02 & 290.B-5113(A) & $M_{\rm nb}$& 472 & 0.42 & $3.09 \pm 0.25$ & $\leq 2.60$ & 0\\
{\bf LHS2397A} & 2013-07-01T00:16:09 & 60.A-9021(A) & $L^{\prime}$& 118 & 0.79 & $30.11 \pm 1.37$ & $\leq 0.68$ & 0\\
{\bf LHS2397A} & 2013-07-01T00:20:16 & 60.A-9021(A) & $M_{\rm nb}$& 118 & 0.85 & $8.74 \pm 0.68$ & $\leq 5.44$ & 0\\
{\bf M87} & 2011-05-18T01:19:07 & 085.B-0639(A) & $L^{\prime}$& 118 & 0.37 & $6.41 \pm 0.24$ & $\leq 0.55$ & 0\\
{\bf M87} & 2011-05-18T01:23:26 & 085.B-0639(A) & $M_{\rm nb}$& 118 & 0.34 & $2.69 \pm 0.24$ & $\leq 4.26$ & 0\\
{\bf MCG+2-4-25} & 2000-11-06T02:56:37 & 65.P-0519(A) & $M_{\rm nb}$& 720 & 0.46 & $7.34 \pm 1.92$ & $\leq 1.88$ & 0\\
MCG-0-29-23 & 2003-06-19T02:04:38 & 70.B-0393(B) & $L^{\prime}$& 600 & 0.54 & $2.74 \pm 0.18$ & $9.36 \pm 0.81$ & 0 \\
MCG-0-29-23 & 2003-06-19T02:27:07 & 70.B-0393(B) & $M_{\rm nb}$& 690 & 0.63 & $3.07 \pm 0.30$ & $\leq 2.15$ & 0 \\
{\bf MCG-0-29-23} & 2003-07-15T23:22:13 & 70.B-0393(B) & $L^{\prime}$& 600 & 0.58 & $4.87 \pm 0.17$ & $13.48 \pm 1.43$ & 0\\
{\bf MCG-0-29-23} & 2003-07-15T23:42:01 & 70.B-0393(B) & $M_{\rm nb}$& 1200 & 0.49 & $6.00 \pm 0.48$ & $\leq 1.79$ & 0\\
{\bf MCG-3-34-64} & 2003-07-15T02:00:36 & 70.B-0393(B) & $L^{\prime}$& 600 & 0.47 & $23.62 \pm 1.68$ & $33.84 \pm 8.19$ & 0\\
{\bf MCG-3-34-64} & 2003-07-15T02:20:20 & 70.B-0393(B) & $M_{\rm nb}$& 1200 & 0.55 & $39.81 \pm 5.47$ & $58.19 \pm 20.87$ & 0\\
MCG-3-34-64 & 2013-07-01T00:39:39 & 290.B-5113(A) & $L^{\prime}$& 118 & 0.79 & $57.09 \pm 3.31$ & $\leq 0.61$ & 0 \\
MCG-3-34-64 & 2013-07-01T00:43:46 & 290.B-5113(A) & $M_{\rm nb}$& 236 & 0.85 & $102.12 \pm 6.80$ & $\leq 3.96$ & 0 \\
{\bf MCG-6-30-15} & 2013-07-04T01:50:12 & 290.B-5113(A) & $L^{\prime}$& 118 & 0.32 & $65.23 \pm 4.07$ & $\leq 0.57$ & 0\\
{\bf MCG-6-30-15} & 2013-07-04T01:54:20 & 290.B-5113(A) & $M_{\rm nb}$& 118 & 0.30 & $56.78 \pm 6.75$ & $54.46 \pm 24.78$ & 0\\
{\bf Mrk 331} & 2000-11-06T23:39:11 & 65.P-0519(A) & $M_{\rm nb}$& 480 & 0.41 & $\leq 2.42$ & $3.67 \pm 1.32$ & 3\\
{\bf Mrk 463} & 2003-07-20T01:09:13 & 70.B-0393(B) & $L^{\prime}$& 600 & 0.58 & $68.53 \pm 9.56$ & $102.17 \pm 45.35$ & 0\\
{\bf Mrk 463} & 2003-07-20T01:29:08 & 70.B-0393(B) & $M_{\rm nb}$& 1200 & 0.49 & $35.55 \pm 7.96$ & $125.28 \pm 27.46$ & 0\\
{\bf Mrk 509} & 2013-07-03T05:25:31 & 290.B-5113(A) & $L^{\prime}$& 118 & 0.41 & $108.47 \pm 4.81$ & $\leq 0.56$ & 0\\
{\bf Mrk 509} & 2013-07-03T05:29:39 & 290.B-5113(A) & $M_{\rm nb}$& 236 & 0.42 & $117.10 \pm 5.40$ & $\leq 3.49$ & 0\\
{\bf Mrk 841} & 2013-07-07T02:37:13 & 290.B-5113(A) & $L^{\prime}$& 118 & 0.46 & $14.58 \pm 1.48$ & $9.97 \pm 4.44$ & 0\\
{\bf Mrk 841} & 2013-07-07T02:41:20 & 290.B-5113(A) & $M_{\rm nb}$& 236 & 0.42 & $23.03 \pm 1.46$ & $\leq 3.56$ & 0\\
Mrk 897 & 2003-06-22T09:25:23 & 71.B-0379(A) & $L^{\prime}$& 480 & 0.69 & $1.74 \pm 0.10$ & $2.79 \pm 0.51$ & 1 \\
{\bf Mrk 897} & 2003-06-22T09:44:18 & 71.B-0379(A) & $L^{\prime}$& 180 & 0.69 & $7.42 \pm 0.47$ & $\leq 0.50$ & 1\\
Mrk 897 & 2003-06-22T09:51:58 & 71.B-0379(A) & $M_{\rm nb}$& 450 & 0.66 & $\leq 6.16$ & $\leq 5.77$ & 1 \\
{\bf Mrk 897} & 2003-06-22T10:11:47 & 71.B-0379(A) & $M_{\rm nb}$& 120 & 0.66 & $\leq 9.10$ & $\leq 8.52$ & 3\\
{\bf NGC 63} & 2000-11-07T00:06:36 & 65.P-0519(A) & $M_{\rm nb}$& 2880 & 0.41 & $0.71 \pm 0.10$ & $\leq 0.82$ & 2\\
{\bf NGC 253} & 2000-09-17T07:53:06 & 65.P-0519(A) & $L^{\prime}$& 144 & 0.57 & $78.77 \pm 2.68$ & $82.78 \pm 10.98$ & 0\\
NGC 253 & 2000-09-17T07:58:33 & 65.P-0519(A) & $M_{\rm nb}$& 288 & 0.54 & $119.08 \pm 4.88$ & $88.08 \pm 41.32$ & 0 \\
{\bf NGC 253} & 2000-09-17T08:09:44 & 65.P-0519(A) & $M_{\rm nb}$& 288 & 0.54 & $121.74 \pm 4.86$ & $88.80 \pm 27.73$ & 0\\
NGC 253 & 2003-06-23T09:58:42 & 71.B-0404(A) & $L^{\prime}$& 330 & 0.69 & $97.21 \pm 5.42$ & $44.23 \pm 12.19$ & 0 \\
NGC 253 & 2003-06-23T10:12:21 & 71.B-0404(A) & $L^{\prime}$& 330 & 0.69 & $95.78 \pm 5.13$ & $48.67 \pm 12.02$ & 1 \\
NGC 253 & 2003-06-23T10:24:29 & 71.B-0404(A) & $M_{\rm nb}$& 90 & 0.35 & $72.42 \pm 5.92$ & $98.43 \pm 25.32$ & 1 \\
NGC 253 & 2003-06-23T10:31:37 & 71.B-0404(A) & $M_{\rm nb}$& 300 & 0.35 & $93.98 \pm 7.70$ & $102.75 \pm 31.10$ & 0 \\
NGC 253 & 2003-06-23T10:45:17 & 71.B-0404(A) & $M_{\rm nb}$& 300 & 0.35 & $68.31 \pm 6.68$ & $117.05 \pm 21.71$ & 0 \\
{\bf NGC 424} & 2003-08-17T06:54:30 & 71.B-0379(A) & $L^{\prime}$& 600 & 0.50 & $203.46 \pm 7.91$ & $\leq 0.27$ & 0\\
{\bf NGC 424} & 2003-08-17T07:14:03 & 71.B-0379(A) & $M_{\rm nb}$& 1200 & 0.53 & $225.63 \pm 8.14$ & $82.03 \pm 33.21$ & 0\\
{\bf NGC 520} & 2000-11-06T03:19:58 & 65.P-0519(A) & $M_{\rm nb}$& 720 & 0.46 & $\leq 2.32$ & $8.50 \pm 0.62$ & 2\\
{\bf NGC 660} & 2000-11-05T04:51:21 & 65.P-0519(A) & $M_{\rm nb}$& 960 & 0.47 & $\leq 2.09$ & $14.42 \pm 0.81$ & 2\\
{\bf NGC 986} & 2000-09-20T04:22:26 & 65.P-0519(A) & $M_{\rm nb}$& 864 & 0.91 & $4.72 \pm 0.53$ & $\leq 3.73$ & 3\\
NGC 1008 & 2001-07-19T09:30:52 & 67.B-0332(A) & $L^{\prime}$& 120 & 0.58 & $\leq 0.59$ & $\leq 0.52$ & 3 \\
{\bf NGC 1008} & 2001-08-15T06:32:11 & 67.B-0332(A) & $L^{\prime}$& 600 & 0.60 & $\leq 0.13$ & $1.69 \pm 0.19$ & 2\\
{\bf NGC 1008} & 2001-08-15T06:48:27 & 67.B-0332(A) & $M_{\rm nb}$& 480 & 0.59 & $\leq 2.94$ & $\leq 2.70$ & 3\\
NGC 1068 & 2001-08-15T07:04:46 & 67.B-0332(A) & $L^{\prime}$& 600 & 0.60 & $0.17 \pm 0.03$ & $\leq 0.23$ & 2 \\
{\bf NGC 1068} & 2001-08-15T07:25:13 & 67.B-0332(A) & $L^{\prime}$& 600 & 0.60 & $1940.88 \pm 241.52$ & $1876.65 \pm 822.79$ & 0\\
{\bf NGC 1068} & 2001-08-15T07:41:27 & 67.B-0332(A) & $M_{\rm nb}$& 480 & 0.59 & $2687.53 \pm 374.86$ & $4267.02 \pm 1471.36$ & 0\\
NGC 1068 & 2004-10-31T03:10:43 & 074.B-0166(A) & $L^{\prime}$& 1534 & 0.45 & $886.54 \pm 124.82$ & $2272.66 \pm 387.63$ & 0 \\
{\bf NGC 1097} & 2003-08-17T08:04:24 & 71.B-0379(A) & $L^{\prime}$& 600 & 0.49 & $5.55 \pm 0.24$ & $8.30 \pm 1.40$ & 0\\
{\bf NGC 1097} & 2003-08-17T08:24:12 & 71.B-0379(A) & $M_{\rm nb}$& 1200 & 0.53 & $5.63 \pm 0.51$ & $\leq 1.84$ & 0\\
NGC 1097 & 2004-10-30T03:50:42 & 074.B-0166(A) & $L^{\prime}$& 1534 & 0.57 & $6.93 \pm 0.28$ & $8.08 \pm 1.15$ & 0 \\
NGC 1097 & 2004-10-31T03:52:44 & 074.B-0166(A) & $L^{\prime}$& 1534 & 0.45 & $4.40 \pm 0.23$ & $8.91 \pm 1.17$ & 0 \\
{\bf NGC 1125} & 2003-08-17T09:12:29 & 71.B-0379(A) & $L^{\prime}$& 600 & 0.43 & $3.51 \pm 0.23$ & $5.95 \pm 1.25$ & 0\\
{\bf NGC 1125} & 2003-08-17T09:32:10 & 71.B-0379(A) & $M_{\rm nb}$& 1200 & 0.53 & $6.76 \pm 0.46$ & $\leq 1.49$ & 0\\
{\bf NGC 1368} & 2004-10-30T04:53:33 & 074.B-0166(A) & $L^{\prime}$& 1534 & 0.57 & $0.67 \pm 0.06$ & $4.34 \pm 0.34$ & 0\\
{\bf NGC 1365} & 2003-12-02T00:31:01 & 072.B-0397(A) & $L^{\prime}$& 590 & 0.65 & $196.44 \pm 6.58$ & $\leq 0.40$ & 0\\
{\bf NGC 1365} & 2003-12-02T00:47:54 & 072.B-0397(A) & $M_{\rm nb}$& 944 & 0.64 & $152.32 \pm 6.74$ & $\leq 1.58$ & 0\\
{\bf NGC 1386} & 2000-09-17T09:45:17 & 65.P-0519(A) & $M_{\rm nb}$& 576 & 0.54 & $29.31 \pm 2.29$ & $18.98 \pm 8.69$ & 0\\
{\bf NGC 1386} & 2010-08-25T07:51:23 & 085.B-0639(A) & $L^{\prime}$& 118 & 0.50 & $18.65 \pm 0.98$ & $21.13 \pm 4.93$ & 0\\
{\bf NGC 1511} & 2000-09-17T08:49:45 & 65.P-0519(A) & $M_{\rm nb}$& 1152 & 0.54 & $2.44 \pm 0.17$ & $\leq 1.41$ & 1\\
{\bf NGC 1566} & 2004-10-30T06:03:36 & 074.B-0166(A) & $L^{\prime}$& 1534 & 0.57 & $10.55 \pm 0.38$ & $11.02 \pm 1.83$ & 0\\
{\bf NGC 1614} & 2000-09-17T08:25:55 & 65.P-0519(A) & $M_{\rm nb}$& 576 & 0.54 & $7.13 \pm 0.65$ & $15.58 \pm 2.04$ & 0\\
{\bf NGC 1667} & 2004-10-30T07:07:28 & 074.B-0166(A) & $L^{\prime}$& 590 & 0.57 & $1.48 \pm 0.08$ & $2.03 \pm 0.39$ & 0\\
NGC 1667 & 2004-10-30T08:14:46 & 074.B-0166(A) & $L^{\prime}$& 944 & 0.57 & $1.47 \pm 0.07$ & $2.87 \pm 0.37$ & 0 \\
{\bf NGC 1808} & 2000-09-15T08:37:55 & 65.P-0519(A) & $L^{\prime}$& 536 & 0.44 & $6.29 \pm 0.51$ & $15.65 \pm 2.15$ & 0\\
NGC 1808 & 2000-09-15T08:50:51 & 65.P-0519(A) & $L^{\prime}$& 144 & 0.44 & $6.40 \pm 0.54$ & $15.28 \pm 2.15$ & 0 \\
NGC 1808 & 2000-09-15T08:56:33 & 65.P-0519(A) & $L^{\prime}$& 144 & 0.44 & $10.09 \pm 0.64$ & $22.79 \pm 2.57$ & 0 \\
NGC 1808 & 2000-09-15T09:08:10 & 65.P-0519(A) & $M_{\rm nb}$& 288 & 0.47 & $9.90 \pm 0.73$ & $10.01 \pm 2.60$ & 0 \\
NGC 1808 & 2003-12-02T02:39:44 & 072.B-0397(A) & $L^{\prime}$& 590 & 0.65 & $19.13 \pm 0.90$ & $23.88 \pm 3.04$ & 0 \\
{\bf NGC 1808} & 2003-12-02T02:56:17 & 072.B-0397(A) & $M_{\rm nb}$& 944 & 0.64 & $10.46 \pm 0.56$ & $8.94 \pm 2.04$ & 0\\
{\bf NGC 3125} & 2010-06-06T23:43:17 & 085.B-0639(A) & $L^{\prime}$& 118 & 0.33 & $\leq 0.75$ & $\leq 0.58$ & 3\\
{\bf NGC 3125} & 2010-06-06T23:47:05 & 085.B-0639(A) & $M_{\rm nb}$& 354 & 0.37 & $\leq 3.47$ & $\leq 3.00$ & 3\\
{\bf NGC 3281} & 2001-06-02T22:52:59 & 65.P-0519(A) & $M_{\rm nb}$& 240 & 0.57 & $116.64 \pm 5.26$ & $\leq 4.67$ & 0\\
{\bf NGC 3660} & 2003-06-19T00:26:28 & 70.B-0393(B) & $L^{\prime}$& 600 & 0.54 & $2.99 \pm 0.13$ & $\leq 0.24$ & 0\\
{\bf NGC 3660} & 2003-06-19T00:49:09 & 70.B-0393(B) & $M_{\rm nb}$& 1200 & 0.63 & $1.46 \pm 0.15$ & $\leq 1.33$ & 0\\
{\bf NGC 4038/9} & 2001-06-03T01:51:39 & 65.P-0519(A) & $M_{\rm nb}$& 2130 & 0.57 & $2.43 \pm 0.26$ & $\leq 1.79$ & 3\\
{\bf NGC 4074} & 2013-06-30T23:31:45 & 290.B-5113(A) & $L^{\prime}$& 118 & 0.79 & $12.53 \pm 0.84$ & $\leq 0.65$ & 0\\
{\bf NGC 4074} & 2013-06-30T23:35:56 & 290.B-5113(A) & $M_{\rm nb}$& 236 & 0.85 & $10.04 \pm 0.72$ & $\leq 4.07$ & 0\\
NGC 4235 & 2011-05-18T00:26:00 & 085.B-0639(A) & $L^{\prime}$& 118 & 1.58 & $17.23 \pm 0.94$ & $\leq 0.61$ & 0 \\
{\bf NGC 4235} & 2011-05-18T00:31:15 & 085.B-0639(A) & $M_{\rm nb}$& 118 & 0.47 & $5.71 \pm 1.15$ & $8.66 \pm 2.07$ & 0\\
{\bf NGC 4235} & 2011-05-18T00:34:53 & 085.B-0639(A) & $L^{\prime}$& 118 & 0.47 & $14.49 \pm 0.97$ & $11.90 \pm 3.98$ & 0\\
{\bf NGC 4253} & 2013-06-30T23:17:03 & 290.B-5113(A) & $L^{\prime}$& 118 & 0.79 & $23.99 \pm 6.38$ & $32.16 \pm 15.01$ & 0\\
{\bf NGC 4253} & 2013-06-30T23:21:10 & 290.B-5113(A) & $M_{\rm nb}$& 236 & 0.85 & $39.53 \pm 2.28$ & $\leq 4.23$ & 0\\
{\bf NGC 4261} & 2011-05-18T00:55:02 & 085.B-0639(A) & $L^{\prime}$& 118 & 0.32 & $0.64 \pm 0.06$ & $5.11 \pm 0.50$ & 0\\
{\bf NGC 4261} & 2011-05-18T00:59:17 & 085.B-0639(A) & $M_{\rm nb}$& 118 & 0.31 & $\leq 5.83$ & $\leq 4.99$ & 3\\
{\bf NGC 4278} & 2013-07-05T23:18:17 & 290.B-5113(A) & $L^{\prime}$& 236 & 0.50 & $\leq 0.57$ & $10.39 \pm 0.49$ & 0\\
{\bf NGC 4278} & 2013-07-05T23:25:00 & 290.B-5113(A) & $M_{\rm nb}$& 708 & 0.45 & $\leq 2.65$ & $\leq 2.37$ & 3\\
NGC 4303 & 2003-05-29T00:50:21 & 71.B-0404(A) & $L^{\prime}$& 590 & 1.41 & $4.01 \pm 0.19$ & $4.37 \pm 0.63$ & 0 \\
{\bf NGC 4303} & 2003-05-29T01:08:30 & 71.B-0404(A) & $M_{\rm nb}$& 600 & 1.37 & $\leq 1.89$ & $\leq 1.99$ & 3\\
{\bf NGC 4303} & 2013-07-06T00:22:55 & 290.B-5113(A) & $L^{\prime}$& 354 & 0.55 & $1.65 \pm 0.13$ & $5.87 \pm 0.67$ & 0\\
{\bf NGC 4303} & 2013-07-06T00:32:36 & 290.B-5113(A) & $M_{\rm nb}$& 472 & 0.64 & $\leq 4.07$ & $11.49 \pm 2.01$ & 1\\
NGC 4374 & 2013-06-27T01:30:13 & 290.B-5113(A) & $L^{\prime}$& 472 & 0.47 & $0.15 \pm 0.05$ & $7.14 \pm 0.39$ & 0 \\
{\bf NGC 4374} & 2013-06-27T01:42:37 & 290.B-5113(A) & $M_{\rm nb}$& 944 & 0.36 & $\leq 2.21$ & $\leq 1.92$ & 3\\
{\bf NGC 4374} & 2013-07-04T00:33:33 & 290.B-5113(A) & $L^{\prime}$& 472 & 0.35 & $0.33 \pm 0.04$ & $7.73 \pm 0.39$ & 0\\
{\bf NGC 4388} & 2013-06-30T23:45:37 & 290.B-5113(A) & $L^{\prime}$& 354 & 0.79 & $34.11 \pm 1.23$ & $\leq 0.41$ & 0\\
{\bf NGC 4388} & 2013-06-30T23:55:22 & 290.B-5113(A) & $M_{\rm nb}$& 708 & 0.85 & $44.11 \pm 2.10$ & $\leq 2.39$ & 0\\
{\bf NGC 4418} & 2013-07-02T01:07:01 & 290.B-5113(A) & $L^{\prime}$& 118 & 0.54 & $2.03 \pm 0.20$ & $1.68 \pm 0.78$ & 0\\
{\bf NGC 4418} & 2013-07-02T01:11:04 & 290.B-5113(A) & $M_{\rm nb}$& 236 & 0.52 & $6.26 \pm 0.61$ & $\leq 4.17$ & 0\\
{\bf NGC 4438} & 2013-03-05T08:53:11 & 290.B-5113(A) & $L^{\prime}$& 118 & 0.54 & $4.04 \pm 0.26$ & $12.56 \pm 1.45$ & 0\\
{\bf NGC 4438} & 2013-03-05T08:58:35 & 290.B-5113(A) & $M_{\rm nb}$& 236 & 0.51 & $2.20 \pm 0.42$ & $\leq 4.84$ & 3\\
{\bf NGC 4457} & 2013-07-07T00:06:51 & 290.B-5113(A) & $L^{\prime}$& 118 & 0.46 & $3.00 \pm 0.23$ & $15.51 \pm 1.23$ & 0\\
{\bf NGC 4457} & 2013-07-07T00:13:54 & 290.B-5113(A) & $M_{\rm nb}$& 236 & 0.42 & $\leq 4.30$ & $\leq 3.62$ & 3\\
{\bf NGC 4472} & 2013-07-03T23:31:28 & 290.B-5113(A) & $L^{\prime}$& 472 & 0.35 & $\leq 0.41$ & $4.97 \pm 0.32$ & 2\\
{\bf NGC 4472} & 2013-07-03T23:43:58 & 290.B-5113(A) & $M_{\rm nb}$& 944 & 0.32 & $\leq 2.09$ & $\leq 1.79$ & 3\\
{\bf NGC 4501} & 2003-07-16T00:30:23 & 70.B-0393(B) & $L^{\prime}$& 600 & 0.58 & $1.50 \pm 0.16$ & $13.68 \pm 1.17$ & 0\\
{\bf NGC 4501} & 2003-07-16T00:50:14 & 70.B-0393(B) & $M_{\rm nb}$& 660 & 0.49 & $\leq 0.76$ & $\leq 2.26$ & 3\\
{\bf NGC 4507} & 2013-06-27T02:25:01 & 290.B-5113(A) & $L^{\prime}$& 118 & 0.47 & $59.67 \pm 5.96$ & $51.83 \pm 24.55$ & 0\\
{\bf NGC 4507} & 2013-06-27T02:29:08 & 290.B-5113(A) & $M_{\rm nb}$& 118 & 0.36 & $62.07 \pm 12.71$ & $\leq 5.72$ & 0\\
NGC 4507 & 2013-07-06T01:37:43 & 290.B-5113(A) & $L^{\prime}$& 118 & 0.55 & $62.47 \pm 6.80$ & $57.66 \pm 23.97$ & 0 \\
NGC 4507 & 2013-07-06T01:41:52 & 290.B-5113(A) & $M_{\rm nb}$& 118 & 0.64 & $88.16 \pm 8.83$ & $60.73 \pm 28.93$ & 0 \\
{\bf NGC 4579} & 2011-05-18T01:43:50 & 085.B-0639(A) & $L^{\prime}$& 118 & 0.31 & $16.91 \pm 0.66$ & $17.10 \pm 4.35$ & 0\\
{\bf NGC 4579} & 2011-05-18T01:48:04 & 085.B-0639(A) & $M_{\rm nb}$& 118 & 0.36 & $14.99 \pm 1.02$ & $7.20 \pm 3.27$ & 0\\
{\bf NGC 4593} & 2013-07-11T00:30:23 & 290.B-5113(A) & $L^{\prime}$& 118 & 0.39 & $32.31 \pm 2.40$ & $22.51 \pm 10.48$ & 0\\
{\bf NGC 4593} & 2013-07-11T00:34:33 & 290.B-5113(A) & $M_{\rm nb}$& 236 & 0.40 & $36.16 \pm 2.54$ & $25.66 \pm 7.17$ & 0\\
{\bf NGC 4594} & 2013-07-07T01:20:48 & 290.B-5113(A) & $L^{\prime}$& 118 & 0.46 & $2.74 \pm 0.32$ & $38.94 \pm 1.97$ & 0\\
{\bf NGC 4594} & 2013-07-07T01:24:53 & 290.B-5113(A) & $M_{\rm nb}$& 236 & 0.42 & $\leq 4.42$ & $\leq 3.73$ & 2\\
{\bf NGC 4746} & 2013-07-06T00:59:50 & 290.B-5113(A) & $L^{\prime}$& 472 & 0.55 & $\leq 0.42$ & $0.56 \pm 0.25$ & 1\\
{\bf NGC 4746} & 2013-07-06T01:12:15 & 290.B-5113(A) & $M_{\rm nb}$& 944 & 0.64 & $\leq 4.66$ & $\leq 4.34$ & 3\\
{\bf NGC 4785} & 2013-07-03T01:15:52 & 290.B-5113(A) & $L^{\prime}$& 472 & 0.53 & $1.11 \pm 0.08$ & $3.99 \pm 0.56$ & 0\\
{\bf NGC 4785} & 2013-07-03T01:28:23 & 290.B-5113(A) & $M_{\rm nb}$& 944 & 0.42 & $\leq 2.17$ & $\leq 1.90$ & 2\\
NGC 4941 & 2003-07-16T01:59:58 & 70.B-0393(B) & $M_{\rm nb}$& 1200 & 0.49 & $1.85 \pm 0.19$ & $\leq 1.91$ & 0 \\
{\bf NGC 4941} & 2003-07-16T02:39:22 & 70.B-0393(B) & $M_{\rm nb}$& 420 & 0.49 & $2.01 \pm 0.34$ & $\leq 3.20$ & 0\\
{\bf NGC 4941} & 2013-07-07T01:33:05 & 290.B-5113(A) & $L^{\prime}$& 118 & 0.46 & $4.54 \pm 0.29$ & $4.20 \pm 1.36$ & 0\\
NGC 4941 & 2013-07-07T01:37:12 & 290.B-5113(A) & $M_{\rm nb}$& 236 & 0.42 & $2.15 \pm 0.35$ & $\leq 3.79$ & 0 \\
NGC 4945 & 2000-06-21T23:22:29 & 65.P-0519(A) & $L^{\prime}$& 120 & 0.50 & $\leq 2.05$ & $65.02 \pm 4.09$ & 0 \\
{\bf NGC 4945} & 2000-06-21T23:26:55 & 65.P-0519(A) & $M_{\rm nb}$& 480 & 0.60 & $\leq 2.58$ & $17.69 \pm 1.47$ & 3\\
NGC 4945 & 2000-06-21T23:38:17 & 65.P-0519(A) & $L^{\prime}$& 120 & 0.50 & $\leq 1.97$ & $67.52 \pm 4.19$ & 0 \\
NGC 4945 & 2000-06-21T23:42:43 & 65.P-0519(A) & $M_{\rm nb}$& 480 & 0.60 & $\leq 2.91$ & $20.34 \pm 1.59$ & 0 \\
{\bf NGC 4945} & 2001-06-08T04:20:48 & 67.B-0332(A) & $L^{\prime}$& 600 & 0.65 & $\leq 2.13$ & $54.70 \pm 3.35$ & 0\\
NGC 4945 & 2001-06-08T04:36:20 & 67.B-0332(A) & $M_{\rm nb}$& 480 & 0.57 & $\leq 4.02$ & $10.93 \pm 2.05$ & 2 \\
NGC 5135 & 2003-07-15T02:59:34 & 70.B-0393(B) & $L^{\prime}$& 600 & 0.47 & $9.10 \pm 0.58$ & $12.31 \pm 4.22$ & 0 \\
NGC 5135 & 2003-07-15T03:19:33 & 70.B-0393(B) & $M_{\rm nb}$& 1200 & 0.55 & $15.16 \pm 2.40$ & $\leq 1.84$ & 0 \\
{\bf NGC 5135} & 2013-07-04T01:36:30 & 290.B-5113(A) & $L^{\prime}$& 118 & 0.32 & $12.63 \pm 0.70$ & $7.95 \pm 3.34$ & 0\\
{\bf NGC 5135} & 2013-07-04T01:40:36 & 290.B-5113(A) & $M_{\rm nb}$& 236 & 0.30 & $19.31 \pm 2.75$ & $\leq 3.34$ & 0\\
{\bf NGC 5252} & 2013-07-11T00:42:28 & 290.B-5113(A) & $L^{\prime}$& 118 & 0.39 & $16.17 \pm 1.20$ & $13.45 \pm 4.88$ & 0\\
{\bf NGC 5252} & 2013-07-11T00:46:33 & 290.B-5113(A) & $M_{\rm nb}$& 236 & 0.40 & $25.19 \pm 1.31$ & $\leq 4.37$ & 0\\
NGC 5252 & 2013-07-24T00:48:30 & 290.B-5113(A) & $L^{\prime}$& 118 & 0.39 & $\leq 0.94$ & $23.62 \pm 1.45$ & 0 \\
NGC 5252 & 2013-07-24T00:52:35 & 290.B-5113(A) & $M_{\rm nb}$& 236 & 0.40 & $\leq 5.00$ & $18.04 \pm 2.50$ & 0 \\
{\bf NGC 5363} & 2013-07-07T01:46:06 & 290.B-5113(A) & $L^{\prime}$& 118 & 0.46 & $0.97 \pm 0.12$ & $8.36 \pm 0.67$ & 0\\
{\bf NGC 5363} & 2013-07-07T01:50:12 & 290.B-5113(A) & $M_{\rm nb}$& 236 & 0.42 & $\leq 4.72$ & $\leq 3.98$ & 3\\
{\bf NGC 5427} & 2013-07-02T01:34:40 & 290.B-5113(A) & $L^{\prime}$& 472 & 0.54 & $2.95 \pm 0.10$ & $\leq 0.31$ & 0\\
{\bf NGC 5427} & 2013-07-02T01:47:12 & 290.B-5113(A) & $M_{\rm nb}$& 944 & 0.52 & $2.09 \pm 0.26$ & $\leq 1.90$ & 2\\
{\bf NGC 5506} & 2003-07-13T23:33:47 & 71.B-0404(A) & $L^{\prime}$& 600 & 0.46 & $343.79 \pm 12.95$ & $124.02 \pm 50.48$ & 0\\
{\bf NGC 5506} & 2003-07-13T23:53:28 & 71.B-0404(A) & $M_{\rm nb}$& 60 & 0.57 & $359.44 \pm 26.41$ & $150.89 \pm 47.07$ & 0\\
{\bf NGC 5548} & 2013-07-24T00:22:19 & 290.B-5113(A) & $L^{\prime}$& 118 & 0.39 & $3.29 \pm 0.89$ & $59.13 \pm 3.82$ & 0\\
{\bf NGC 5548} & 2013-07-24T00:26:23 & 290.B-5113(A) & $M_{\rm nb}$& 236 & 0.40 & $\leq 4.99$ & $51.91 \pm 2.65$ & 0\\
{\bf NGC 5643} & 2013-07-04T02:16:55 & 290.B-5113(A) & $L^{\prime}$& 118 & 0.32 & $8.56 \pm 0.45$ & $10.99 \pm 2.42$ & 0\\
{\bf NGC 5643} & 2013-07-04T02:21:04 & 290.B-5113(A) & $M_{\rm nb}$& 236 & 0.30 & $14.11 \pm 1.55$ & $\leq 3.35$ & 0\\
{\bf NGC 5728} & 2013-07-06T02:39:55 & 290.B-5113(A) & $L^{\prime}$& 118 & 0.73 & $2.89 \pm 0.22$ & $2.93 \pm 1.12$ & 0\\
{\bf NGC 5728} & 2013-07-06T02:44:04 & 290.B-5113(A) & $M_{\rm nb}$& 236 & 0.74 & $2.76 \pm 0.33$ & $\leq 4.33$ & 0\\
{\bf NGC 5813} & 2013-07-07T01:58:27 & 290.B-5113(A) & $L^{\prime}$& 472 & 0.46 & $\leq 0.38$ & $6.57 \pm 0.34$ & 0\\
{\bf NGC 5813} & 2013-07-07T02:10:54 & 290.B-5113(A) & $M_{\rm nb}$& 944 & 0.42 & $\leq 2.06$ & $\leq 1.74$ & 1\\
NGC 5953 & 2001-06-10T02:46:05 & 67.B-0332(A) & $L^{\prime}$& 600 & 0.95 & $1.60 \pm 0.17$ & $2.71 \pm 0.59$ & 0 \\
NGC 5953 & 2001-06-10T03:02:24 & 67.B-0332(A) & $M_{\rm nb}$& 480 & 0.81 & $\leq 4.43$ & $\leq 3.91$ & 3 \\
NGC 5953 & 2003-05-31T05:14:11 & 70.B-0393(B) & $L^{\prime}$& 590 & 1.41 & $6.01 \pm 0.63$ & $\leq 0.26$ & 1 \\
NGC 5953 & 2003-05-31T05:32:20 & 70.B-0393(B) & $M_{\rm nb}$& 450 & 1.37 & $\leq 6.36$ & $\leq 6.69$ & 3 \\
NGC 5953 & 2003-06-01T05:31:44 & 70.B-0393(B) & $L^{\prime}$& 590 & 1.41 & $4.39 \pm 0.33$ & $3.67 \pm 1.07$ & 1 \\
NGC 5953 & 2003-06-01T05:49:46 & 70.B-0393(B) & $M_{\rm nb}$& 1200 & 1.37 & $\leq 1.65$ & $\leq 1.74$ & 3 \\
NGC 5953 & 2003-07-16T03:18:09 & 70.B-0393(B) & $L^{\prime}$& 600 & 0.58 & $0.68 \pm 0.12$ & $9.63 \pm 0.84$ & 1 \\
NGC 5953 & 2003-07-16T03:37:56 & 70.B-0393(B) & $M_{\rm nb}$& 1200 & 0.49 & $\leq 1.59$ & $\leq 1.66$ & 3 \\
{\bf NGC 5953} & 2013-07-07T02:51:22 & 290.B-5113(A) & $L^{\prime}$& 472 & 0.46 & $0.54 \pm 0.09$ & $7.25 \pm 0.50$ & 1\\
{\bf NGC 5953} & 2013-07-07T03:03:48 & 290.B-5113(A) & $M_{\rm nb}$& 944 & 0.42 & $\leq 2.29$ & $\leq 1.93$ & 3\\
{\bf NGC 5995} & 2003-06-19T04:49:17 & 70.B-0393(B) & $L^{\prime}$& 600 & 0.56 & $60.09 \pm 3.02$ & $44.62 \pm 13.35$ & 0\\
{\bf NGC 5995} & 2003-06-19T05:11:47 & 70.B-0393(B) & $M_{\rm nb}$& 1200 & 0.46 & $54.06 \pm 4.99$ & $53.44 \pm 19.25$ & 0\\
{\bf NGC 6000} & 2000-08-15T00:21:24 & 65.P-0519(A) & $M_{\rm nb}$& 960 & 0.53 & $2.72 \pm 0.46$ & $13.36 \pm 1.71$ & 0\\
{\bf NGC 6221} & 2013-07-03T04:48:31 & 290.B-5113(A) & $L^{\prime}$& 118 & 0.41 & $12.53 \pm 0.47$ & $12.80 \pm 2.95$ & 0\\
{\bf NGC 6221} & 2013-07-03T04:52:39 & 290.B-5113(A) & $M_{\rm nb}$& 236 & 0.42 & $11.30 \pm 0.62$ & $\leq 3.67$ & 0\\
{\bf NGC 6240N} & 2000-07-13T03:00:50 & 65.P-0519(A) & $L^{\prime}$& 240 & 0.99 & $16.65 \pm 0.56$ & $12.41 \pm 1.48$ & 0\\
{\bf NGC 6240N} & 2000-07-13T03:08:55 & 65.P-0519(A) & $M_{\rm nb}$& 720 & 0.68 & $18.47 \pm 0.68$ & $\leq 1.79$ & 0\\
{\bf NGC 6300} & 2013-07-02T04:57:16 & 290.B-5113(A) & $L^{\prime}$& 118 & 0.54 & $31.34 \pm 1.73$ & $19.64 \pm 8.69$ & 0\\
{\bf NGC 6300} & 2013-07-02T05:01:22 & 290.B-5113(A) & $M_{\rm nb}$& 118 & 0.52 & $49.80 \pm 2.54$ & $\leq 5.92$ & 0\\
{\bf NGC 6810} & 2003-07-14T04:22:03 & 71.B-0379(A) & $L^{\prime}$& 150 & 0.36 & $7.10 \pm 0.56$ & $28.41 \pm 2.52$ & 0\\
NGC 6810 & 2003-07-14T04:29:43 & 71.B-0379(A) & $L^{\prime}$& 600 & 0.36 & $10.10 \pm 0.55$ & $26.71 \pm 2.84$ & 0 \\
{\bf NGC 6810} & 2003-07-14T04:47:59 & 71.B-0379(A) & $M_{\rm nb}$& 1200 & 0.43 & $4.85 \pm 0.51$ & $10.06 \pm 1.98$ & 0\\
NGC 6814 & 2000-08-13T04:25:33 & 65.P-0519(A) & $M_{\rm nb}$& 720 & 0.63 & $23.15 \pm 1.92$ & $\leq 1.76$ & 0 \\
{\bf NGC 6814} & 2013-07-03T05:12:02 & 290.B-5113(A) & $L^{\prime}$& 118 & 0.41 & $10.57 \pm 0.40$ & $5.39 \pm 1.44$ & 0\\
{\bf NGC 6814} & 2013-07-03T05:16:10 & 290.B-5113(A) & $M_{\rm nb}$& 236 & 0.42 & $12.04 \pm 0.69$ & $\leq 3.35$ & 0\\
{\bf NGC 6860} & 2013-06-26T08:39:54 & 290.B-5113(A) & $L^{\prime}$& 118 & 0.78 & $35.24 \pm 1.50$ & $15.08 \pm 4.98$ & 0\\
{\bf NGC 6860} & 2013-06-26T08:44:01 & 290.B-5113(A) & $M_{\rm nb}$& 236 & 0.72 & $39.72 \pm 2.47$ & $\leq 4.07$ & 0\\
{\bf NGC 6890} & 2003-07-14T06:50:47 & 71.B-0379(A) & $L^{\prime}$& 600 & 0.50 & $8.12 \pm 0.35$ & $6.79 \pm 1.56$ & 0\\
{\bf NGC 6890} & 2003-07-14T07:10:58 & 71.B-0379(A) & $M_{\rm nb}$& 1200 & 0.50 & $9.45 \pm 0.66$ & $\leq 1.23$ & 0\\
NGC 6890 & 2004-10-30T00:07:52 & 074.B-0166(A) & $L^{\prime}$& 1534 & 0.57 & $6.14 \pm 0.61$ & $7.98 \pm 2.20$ & 0 \\
{\bf NGC 7130} & 2003-08-12T03:10:03 & 71.B-0379(A) & $L^{\prime}$& 600 & 0.60 & $6.91 \pm 0.30$ & $7.75 \pm 1.15$ & 0\\
{\bf NGC 7130} & 2003-08-12T03:29:58 & 71.B-0379(A) & $M_{\rm nb}$& 1200 & 0.55 & $14.39 \pm 0.75$ & $\leq 1.80$ & 0\\
NGC 7172 & 2001-06-09T09:38:20 & 67.B-0332(A) & $L^{\prime}$& 210 & 0.95 & $37.80 \pm 7.16$ & $46.56 \pm 19.09$ & 1 \\
NGC 7172 & 2001-06-09T09:48:16 & 67.B-0332(A) & $L^{\prime}$& 600 & 0.95 & $54.89 \pm 3.02$ & $37.55 \pm 11.43$ & 1 \\
{\bf NGC 7172} & 2001-06-09T10:03:39 & 67.B-0332(A) & $M_{\rm nb}$& 480 & 0.81 & $61.82 \pm 3.30$ & $\leq 3.30$ & 1\\
{\bf NGC 7172} & 2004-10-30T01:27:32 & 074.B-0166(A) & $L^{\prime}$& 1534 & 0.57 & $48.96 \pm 1.92$ & $31.79 \pm 7.59$ & 0\\
{\bf NGC 7213} & 2000-08-13T05:44:26 & 65.P-0519(A) & $M_{\rm nb}$& 240 & 0.72 & $\leq 3.39$ & $\leq 2.98$ & 3\\
{\bf NGC 7314} & 2000-08-13T04:53:57 & 65.P-0519(A) & $M_{\rm nb}$& 480 & 0.63 & $18.47 \pm 0.83$ & $\leq 2.39$ & 0\\
{\bf NGC 7314} & 2004-10-31T01:17:55 & 074.B-0166(A) & $L^{\prime}$& 1534 & 0.45 & $7.94 \pm 0.43$ & $6.82 \pm 1.46$ & 0\\
{\bf NGC 7479} & 2004-10-30T02:40:56 & 074.B-0166(A) & $L^{\prime}$& 1534 & 0.57 & $12.84 \pm 0.58$ & $10.03 \pm 2.48$ & 0\\
NGC 7496 & 2003-08-12T04:49:44 & 71.B-0379(A) & $L^{\prime}$& 90 & 0.60 & $4.20 \pm 0.36$ & $4.86 \pm 1.35$ & 0 \\
{\bf NGC 7496} & 2003-08-12T04:59:19 & 71.B-0379(A) & $L^{\prime}$& 600 & 0.60 & $6.59 \pm 0.26$ & $4.67 \pm 1.12$ & 0\\
{\bf NGC 7496} & 2003-08-12T05:19:17 & 71.B-0379(A) & $M_{\rm nb}$& 1200 & 0.55 & $6.89 \pm 0.64$ & $\leq 2.10$ & 0\\
NGC 7552 & 2000-06-15T08:36:37 & 65.P-0519(A) & $M_{\rm nb}$& 1440 & 0.78 & $\leq 1.66$ & $2.44 \pm 0.42$ & 0 \\
{\bf NGC 7552} & 2000-07-13T05:46:58 & 65.P-0519(A) & $M_{\rm nb}$& 1404 & 0.55 & $\leq 2.13$ & $\leq 1.86$ & 0\\
NGC 7582 & 2000-08-13T05:55:46 & 65.P-0519(A) & $M_{\rm nb}$& 240 & 0.72 & $69.95 \pm 4.49$ & $92.92 \pm 21.84$ & 0 \\
NGC 7582 & 2001-07-19T08:06:45 & 67.B-0332(A) & $L^{\prime}$& 600 & 0.58 & $196.65 \pm 7.49$ & $59.53 \pm 26.95$ & 0 \\
{\bf NGC 7582} & 2001-07-19T08:22:58 & 67.B-0332(A) & $M_{\rm nb}$& 480 & 0.39 & $91.64 \pm 5.08$ & $81.68 \pm 21.87$ & 0\\
{\bf NGC 7582} & 2001-08-17T05:06:52 & 67.B-0332(A) & $L^{\prime}$& 600 & 0.40 & $107.81 \pm 3.99$ & $84.13 \pm 22.80$ & 0\\
NGC 7582 & 2001-08-17T05:23:17 & 67.B-0332(A) & $M_{\rm nb}$& 480 & 0.37 & $96.60 \pm 5.08$ & $80.12 \pm 24.98$ & 0 \\
NGC 7582 & 2003-08-17T04:30:43 & 71.B-0379(A) & $L^{\prime}$& 600 & 0.50 & $111.83 \pm 8.31$ & $88.53 \pm 39.71$ & 0 \\
NGC 7582 & 2003-08-17T04:50:28 & 71.B-0379(A) & $M_{\rm nb}$& 1200 & 0.53 & $101.52 \pm 5.43$ & $96.47 \pm 26.76$ & 0 \\
NGC 7582 & 2004-10-31T02:27:17 & 074.B-0166(A) & $L^{\prime}$& 1534 & 0.45 & $85.10 \pm 5.22$ & $101.69 \pm 28.71$ & 0 \\
{\bf NGC 7590} & 2003-07-15T06:11:02 & 71.B-0379(A) & $L^{\prime}$& 600 & 0.58 & $0.68 \pm 0.05$ & $3.99 \pm 0.29$ & 0\\
{\bf NGC 7590} & 2003-07-15T06:30:44 & 71.B-0379(A) & $M_{\rm nb}$& 1200 & 0.49 & $\leq 1.79$ & $\leq 1.62$ & 3\\
{\bf NGC 7679} & 2003-06-23T09:14:46 & 71.B-0379(A) & $M_{\rm nb}$& 600 & 0.35 & $\leq 2.84$ & $\leq 2.36$ & 3\\
{\bf PG 2130+099} & 2013-06-27T09:42:30 & 290.B-5113(A) & $L^{\prime}$& 118 & 0.68 & $17.64 \pm 3.15$ & $20.29 \pm 9.73$ & 0\\
PG 2130+099 & 2013-06-27T09:46:41 & 290.B-5113(A) & $M_{\rm nb}$& 236 & 0.66 & $42.59 \pm 1.76$ & $\leq 4.37$ & 0 \\
PG 2130+099 & 2013-06-29T09:41:27 & 290.B-5113(A) & $L^{\prime}$& 118 & 0.78 & $30.34 \pm 2.23$ & $\leq 0.56$ & 3 \\
PG 2130+099 & 2013-06-29T09:48:09 & 290.B-5113(A) & $M_{\rm nb}$& 236 & 0.80 & $39.94 \pm 2.89$ & $\leq 4.13$ & 0 \\
PG 2130+099 & 2013-06-29T09:58:29 & 290.B-5113(A) & $L^{\prime}$& 118 & 0.78 & $41.19 \pm 2.51$ & $\leq 0.62$ & 0 \\
PG 2130+099 & 2013-06-29T10:02:41 & 290.B-5113(A) & $M_{\rm nb}$& 236 & 0.80 & $39.00 \pm 2.61$ & $\leq 3.99$ & 0 \\
PG 2130+099 & 2013-06-30T09:42:35 & 290.B-5113(A) & $L^{\prime}$& 118 & 0.78 & $42.95 \pm 1.29$ & $\leq 0.64$ & 0 \\
{\bf PG 2130+099} & 2013-06-30T09:46:46 & 290.B-5113(A) & $M_{\rm nb}$& 236 & 0.80 & $44.28 \pm 1.44$ & $\leq 3.97$ & 0\\
{\bf PKS 1417-19} & 2013-07-06T02:24:43 & 290.B-5113(A) & $L^{\prime}$& 118 & 0.73 & $6.20 \pm 0.22$ & $\leq 0.63$ & 0\\
{\bf PKS 1417-19} & 2013-07-06T02:28:50 & 290.B-5113(A) & $M_{\rm nb}$& 236 & 0.74 & $3.04 \pm 0.40$ & $\leq 4.02$ & 2\\
{\bf PKS 1814-63} & 2013-07-02T05:22:22 & 290.B-5113(A) & $L^{\prime}$& 472 & 0.54 & $3.71 \pm 0.12$ & $\leq 0.36$ & 0\\
{\bf PKS 1814-63} & 2013-07-02T05:34:51 & 290.B-5113(A) & $M_{\rm nb}$& 944 & 0.52 & $2.09 \pm 0.19$ & $\leq 2.08$ & 0\\
{\bf PKS 1932-46} & 2013-06-26T09:06:59 & 290.B-5113(A) & $L^{\prime}$& 472 & 0.78 & $\leq 0.33$ & $\leq 0.29$ & 3\\
{\bf PKS 1932-46} & 2013-06-26T09:19:25 & 290.B-5113(A) & $M_{\rm nb}$& 944 & 0.72 & $\leq 2.02$ & $\leq 1.94$ & 3\\
{\bf Superantennae S} & 2003-06-19T08:22:54 & 71.B-0379(A) & $L^{\prime}$& 600 & 0.56 & $16.72 \pm 0.81$ & $9.34 \pm 2.82$ & 0\\
Superantennae S & 2003-06-19T08:45:13 & 71.B-0379(A) & $M_{\rm nb}$& 60 & 0.46 & $36.60 \pm 1.69$ & $\leq 5.93$ & 0 \\
Superantennae S & 2003-06-21T08:19:13 & 71.B-0379(A) & $M_{\rm nb}$& 780 & 0.66 & $30.18 \pm 2.02$ & $\leq 1.96$ & 0 \\
Superantennae S & 2003-06-21T08:53:44 & 71.B-0379(A) & $M_{\rm nb}$& 600 & 0.66 & $28.62 \pm 1.78$ & $19.23 \pm 6.46$ & 0 \\
{\bf Superantennae S} & 2003-06-23T07:42:39 & 71.B-0379(A) & $M_{\rm nb}$& 630 & 0.49 & $26.91 \pm 1.52$ & $17.36 \pm 4.87$ & 0\\
Superantennae S & 2003-06-23T08:06:49 & 71.B-0379(A) & $M_{\rm nb}$& 630 & 0.35 & $19.43 \pm 2.37$ & $23.88 \pm 8.28$ & 0 \\
{\bf UGC 2369 S} & 2000-11-05T04:27:27 & 65.P-0519(A) & $M_{\rm nb}$& 720 & 0.47 & $1.18 \pm 0.11$ & $\leq 1.85$ & 3\\
{\bf Z 41-20} & 2013-07-03T00:47:57 & 290.B-5113(A) & $L^{\prime}$& 118 & 0.53 & $2.87 \pm 0.13$ & $\leq 0.66$ & 0\\
{\bf Z 41-20} & 2013-07-03T00:52:05 & 290.B-5113(A) & $M_{\rm nb}$& 236 & 0.43 & $2.18 \pm 0.34$ & $\leq 3.92$ & 2\\
\enddata
\tablecomments{Epochs marked in bold are used in the analysis above, typically because of better seeing conditions. Flags: 0, no issue; 1, possible error in the reduction; 2, likely error in reduction (chopping/nodding); 3, failure in full chop/nod reduction.   }
\end{deluxetable*}

\subsection{Special Cases}

\subsubsection{Arp 220 -- A Known Double Nucleus}
Special care was taken for Arp 220 (a.k.a. IC4553) which is known to have a double nucleus \citep[e.g.,][]{soifer1999, aalto2009}. This is the only galaxy in which we fit three elliptical Gaussians to the emission, rather than the usual two; one for each nucleus, and one for any background/extended emission. We report the flux of each nucleus separately in Table 2. 

\subsubsection{NGC 7552 -- An AGN with a Starburst Ring}
This (possibly dormant) AGN is located at the center of a bright starburst ring \citep{forbes1994}. To avoid fitting starburst regions rather than the AGN, we limit the fit to within the central 1'' of the galaxy. 

\subsubsection{Faint Sources}
There were 13 sources which are marginal detections with our two-Gaussian approach, but which are clear detections "by-eye." To reduce the number of free parameters and increase the flux significance of the results, we fit the following with only one Gaussian. The FWHM of this Gaussian is set equal to that of the PSF calibrator closest in time, as in the two-Gaussian case.
The AGN: 3C321, 3C327, 3C424, ESO 323-32, M87, NGC 63, NGC 986, NGC 3660, NGC 4038/9, NGC 5427, PKS 1814-63, UGC 2369 S, Z 41-20.

\section{Calibration Strategy}
\begin{figure*}[ht]
    \centering
    \includegraphics[width=0.75\textwidth]{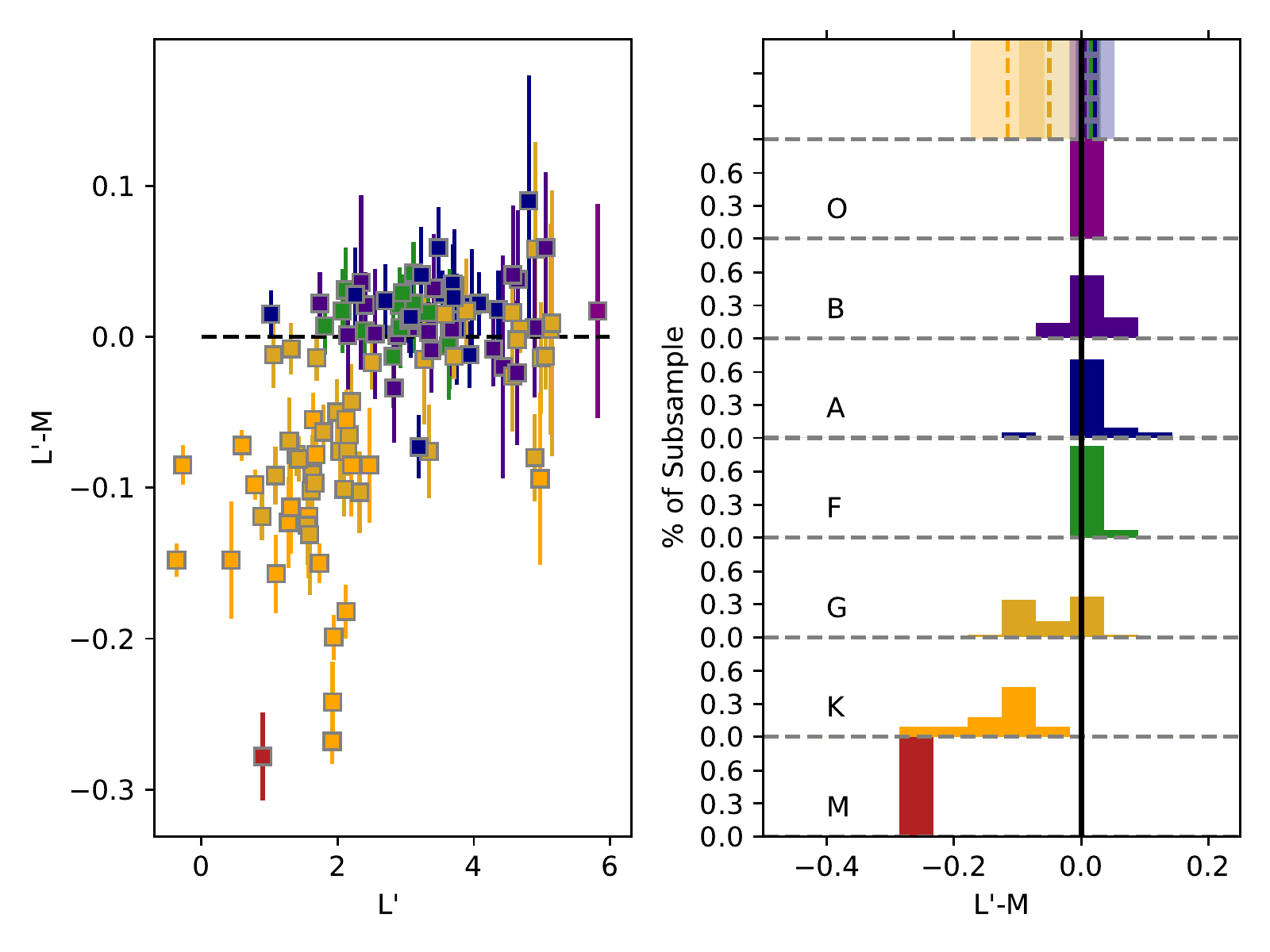}
    \caption{The L-M colors for all calibration stars in \citet{vanderbliek1996}, separated by spectral type. For stars of type earlier than G, we find that L-M$\approx 0$. Bins are spaced every 0.05 mag. }
    \label{fig:app_calibs}
\end{figure*}

As we took many of these sources from archival programs with various setups, we had to define a consistent calibration strategy. While most of the calibrators chosen have \lprime~magnitudes in the catalog of \citet{vanderbliek1996}, many do not have M magnitudes, and none have M$_{\rm nb}$. We therefore first make the assumption that ${\rm M}-{\rm M}_{\rm nb} = 0$. Secondly, we investigate the relationships between the \lprime$-$M$_{\rm nb}$ color, \lprime magnitude, and spectral type of the calibrator, shown in Fig \ref{fig:app_calibs} . We find that for stars of type O, B, A, and F the color \lprime$-$M$_{\rm nb} \approx 0$ with very little scatter ($\sigma_{\rm L-M} \lesssim 0.05$). This results in the calibration strategy below:

\begin{algorithmic}
\Function {Calibrate Targets}{}
\For{each {\it targ} observation}
    \For {each {\it band} $\in$ [\lprime, M$_{\rm nb}$] }
    
        \State 1) Find the calibrator ({\it calib}) observed closest in time which
        
        \State   \hskip1.5em a) was observed within 6hrs of the target
        \State   \hskip1.5em b) (has both L' and M band catalog values) $\lor$ (is spectral type $\in$ [O, B, A, F])
        
        \State 2) Read $F_{calib, real}$ from {\it band} or calculate from $\neg${\it band} if necessary
        
        \State 3) Compute $F_{\rm targ,cal} = F_{\rm targ, raw} / F_{\rm calib, raw} \times F_{calib, real}$
        
        \State 4) Compute the error from the relative errors on the individual fits
    \EndFor
\EndFor
\EndFunction
\label{lst:cal_algo}
\end{algorithmic}
where catalog values refers to the NIR catalog of ESO calibrators from \citet{vanderbliek1996}; and $F_{\rm targ, raw}$ and $F_{\rm calib, raw}$ are the fitted integrated fluxes in counts for the target and calibrator, respectively. Finally, we list each target and its calibrator in Table A.1. 

\clearpage
\section{Source Cutouts}
Here we present the cutouts of all sources included in this work in each band. We present each image with ``slices'' across the image to show the quality of the model fits. 
%See ``supplementary\_figures.pdf'', as they will be in the online version only.

\begin{figure*}
\centering
\subfloat{\includegraphics[width=0.25\hsize]{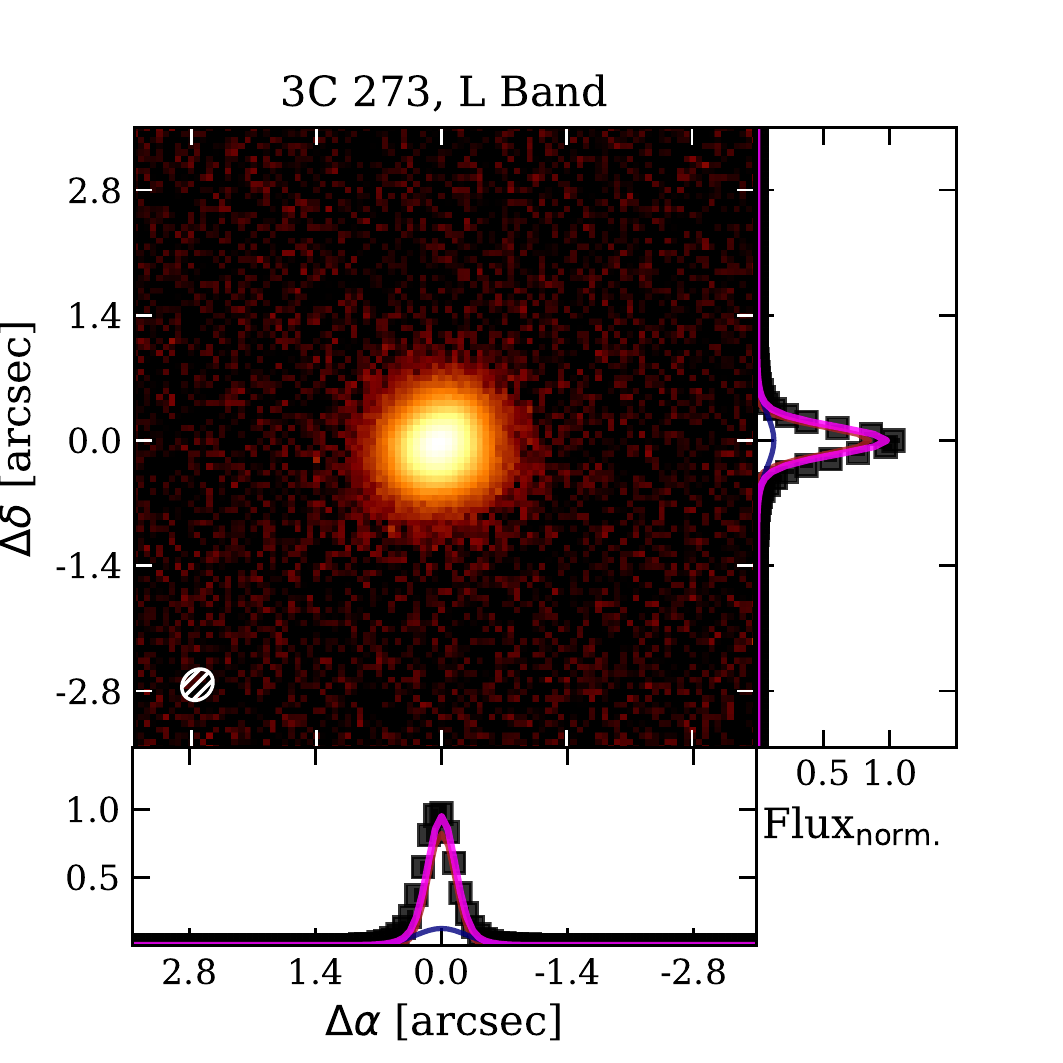}}
\subfloat{\includegraphics[width=0.25\hsize]{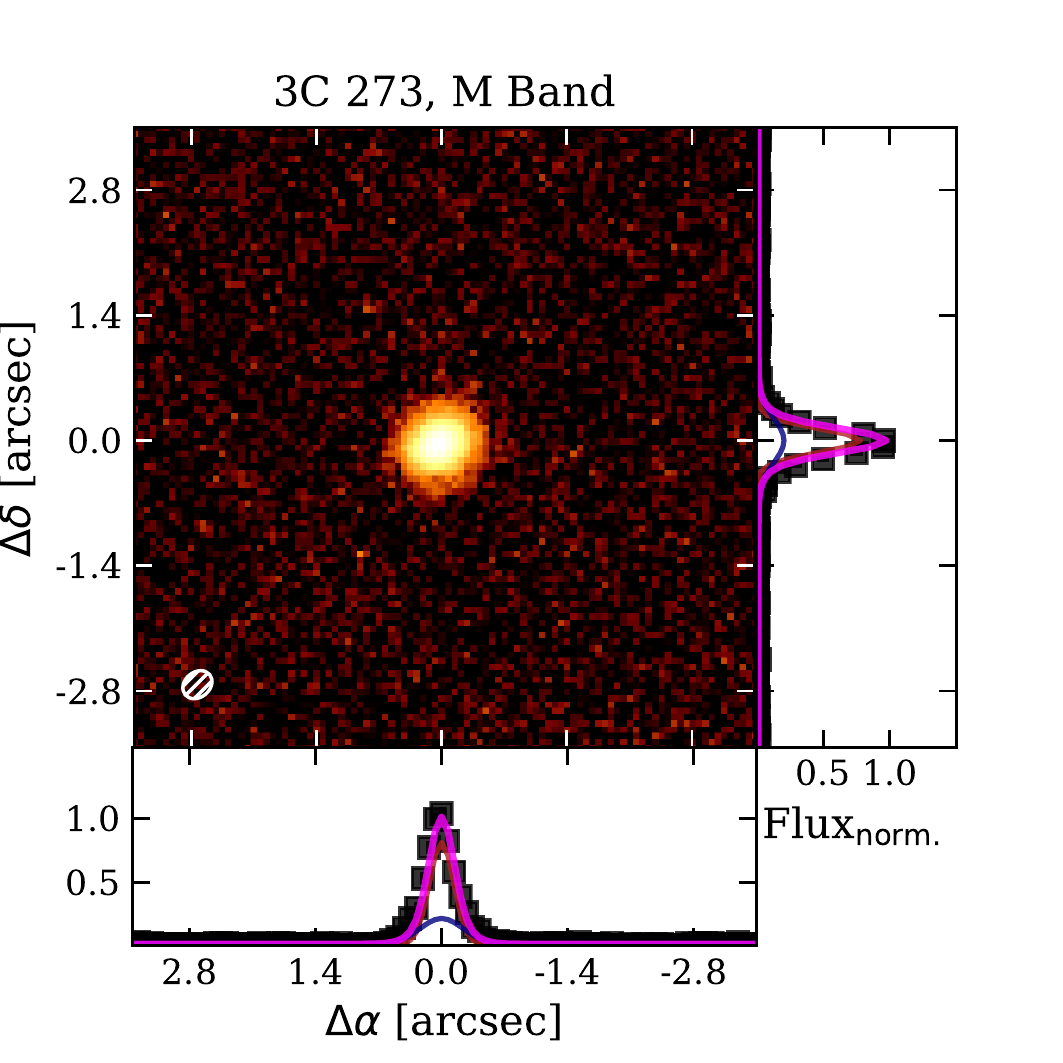}}
\subfloat{\includegraphics[width=0.25\hsize]{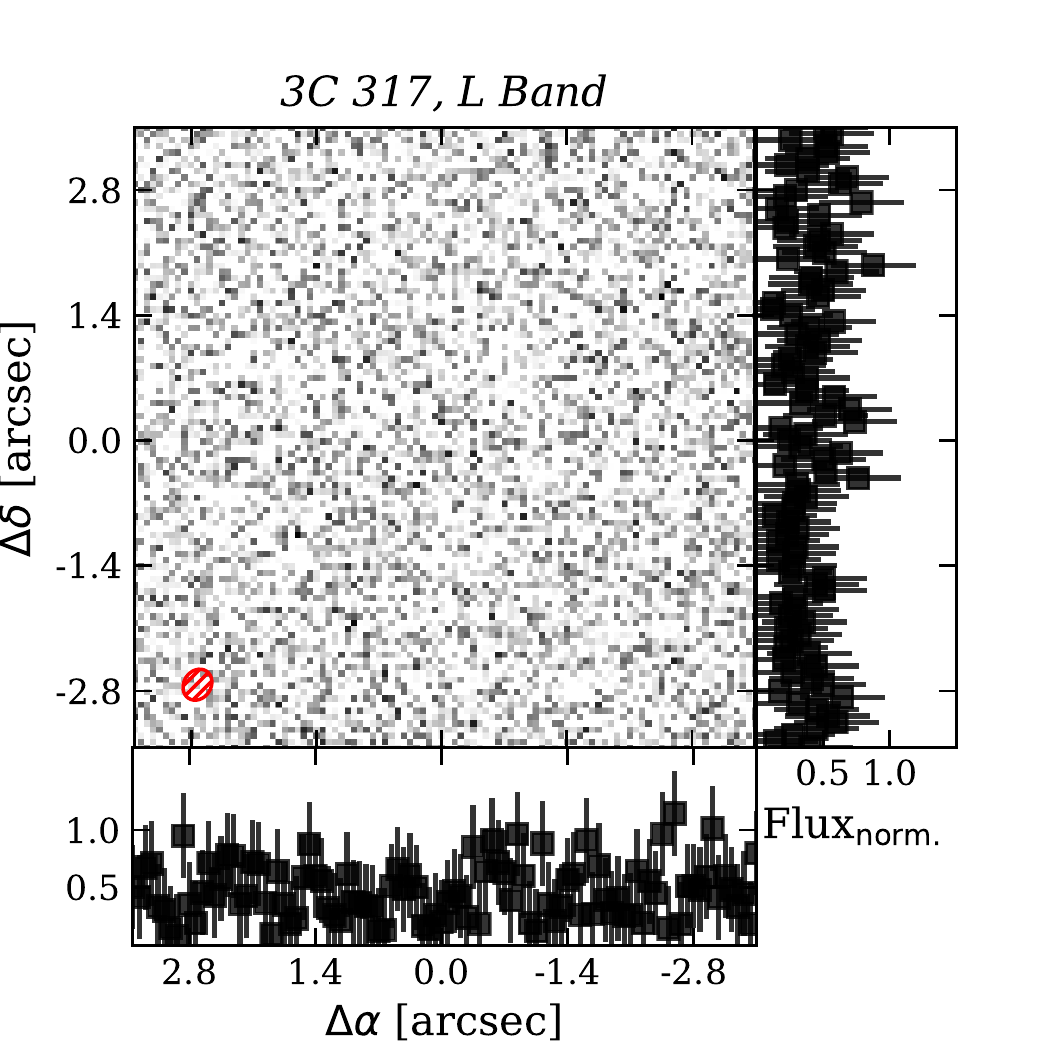}}
\subfloat{\includegraphics[width=0.25\hsize]{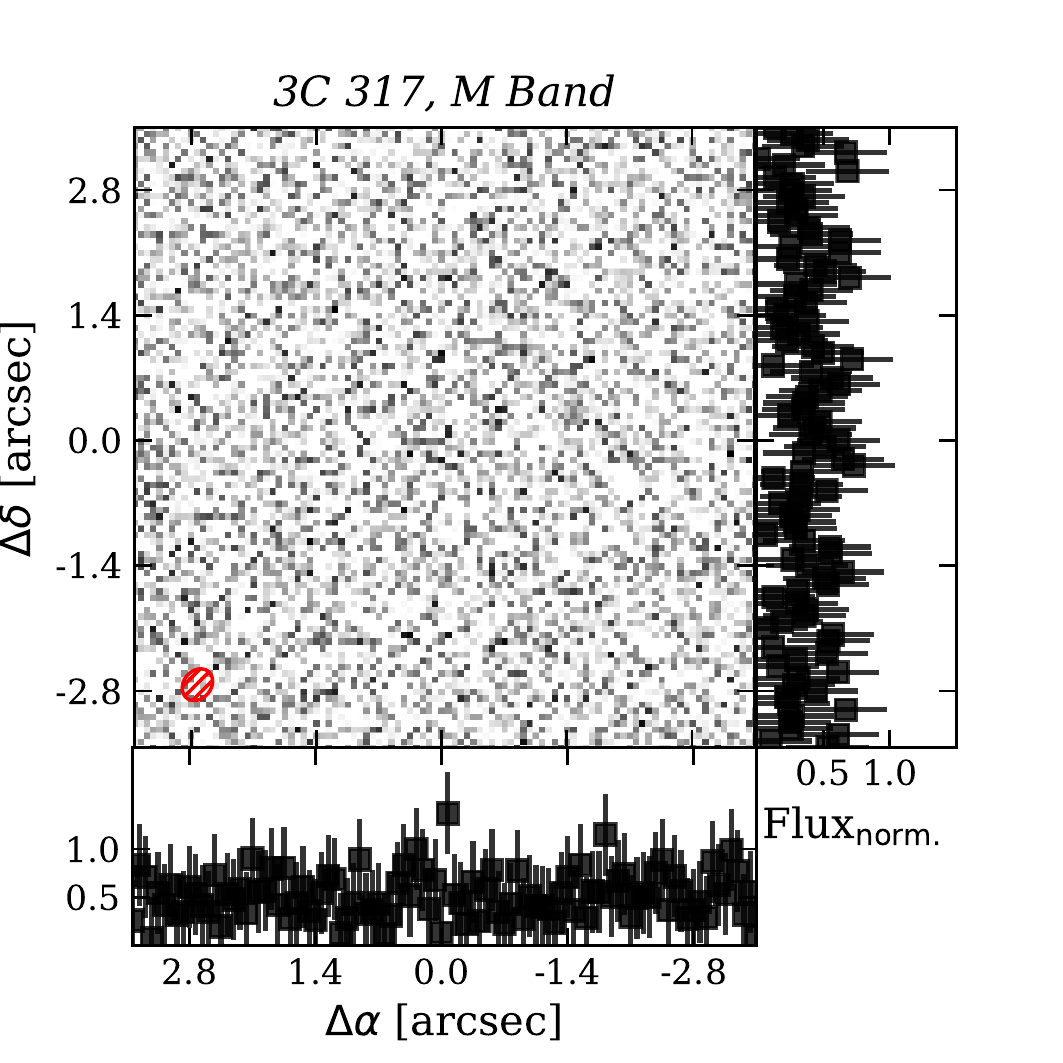}} \\
\subfloat{\includegraphics[width=0.25\hsize]{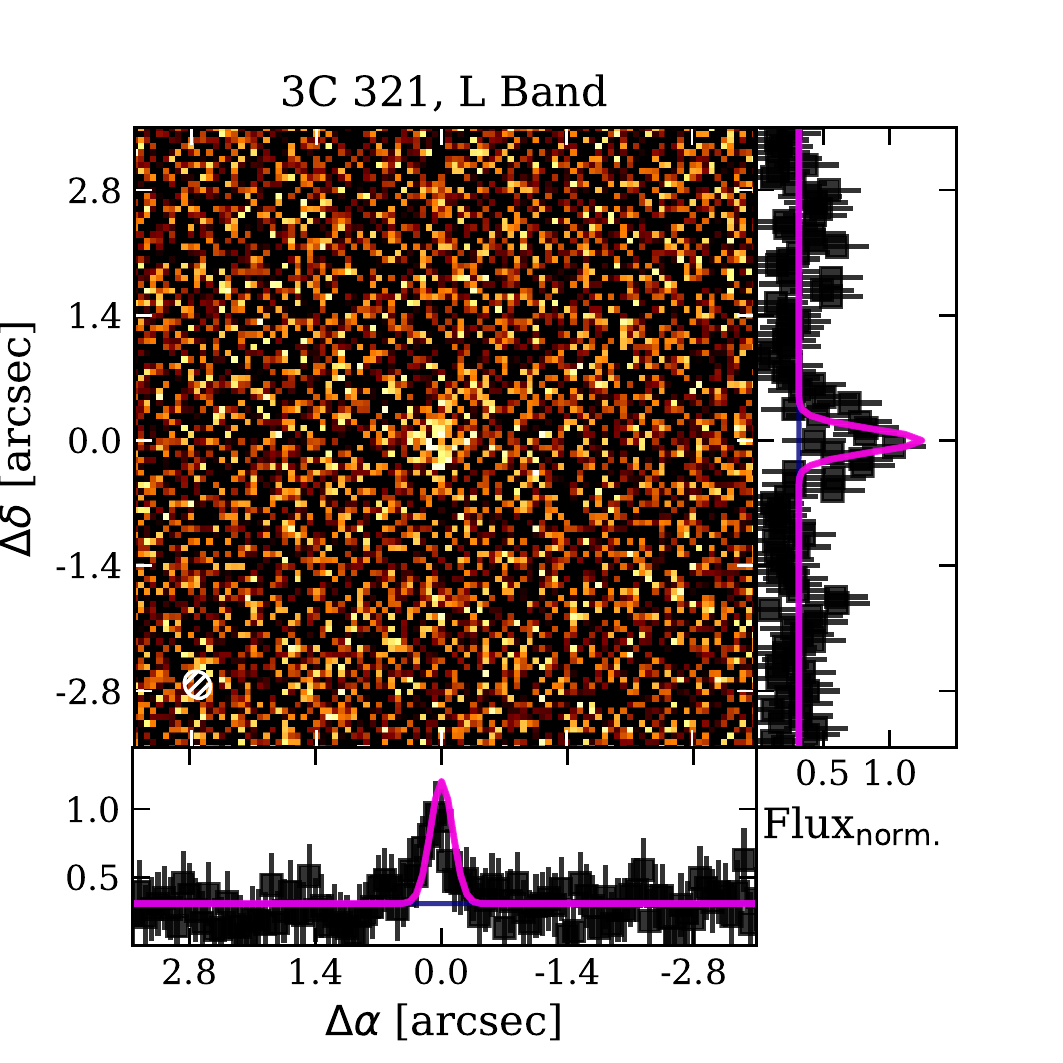}}
\subfloat{\includegraphics[width=0.25\hsize]{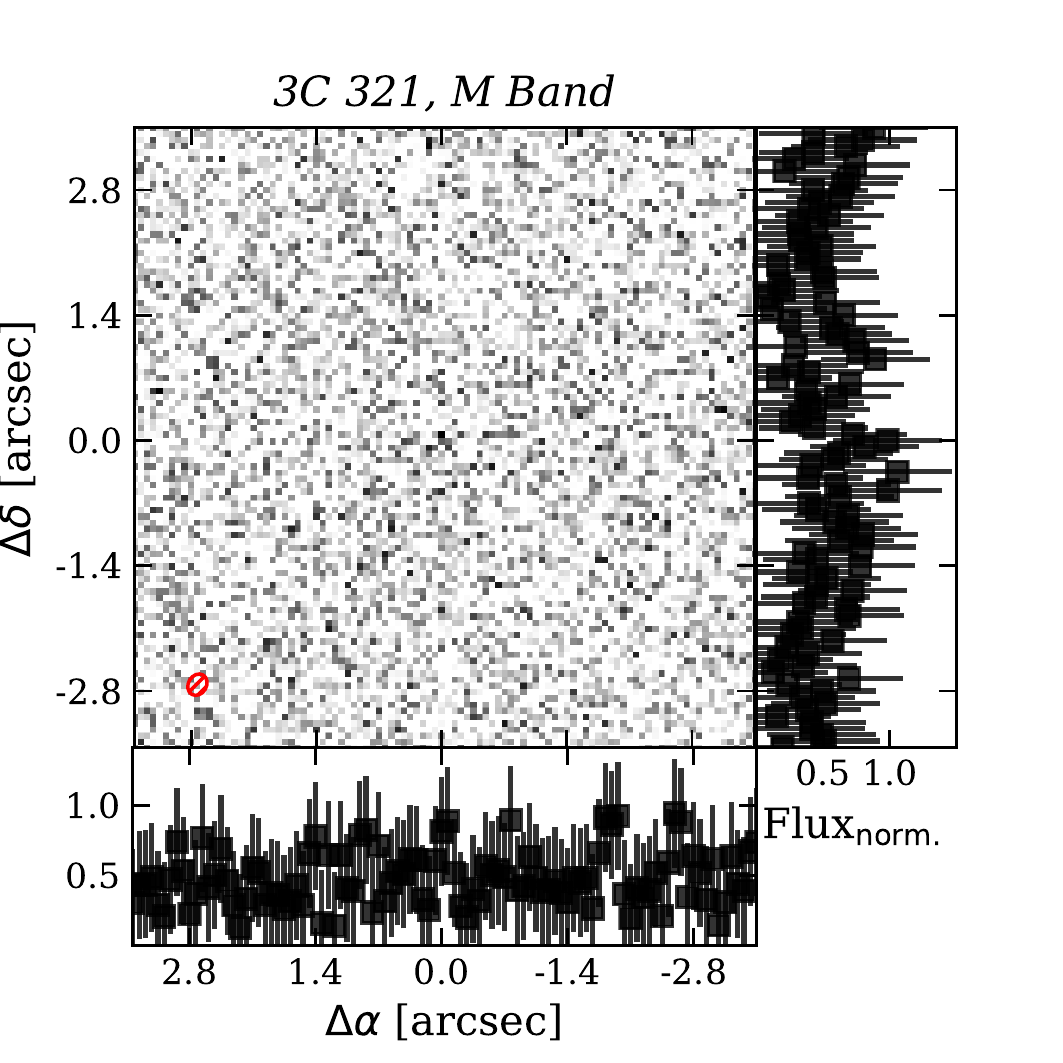}}
\subfloat{\includegraphics[width=0.25\hsize]{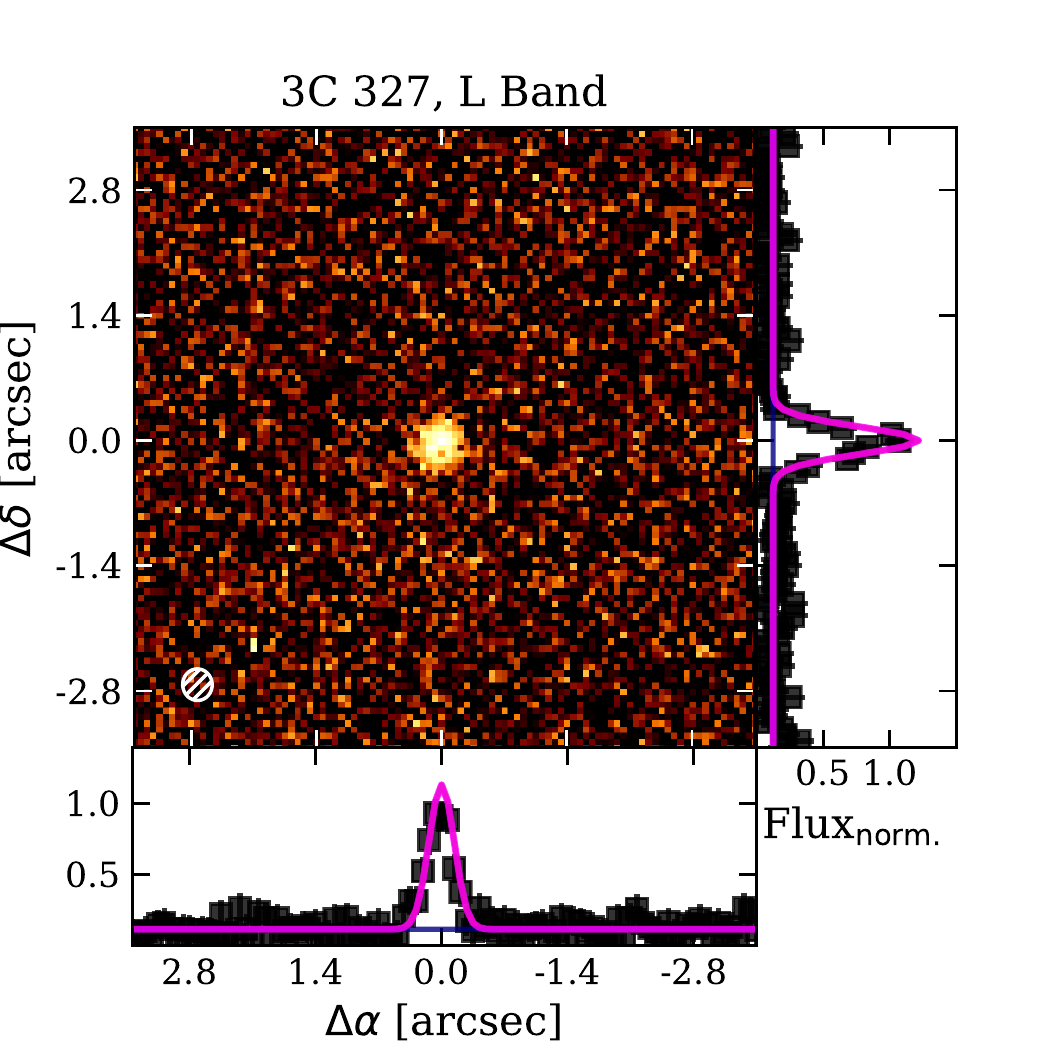}}
\subfloat{\includegraphics[width=0.25\hsize]{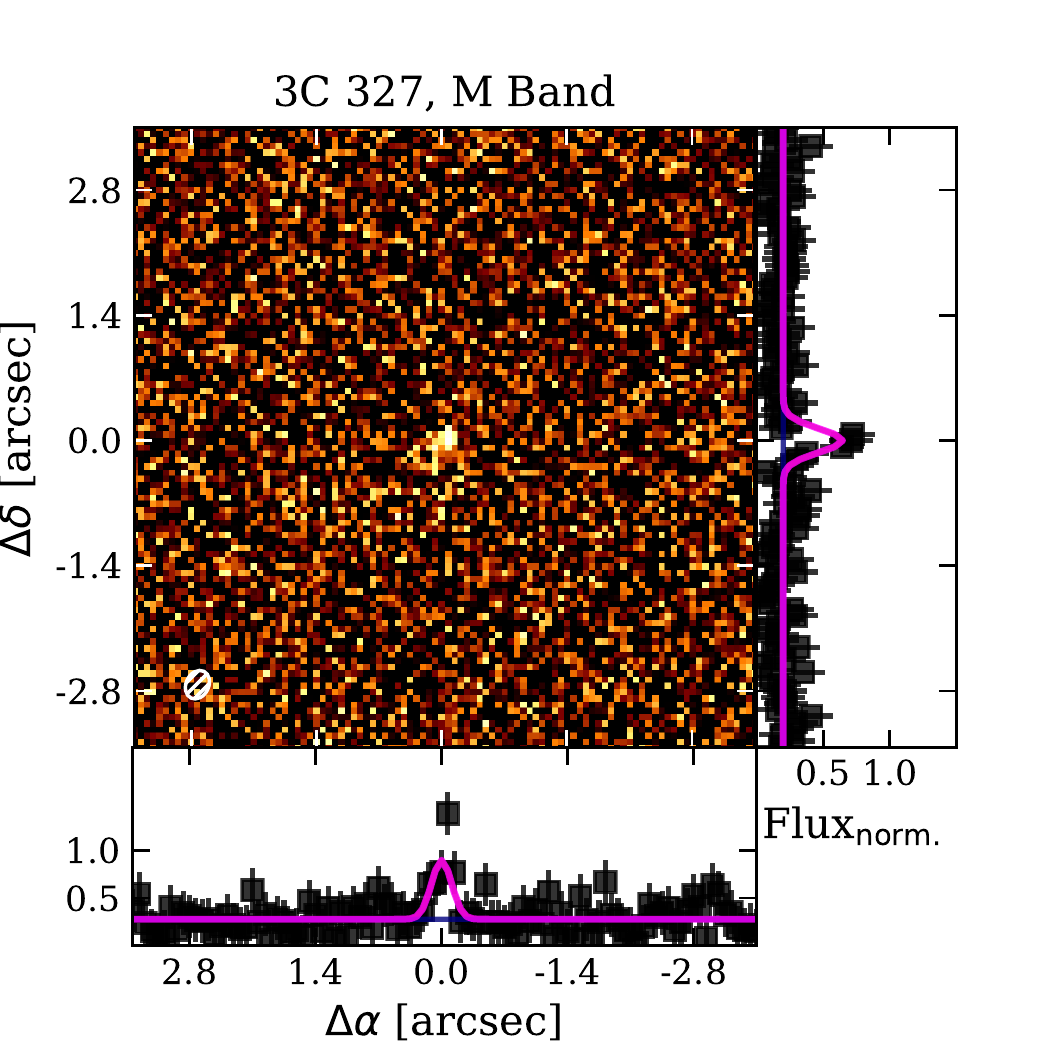}} \\
\subfloat{\includegraphics[width=0.25\hsize]{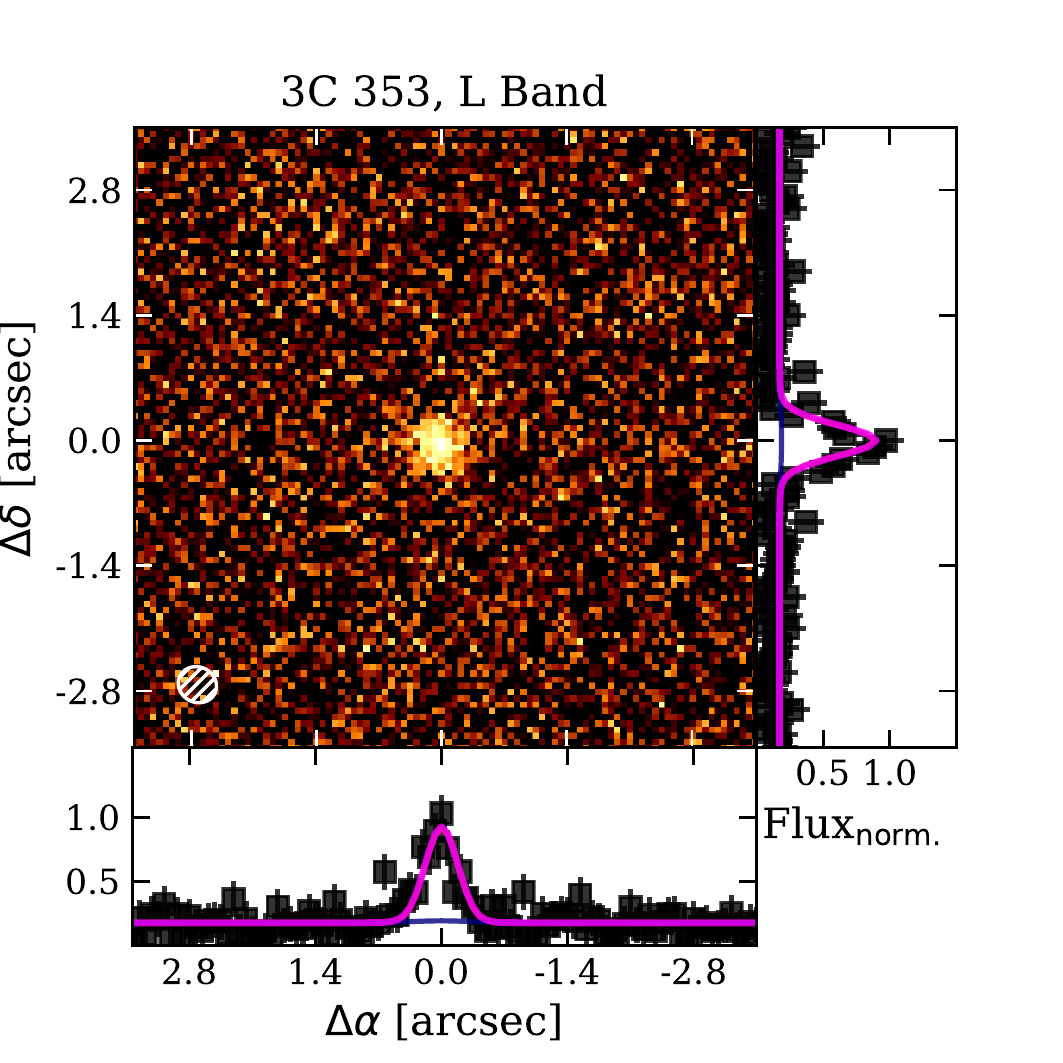}}
\subfloat{\includegraphics[width=0.25\hsize]{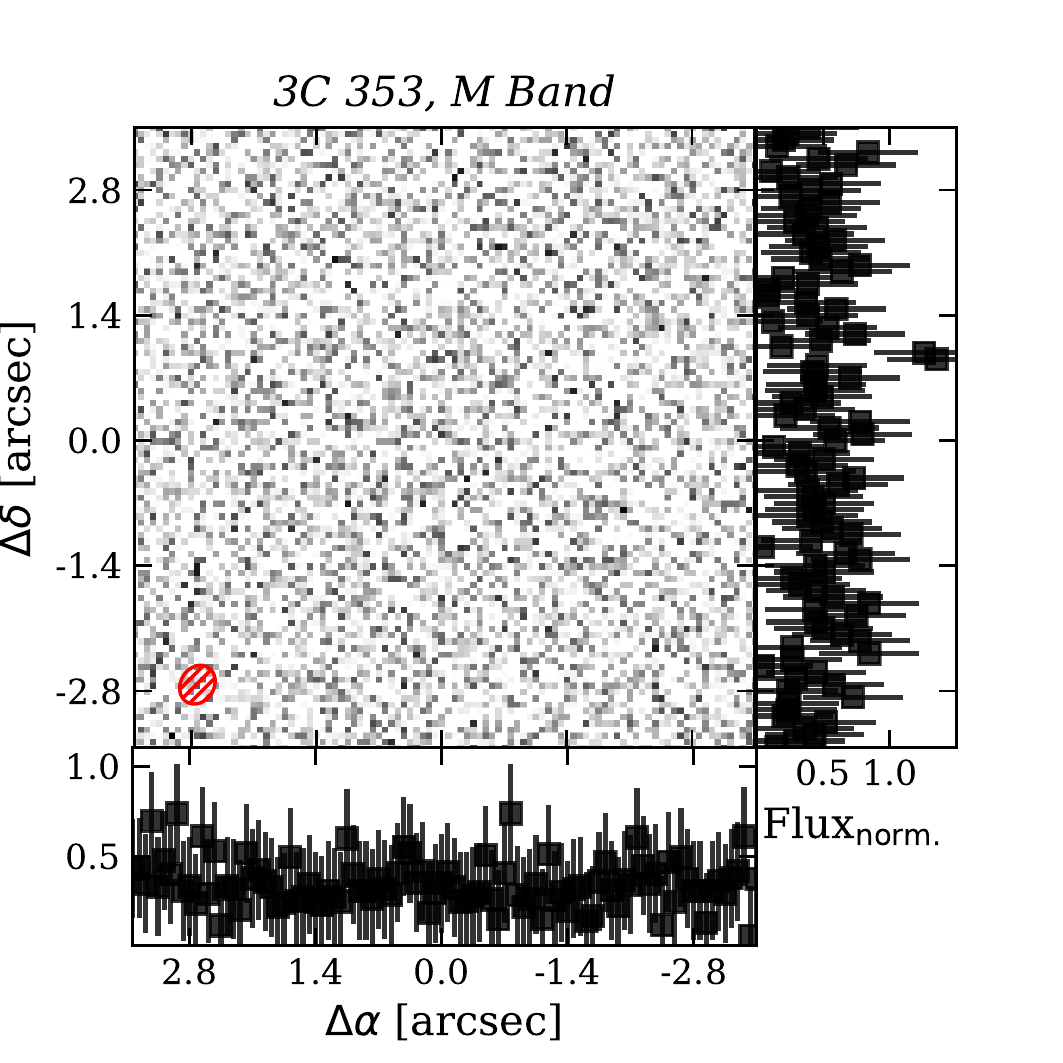}}
\subfloat{\includegraphics[width=0.25\hsize]{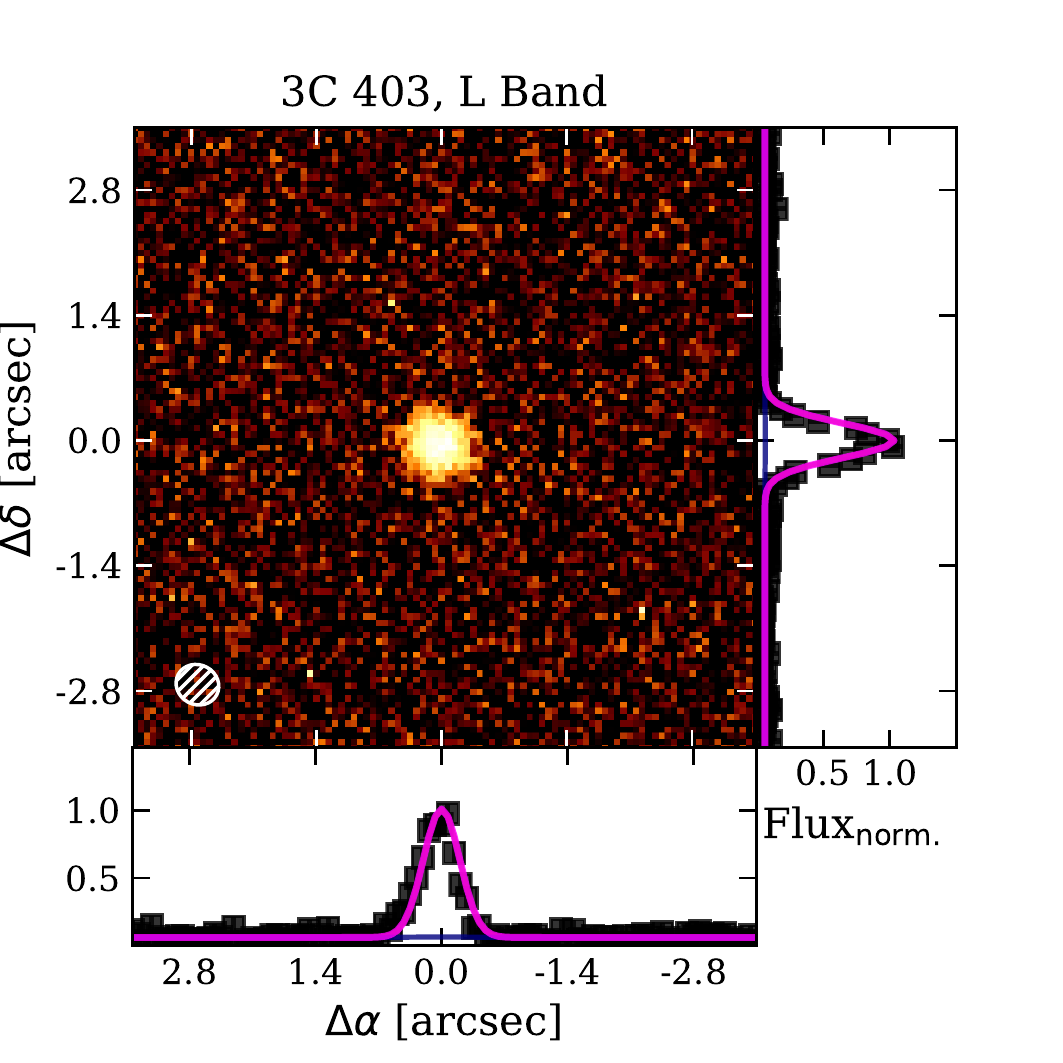}}
\subfloat{\includegraphics[width=0.25\hsize]{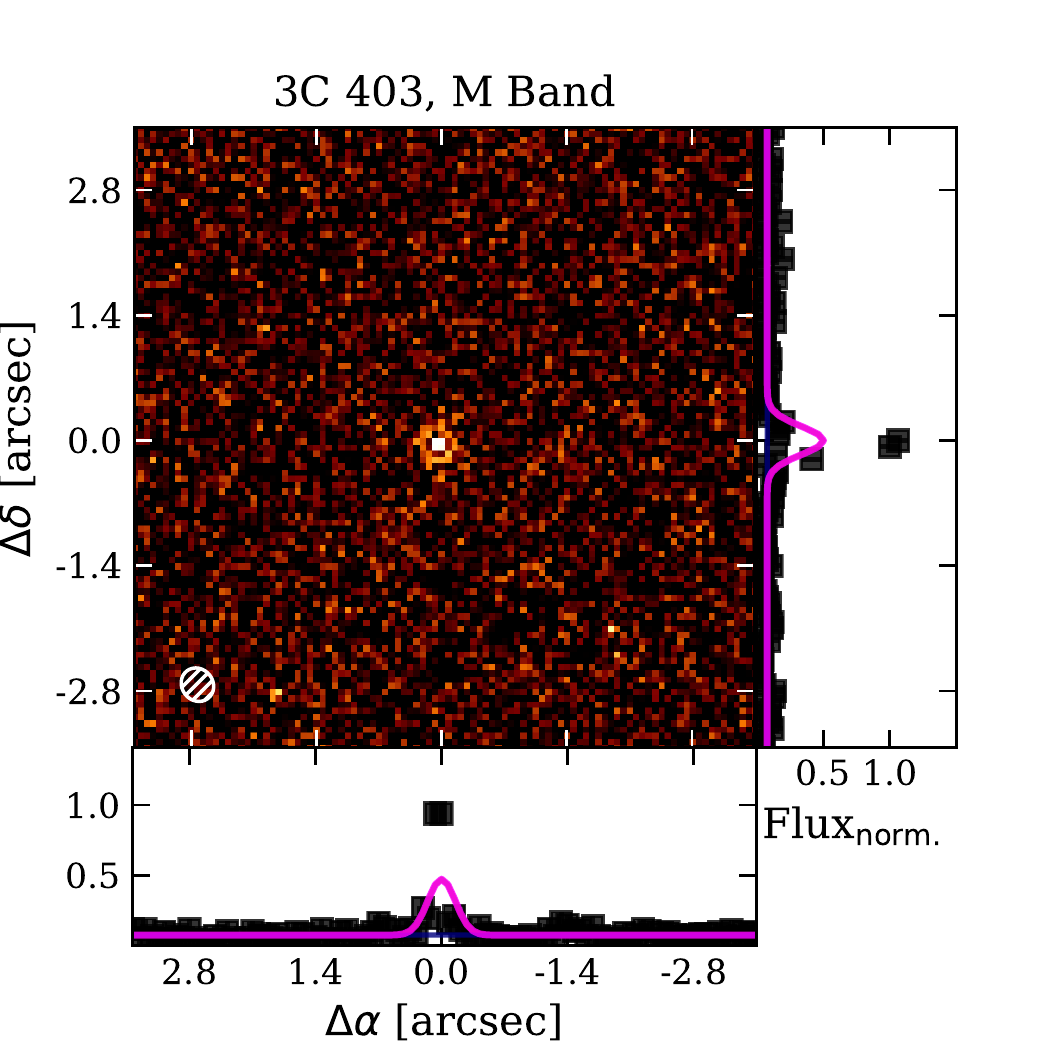}} \\
\subfloat{\includegraphics[width=0.25\hsize]{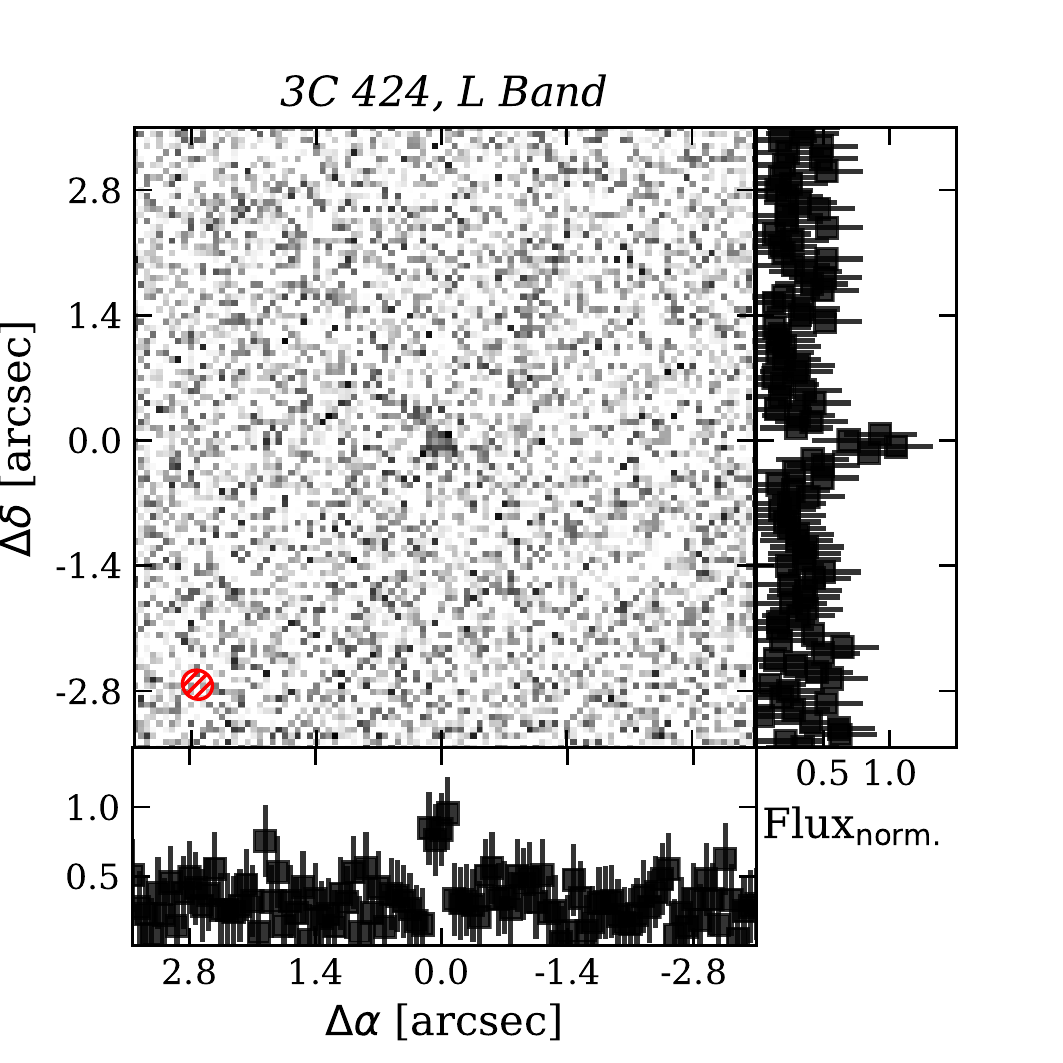}}
\subfloat{\includegraphics[width=0.25\hsize]{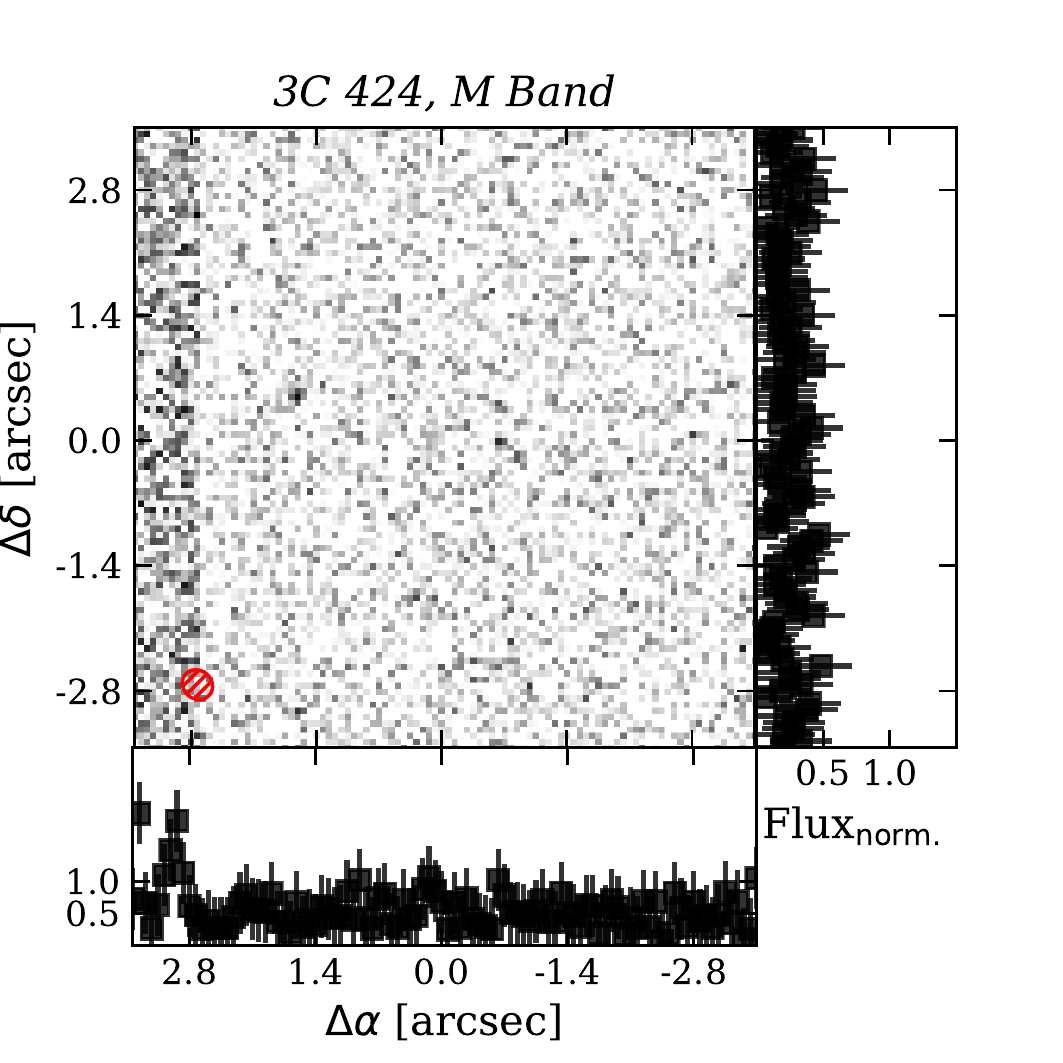}}
\subfloat{\includegraphics[width=0.25\hsize]{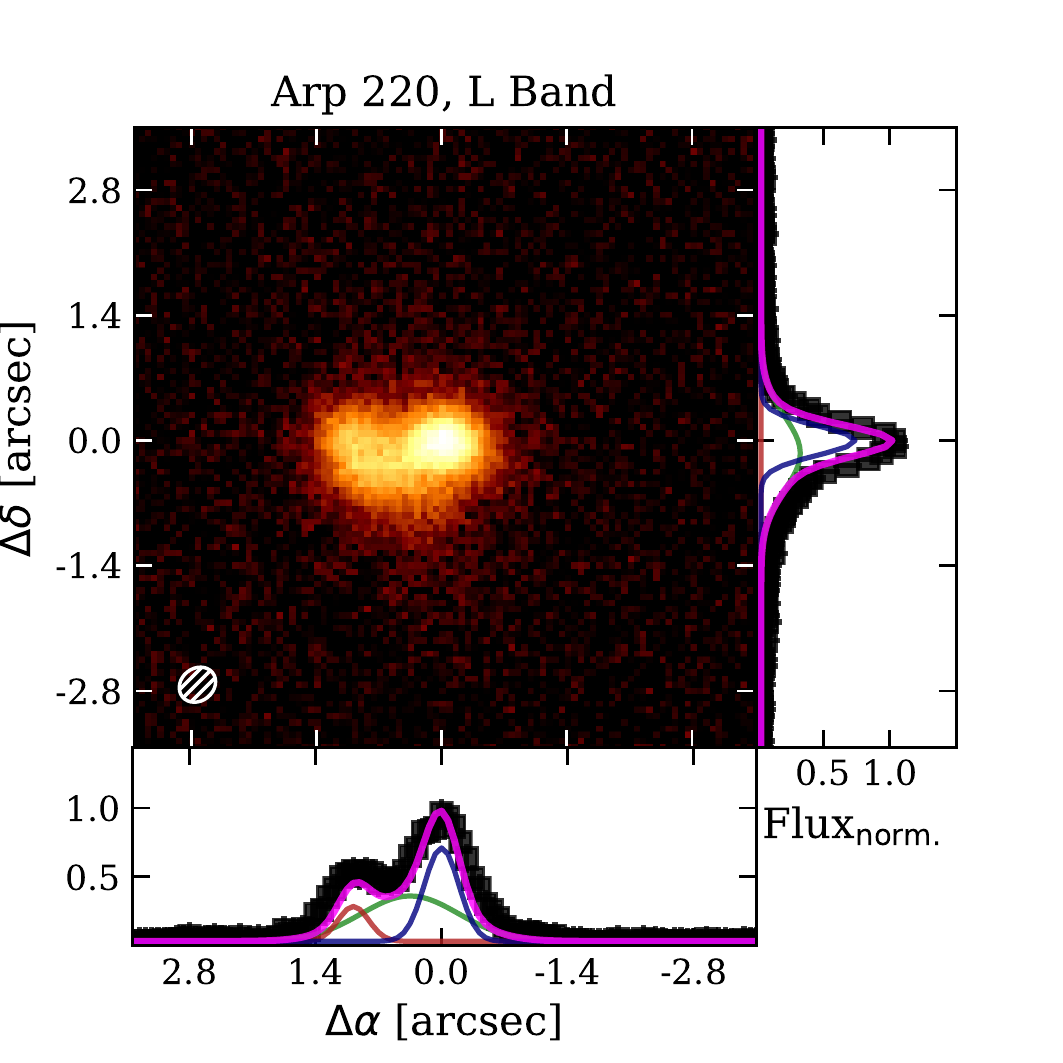}}
\subfloat{\includegraphics[width=0.25\hsize]{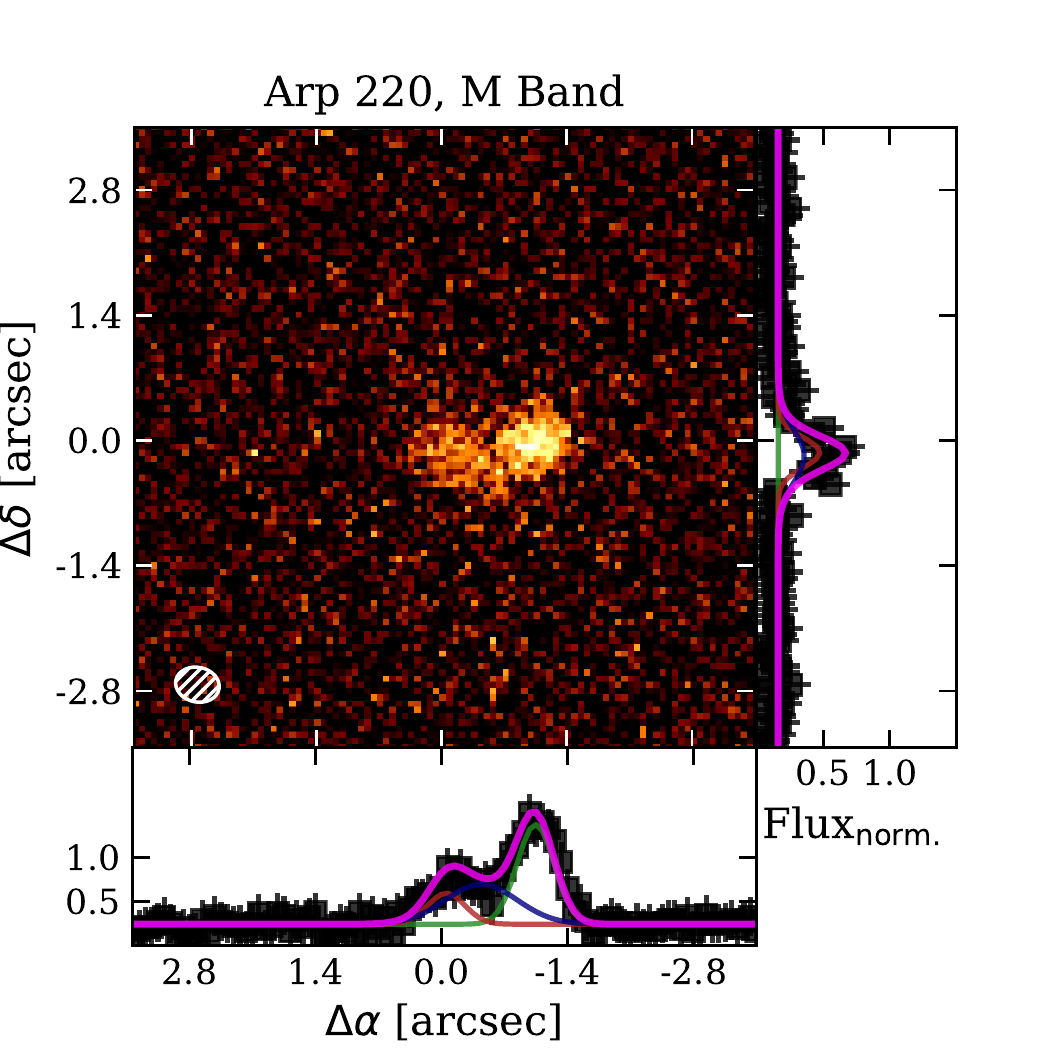}} \\
\subfloat{\includegraphics[width=0.25\hsize]{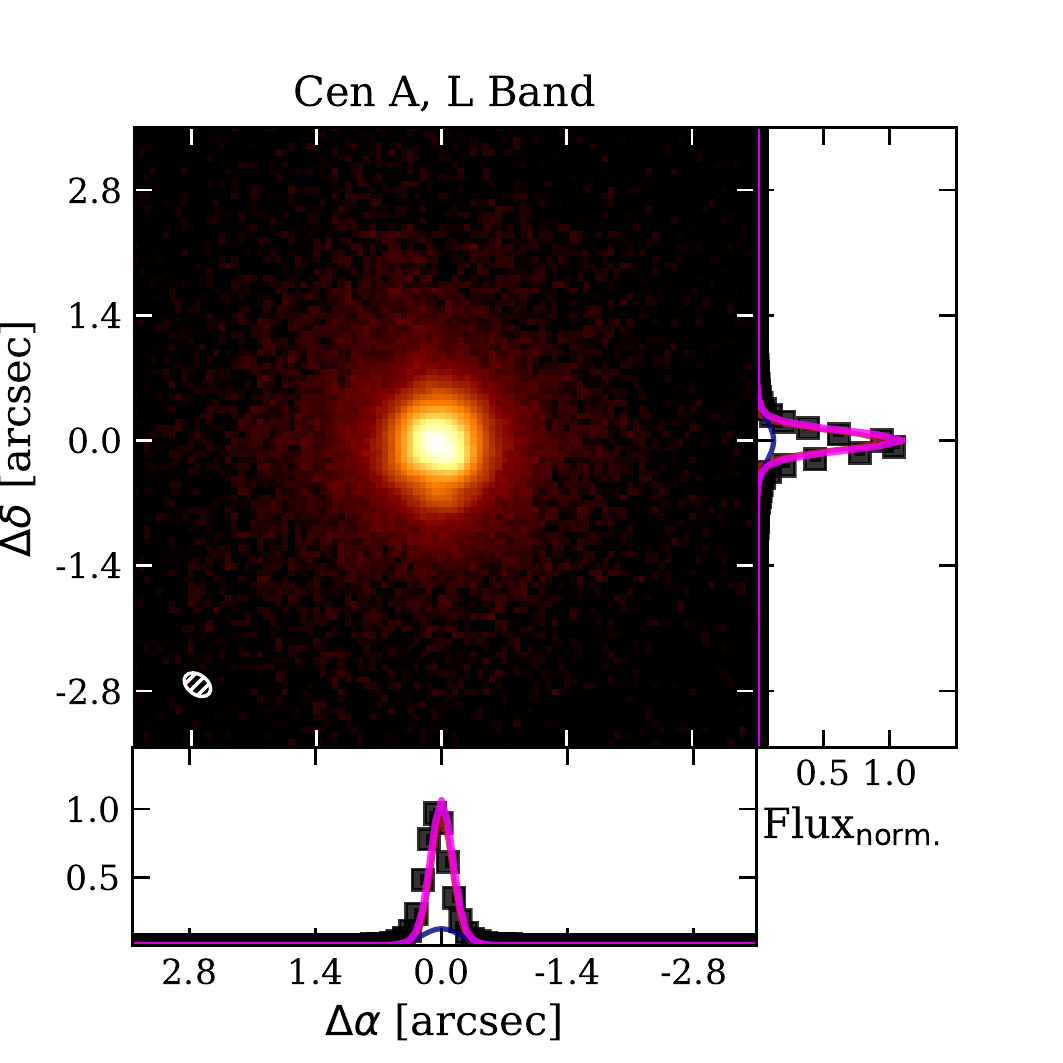}}
\subfloat{\includegraphics[width=0.25\hsize]{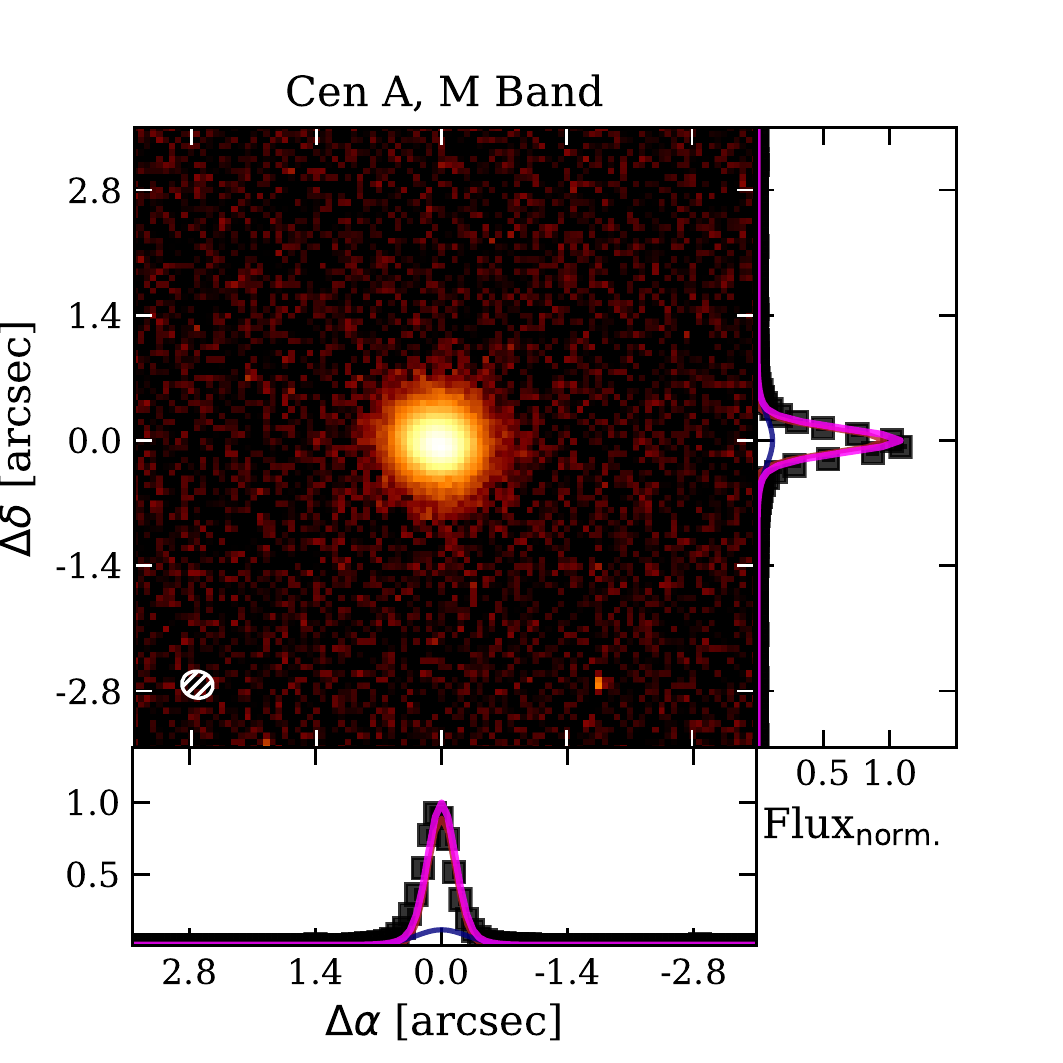}}
\subfloat{\includegraphics[width=0.25\hsize]{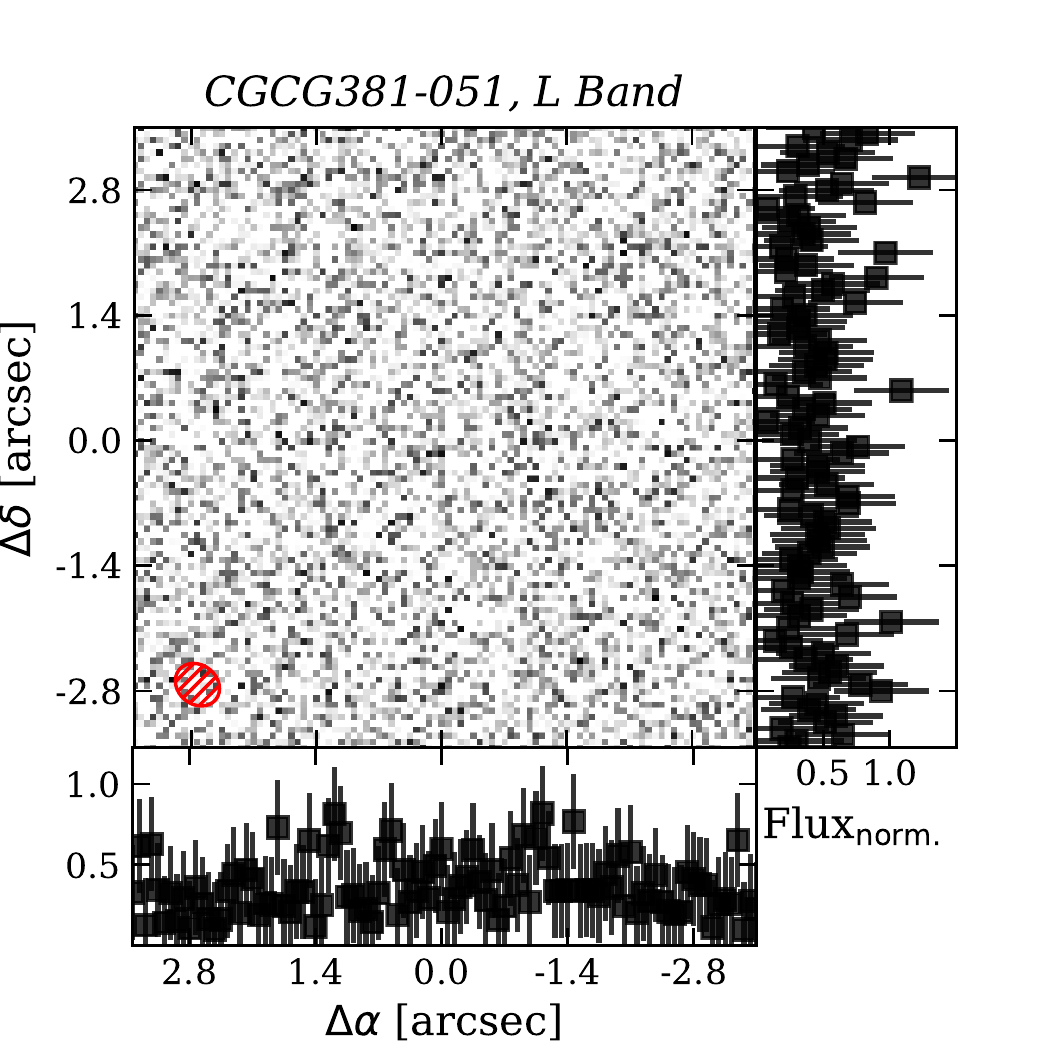}}
\subfloat{\includegraphics[width=0.25\hsize]{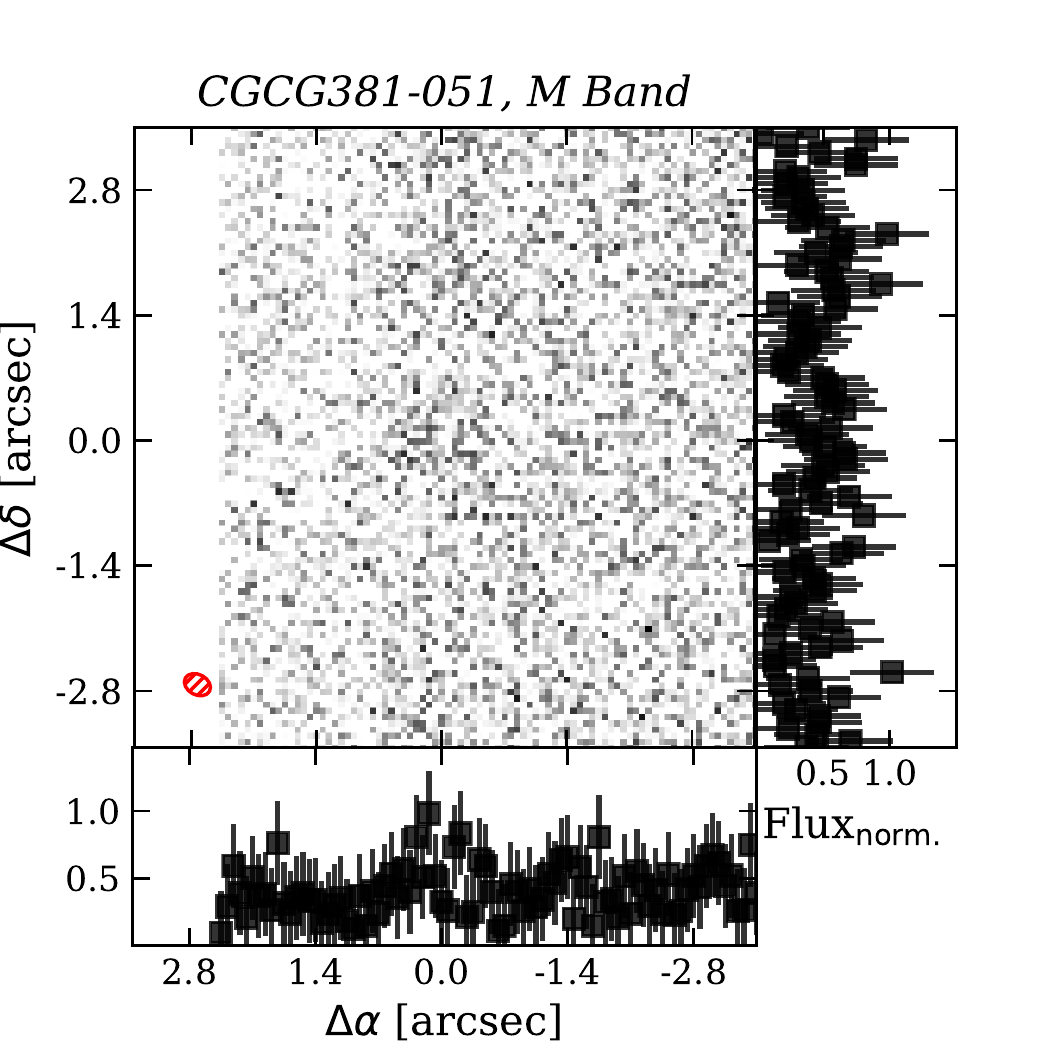}} \\
\caption{ As Fig \ref{fig:cutouts_one} but for all sources.}
\end{figure*}
\begin{figure*}
\subfloat{\includegraphics[width=0.25\hsize]{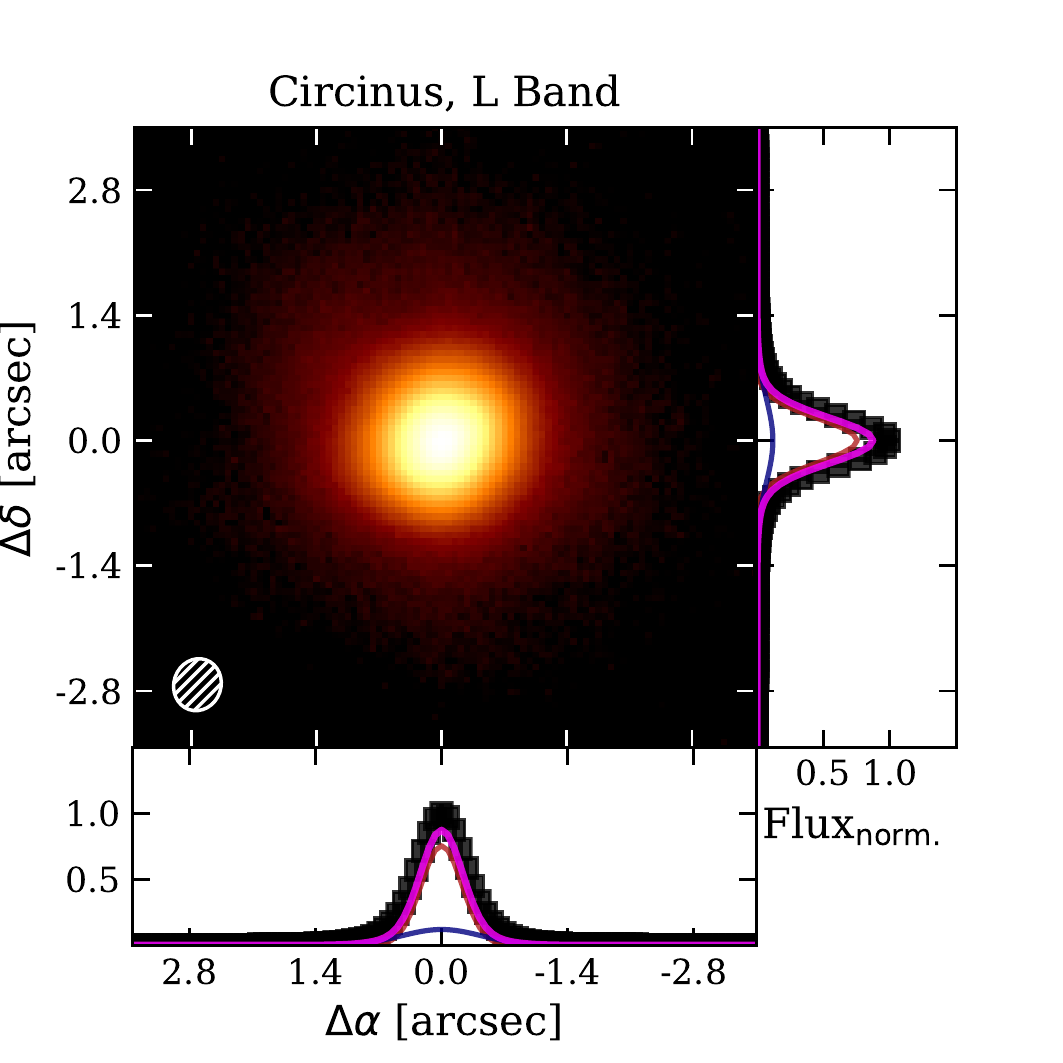}}
\subfloat{\includegraphics[width=0.25\hsize]{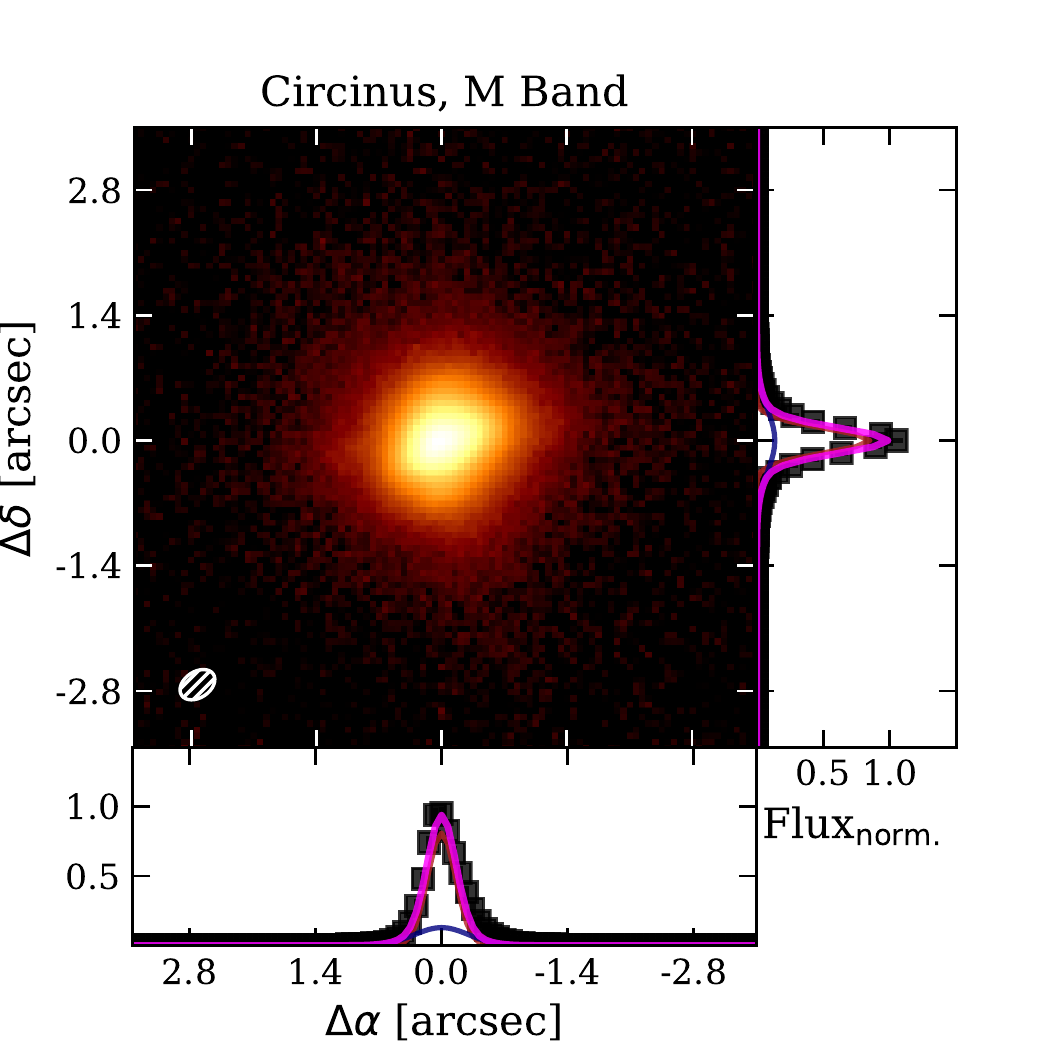}}
\subfloat{\includegraphics[width=0.25\hsize]{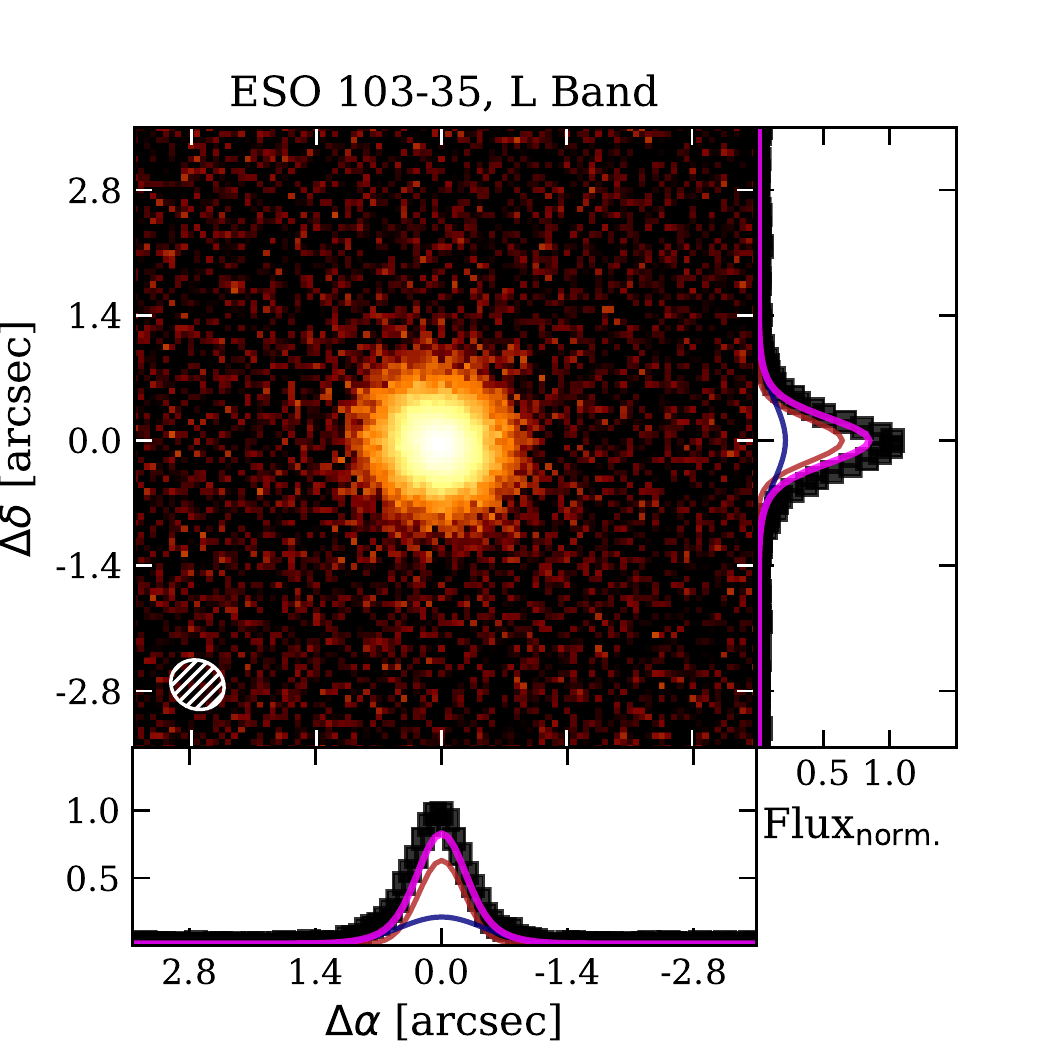}}
\subfloat{\includegraphics[width=0.25\hsize]{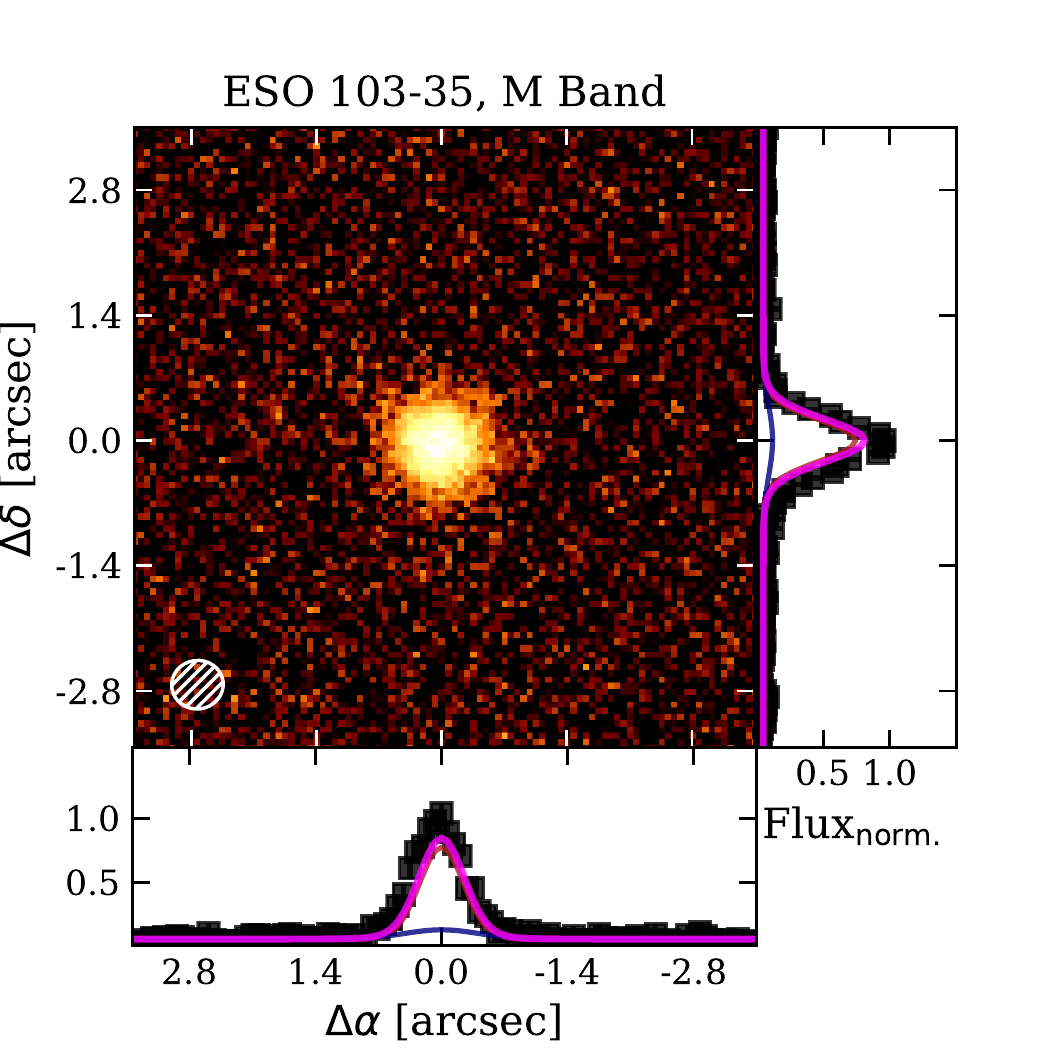}} \\
\subfloat{\includegraphics[width=0.25\hsize]{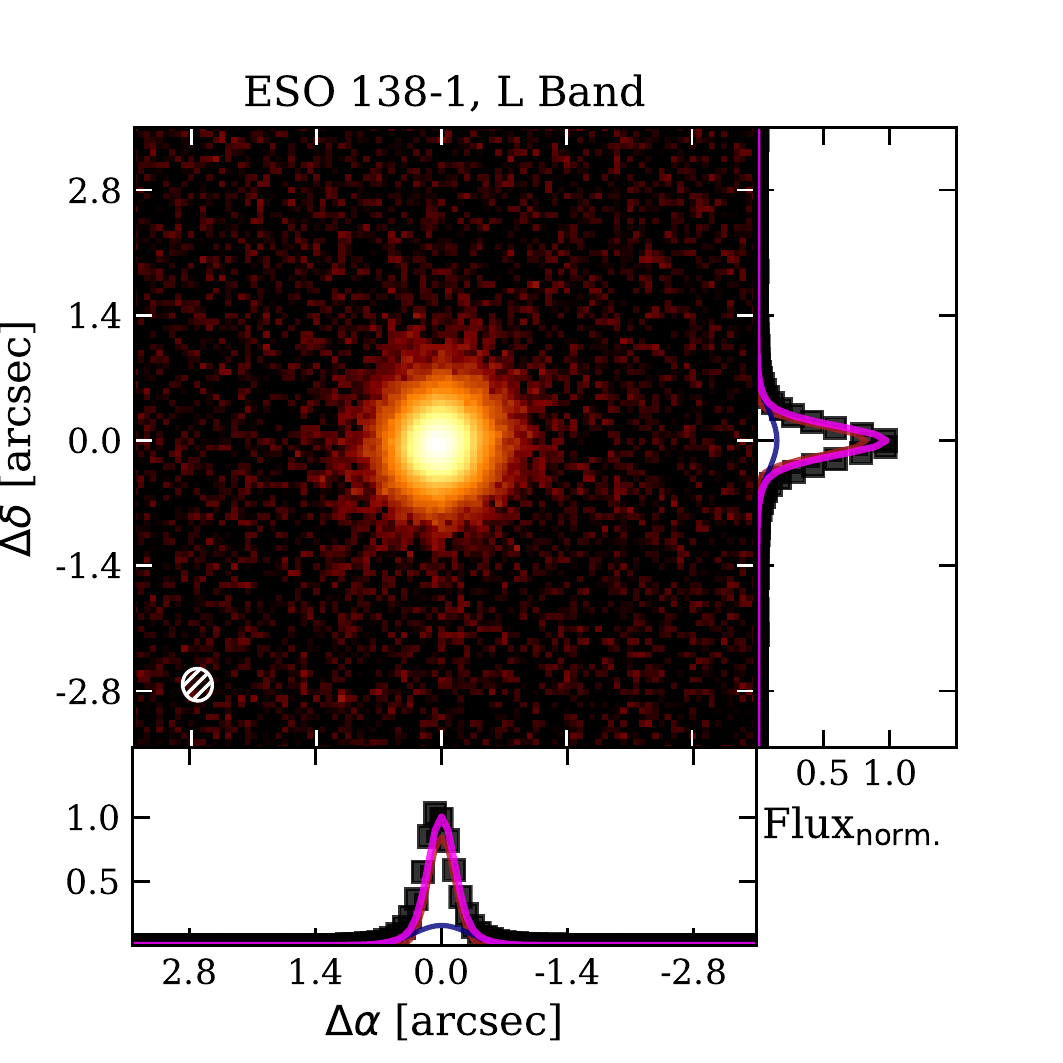}}
\subfloat{\includegraphics[width=0.25\hsize]{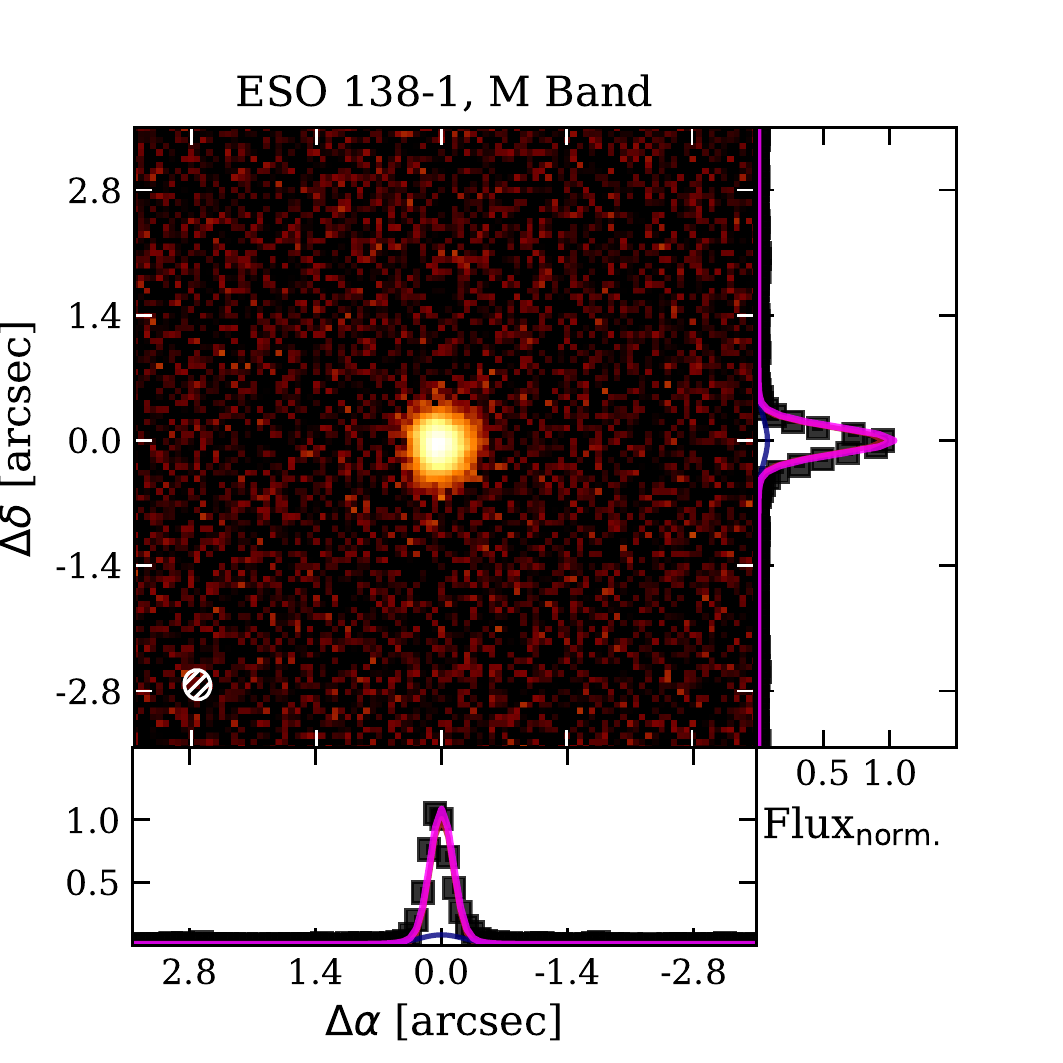}}
\subfloat{\includegraphics[width=0.25\hsize]{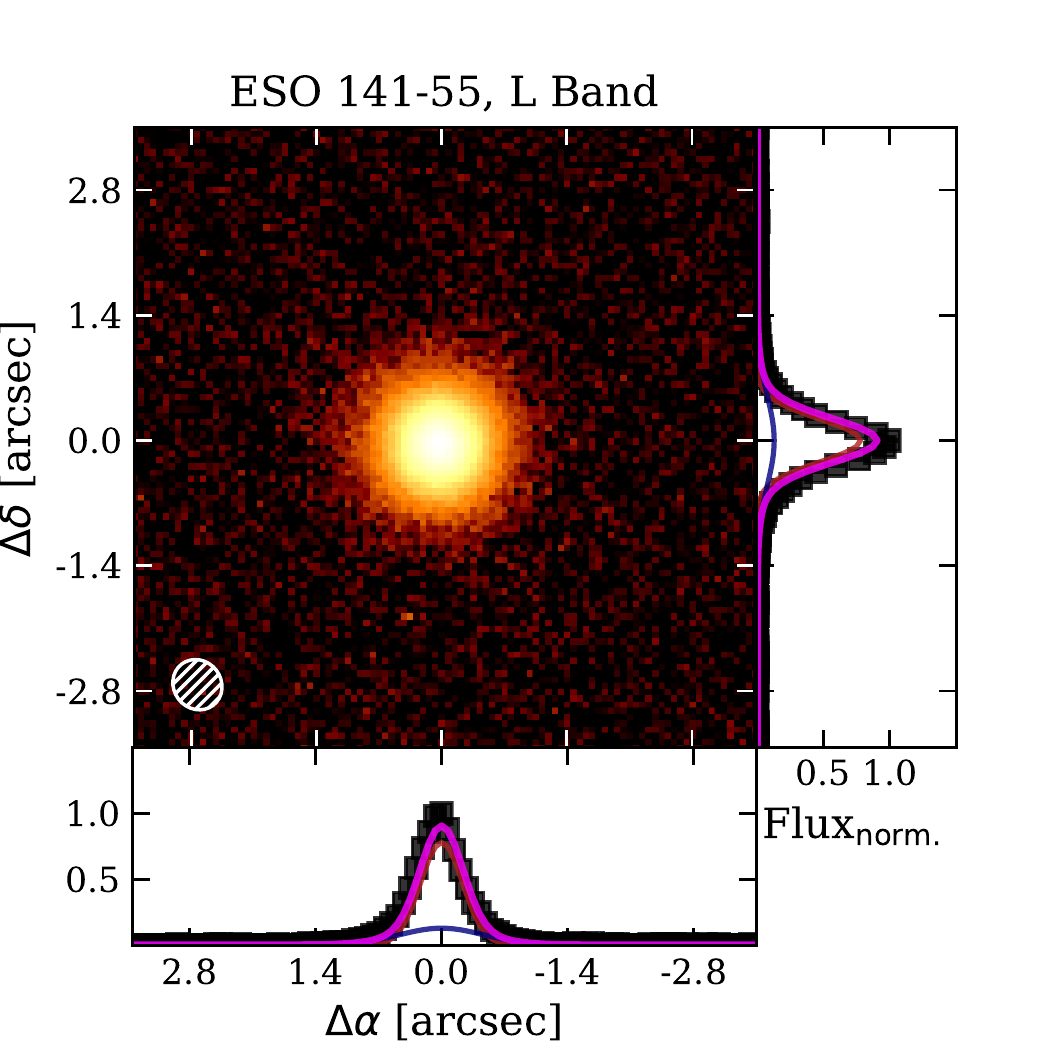}}
\subfloat{\includegraphics[width=0.25\hsize]{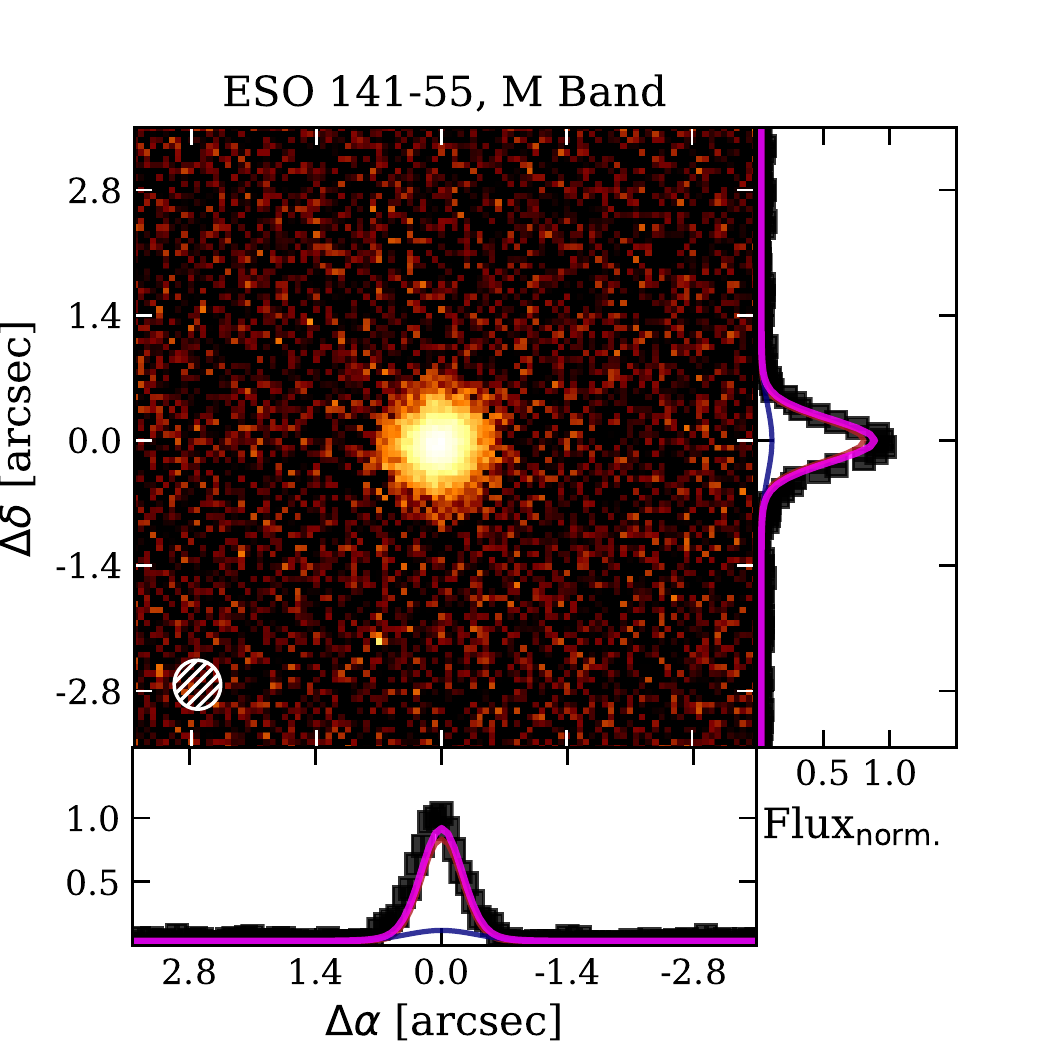}} \\
\subfloat{\includegraphics[width=0.25\hsize]{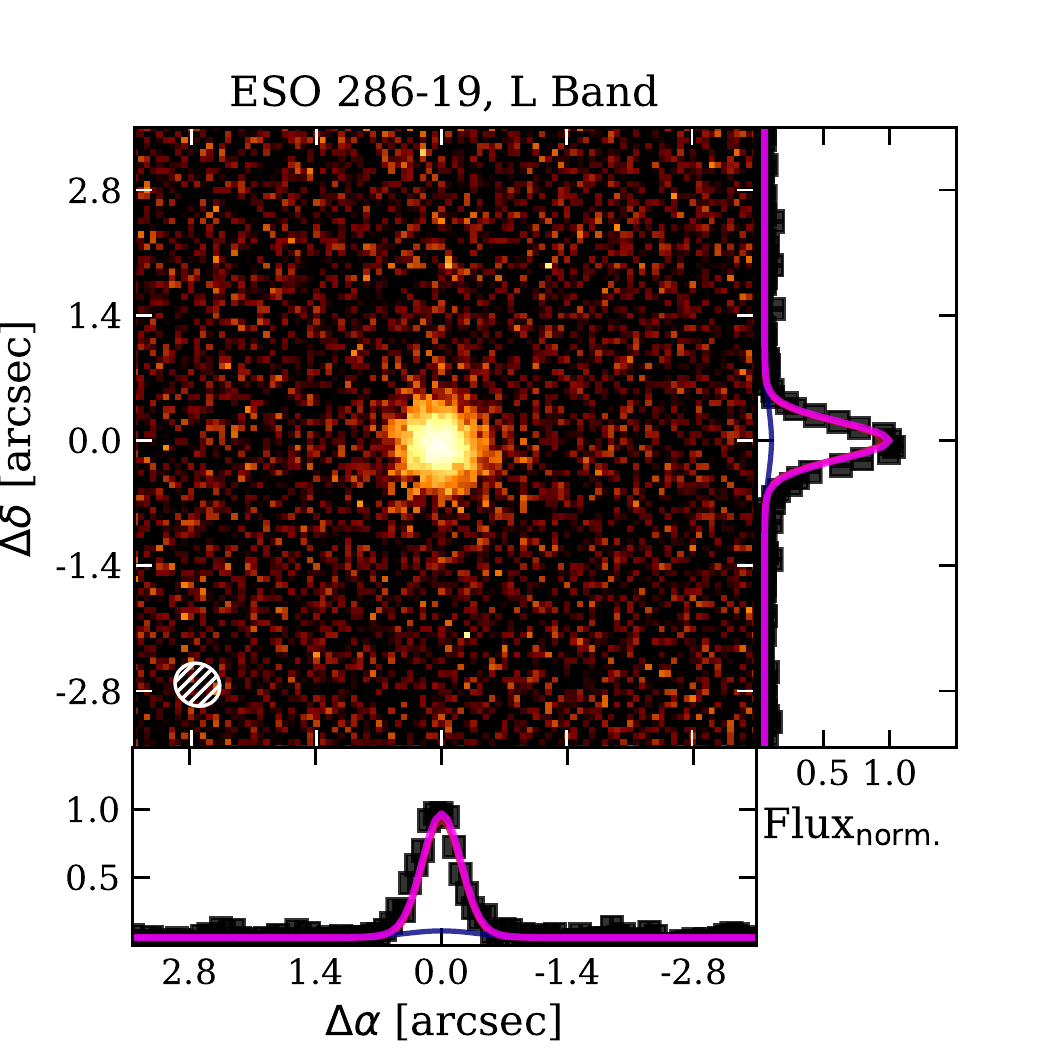}}
\subfloat{\includegraphics[width=0.25\hsize]{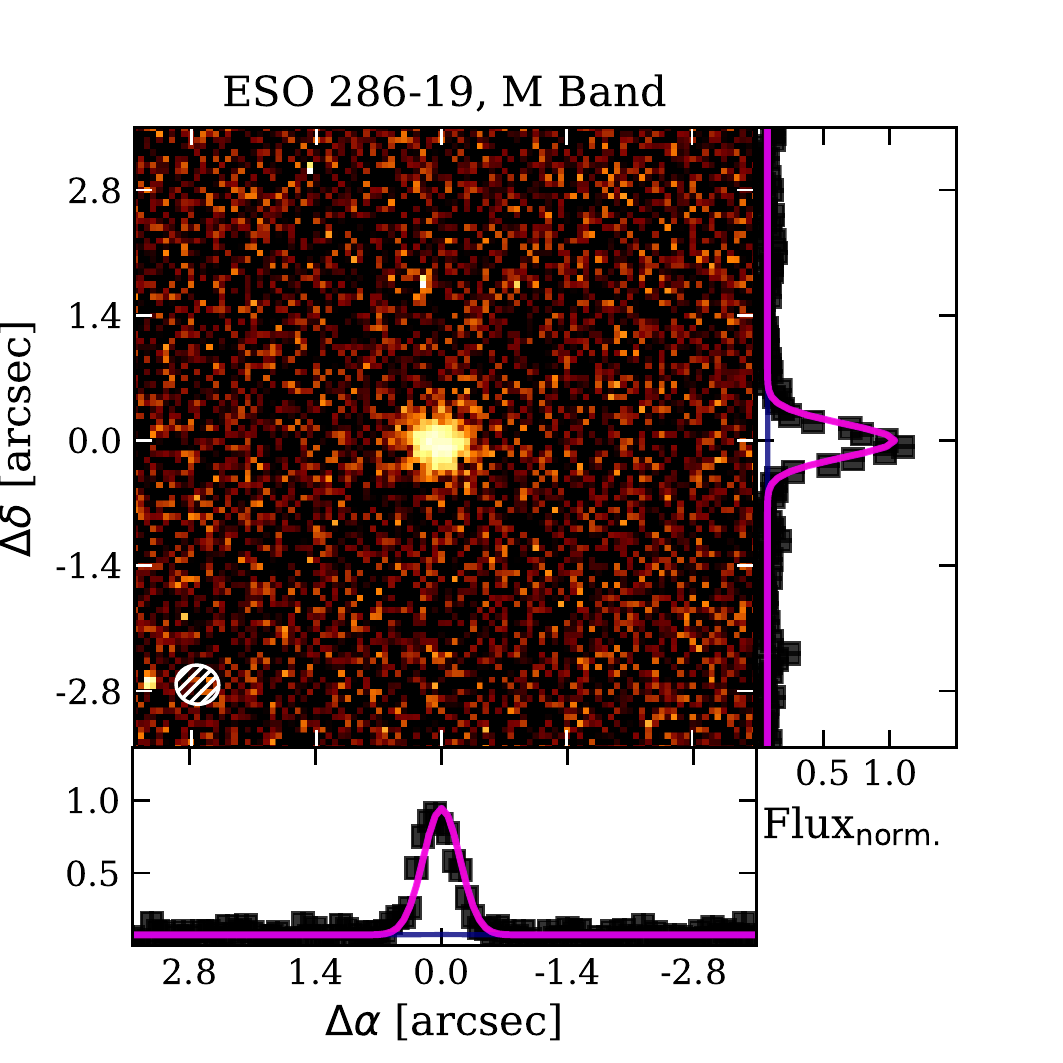}}
\subfloat{\includegraphics[width=0.25\hsize]{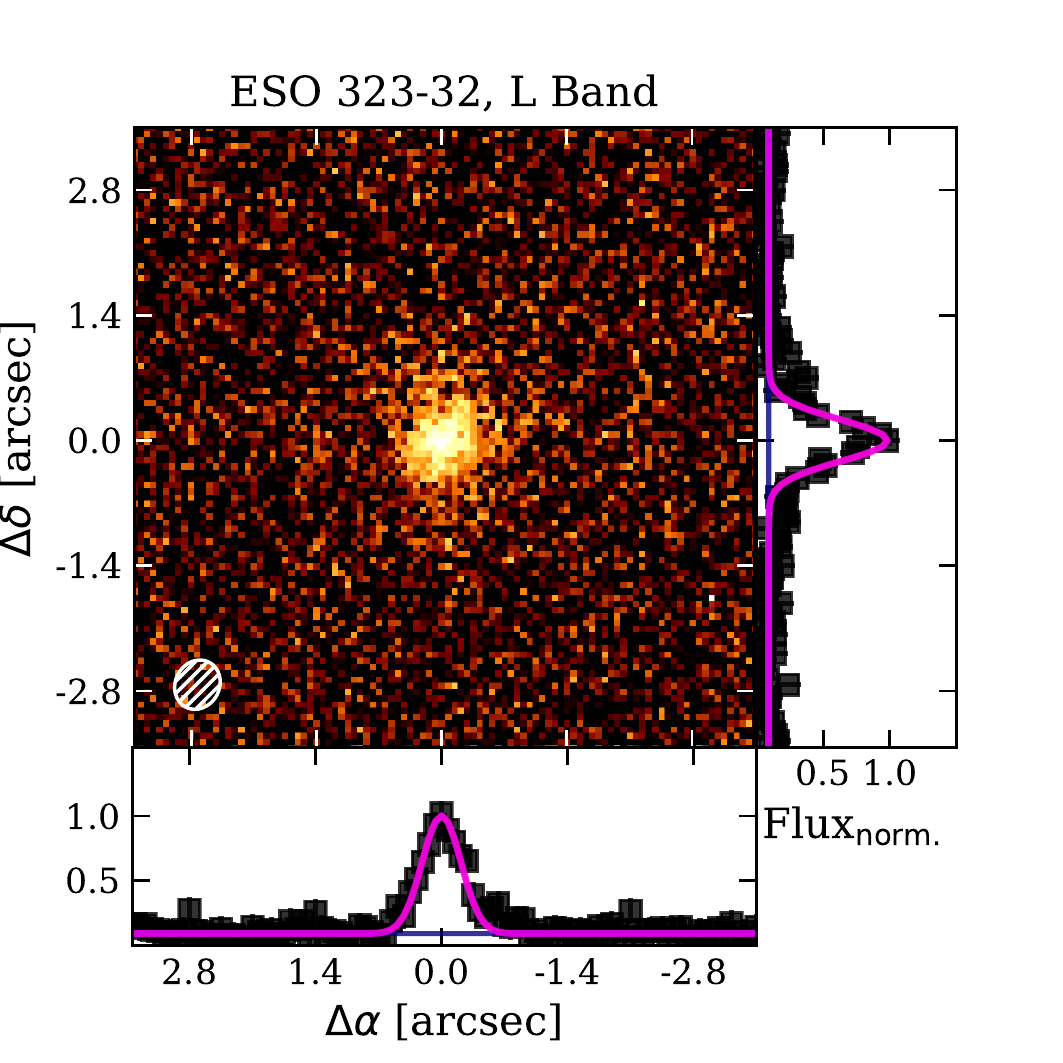}}
\subfloat{\includegraphics[width=0.25\hsize]{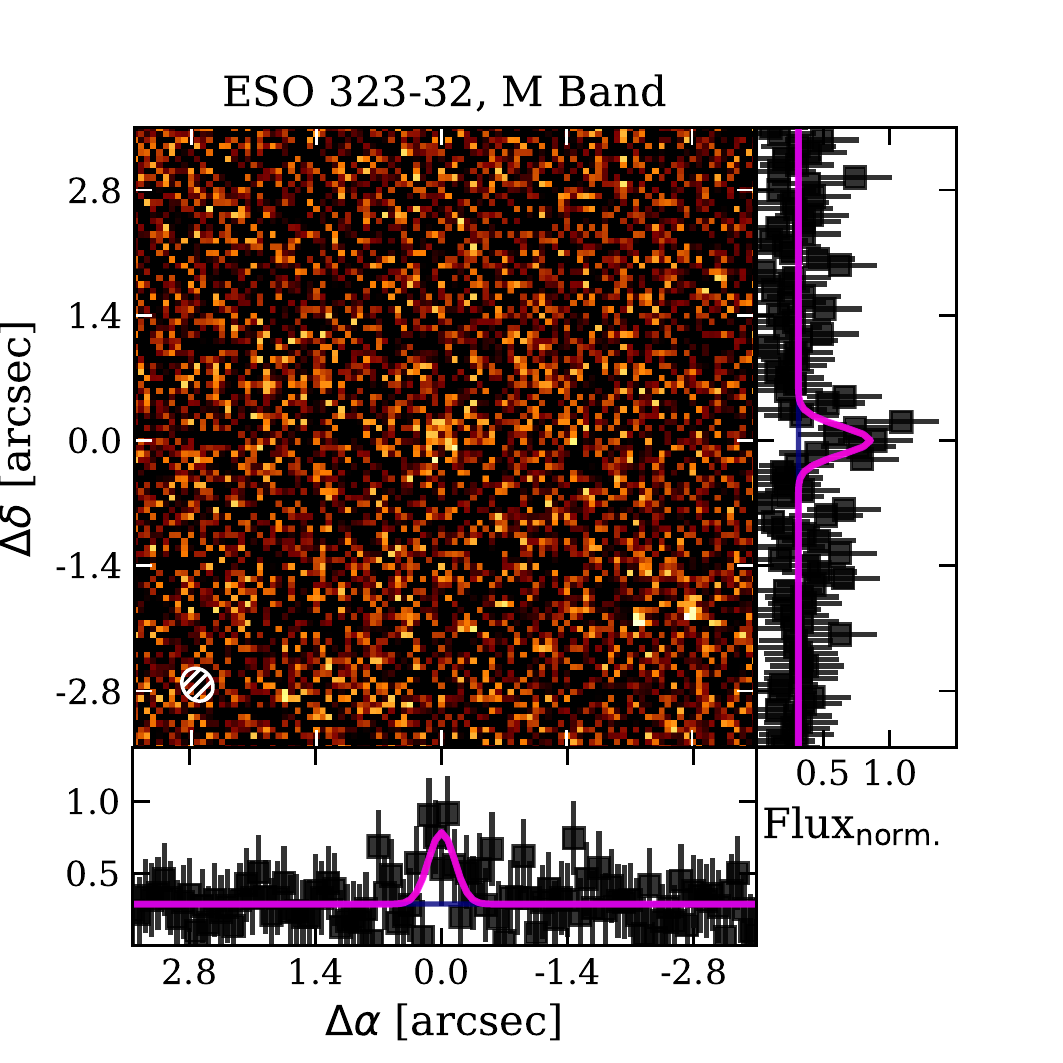}} \\
\subfloat{\includegraphics[width=0.25\hsize]{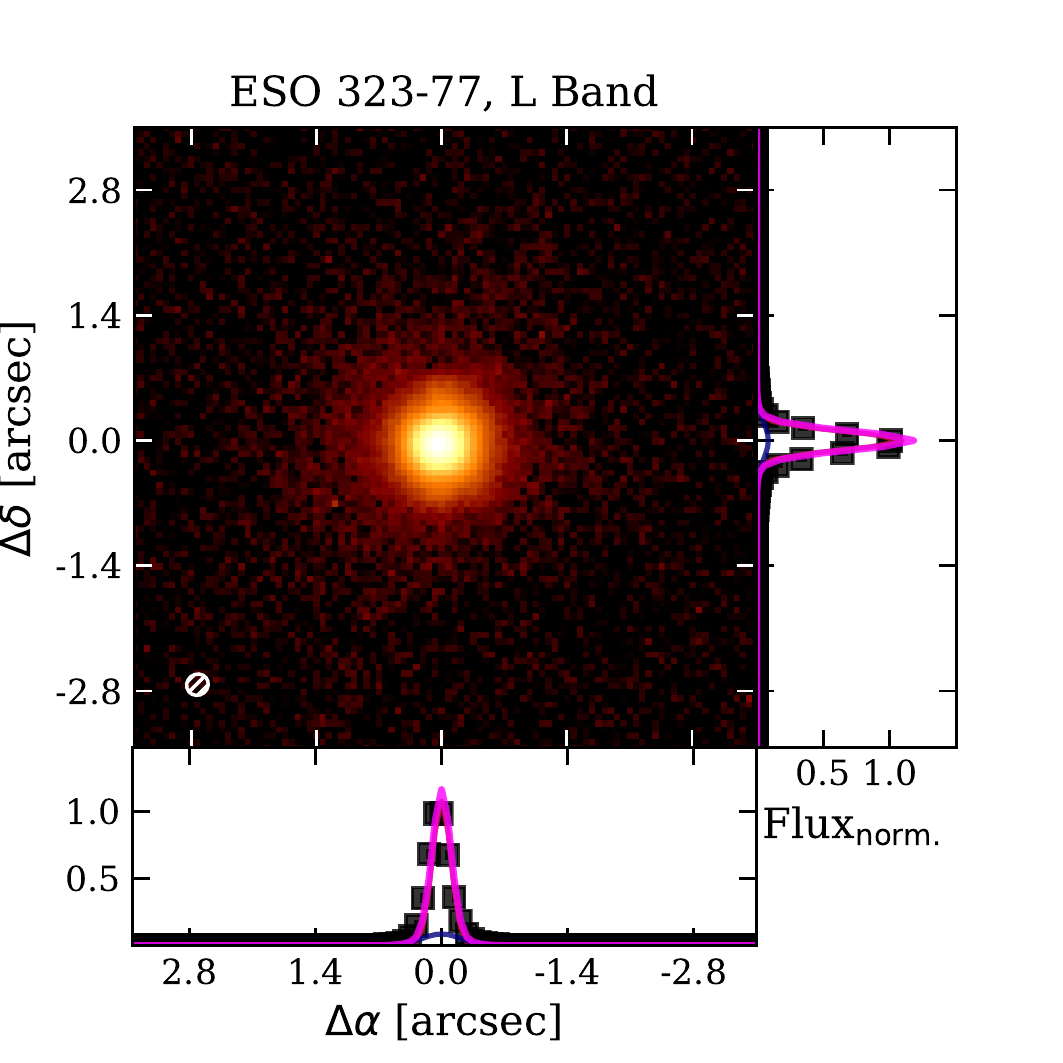}}
\subfloat{\includegraphics[width=0.25\hsize]{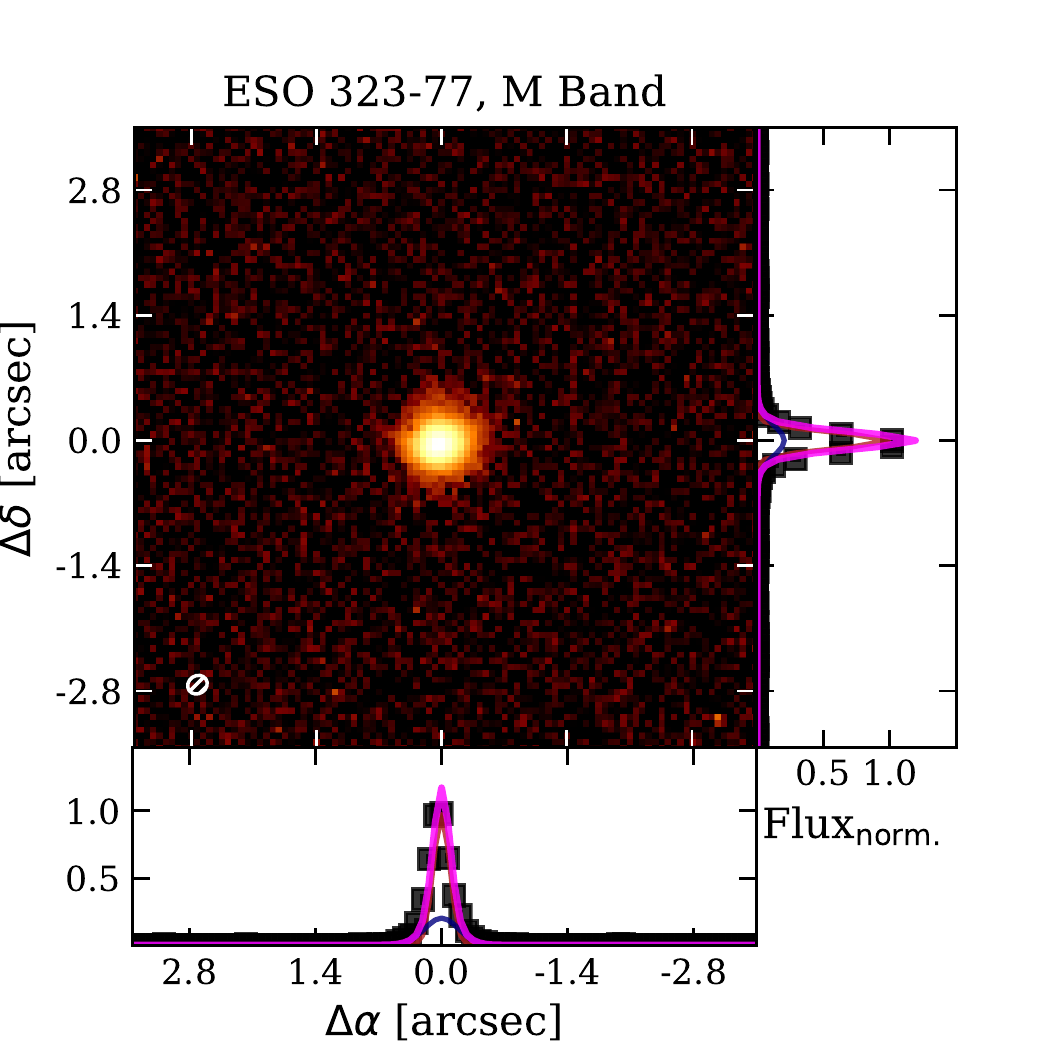}}
\subfloat{\includegraphics[width=0.25\hsize]{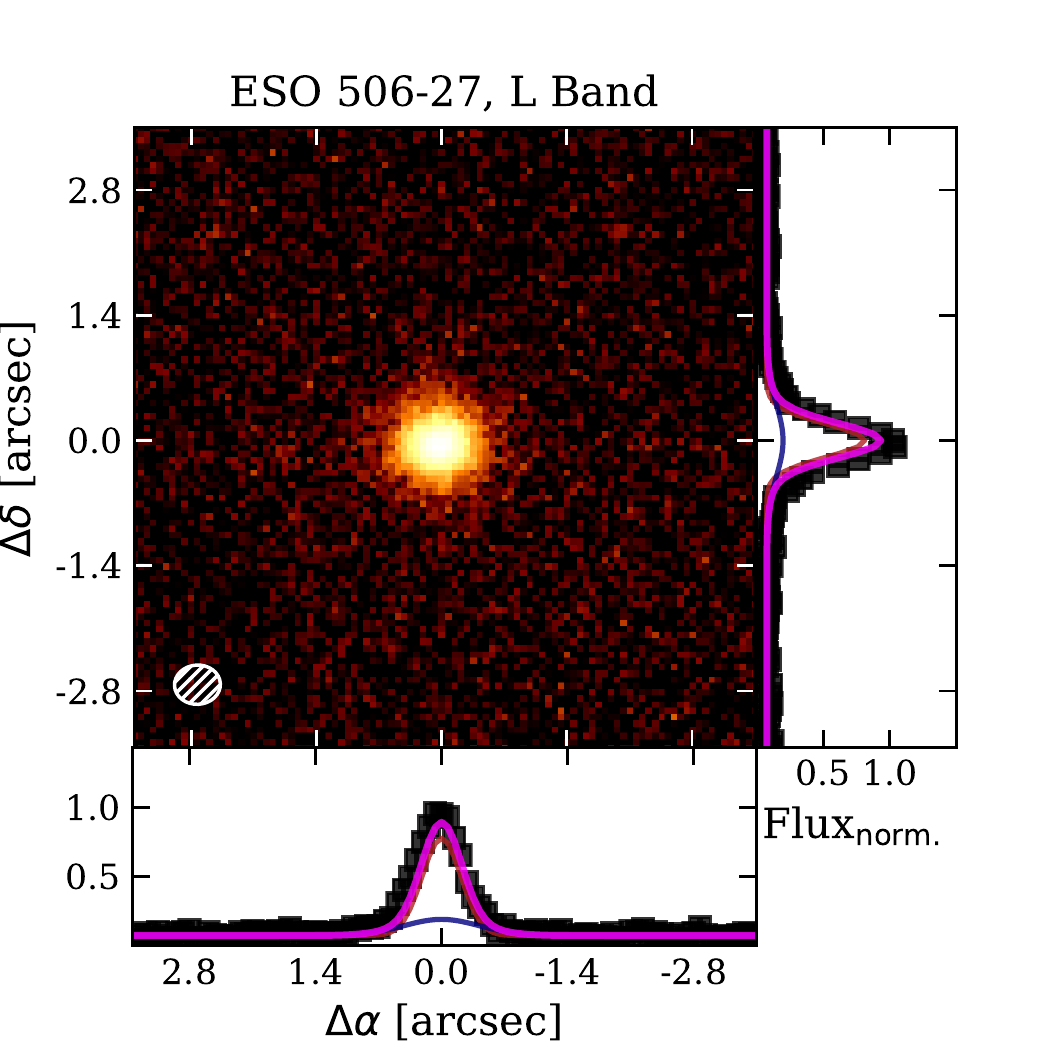}}
\subfloat{\includegraphics[width=0.25\hsize]{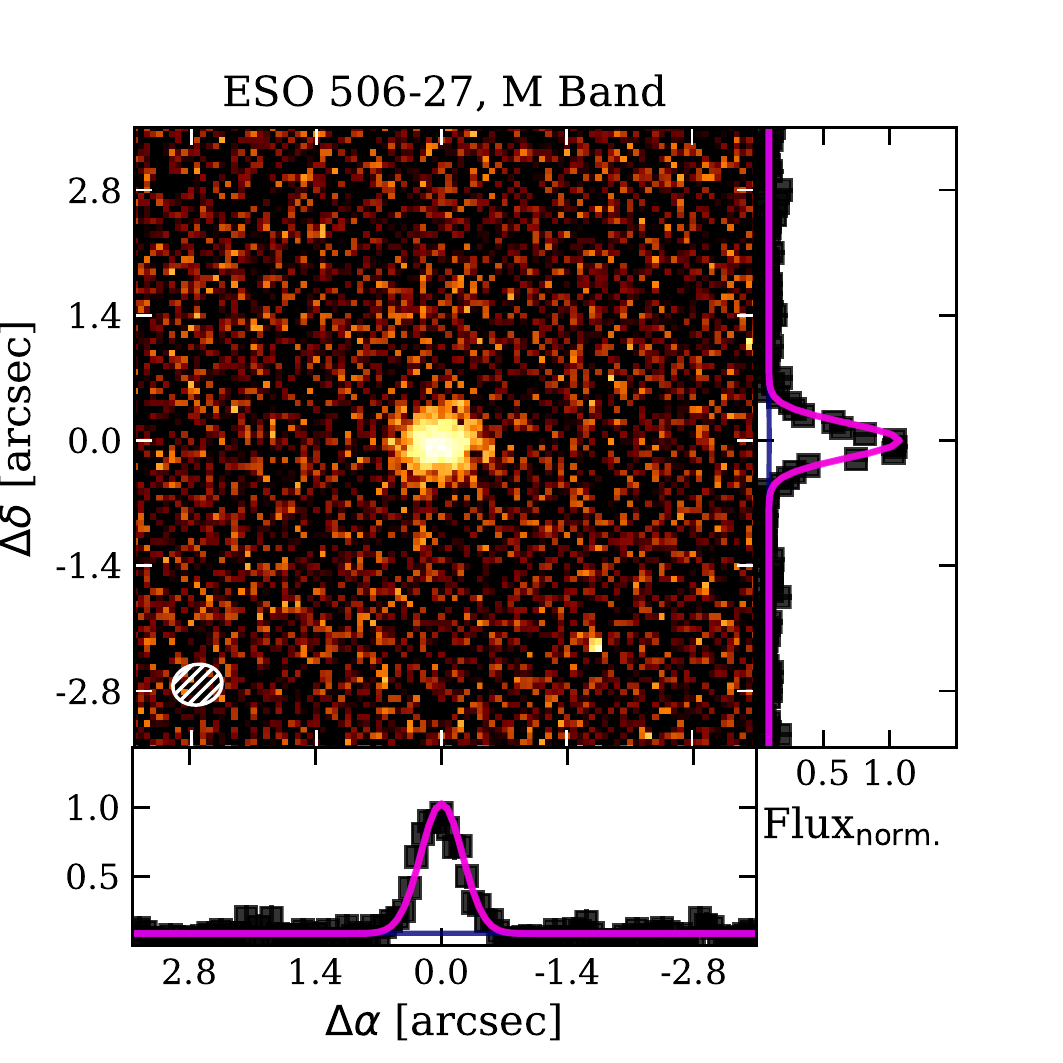}} \\
\subfloat{\includegraphics[width=0.25\hsize]{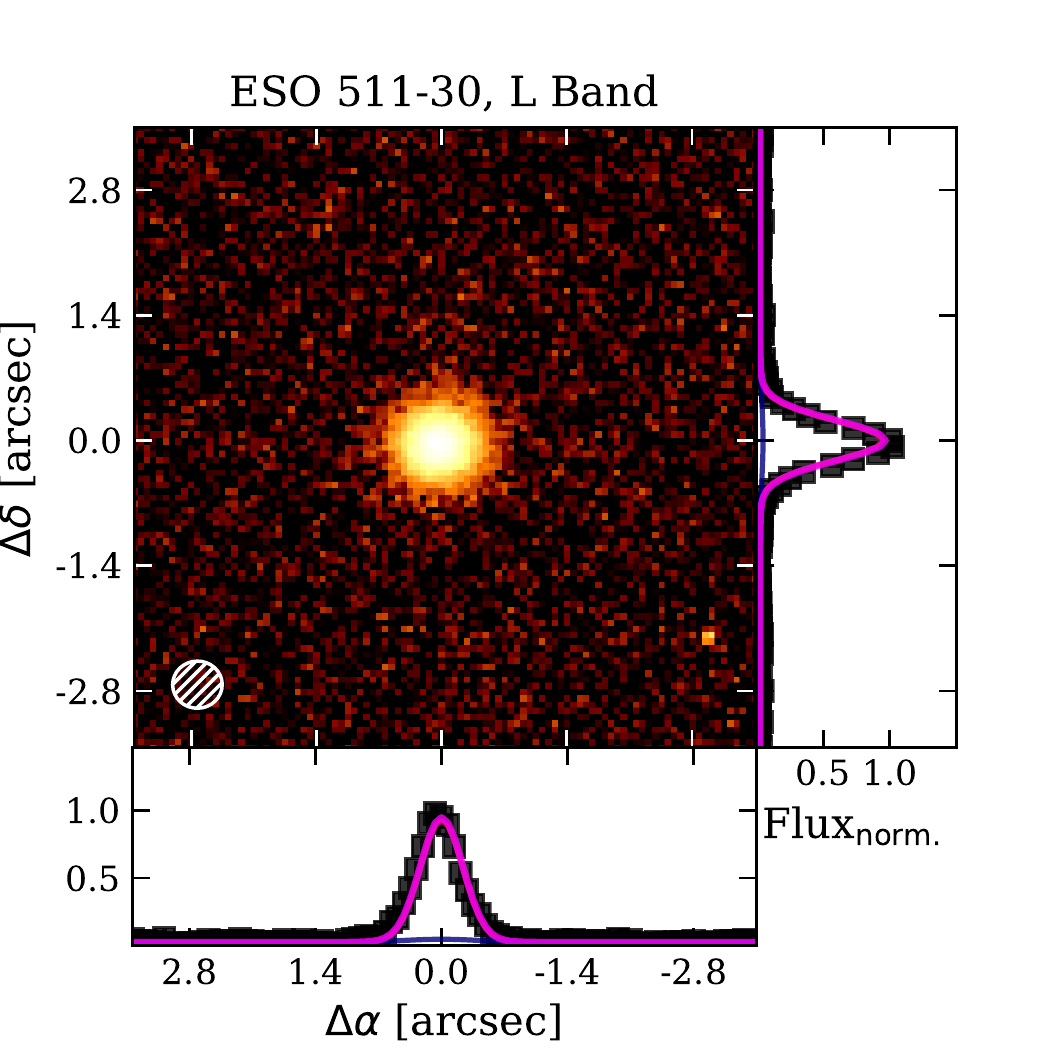}}
\subfloat{\includegraphics[width=0.25\hsize]{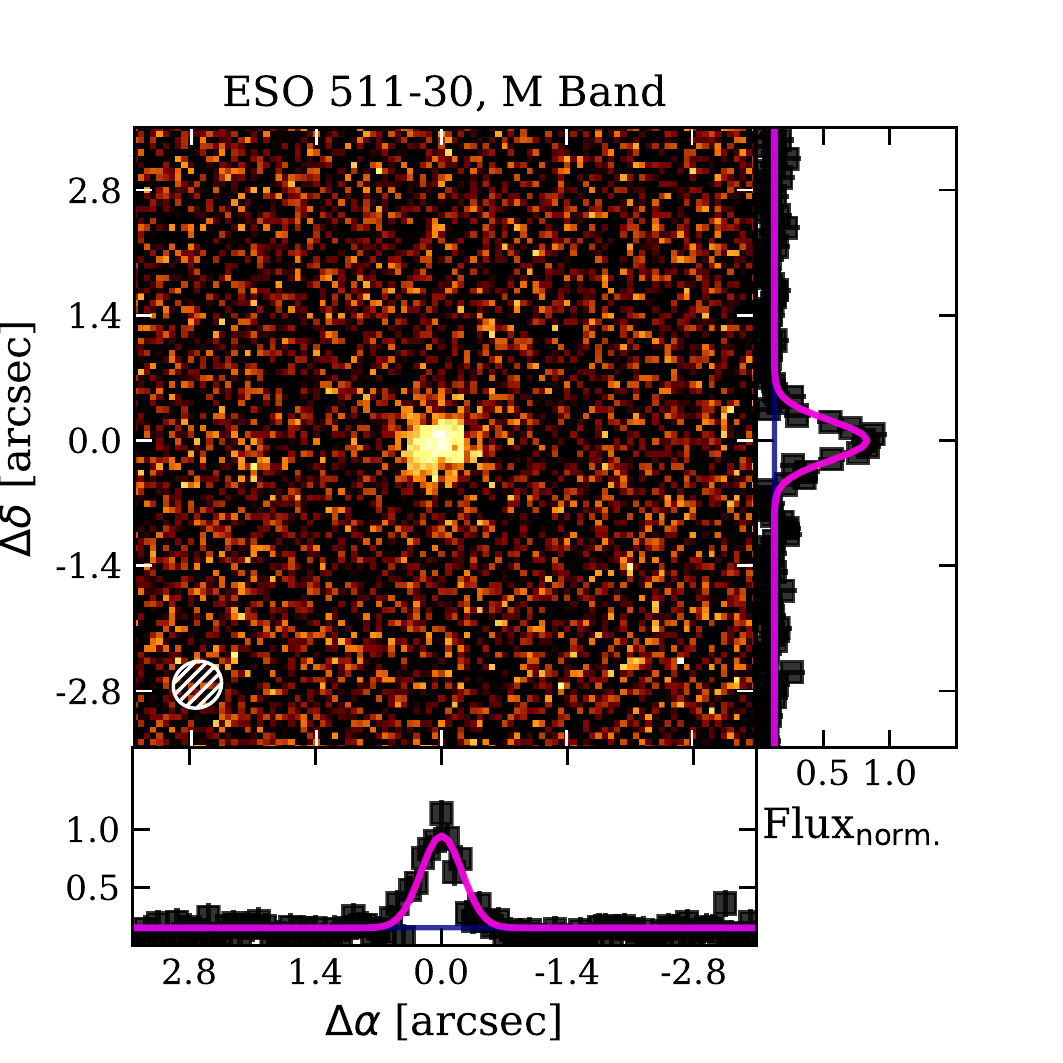}}
\subfloat{\includegraphics[width=0.25\hsize]{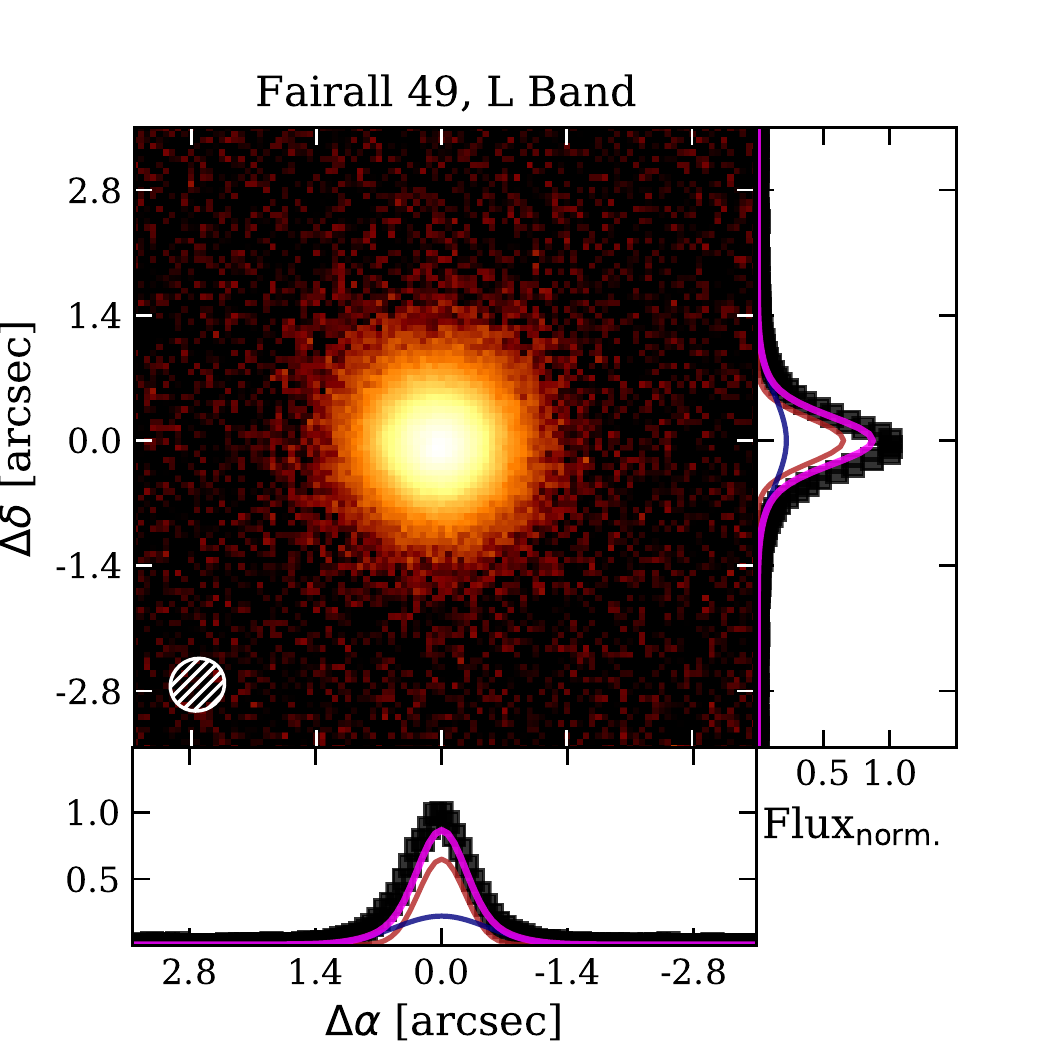}}
\subfloat{\includegraphics[width=0.25\hsize]{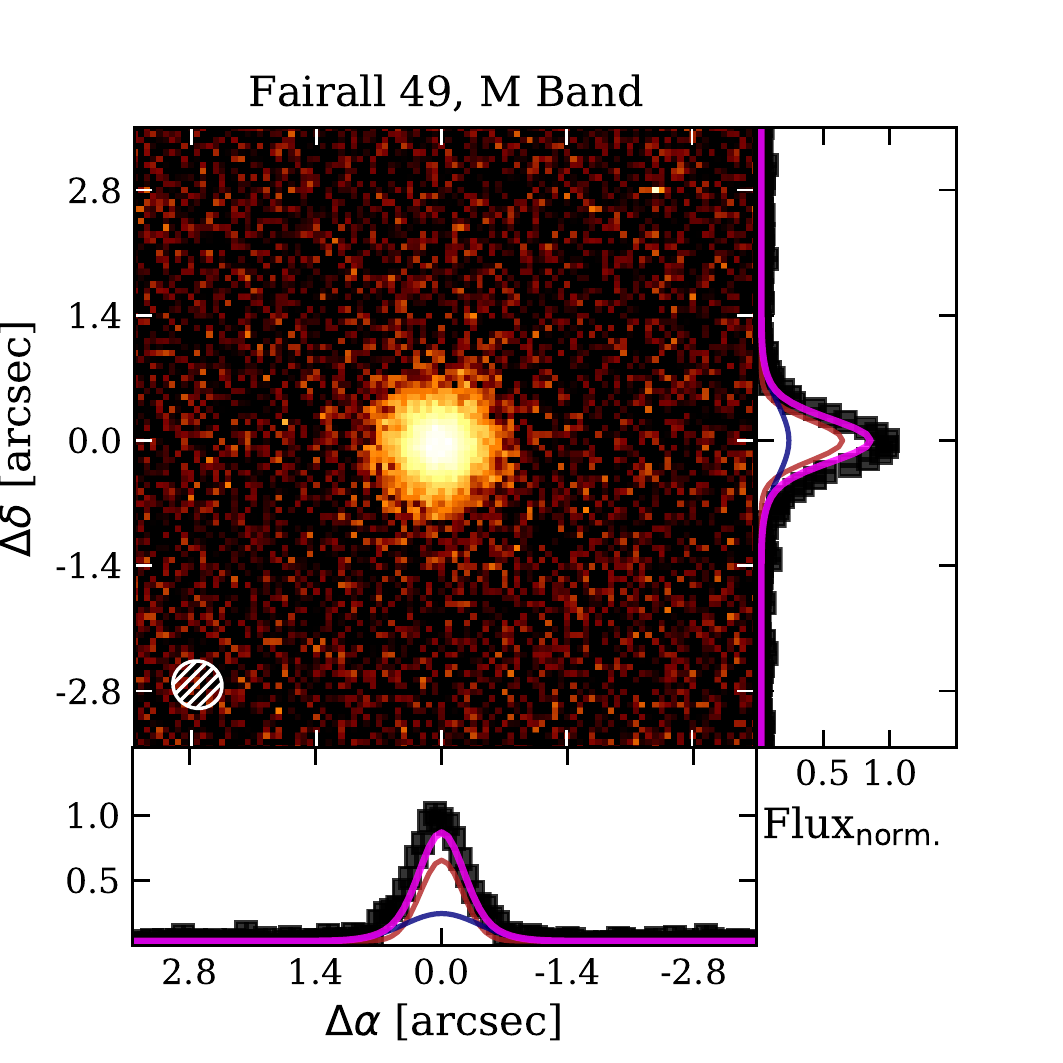}} \\
\caption{ As Fig \ref{fig:cutouts_one} but for all sources.}
\end{figure*}
\begin{figure*}
\subfloat{\includegraphics[width=0.25\hsize]{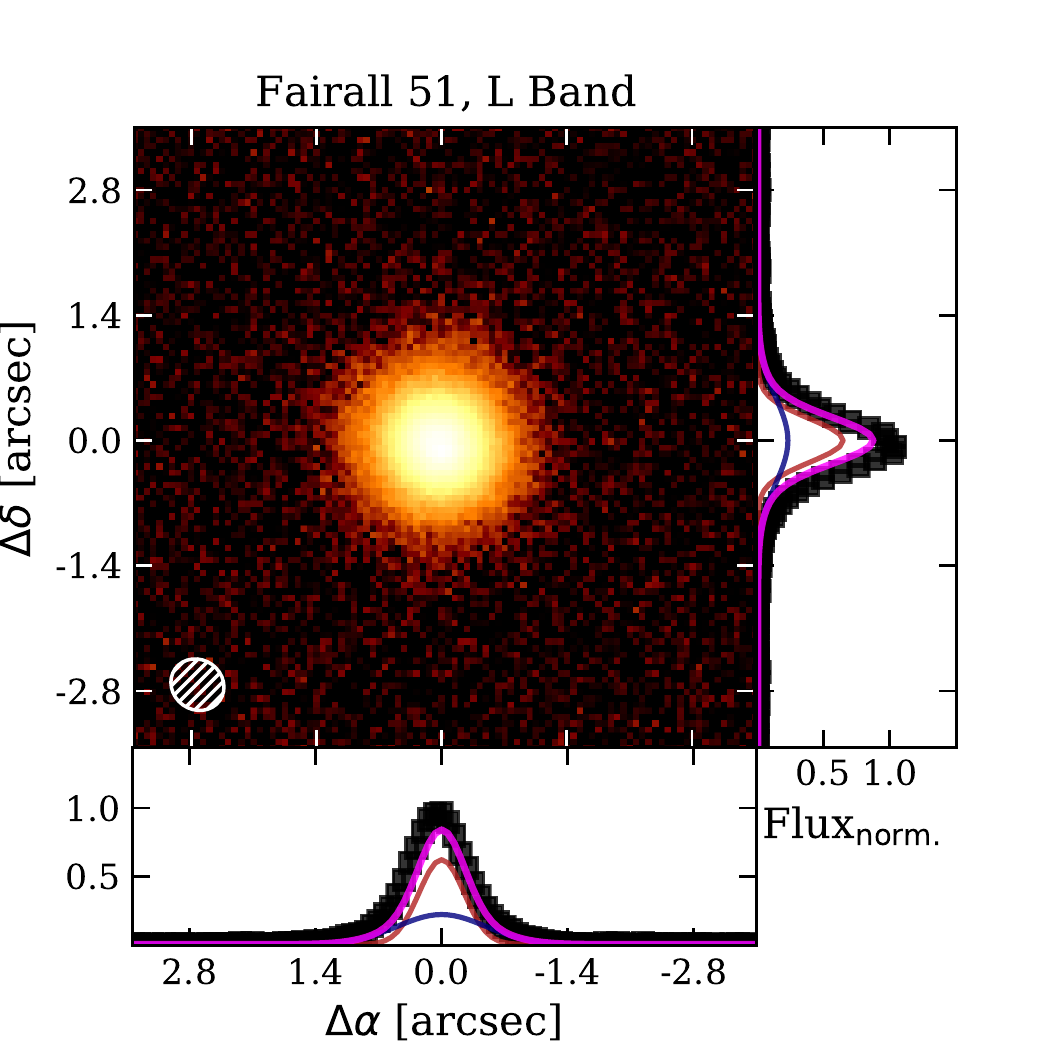}}
\subfloat{\includegraphics[width=0.25\hsize]{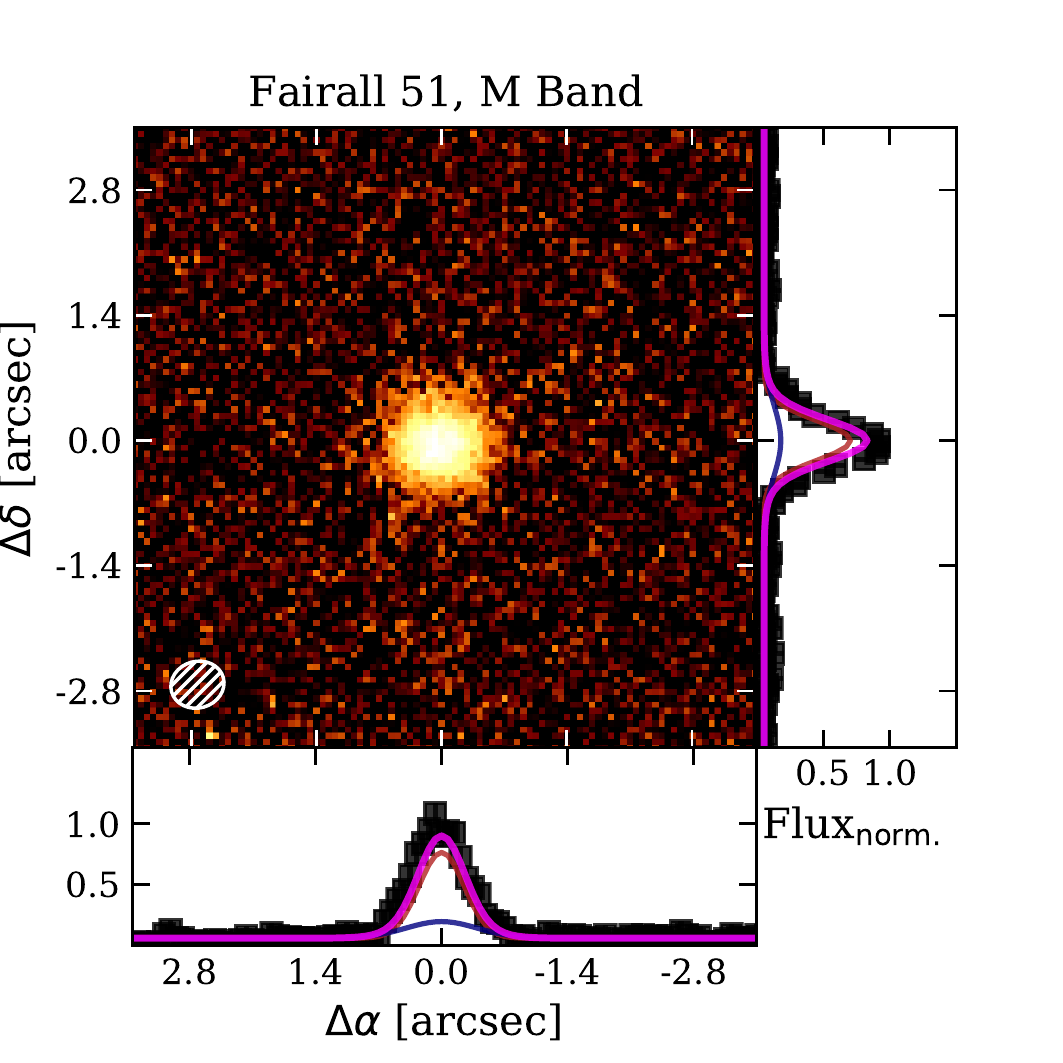}}
\subfloat{\includegraphics[width=0.25\hsize]{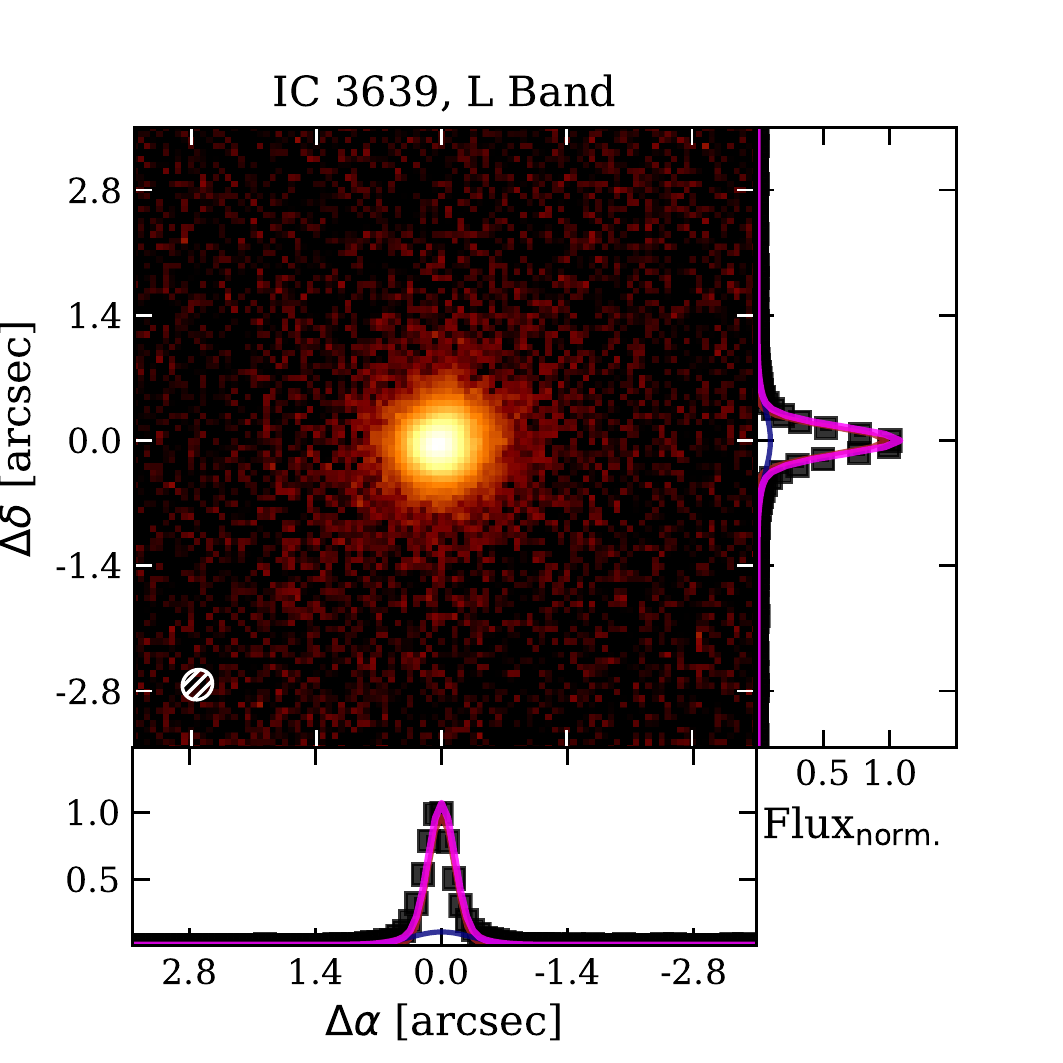}}
\subfloat{\includegraphics[width=0.25\hsize]{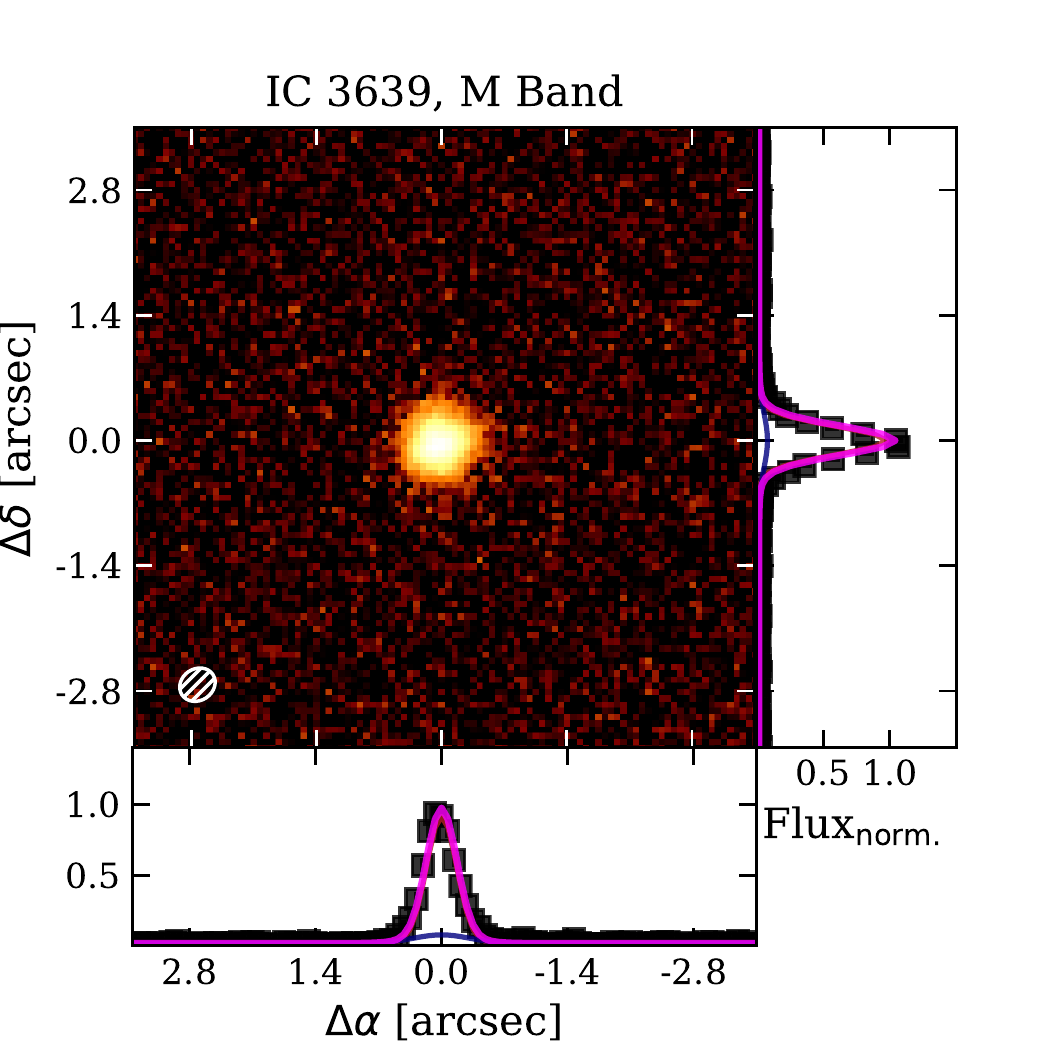}} \\
\subfloat{\includegraphics[width=0.25\hsize]{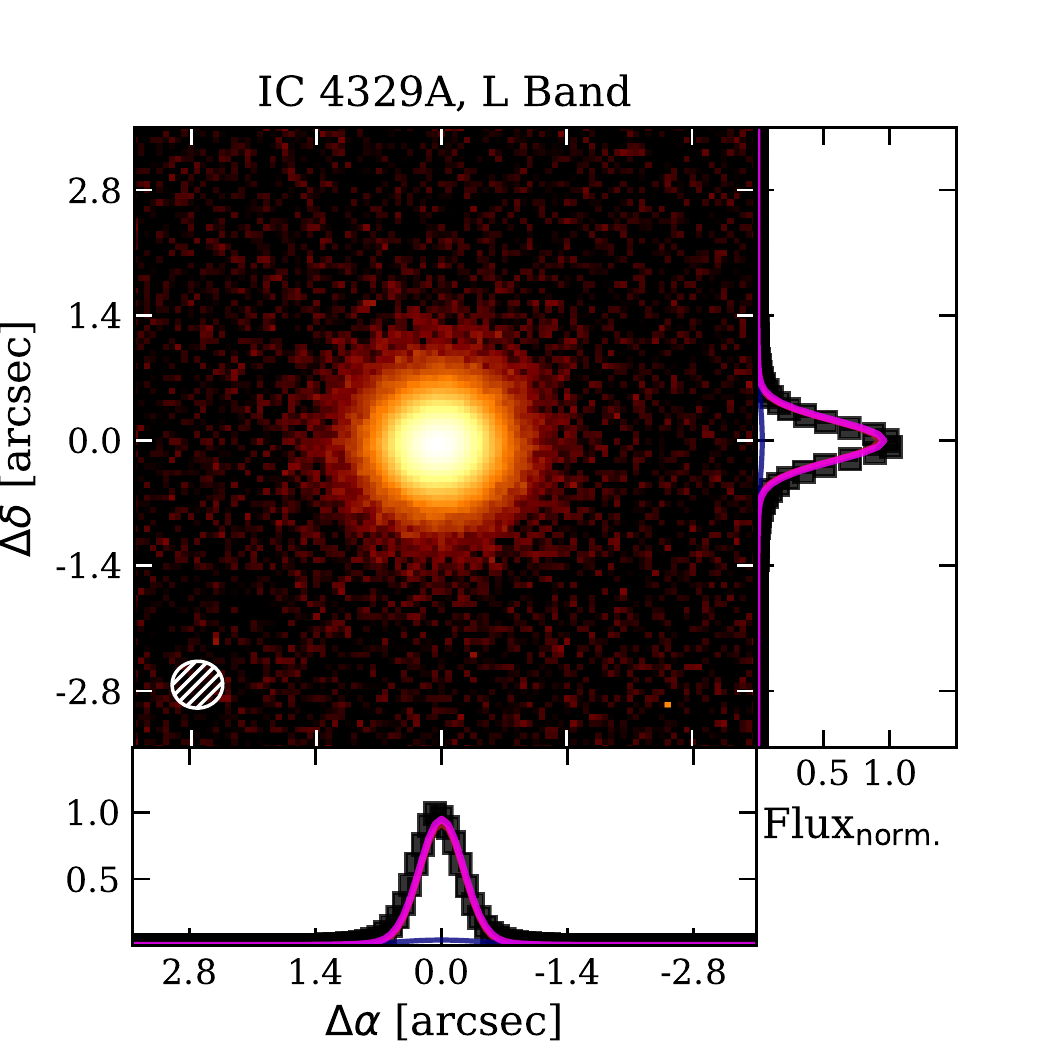}}
\subfloat{\includegraphics[width=0.25\hsize]{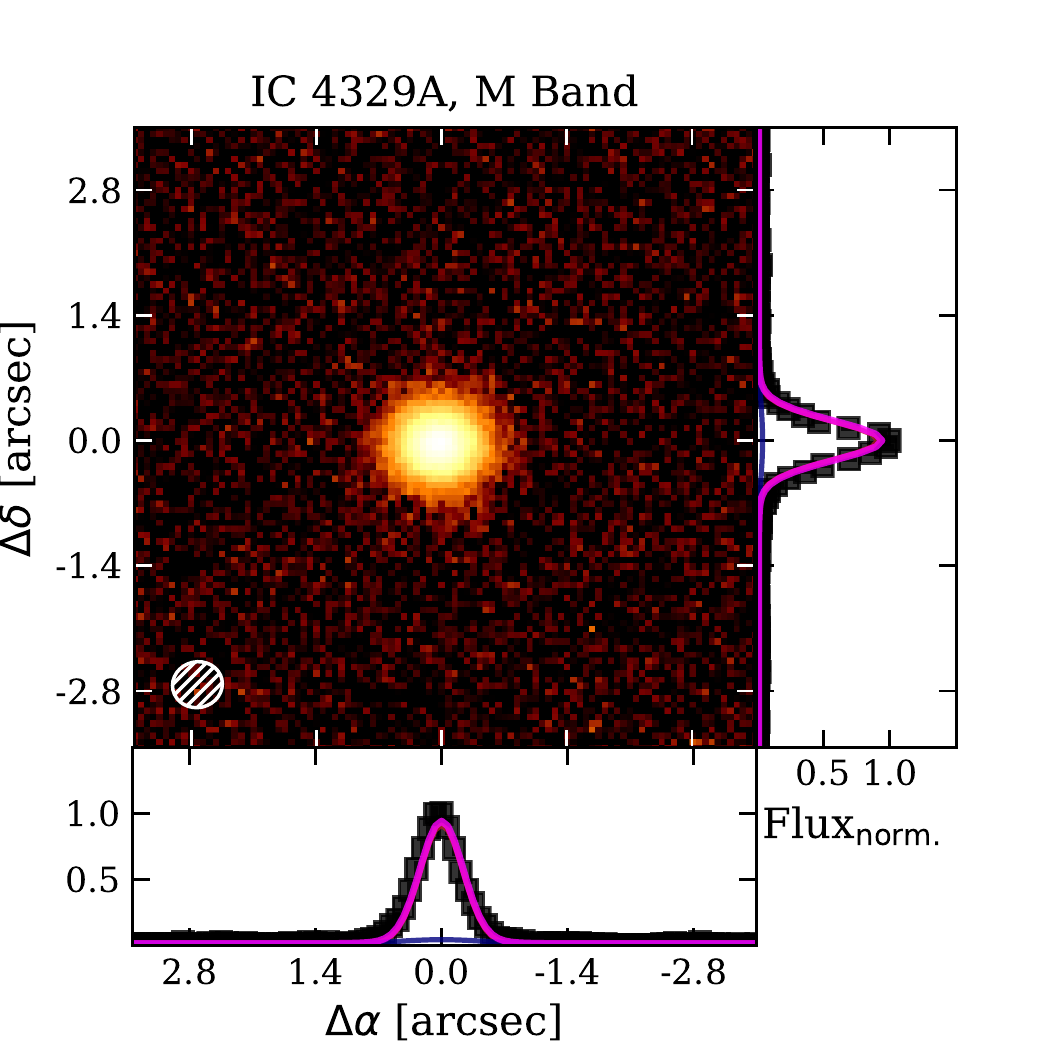}}
\subfloat{\includegraphics[width=0.25\hsize]{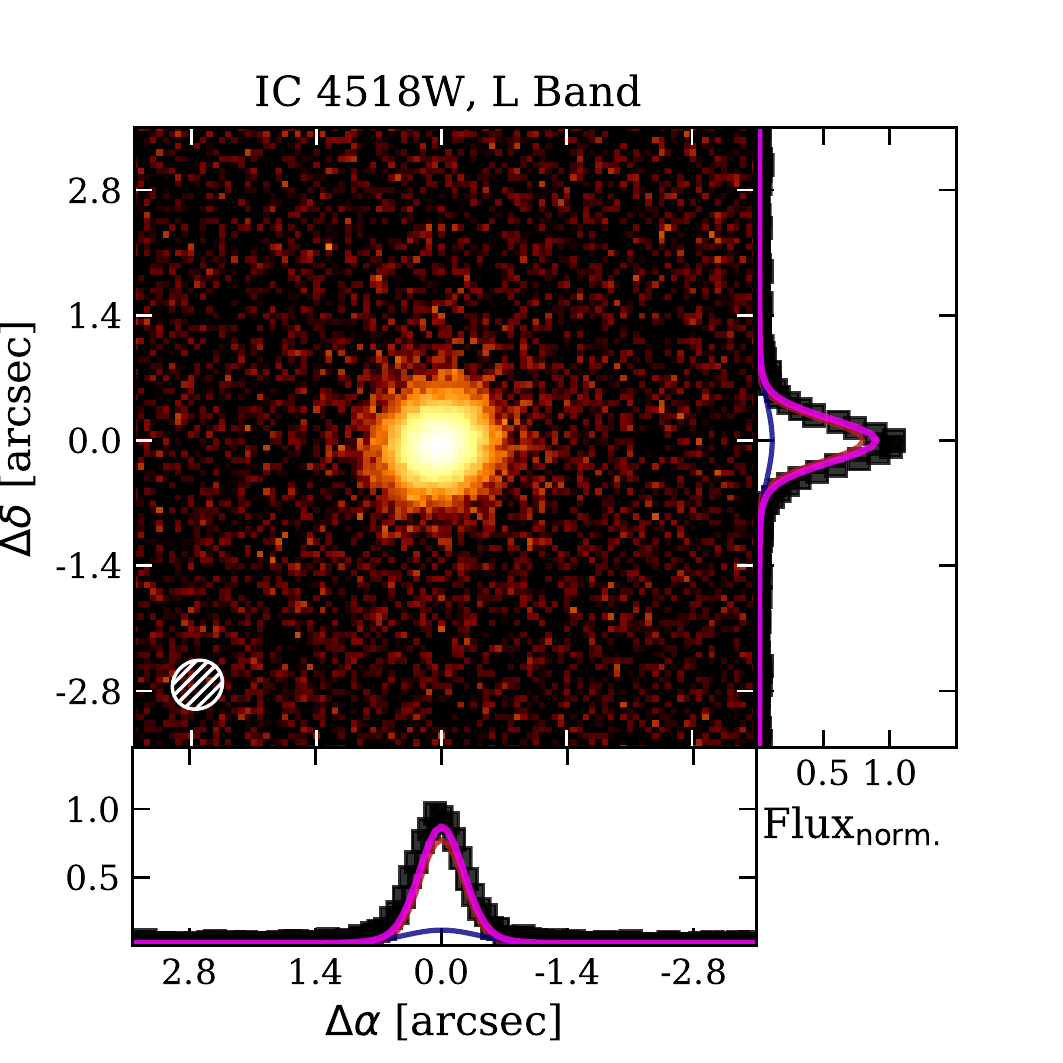}}
\subfloat{\includegraphics[width=0.25\hsize]{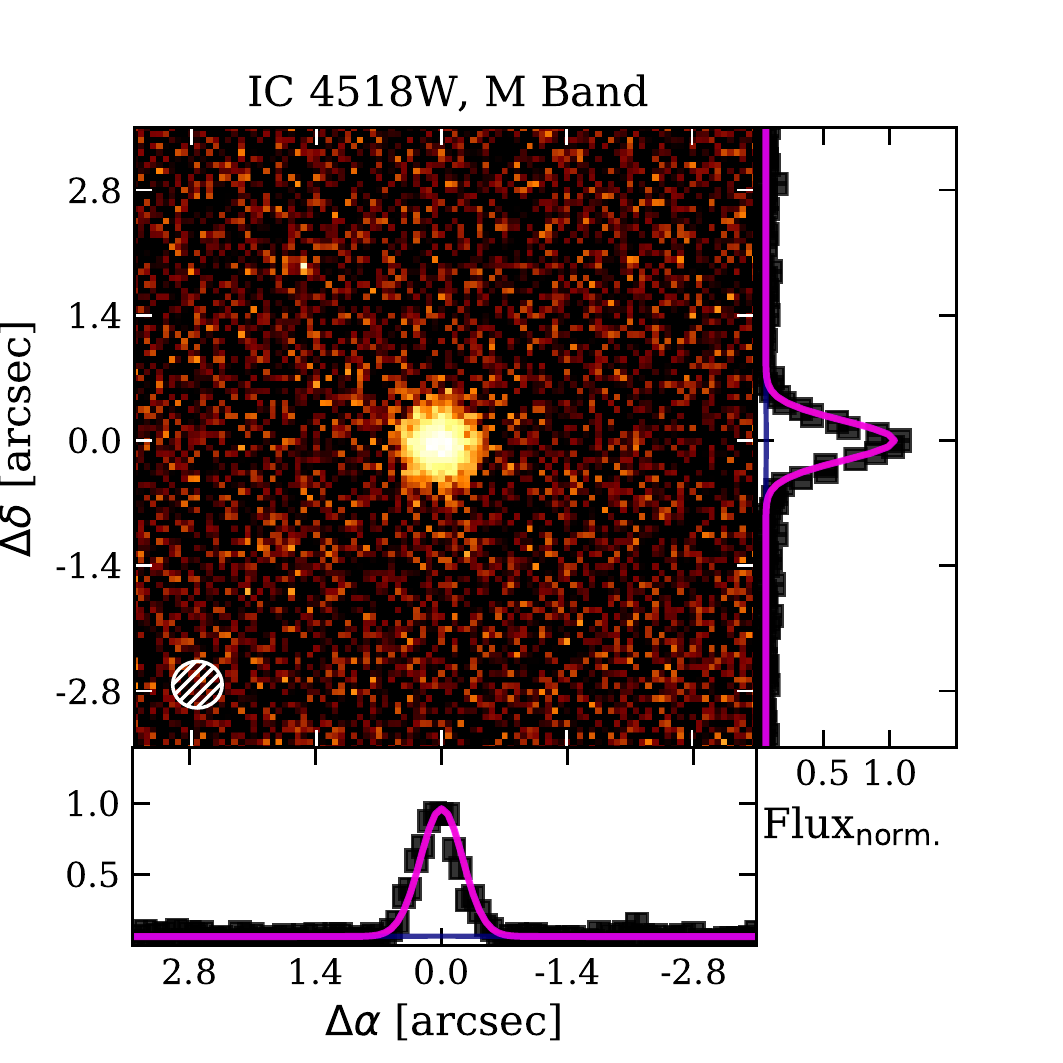}}\\
\subfloat{\includegraphics[width=0.25\hsize]{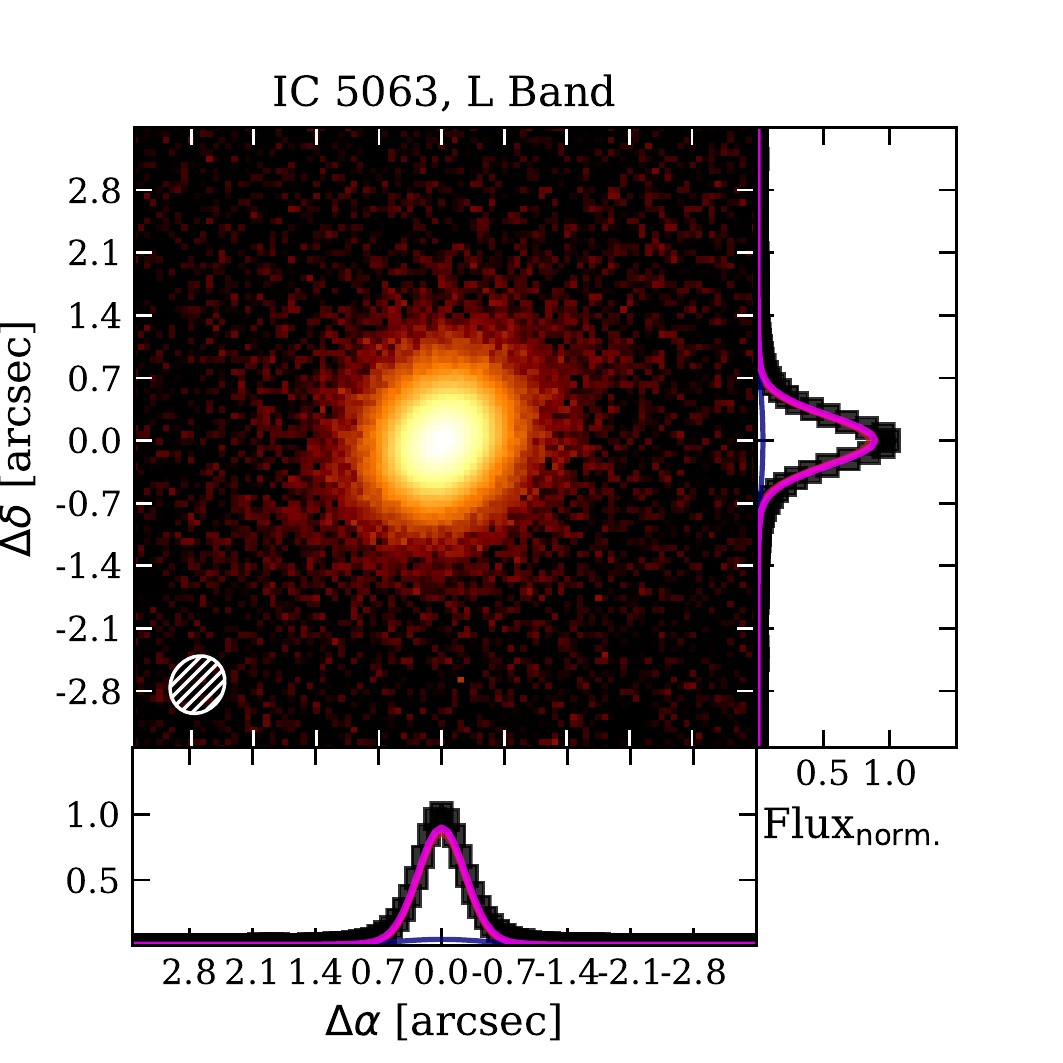}}
\subfloat{\includegraphics[width=0.25\hsize]{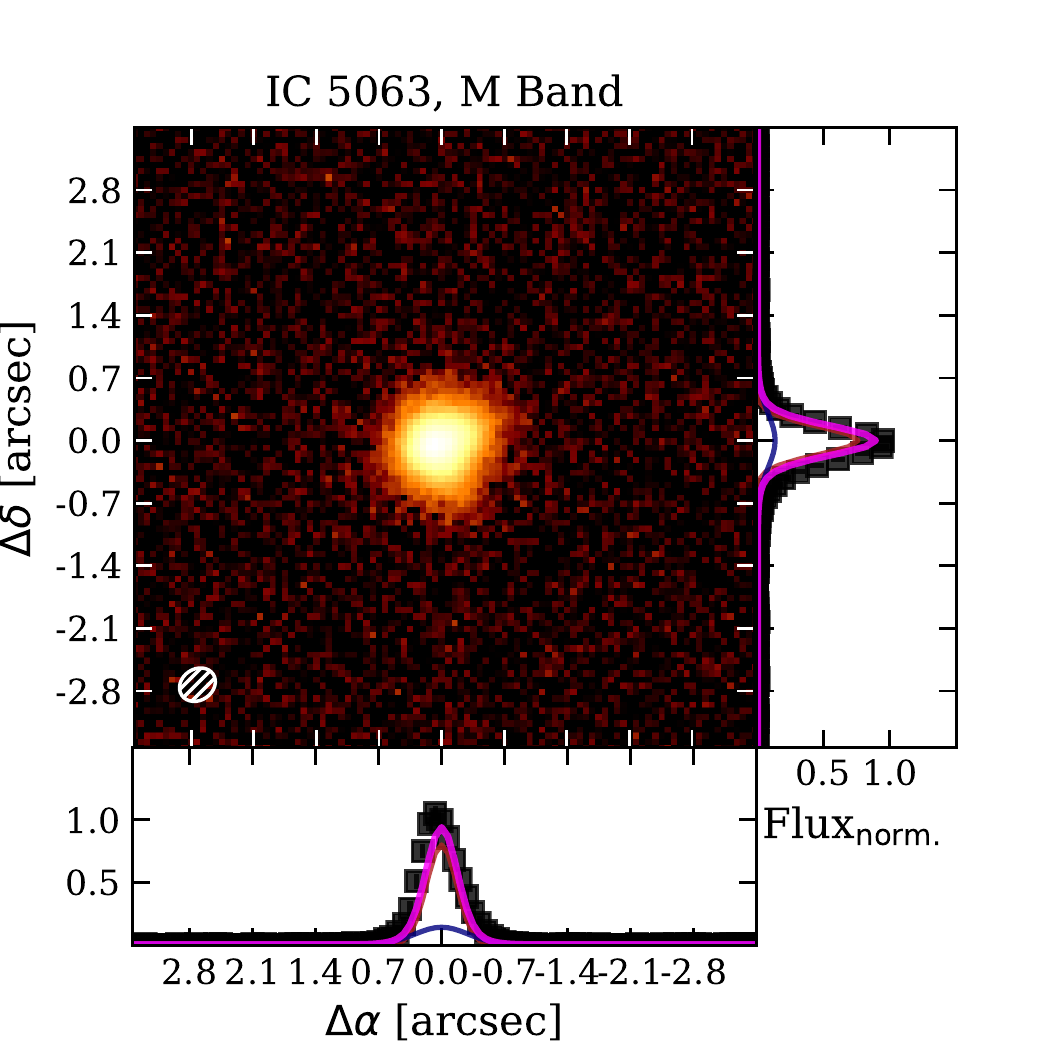}}
\subfloat{\includegraphics[width=0.25\hsize]{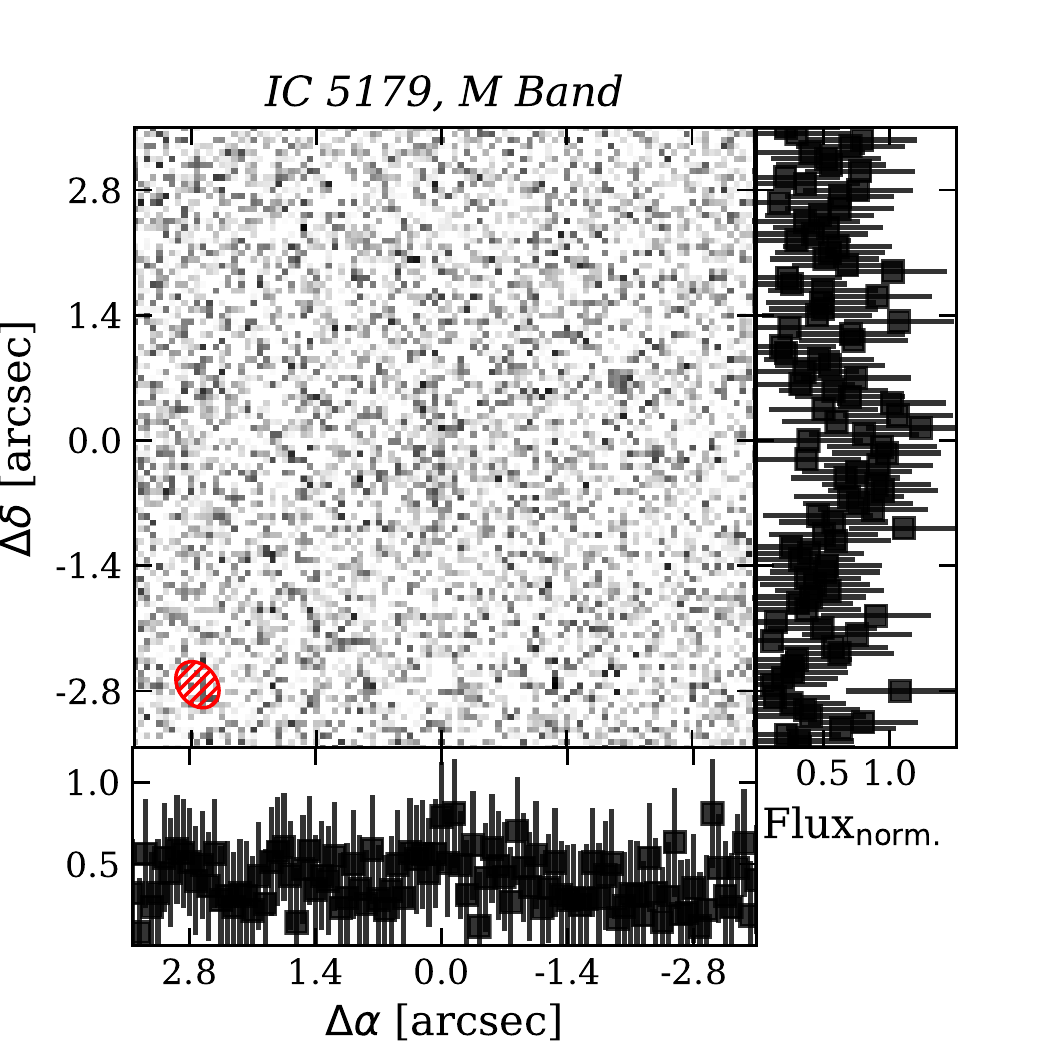}}
\subfloat{\includegraphics[width=0.25\hsize]{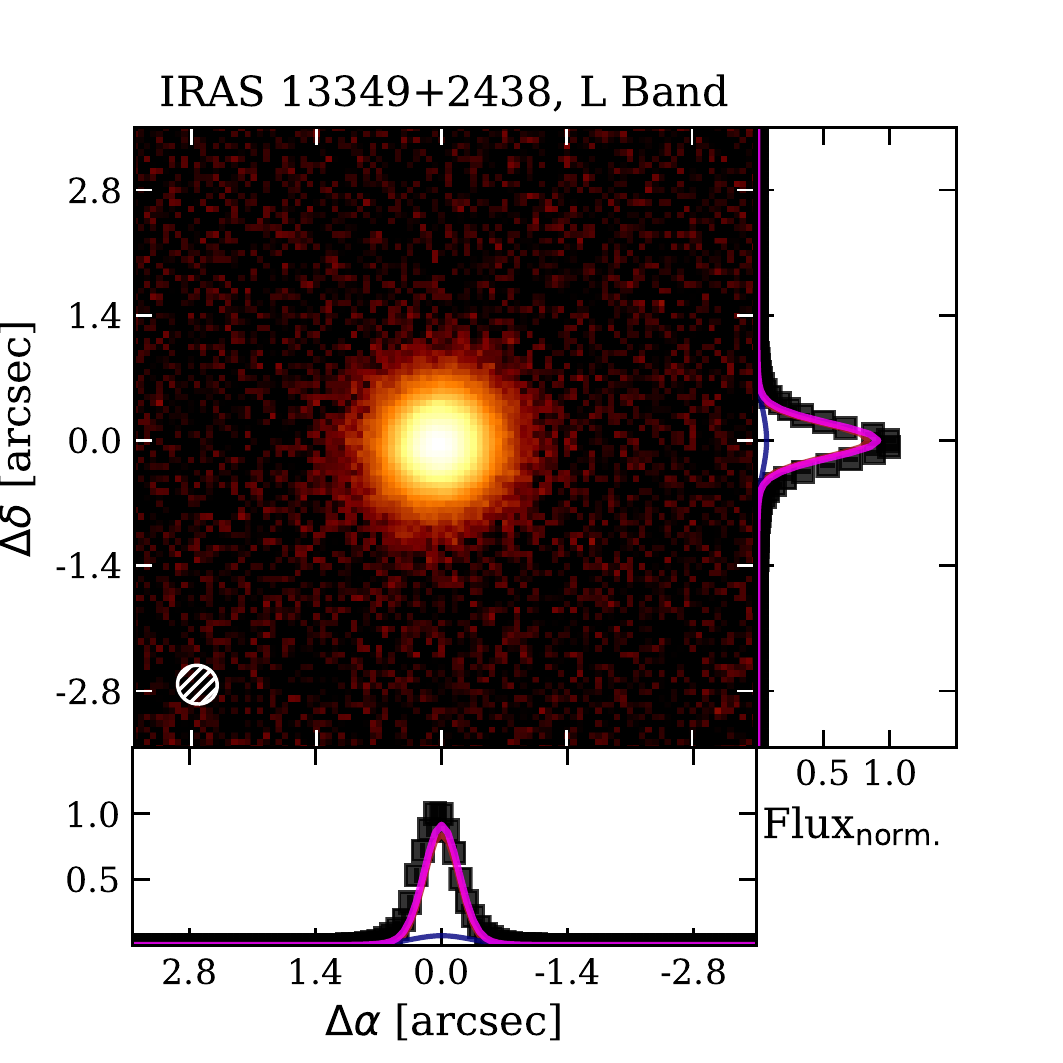}}\\
\subfloat{\includegraphics[width=0.25\hsize]{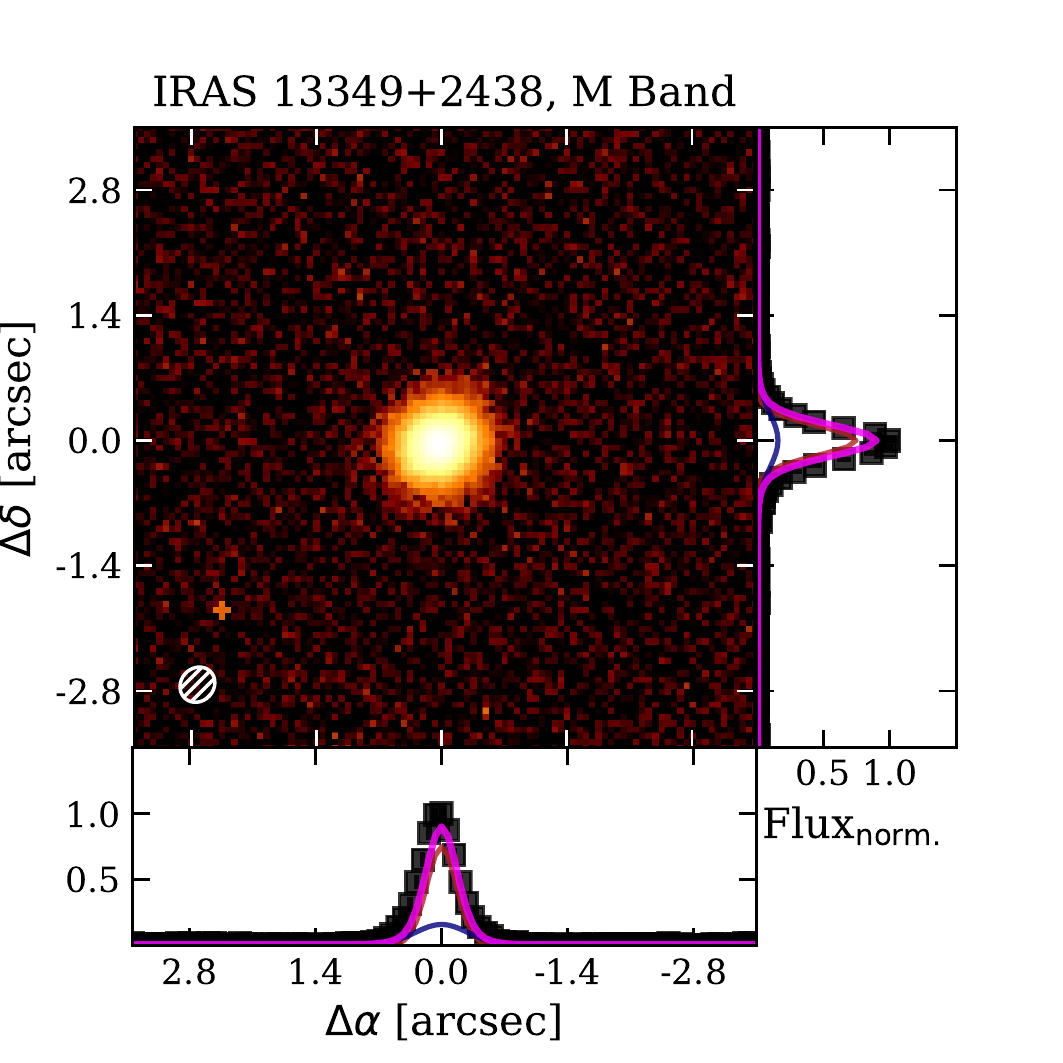}}
\subfloat{\includegraphics[width=0.25\hsize]{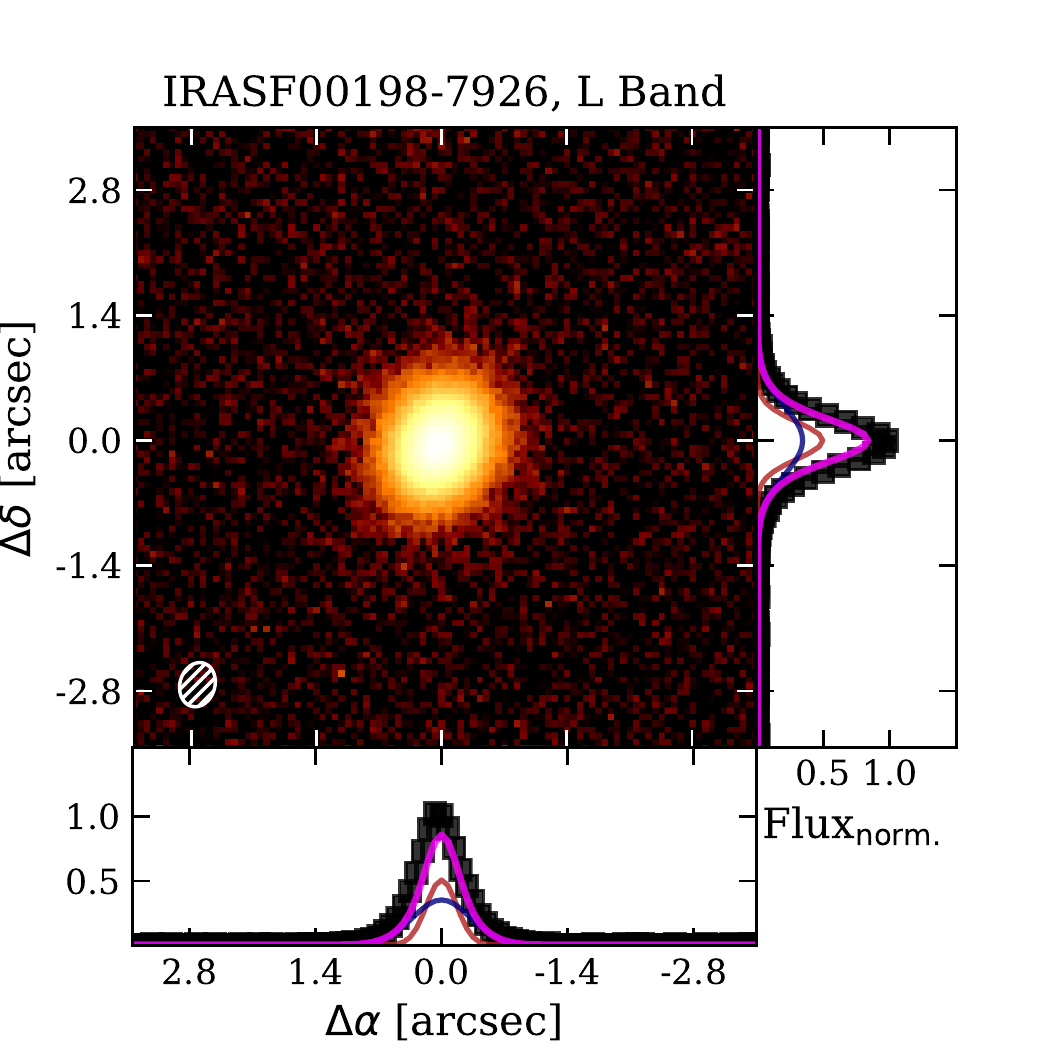}} 
\subfloat{\includegraphics[width=0.25\hsize]{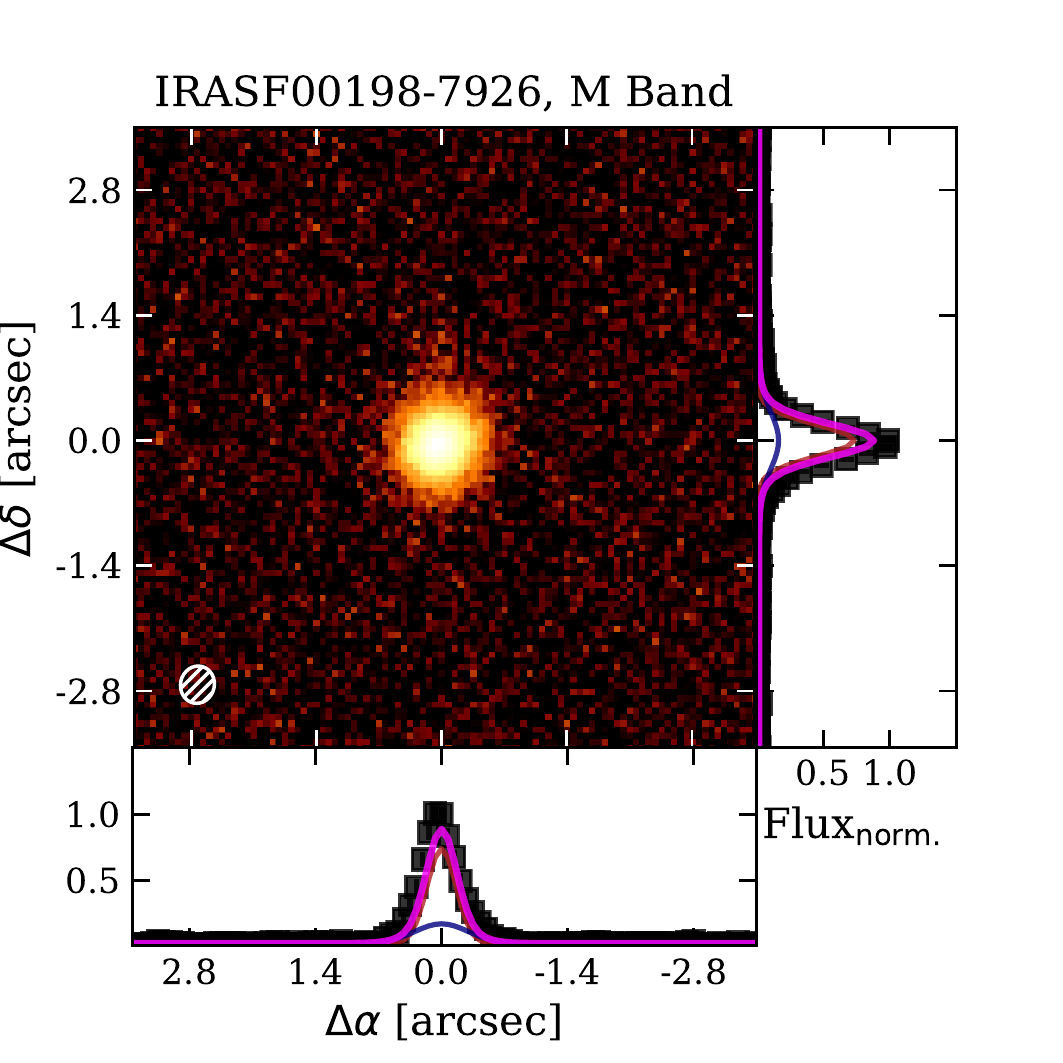}}
\subfloat{\includegraphics[width=0.25\hsize]{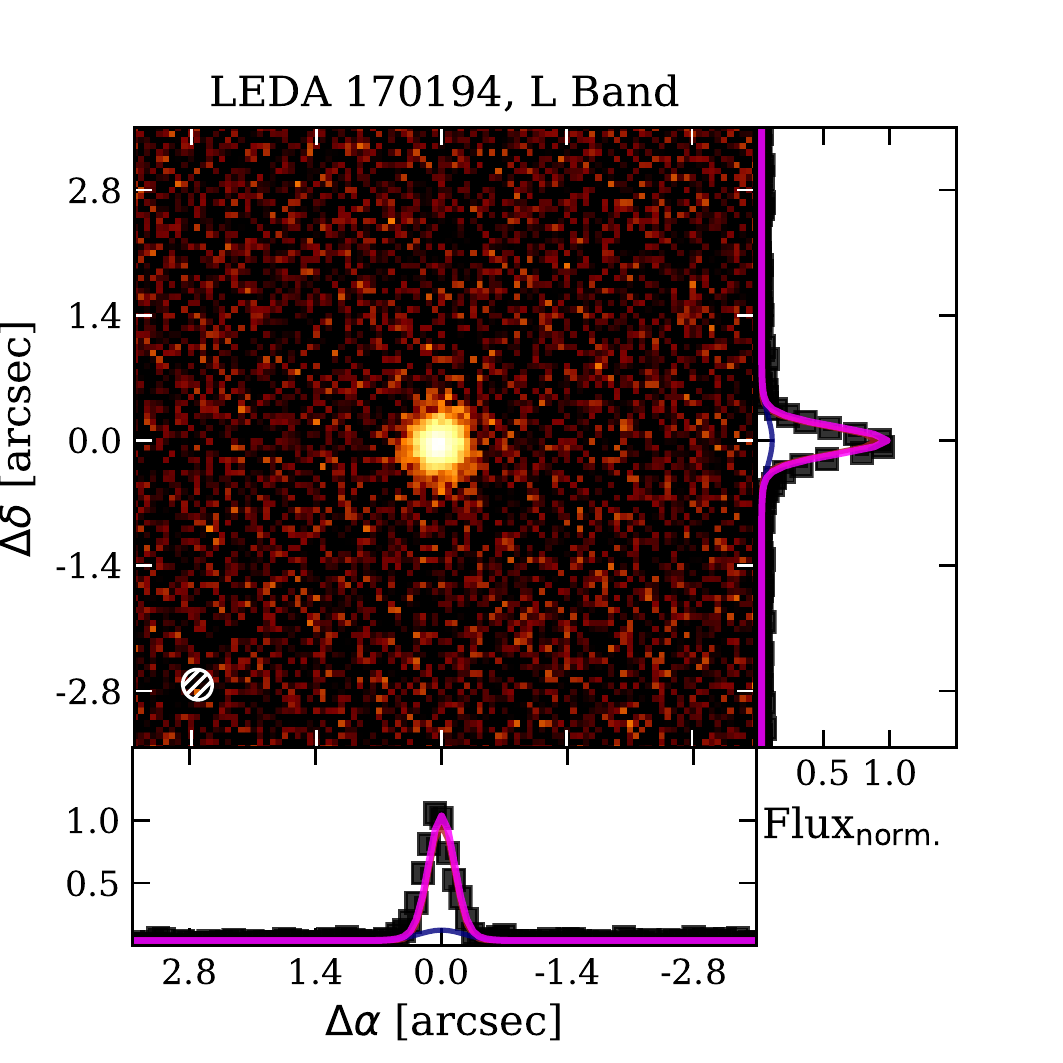}}\\
\subfloat{\includegraphics[width=0.25\hsize]{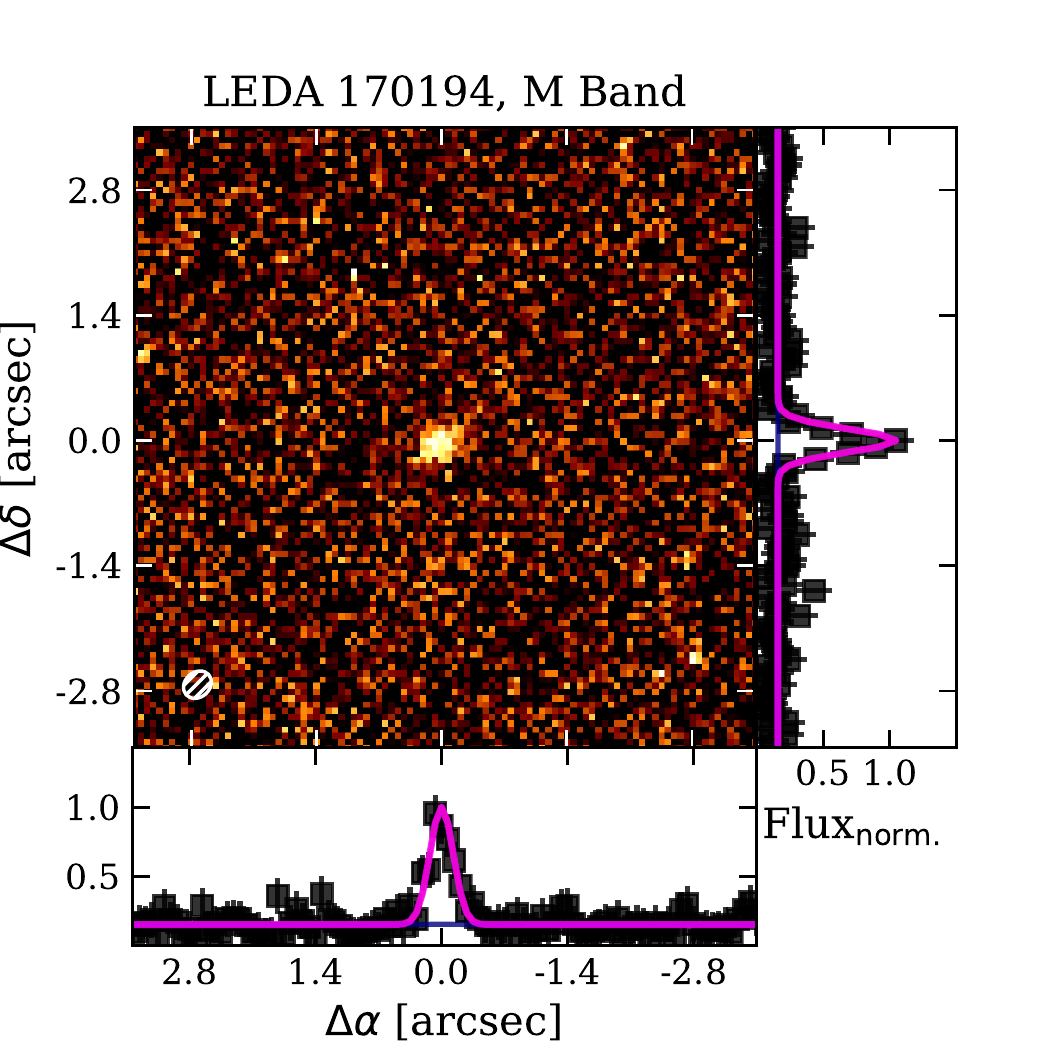}}
\subfloat{\includegraphics[width=0.25\hsize]{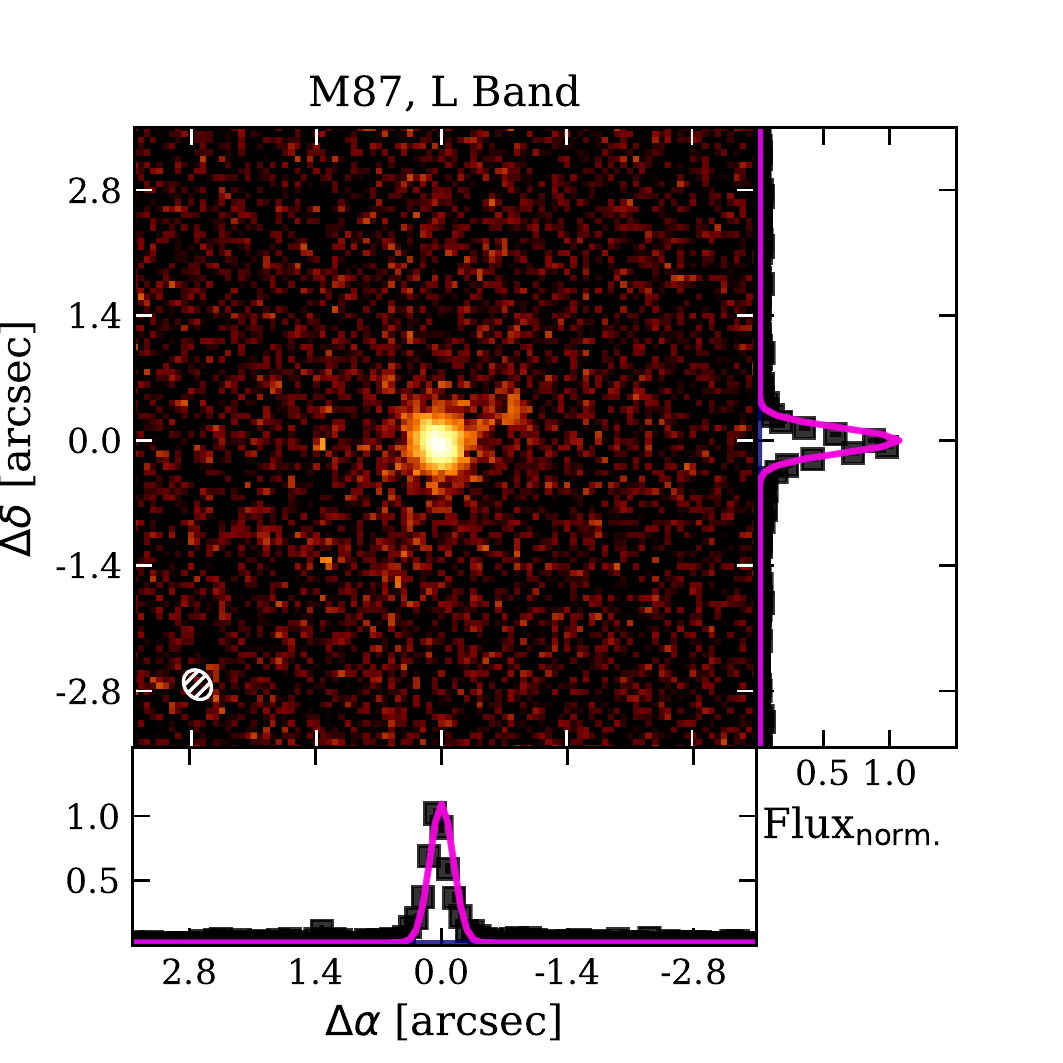}} 
\subfloat{\includegraphics[width=0.25\hsize]{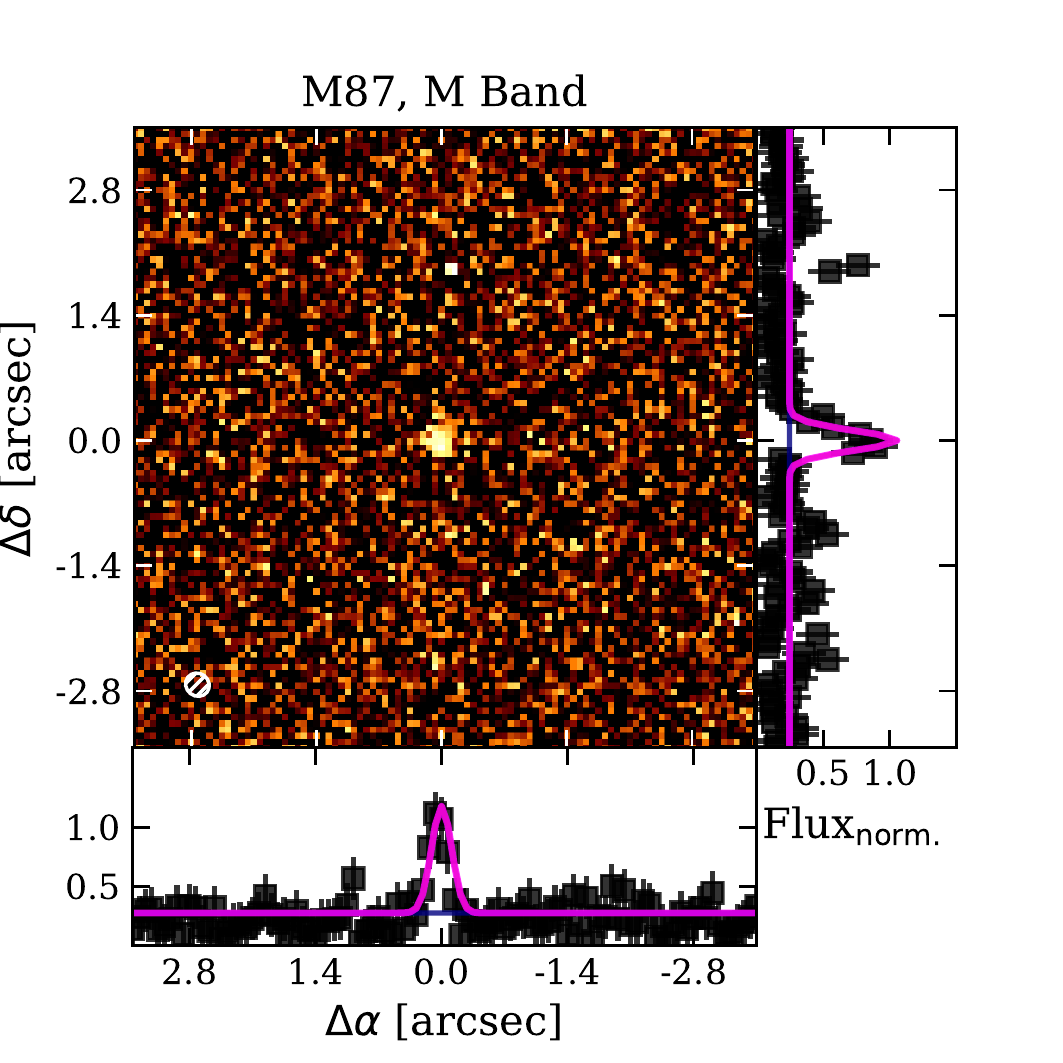}}
\subfloat{\includegraphics[width=0.25\hsize]{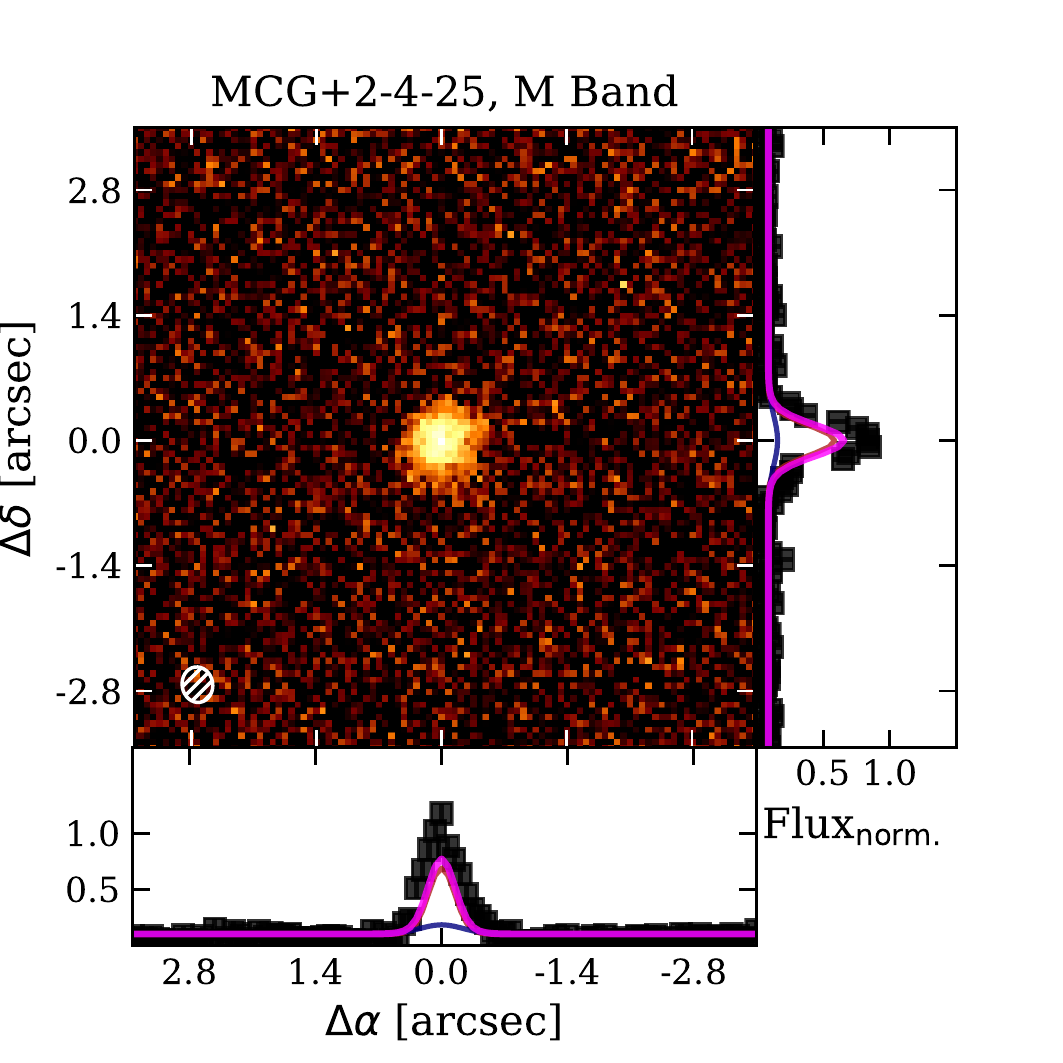}} \\
\caption{ As Fig \ref{fig:cutouts_one} but for all sources.}
\end{figure*}
\begin{figure*}
\subfloat{\includegraphics[width=0.25\hsize]{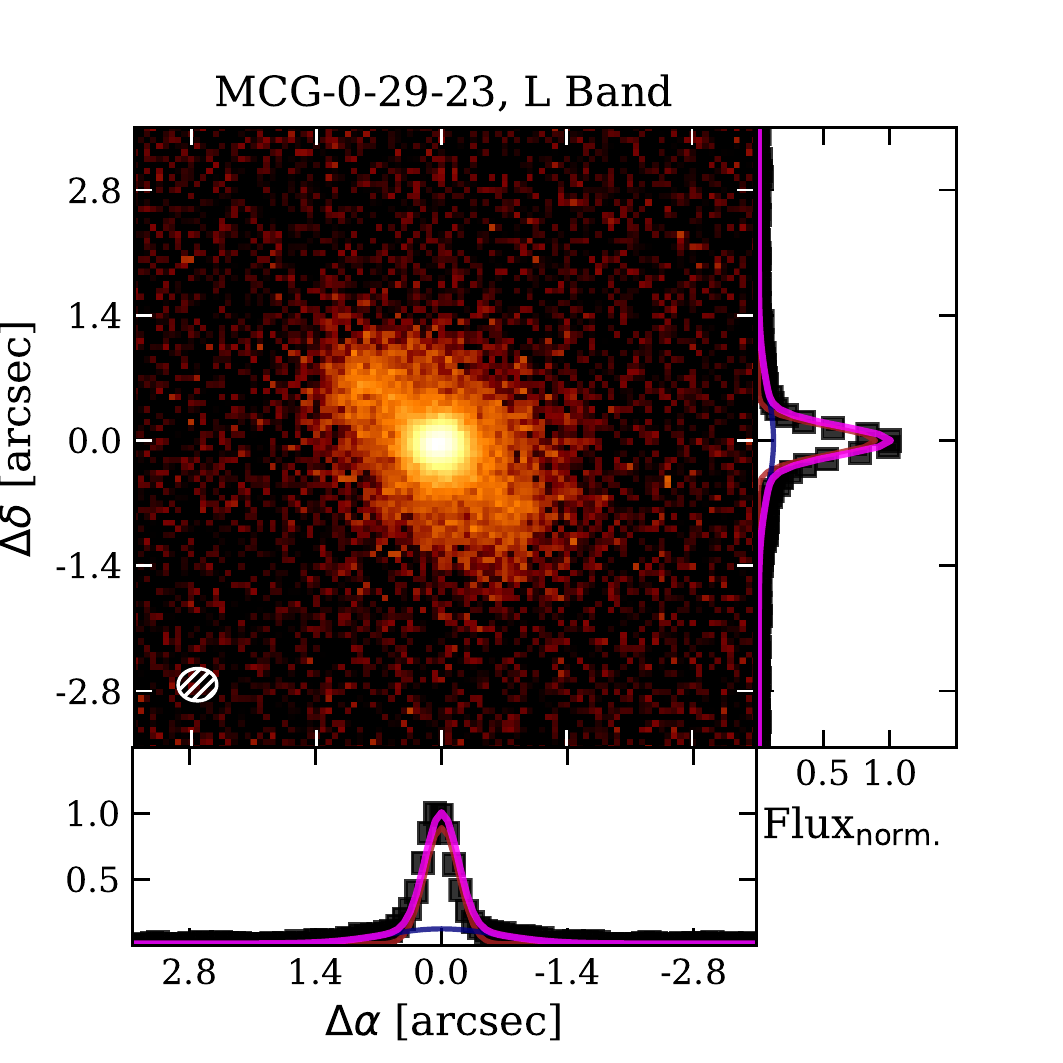}}
\subfloat{\includegraphics[width=0.25\hsize]{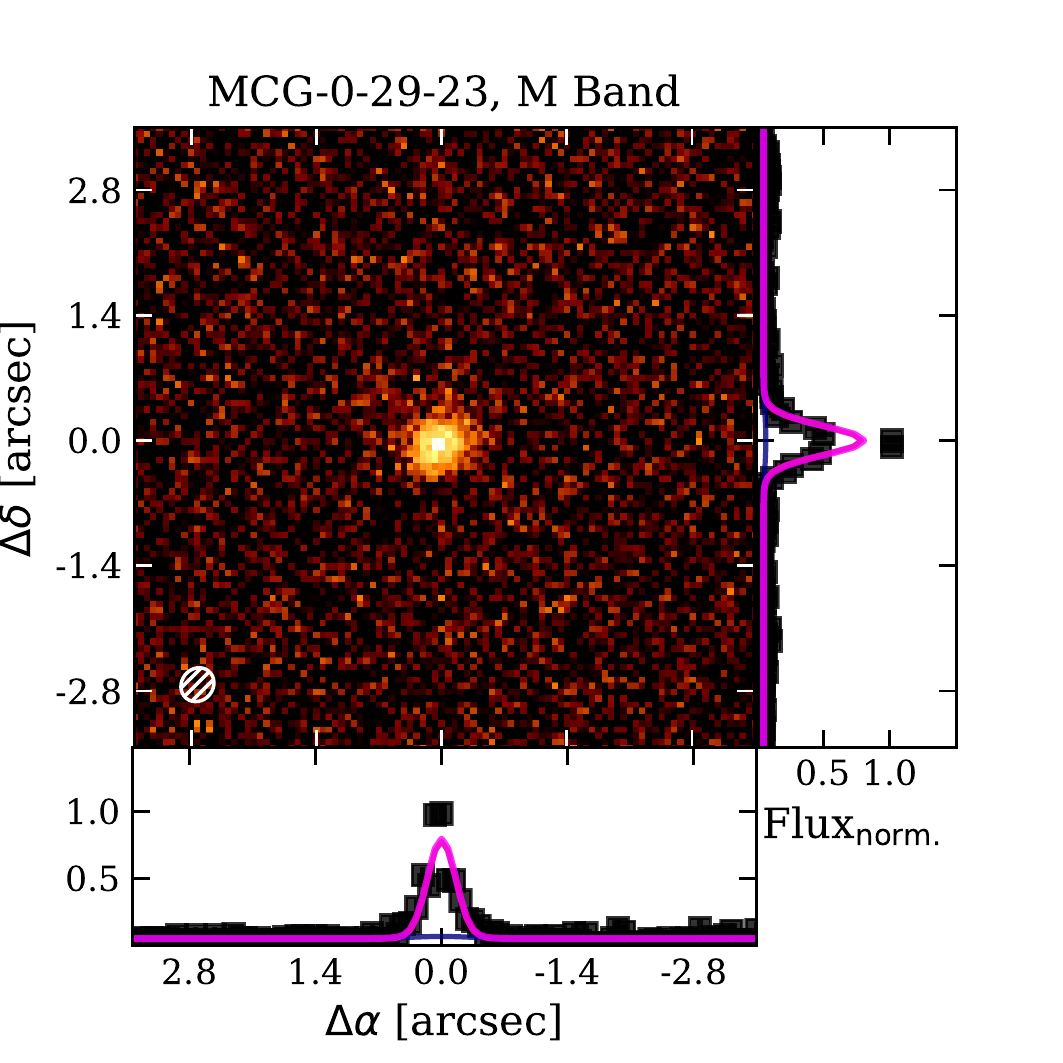}} 
\subfloat{\includegraphics[width=0.25\hsize]{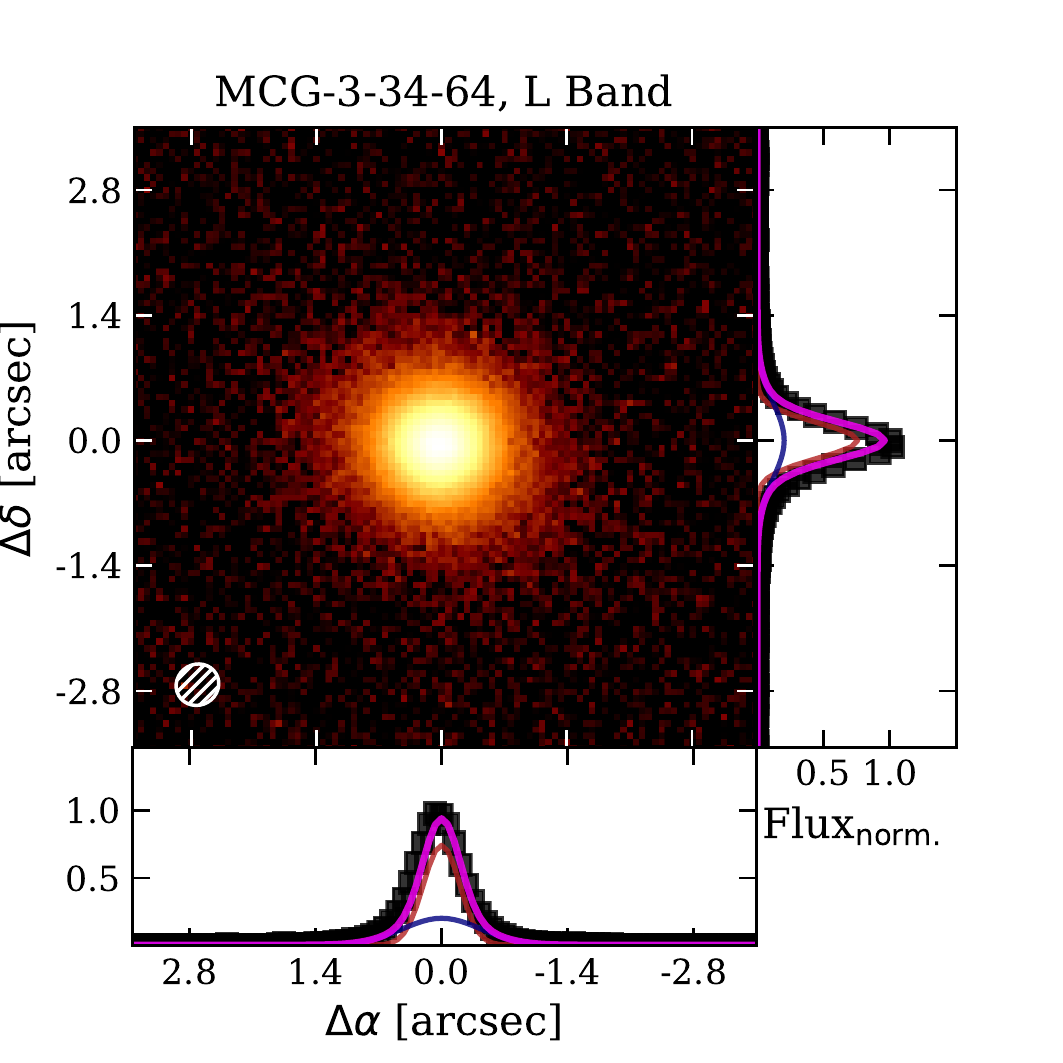}}
\subfloat{\includegraphics[width=0.25\hsize]{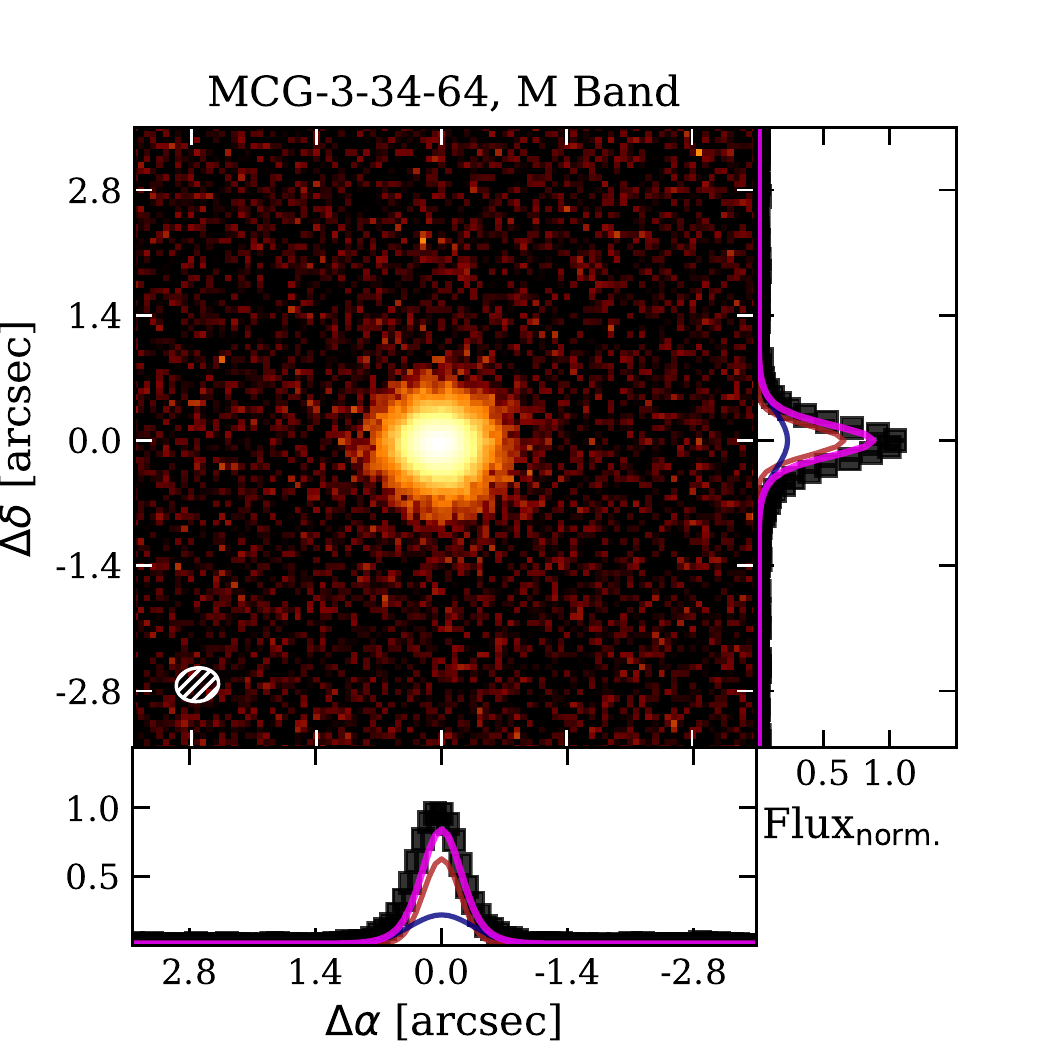}} \\
\subfloat{\includegraphics[width=0.25\hsize]{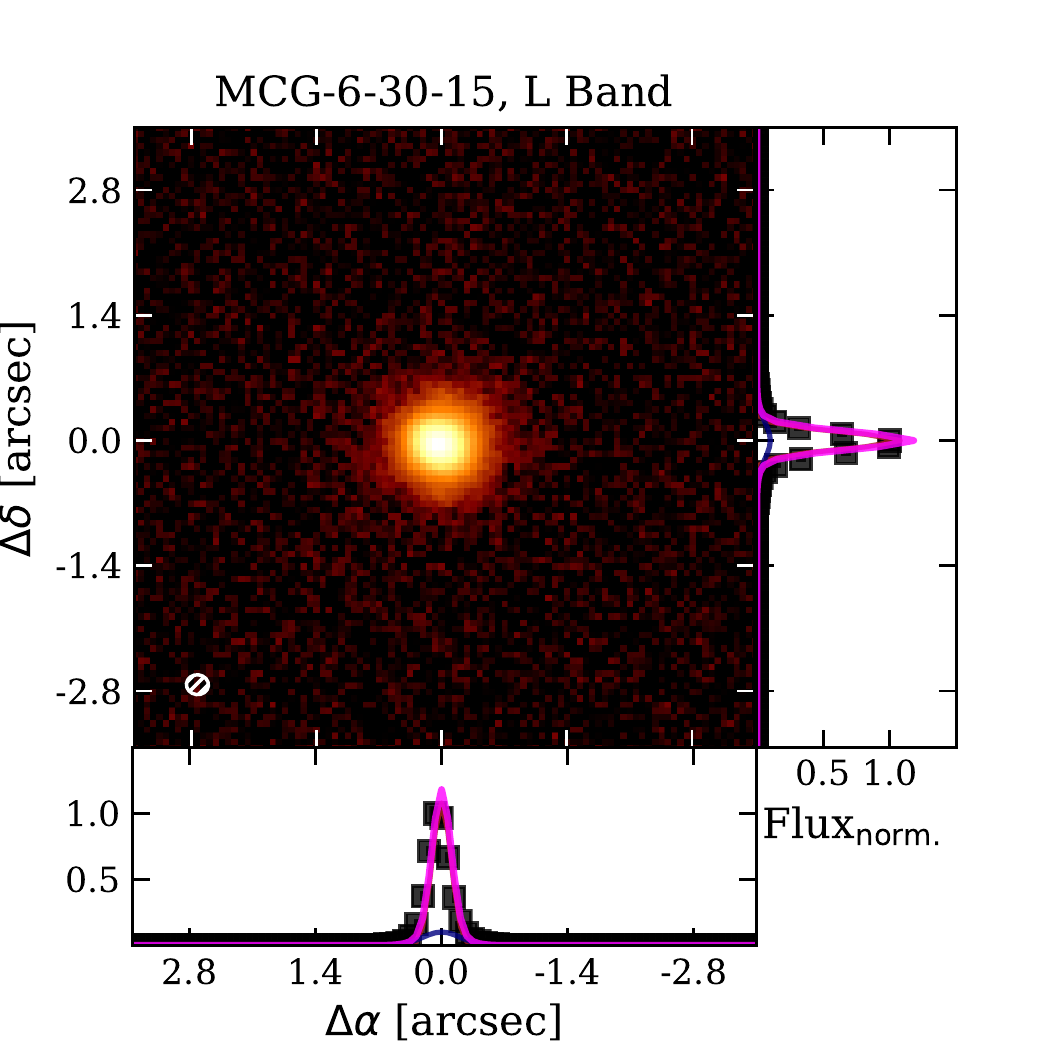}}
\subfloat{\includegraphics[width=0.25\hsize]{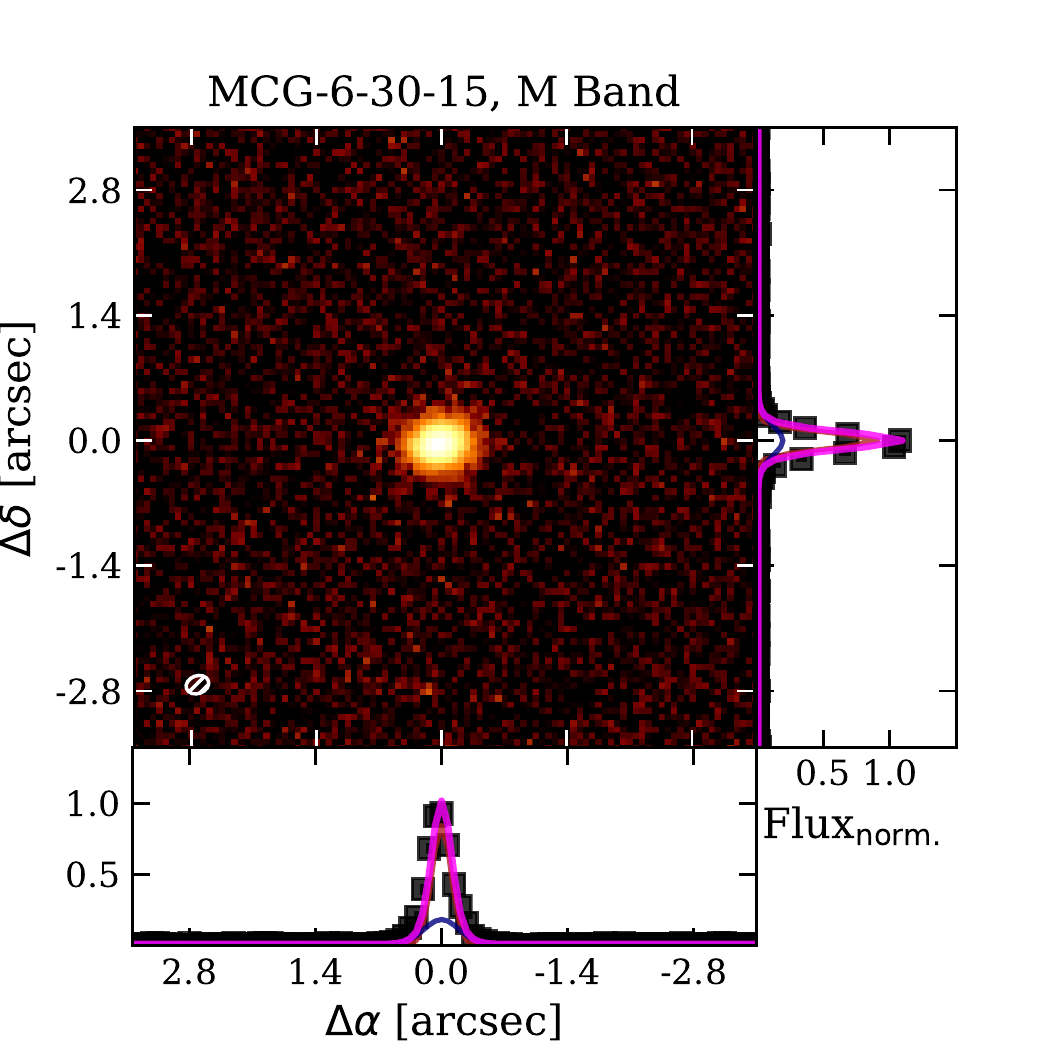}} 
\subfloat{\includegraphics[width=0.25\hsize]{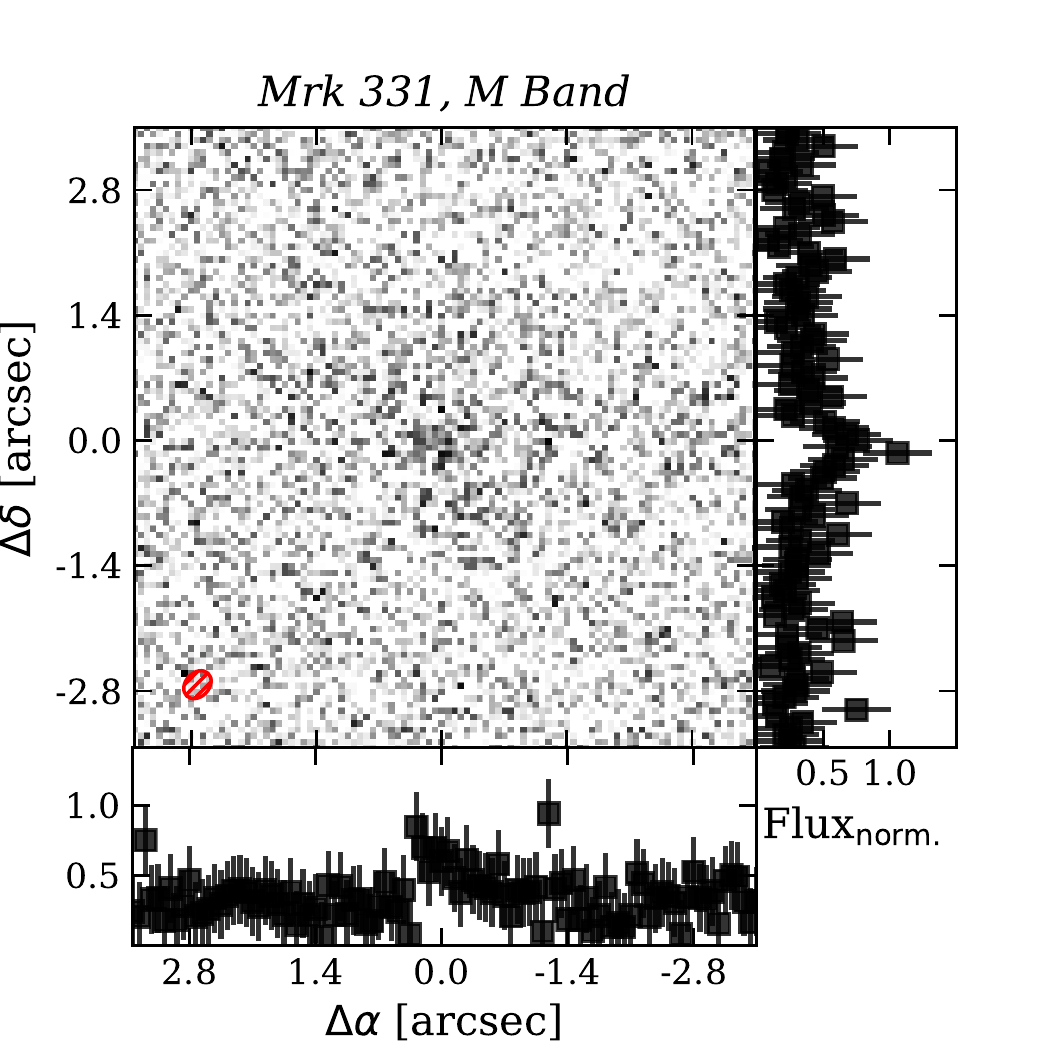}}
\subfloat{\includegraphics[width=0.25\hsize]{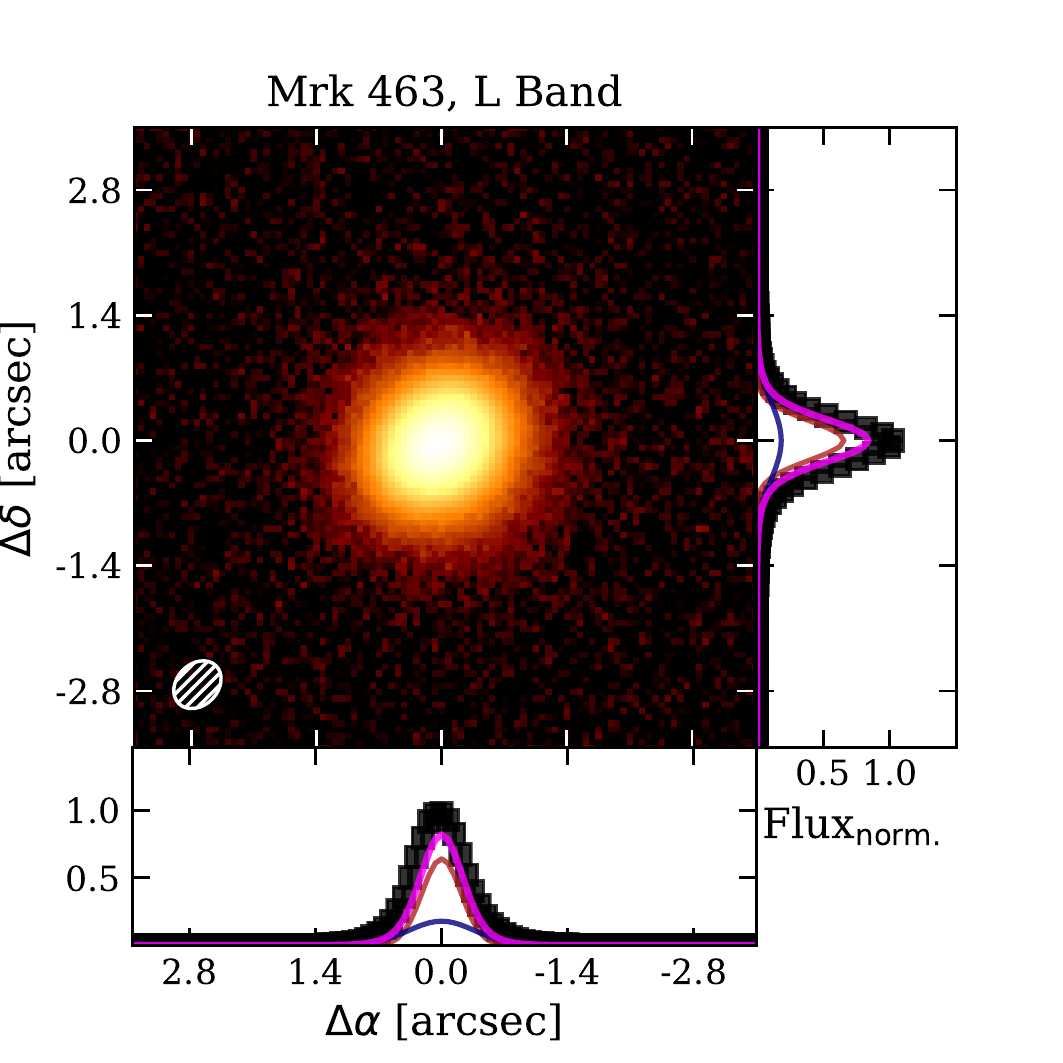}}\\
\subfloat{\includegraphics[width=0.25\hsize]{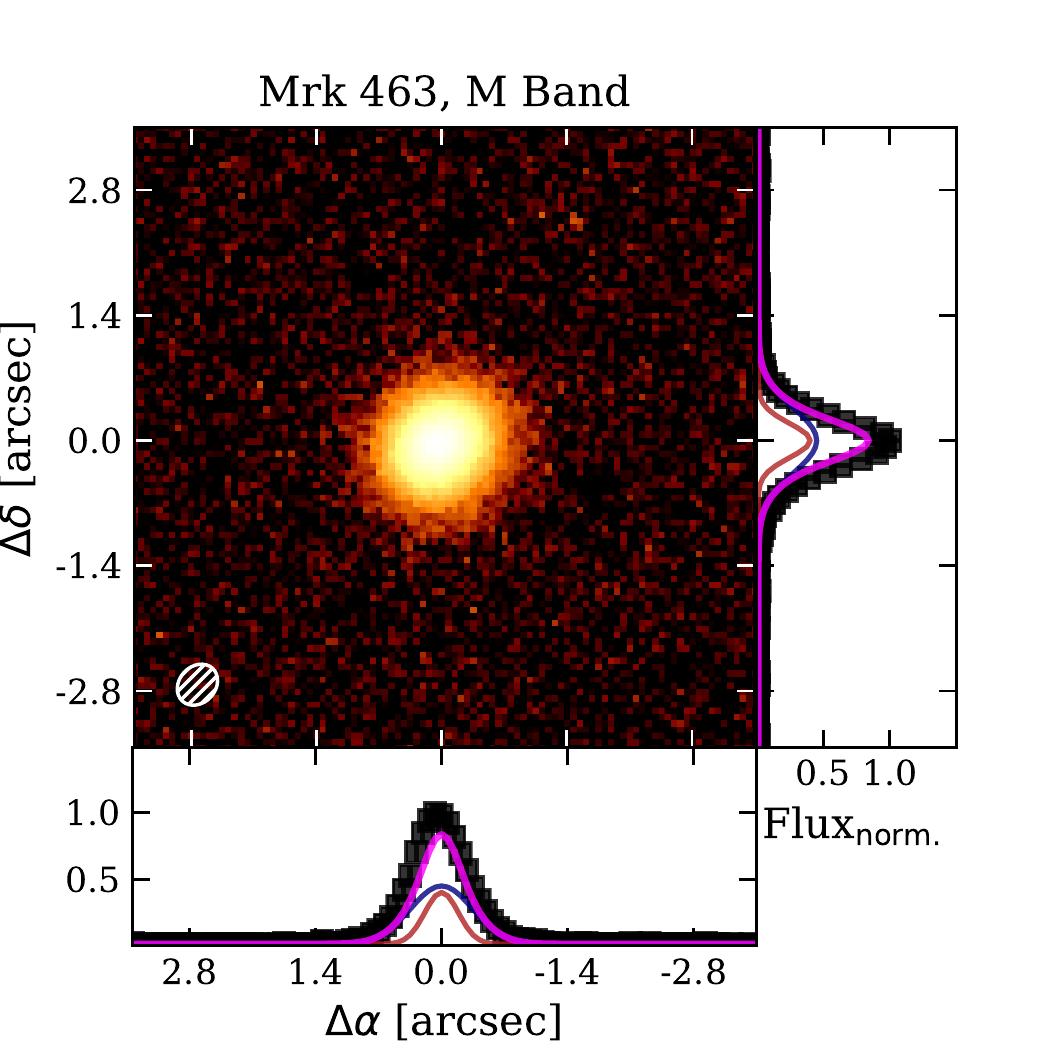}}
\subfloat{\includegraphics[width=0.25\hsize]{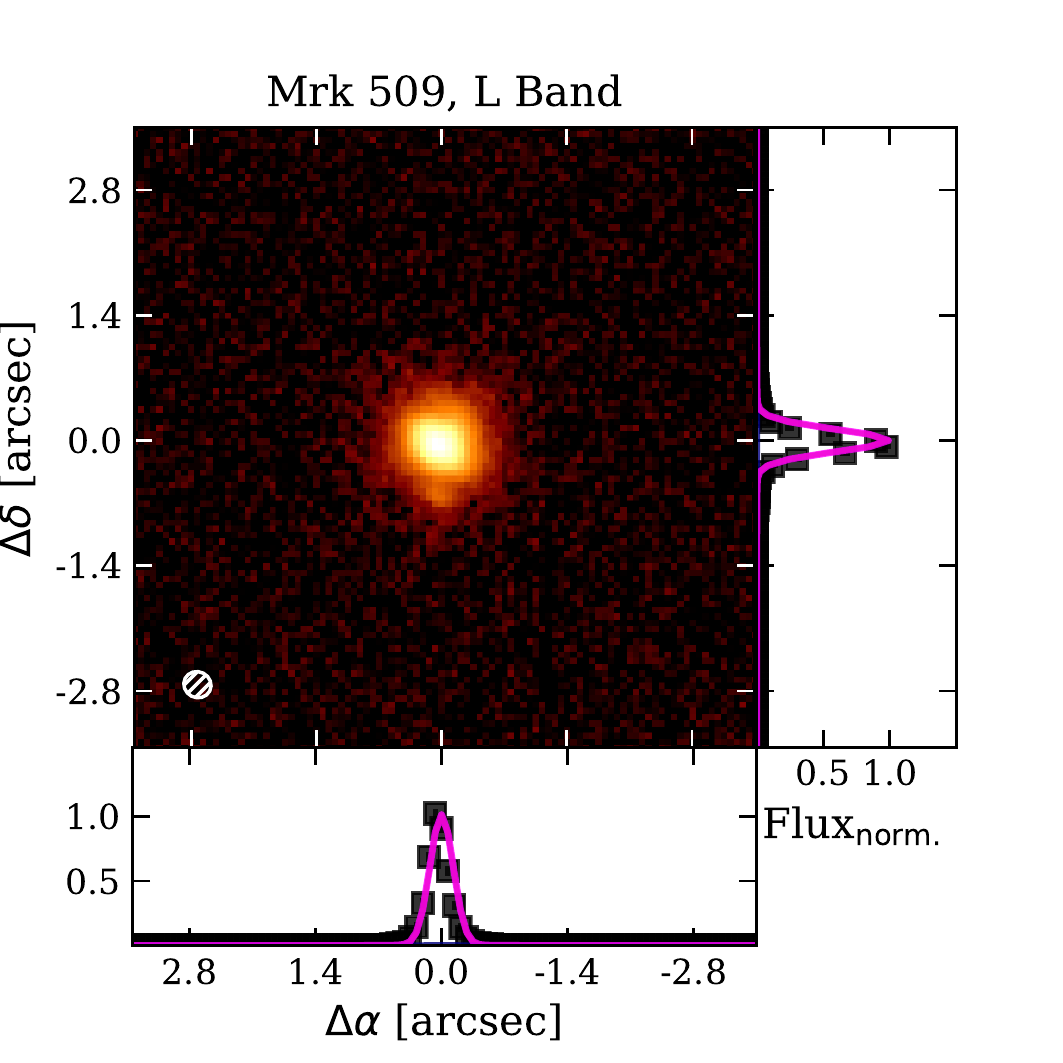}} 
\subfloat{\includegraphics[width=0.25\hsize]{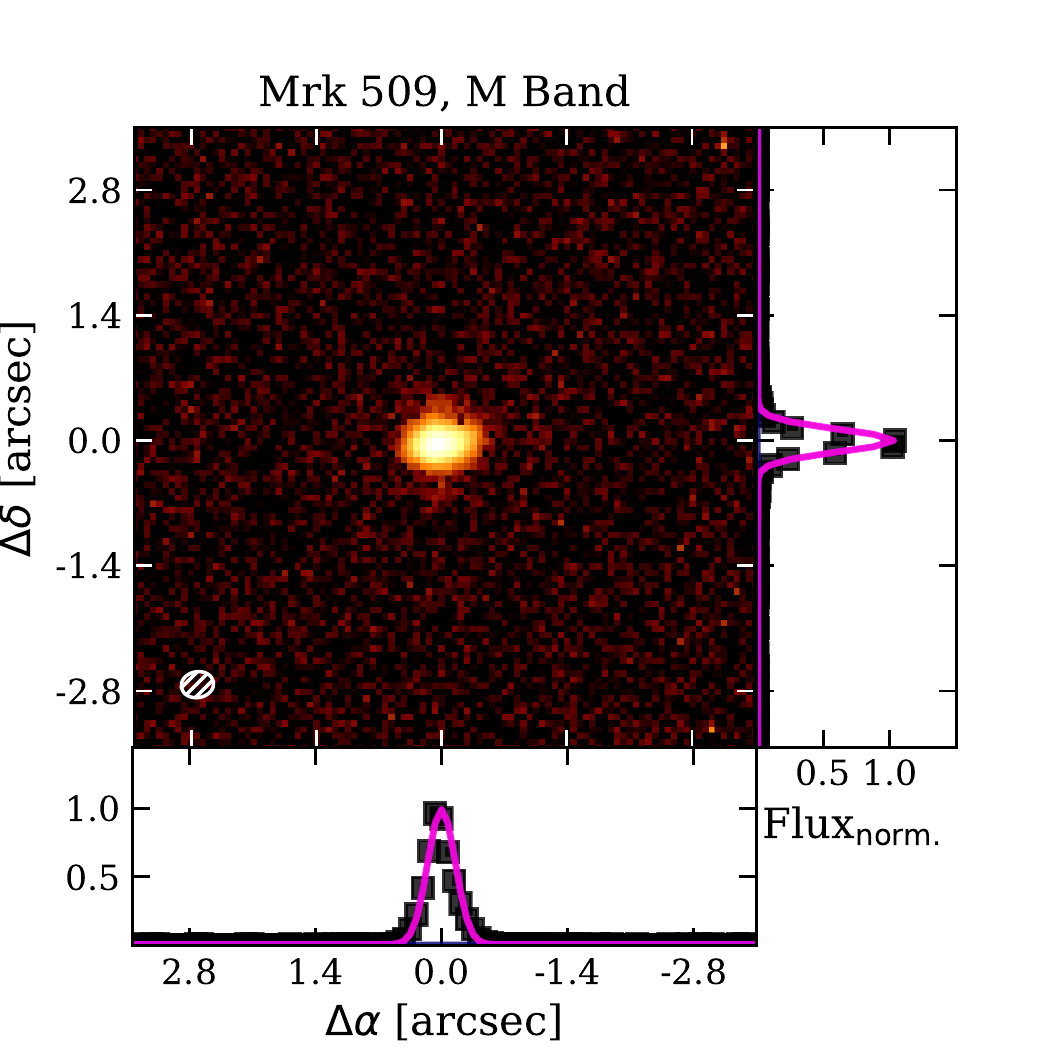}}
\subfloat{\includegraphics[width=0.25\hsize]{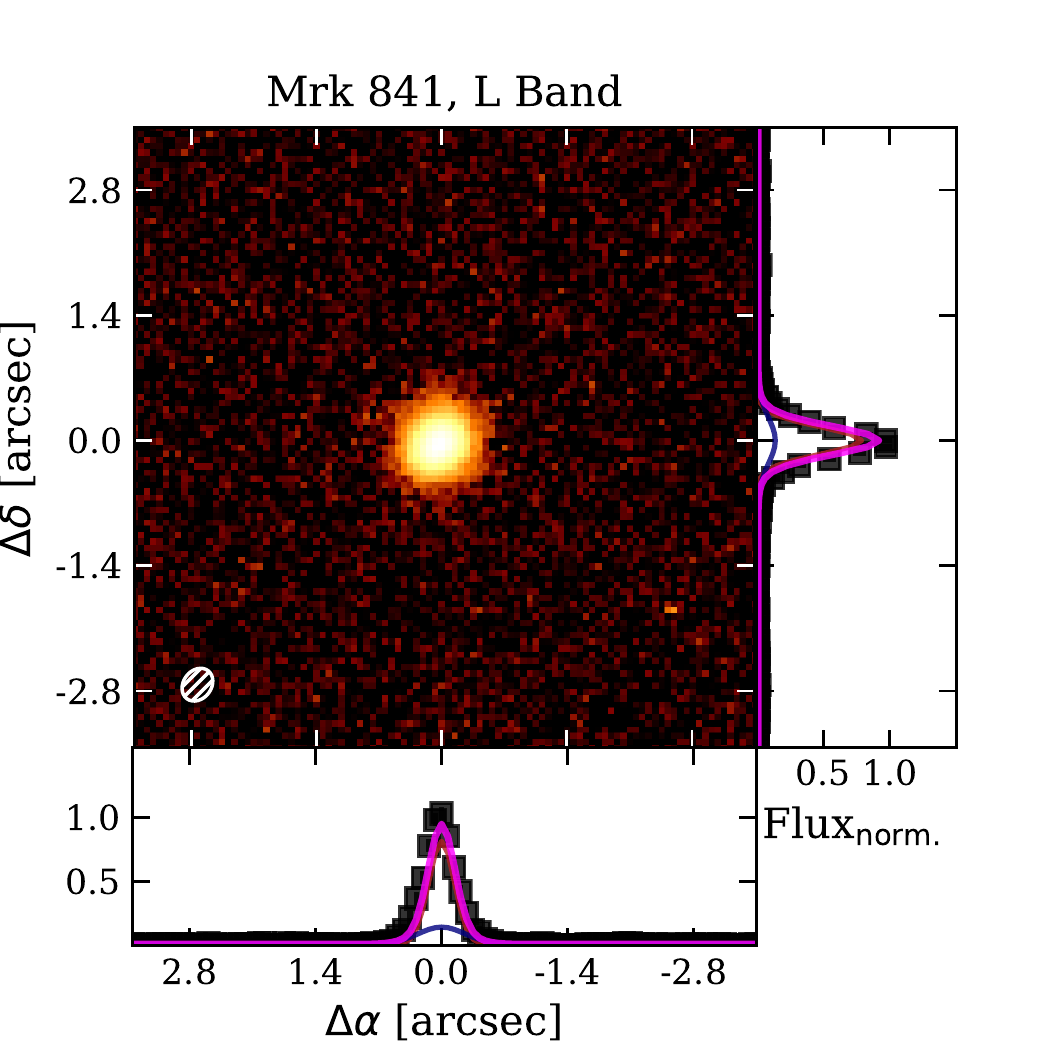}} \\
\subfloat{\includegraphics[width=0.25\hsize]{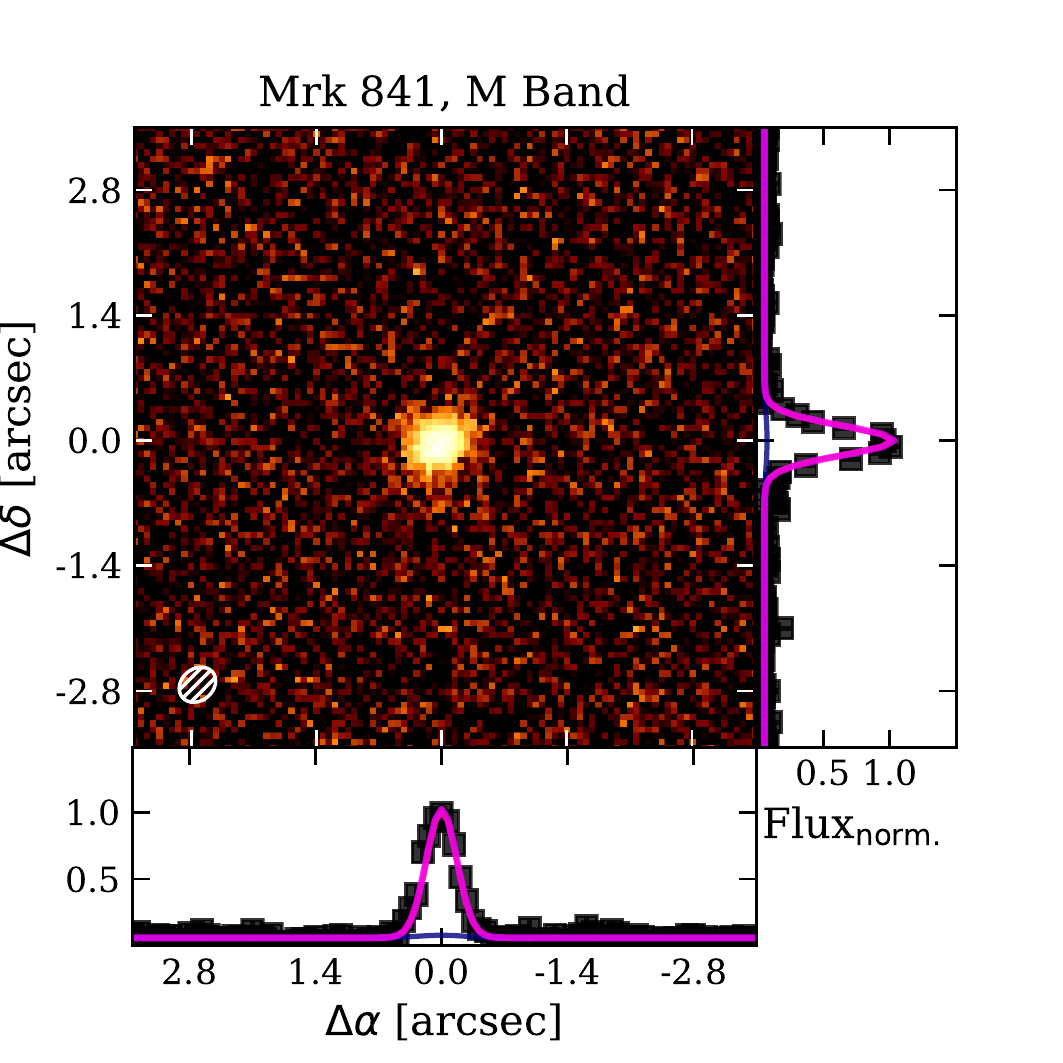}}
\subfloat{\includegraphics[width=0.25\hsize]{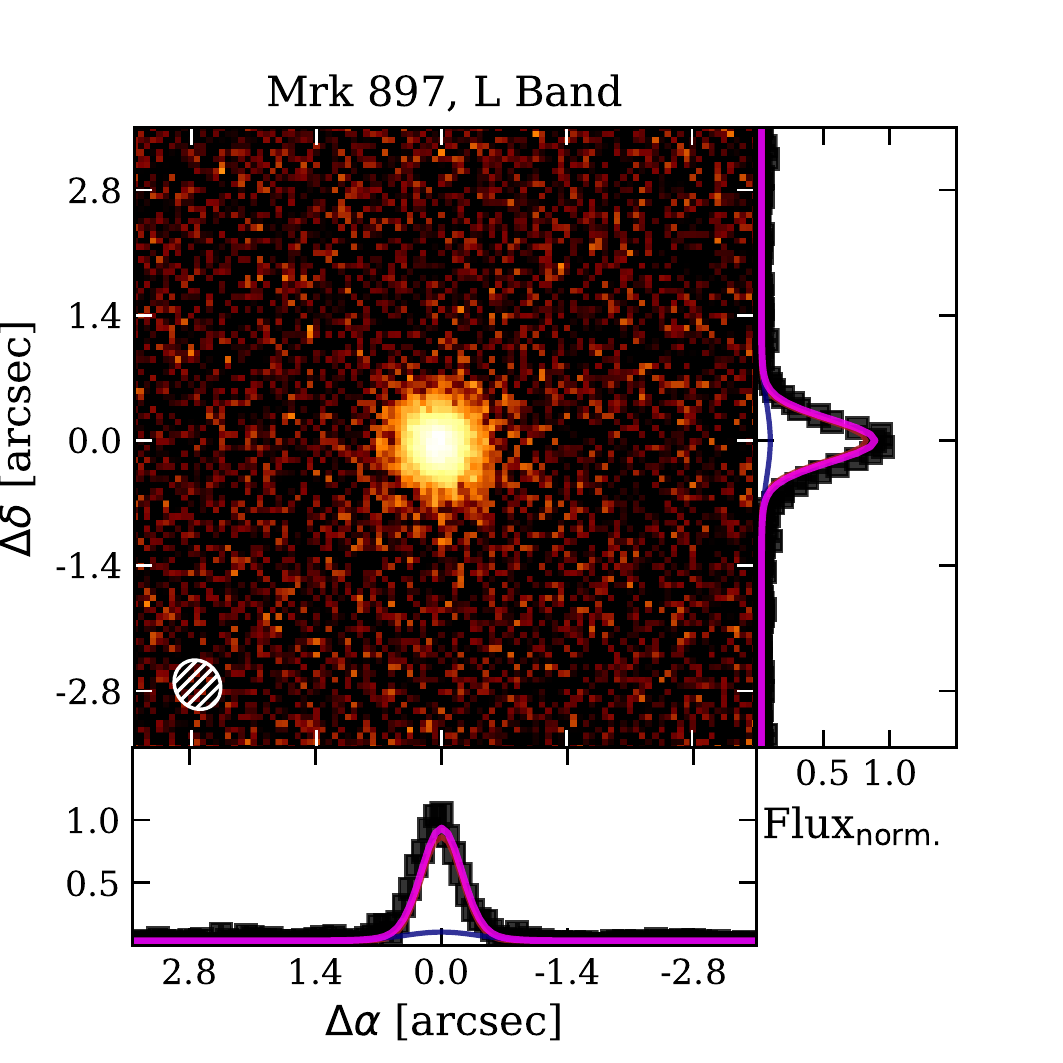}} 
\subfloat{\includegraphics[width=0.25\hsize]{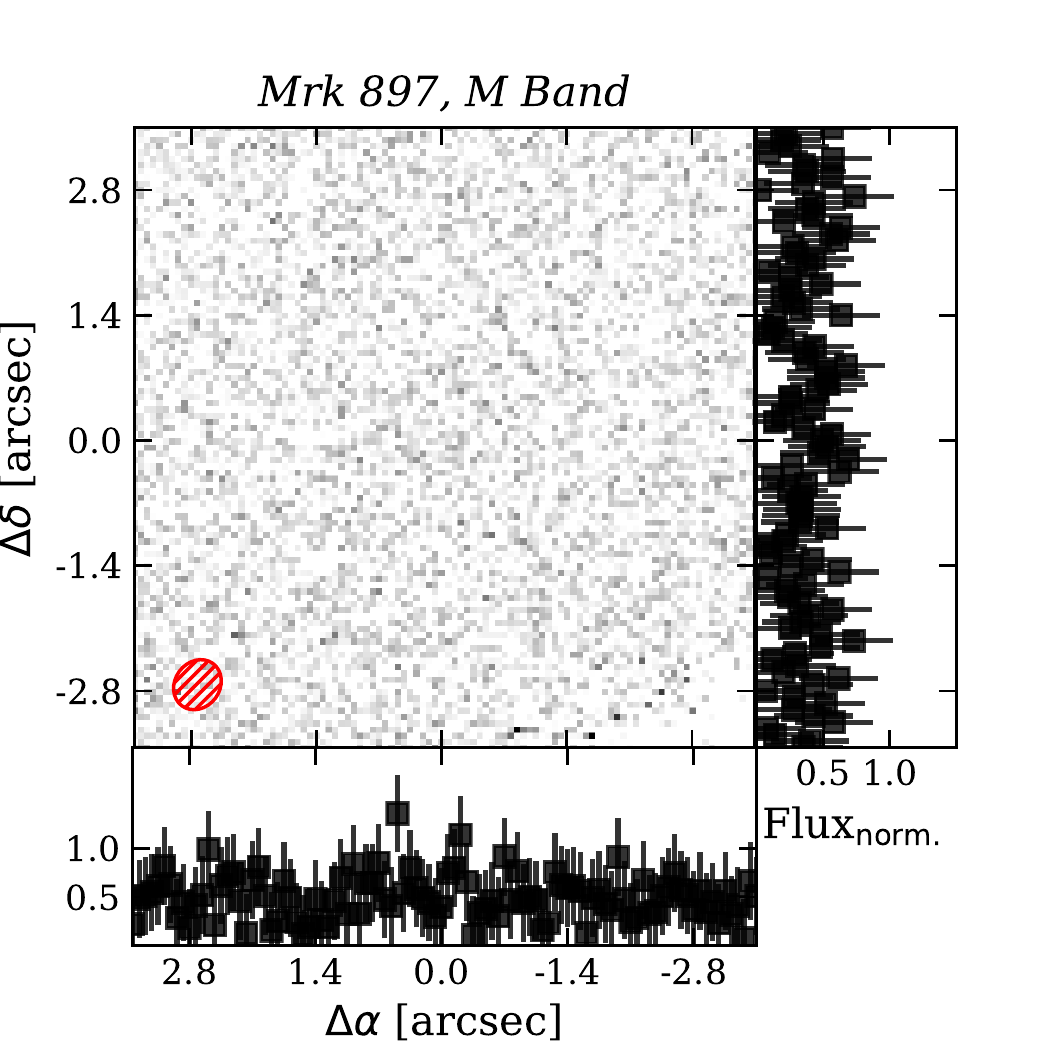}}
\subfloat{\includegraphics[width=0.25\hsize]{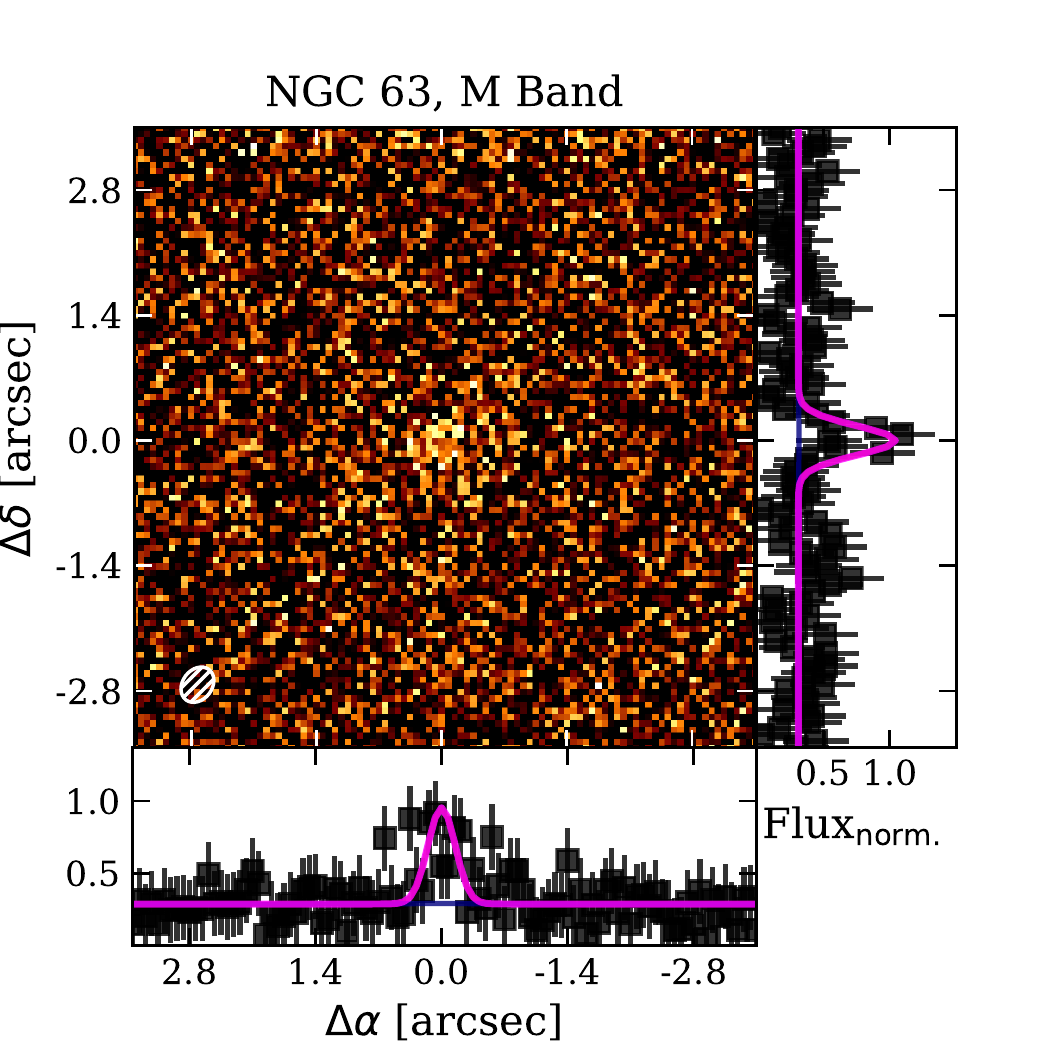}} \\
\subfloat{\includegraphics[width=0.25\hsize]{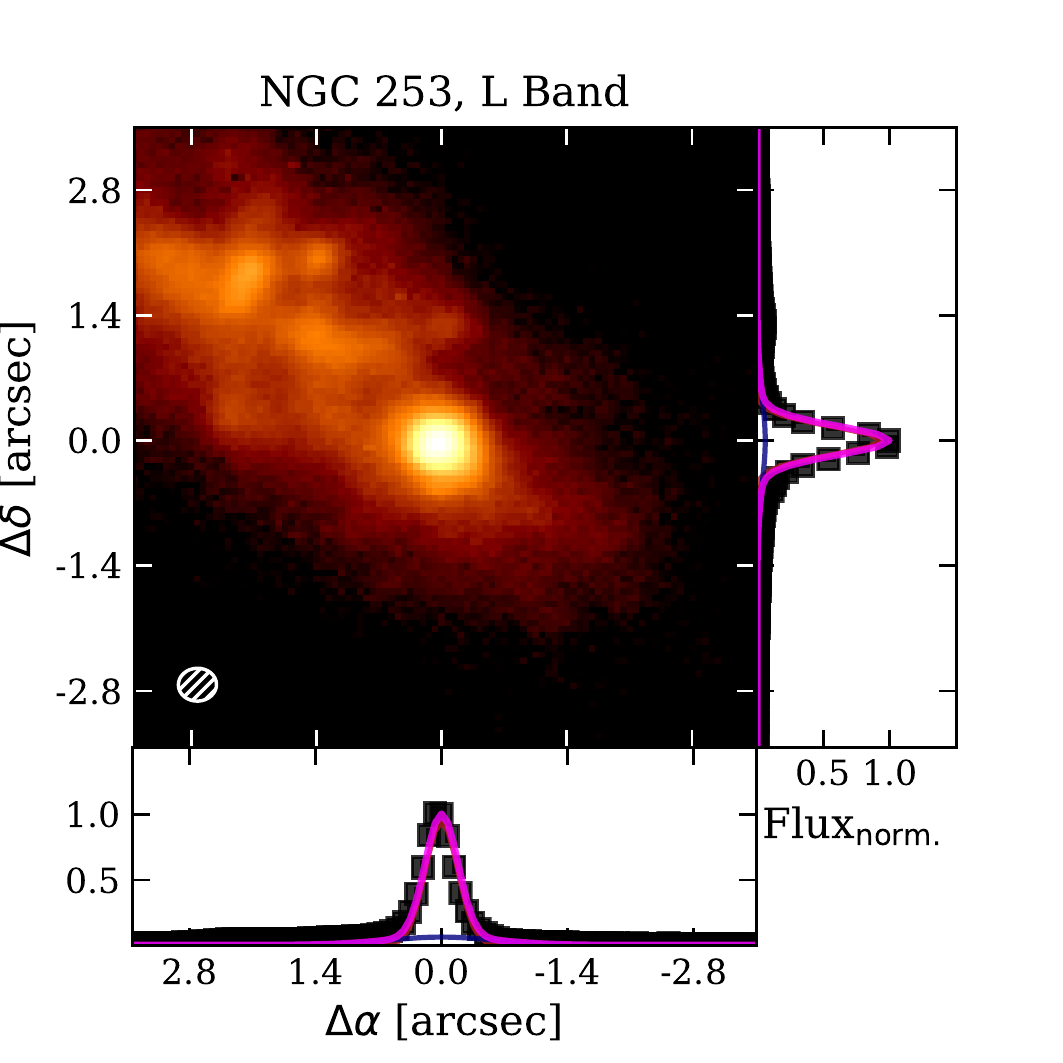}}
\subfloat{\includegraphics[width=0.25\hsize]{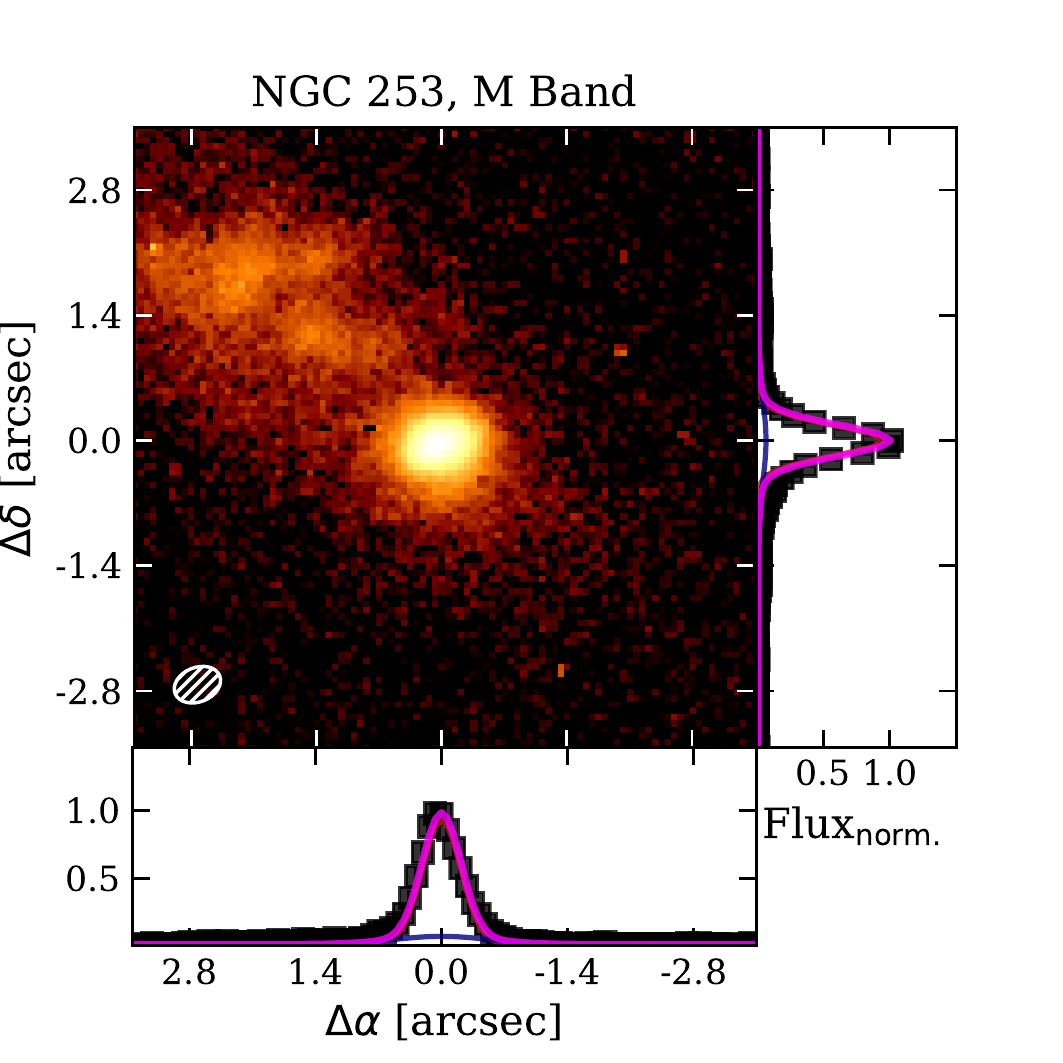}} 
\subfloat{\includegraphics[width=0.25\hsize]{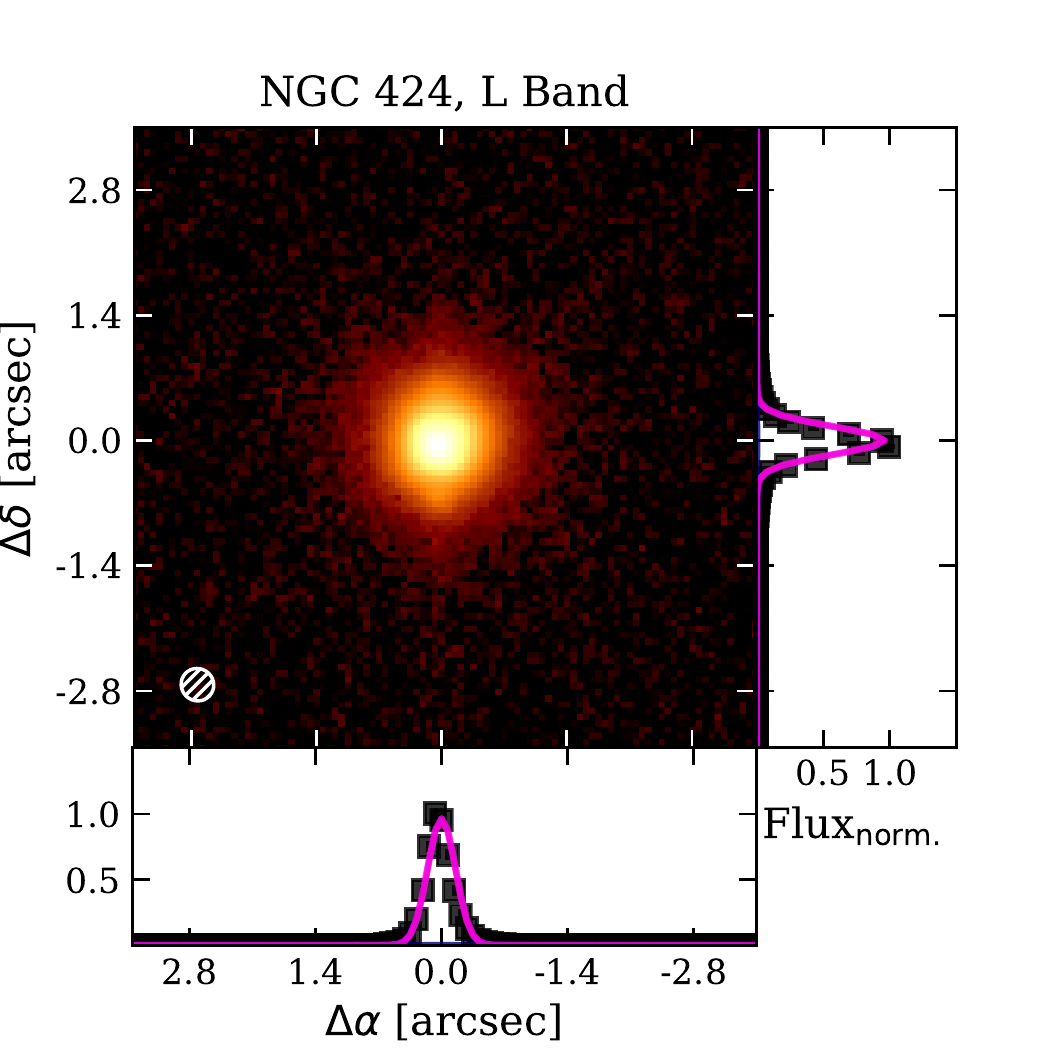}}
\subfloat{\includegraphics[width=0.25\hsize]{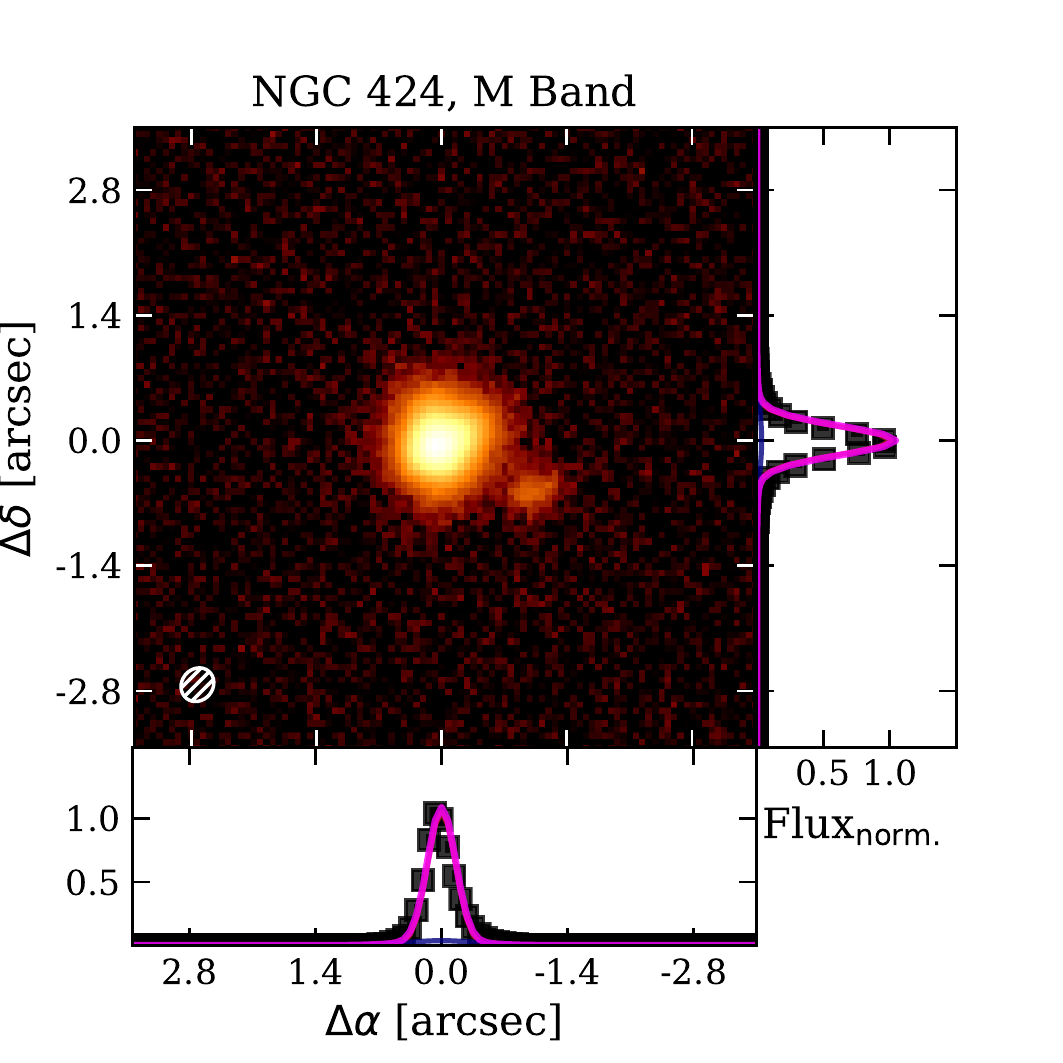}}\\
\caption{ As Fig \ref{fig:cutouts_one} but for all sources.}
\end{figure*}
\begin{figure*}
\subfloat{\includegraphics[width=0.25\hsize]{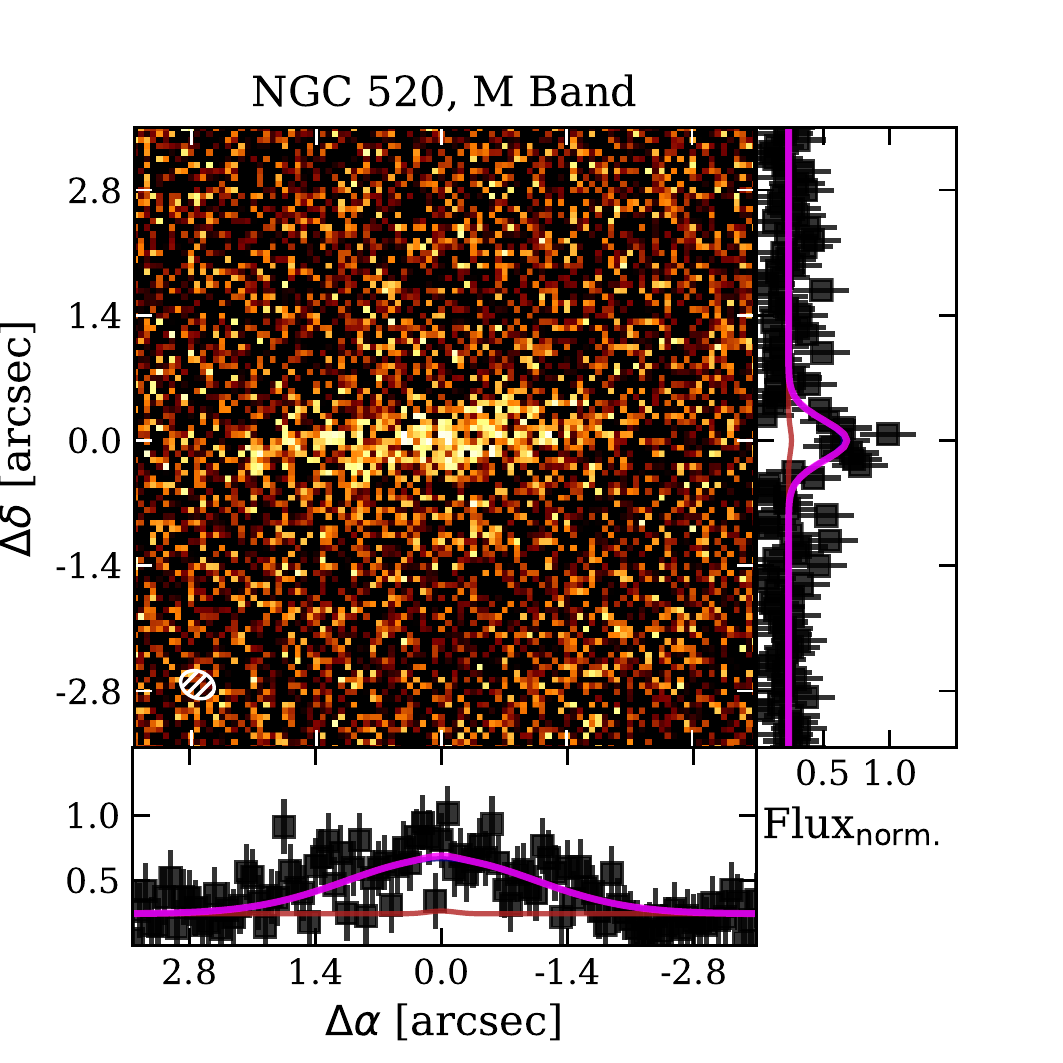}}
\subfloat{\includegraphics[width=0.25\hsize]{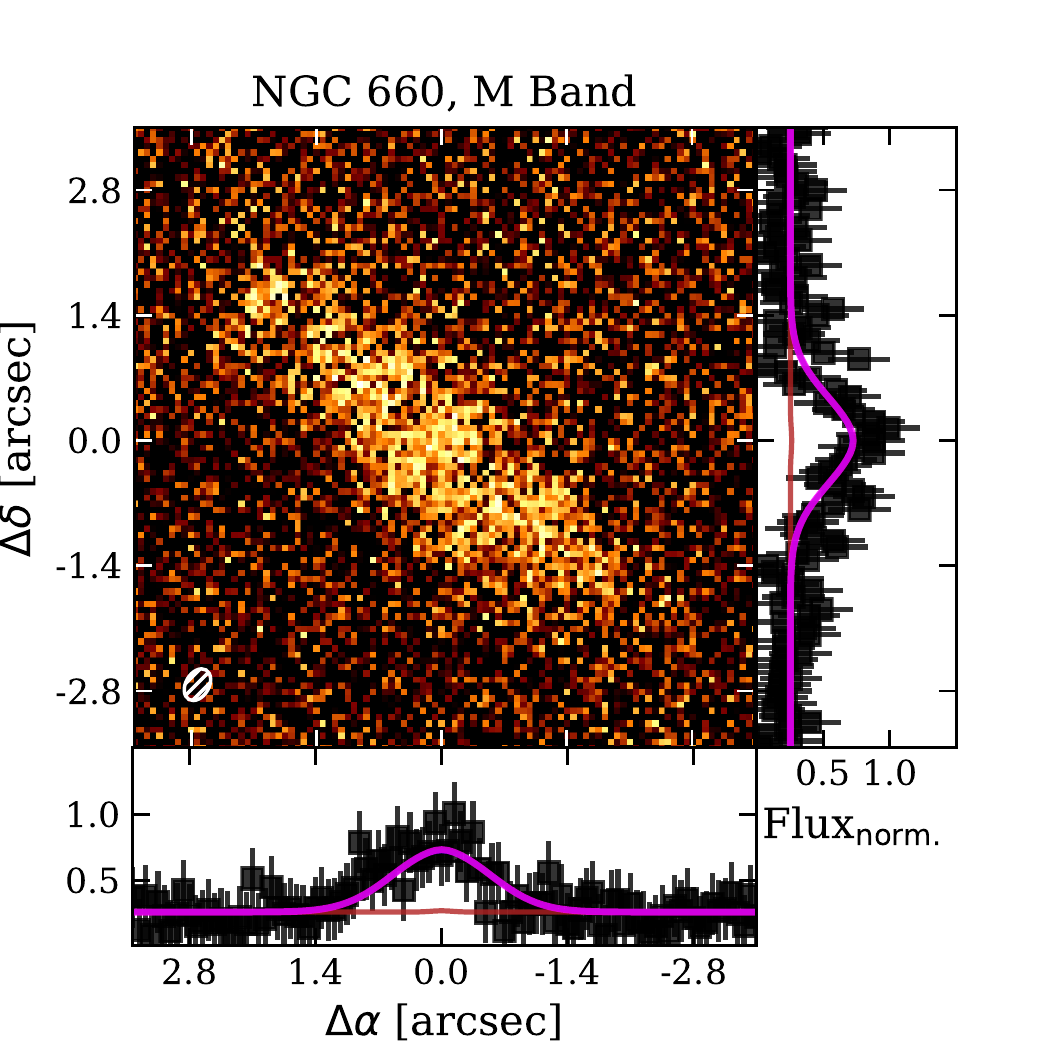}} 
\subfloat{\includegraphics[width=0.25\hsize]{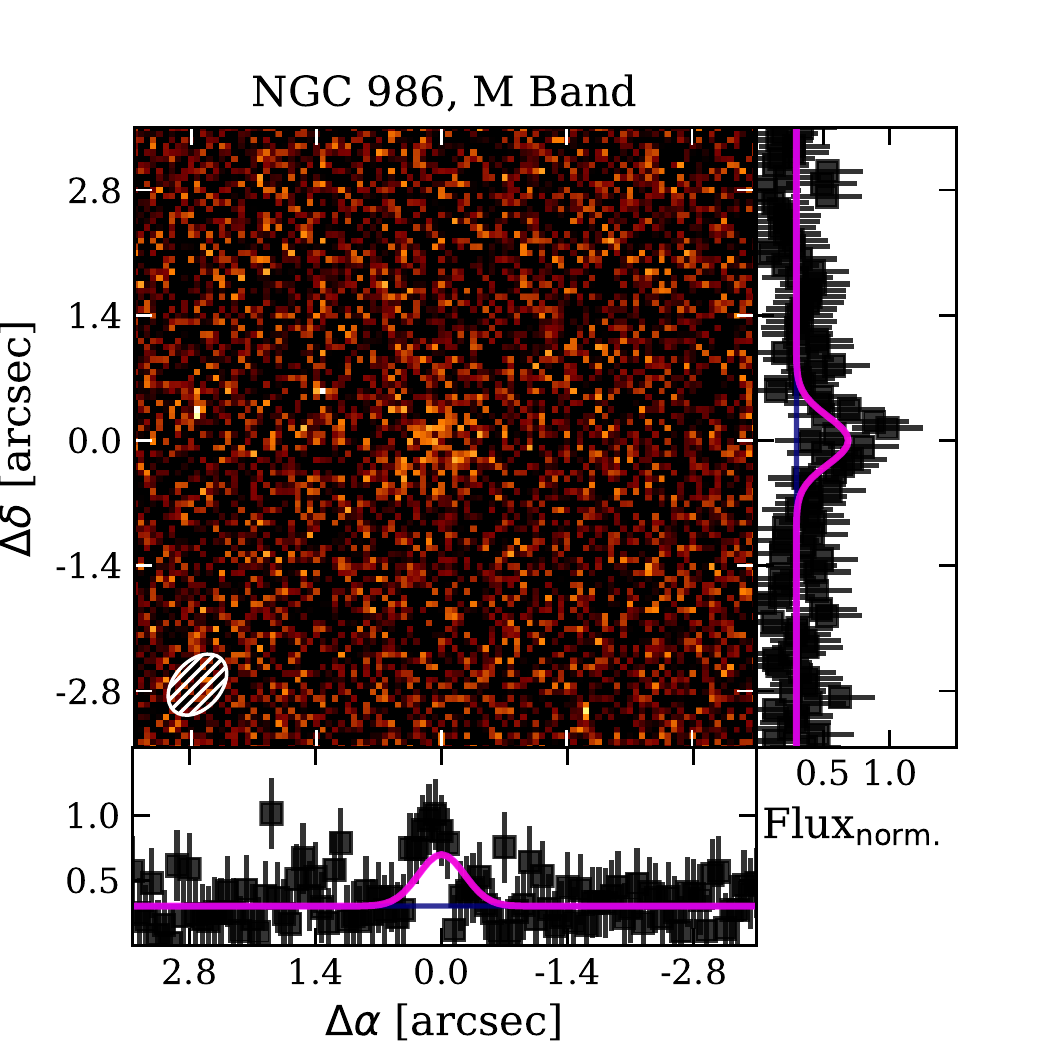}}
\subfloat{\includegraphics[width=0.25\hsize]{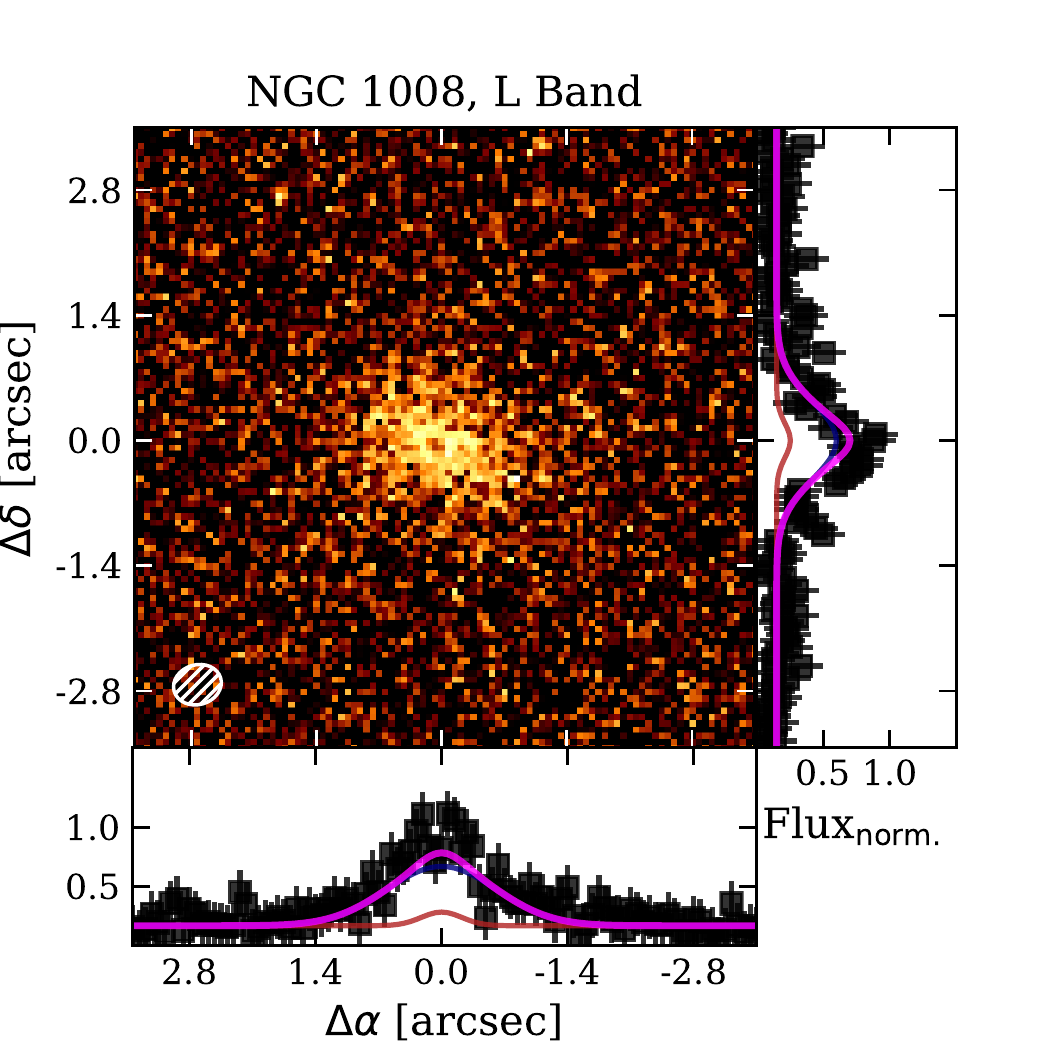}}\\
\subfloat{\includegraphics[width=0.25\hsize]{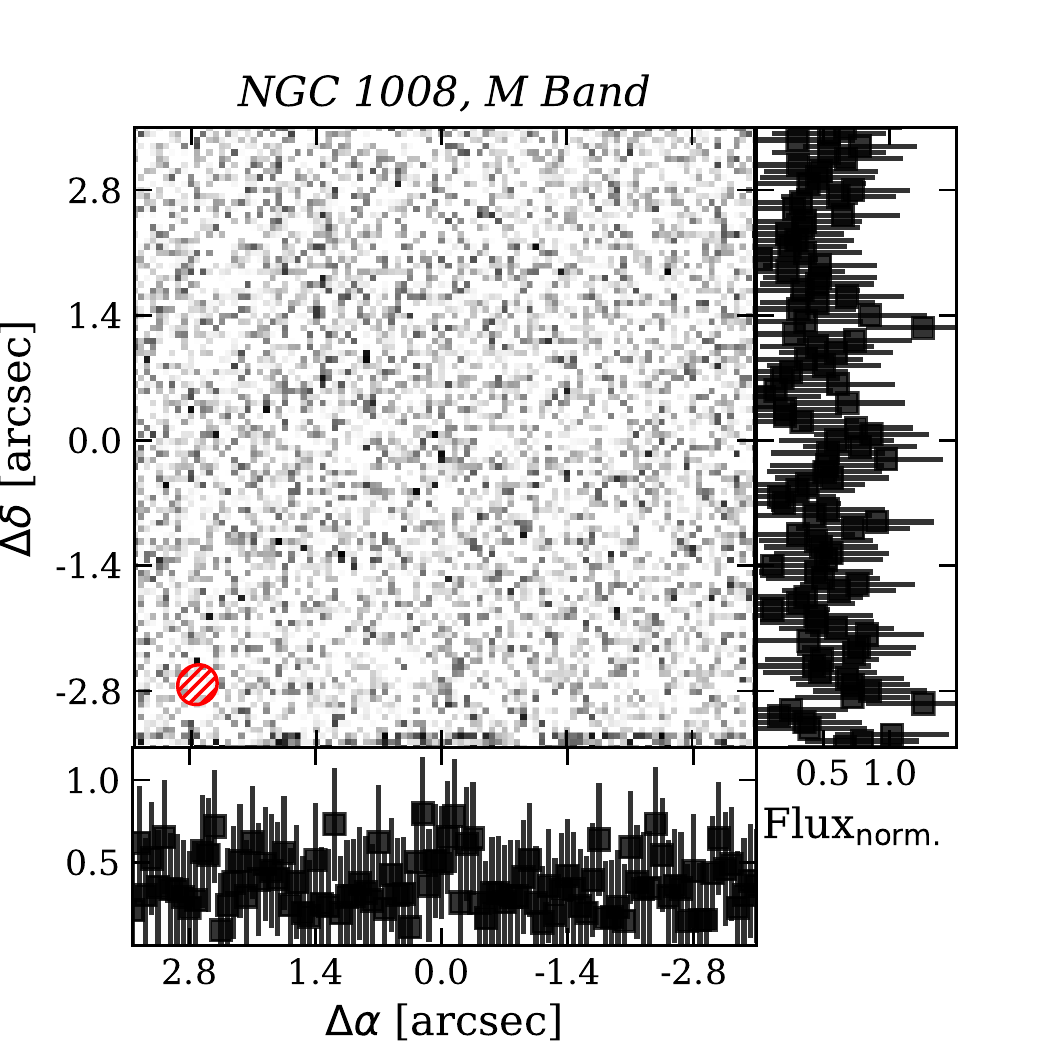}}
\subfloat{\includegraphics[width=0.25\hsize]{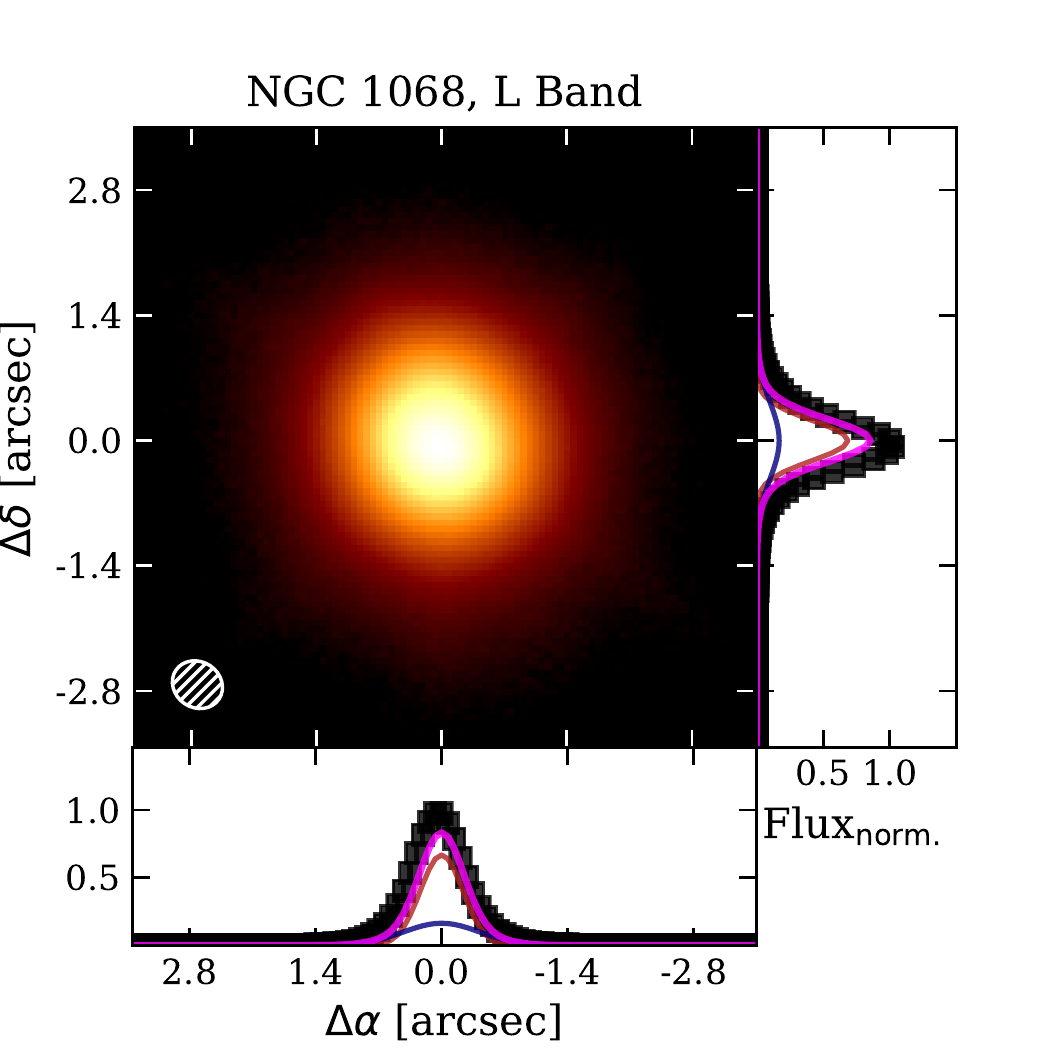}} 
\subfloat{\includegraphics[width=0.25\hsize]{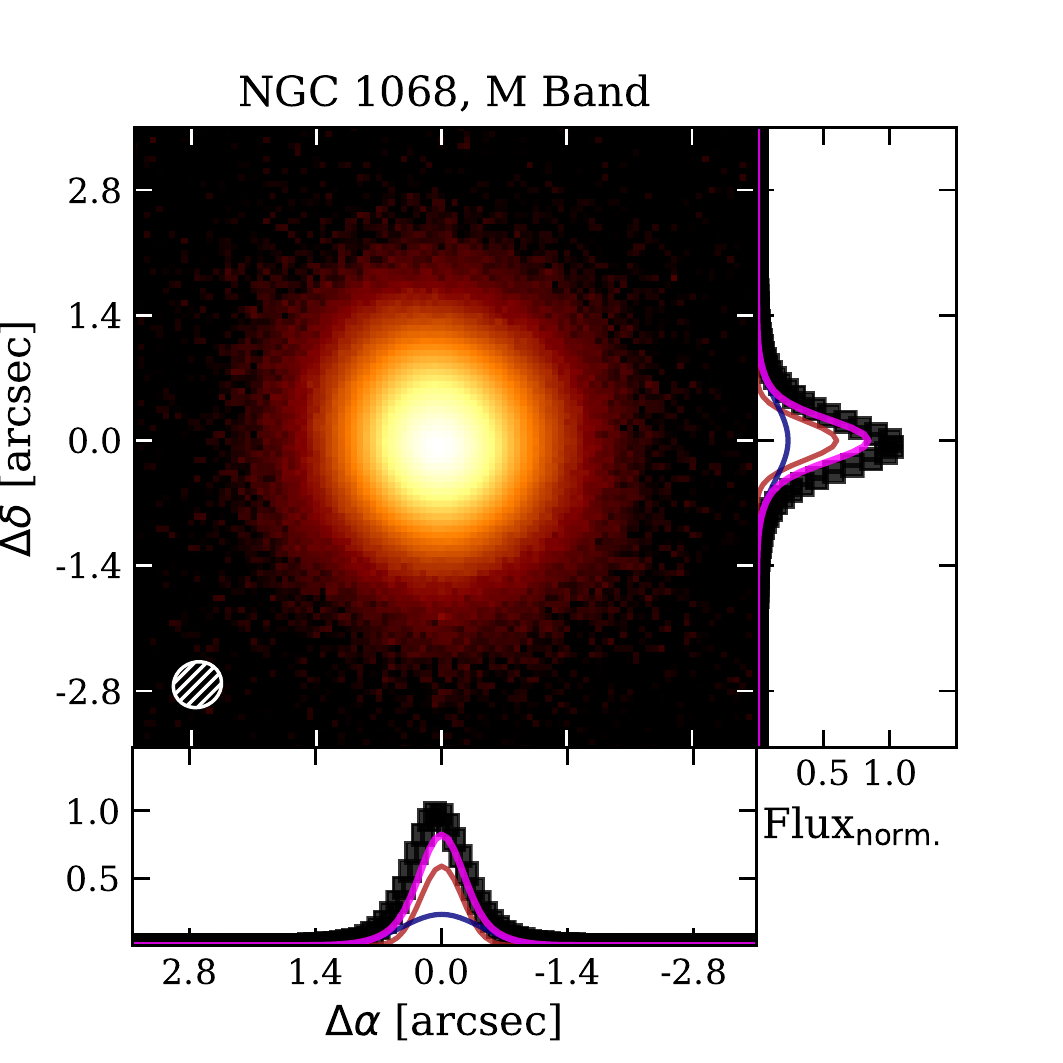}}
\subfloat{\includegraphics[width=0.25\hsize]{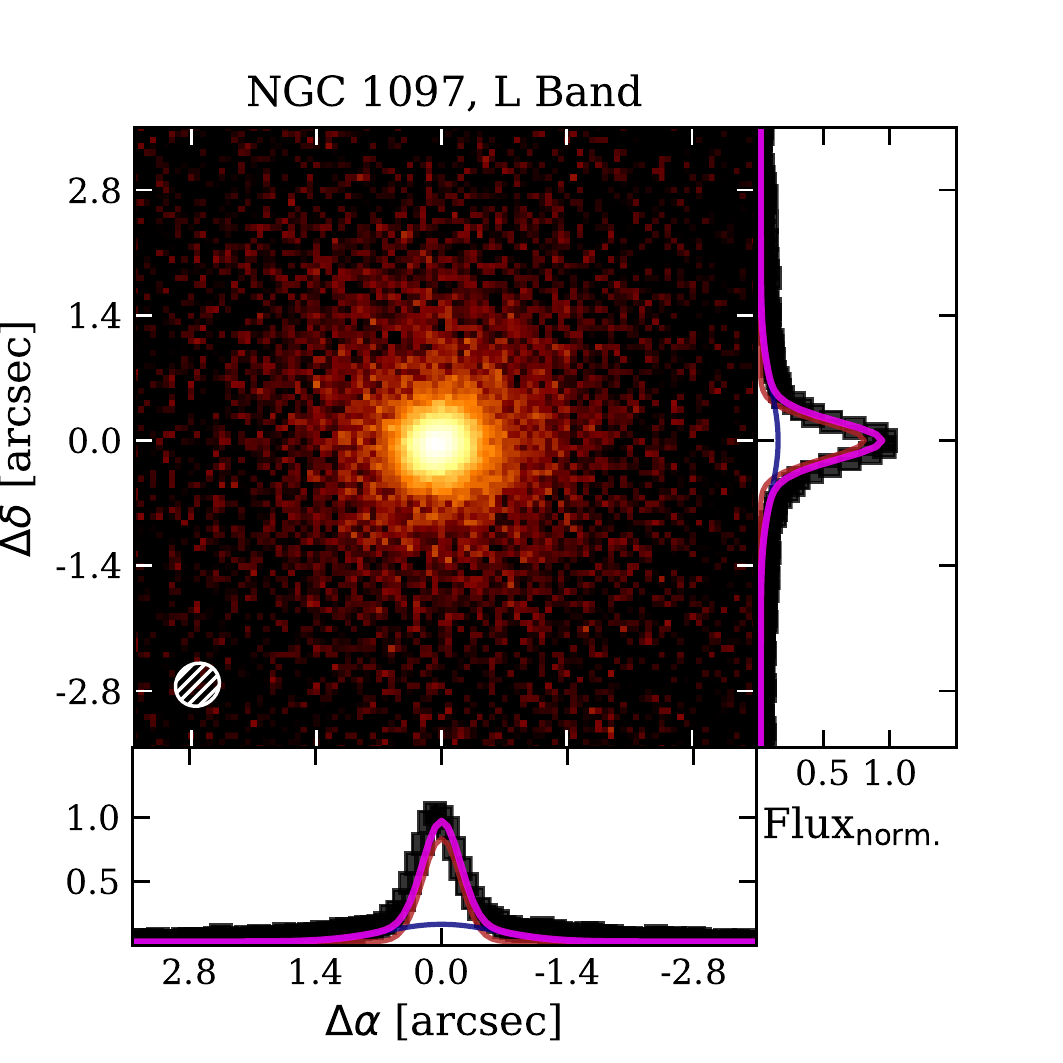}}\\
\subfloat{\includegraphics[width=0.25\hsize]{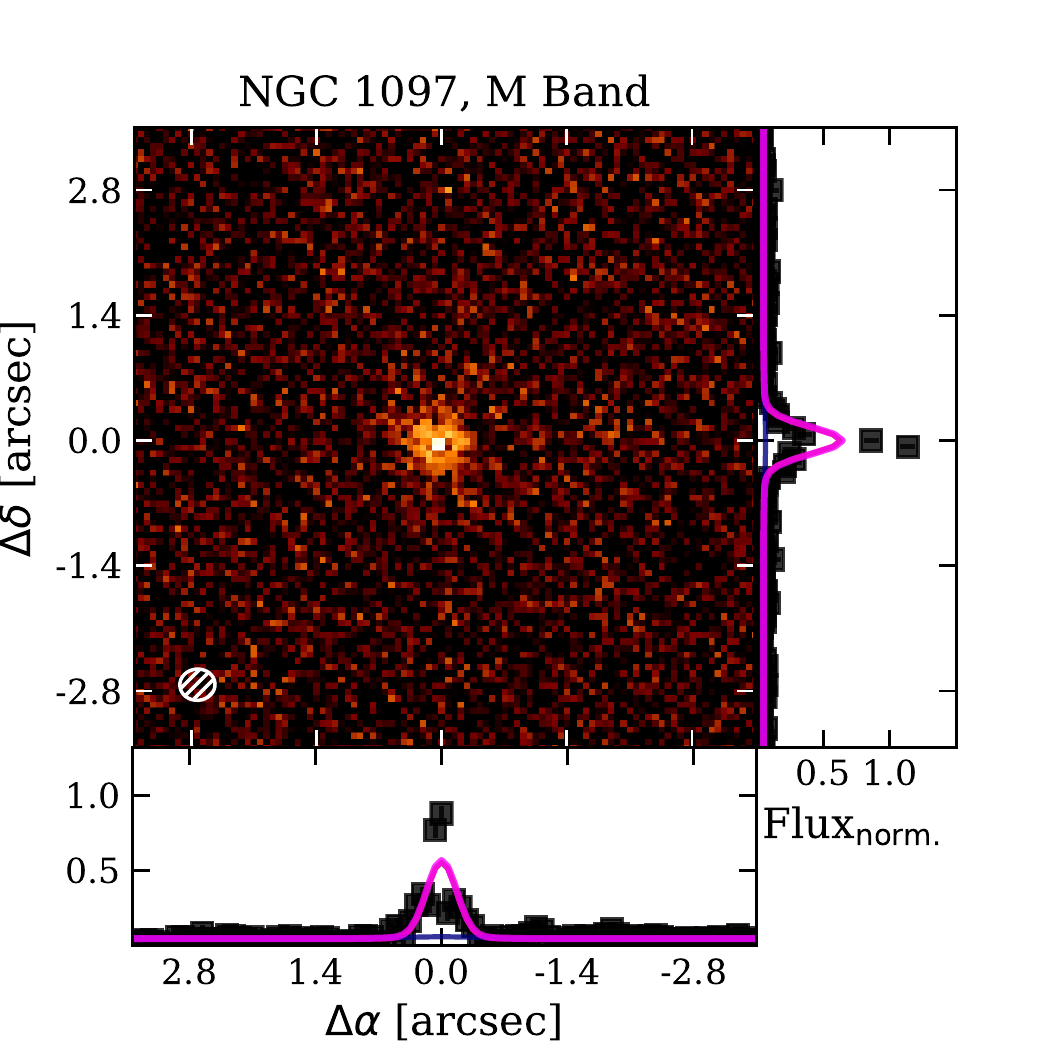}}
\subfloat{\includegraphics[width=0.25\hsize]{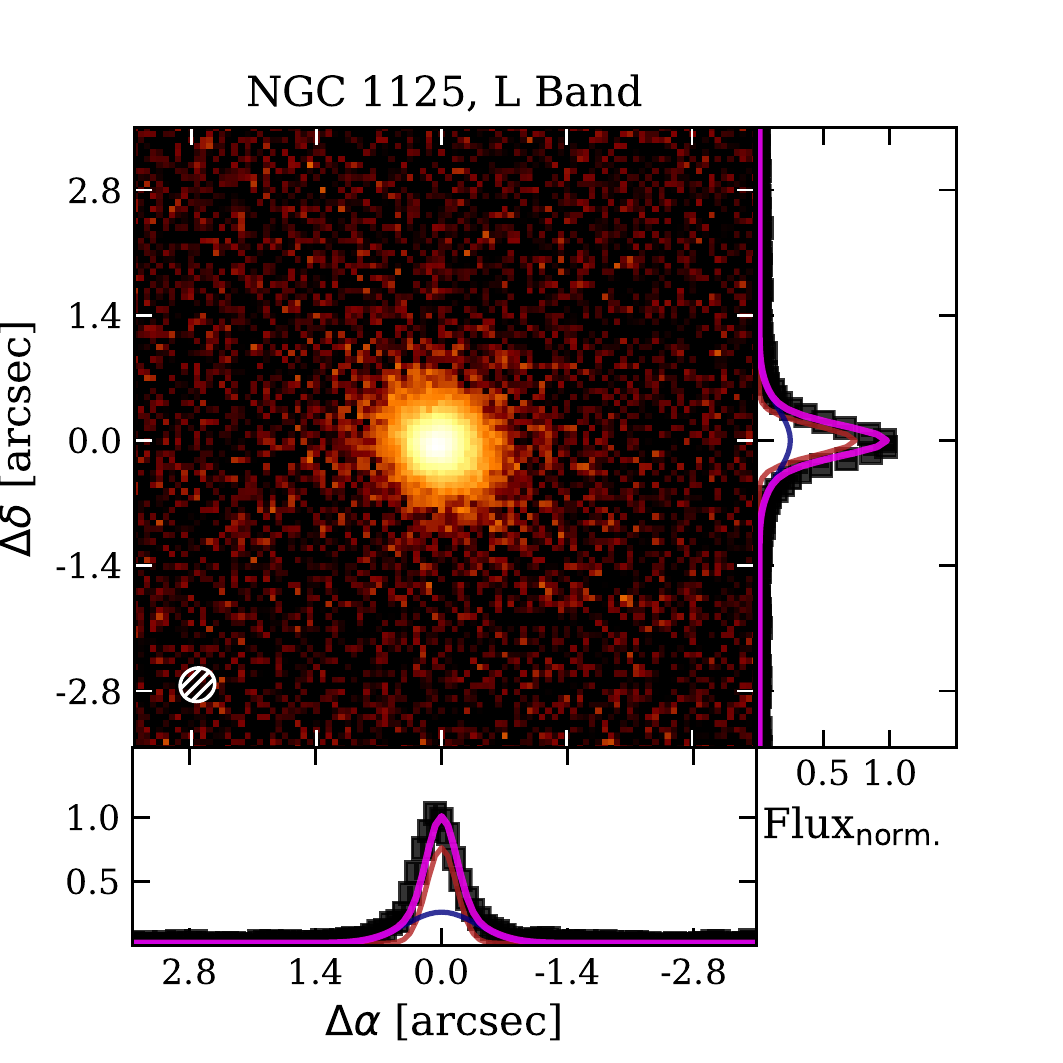}} 
\subfloat{\includegraphics[width=0.25\hsize]{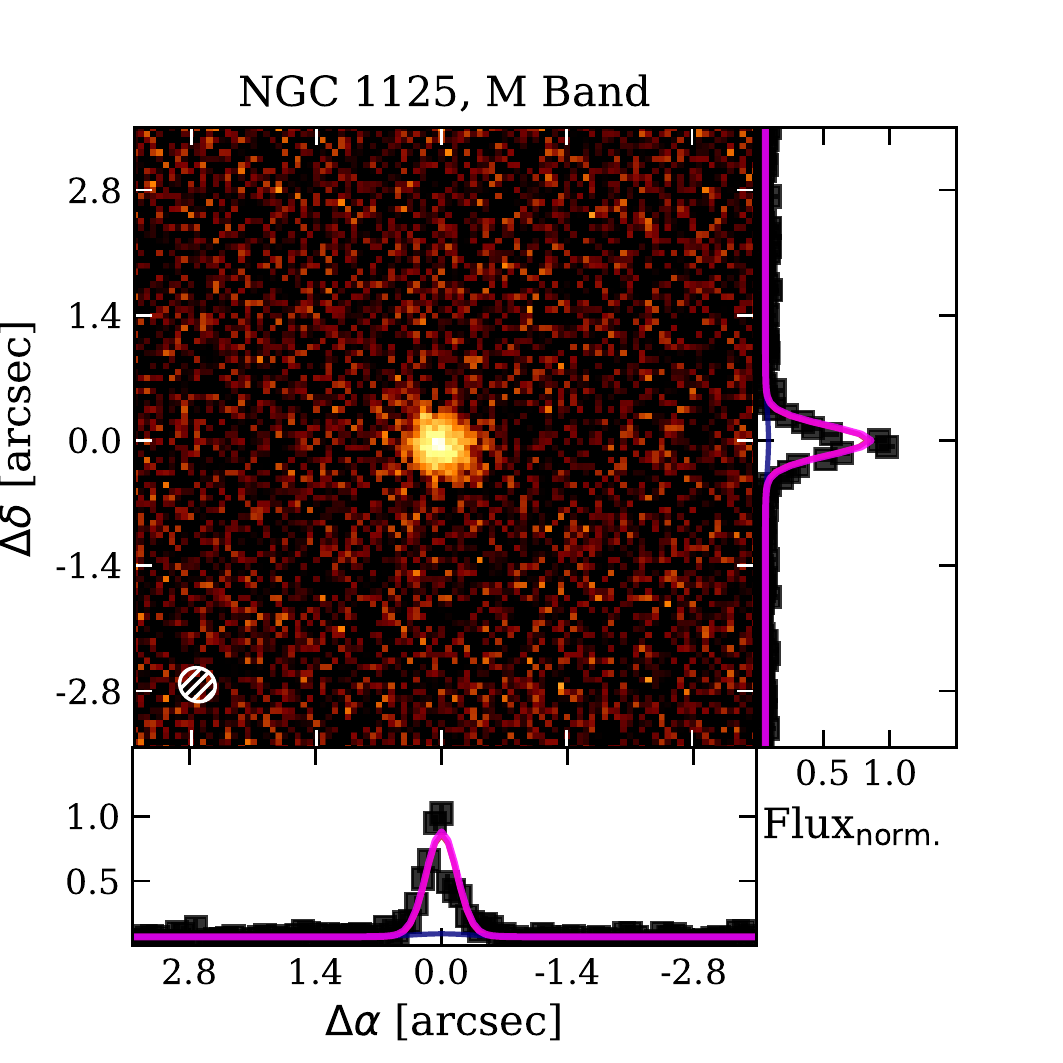}}
\subfloat{\includegraphics[width=0.25\hsize]{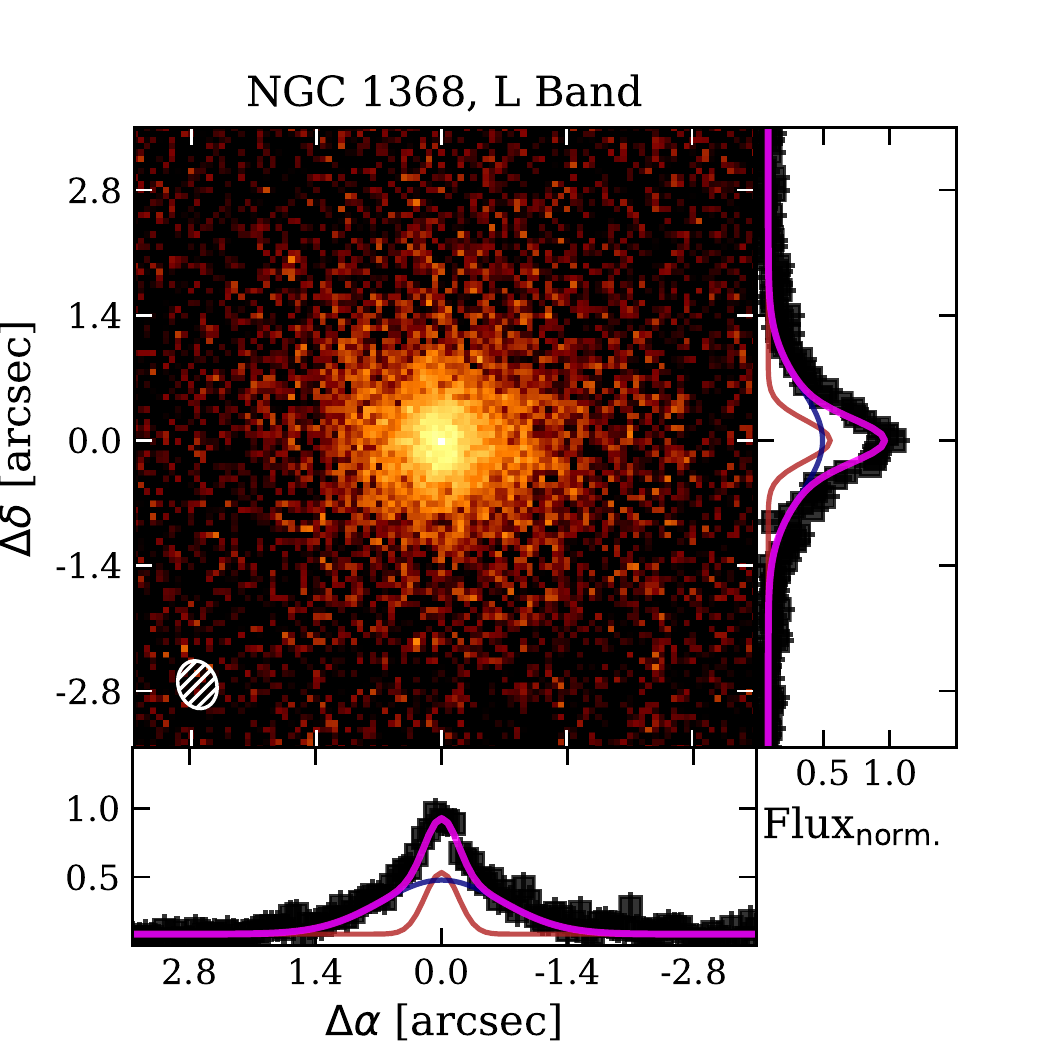}}\\
\subfloat{\includegraphics[width=0.25\hsize]{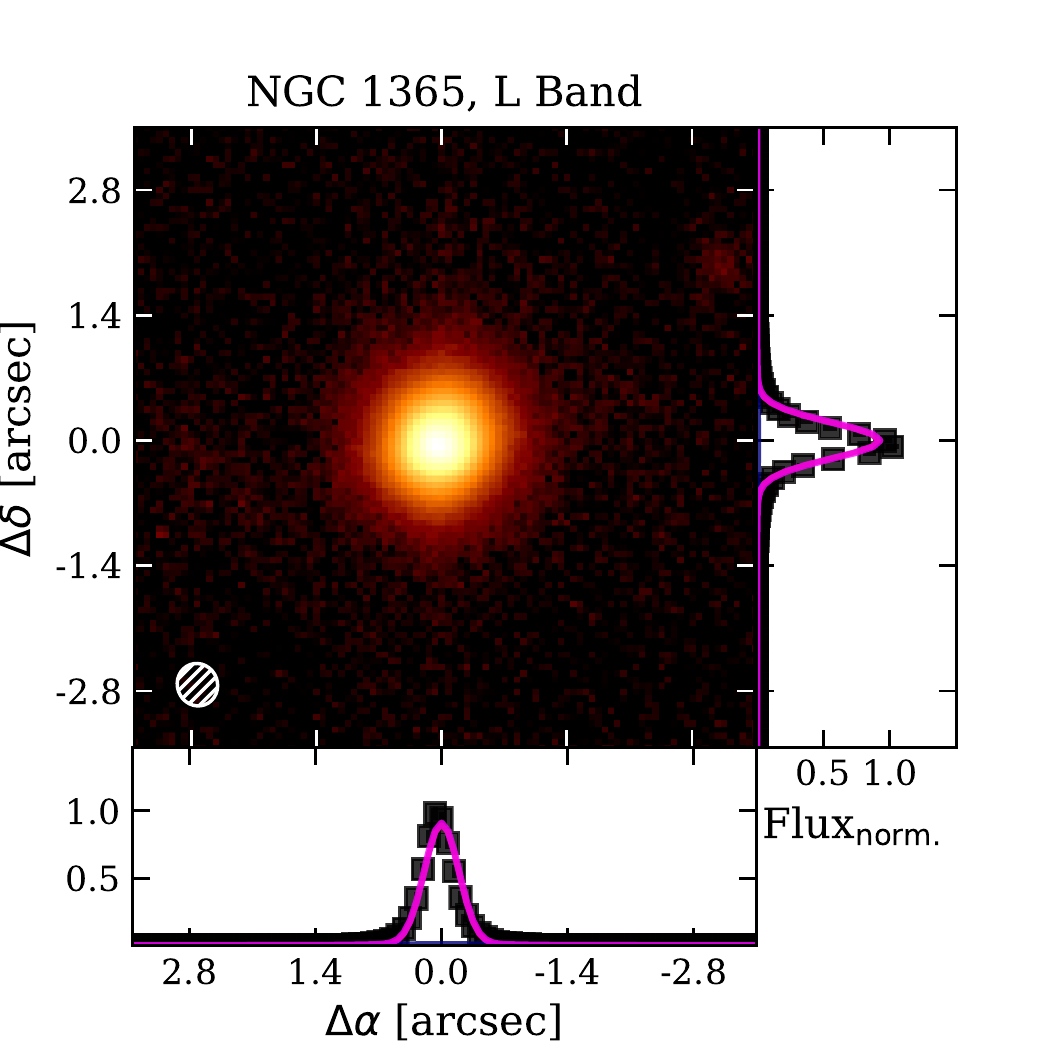}}
\subfloat{\includegraphics[width=0.25\hsize]{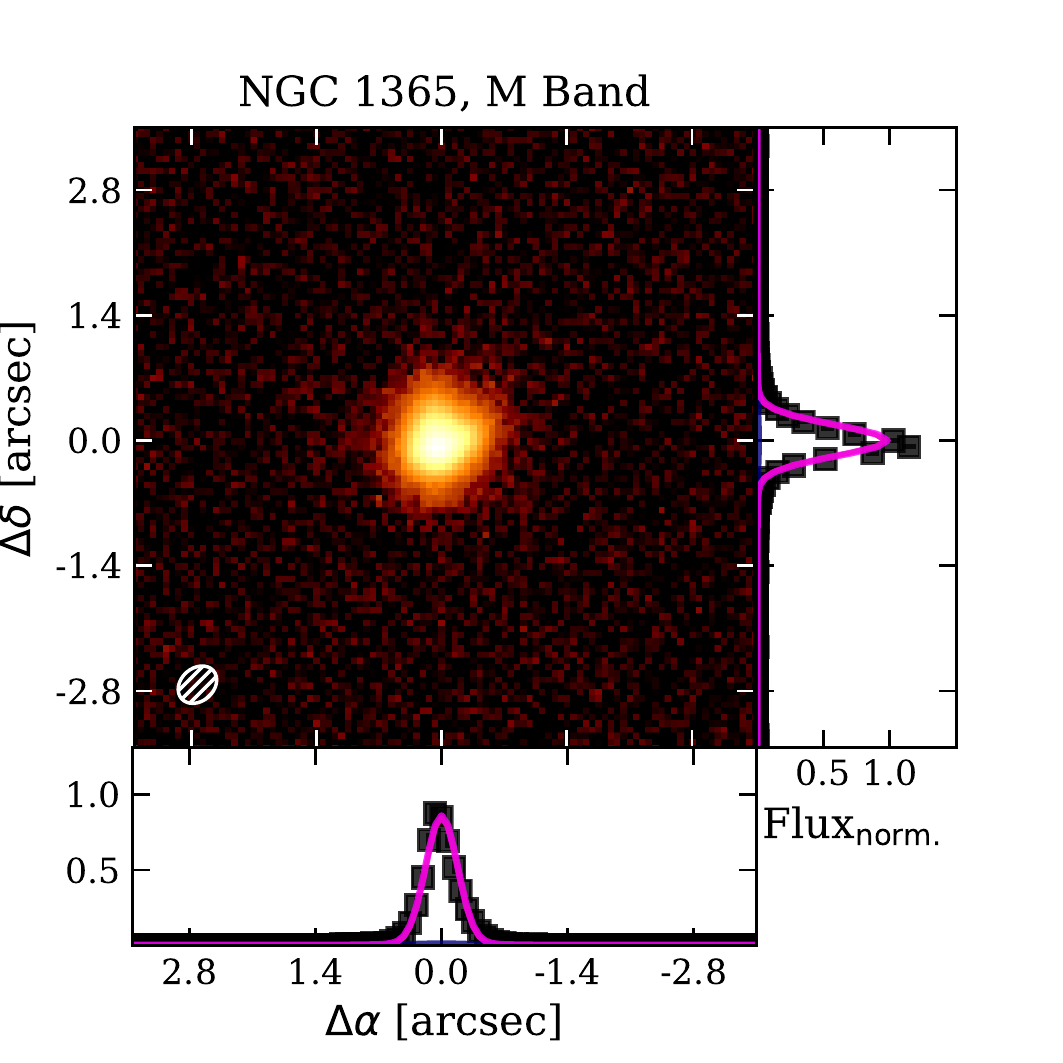}} 
\subfloat{\includegraphics[width=0.25\hsize]{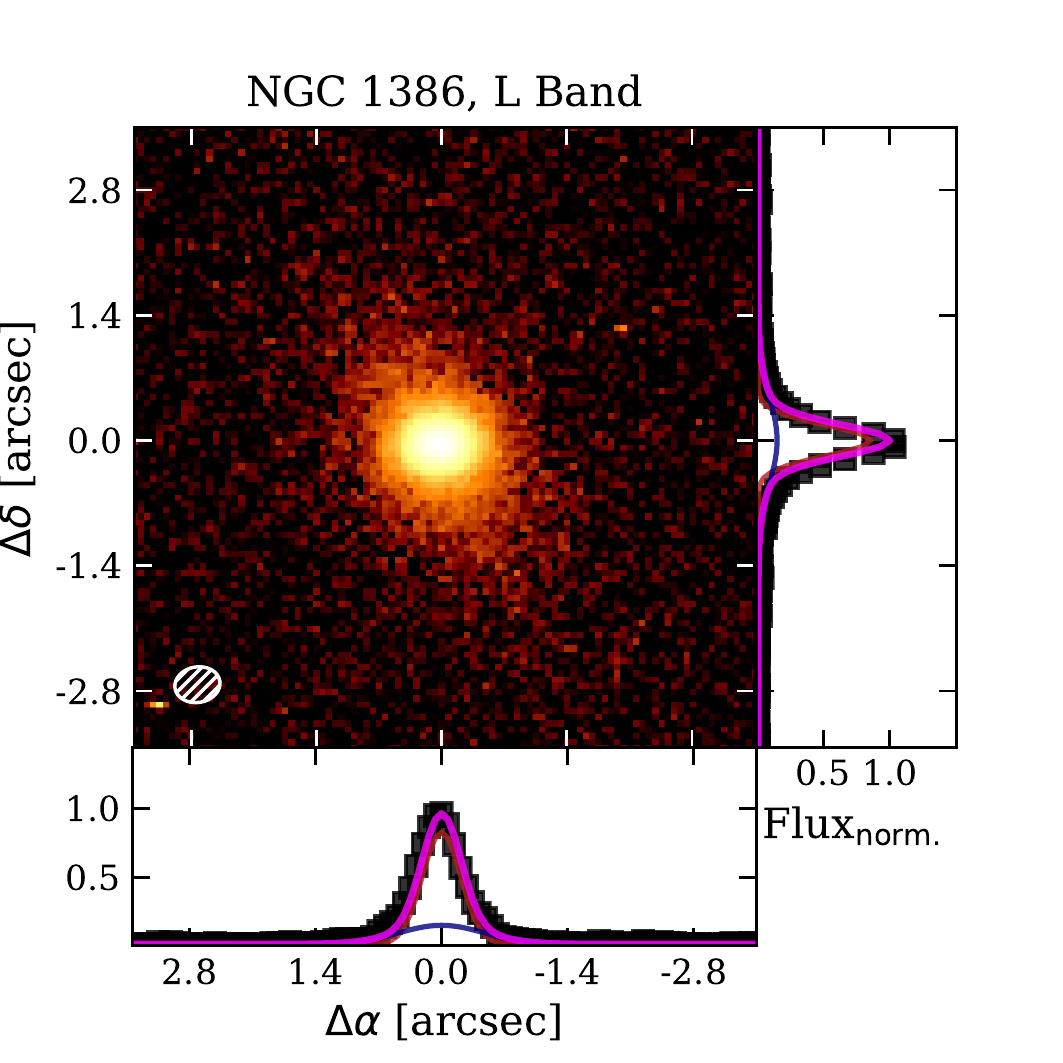}}
\subfloat{\includegraphics[width=0.25\hsize]{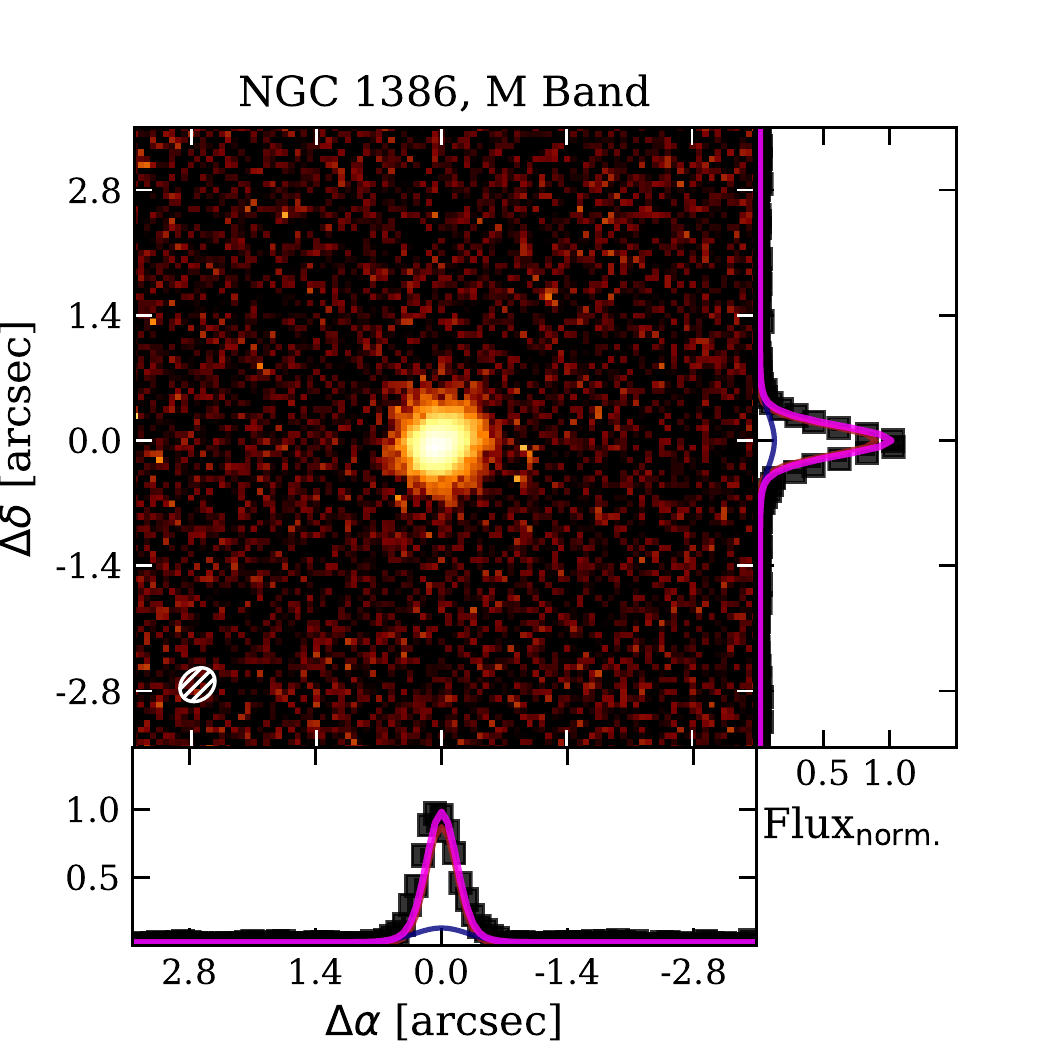}} \\
\subfloat{\includegraphics[width=0.25\hsize]{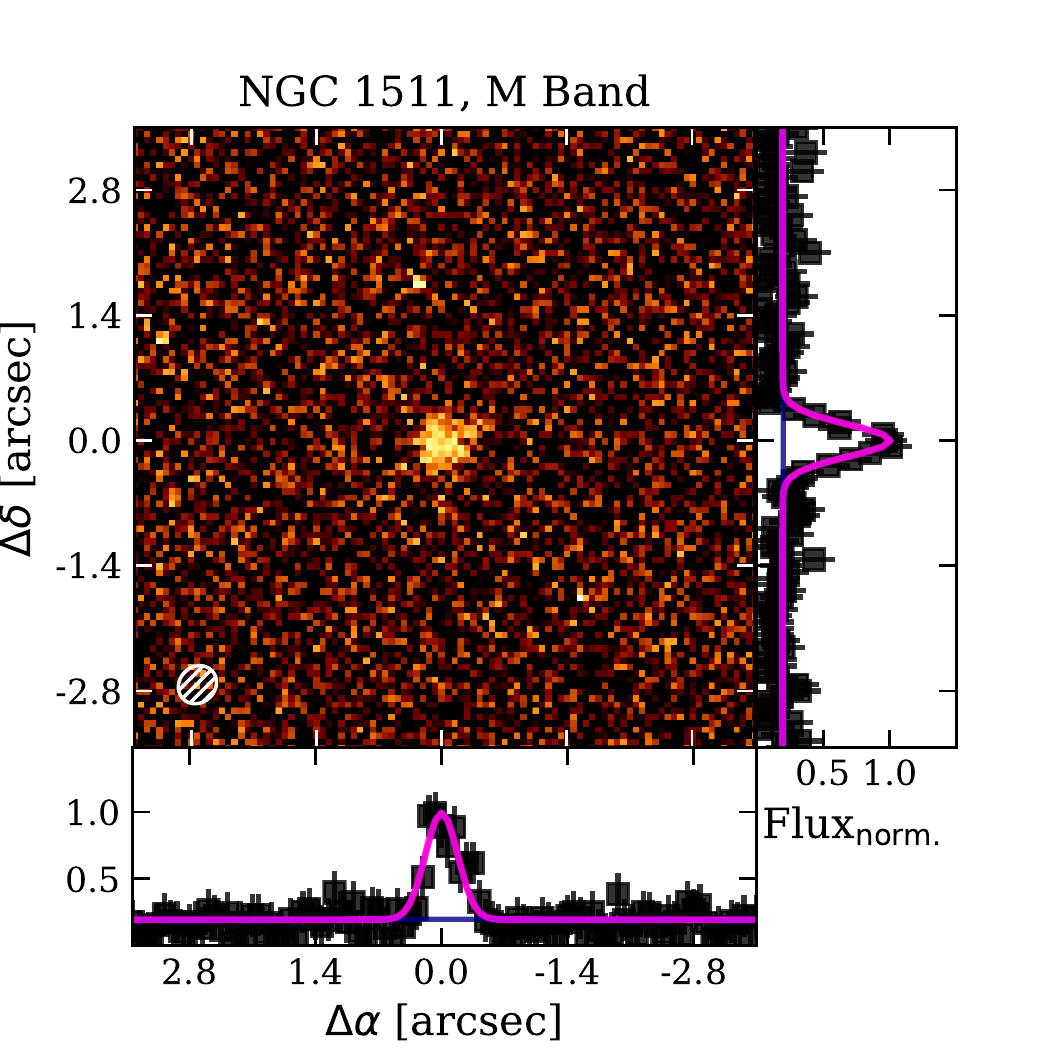}}
\subfloat{\includegraphics[width=0.25\hsize]{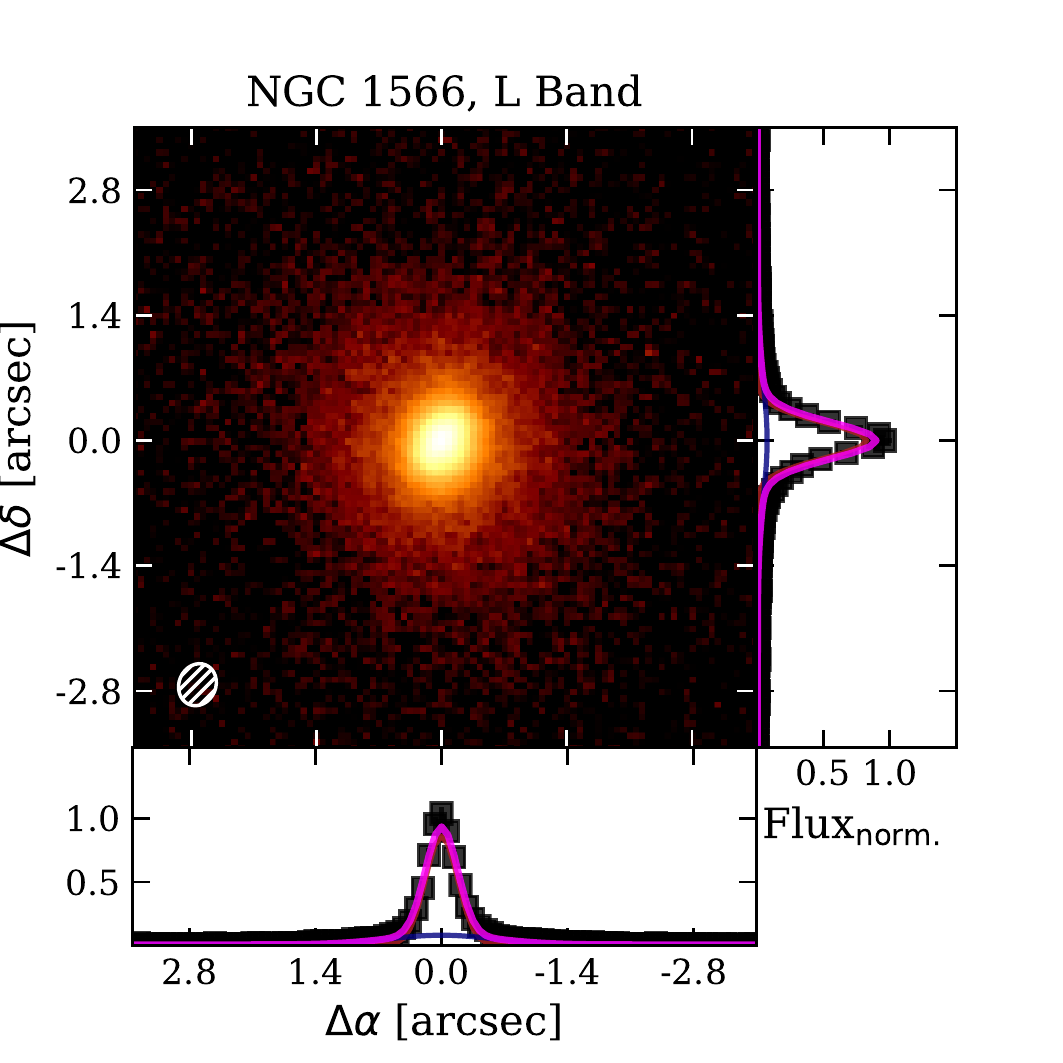}} 
\subfloat{\includegraphics[width=0.25\hsize]{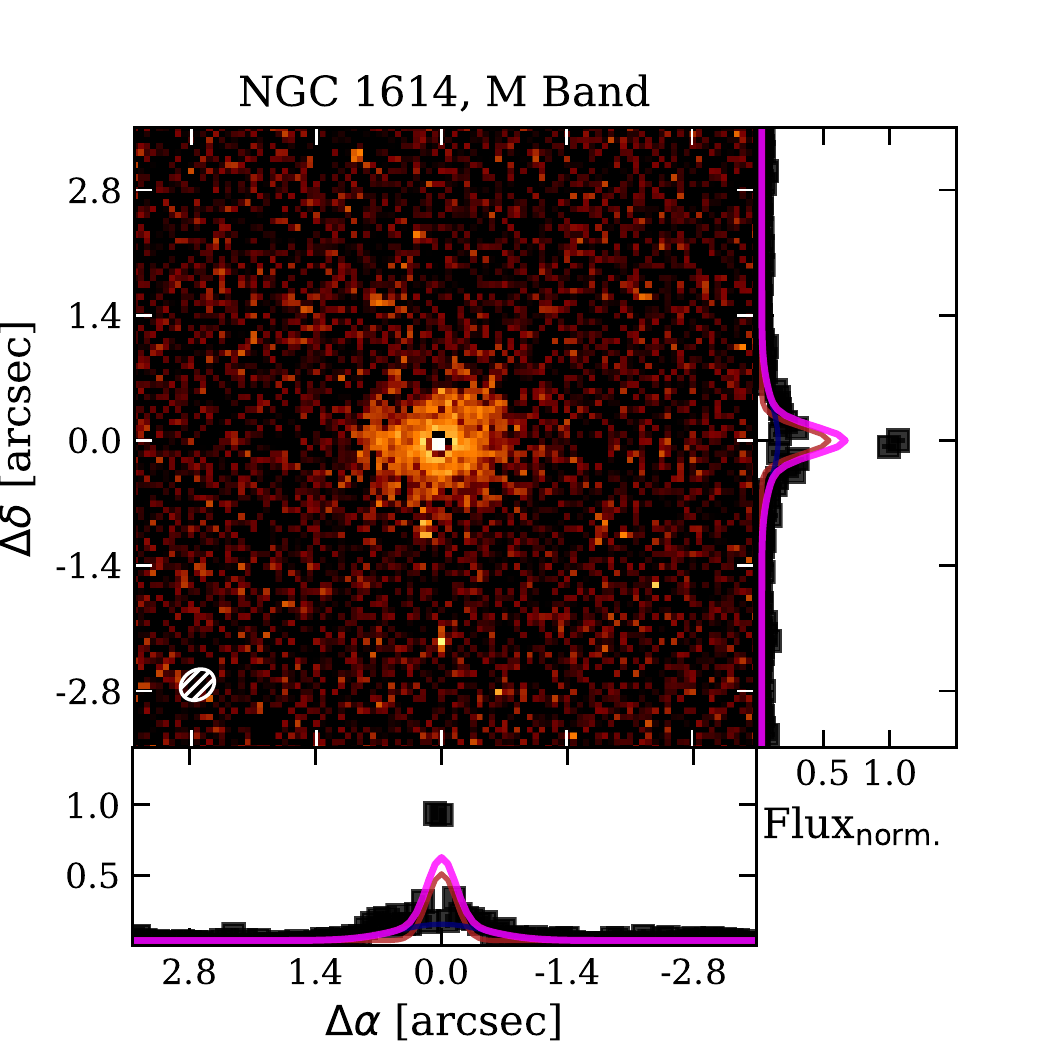}}
\subfloat{\includegraphics[width=0.25\hsize]{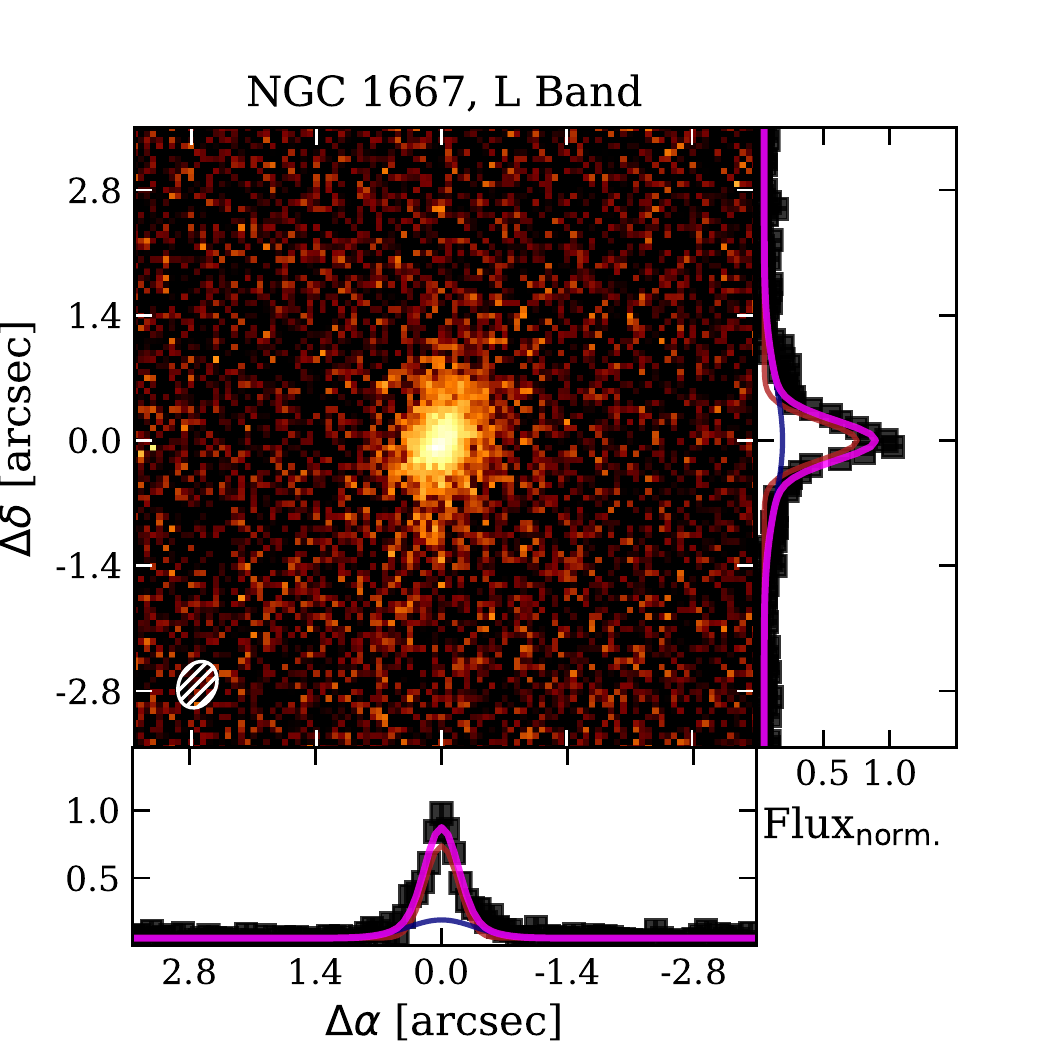}}\\
\caption{ As Fig \ref{fig:cutouts_one} but for all sources.}
\end{figure*}
\begin{figure*}
\subfloat{\includegraphics[width=0.25\hsize]{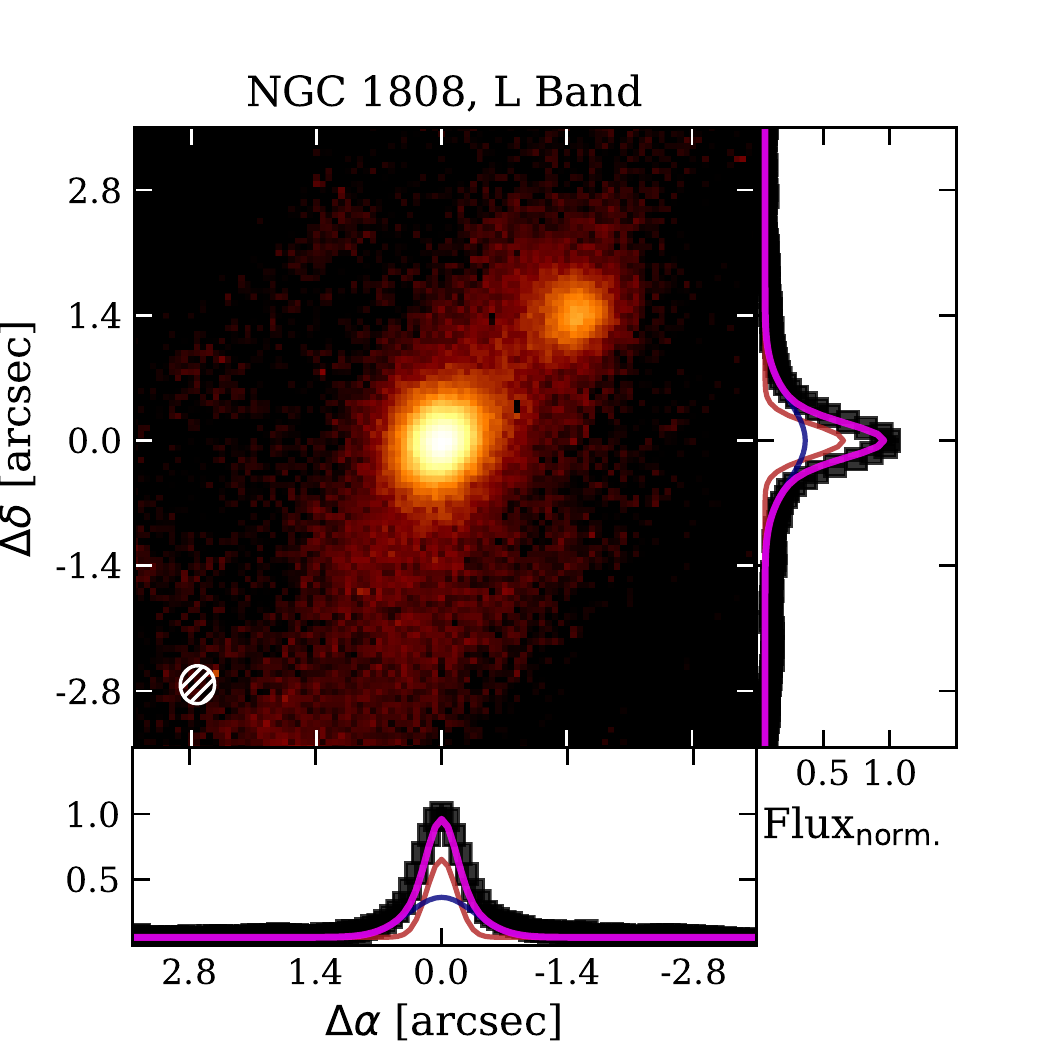}}
\subfloat{\includegraphics[width=0.25\hsize]{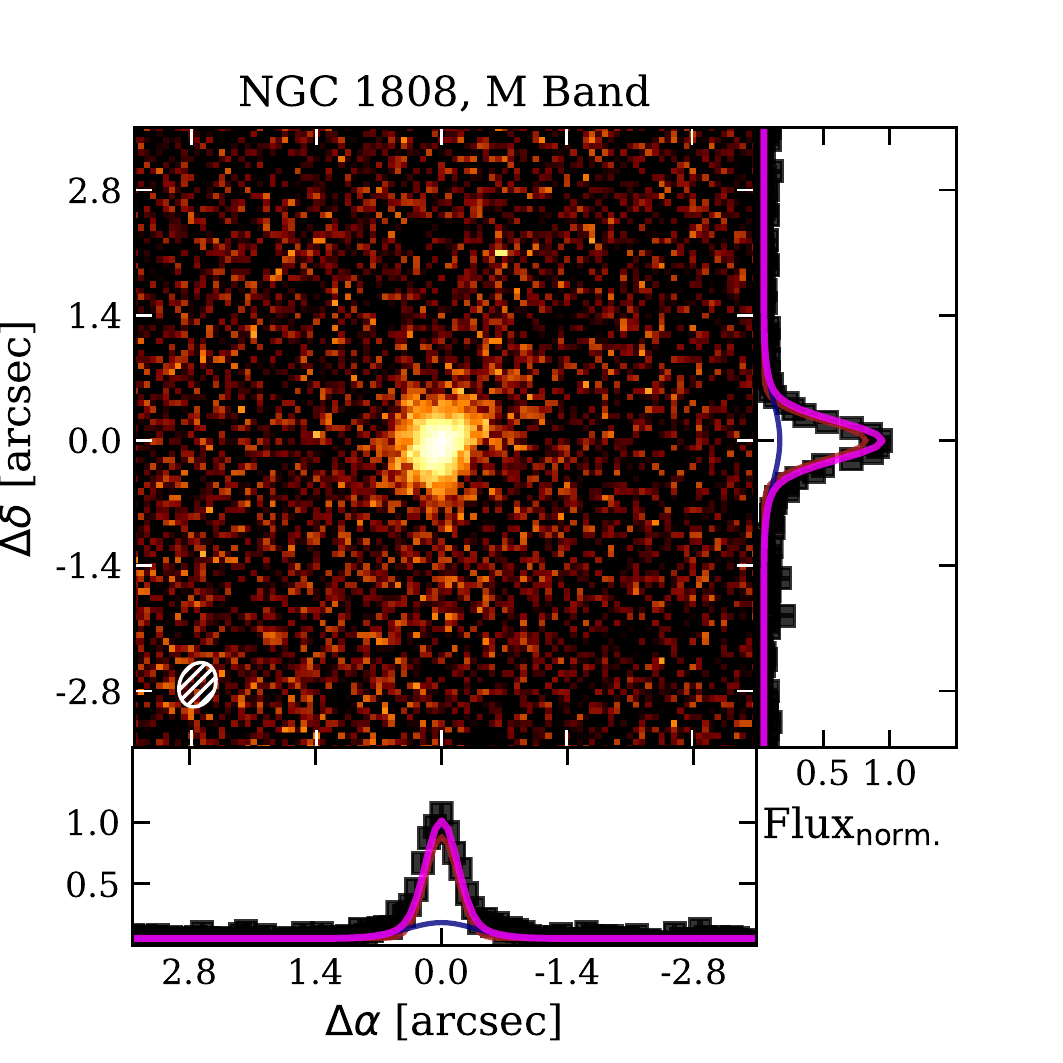}} 
\subfloat{\includegraphics[width=0.25\hsize]{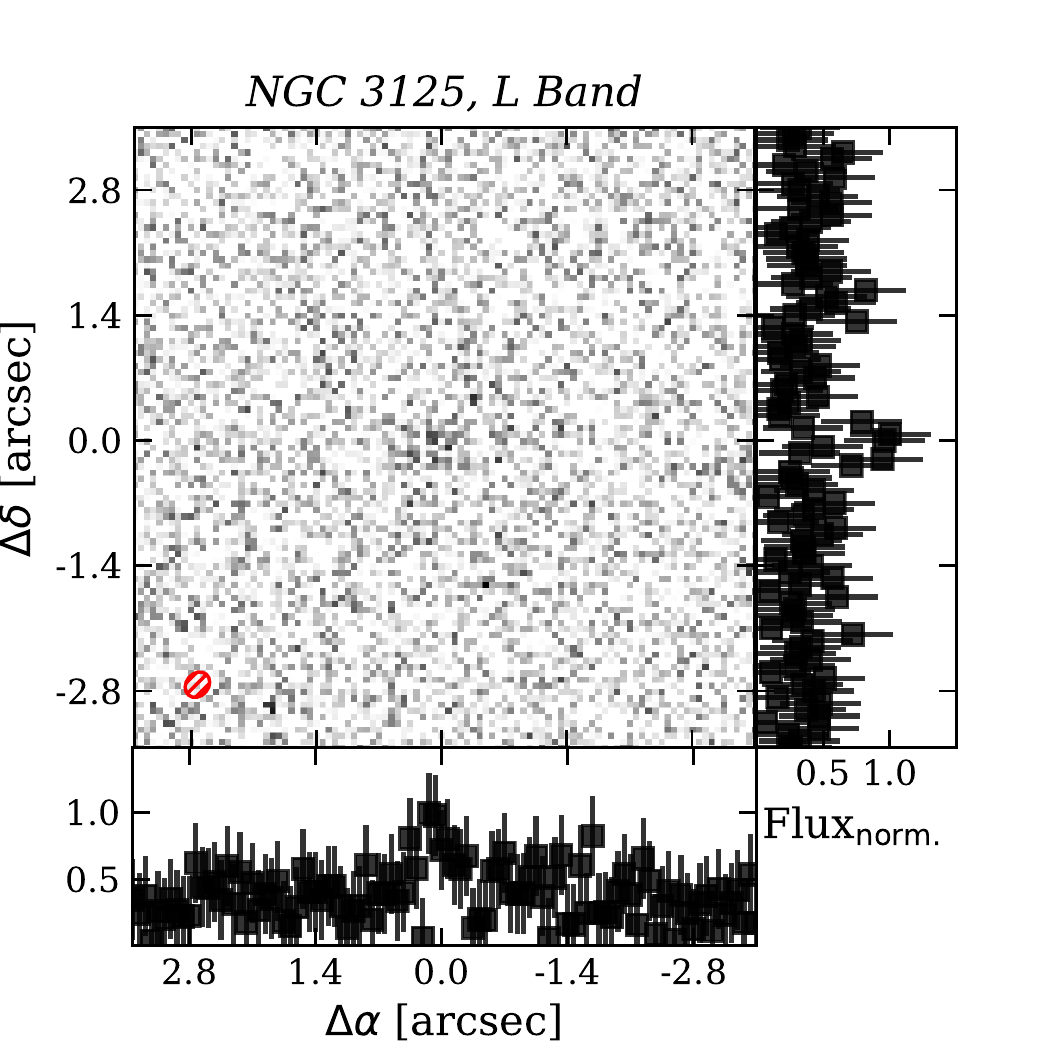}}
\subfloat{\includegraphics[width=0.25\hsize]{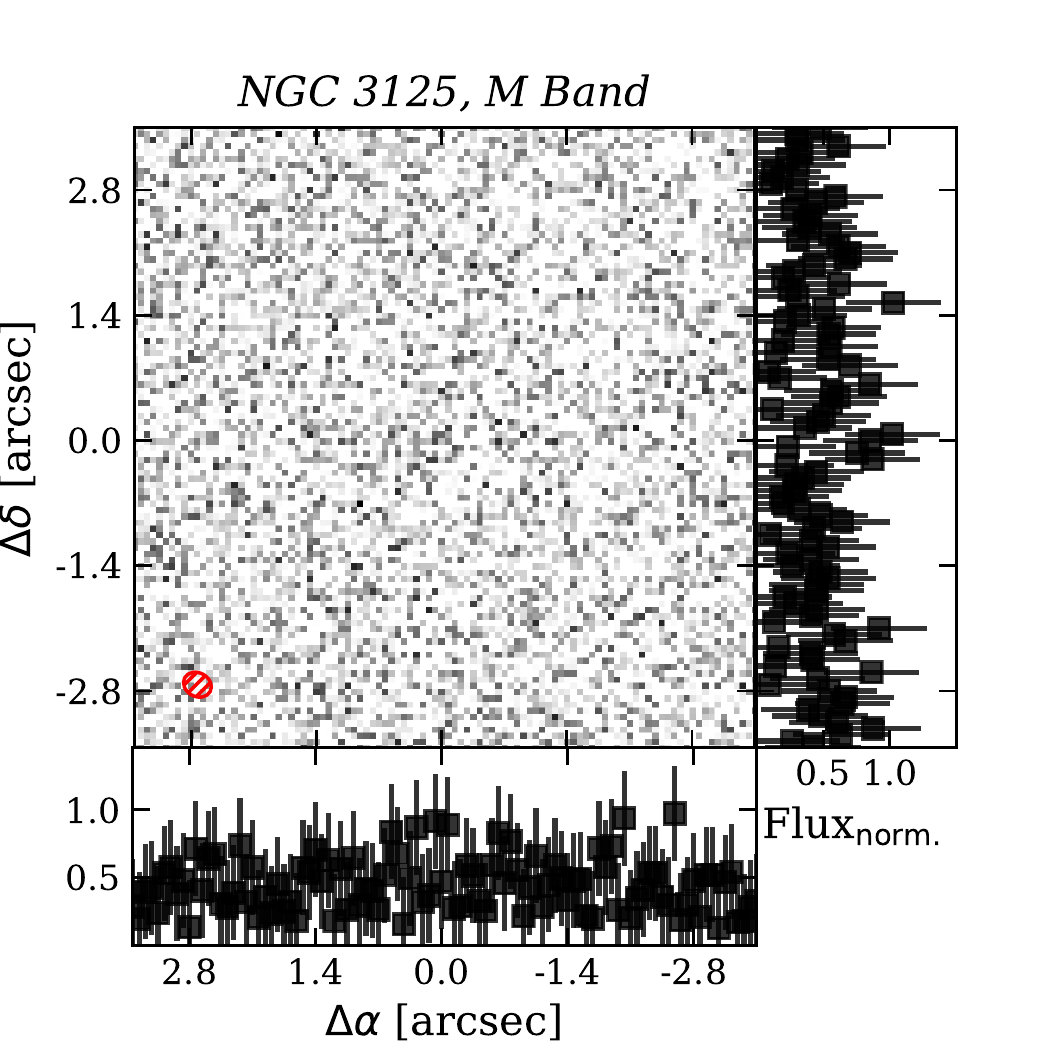}}\\
\subfloat{\includegraphics[width=0.25\hsize]{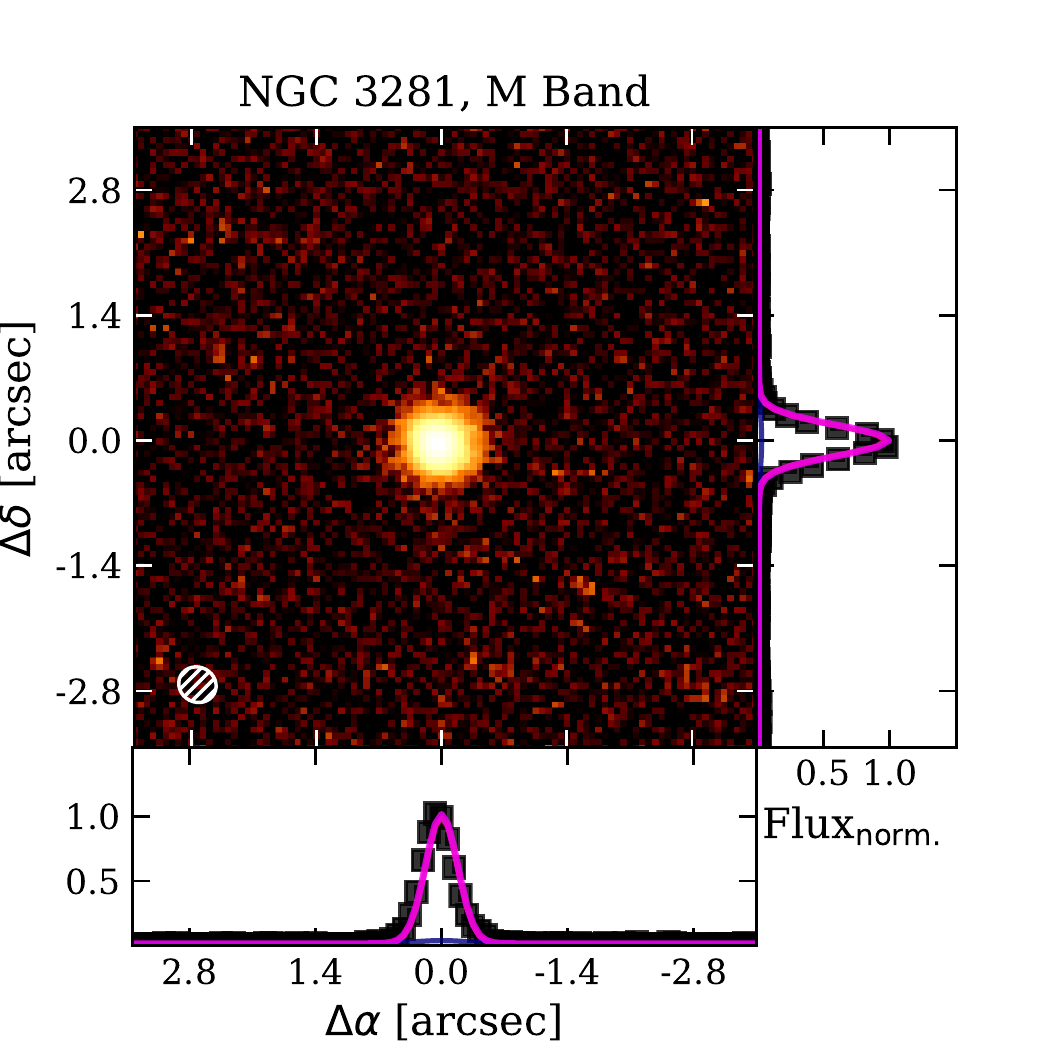}}
\subfloat{\includegraphics[width=0.25\hsize]{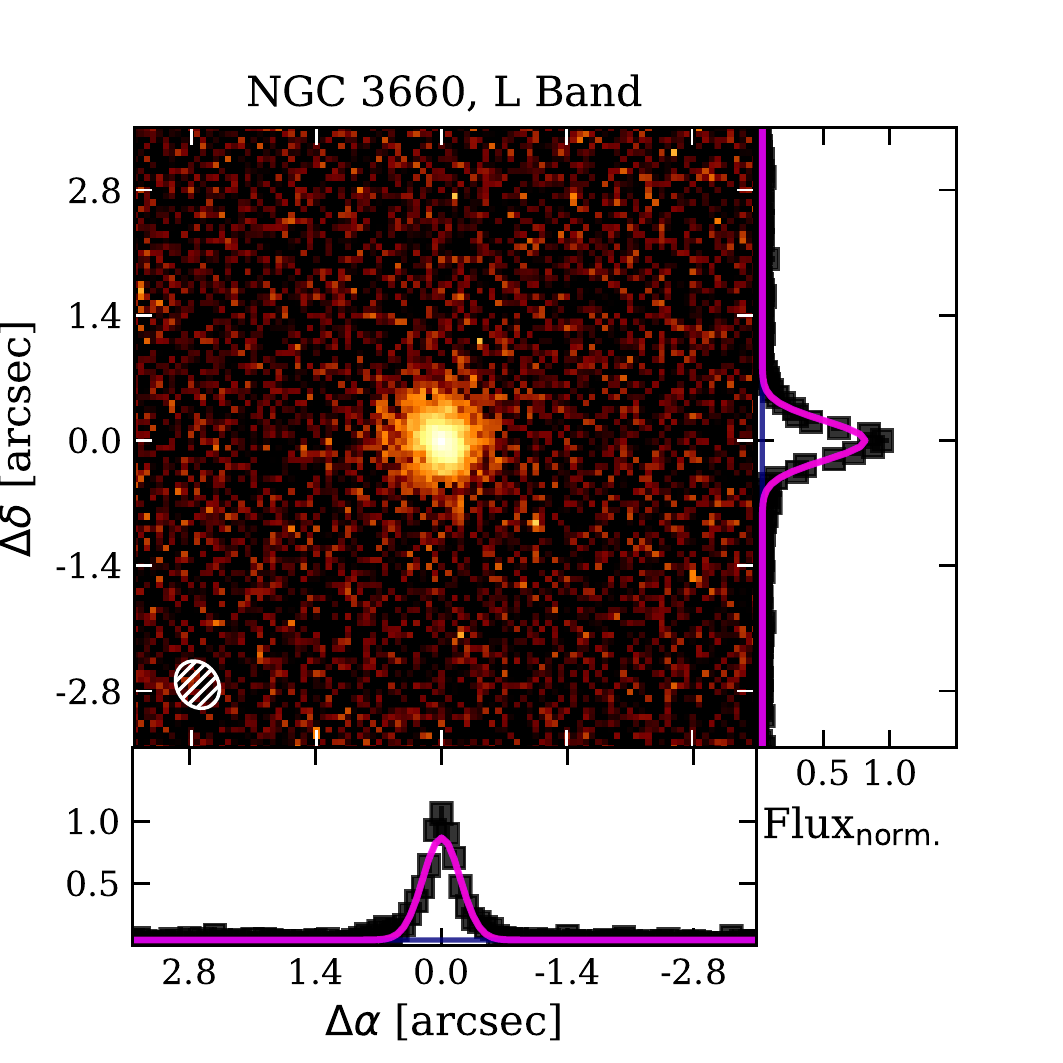}} 
\subfloat{\includegraphics[width=0.25\hsize]{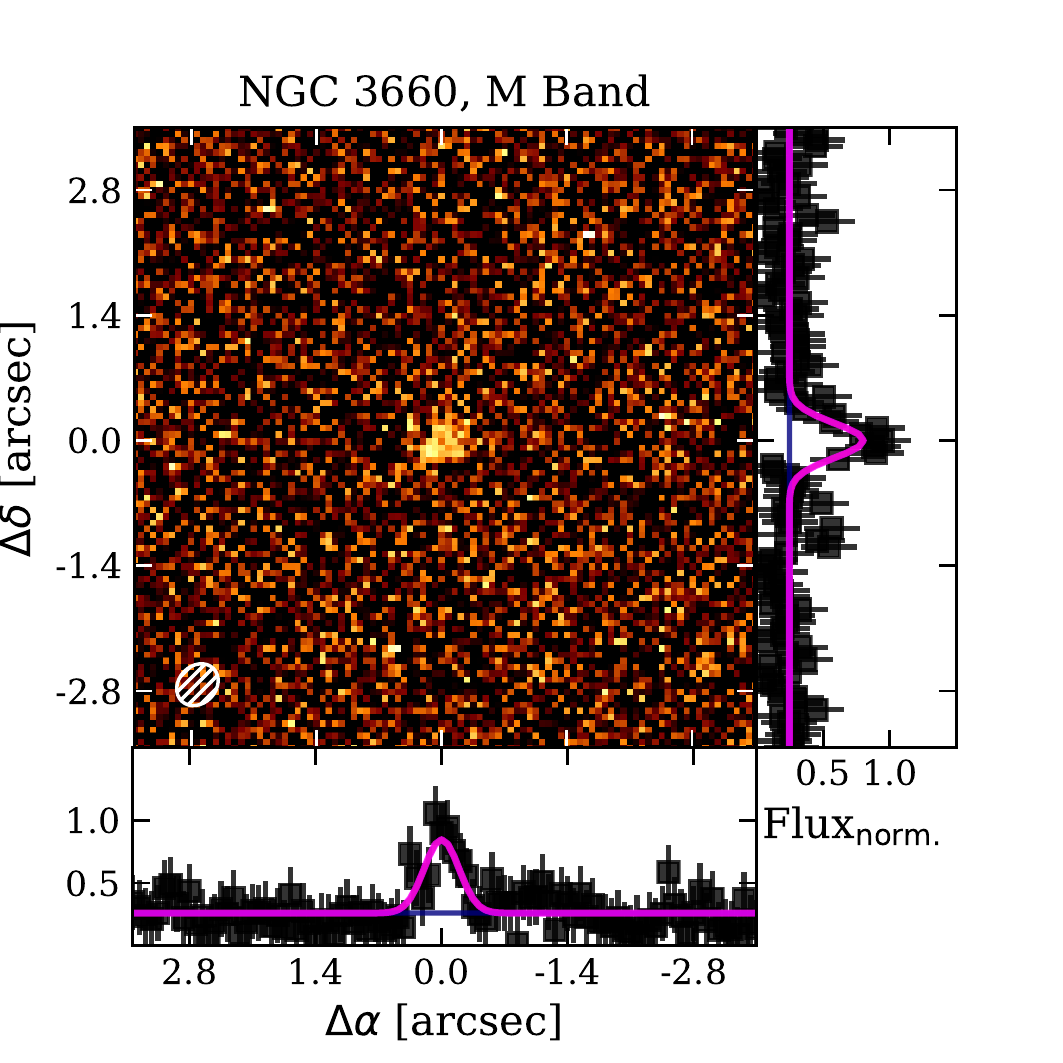}}
\subfloat{\includegraphics[width=0.25\hsize]{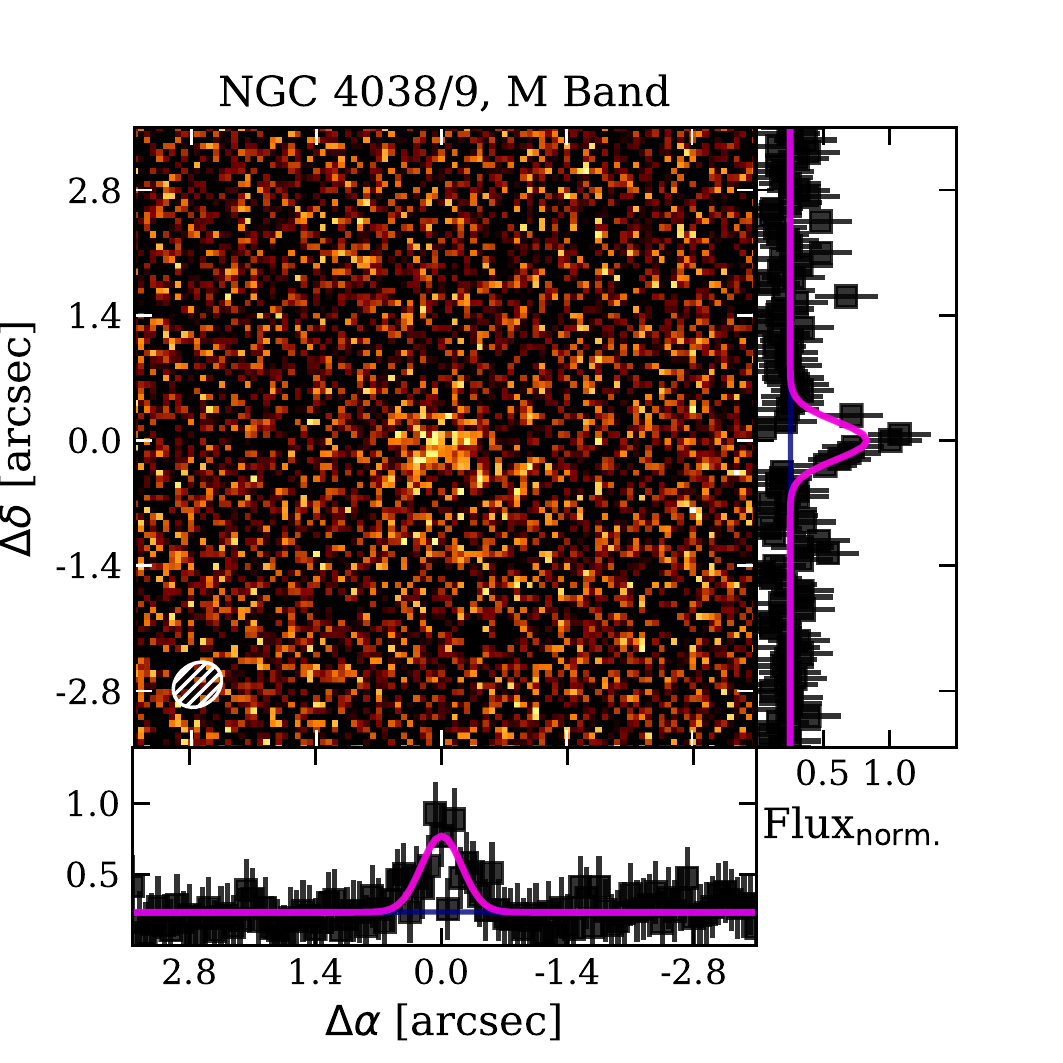}}\\
\subfloat{\includegraphics[width=0.25\hsize]{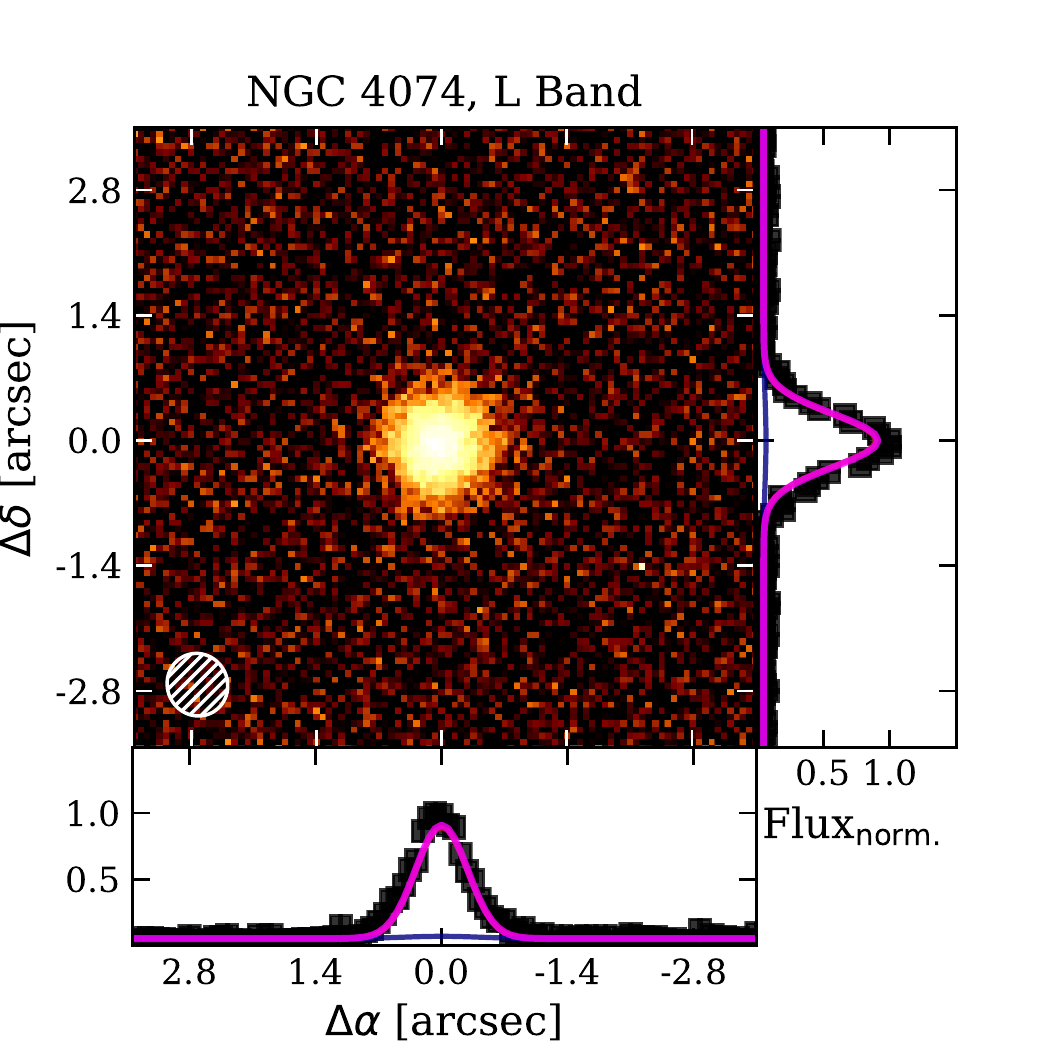}}
\subfloat{\includegraphics[width=0.25\hsize]{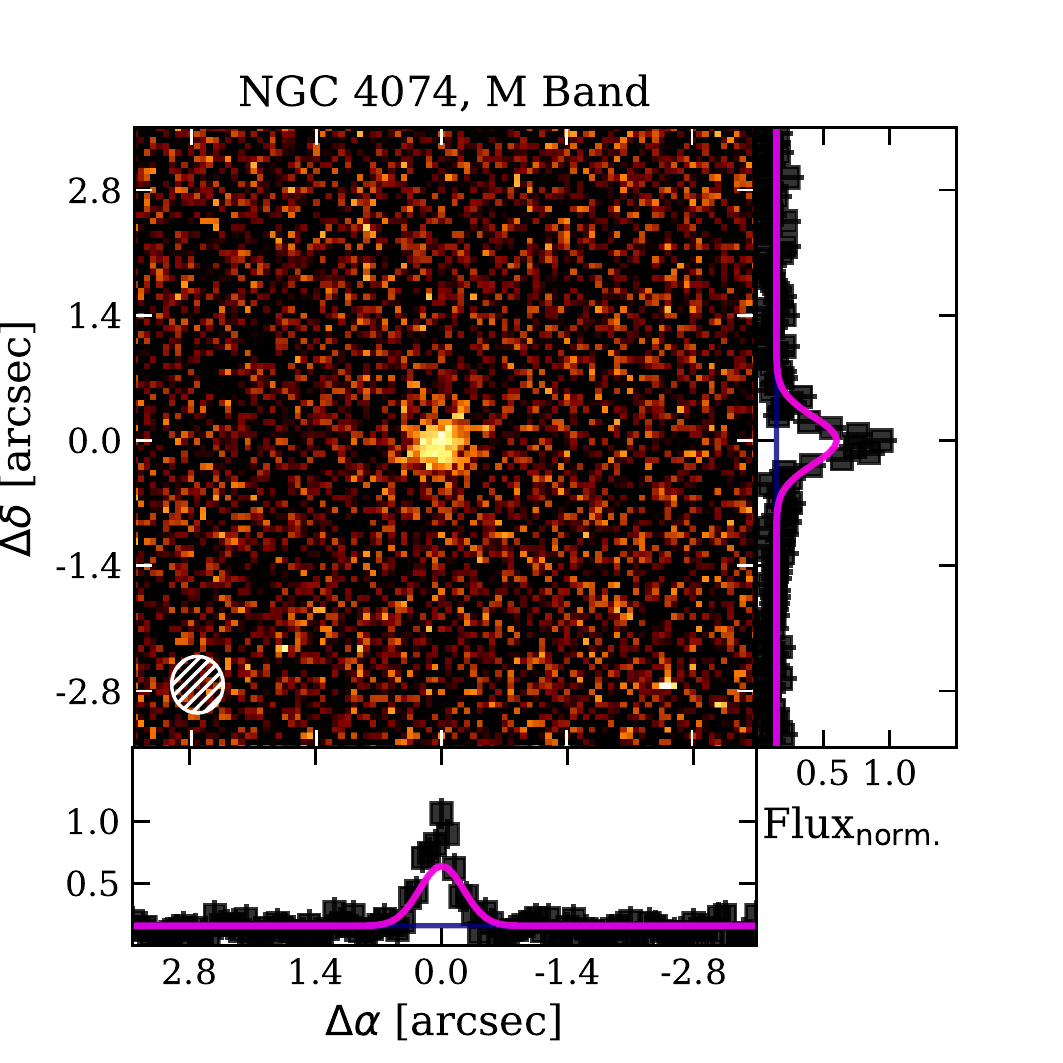}} 
\subfloat{\includegraphics[width=0.25\hsize]{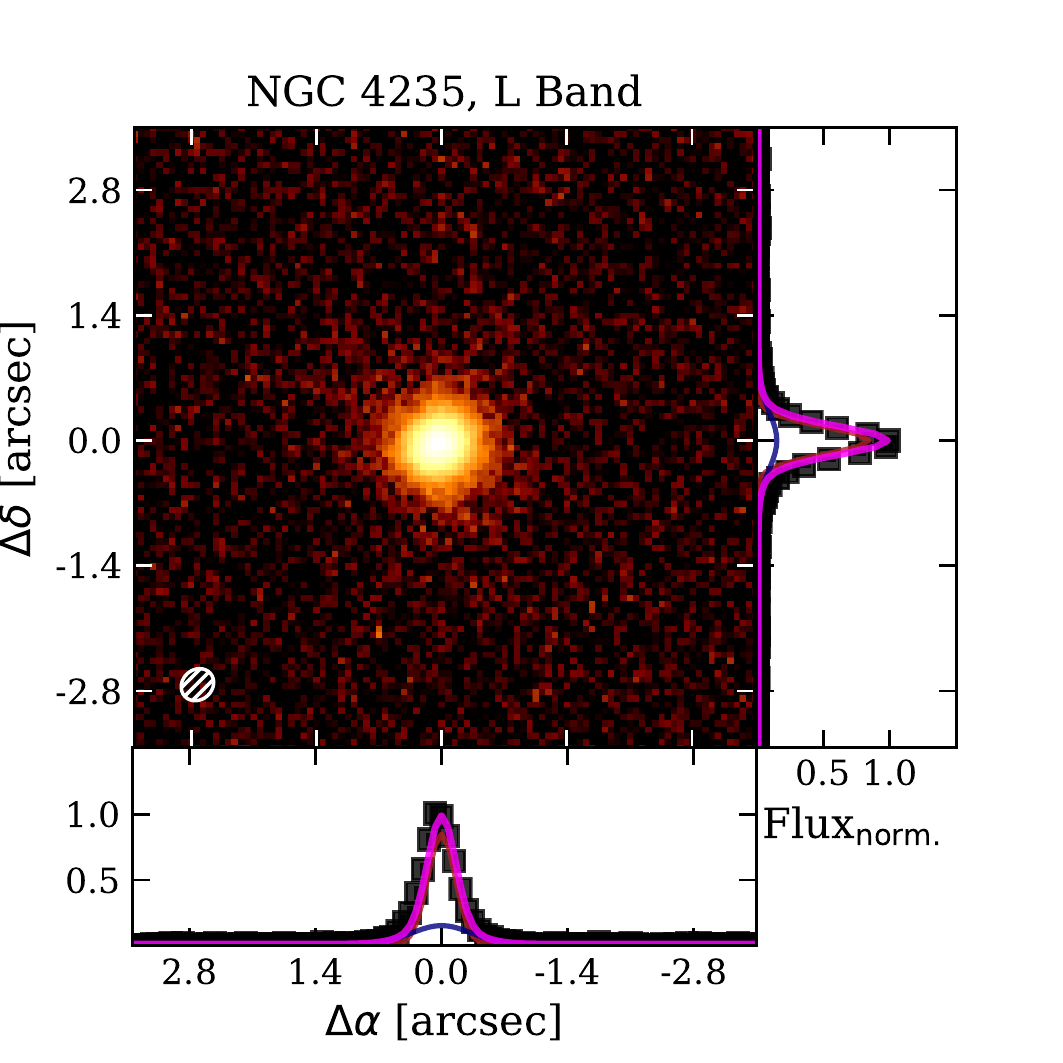}}
\subfloat{\includegraphics[width=0.25\hsize]{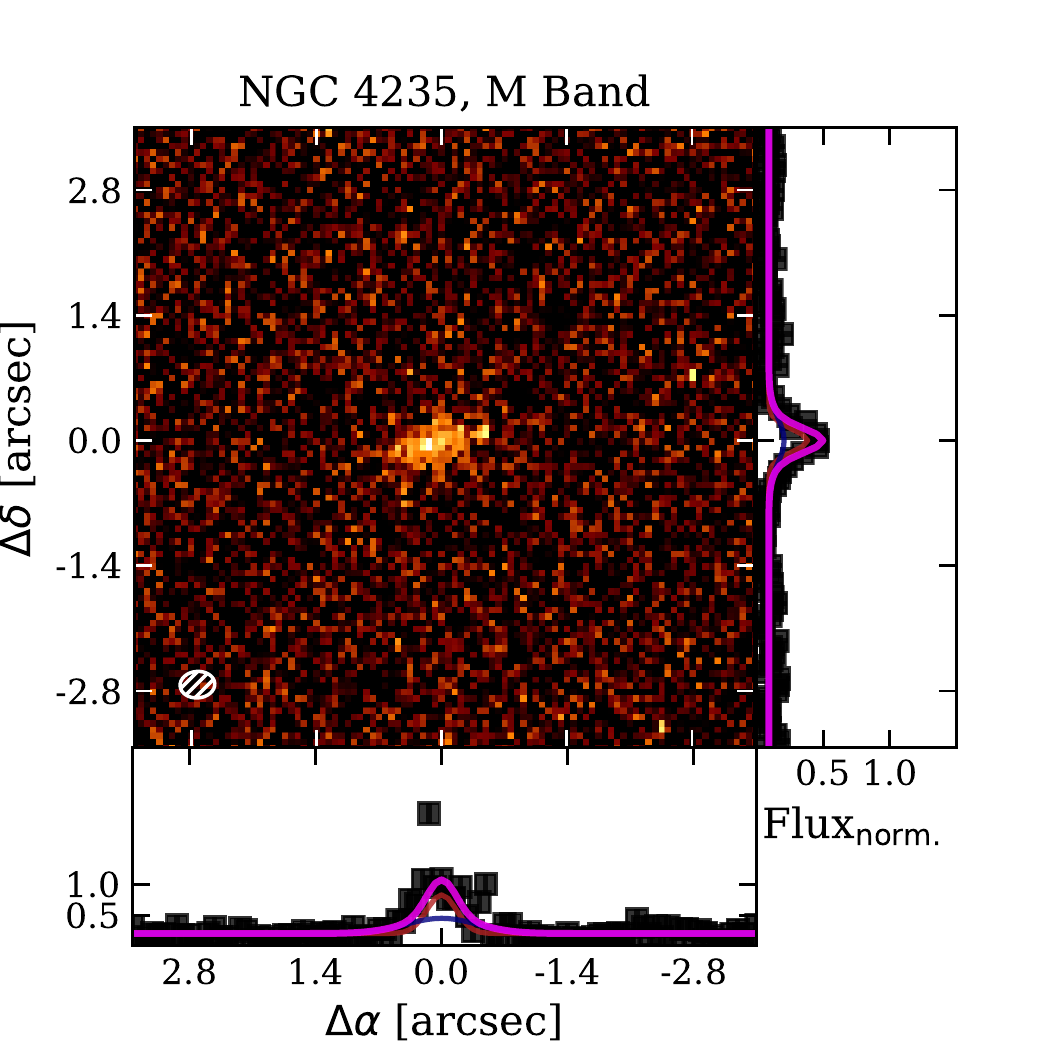}}\\
\subfloat{\includegraphics[width=0.25\hsize]{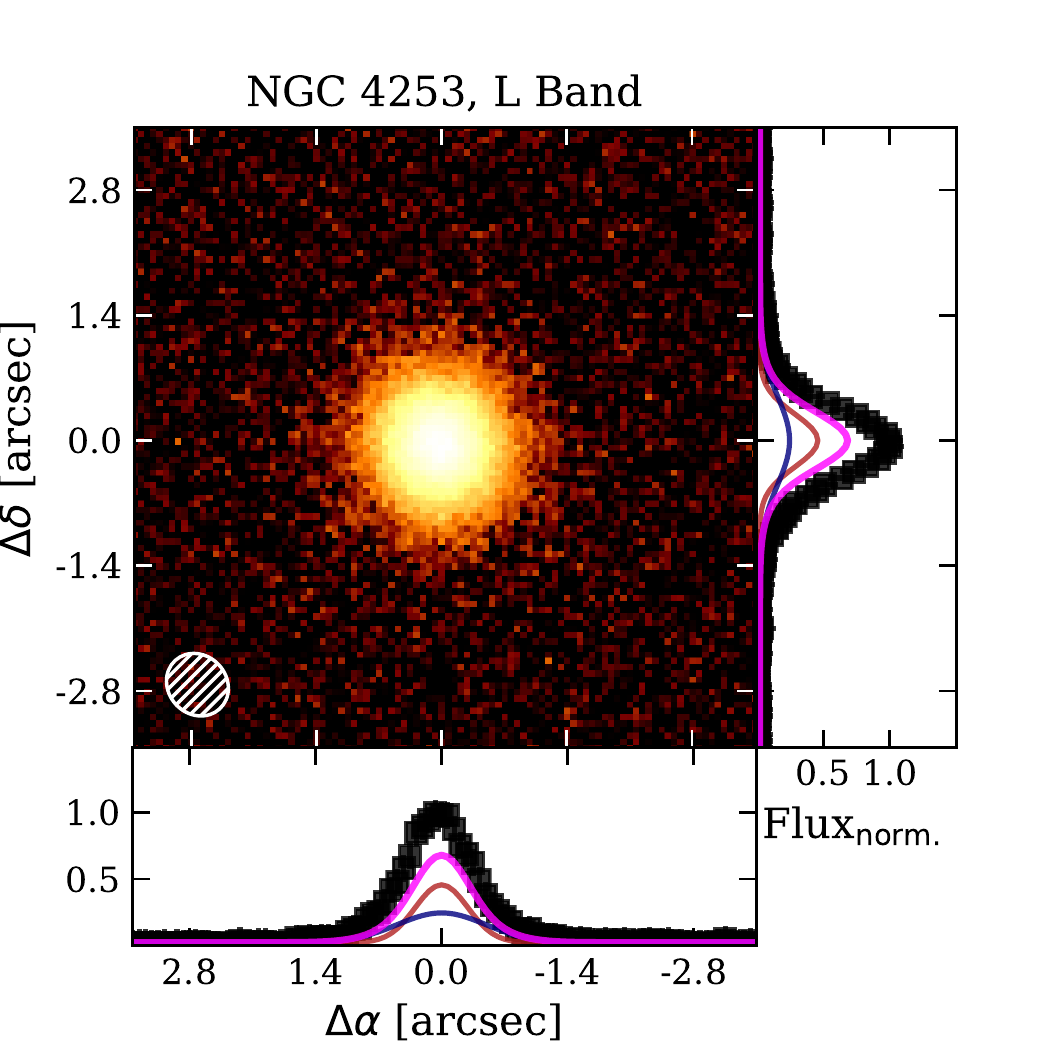}}
\subfloat{\includegraphics[width=0.25\hsize]{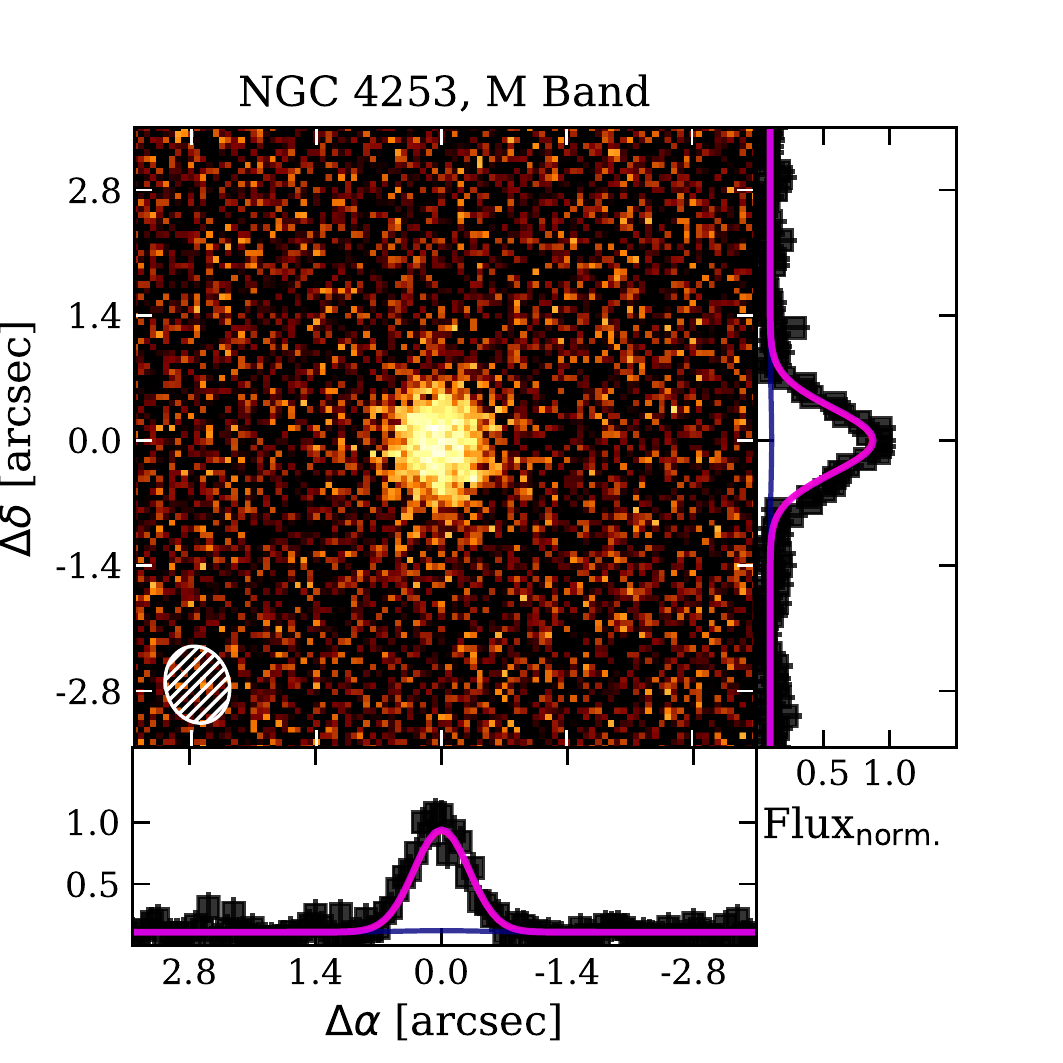}} 
\subfloat{\includegraphics[width=0.25\hsize]{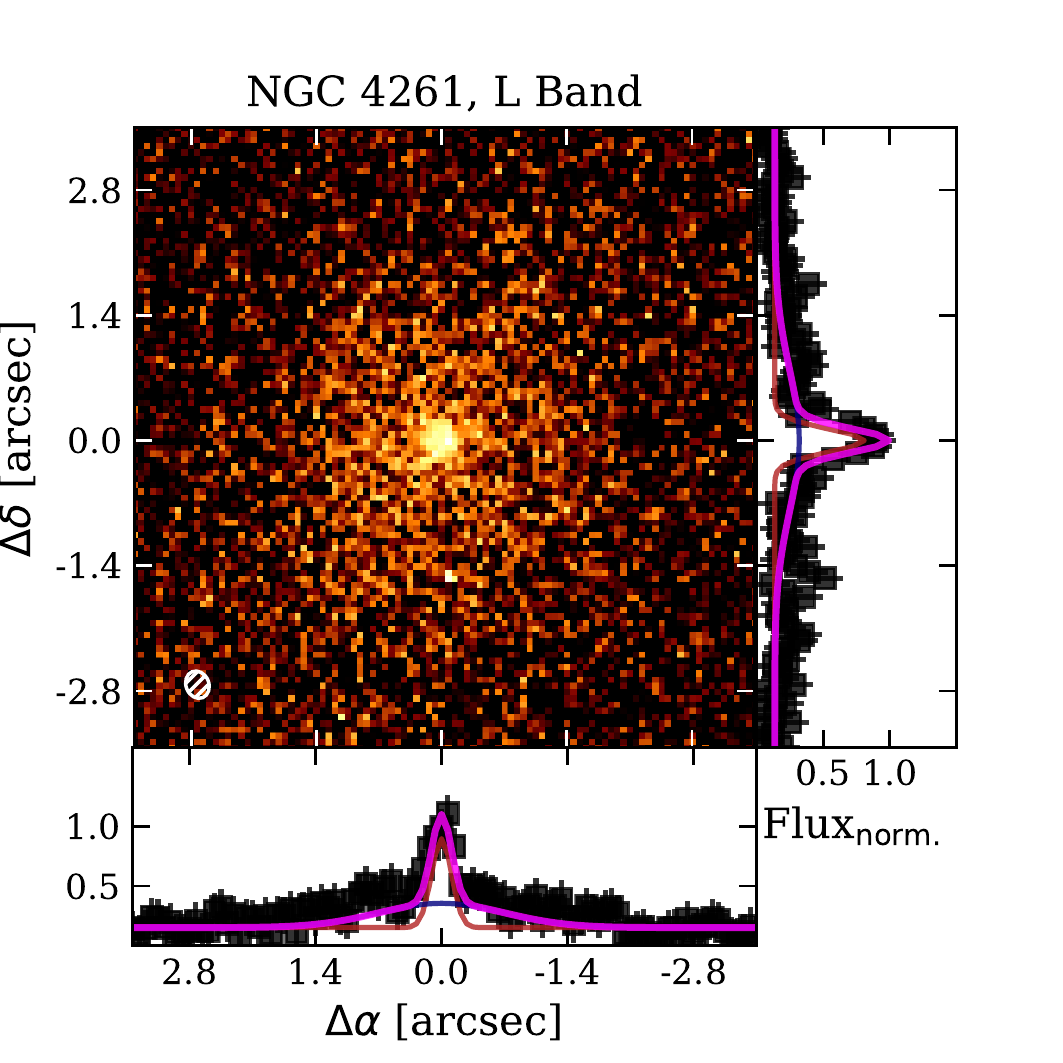}}
\subfloat{\includegraphics[width=0.25\hsize]{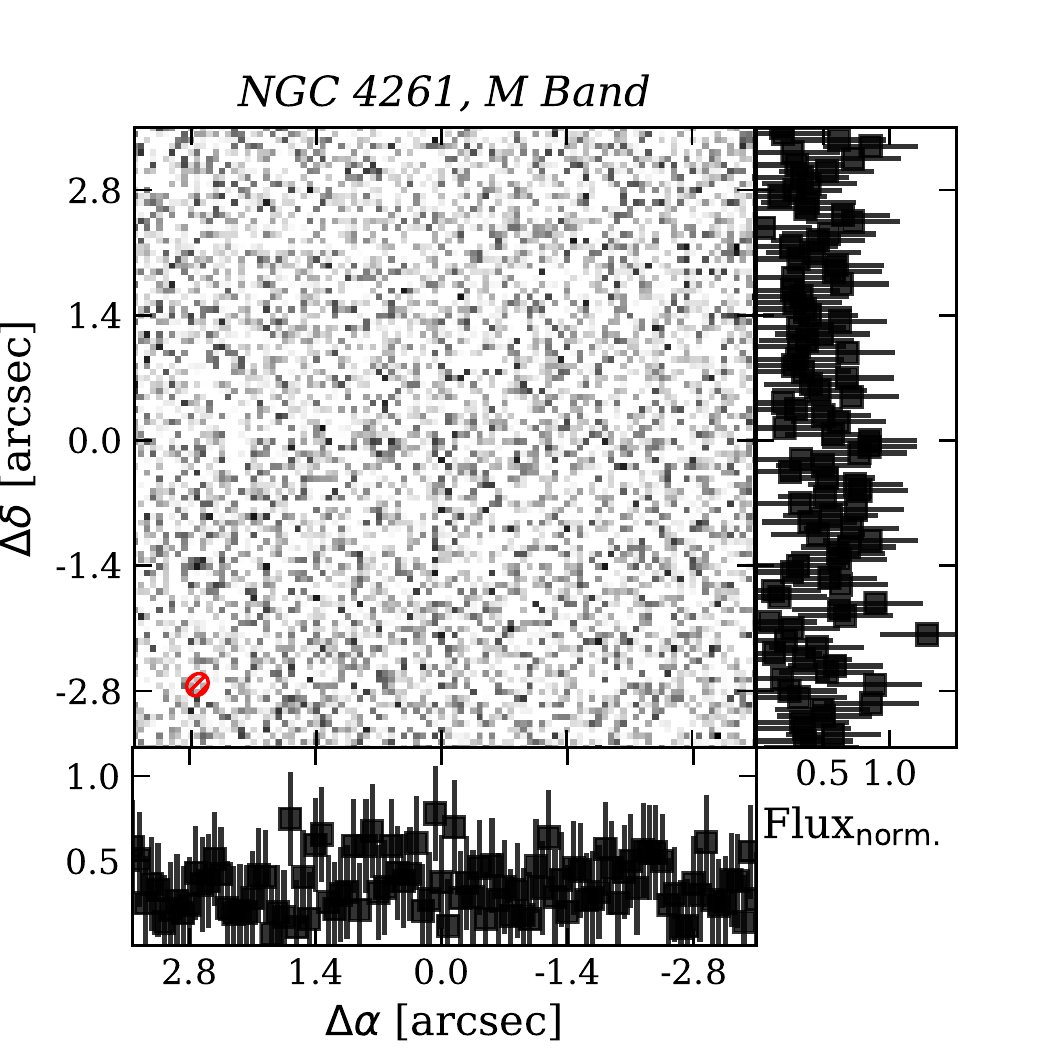}}\\
\subfloat{\includegraphics[width=0.25\hsize]{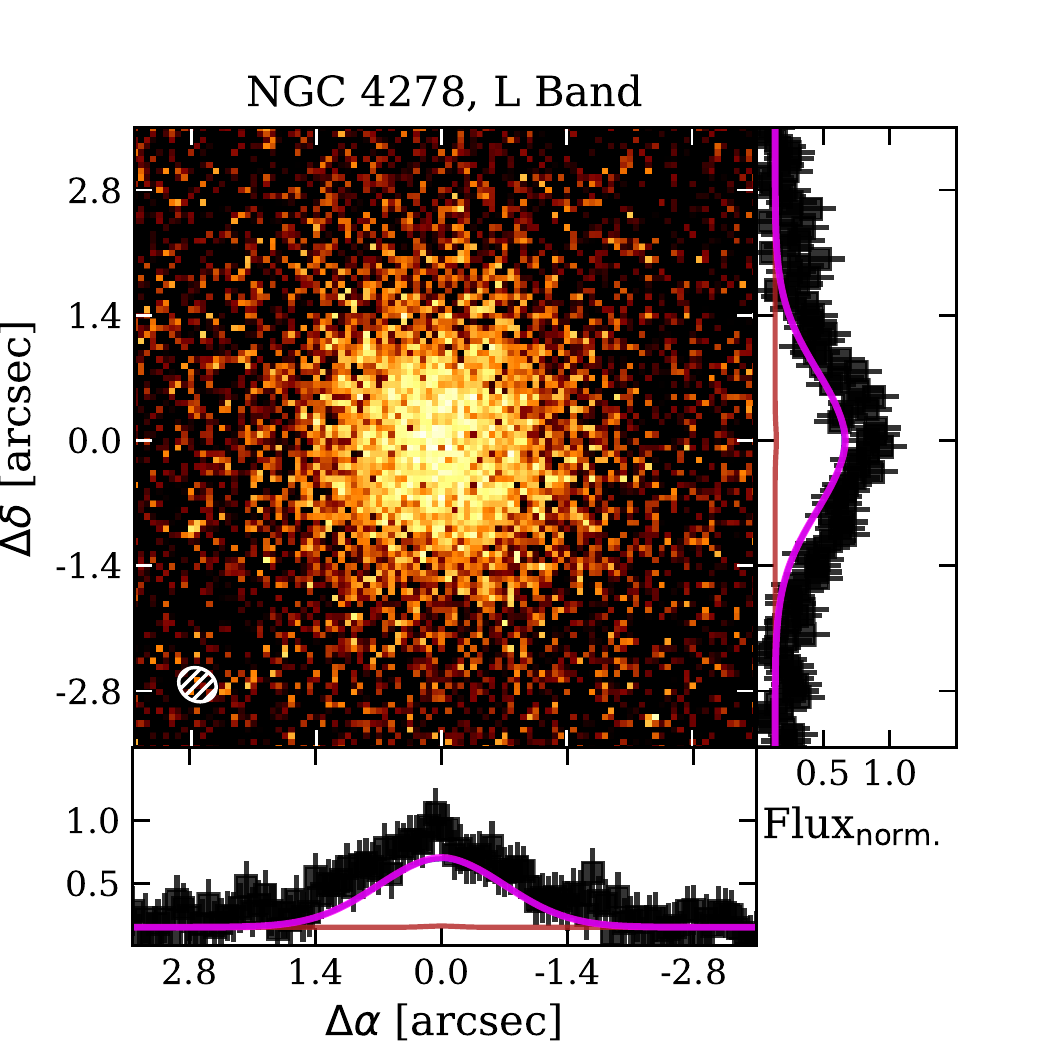}}
\subfloat{\includegraphics[width=0.25\hsize]{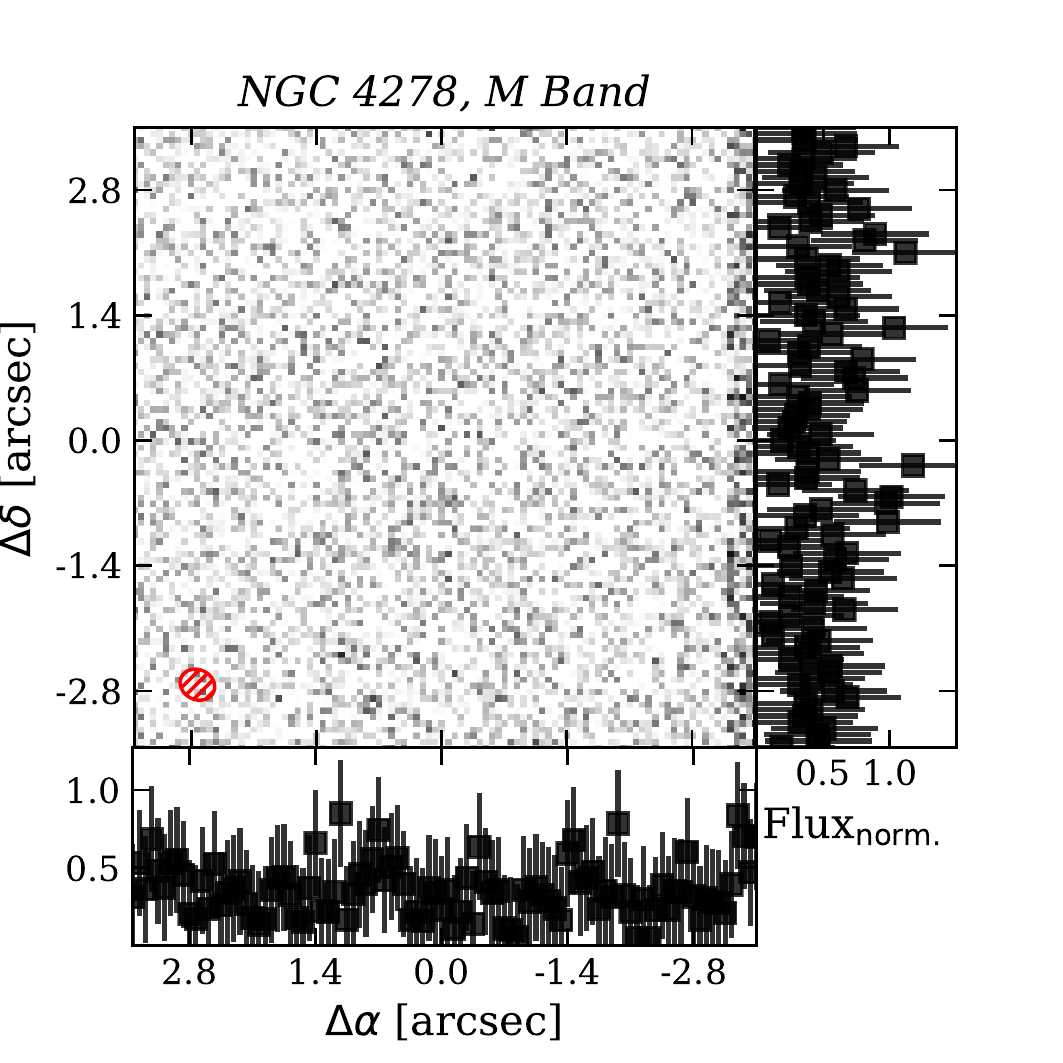}} 
\subfloat{\includegraphics[width=0.25\hsize]{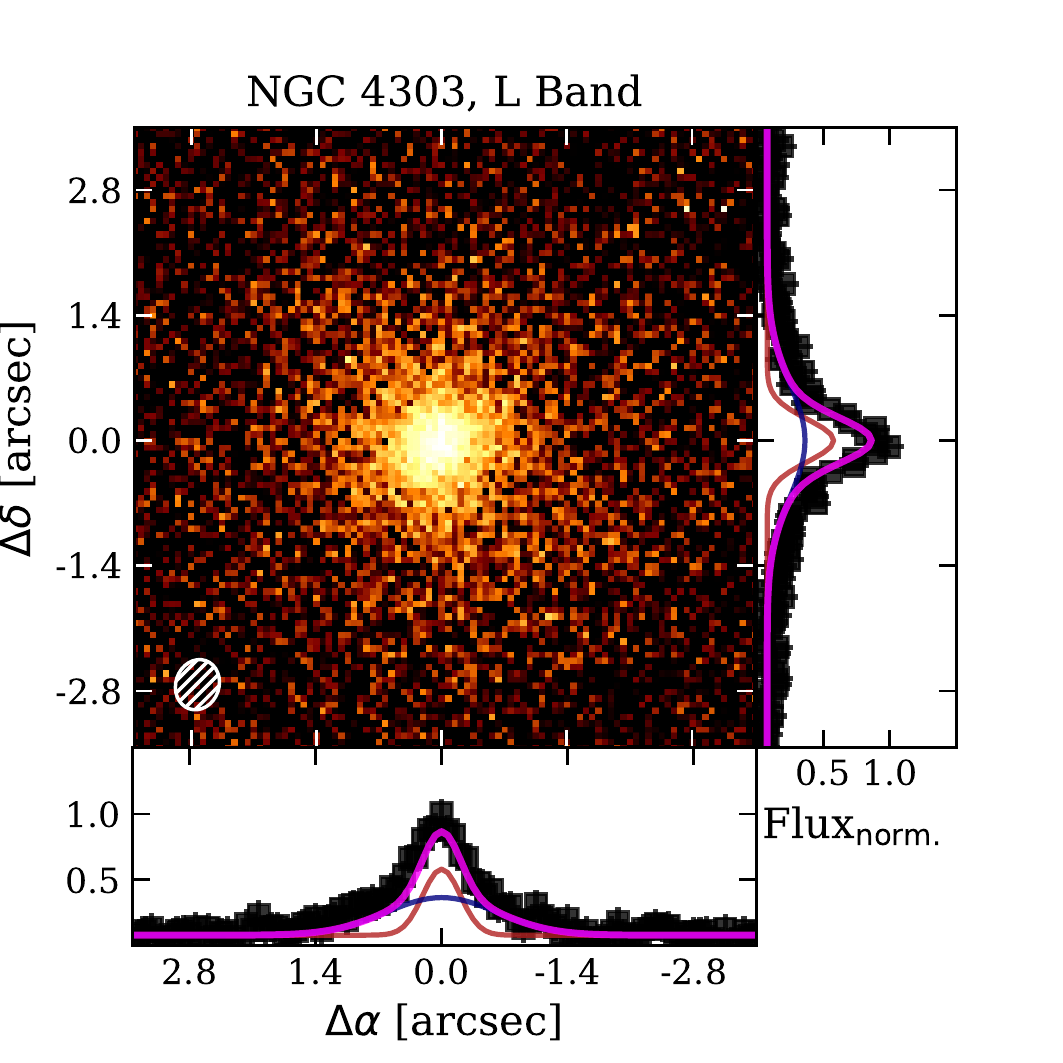}}
\subfloat{\includegraphics[width=0.25\hsize]{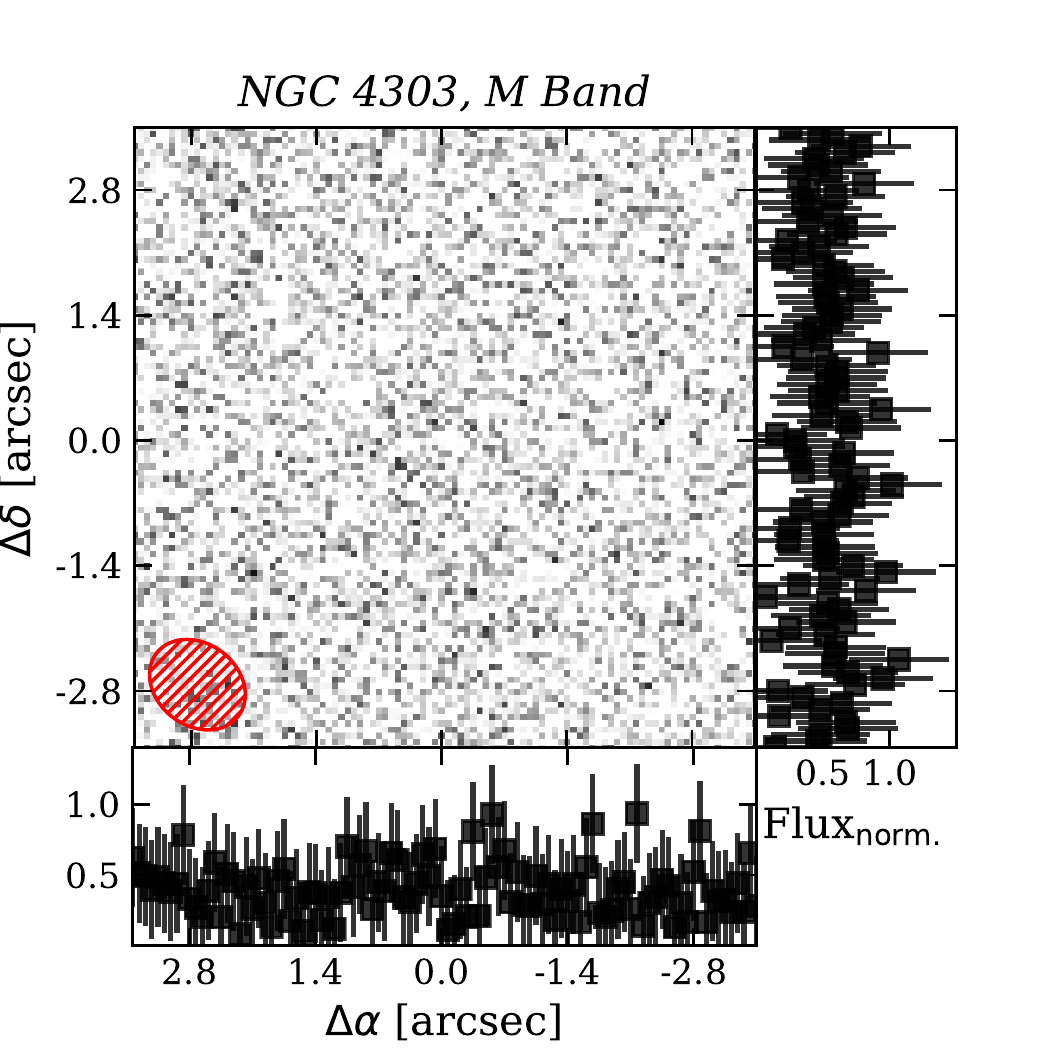}}\\
\caption{ As Fig \ref{fig:cutouts_one} but for all sources.}
\end{figure*}
\begin{figure*}
\subfloat{\includegraphics[width=0.25\hsize]{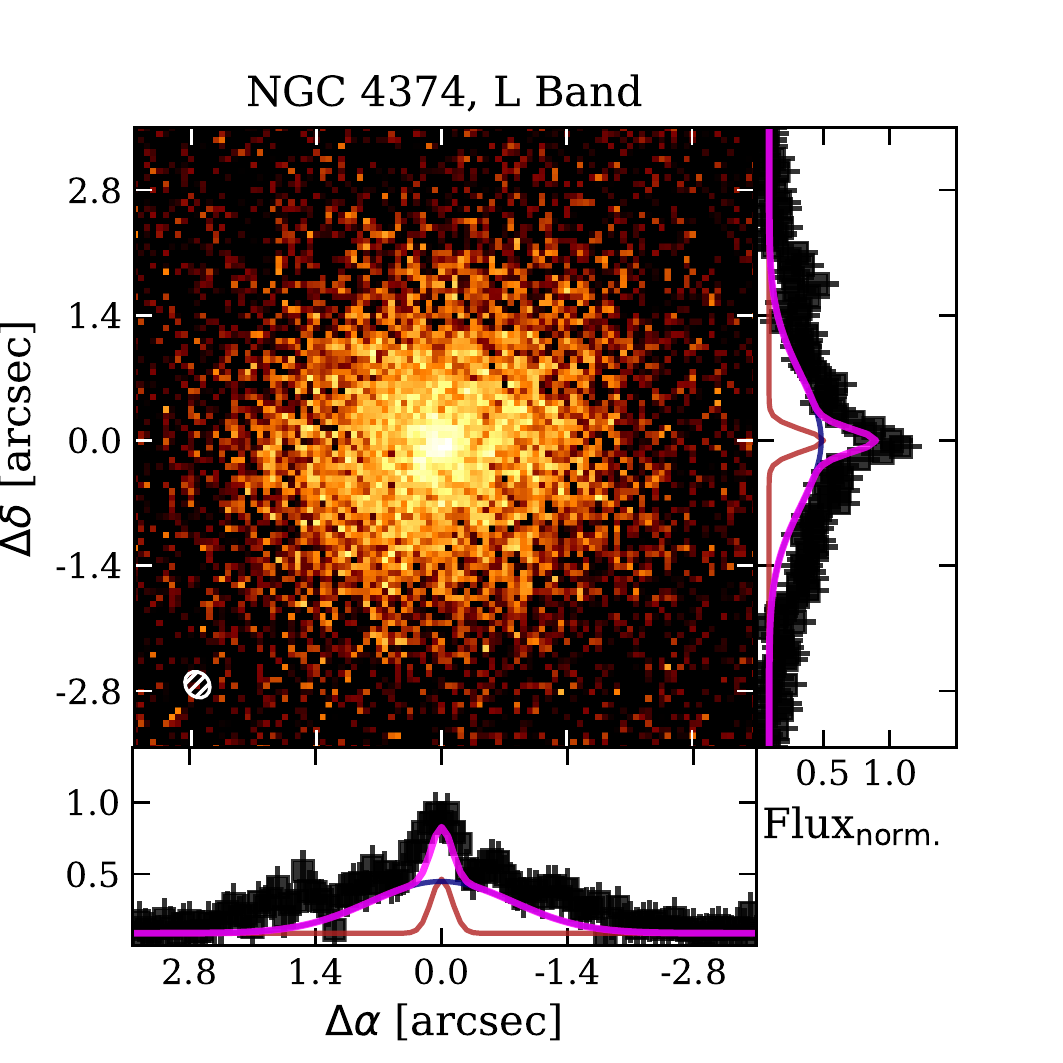}}
\subfloat{\includegraphics[width=0.25\hsize]{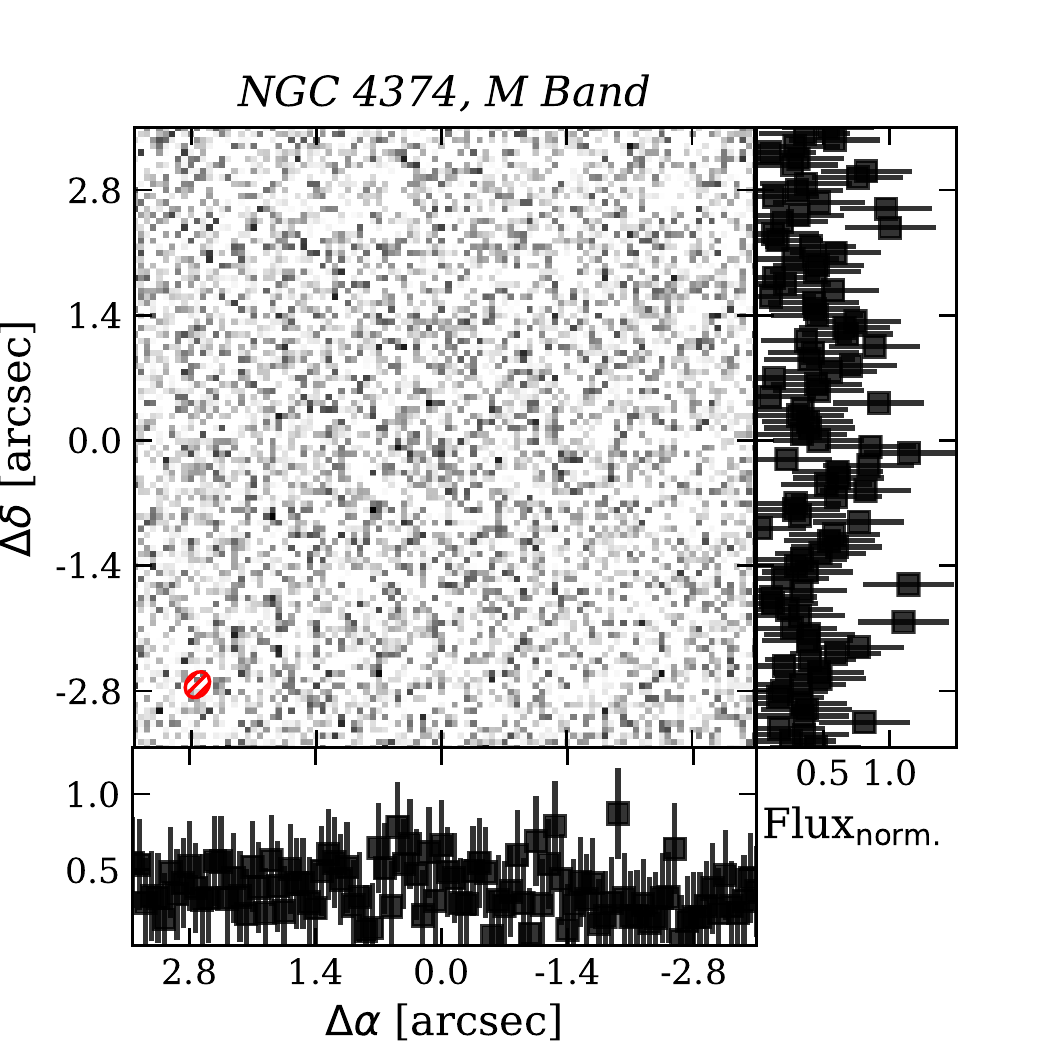}} 
\subfloat{\includegraphics[width=0.25\hsize]{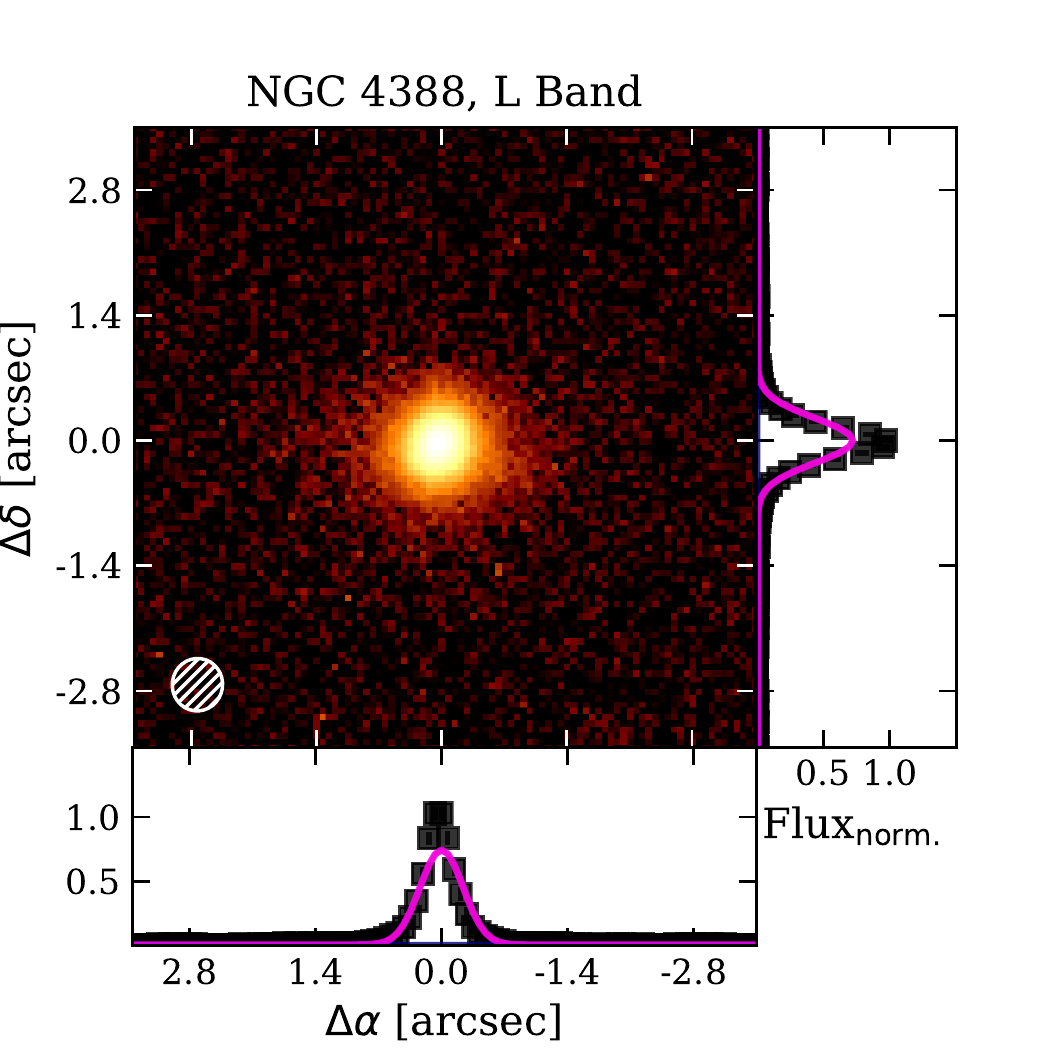}}
\subfloat{\includegraphics[width=0.25\hsize]{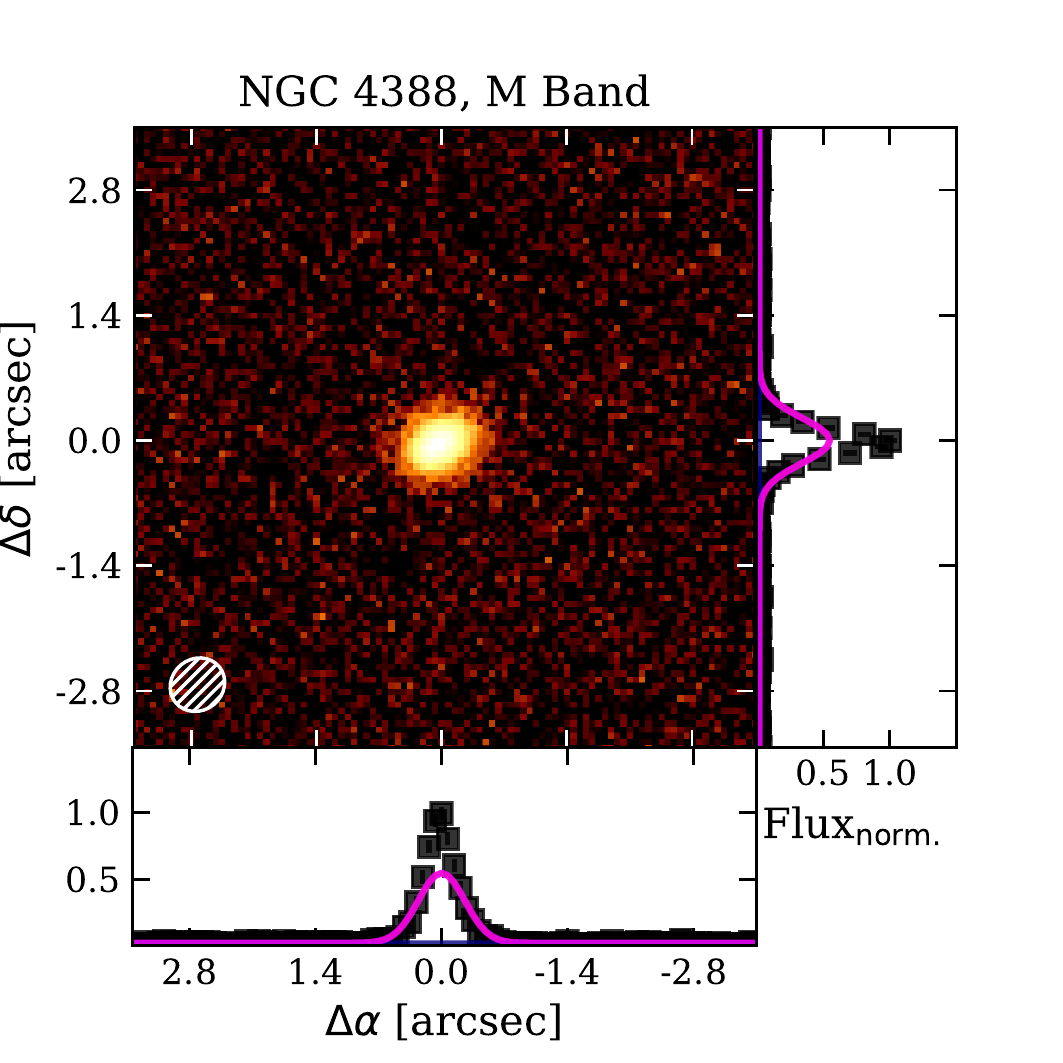}}\\
\subfloat{\includegraphics[width=0.25\hsize]{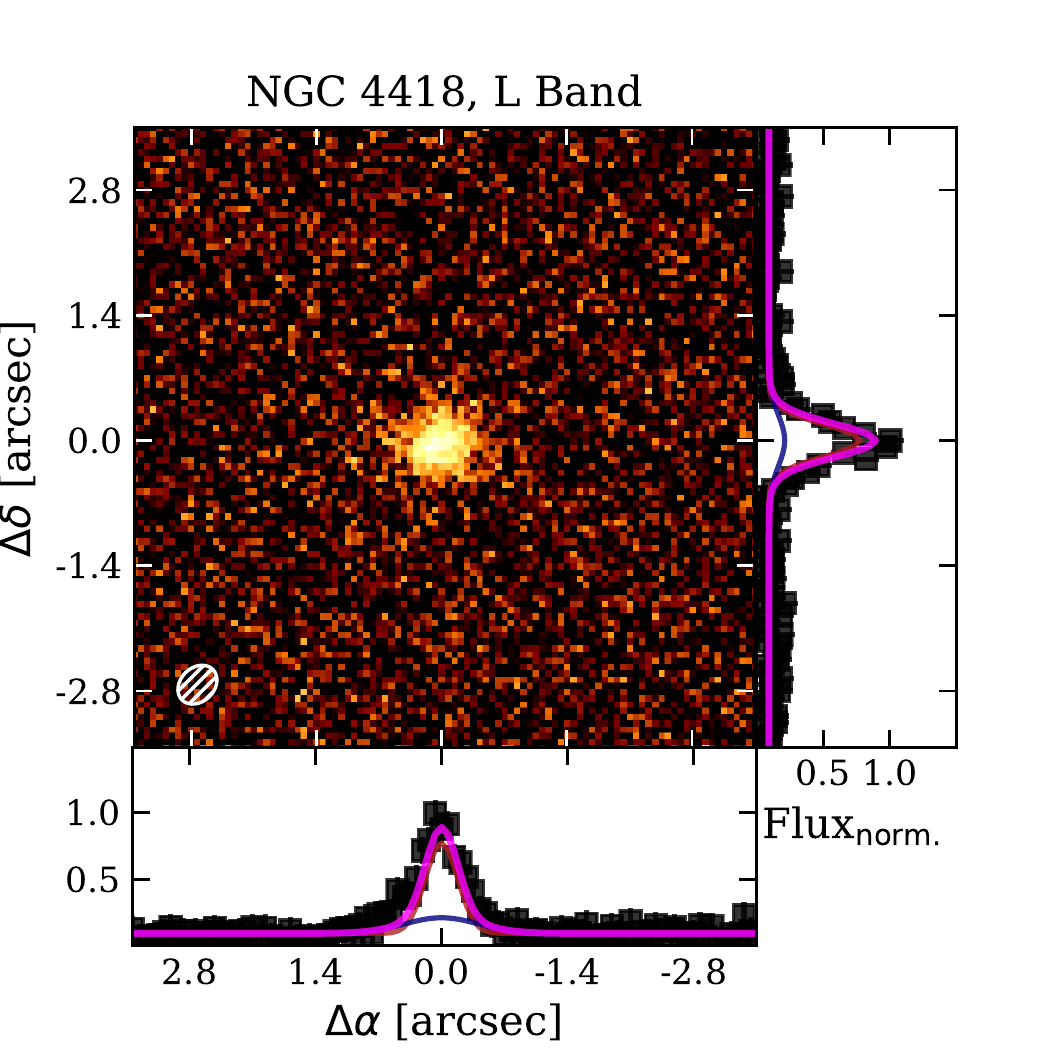}}
\subfloat{\includegraphics[width=0.25\hsize]{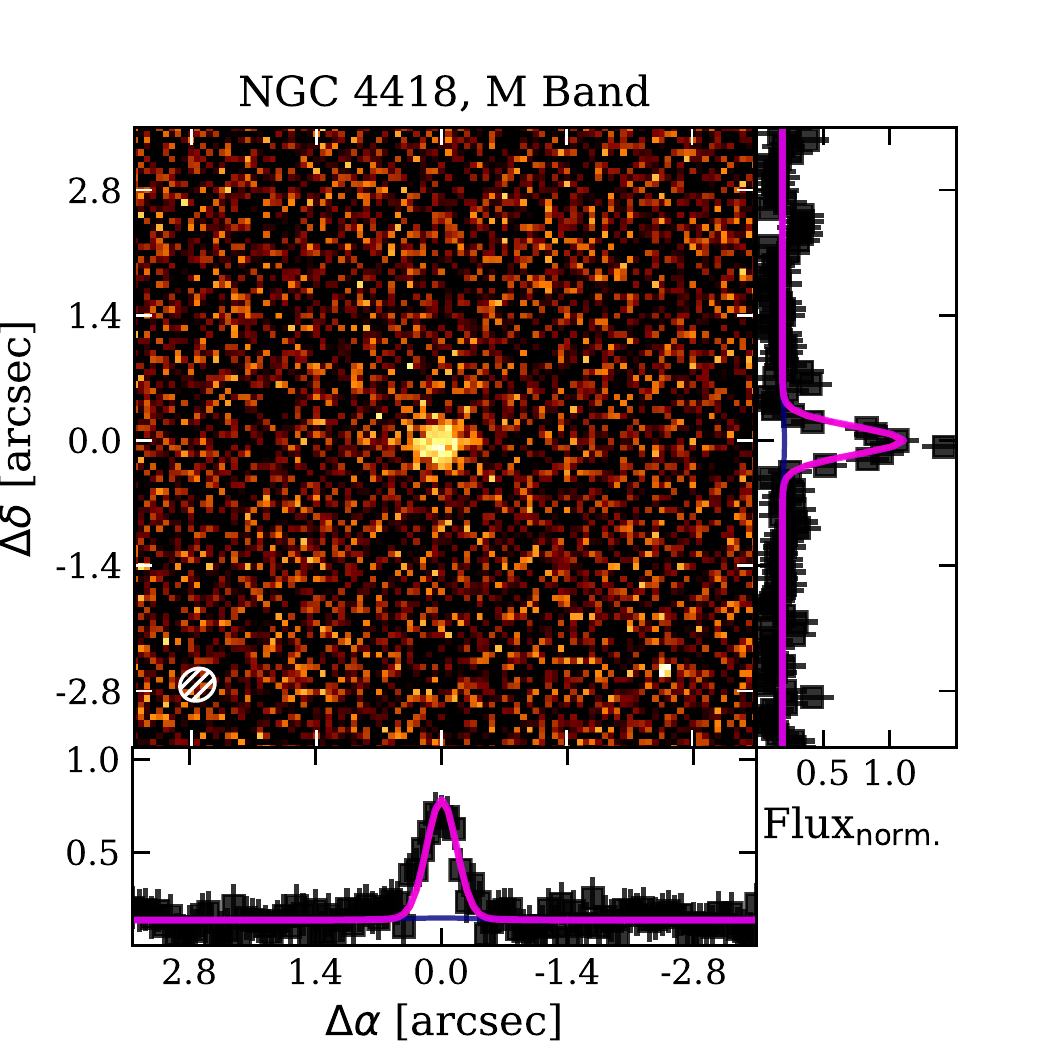}} 
\subfloat{\includegraphics[width=0.25\hsize]{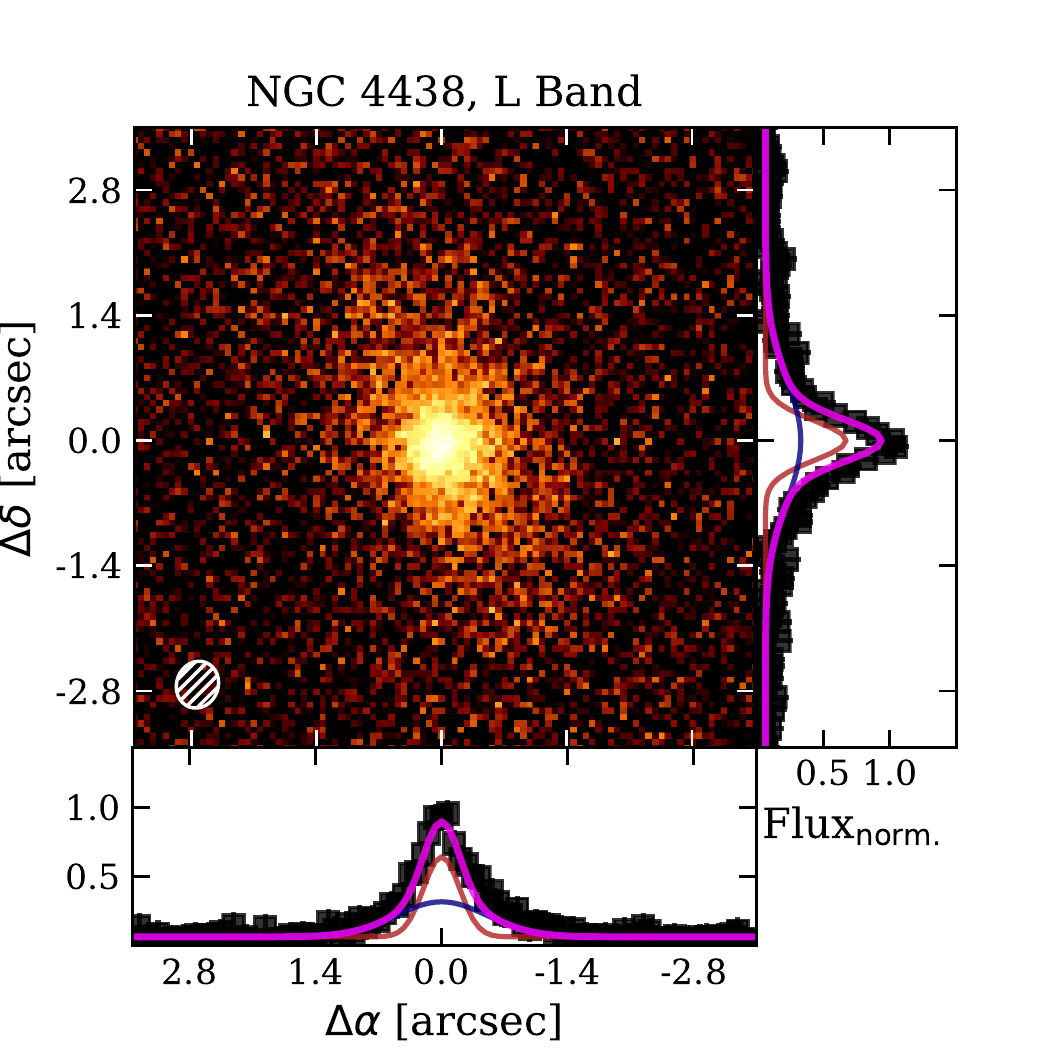}}
\subfloat{\includegraphics[width=0.25\hsize]{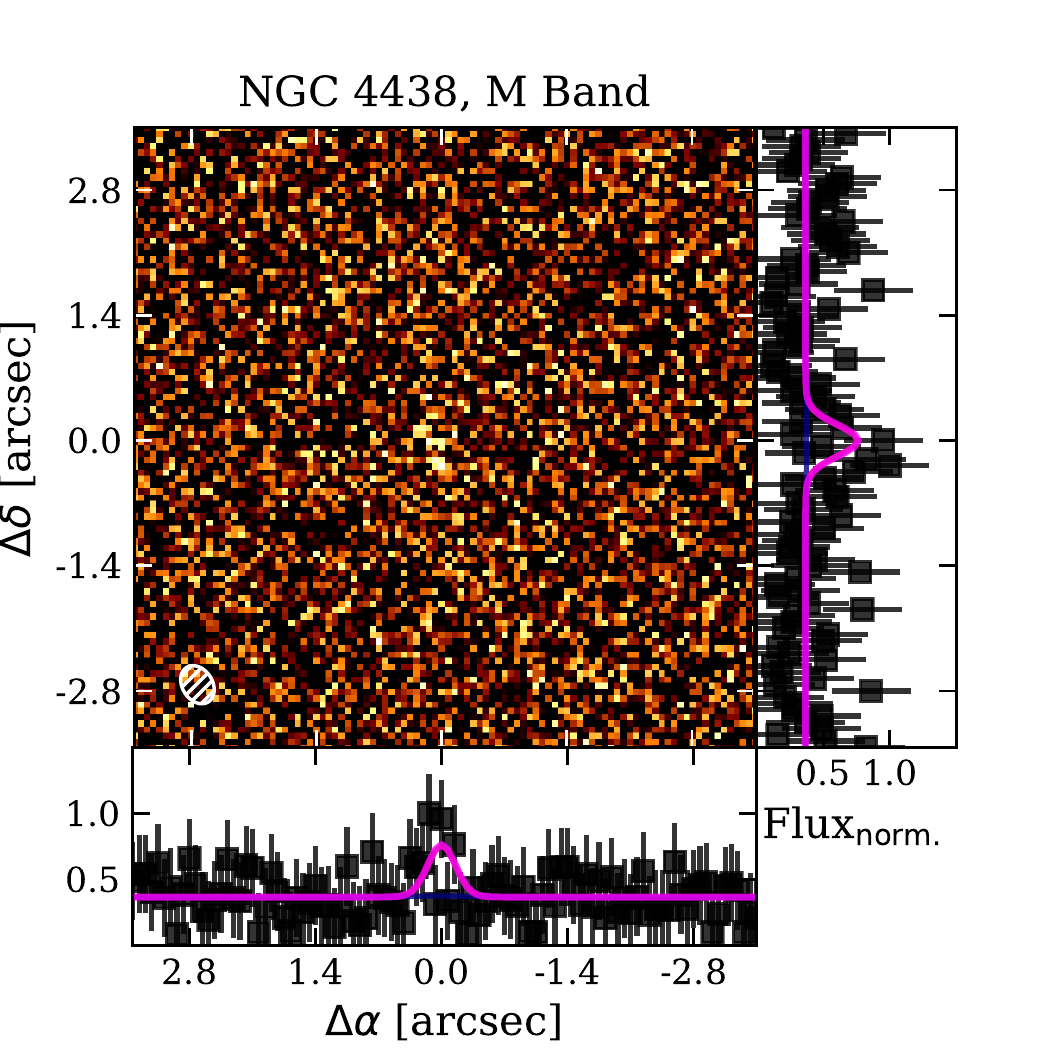}} \\
\subfloat{\includegraphics[width=0.25\hsize]{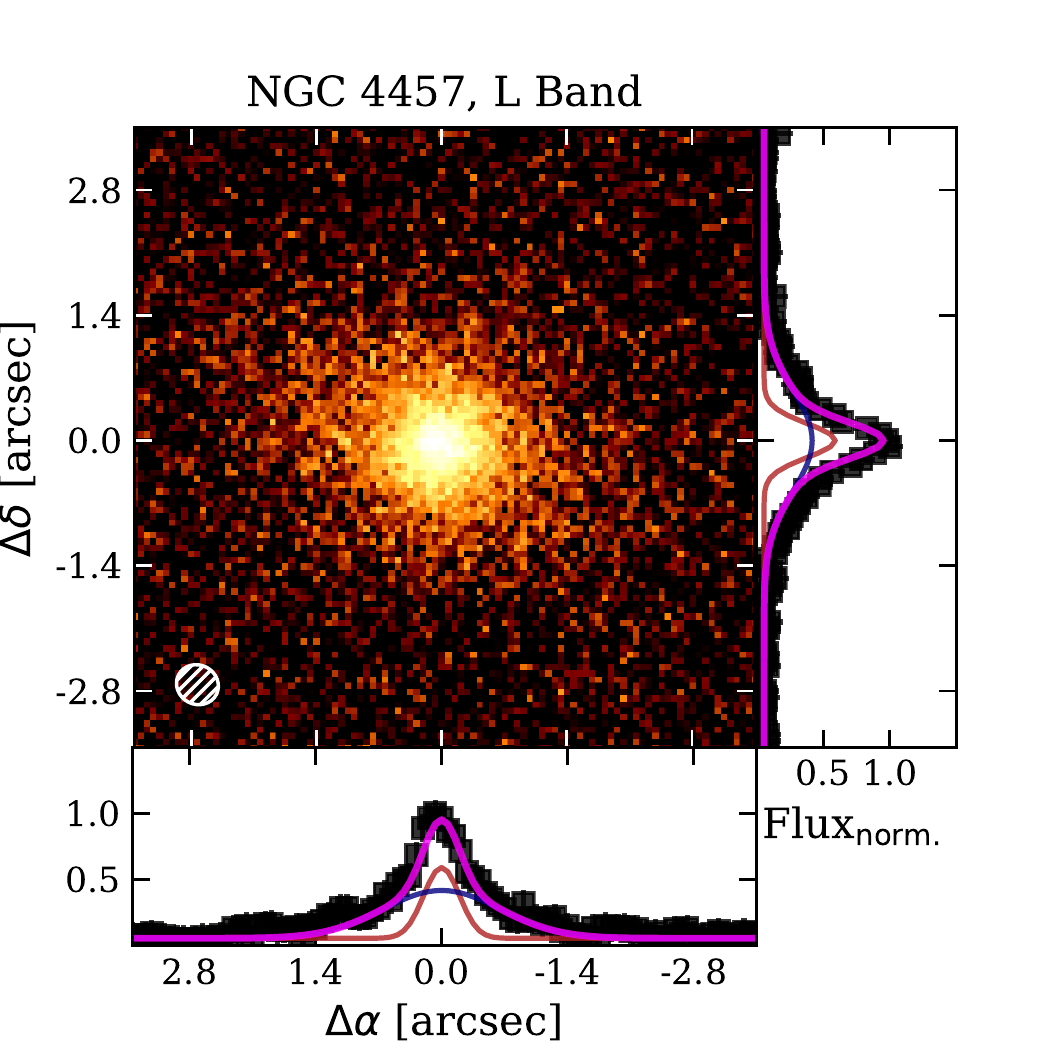}}
\subfloat{\includegraphics[width=0.25\hsize]{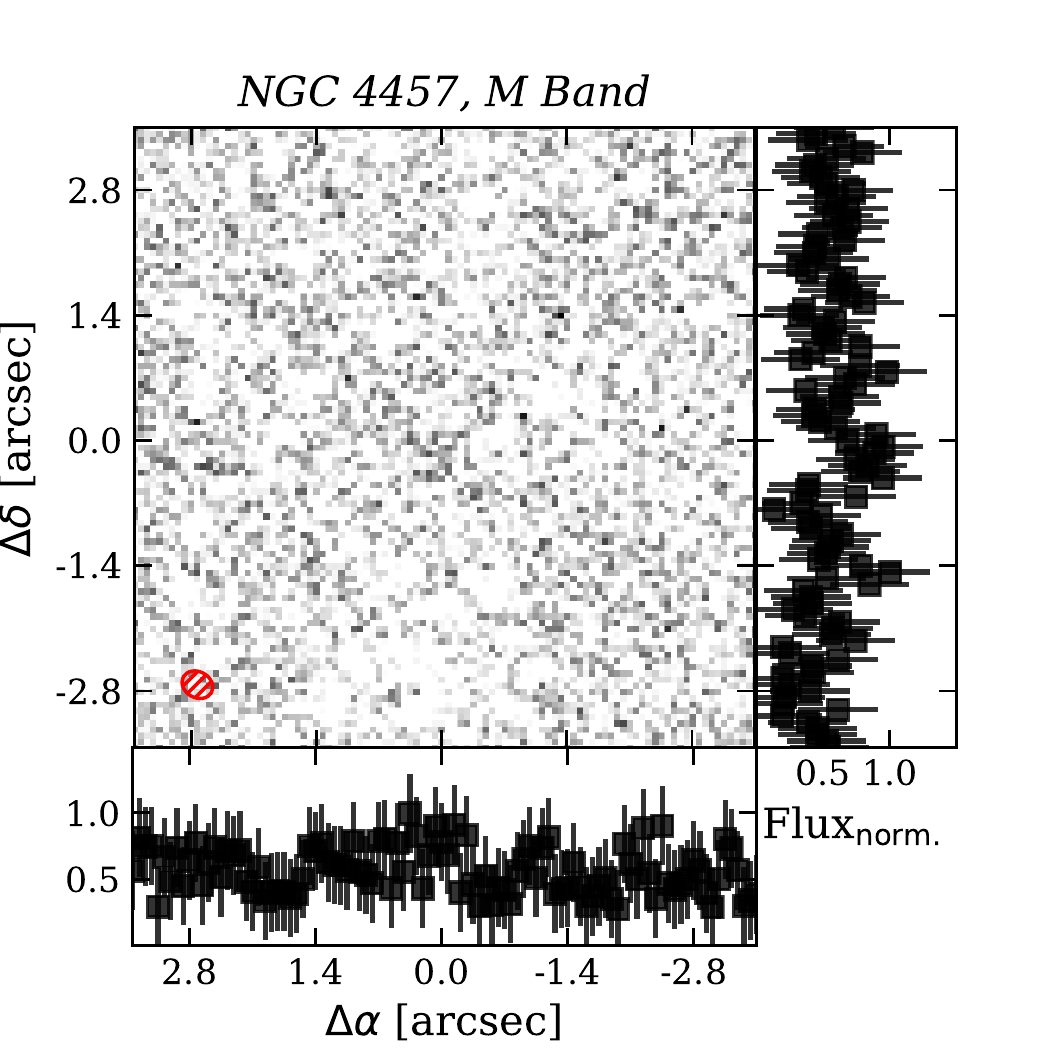}} 
\subfloat{\includegraphics[width=0.25\hsize]{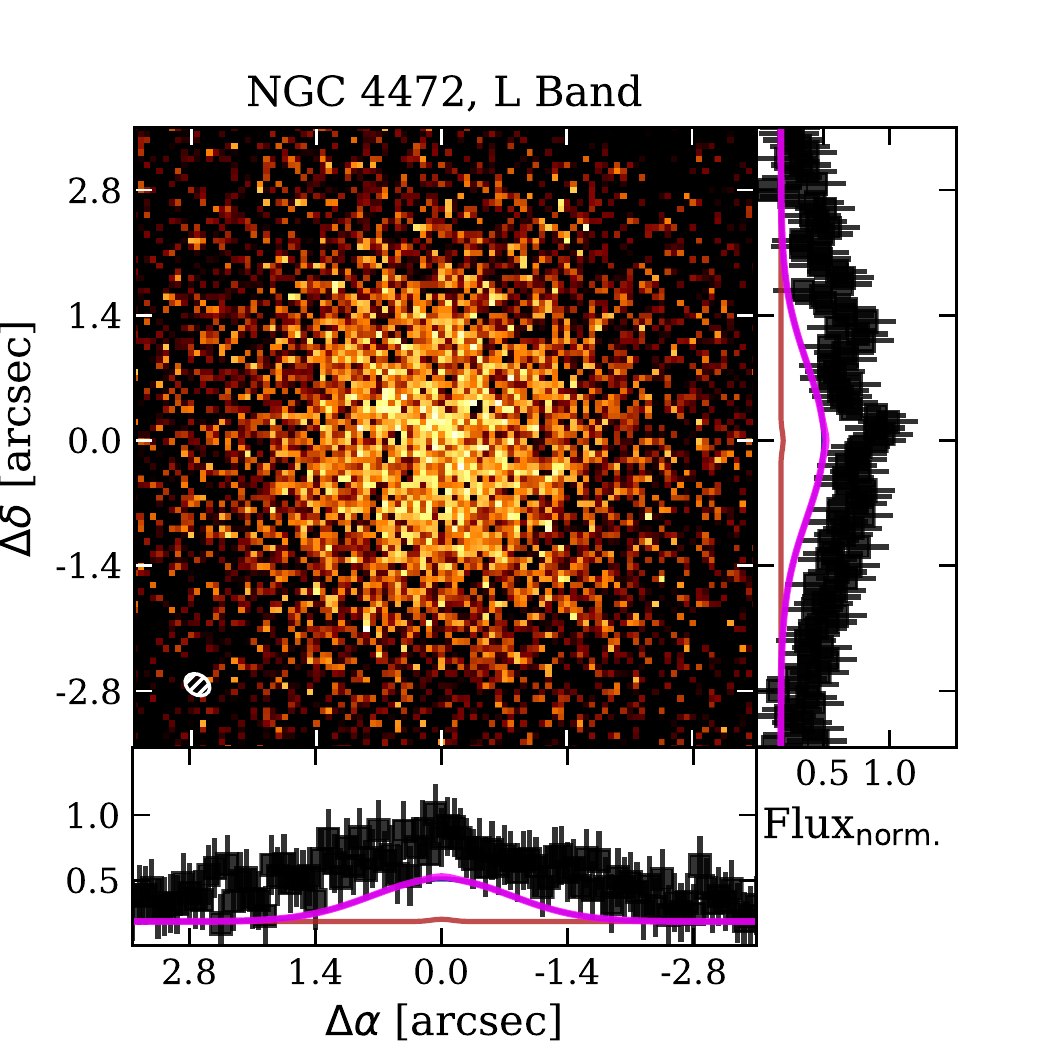}}
\subfloat{\includegraphics[width=0.25\hsize]{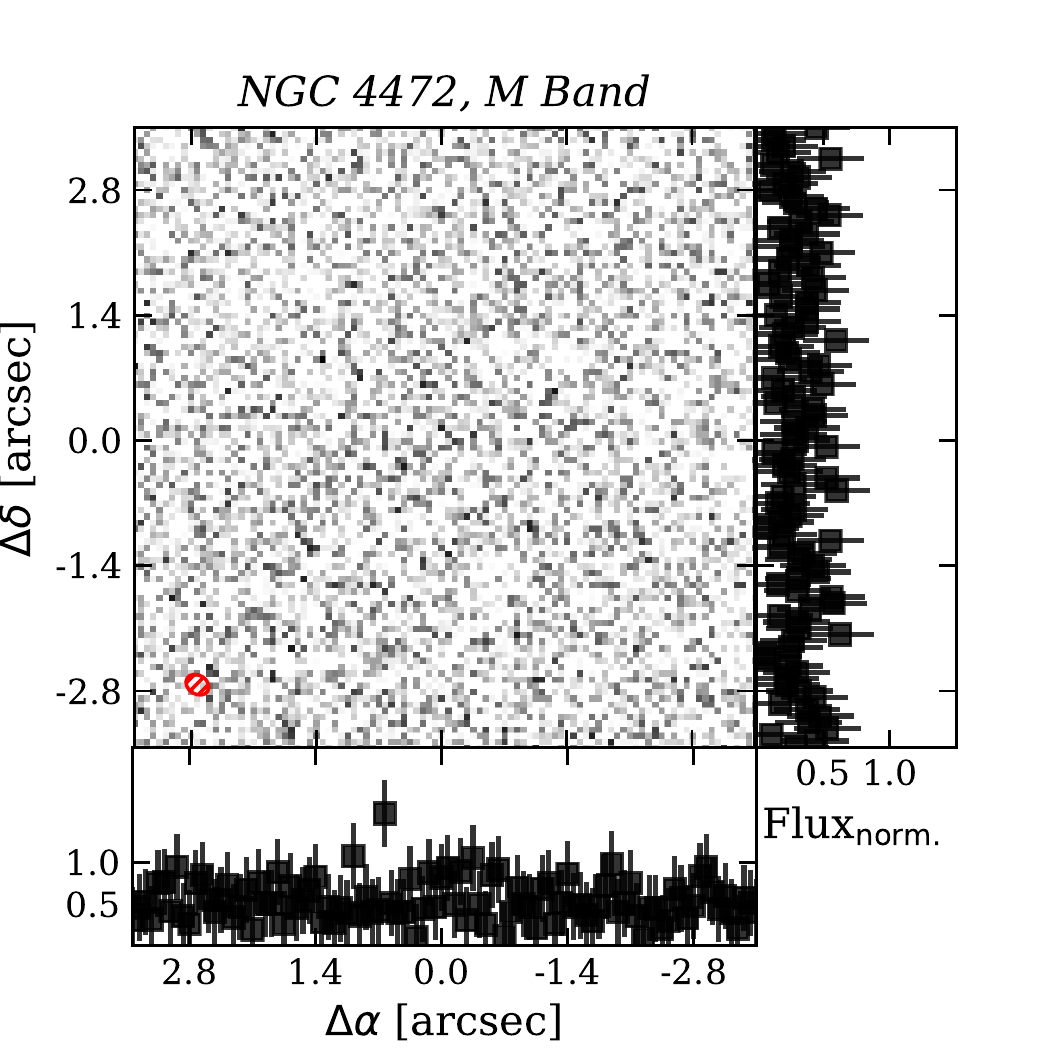}} \\
\subfloat{\includegraphics[width=0.25\hsize]{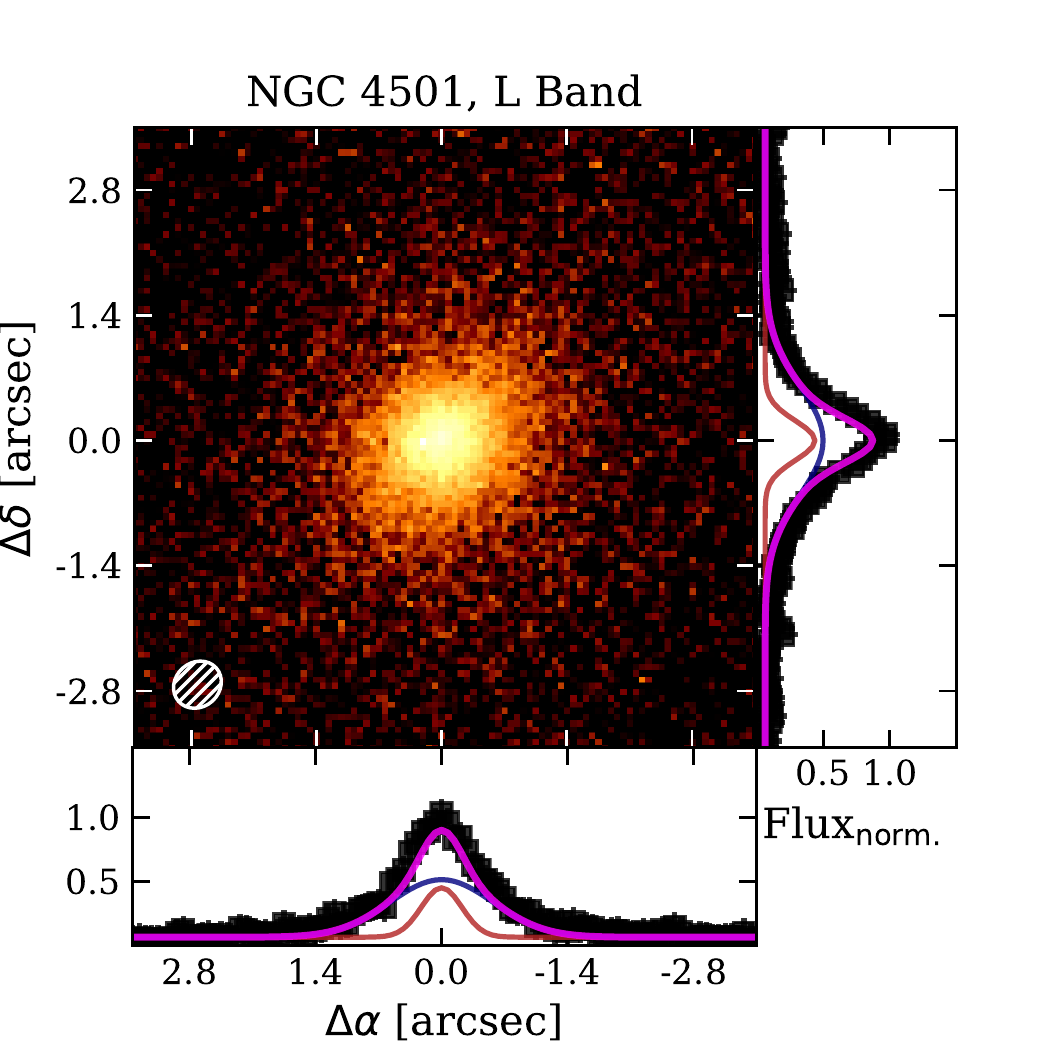}}
\subfloat{\includegraphics[width=0.25\hsize]{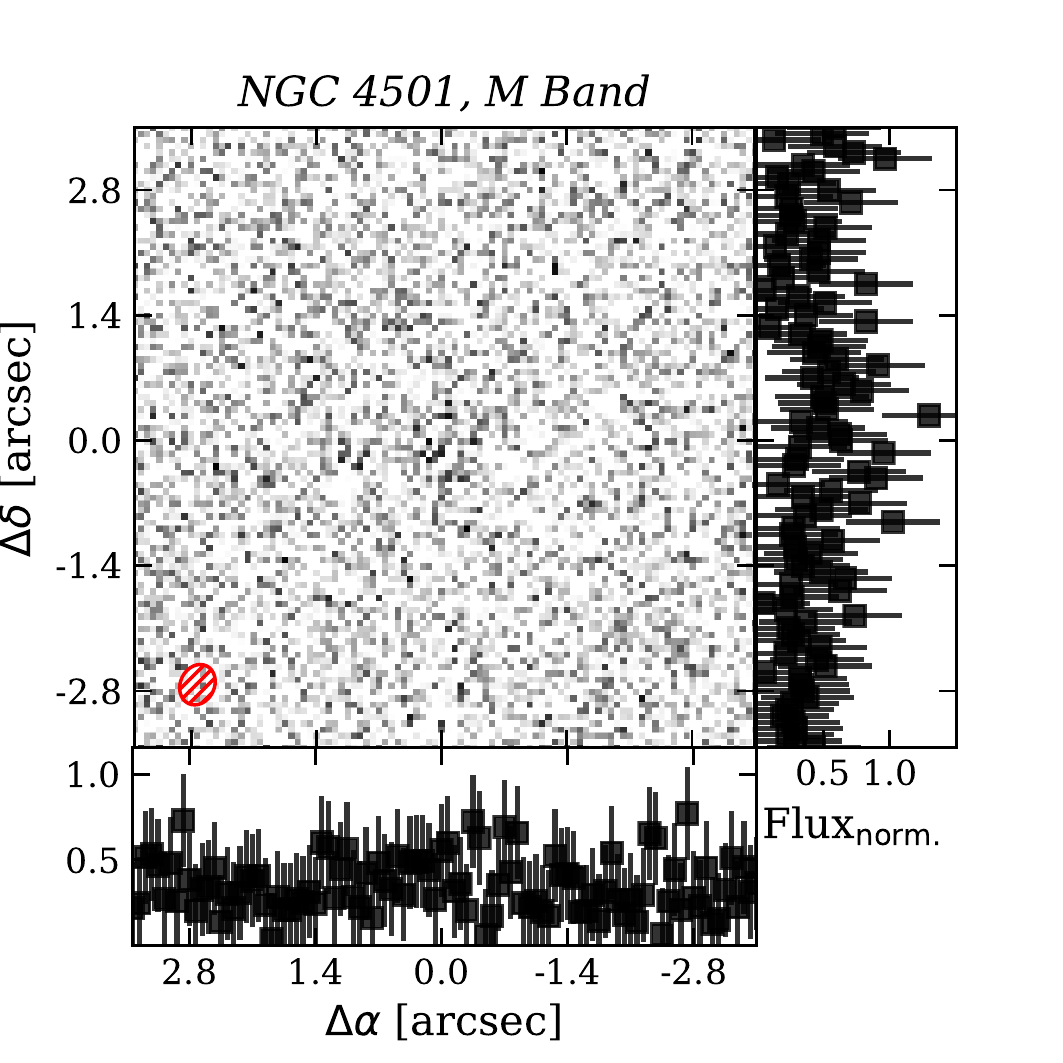}} 
\subfloat{\includegraphics[width=0.25\hsize]{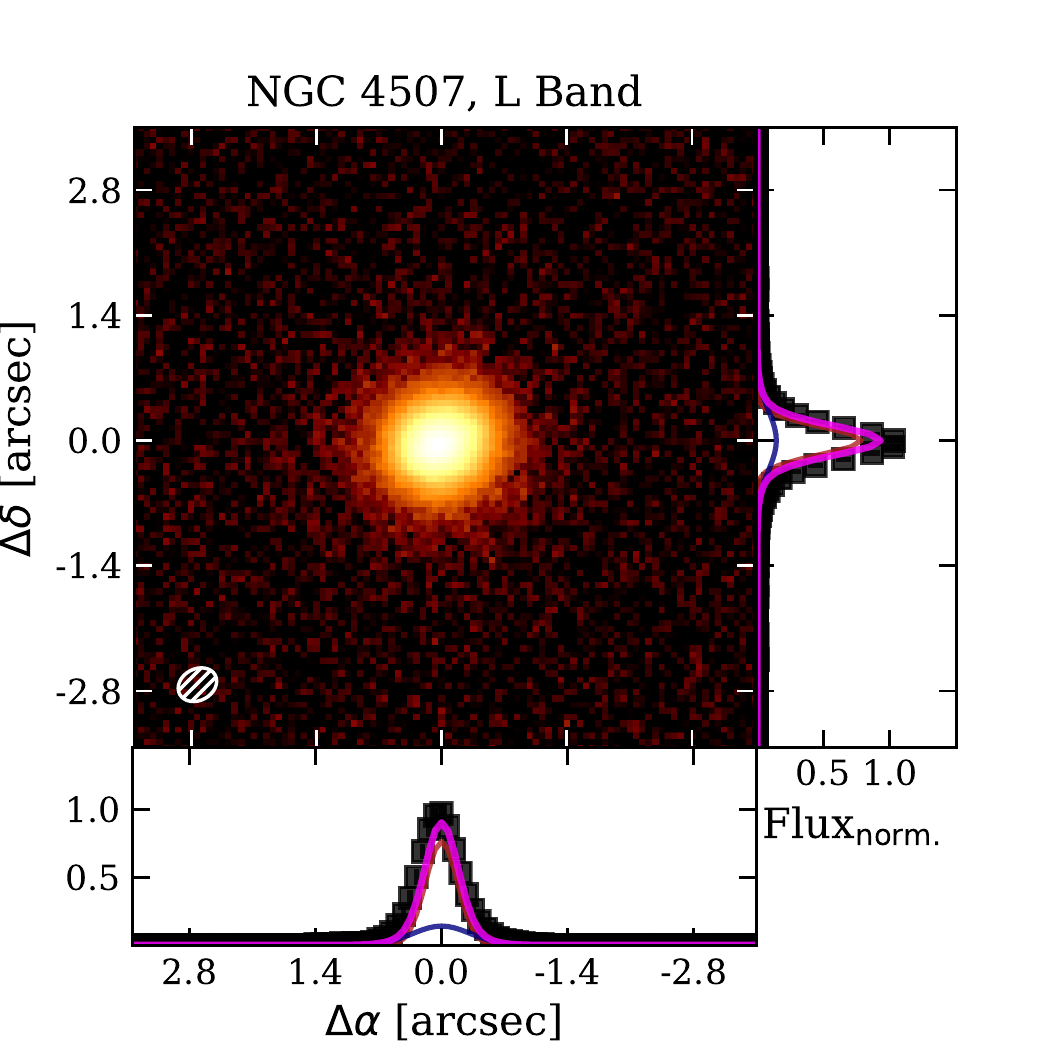}}
\subfloat{\includegraphics[width=0.25\hsize]{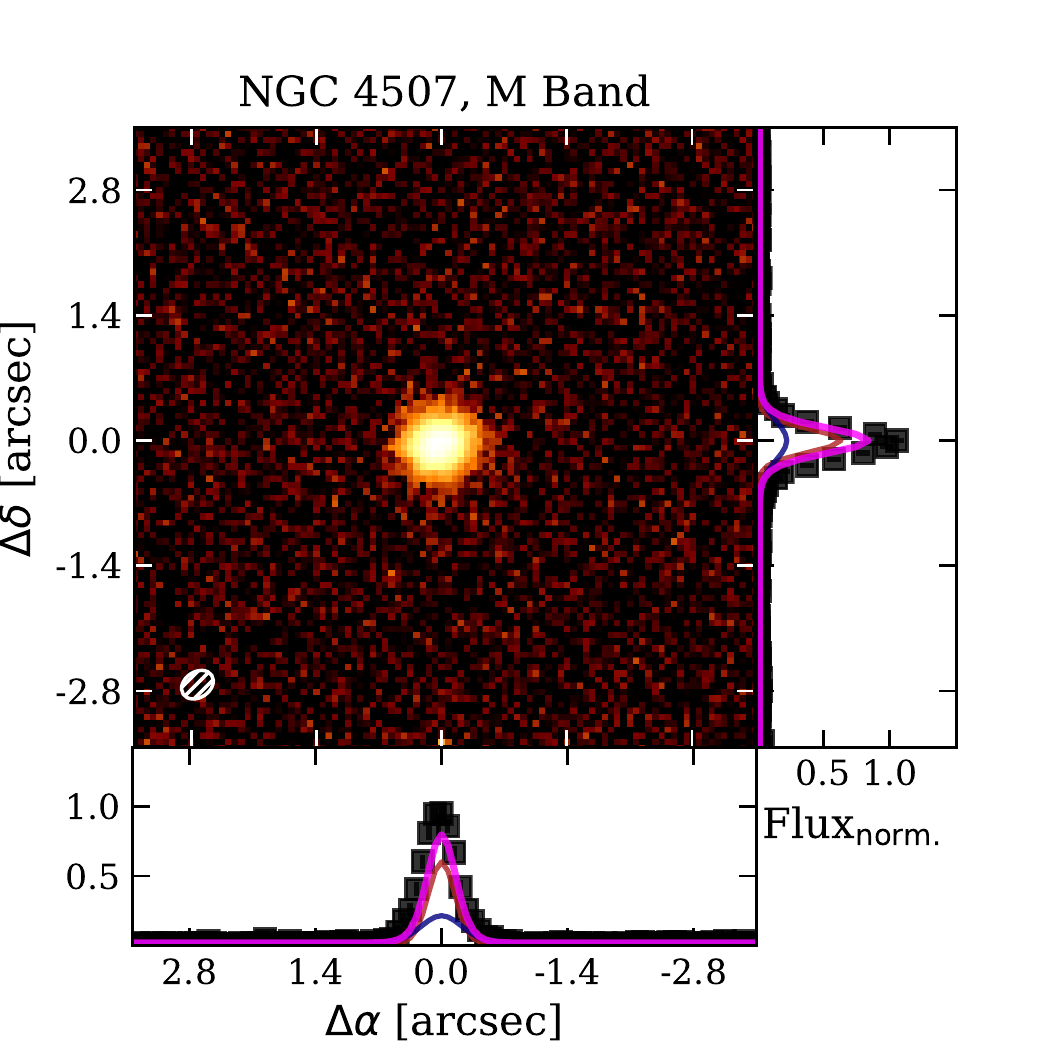}}\\
\subfloat{\includegraphics[width=0.25\hsize]{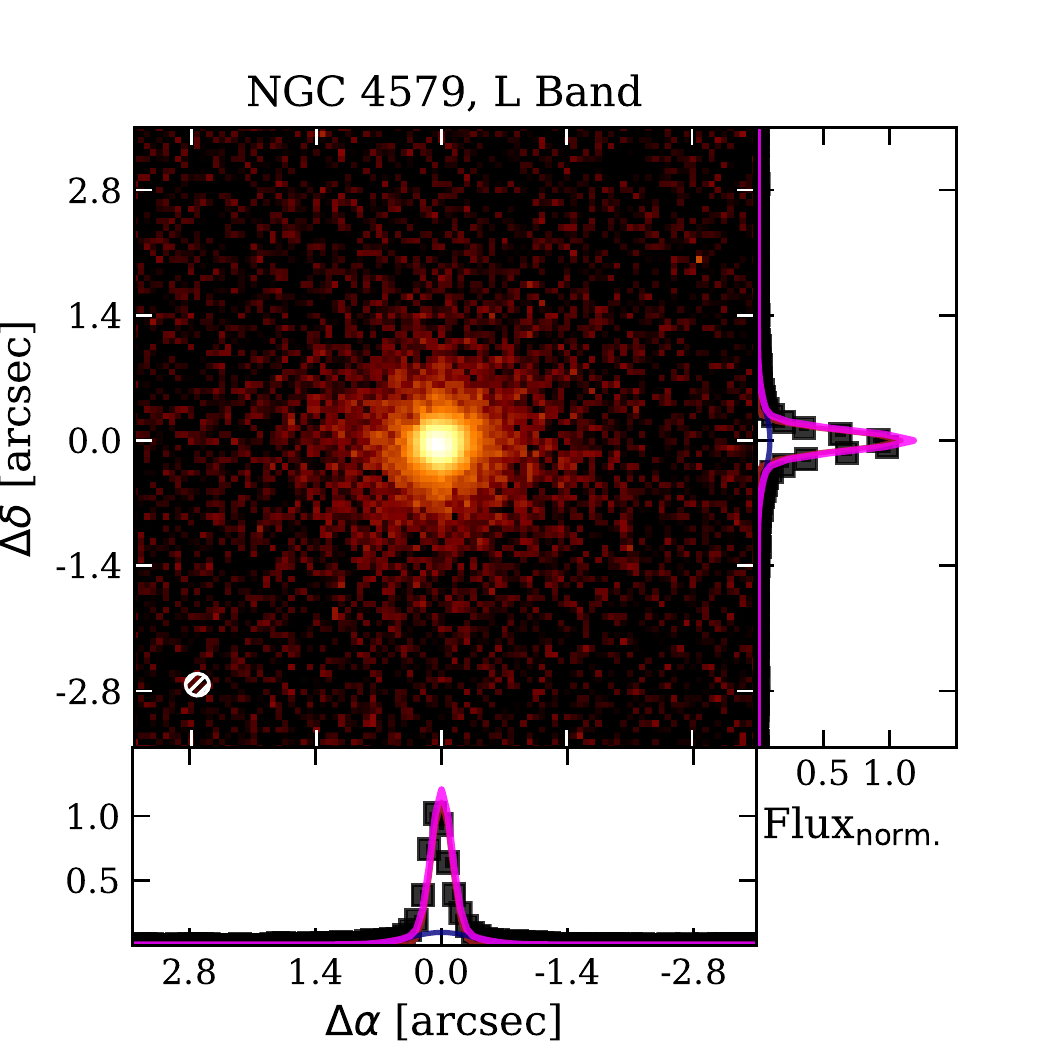}}
\subfloat{\includegraphics[width=0.25\hsize]{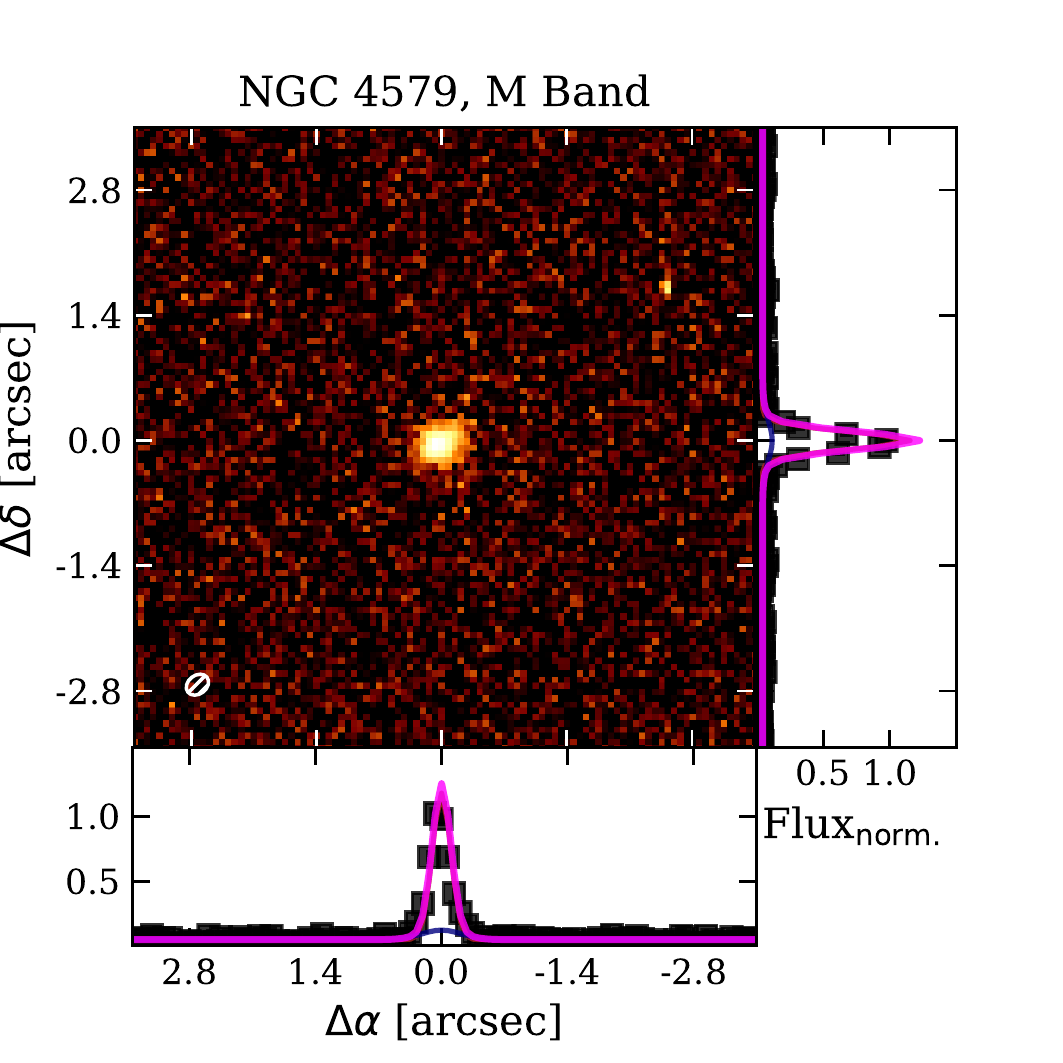}} 
\subfloat{\includegraphics[width=0.25\hsize]{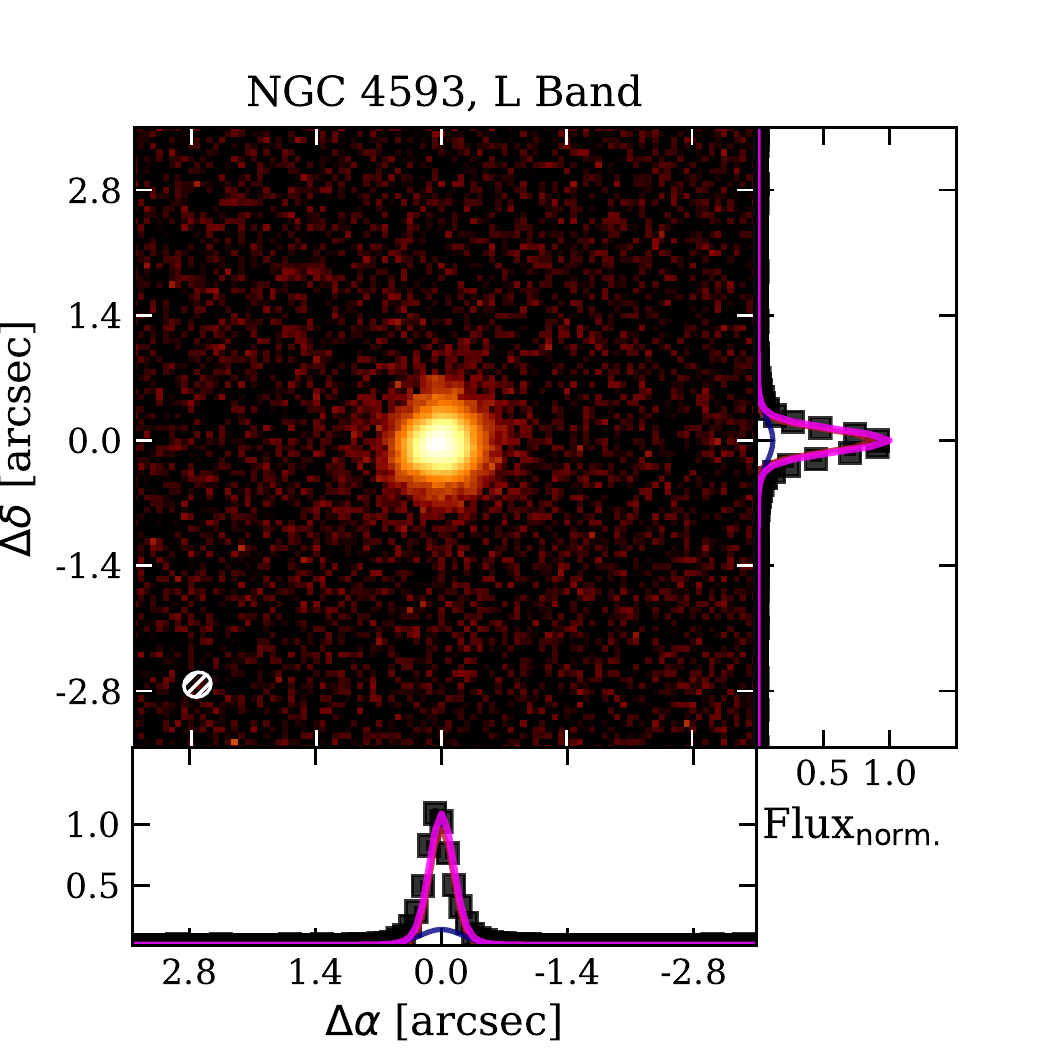}}
\subfloat{\includegraphics[width=0.25\hsize]{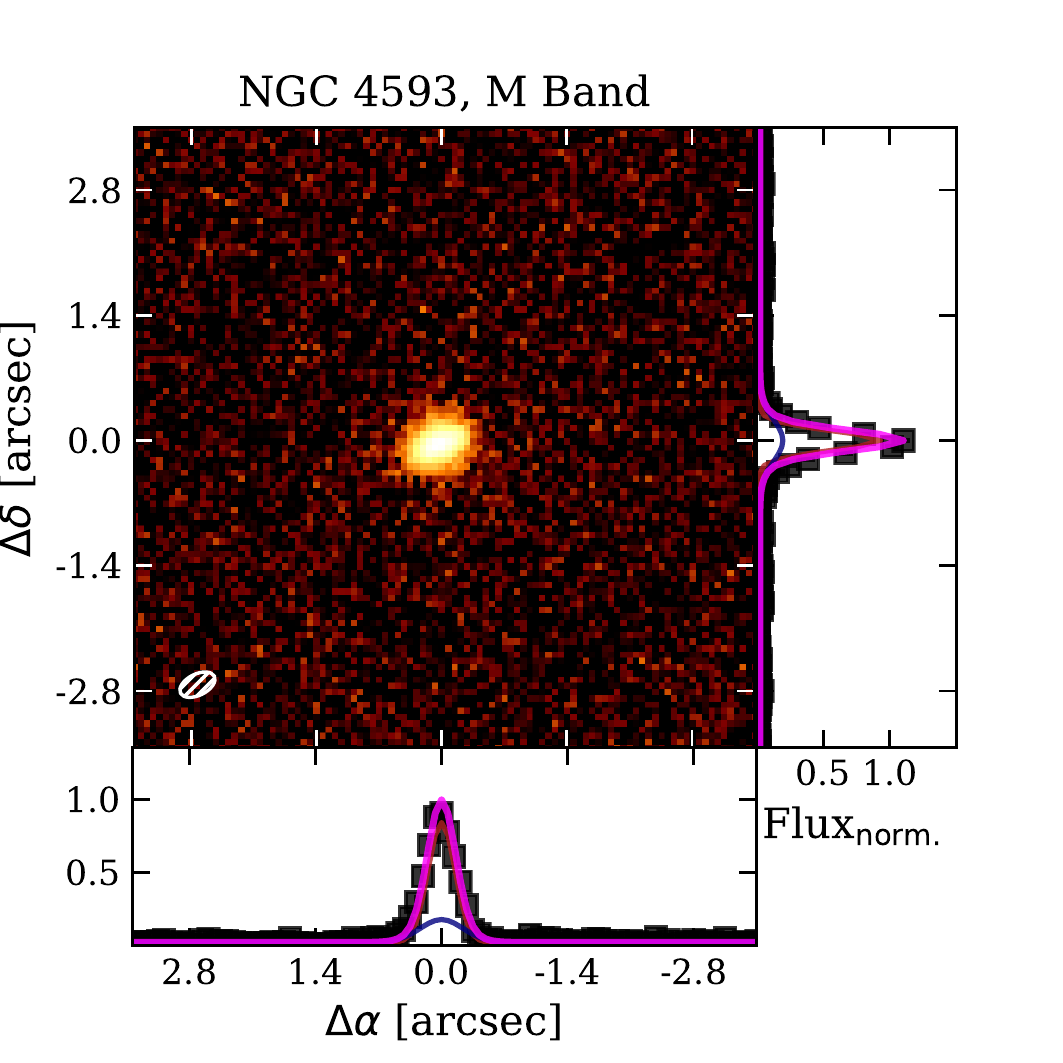}}\\
\caption{ As Fig \ref{fig:cutouts_one} but for all sources.}
\end{figure*}
\begin{figure*}
\subfloat{\includegraphics[width=0.25\hsize]{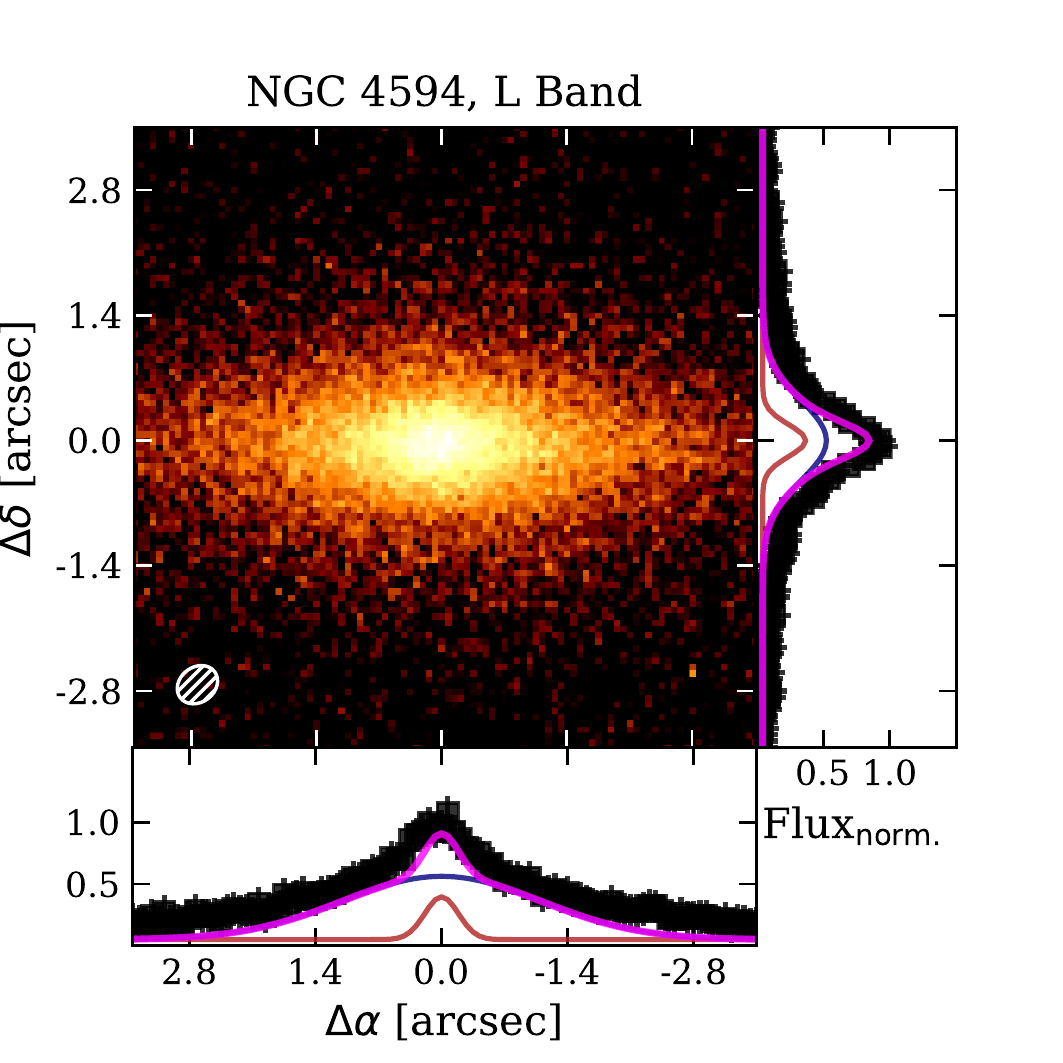}}
\subfloat{\includegraphics[width=0.25\hsize]{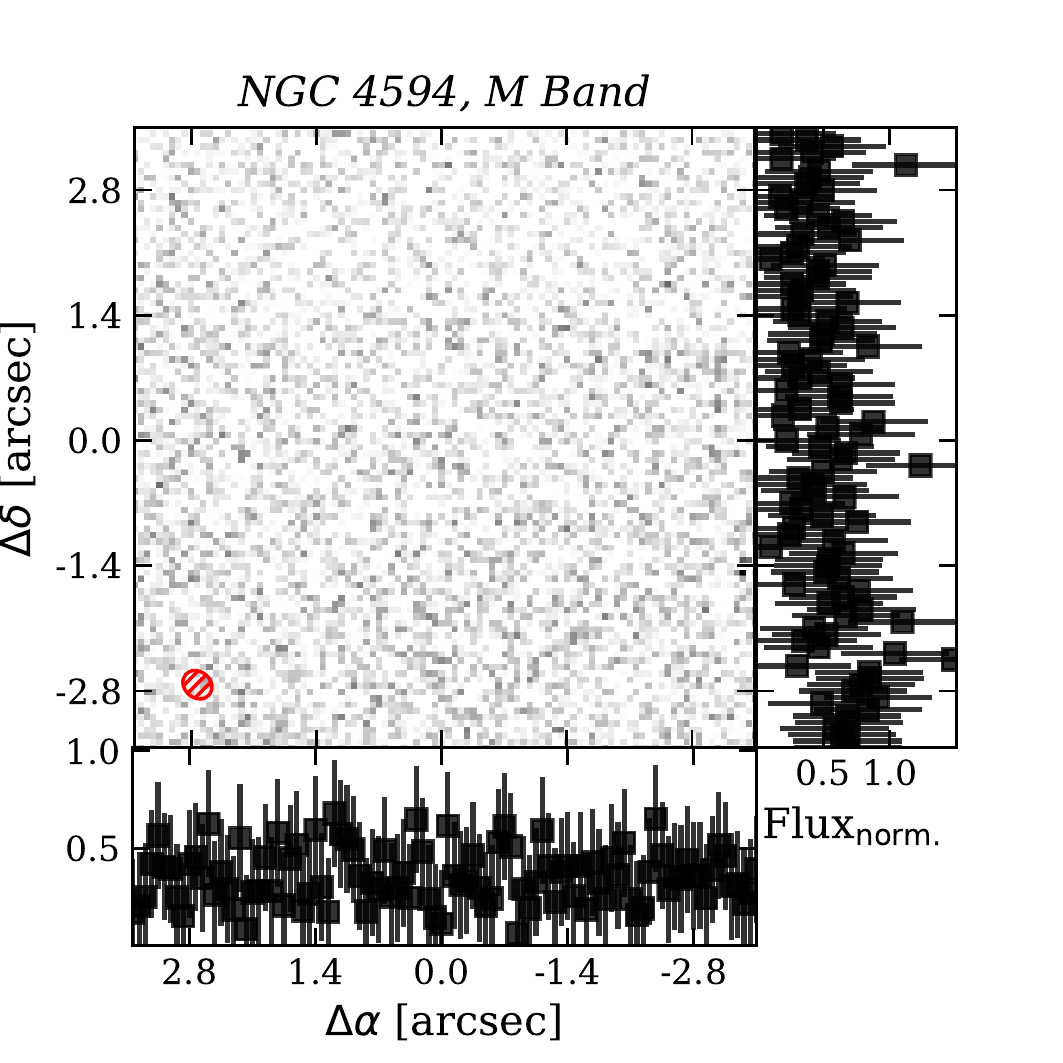}} 
\subfloat{\includegraphics[width=0.25\hsize]{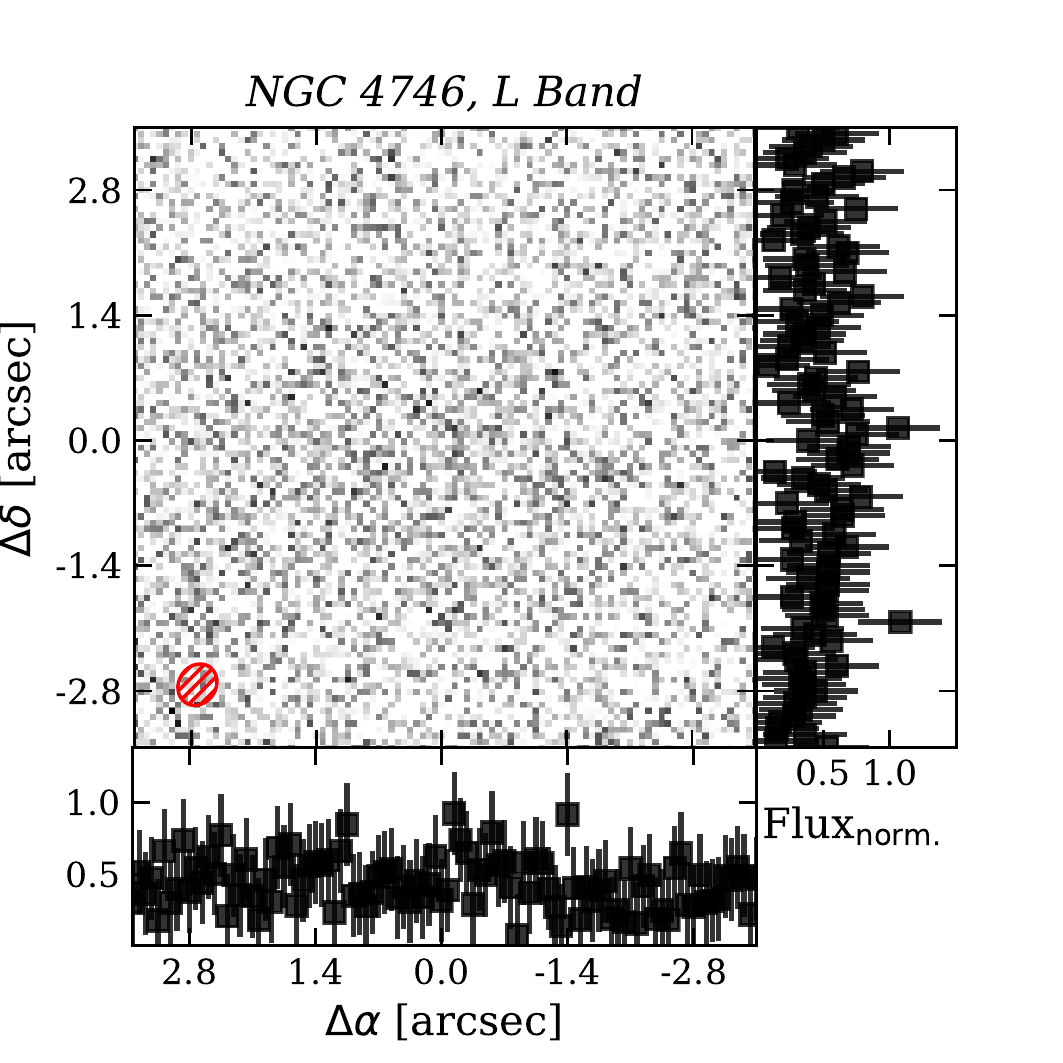}}
\subfloat{\includegraphics[width=0.25\hsize]{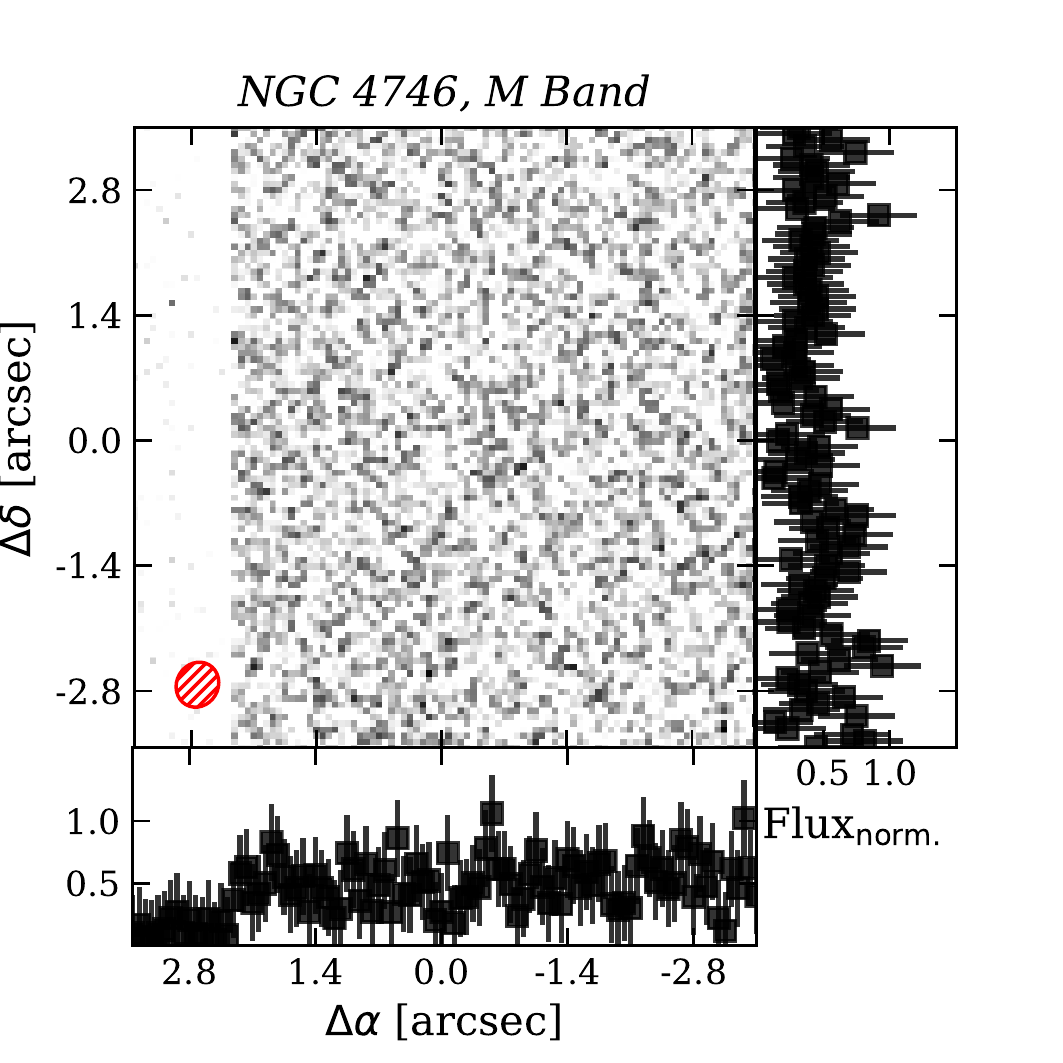}} \\
\subfloat{\includegraphics[width=0.25\hsize]{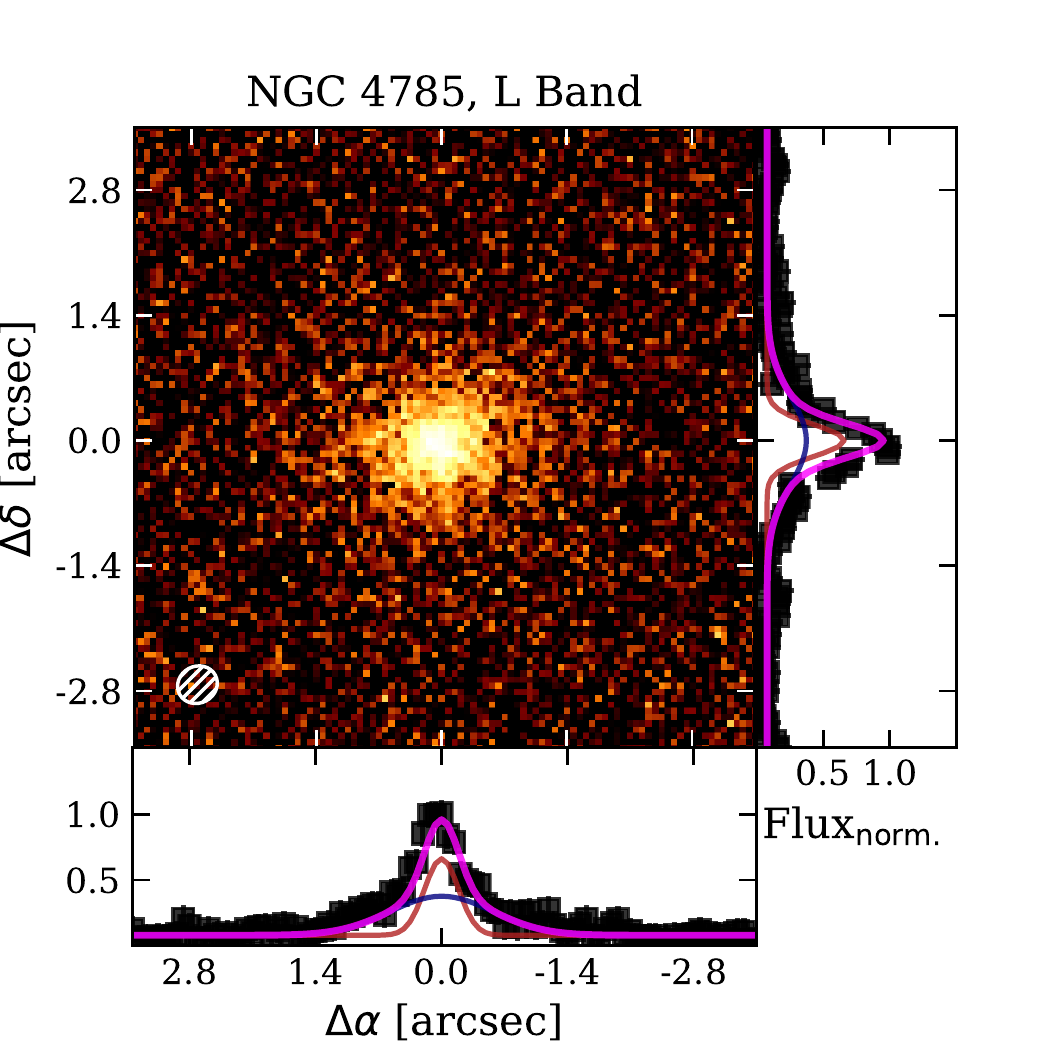}}
\subfloat{\includegraphics[width=0.25\hsize]{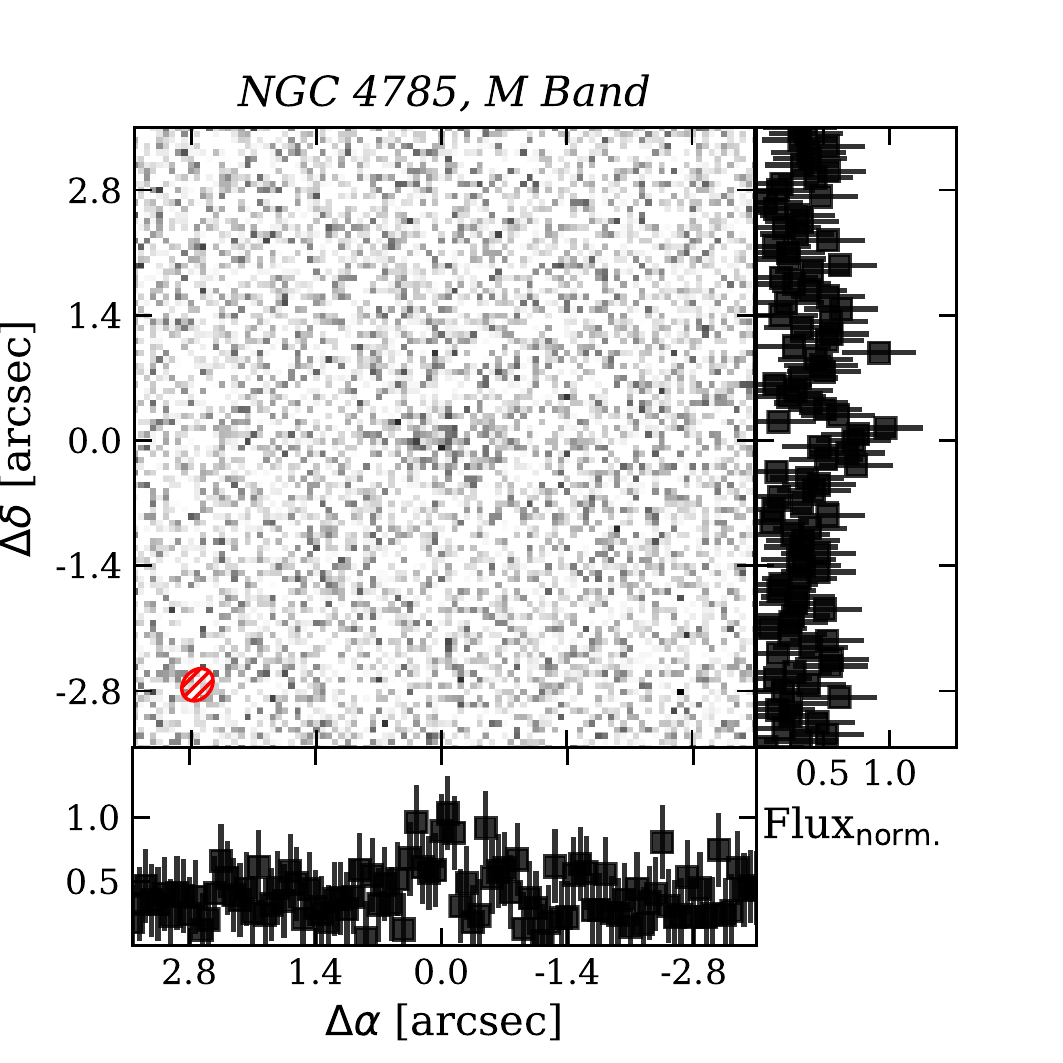}} 
\subfloat{\includegraphics[width=0.25\hsize]{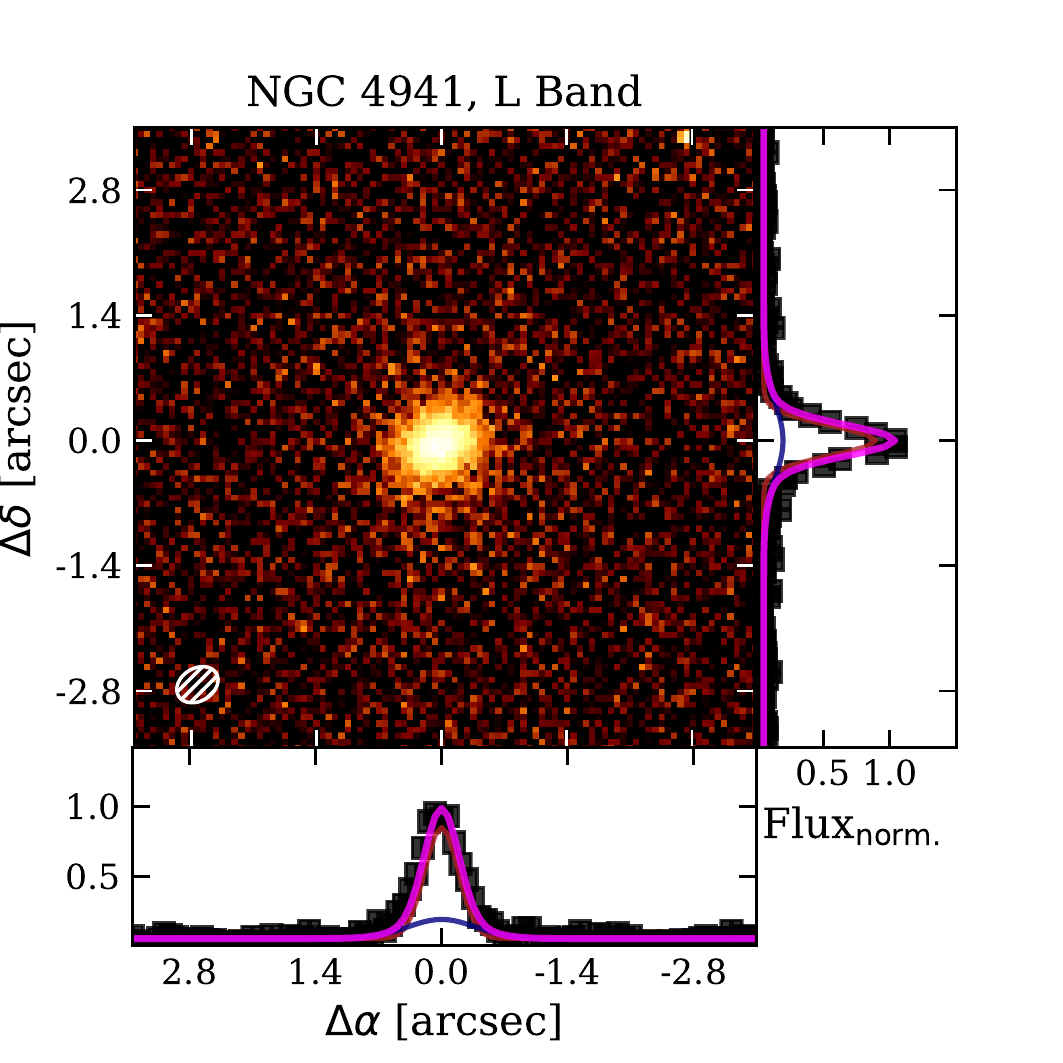}}
\subfloat{\includegraphics[width=0.25\hsize]{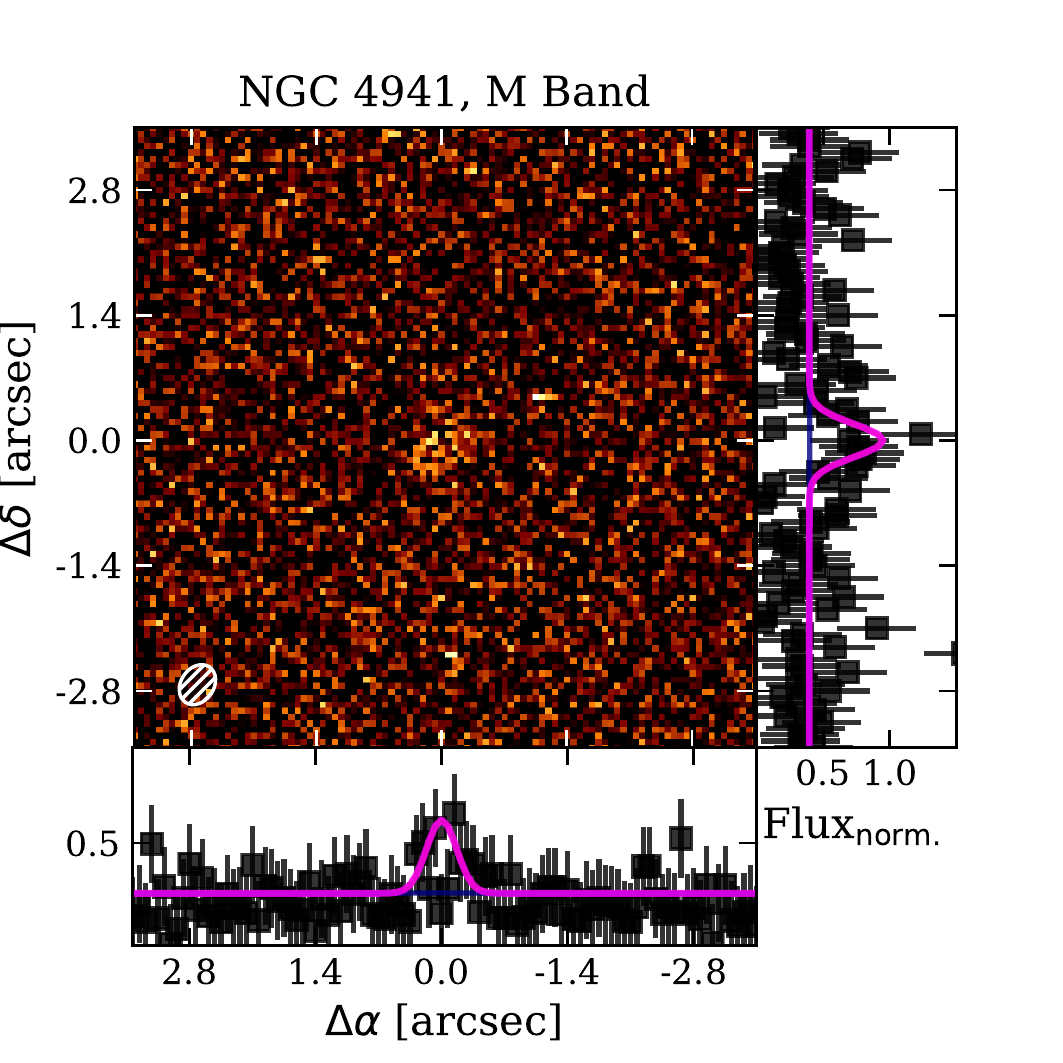}} \\
\subfloat{\includegraphics[width=0.25\hsize]{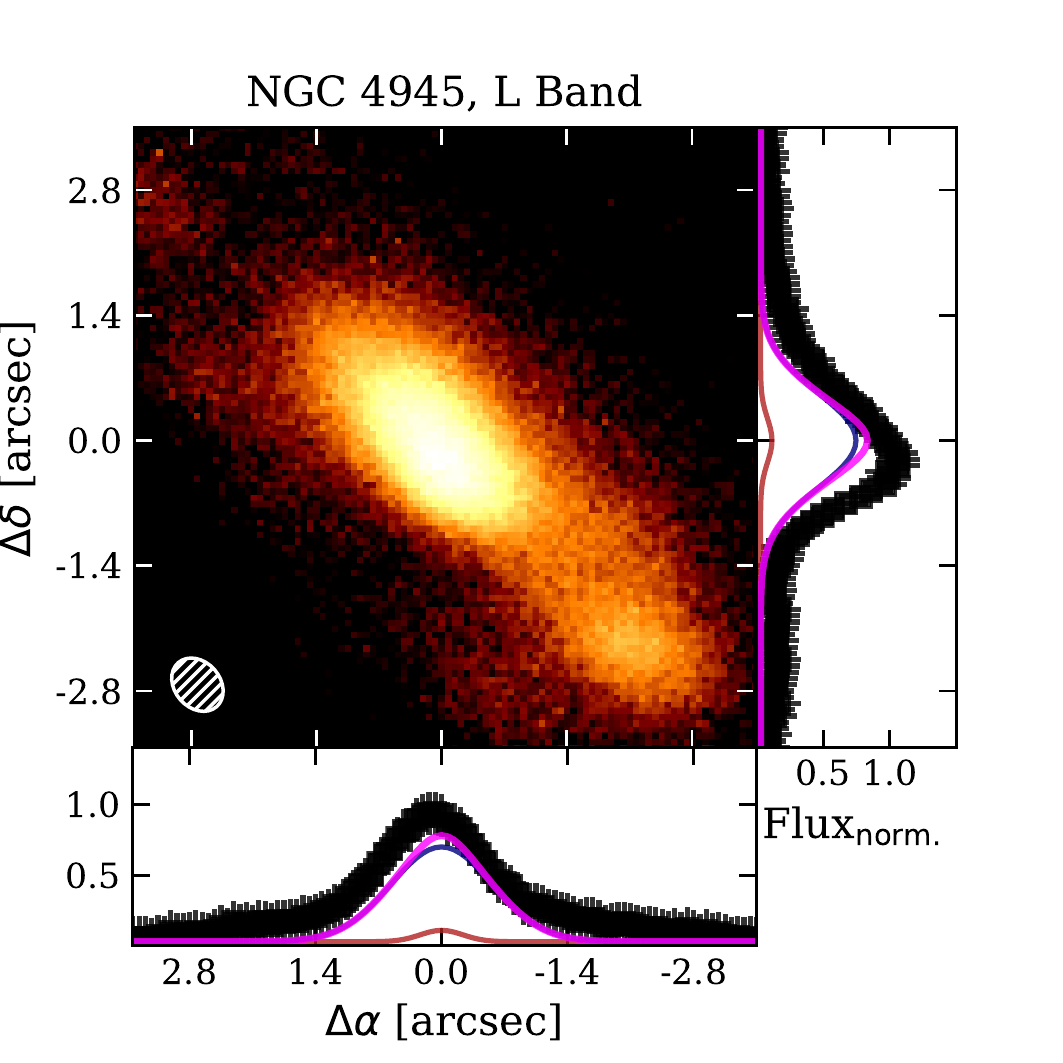}}
\subfloat{\includegraphics[width=0.25\hsize]{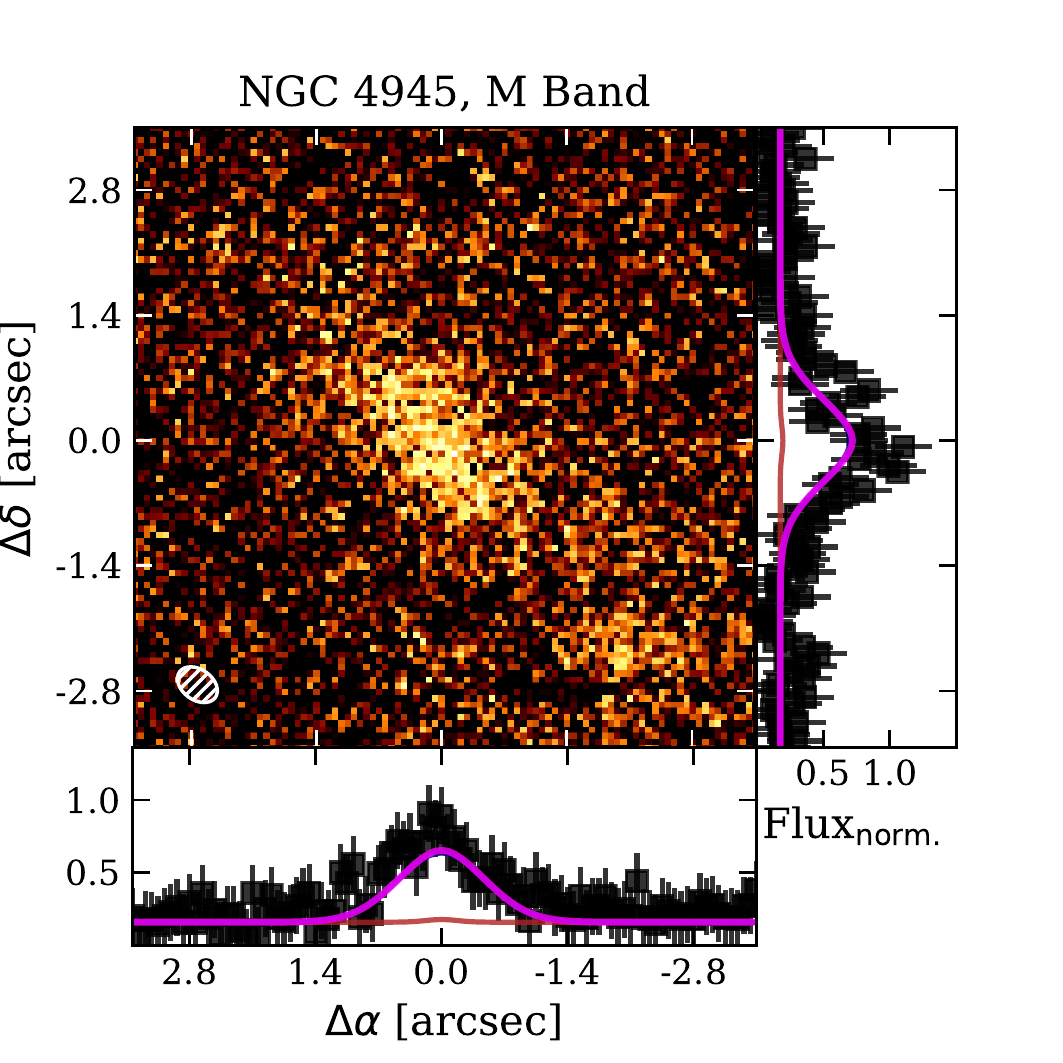}} 
\subfloat{\includegraphics[width=0.25\hsize]{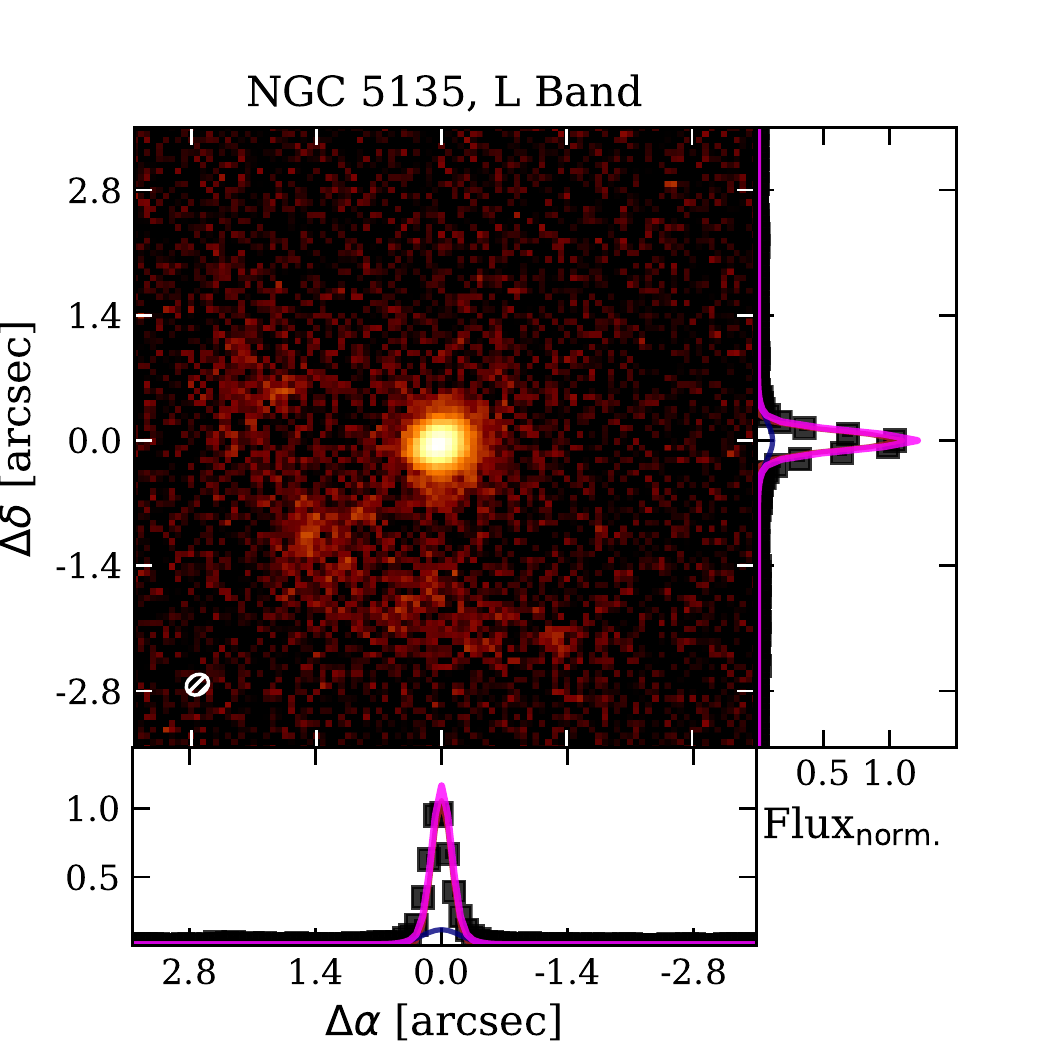}}
\subfloat{\includegraphics[width=0.25\hsize]{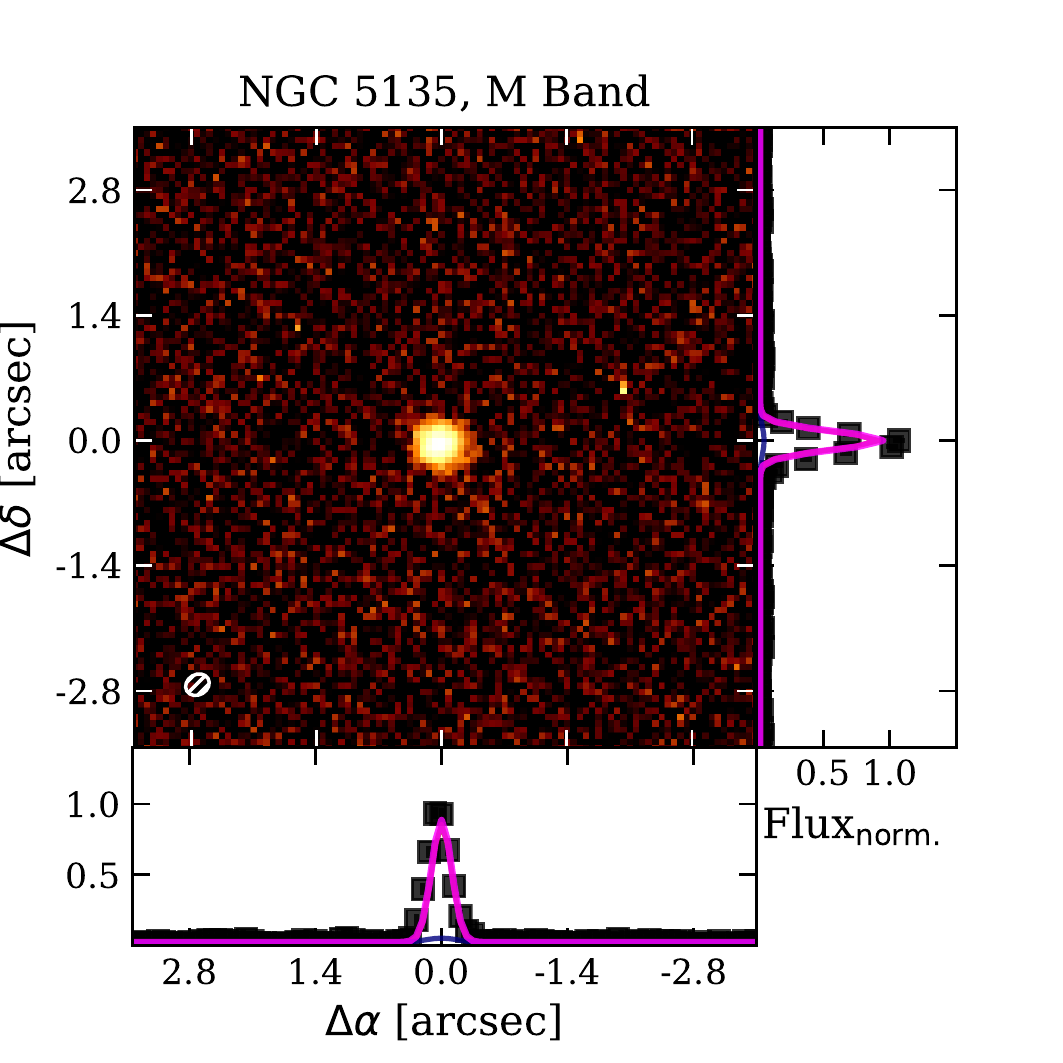}} \\
\subfloat{\includegraphics[width=0.25\hsize]{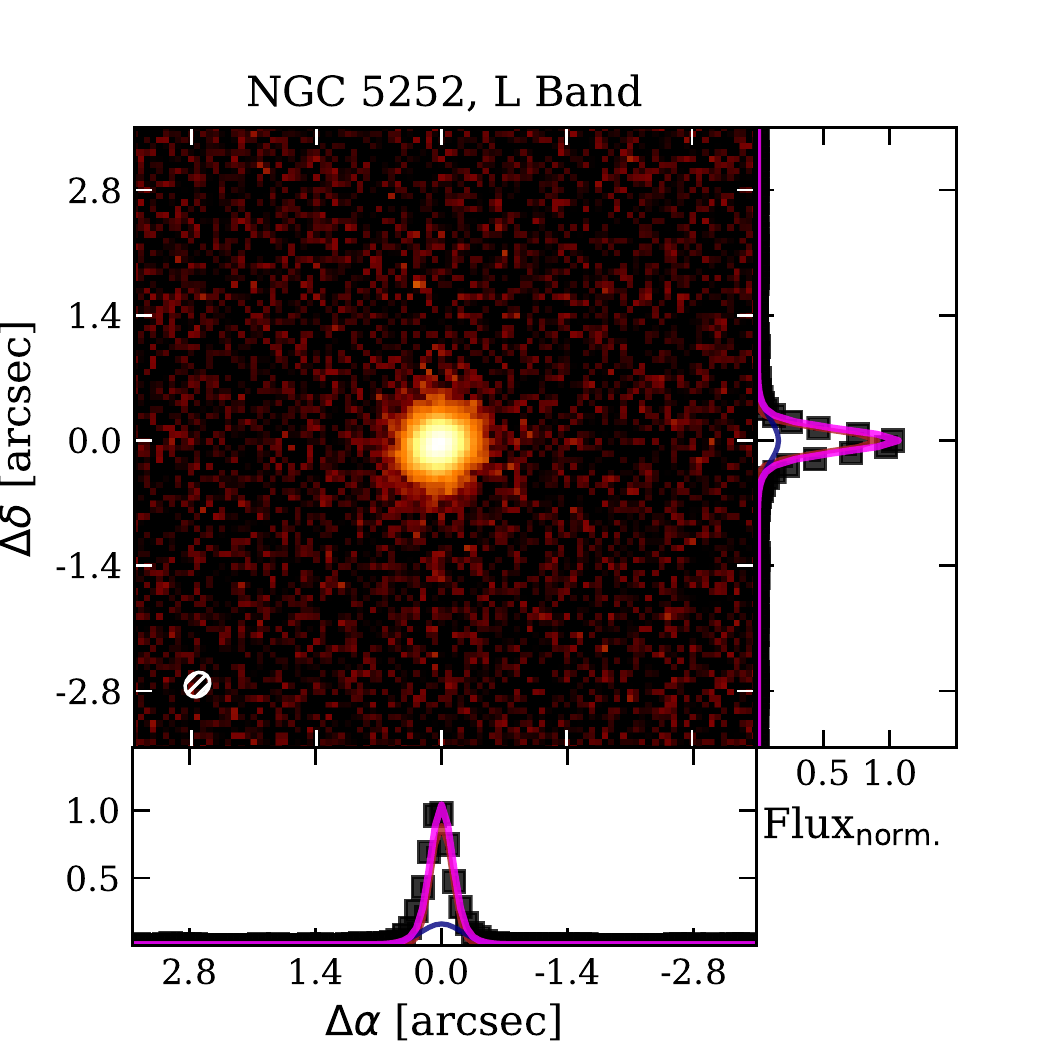}}
\subfloat{\includegraphics[width=0.25\hsize]{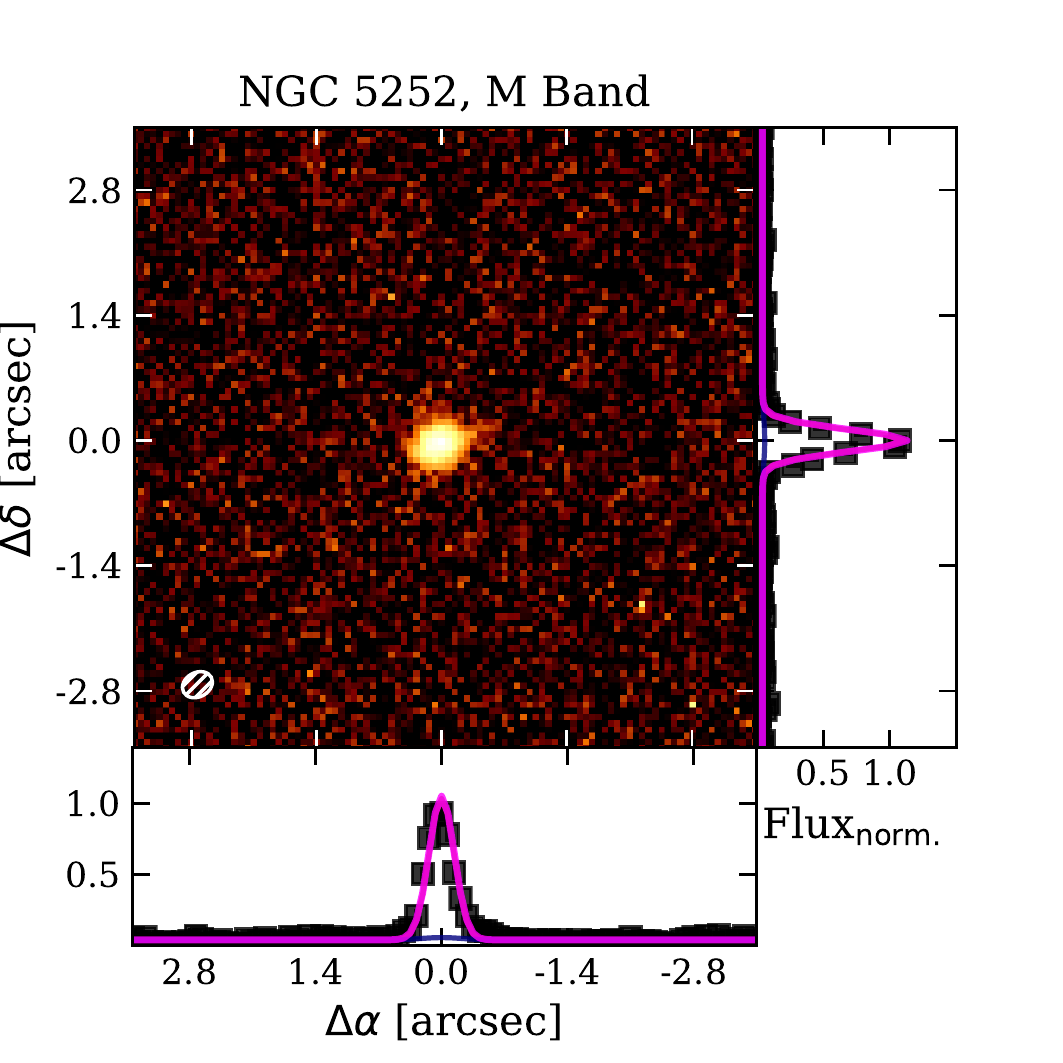}} 
\subfloat{\includegraphics[width=0.25\hsize]{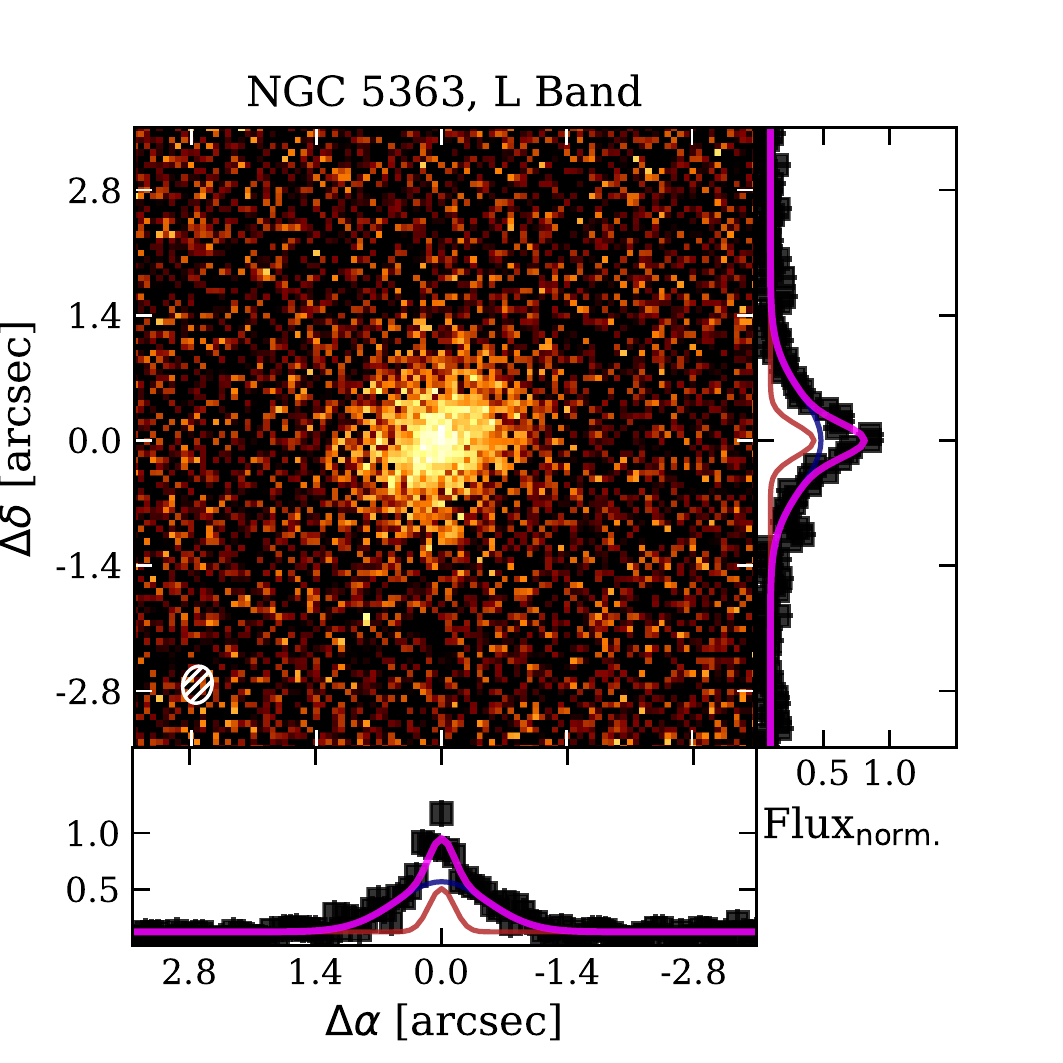}}
\subfloat{\includegraphics[width=0.25\hsize]{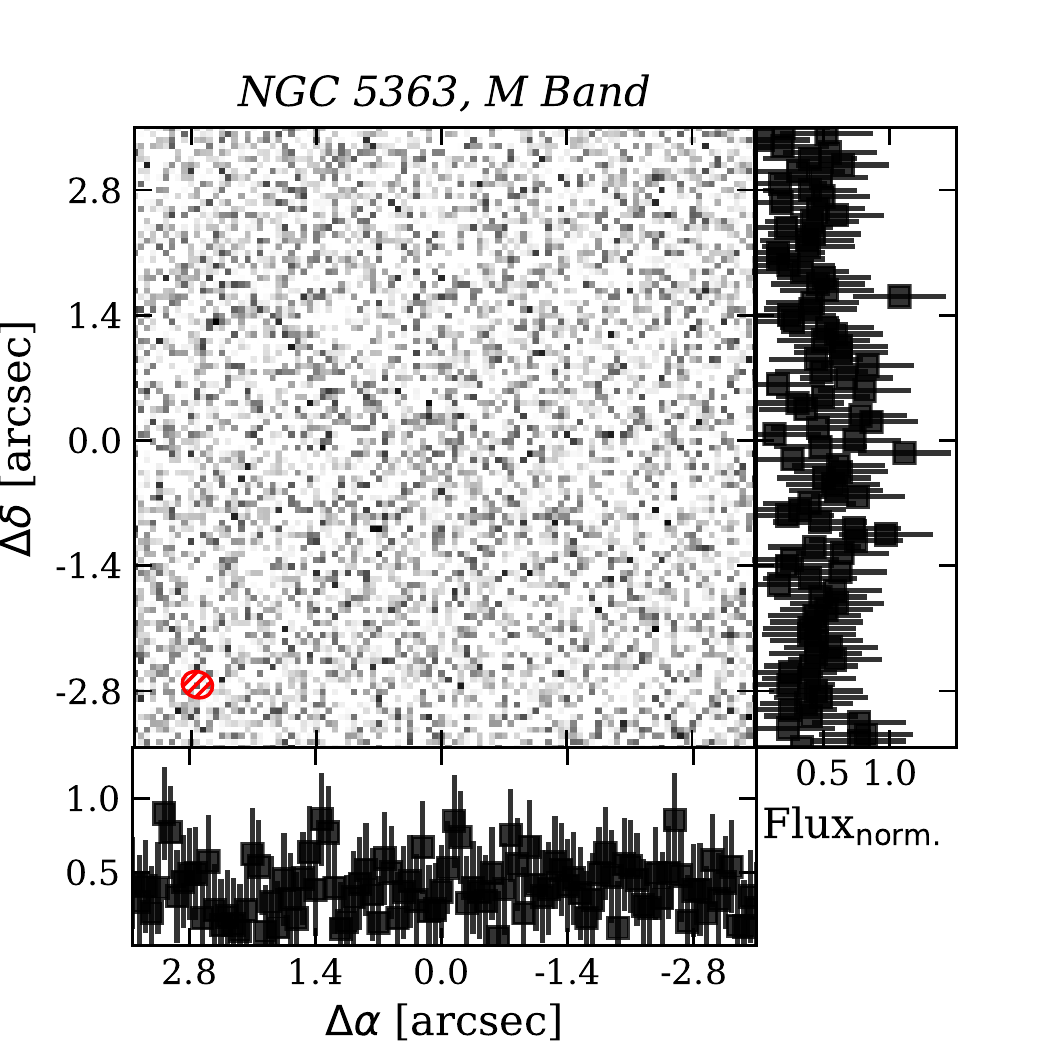}} \\
\subfloat{\includegraphics[width=0.25\hsize]{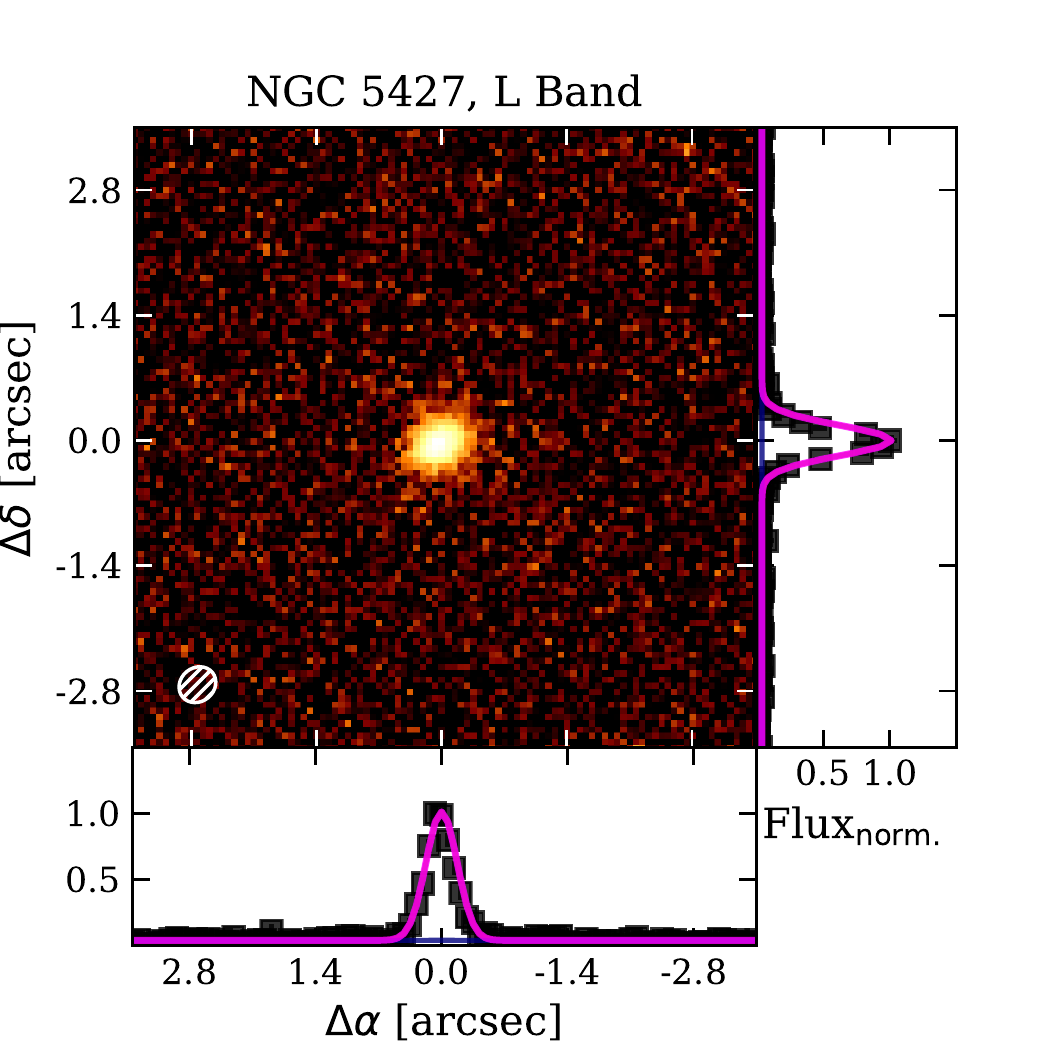}}
\subfloat{\includegraphics[width=0.25\hsize]{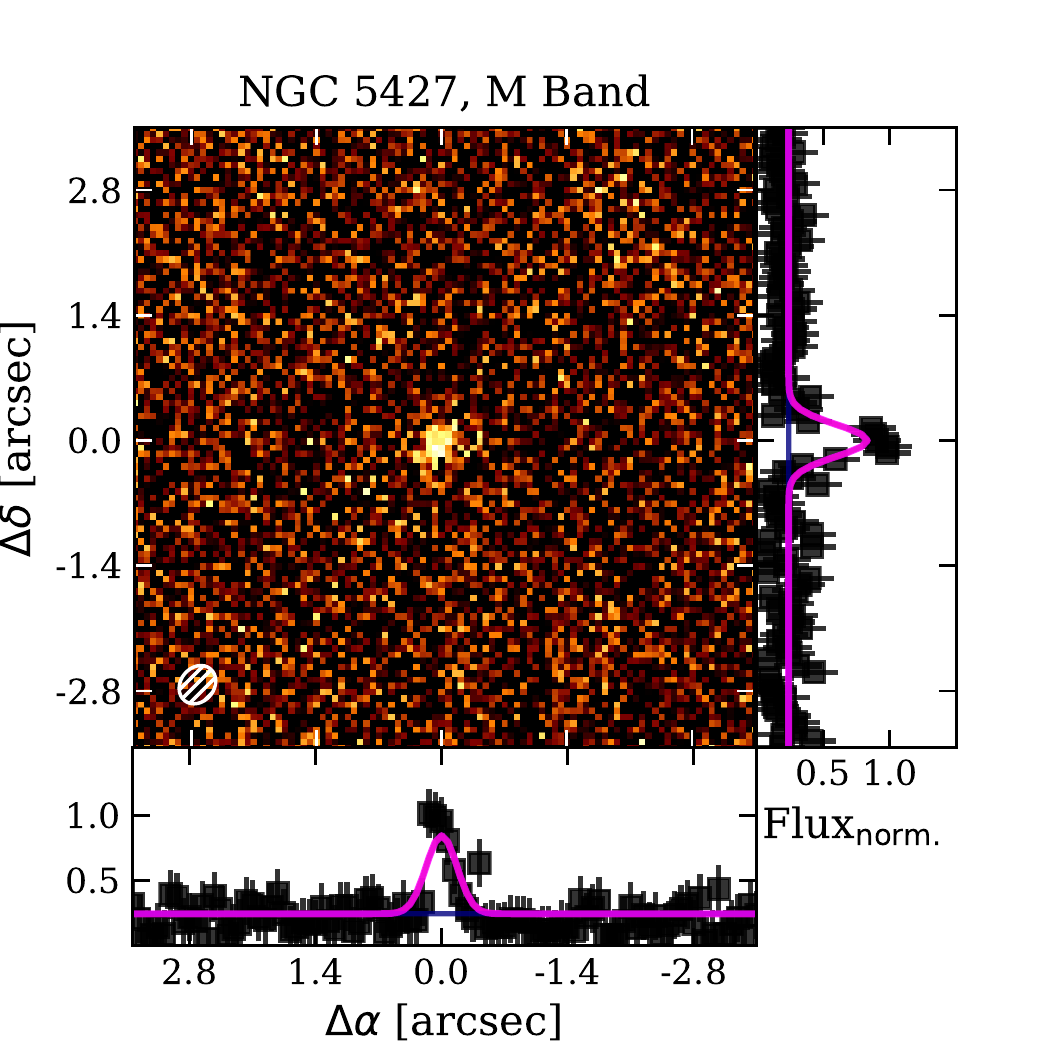}} 
\subfloat{\includegraphics[width=0.25\hsize]{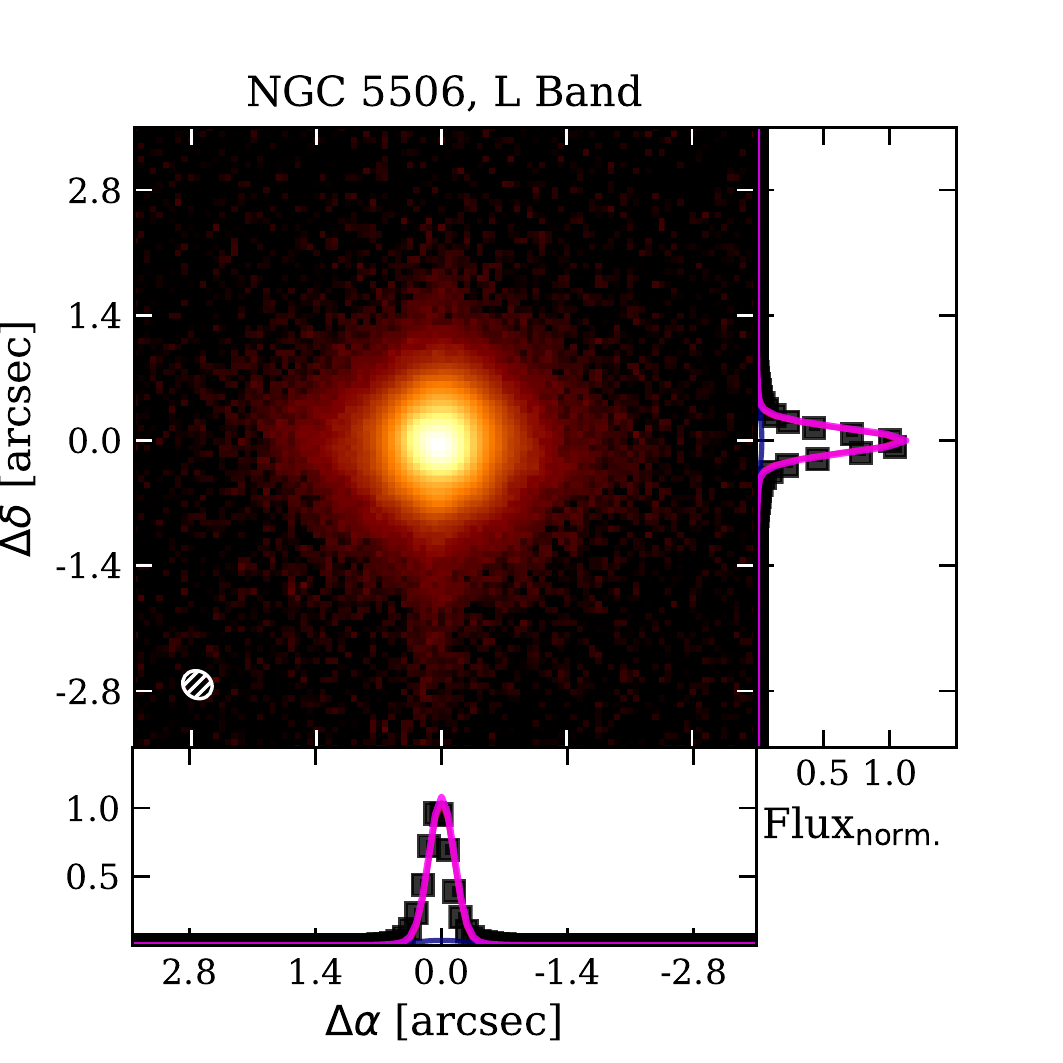}}
\subfloat{\includegraphics[width=0.25\hsize]{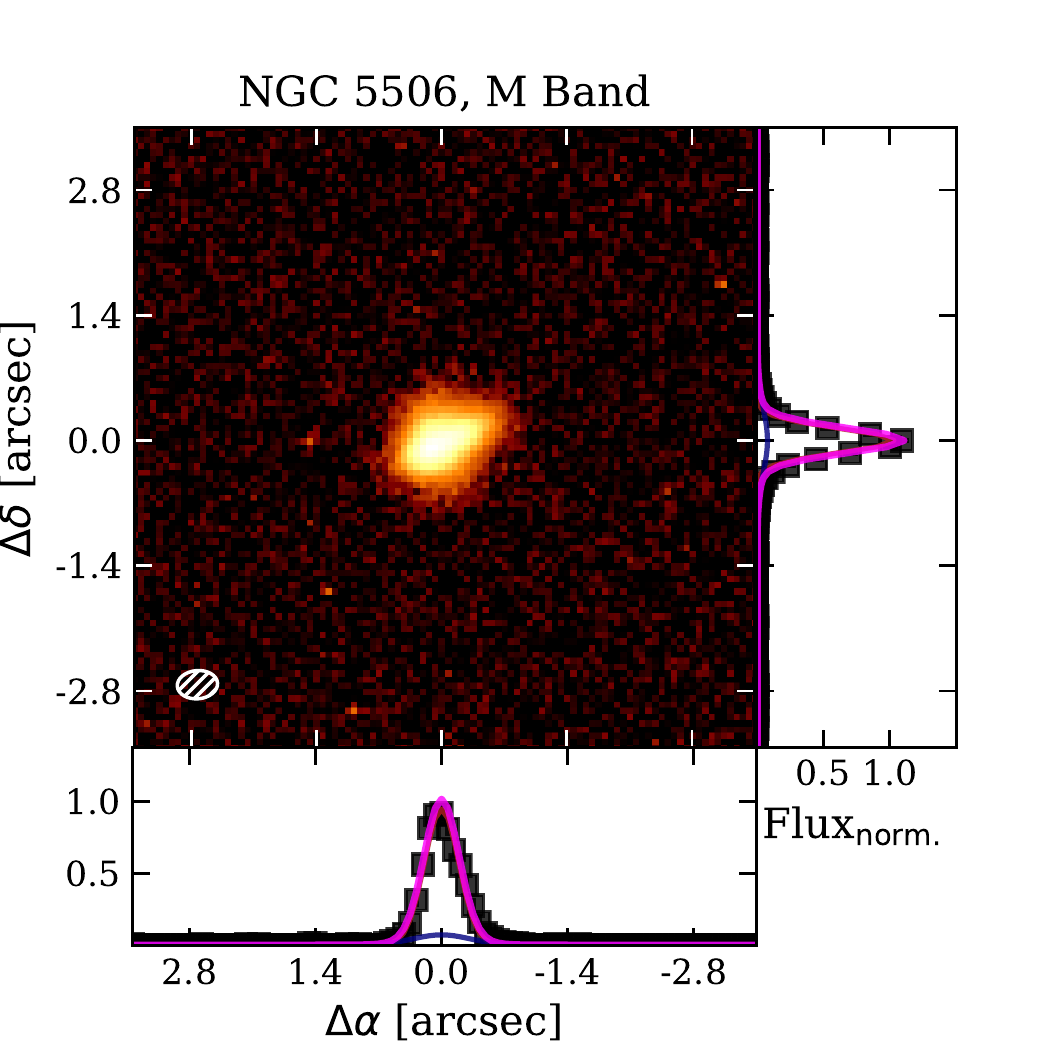}} \\
\caption{ As Fig \ref{fig:cutouts_one} but for all sources.}
\end{figure*}
\begin{figure*}
\subfloat{\includegraphics[width=0.25\hsize]{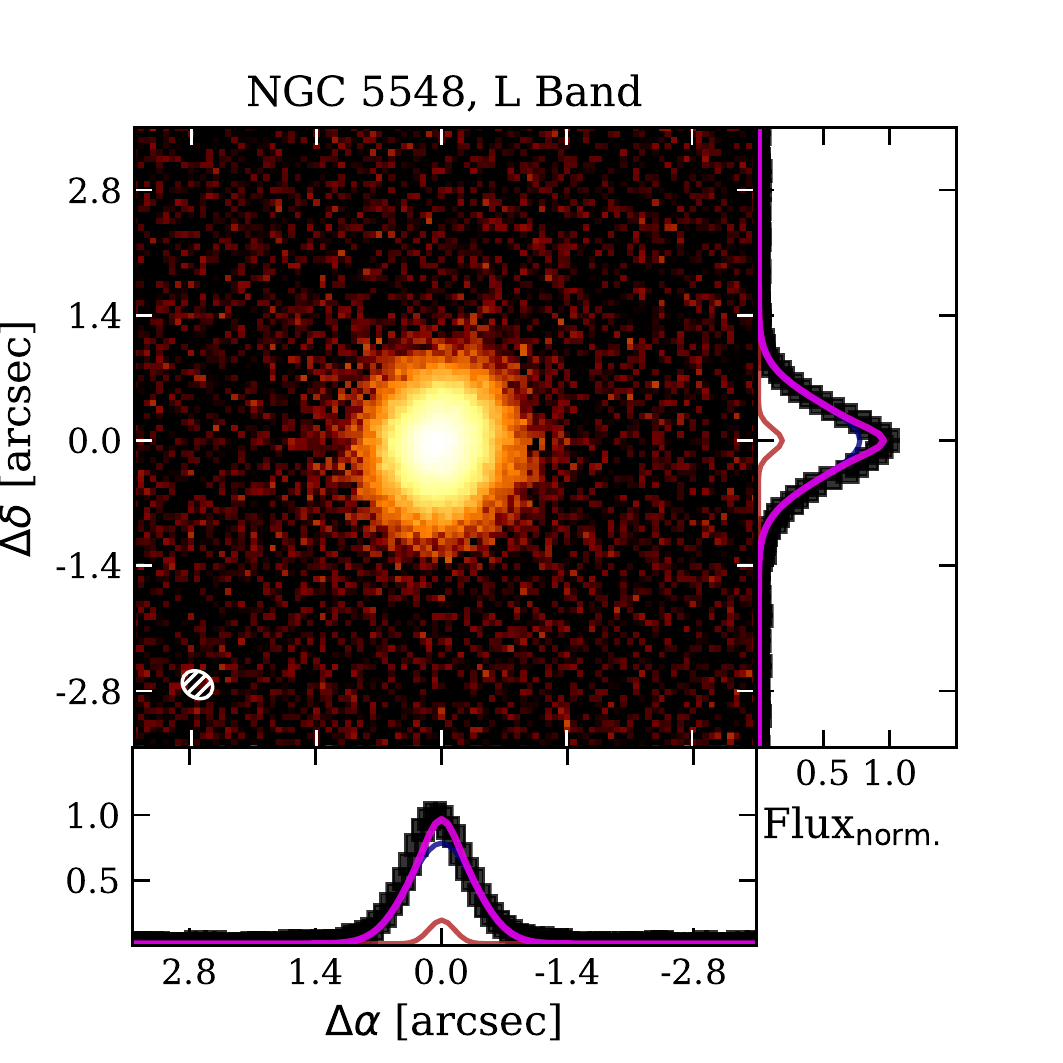}}
\subfloat{\includegraphics[width=0.25\hsize]{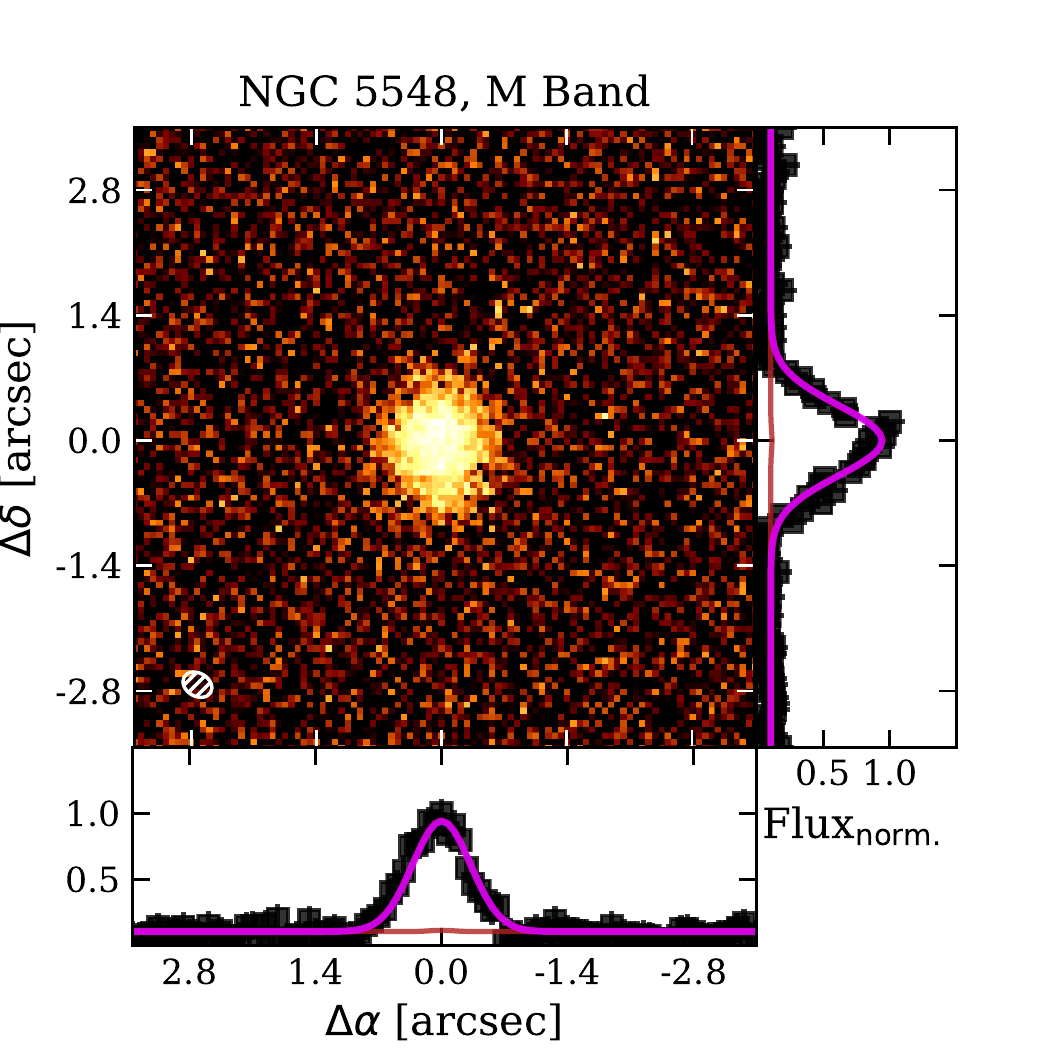}} 
\subfloat{\includegraphics[width=0.25\hsize]{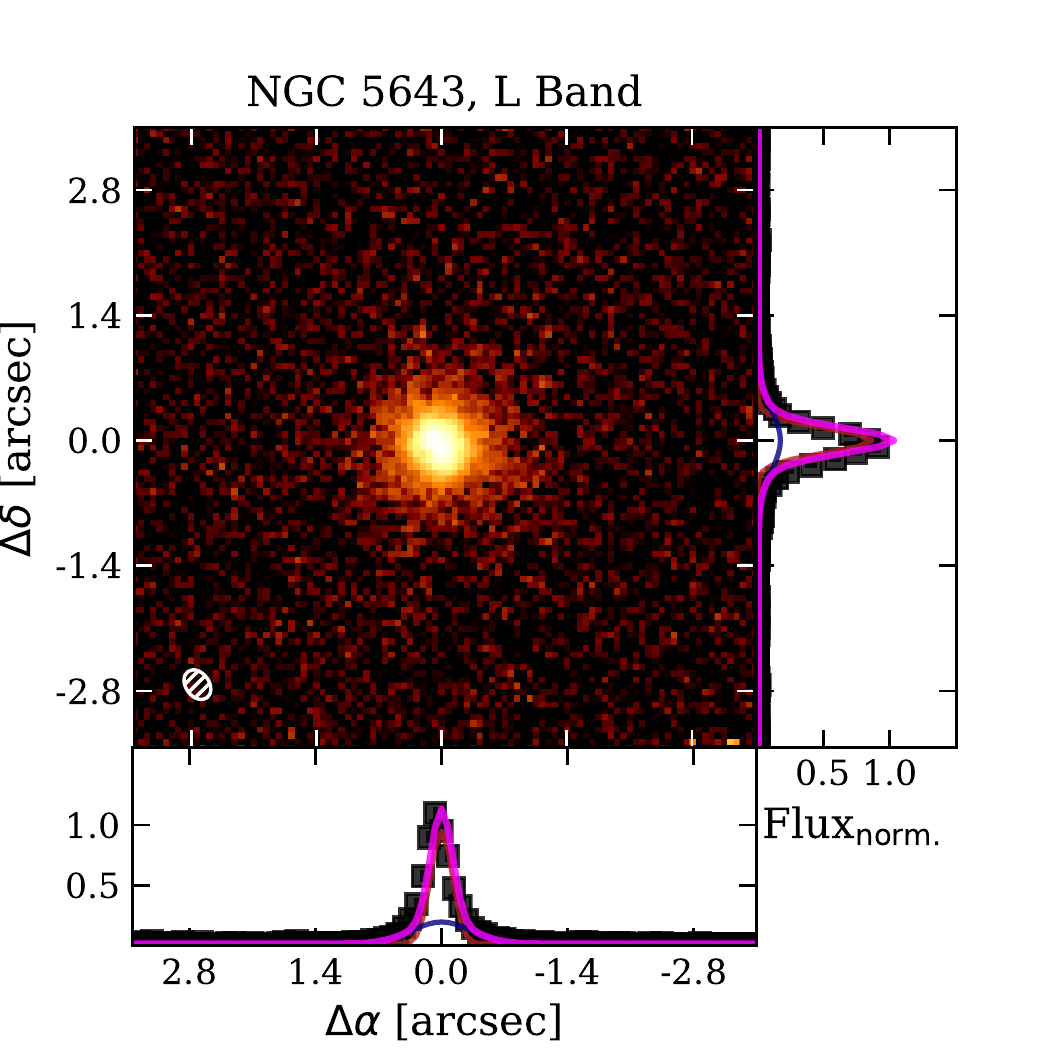}}
\subfloat{\includegraphics[width=0.25\hsize]{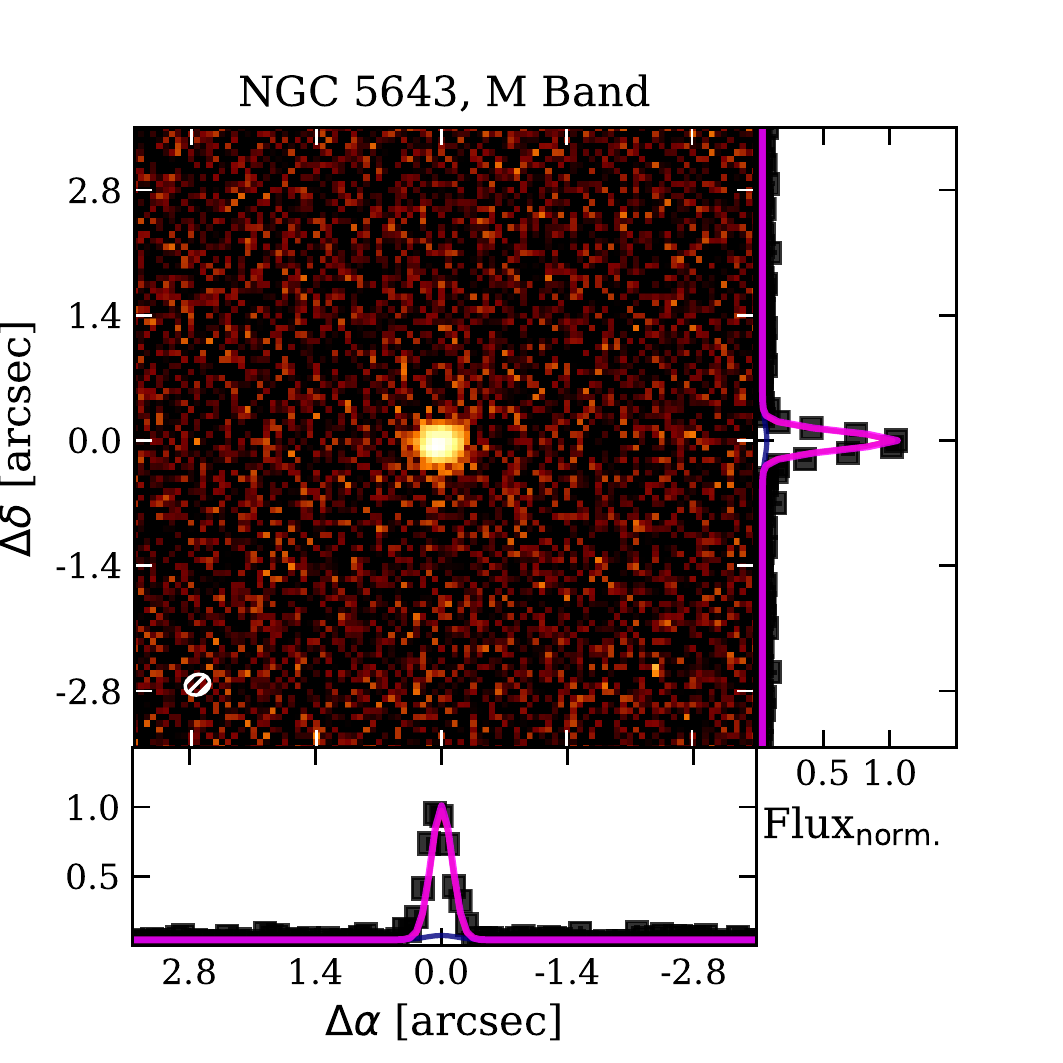}} \\
\subfloat{\includegraphics[width=0.25\hsize]{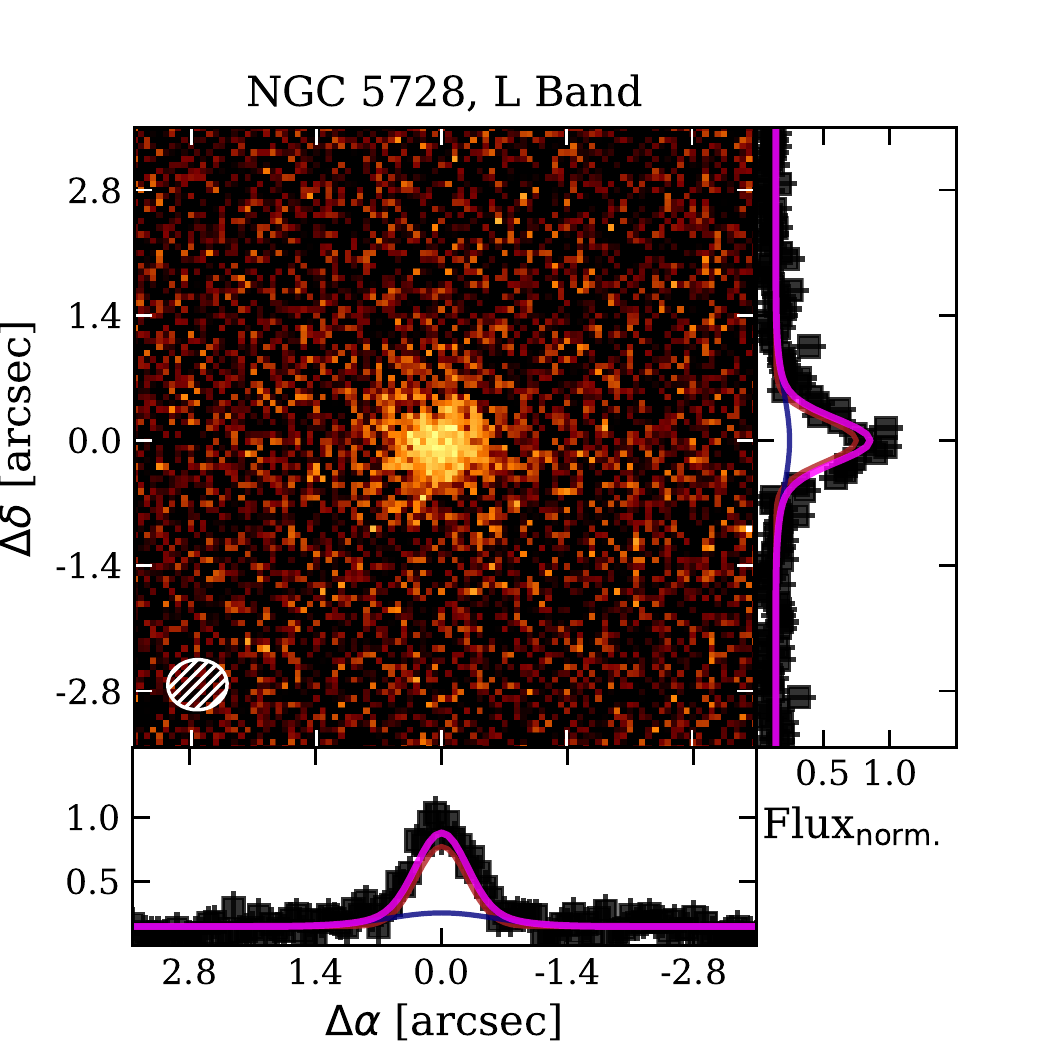}}
\subfloat{\includegraphics[width=0.25\hsize]{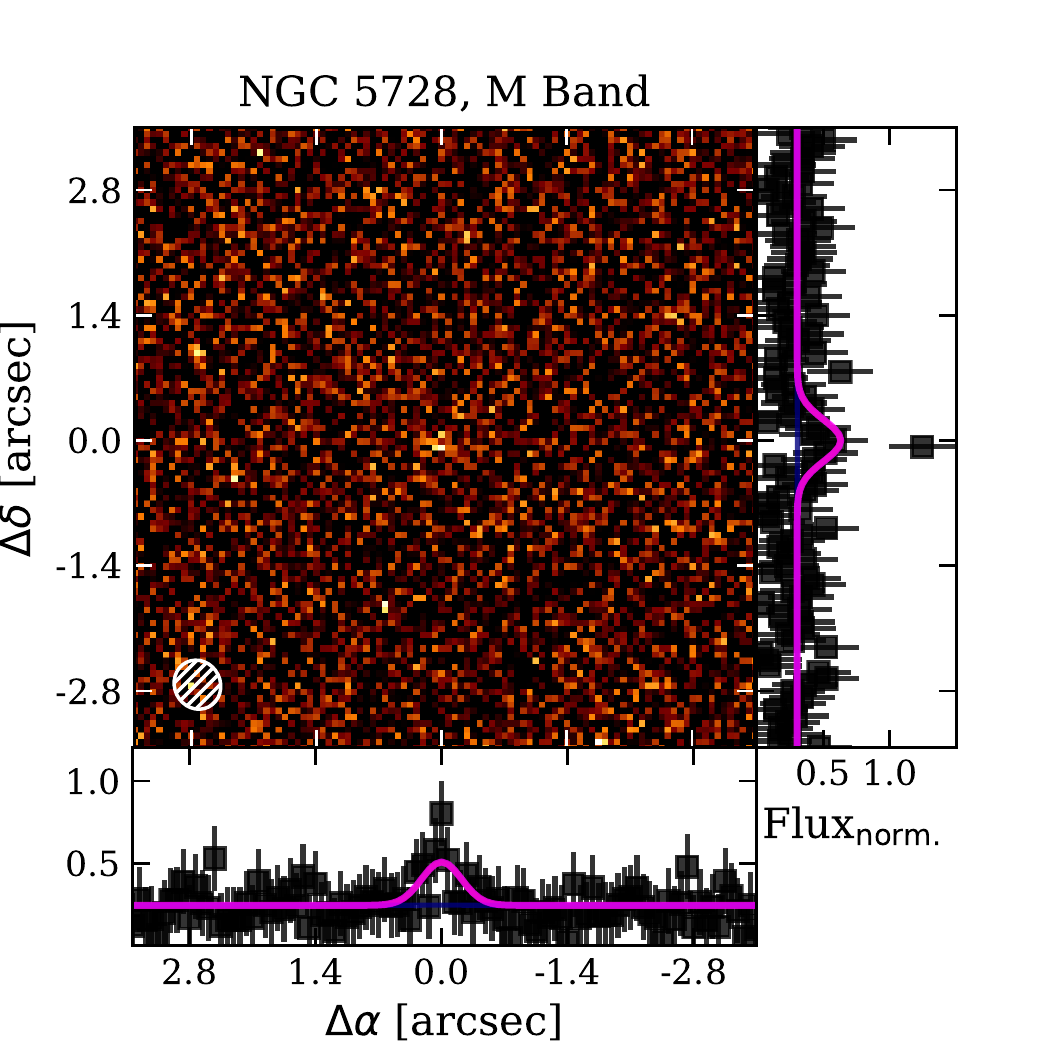}} 
\subfloat{\includegraphics[width=0.25\hsize]{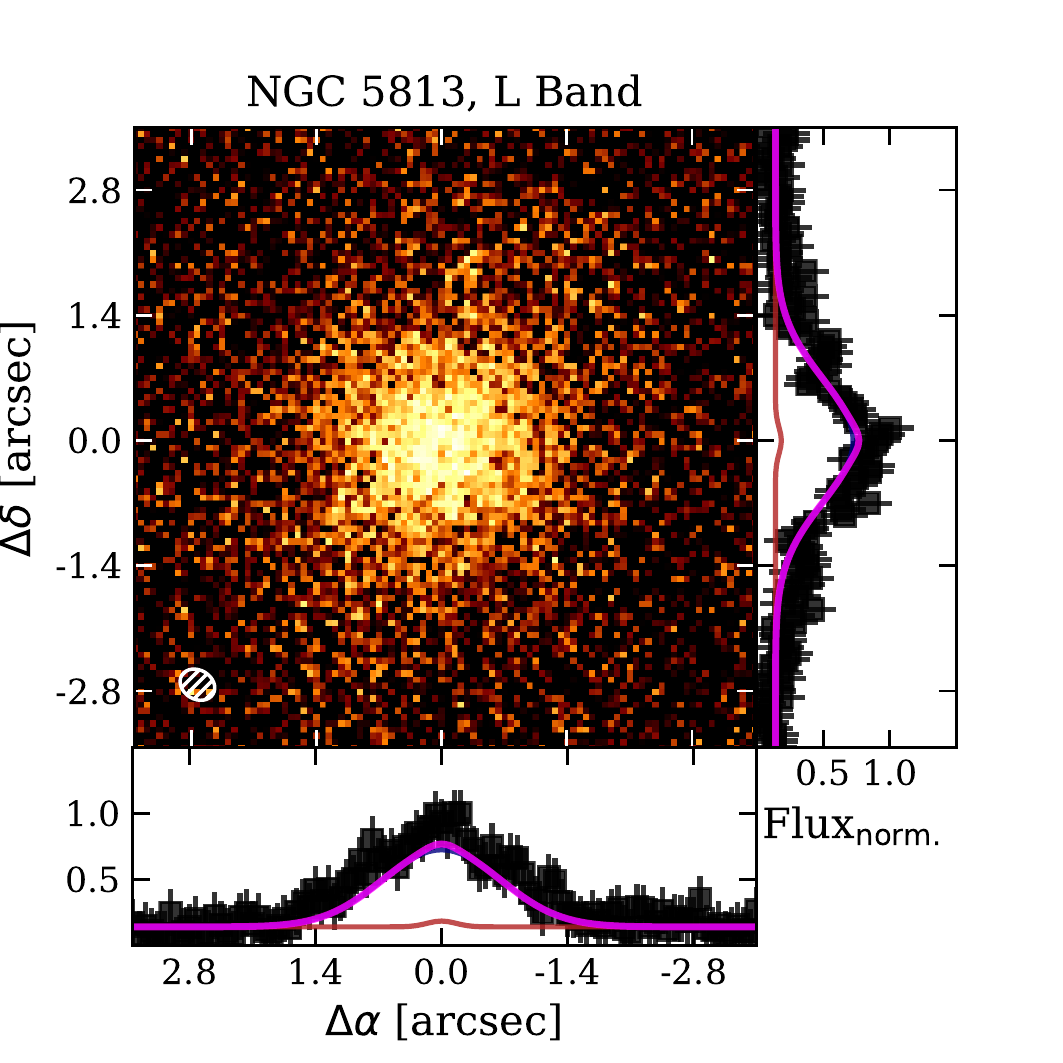}}
\subfloat{\includegraphics[width=0.25\hsize]{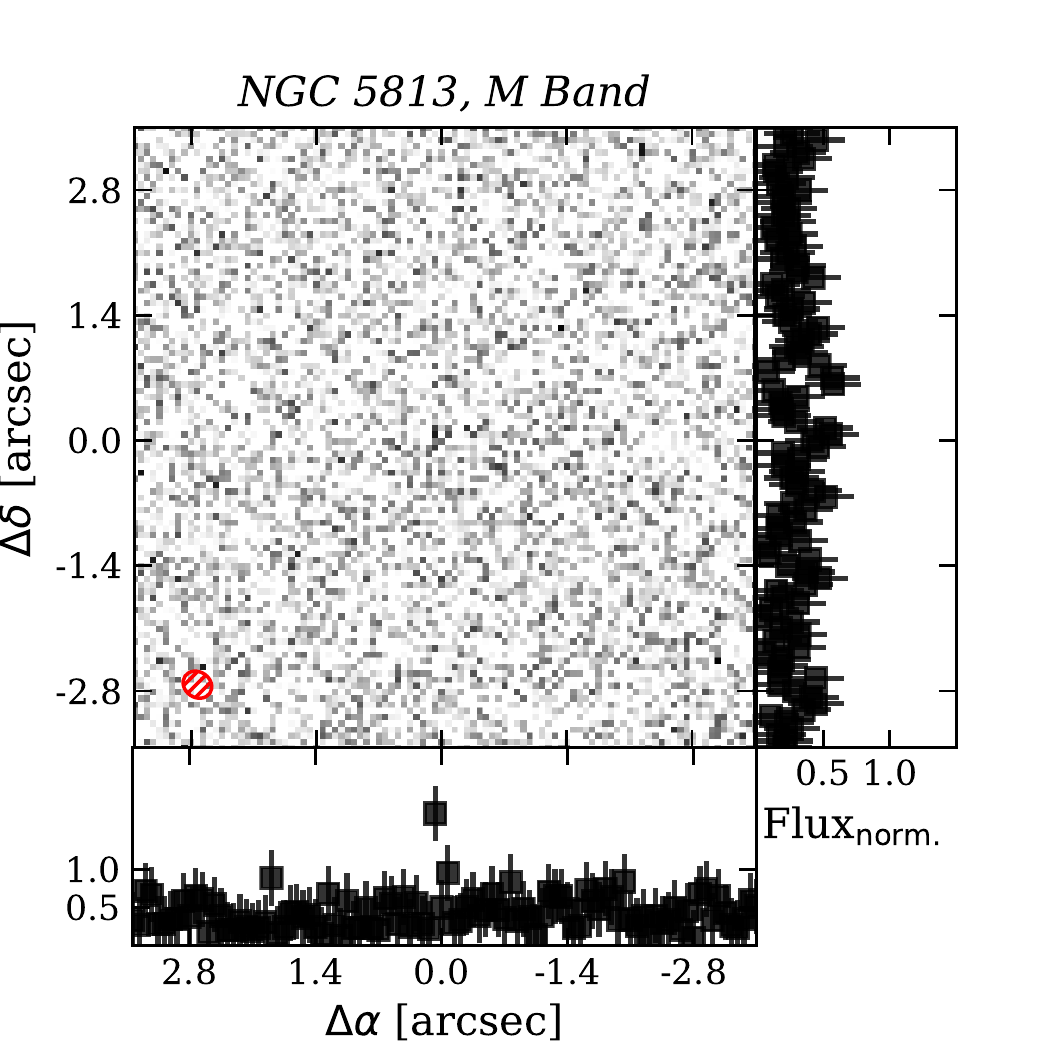}} \\
\subfloat{\includegraphics[width=0.25\hsize]{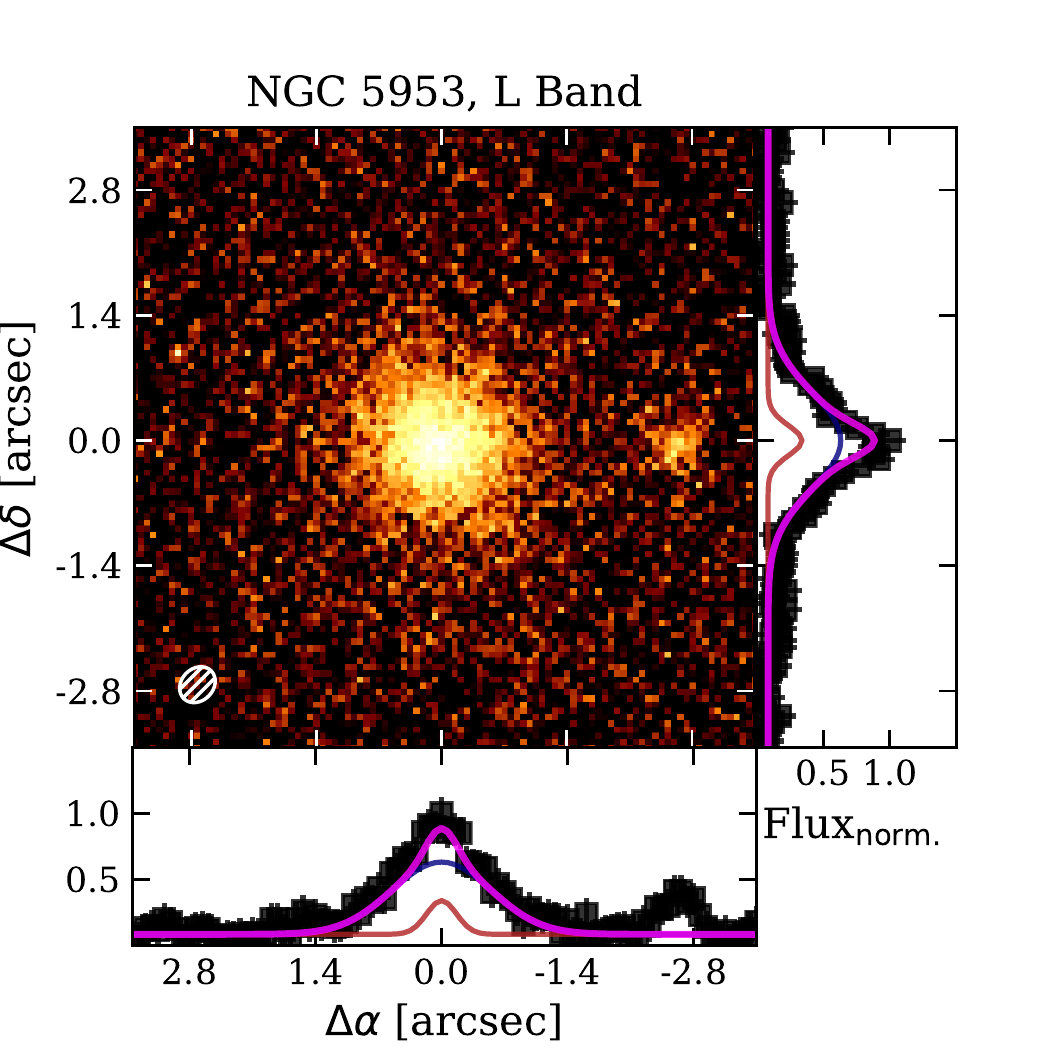}}
\subfloat{\includegraphics[width=0.25\hsize]{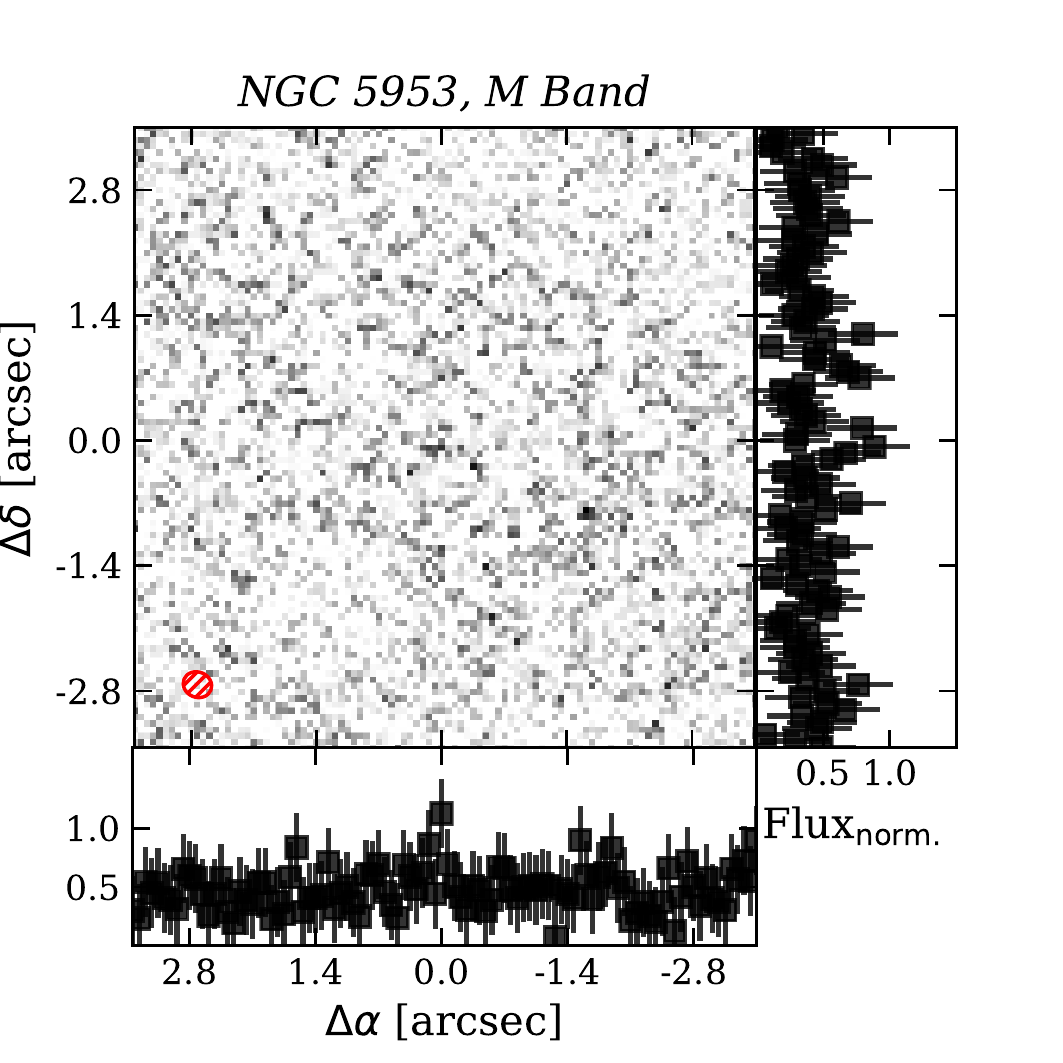}} 
\subfloat{\includegraphics[width=0.25\hsize]{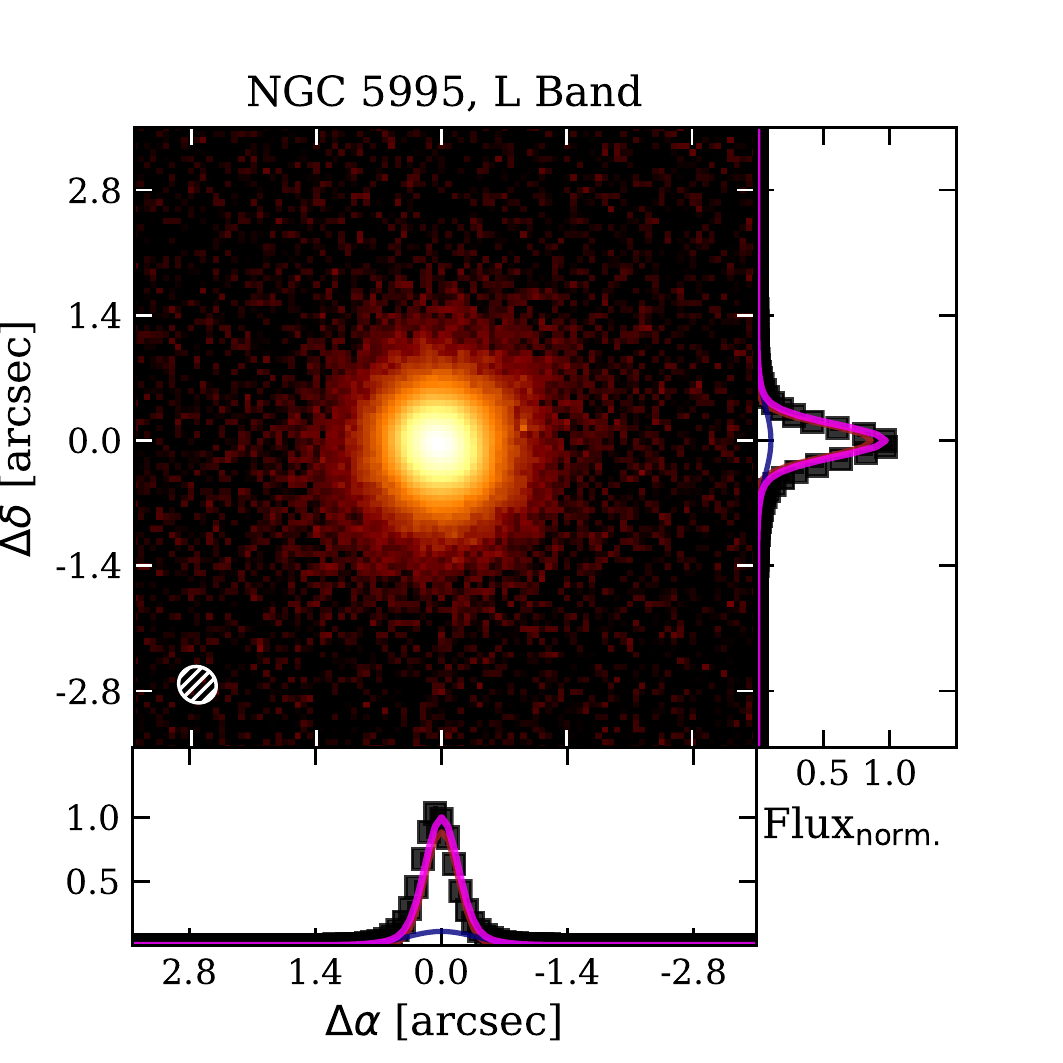}}
\subfloat{\includegraphics[width=0.25\hsize]{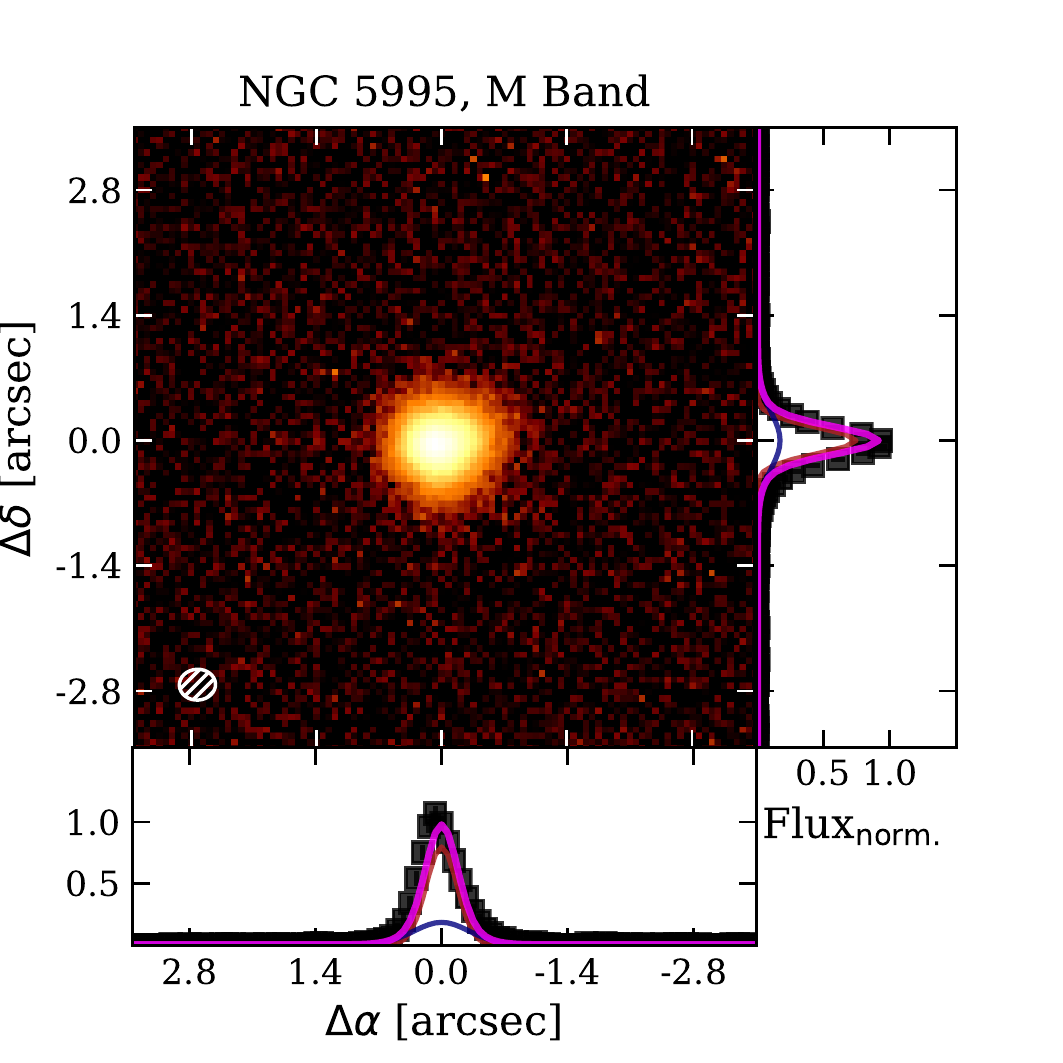}} \\
\subfloat{\includegraphics[width=0.25\hsize]{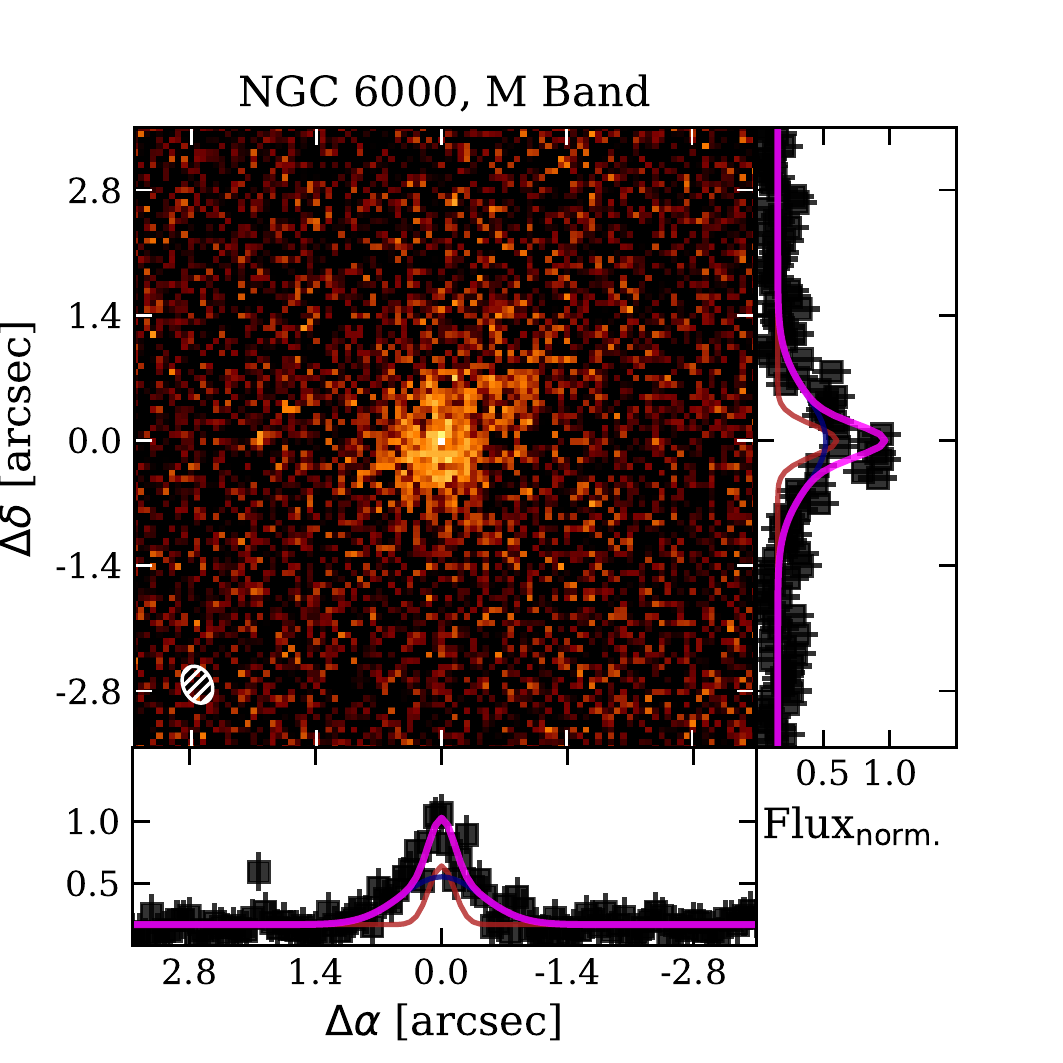}}
\subfloat{\includegraphics[width=0.25\hsize]{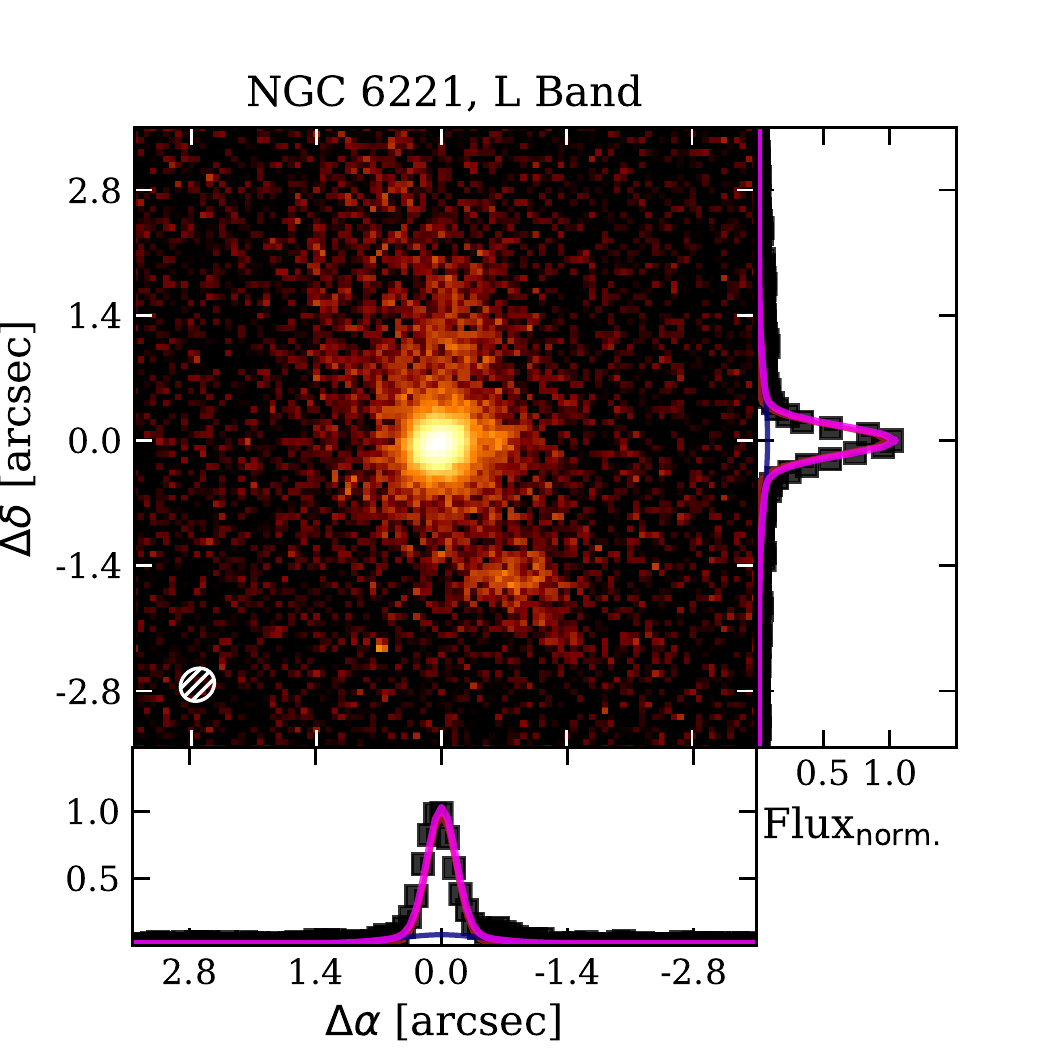}} 
\subfloat{\includegraphics[width=0.25\hsize]{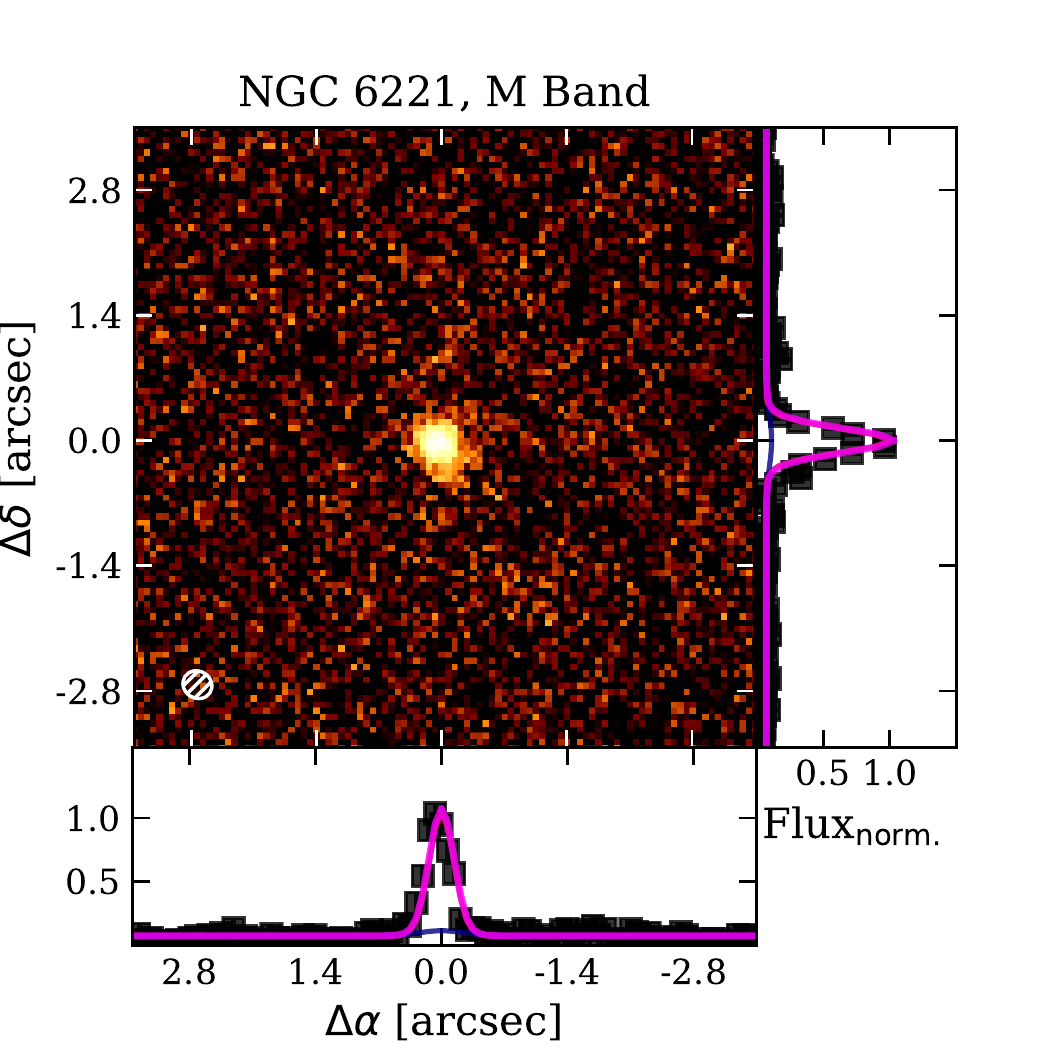}}
\subfloat{\includegraphics[width=0.25\hsize]{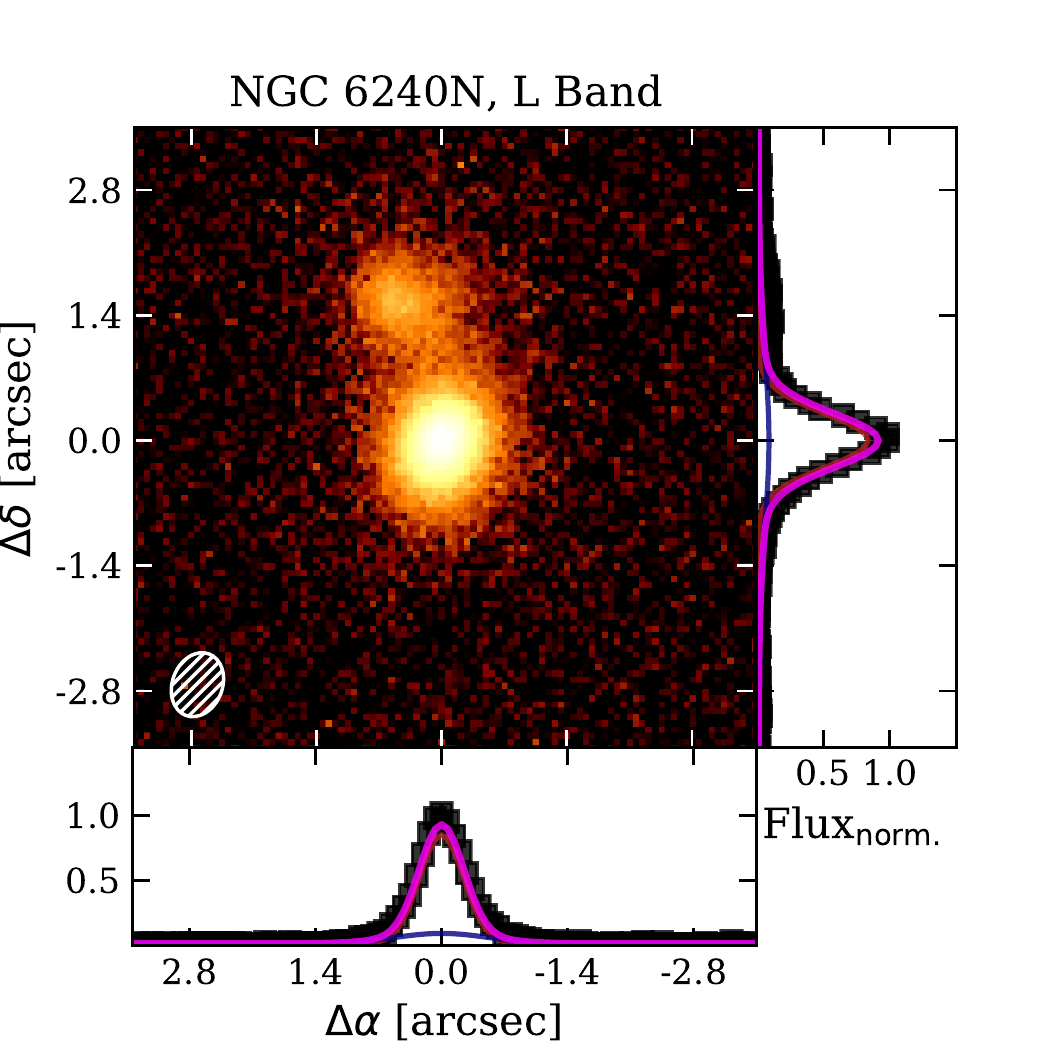}} \\
\subfloat{\includegraphics[width=0.25\hsize]{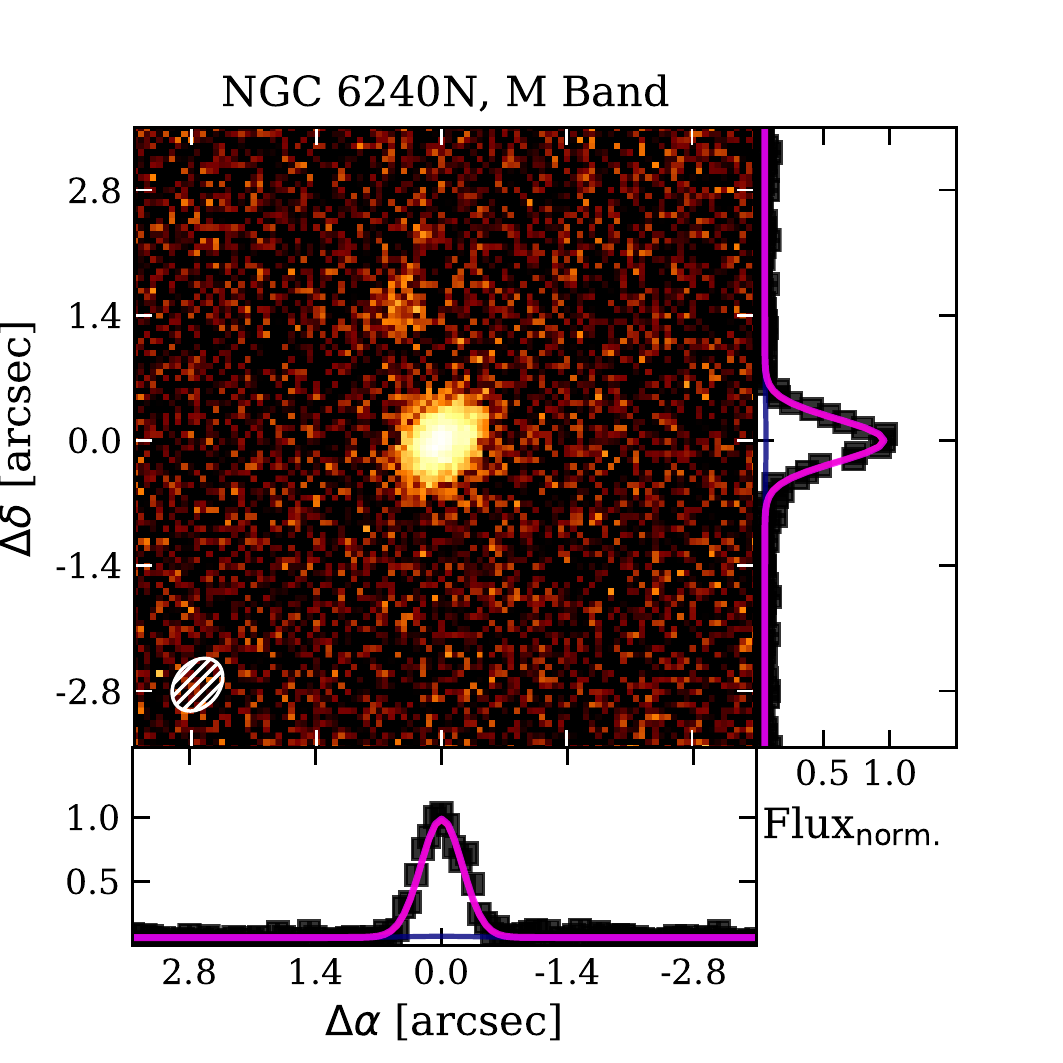}}
\subfloat{\includegraphics[width=0.25\hsize]{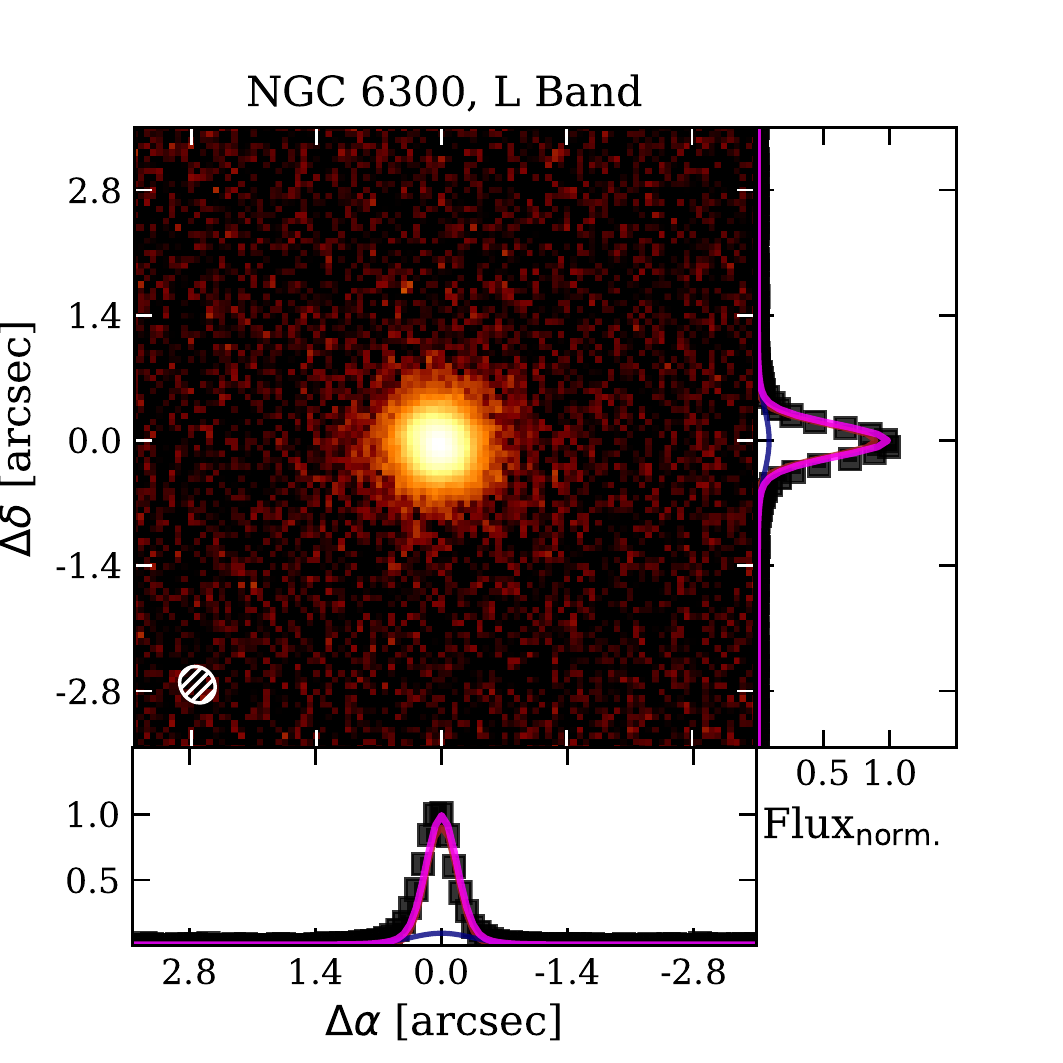}} 
\subfloat{\includegraphics[width=0.25\hsize]{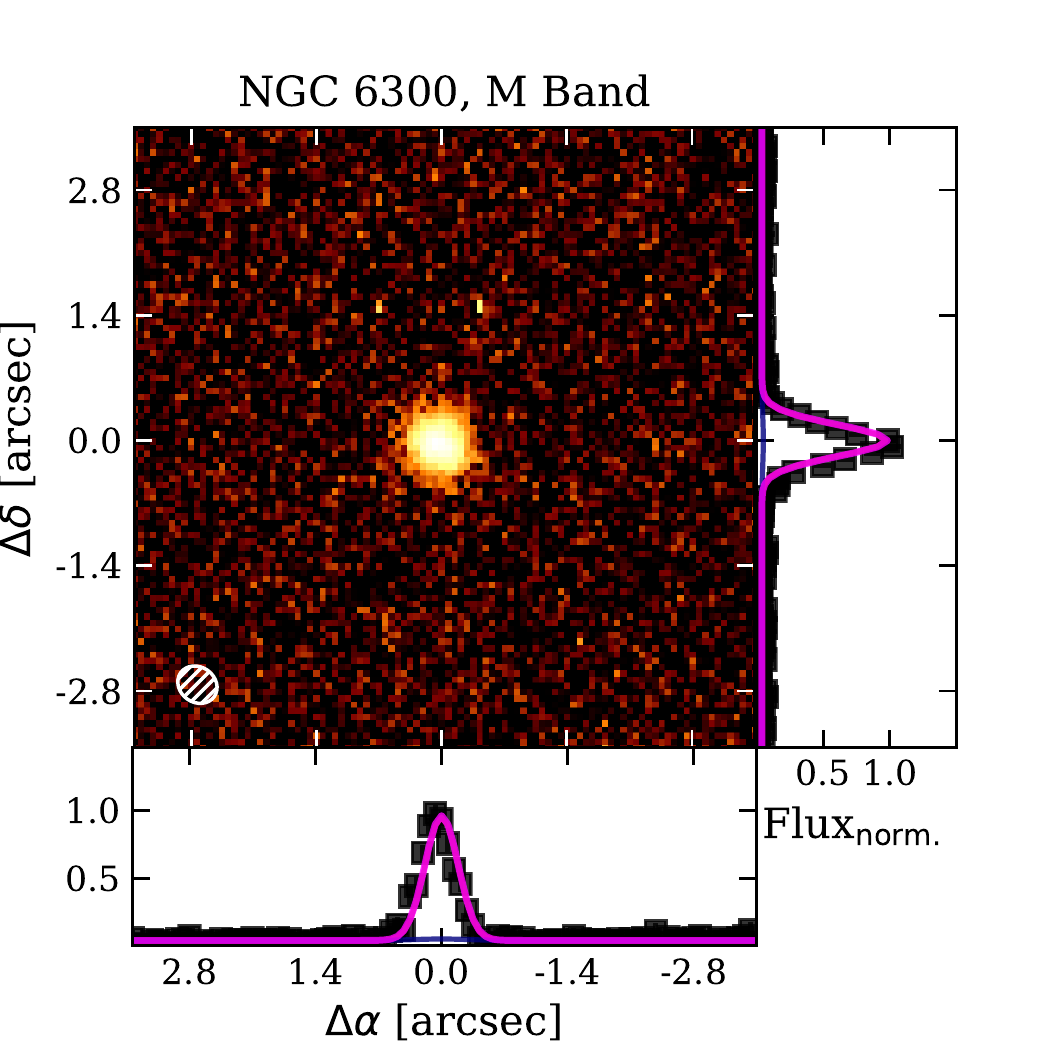}}
\subfloat{\includegraphics[width=0.25\hsize]{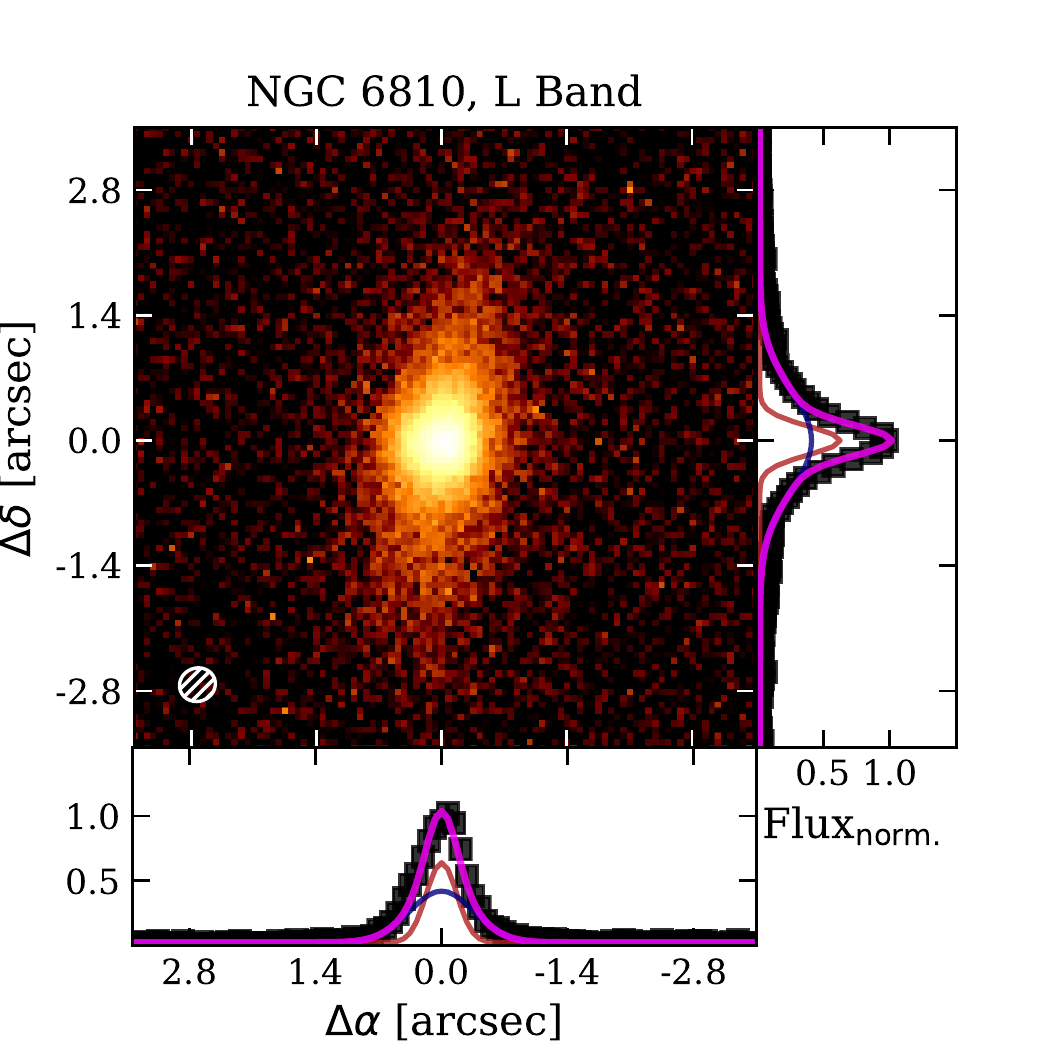}} \\
\caption{ As Fig \ref{fig:cutouts_one} but for all sources.}
\end{figure*}
\begin{figure*}
\subfloat{\includegraphics[width=0.25\hsize]{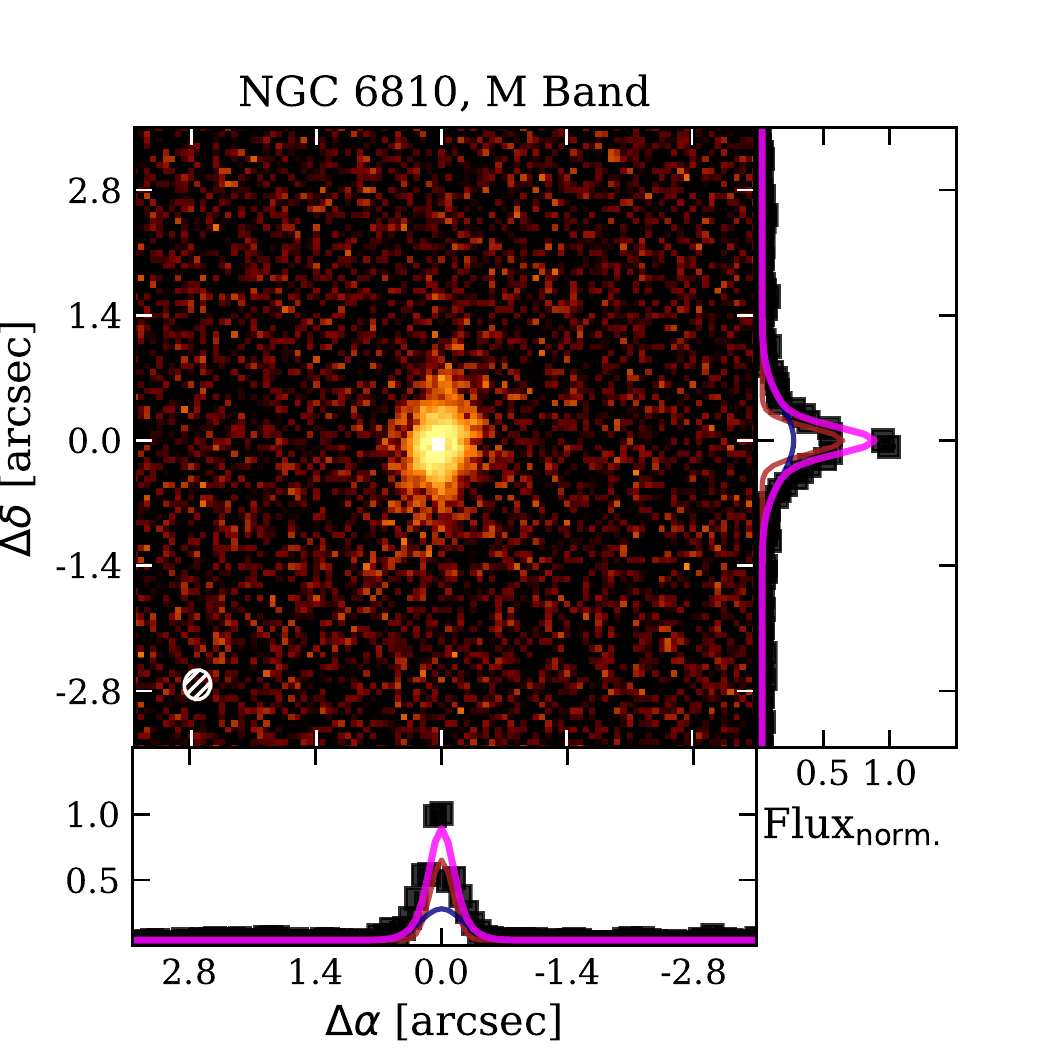}}
\subfloat{\includegraphics[width=0.25\hsize]{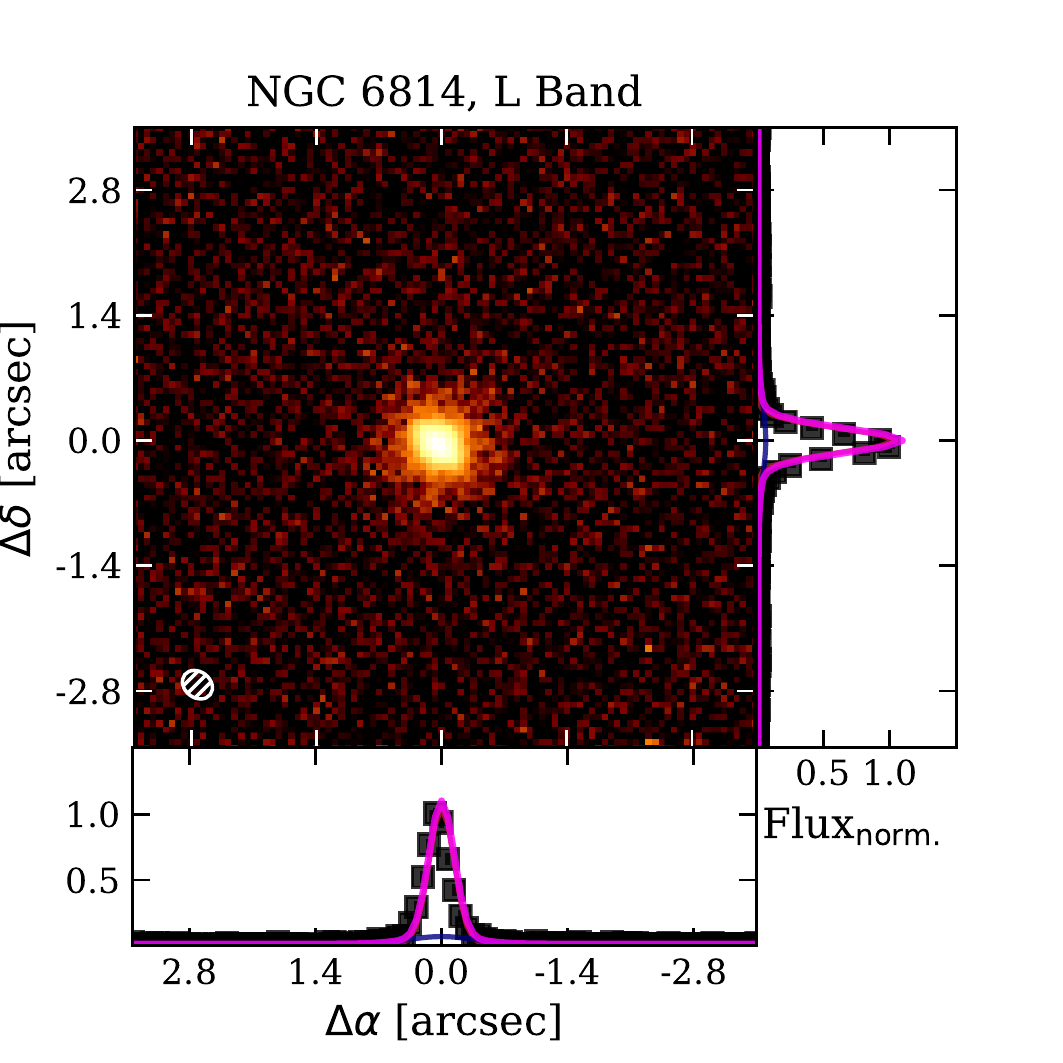}} 
\subfloat{\includegraphics[width=0.25\hsize]{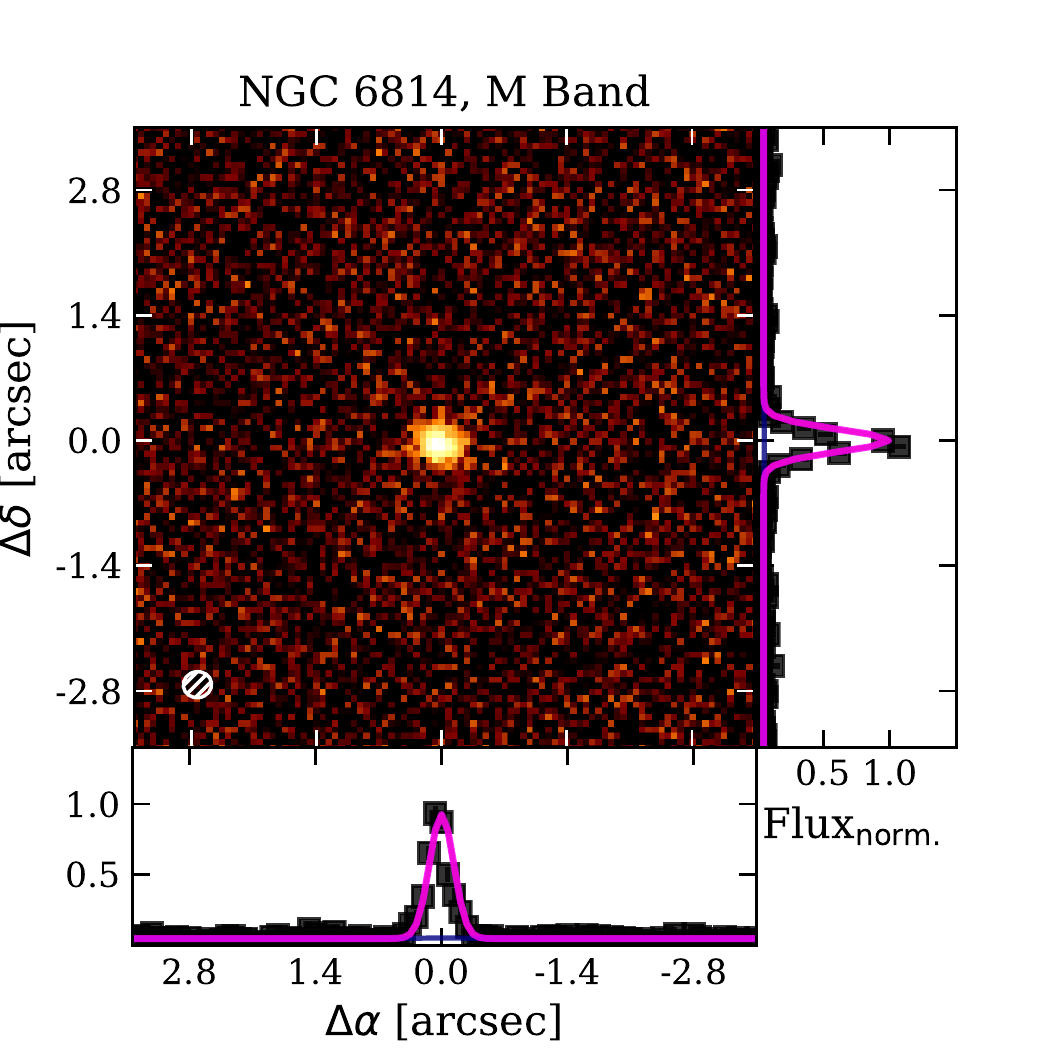}}
\subfloat{\includegraphics[width=0.25\hsize]{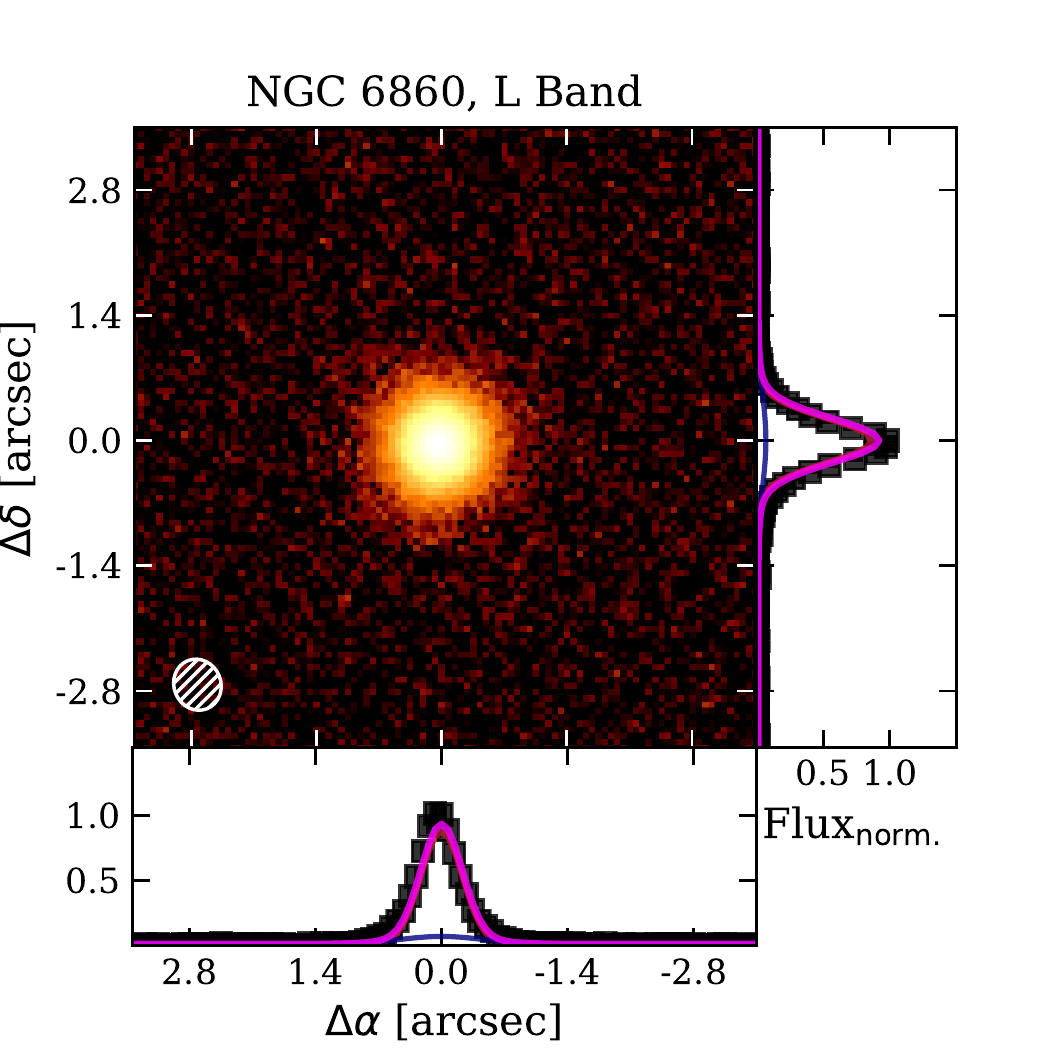}} \\
\subfloat{\includegraphics[width=0.25\hsize]{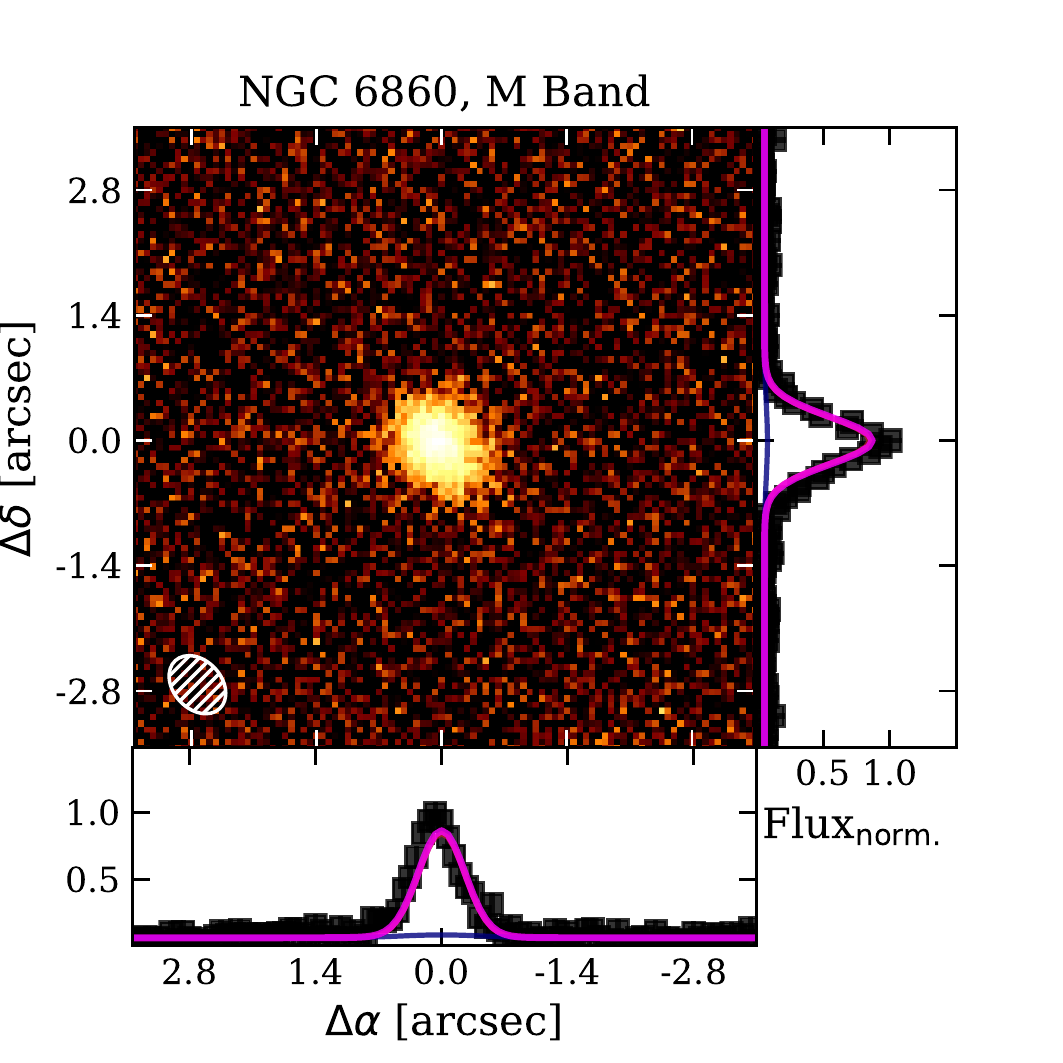}}
\subfloat{\includegraphics[width=0.25\hsize]{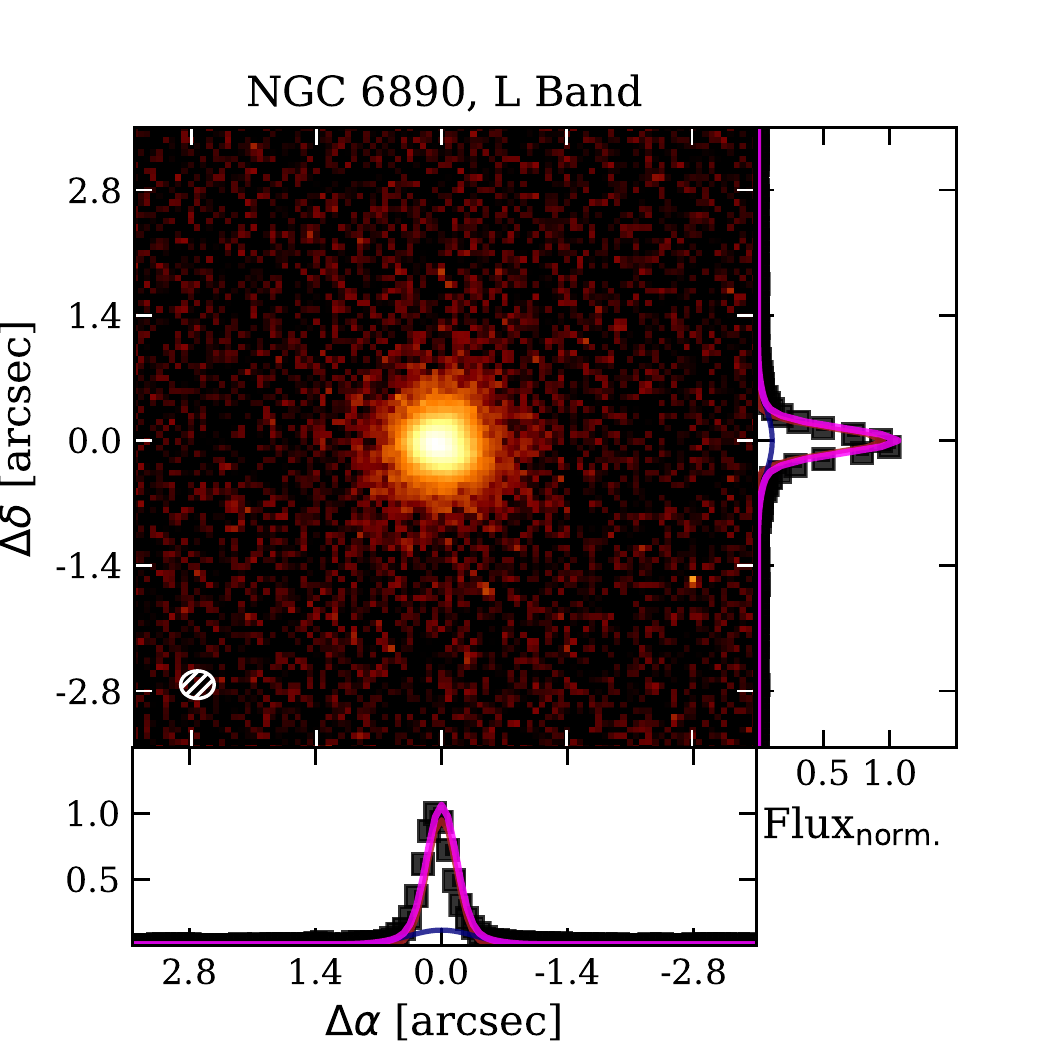}} 
\subfloat{\includegraphics[width=0.25\hsize]{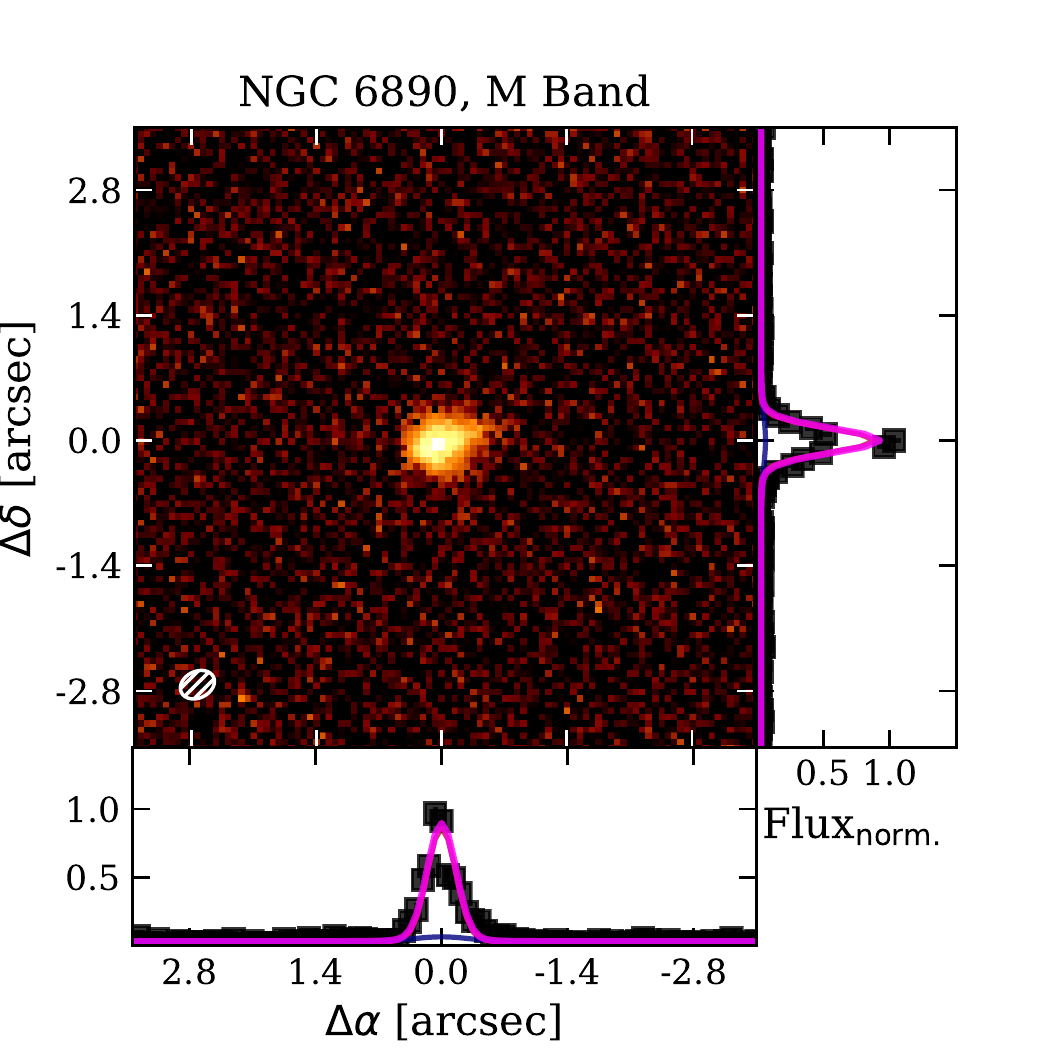}}
\subfloat{\includegraphics[width=0.25\hsize]{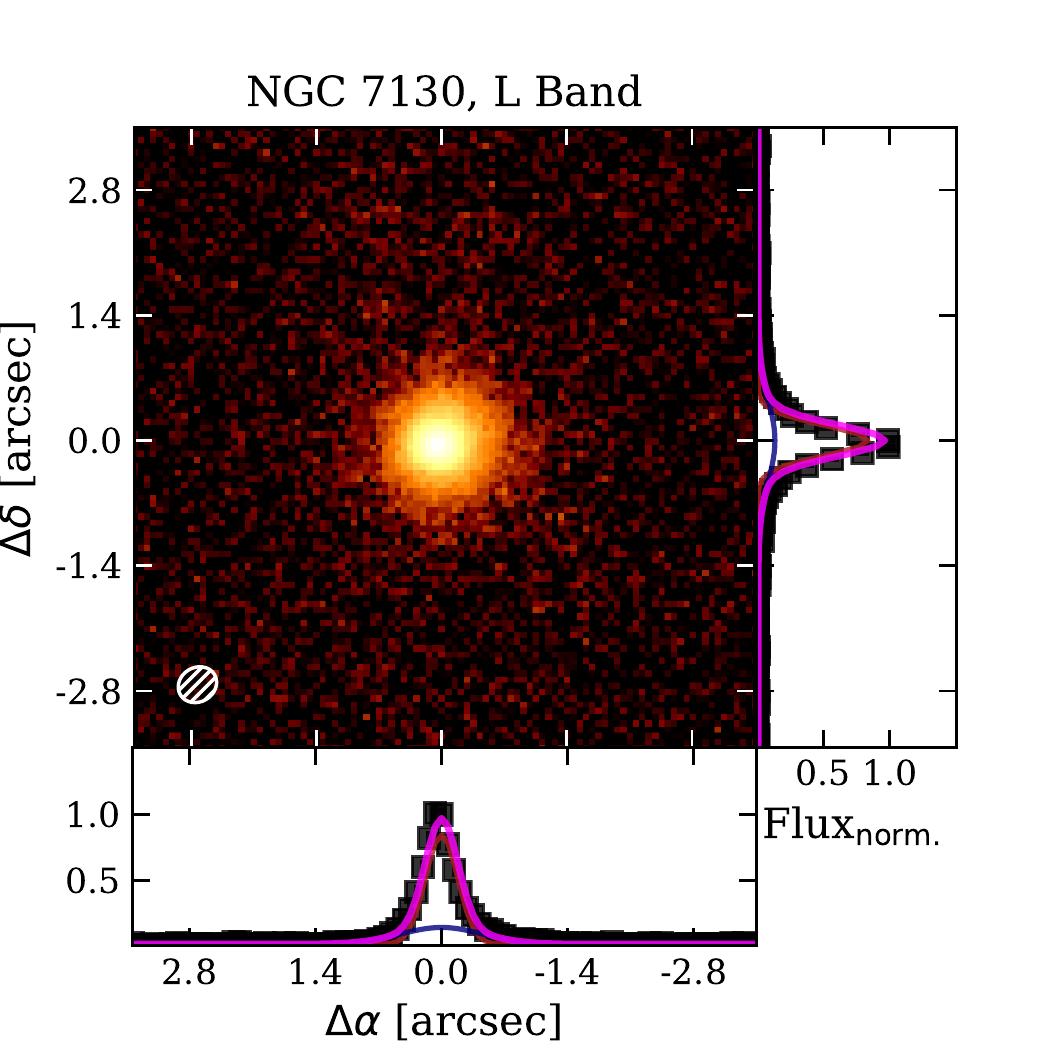}} \\
\subfloat{\includegraphics[width=0.25\hsize]{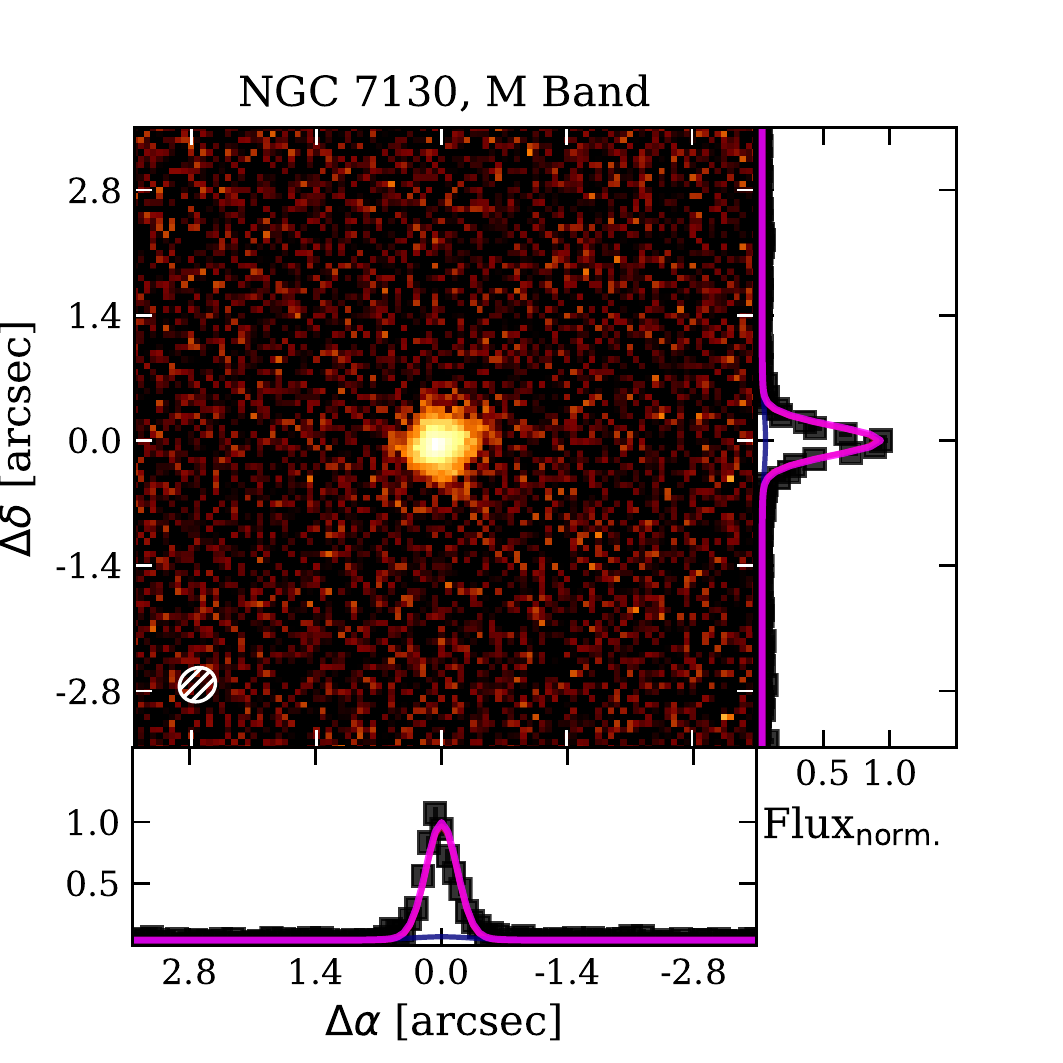}}
\subfloat{\includegraphics[width=0.25\hsize]{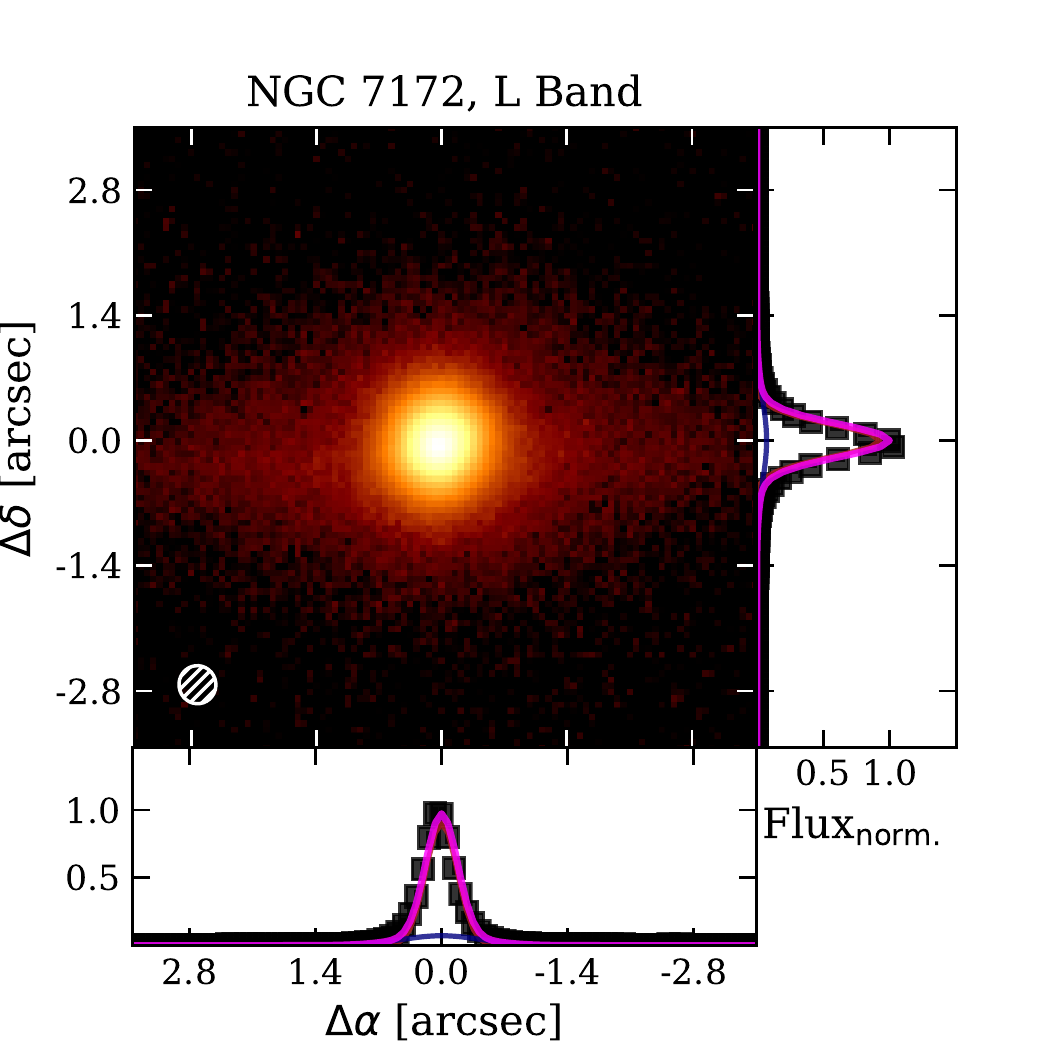}} 
\subfloat{\includegraphics[width=0.25\hsize]{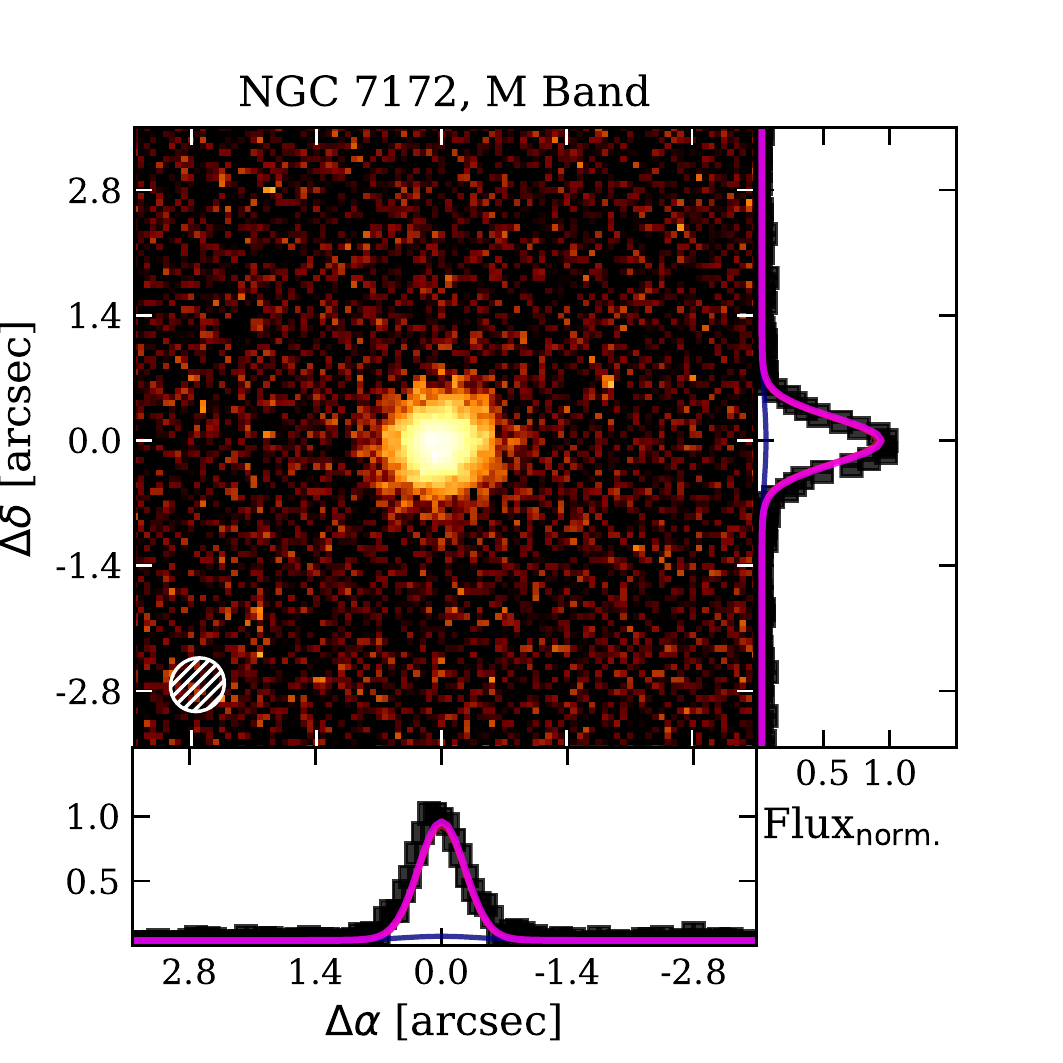}}
\subfloat{\includegraphics[width=0.25\hsize]{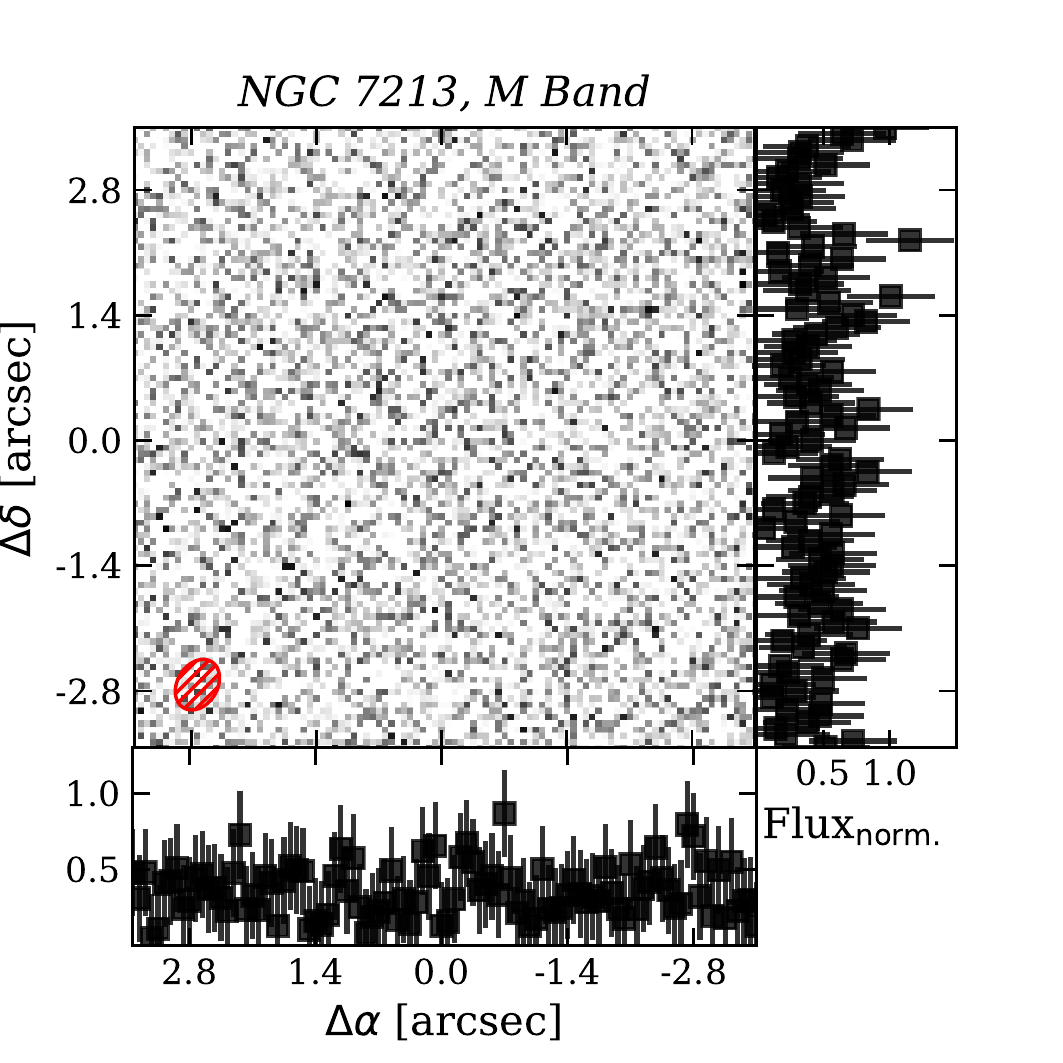}} \\
\subfloat{\includegraphics[width=0.25\hsize]{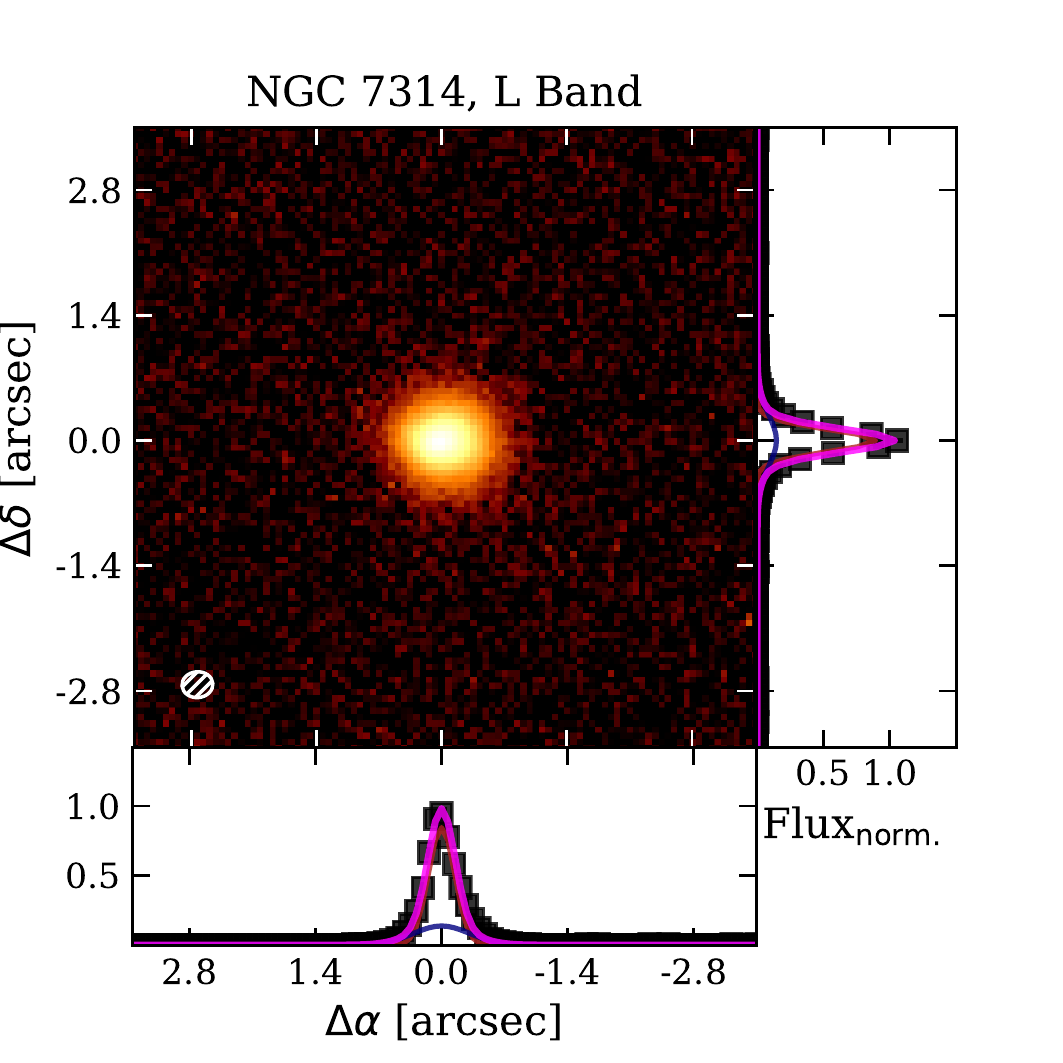}}
\subfloat{\includegraphics[width=0.25\hsize]{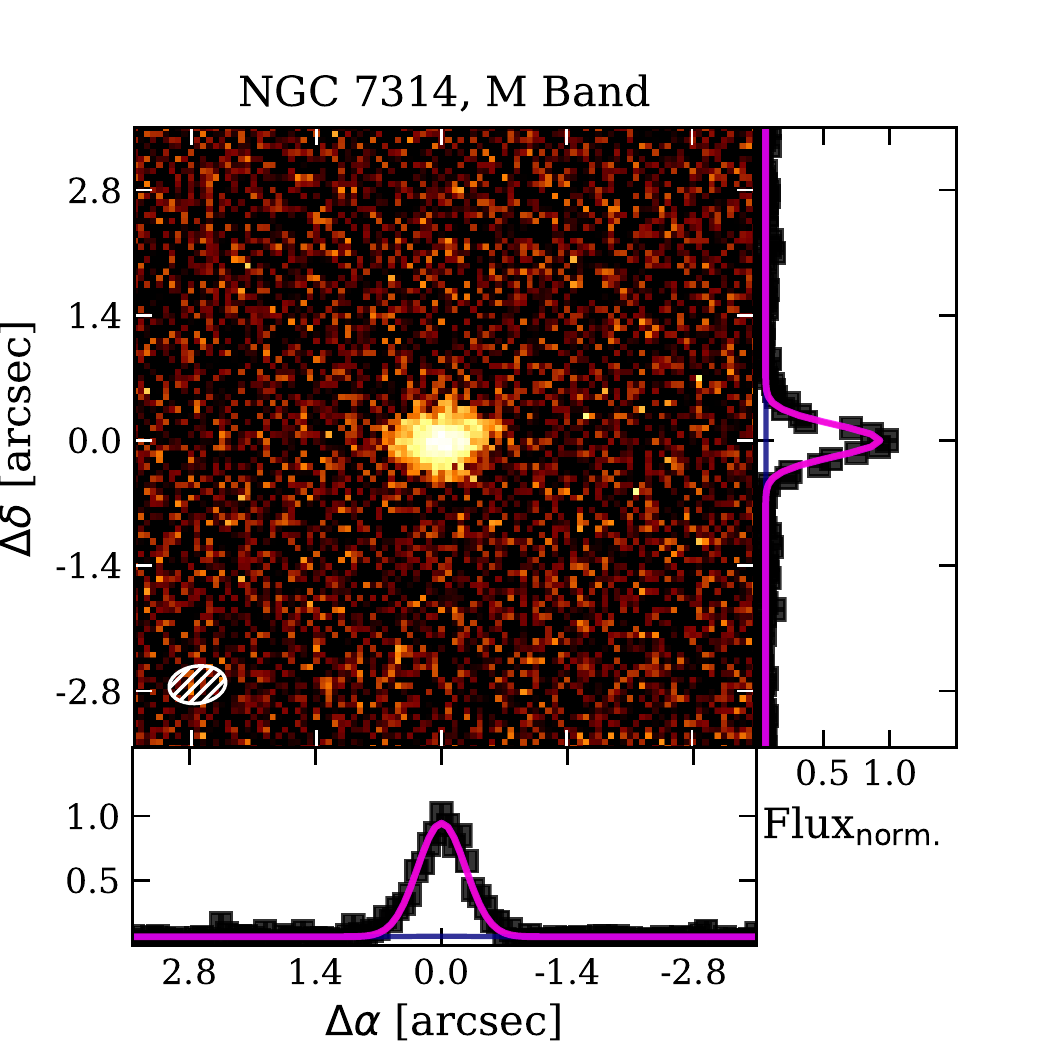}} 
\subfloat{\includegraphics[width=0.25\hsize]{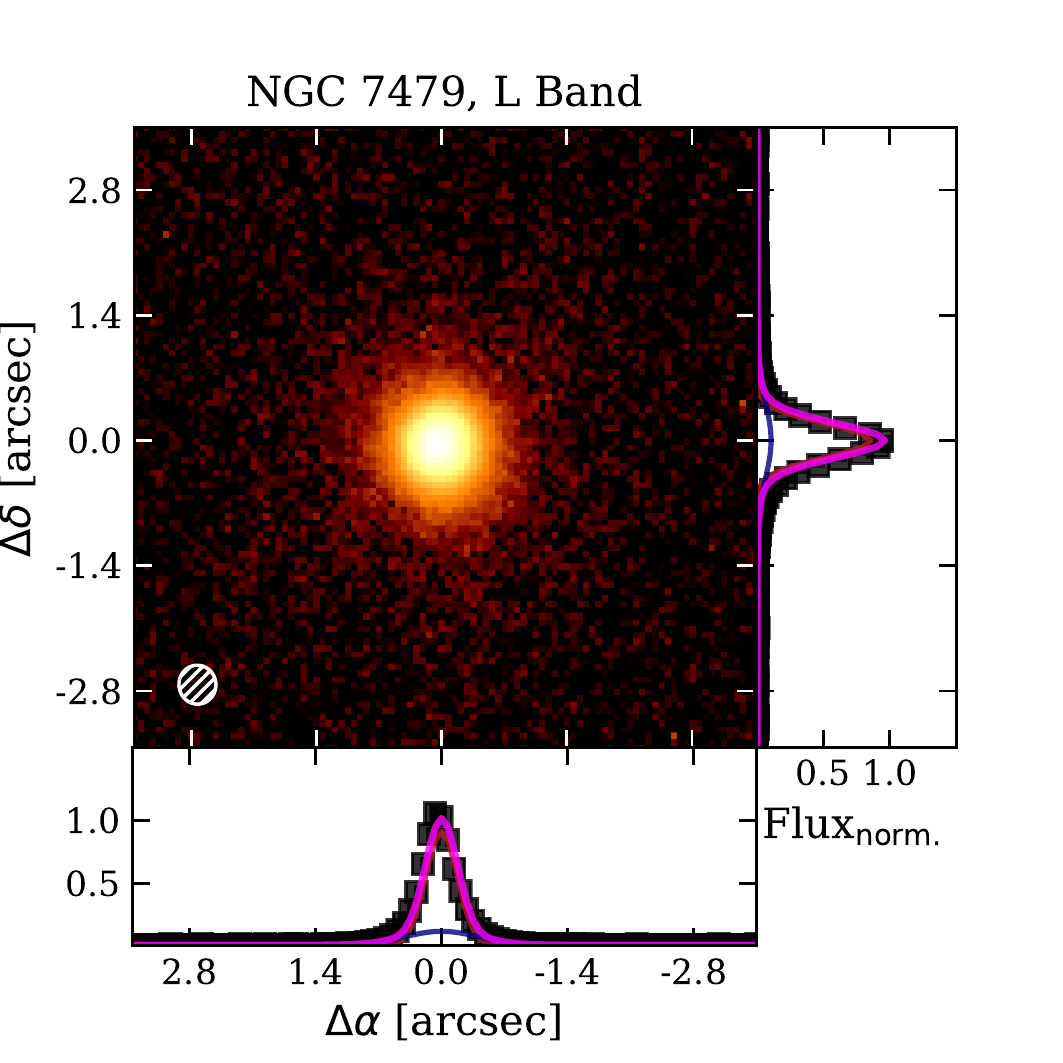}}
\subfloat{\includegraphics[width=0.25\hsize]{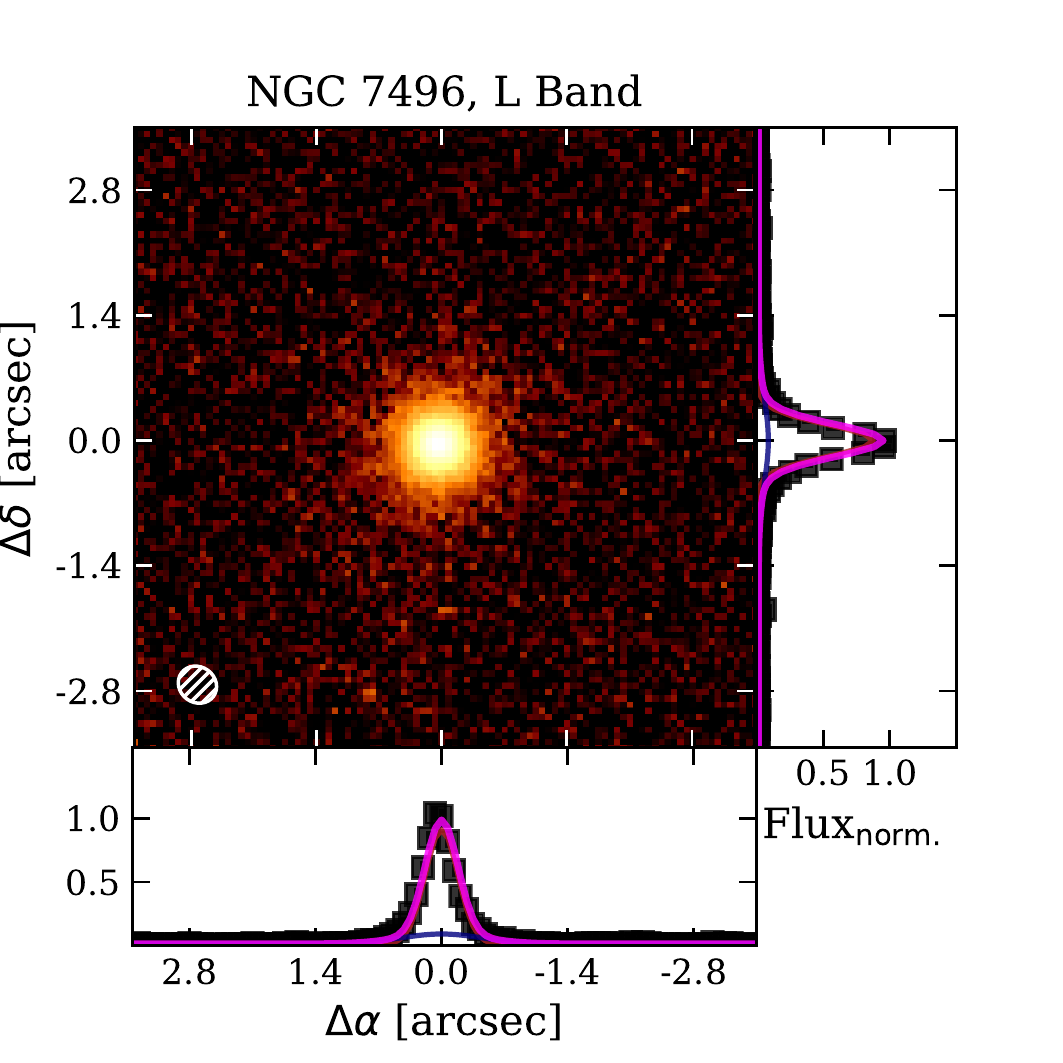}} \\
\subfloat{\includegraphics[width=0.25\hsize]{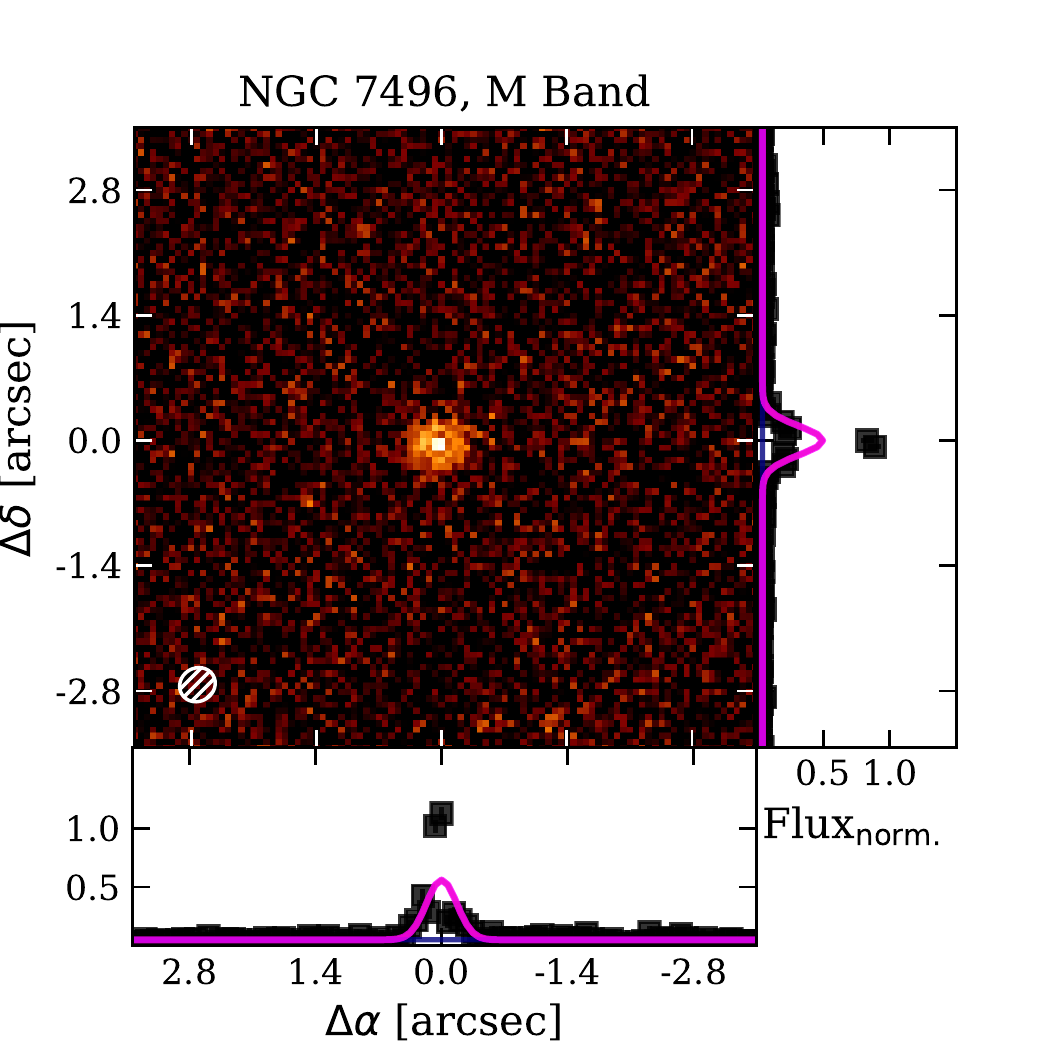}}
\subfloat{\includegraphics[width=0.25\hsize]{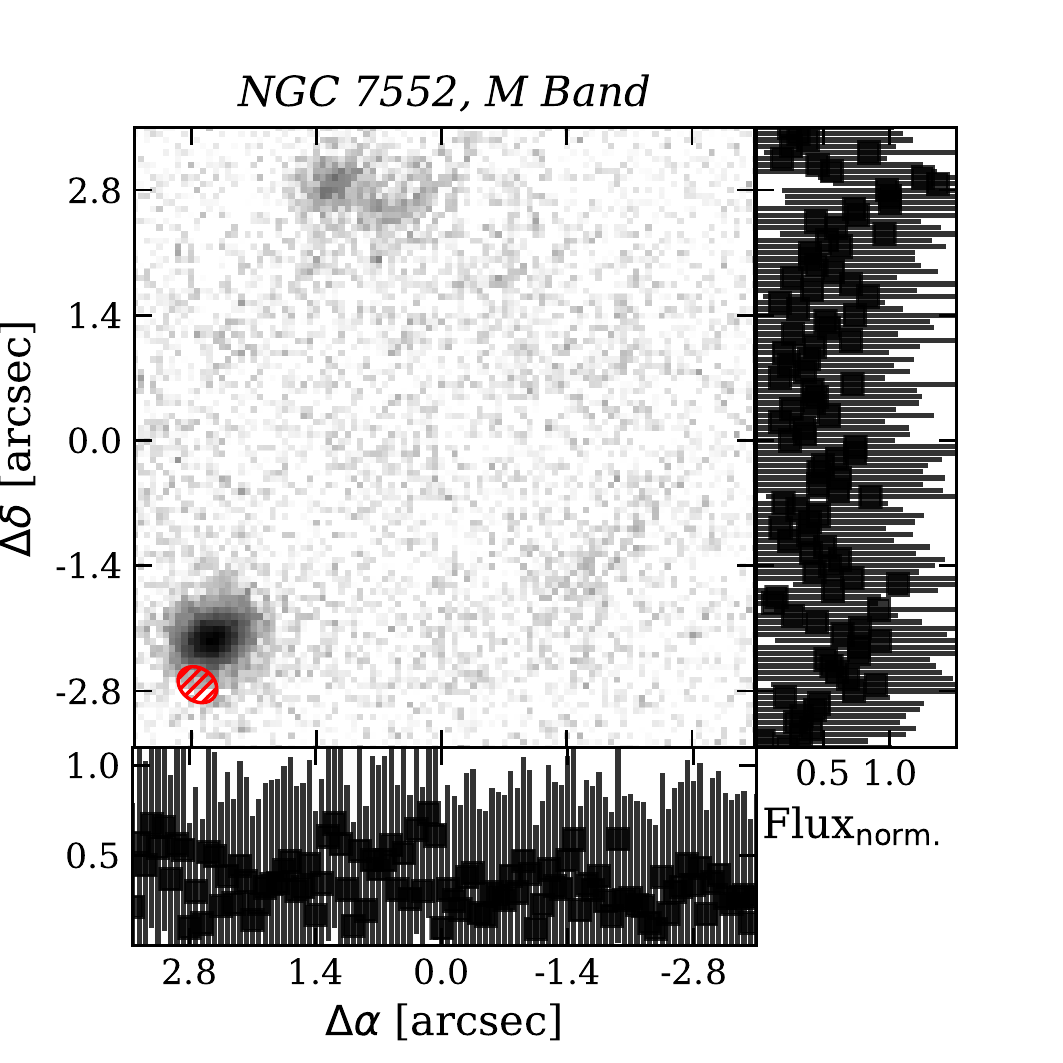}} 
\subfloat{\includegraphics[width=0.25\hsize]{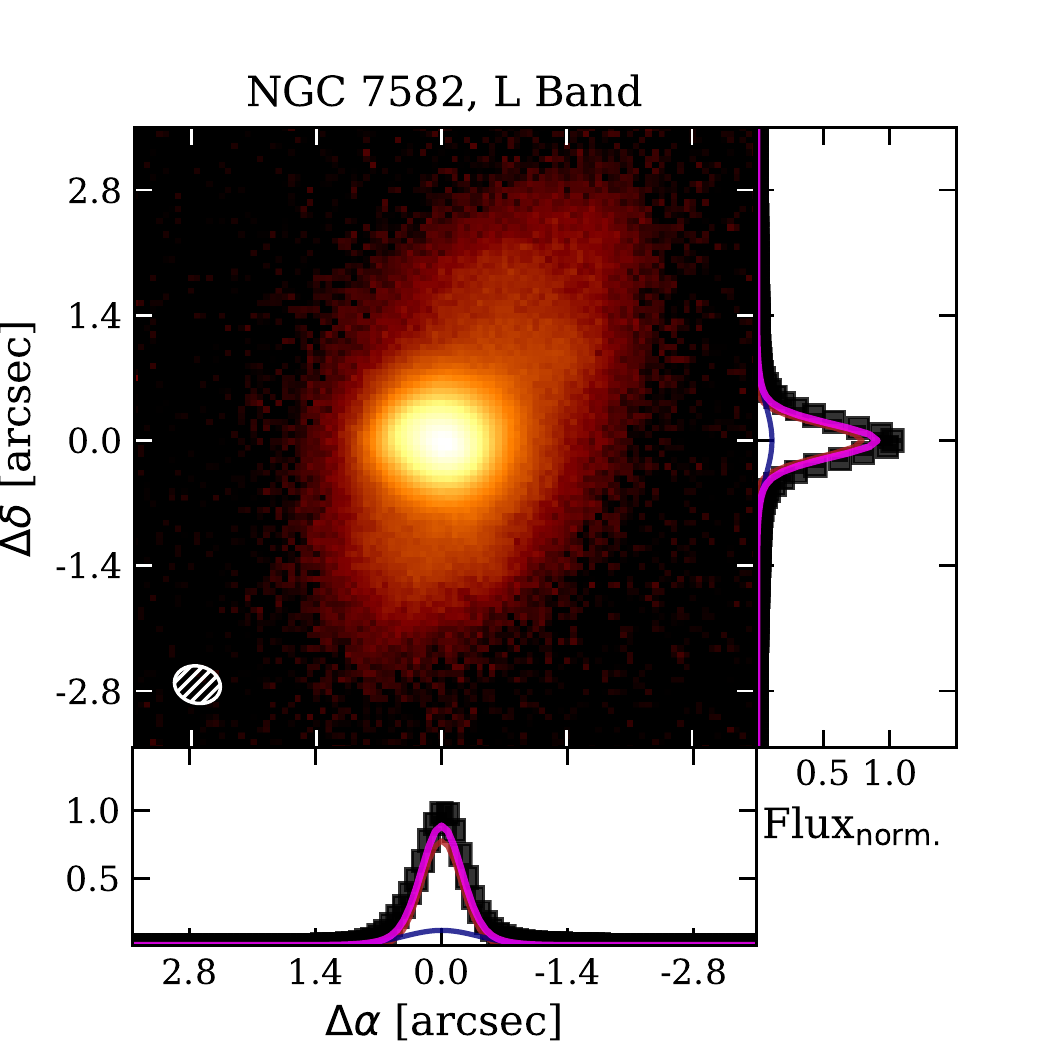}}
\subfloat{\includegraphics[width=0.25\hsize]{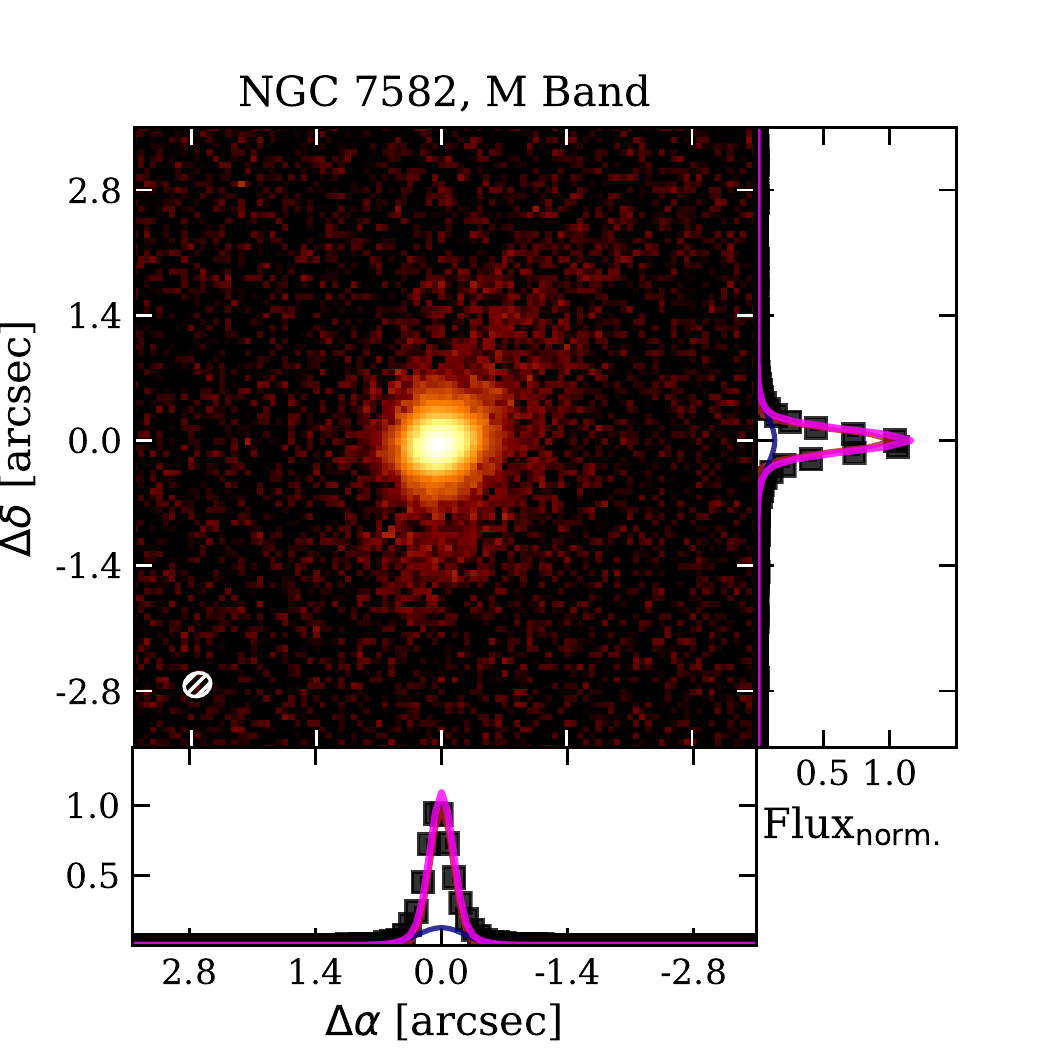}} \\
\caption{ As Fig \ref{fig:cutouts_one} but for all sources.}
\end{figure*}
\begin{figure*}
\subfloat{\includegraphics[width=0.25\hsize]{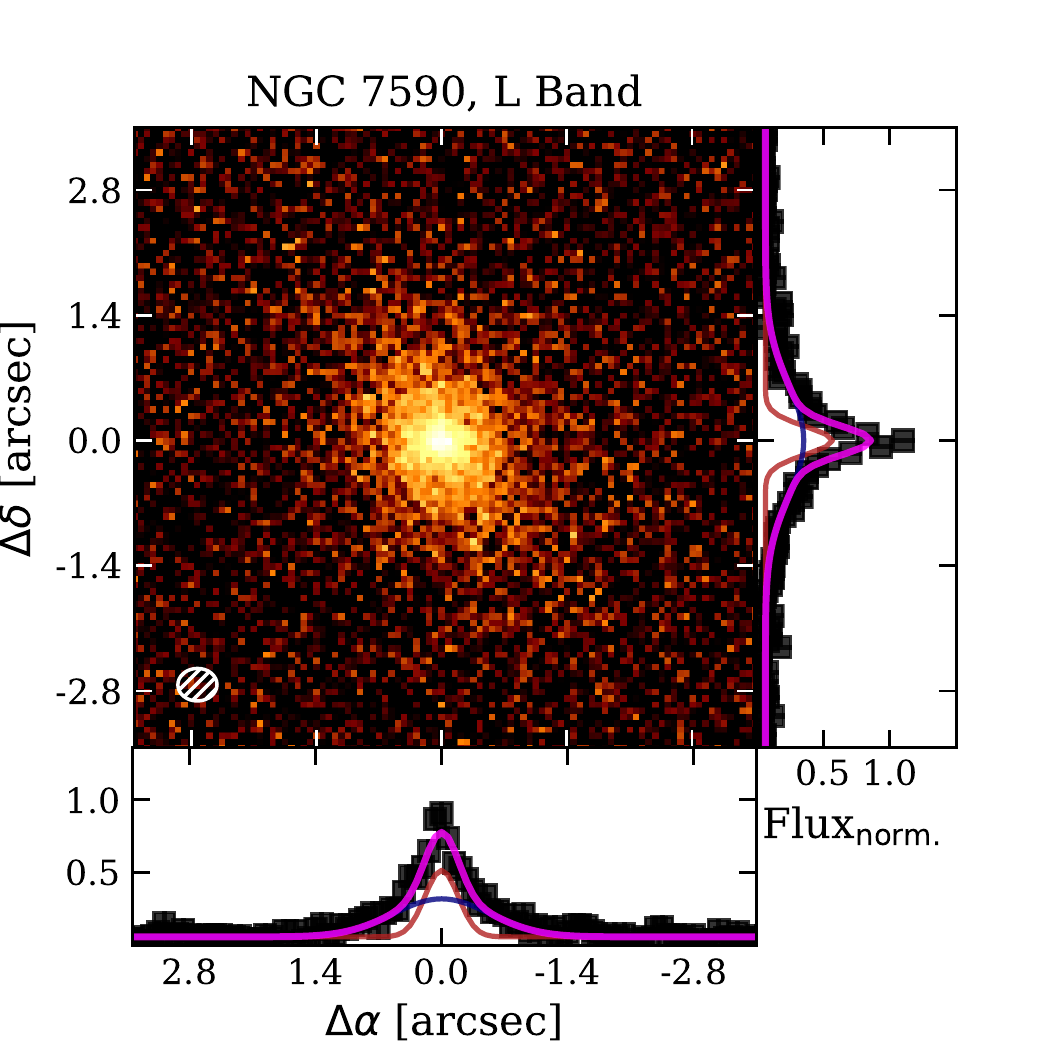}}
\subfloat{\includegraphics[width=0.25\hsize]{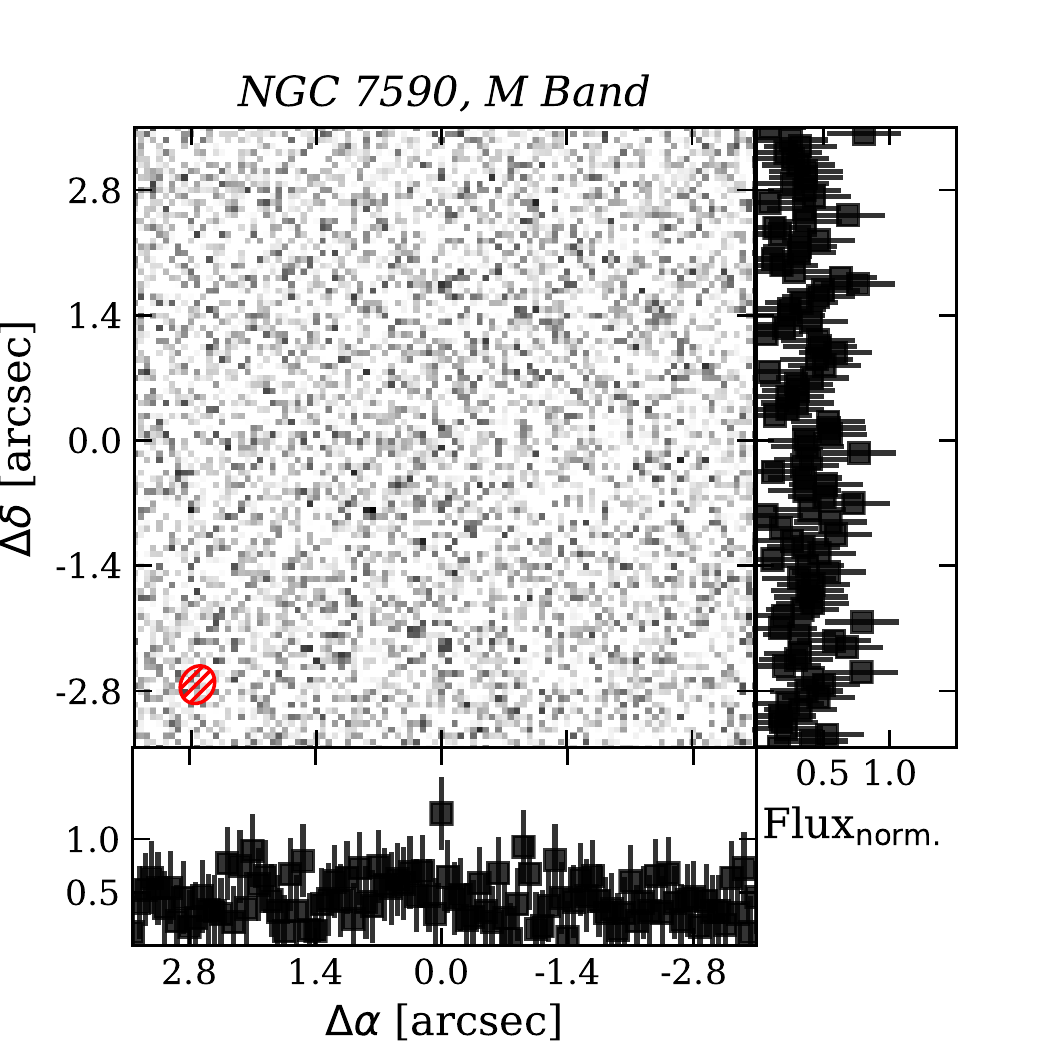}} 
\subfloat{\includegraphics[width=0.25\hsize]{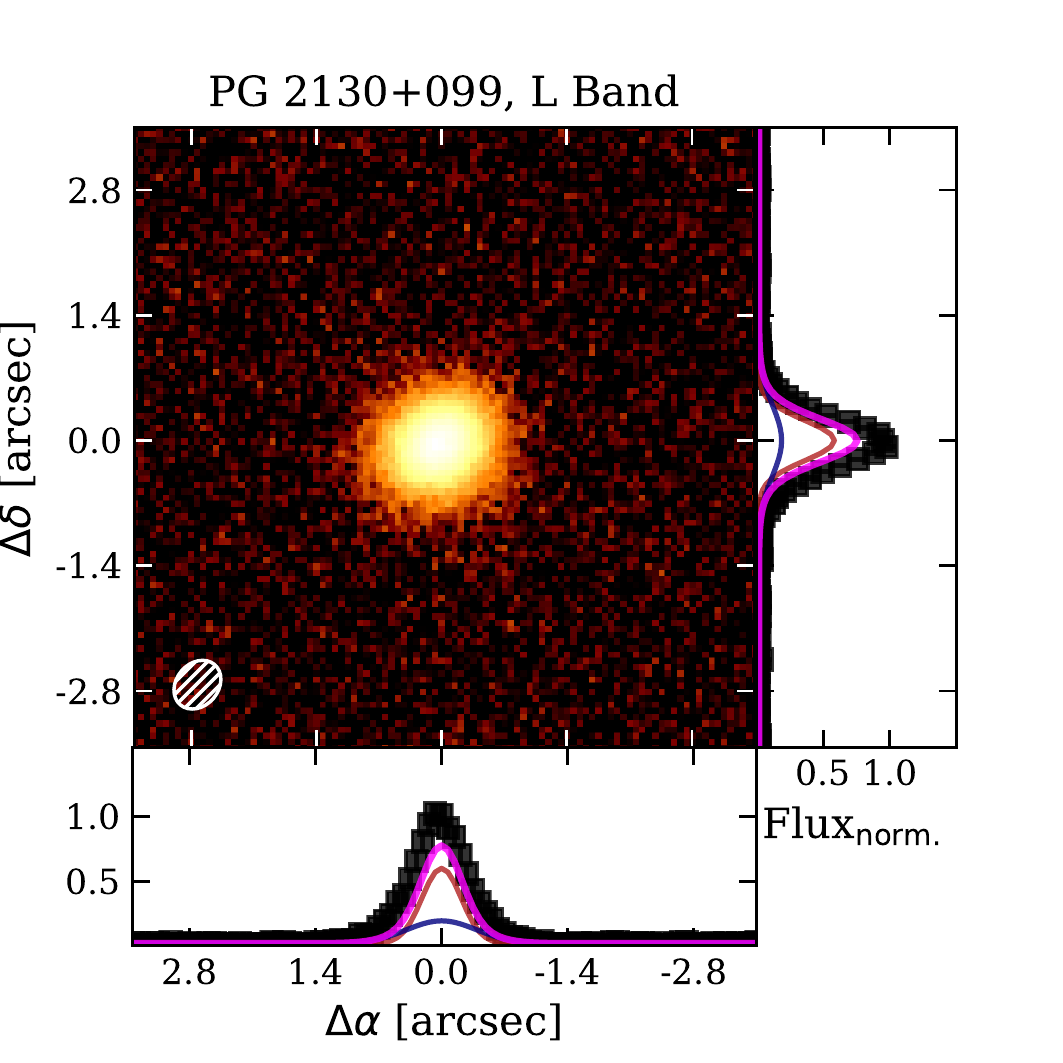}}
\subfloat{\includegraphics[width=0.25\hsize]{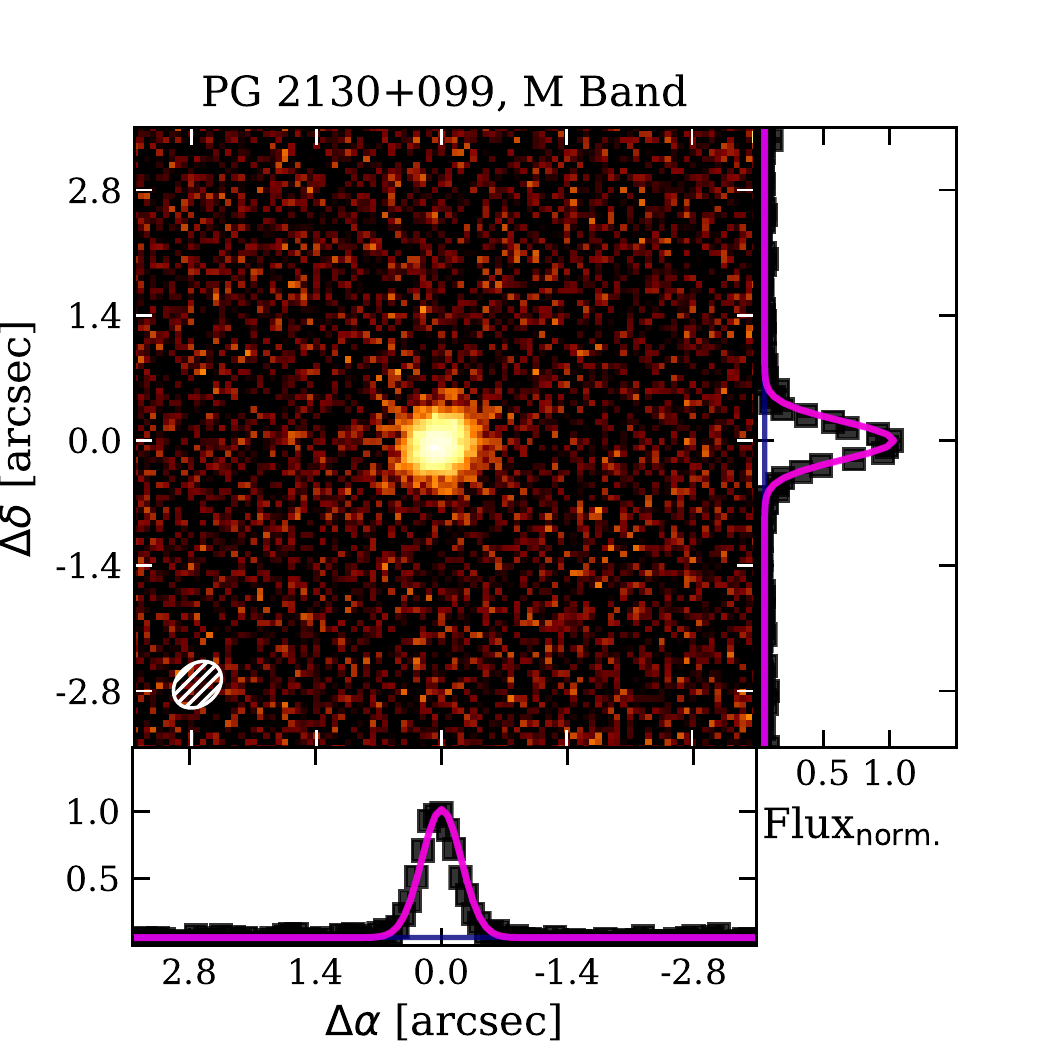}}\\
\subfloat{\includegraphics[width=0.25\hsize]{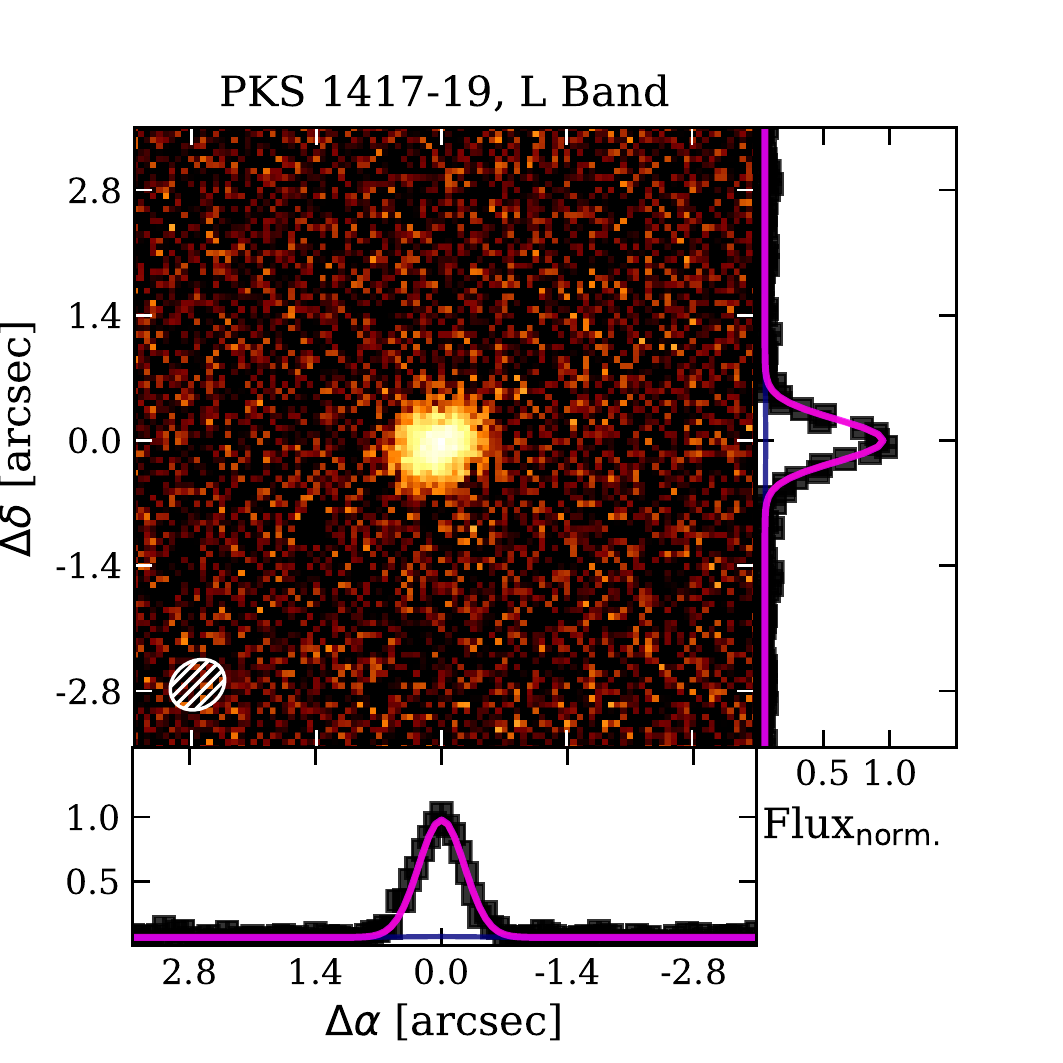}}
\subfloat{\includegraphics[width=0.25\hsize]{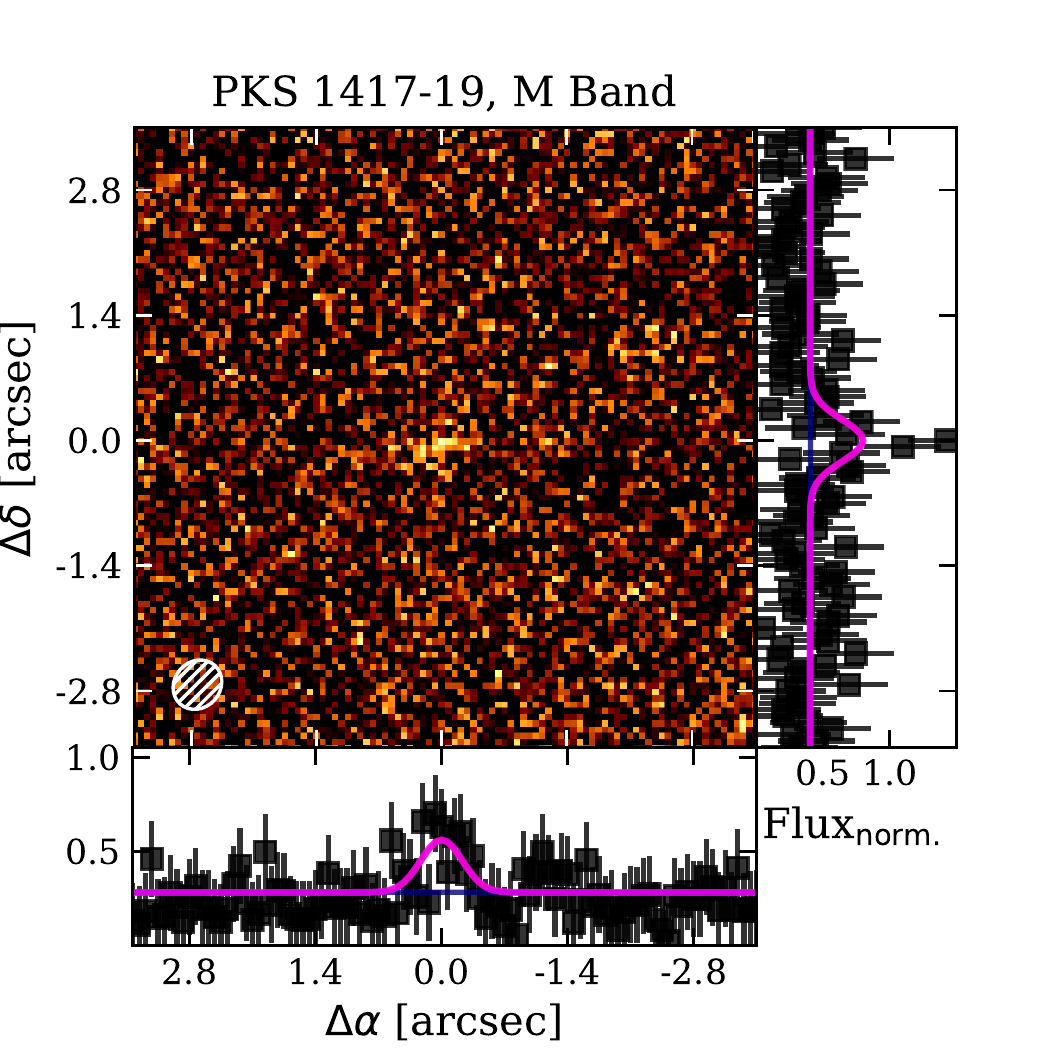}} 
\subfloat{\includegraphics[width=0.25\hsize]{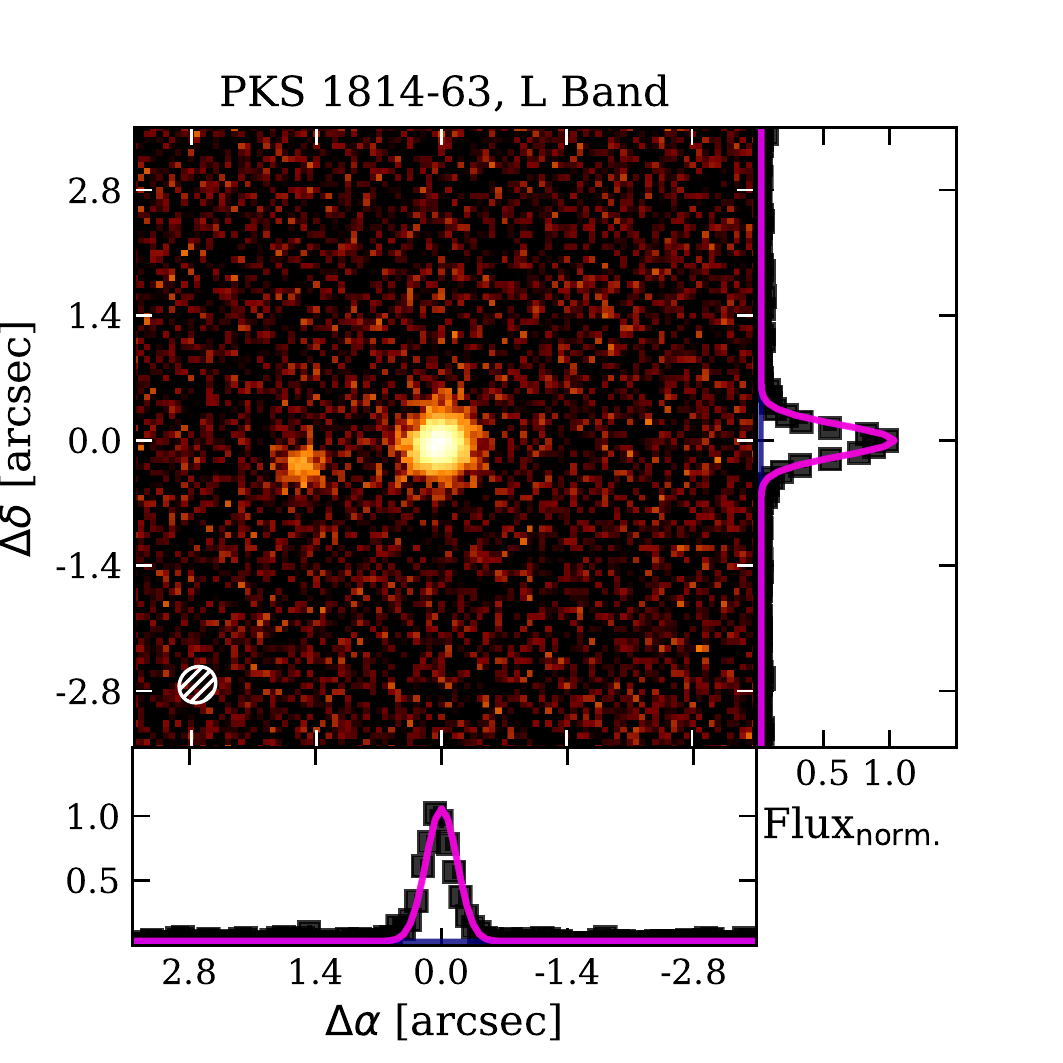}}
\subfloat{\includegraphics[width=0.25\hsize]{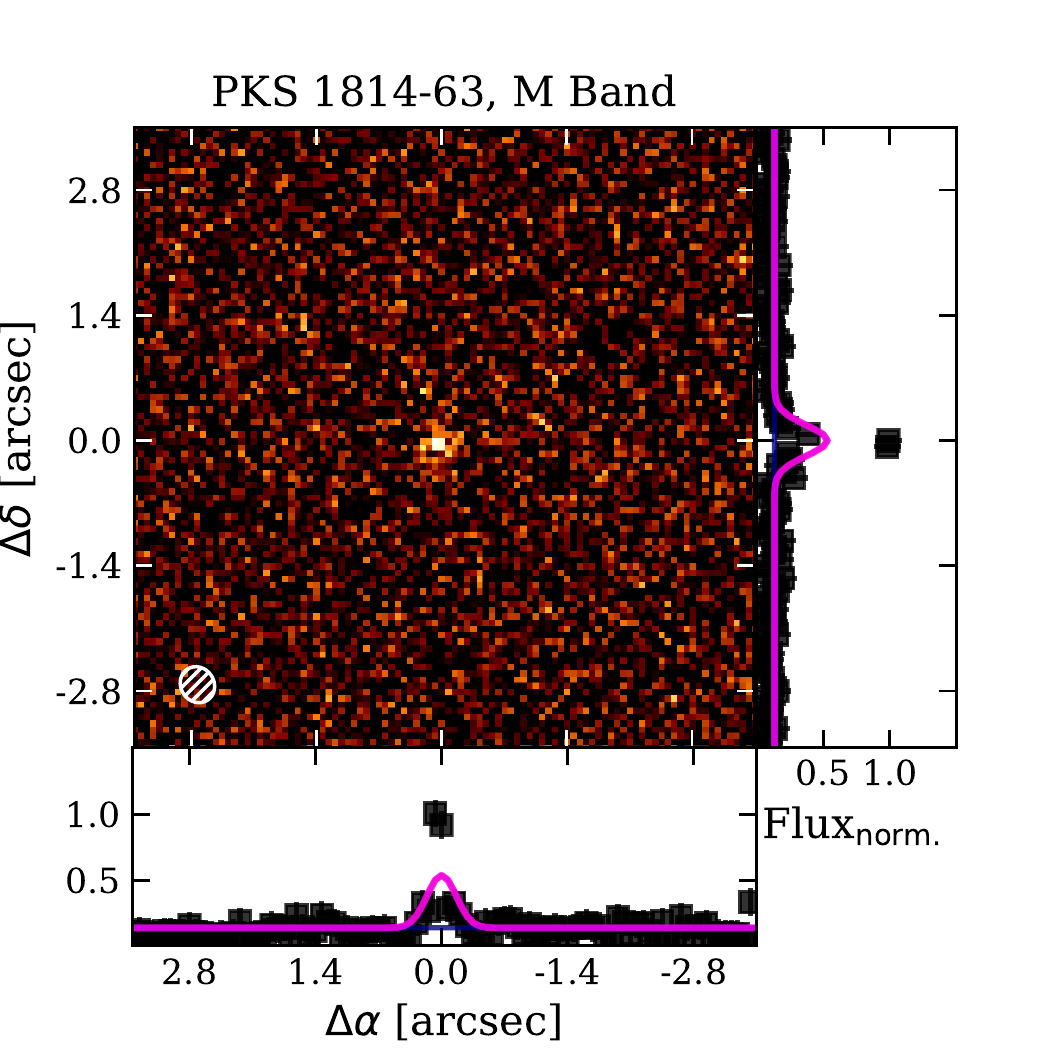}}\\
\subfloat{\includegraphics[width=0.25\hsize]{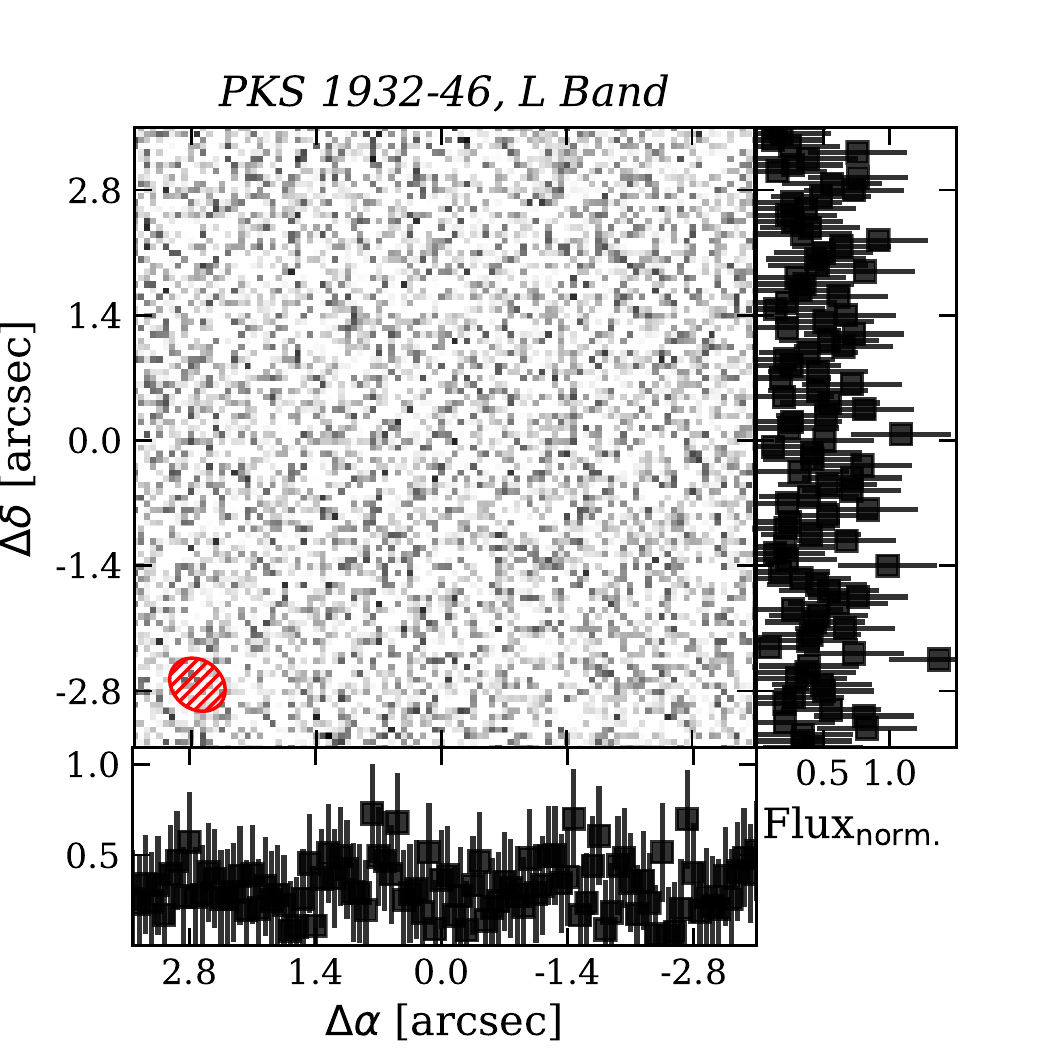}}
\subfloat{\includegraphics[width=0.25\hsize]{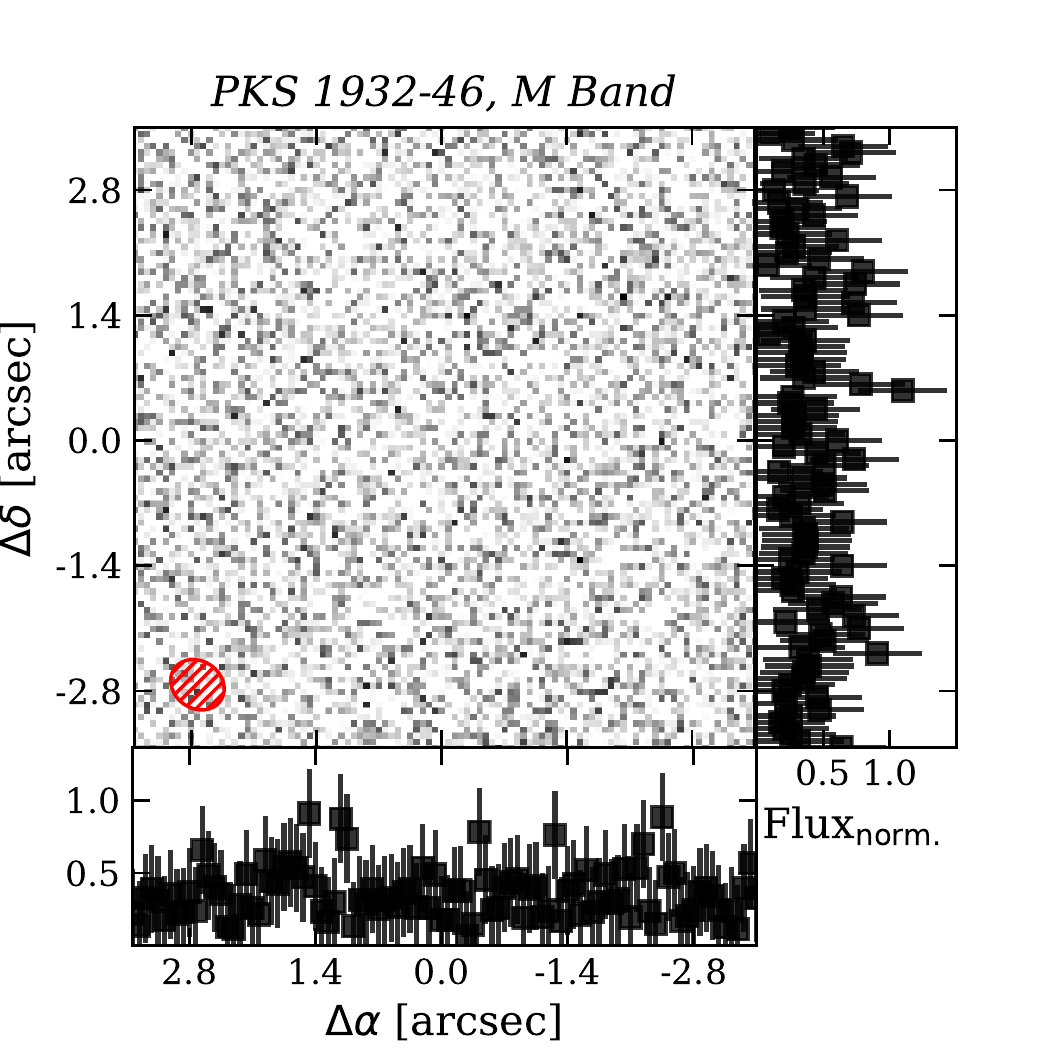}} 
\subfloat{\includegraphics[width=0.25\hsize]{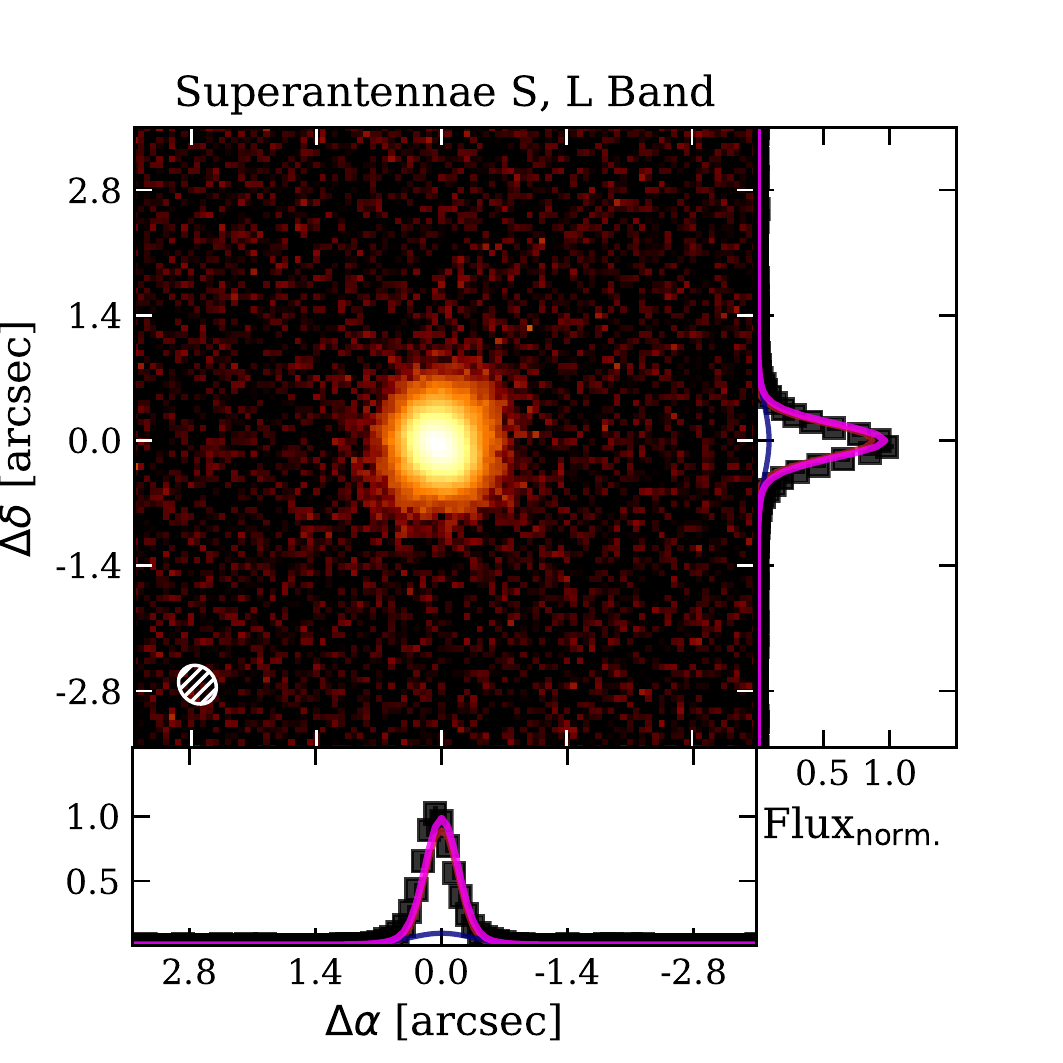}}
\subfloat{\includegraphics[width=0.25\hsize]{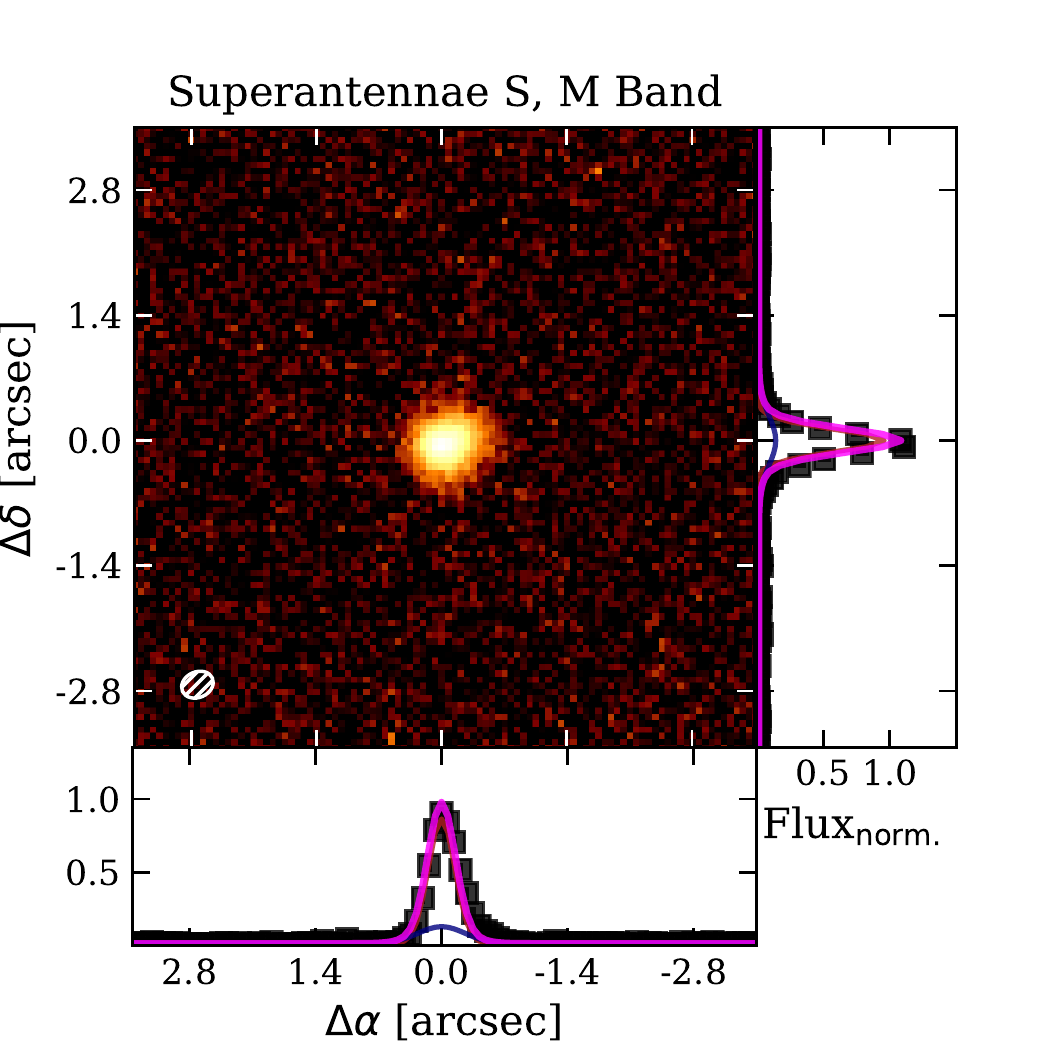}}\\
\subfloat{\includegraphics[width=0.25\hsize]{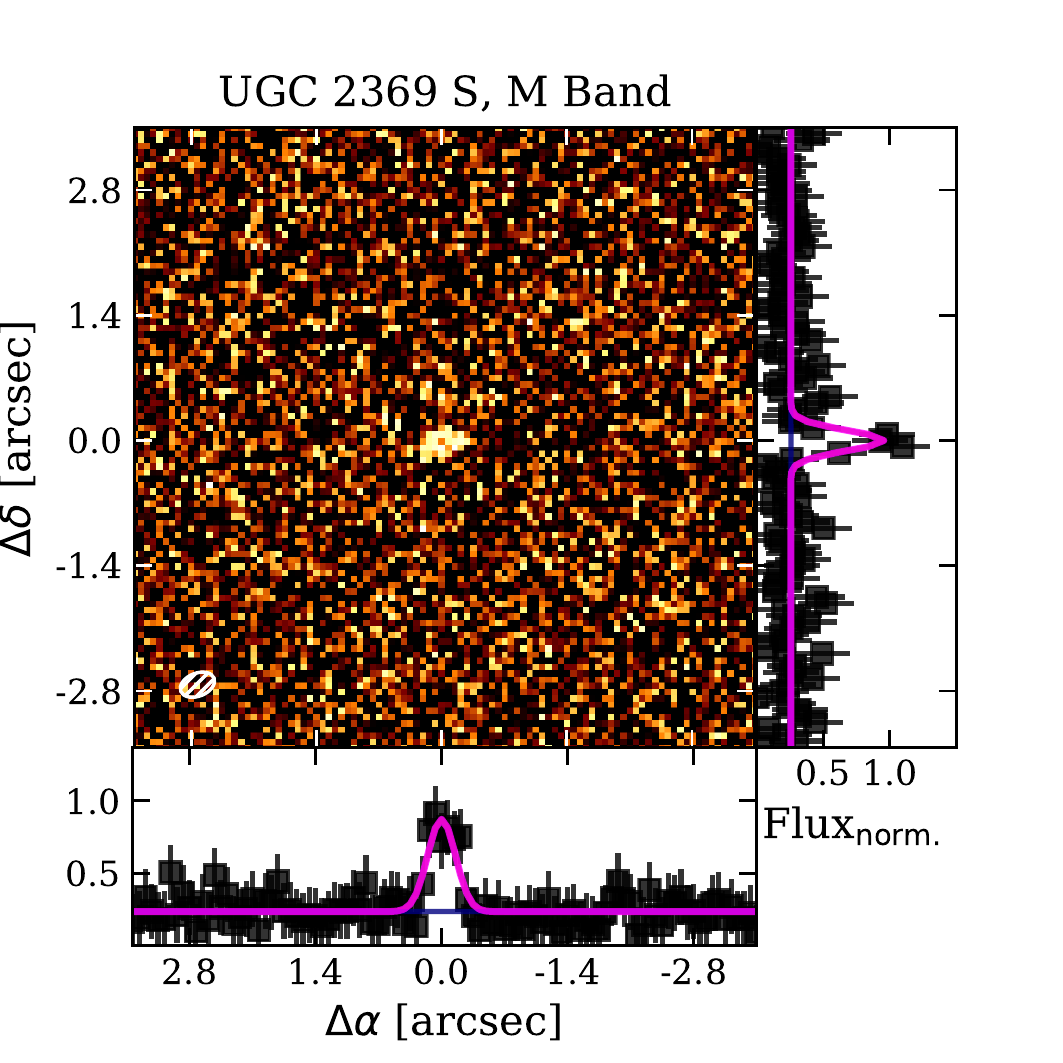}}
\subfloat{\includegraphics[width=0.25\hsize]{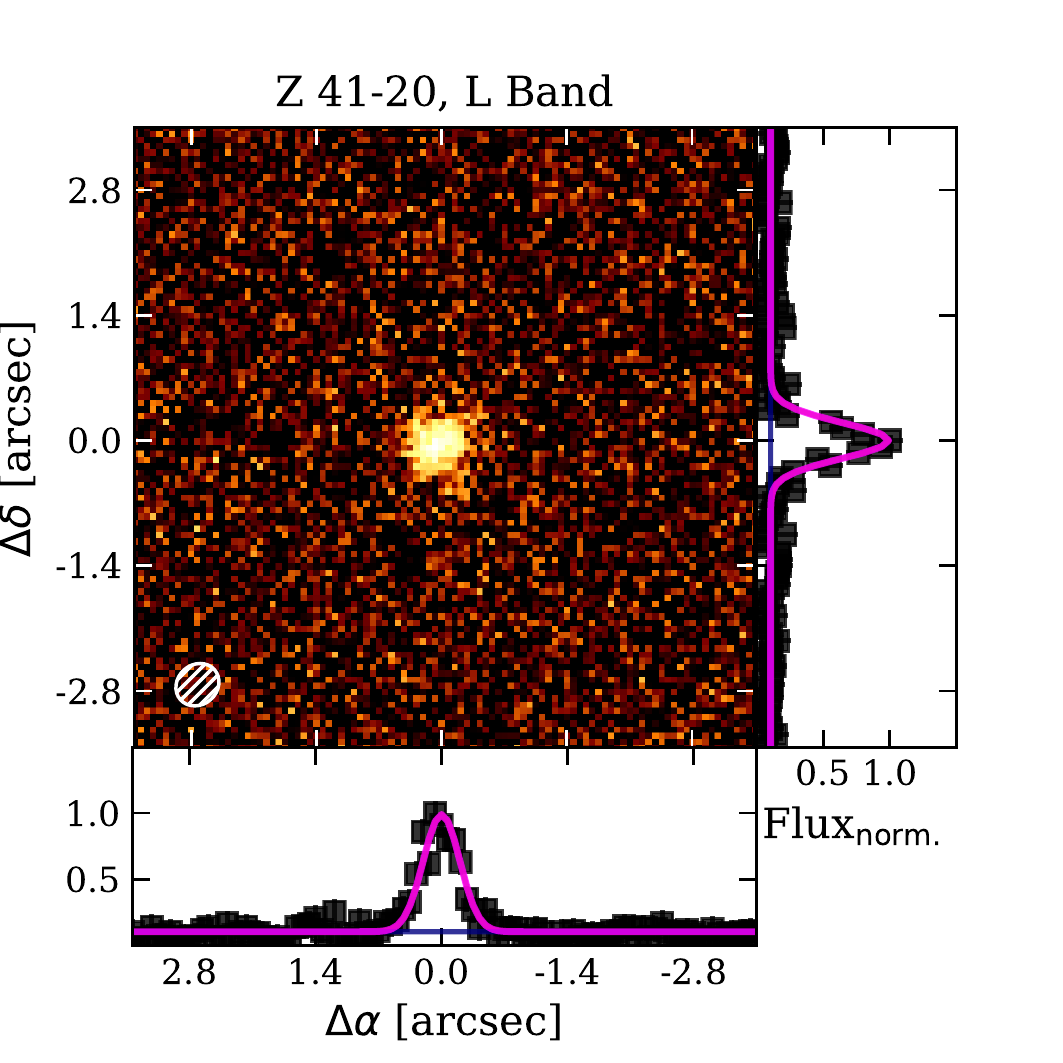}} 
\subfloat{\includegraphics[width=0.25\hsize]{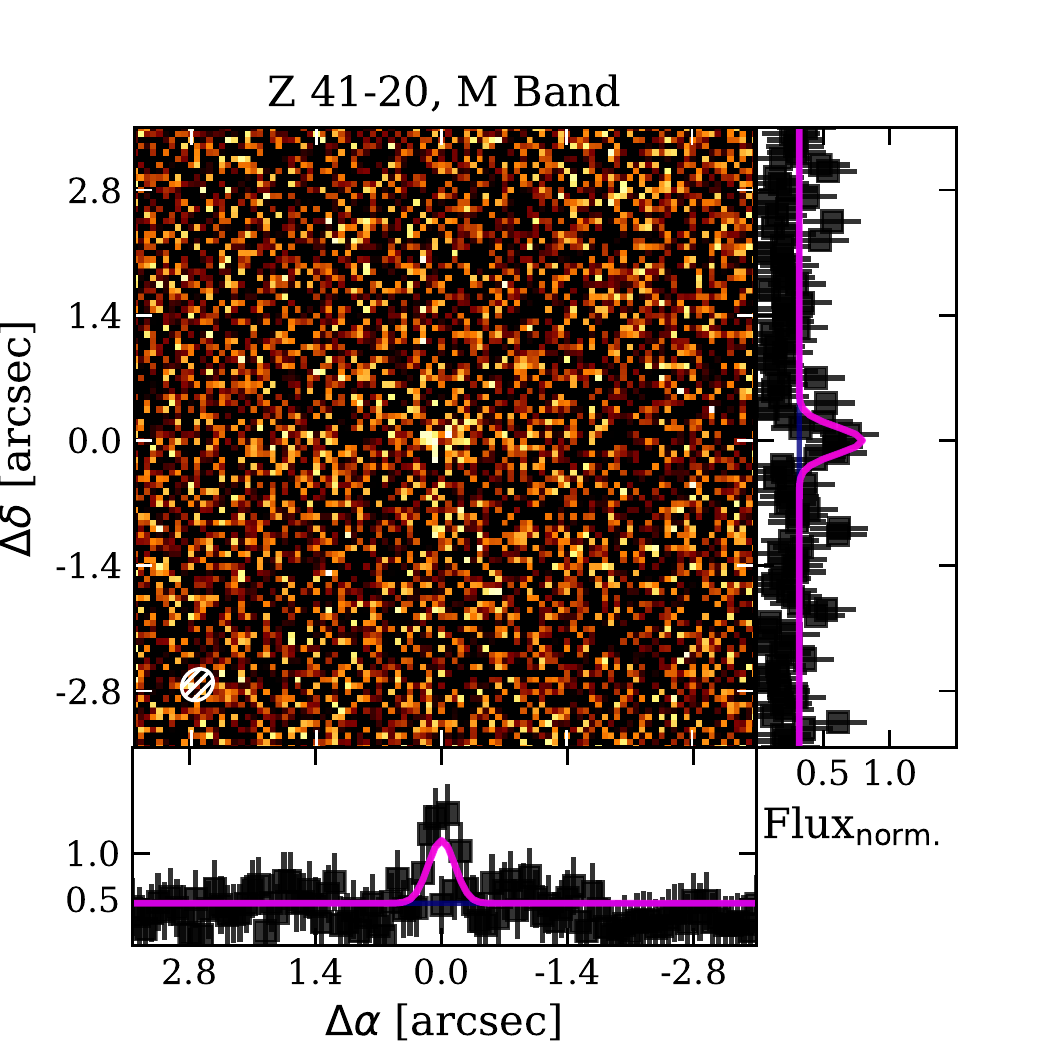}}\\
\caption{ As Fig \ref{fig:cutouts_one} but for all sources.}
\end{figure*}

\end{document}

\begin{figure*}
\centering
\subfloat{\includegraphics[width=0.25\hsize]{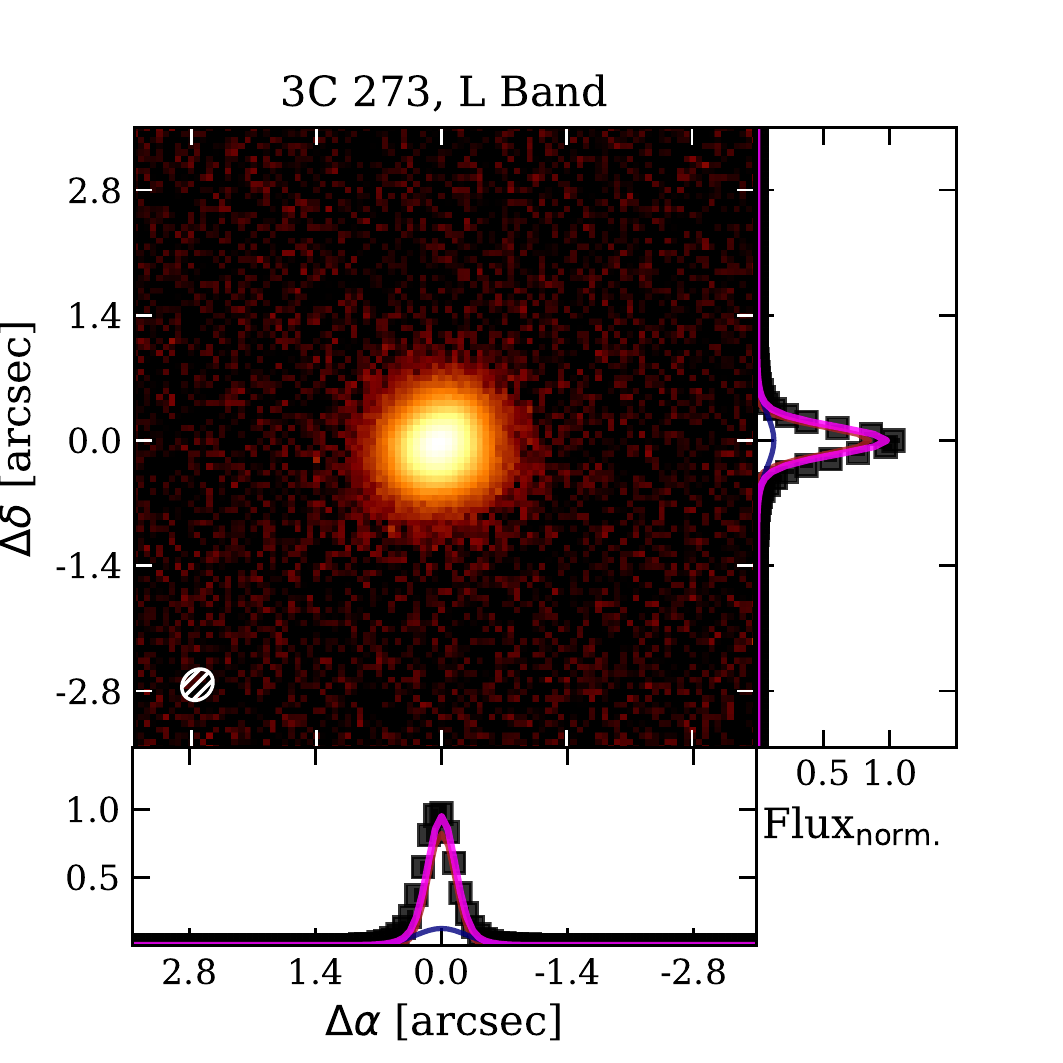}}
\subfloat{\includegraphics[width=0.25\hsize]{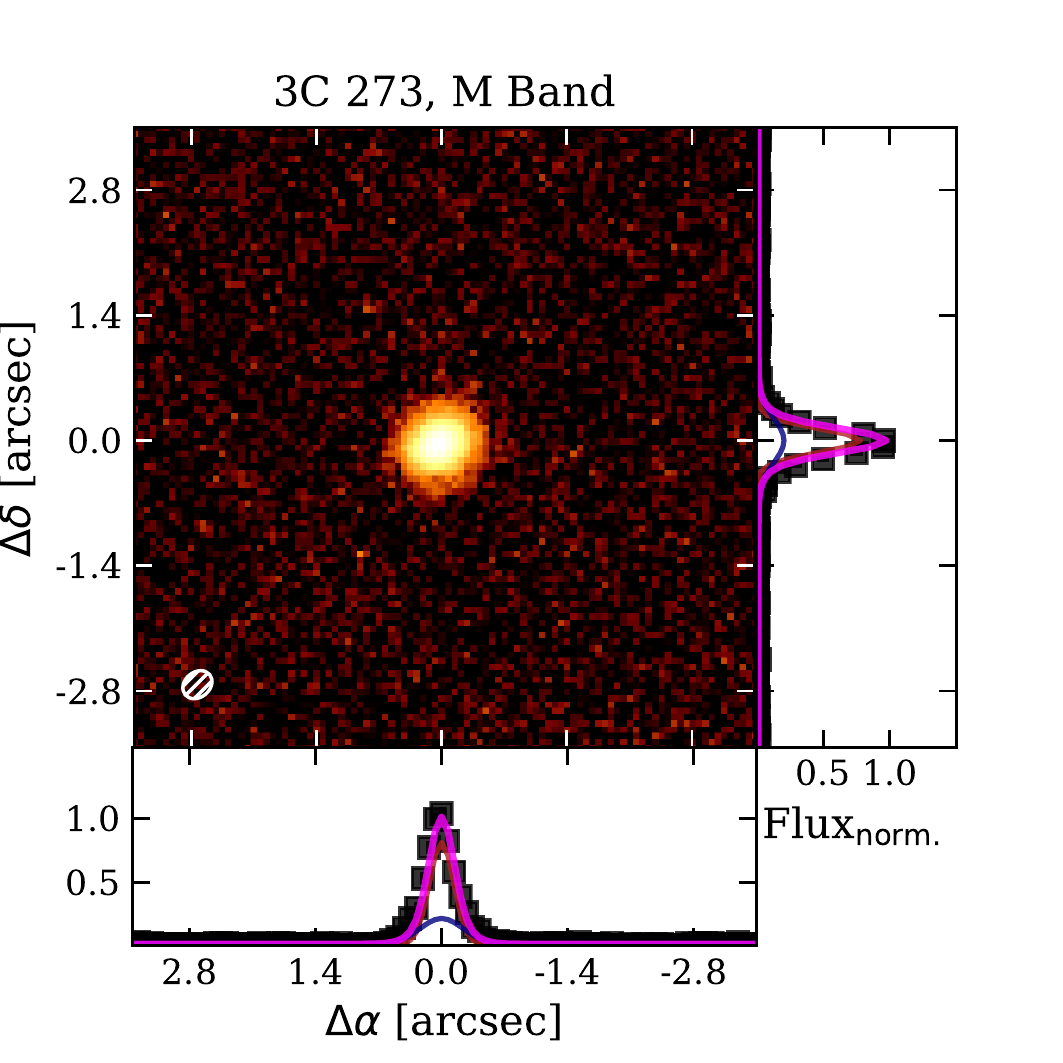}}
\subfloat{\includegraphics[width=0.25\hsize]{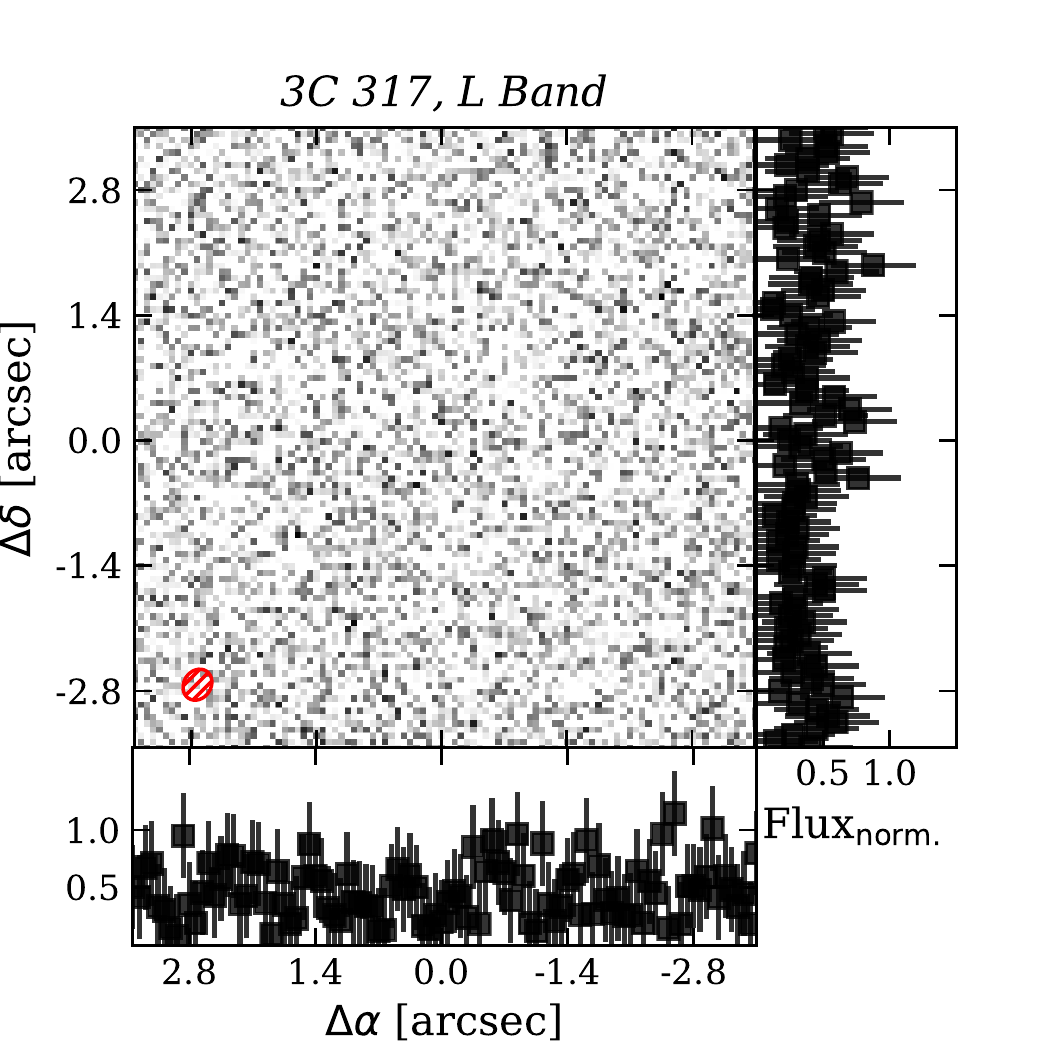}}
\subfloat{\includegraphics[width=0.25\hsize]{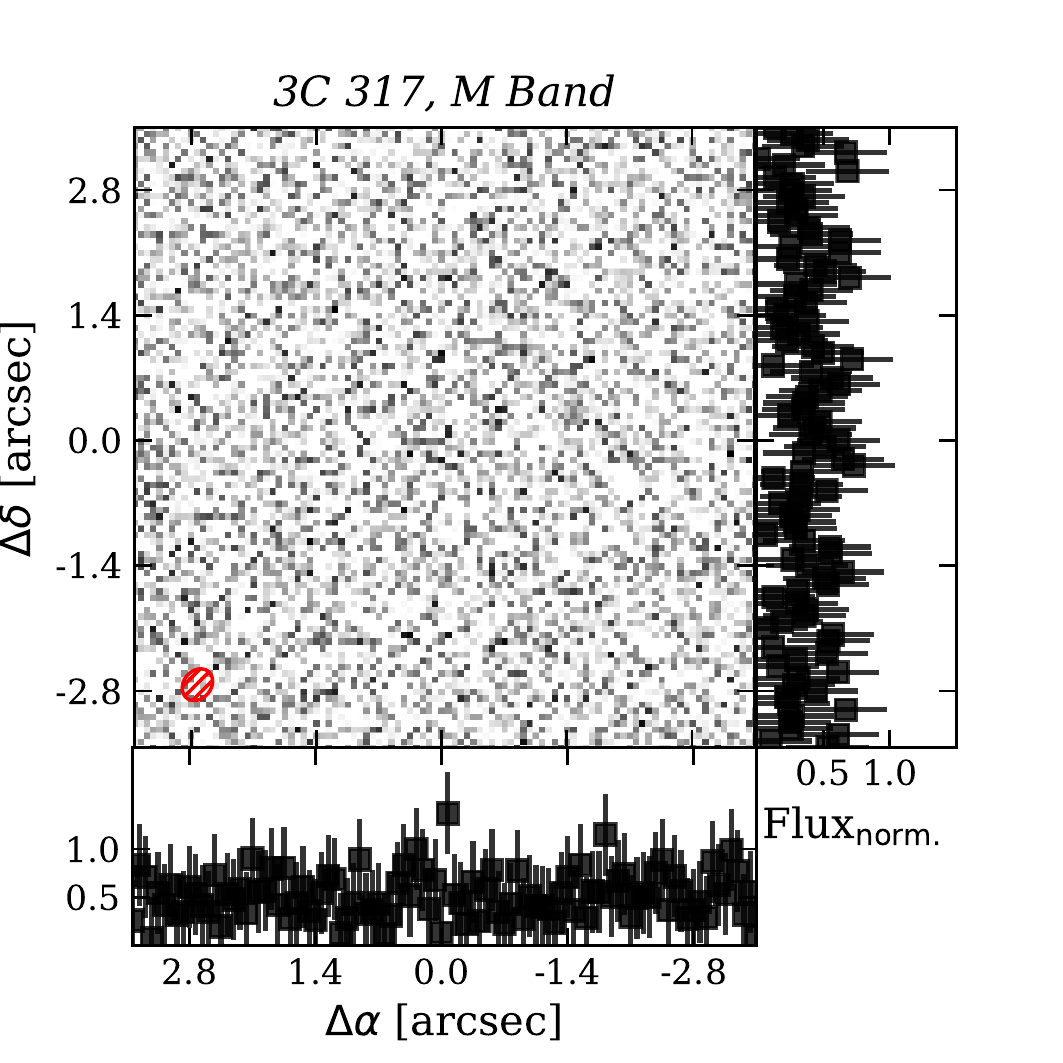}} \\
\subfloat{\includegraphics[width=0.25\hsize]{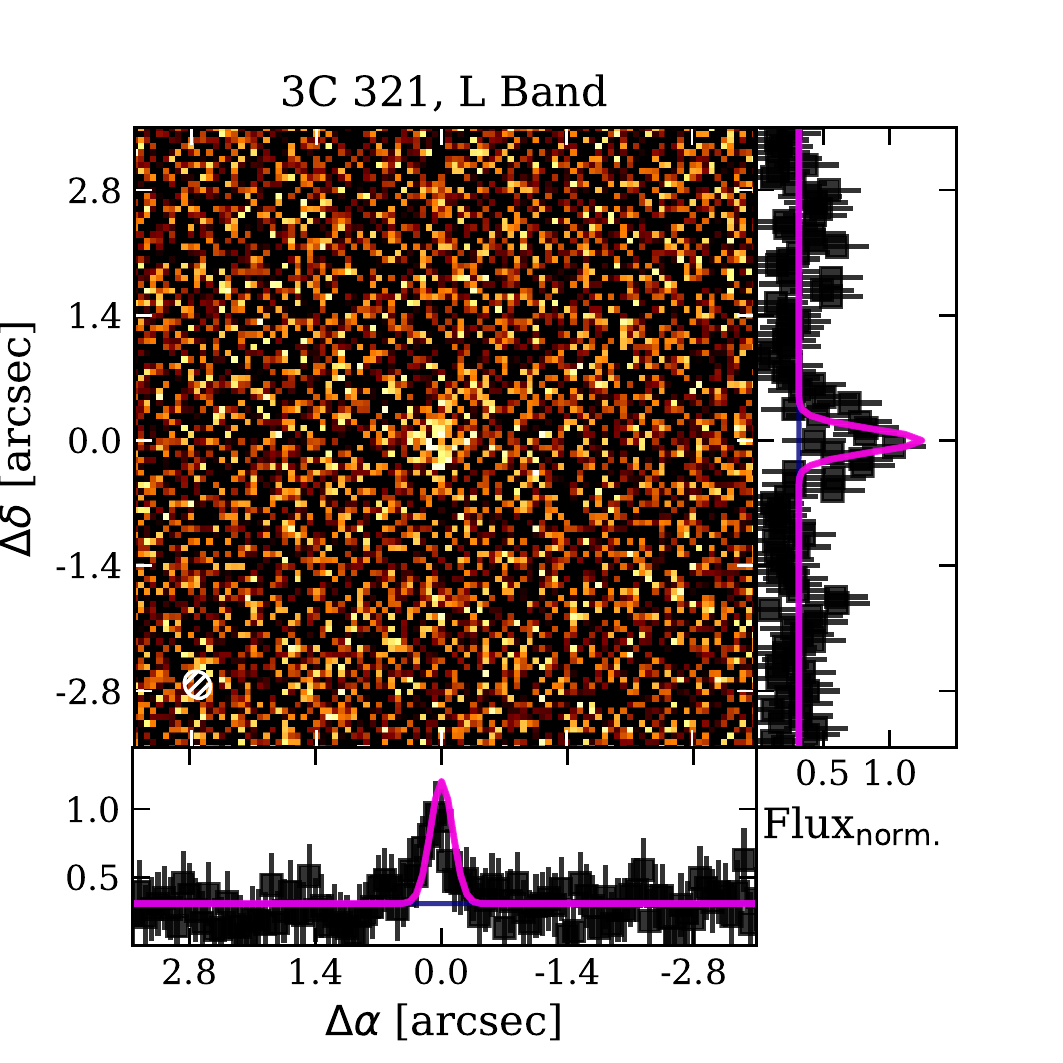}}
\subfloat{\includegraphics[width=0.25\hsize]{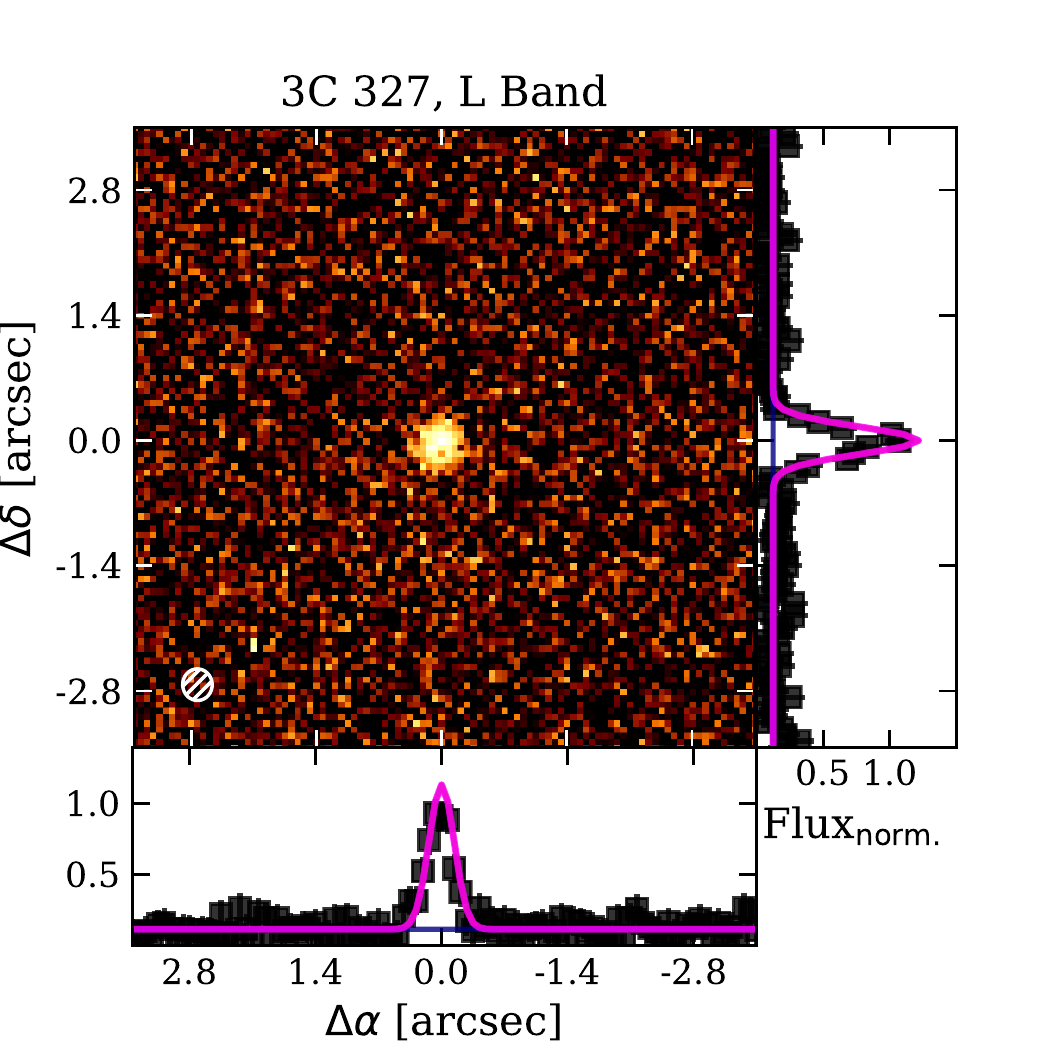}}
\subfloat{\includegraphics[width=0.25\hsize]{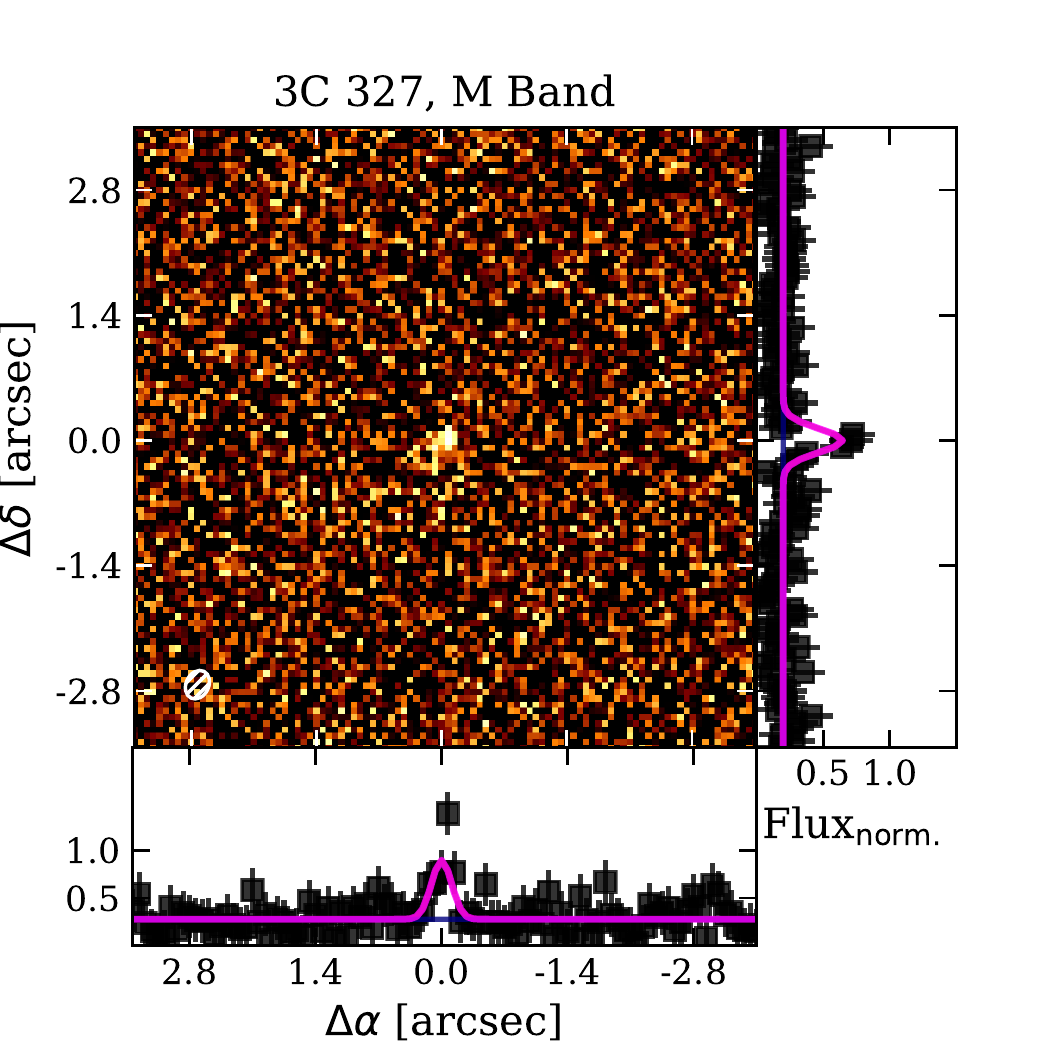}}
\subfloat{\includegraphics[width=0.25\hsize]{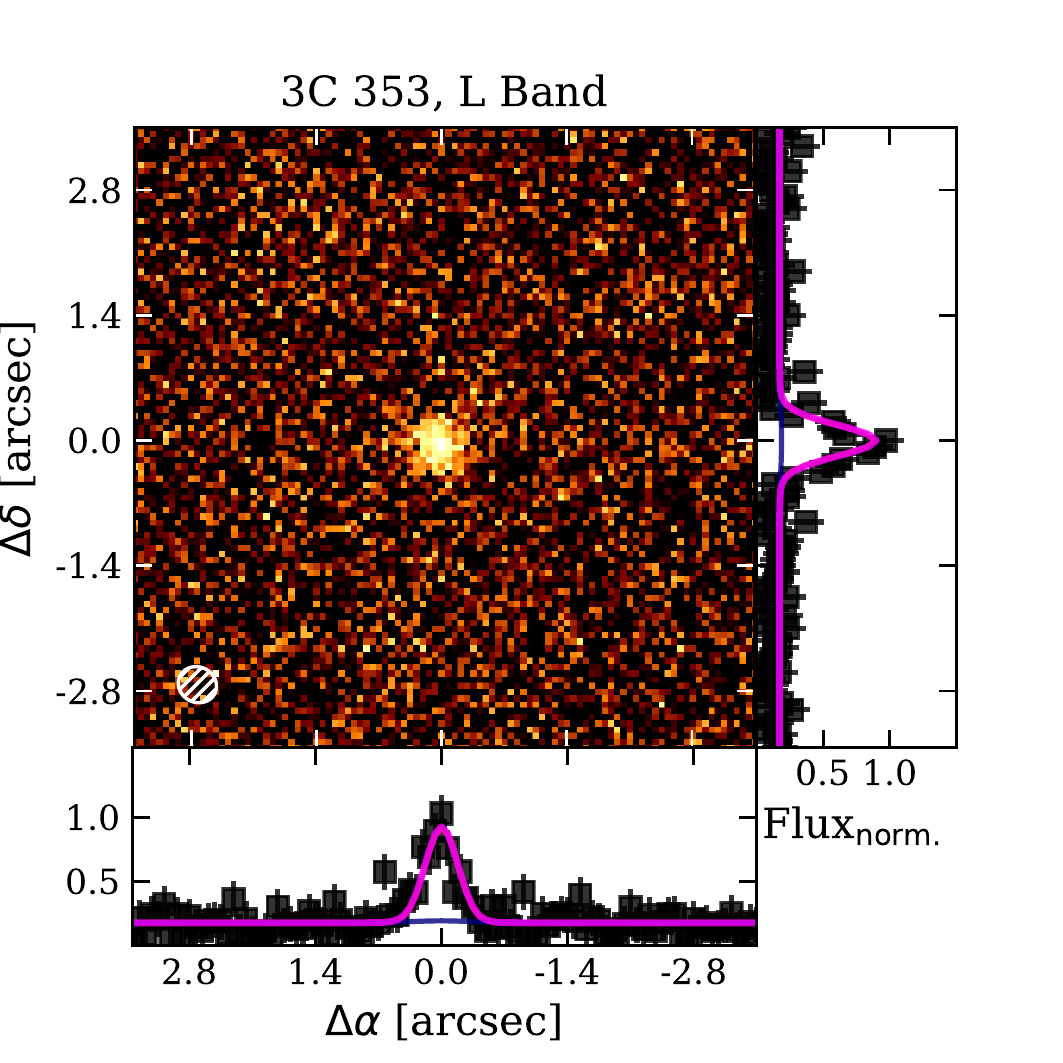}} \\
\subfloat{\includegraphics[width=0.25\hsize]{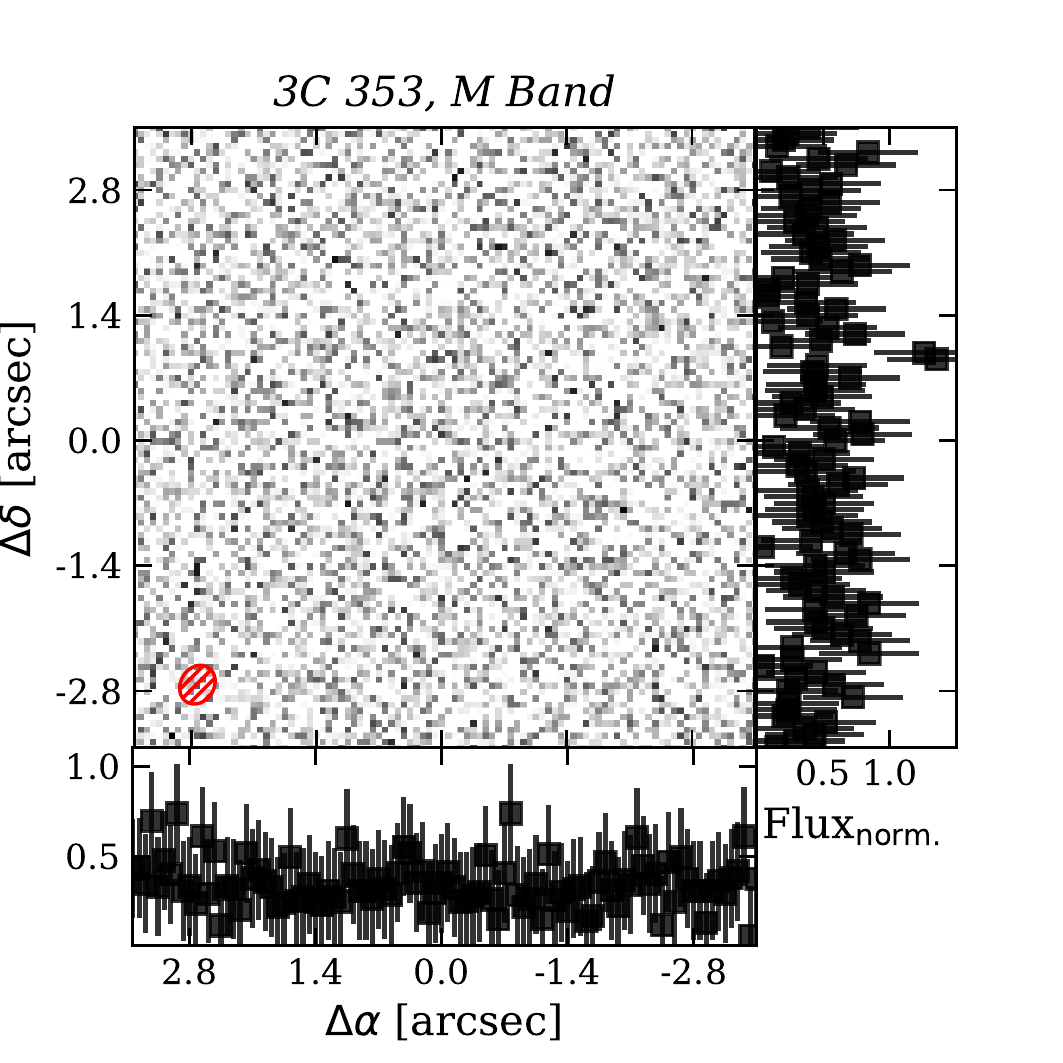}}
\subfloat{\includegraphics[width=0.25\hsize]{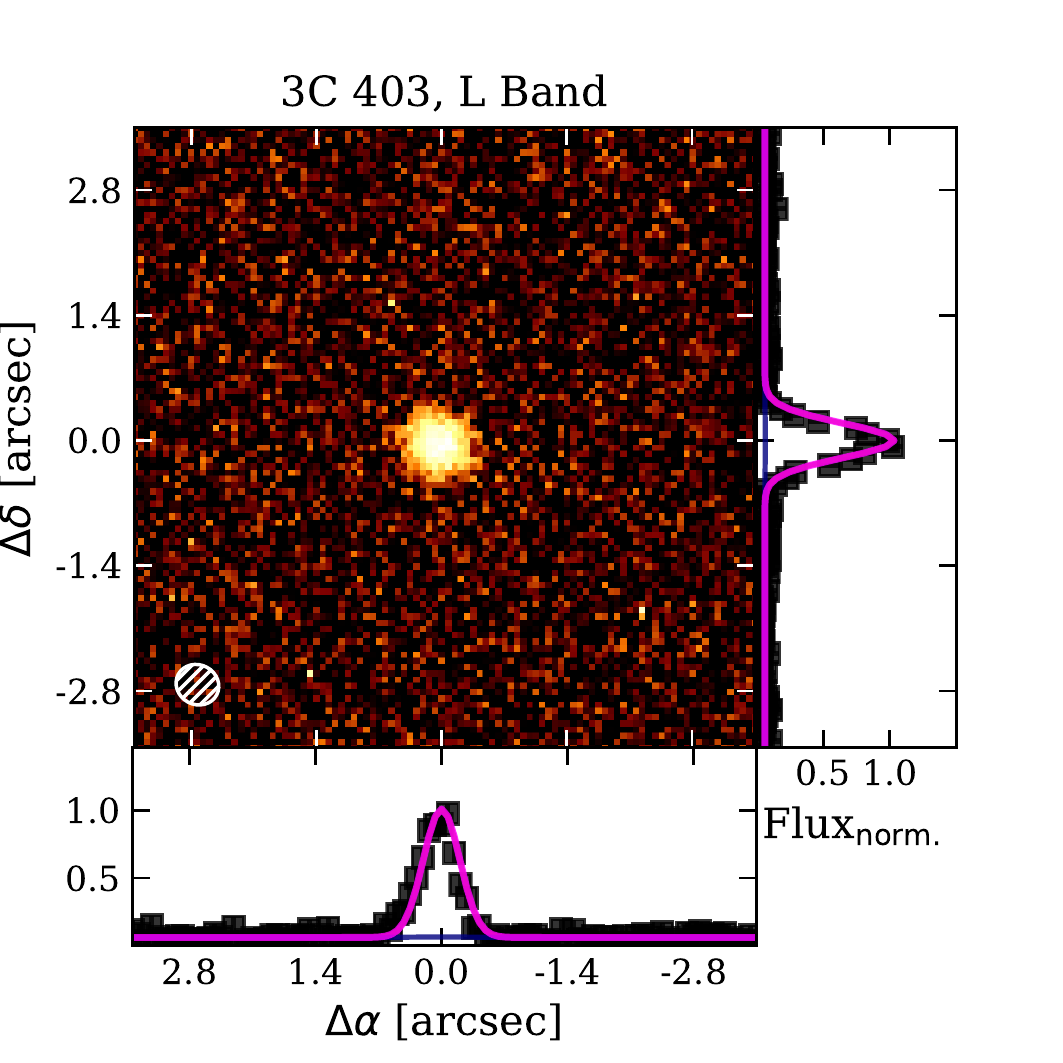}}
\subfloat{\includegraphics[width=0.25\hsize]{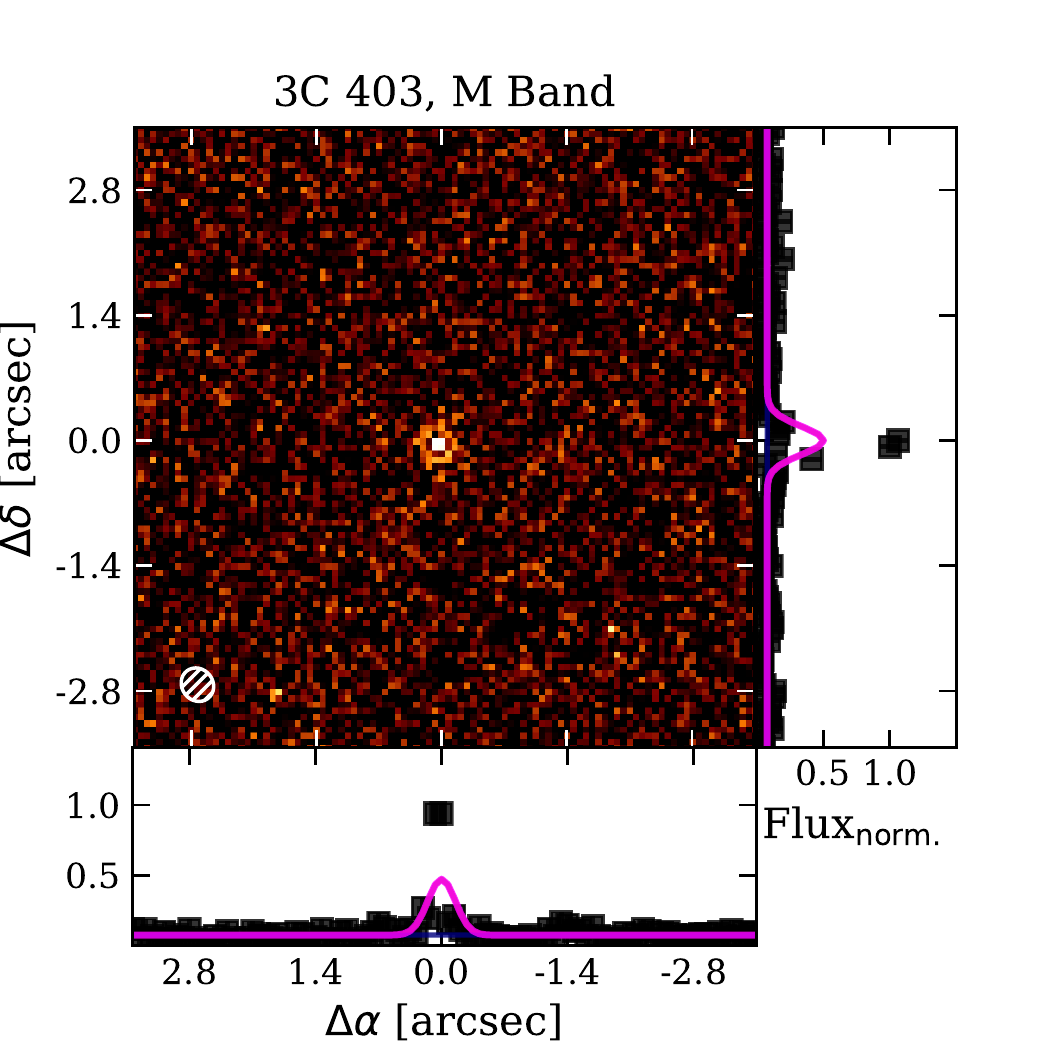}}
\subfloat{\includegraphics[width=0.25\hsize]{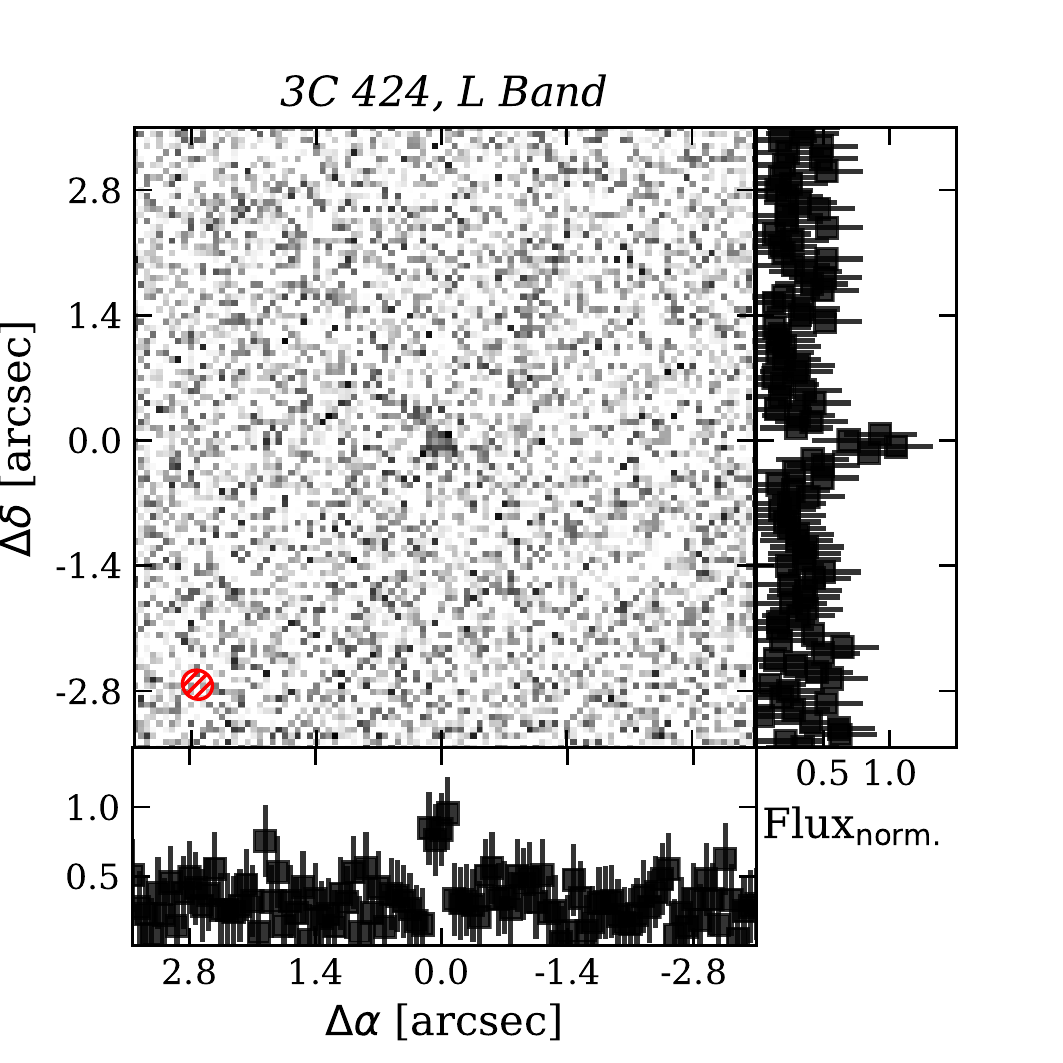}} \\
\subfloat{\includegraphics[width=0.25\hsize]{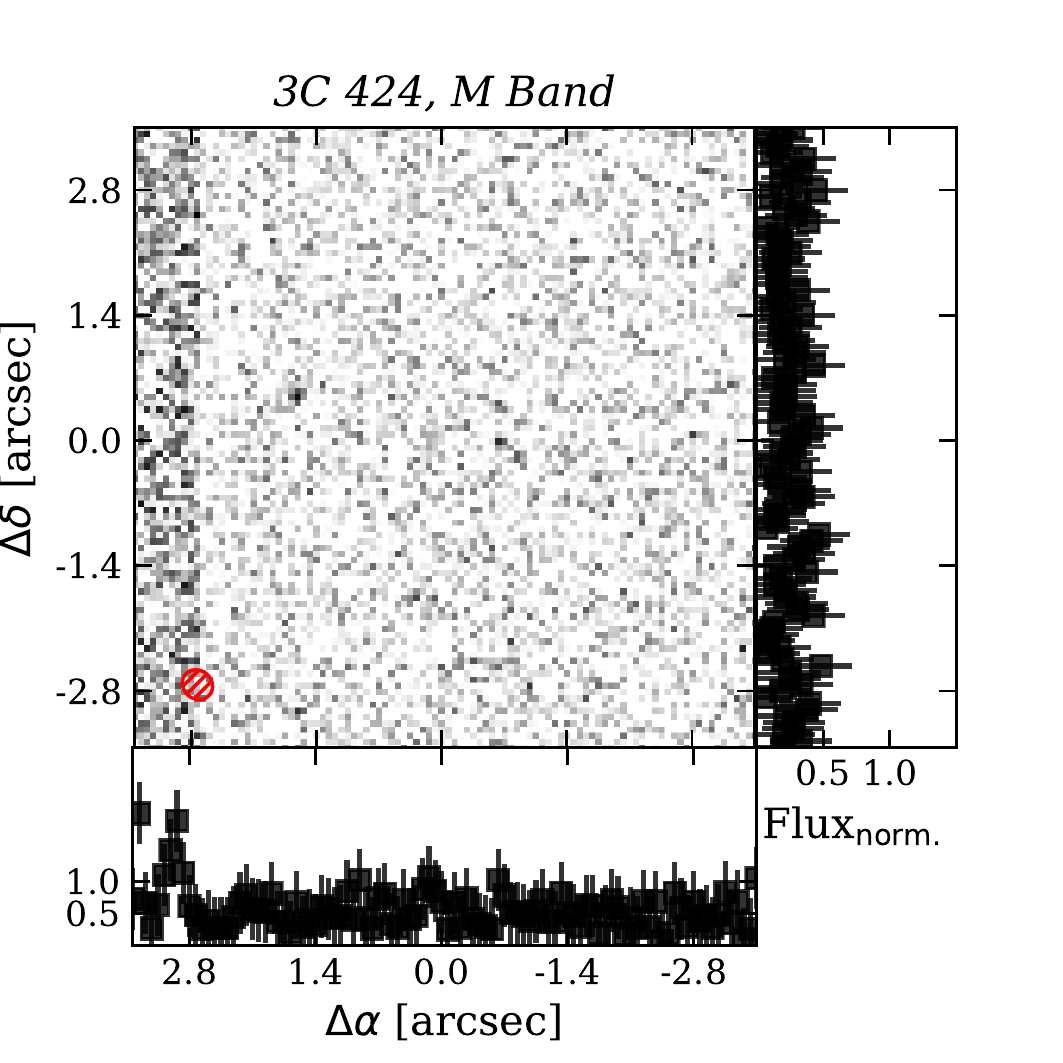}}
\subfloat{\includegraphics[width=0.25\hsize]{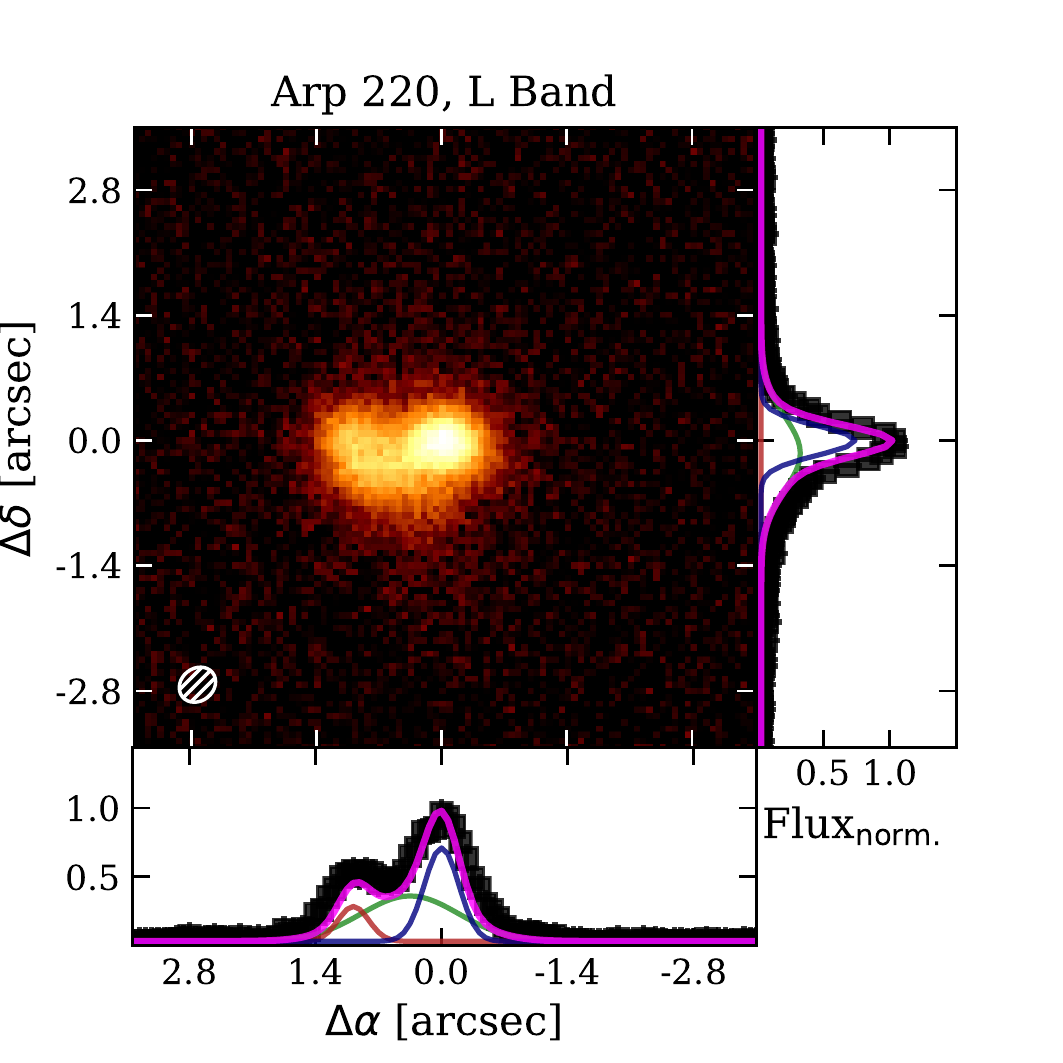}}
\subfloat{\includegraphics[width=0.25\hsize]{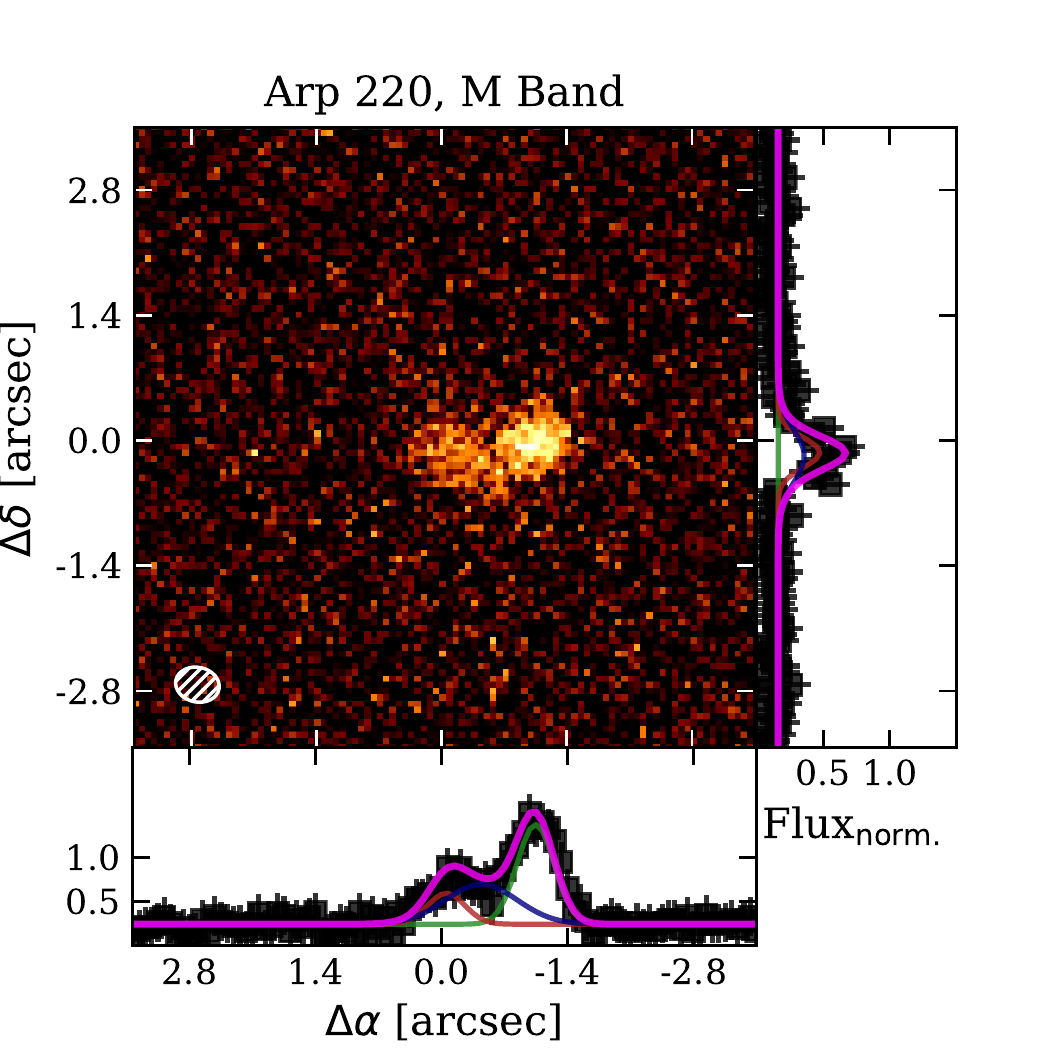}}
\subfloat{\includegraphics[width=0.25\hsize]{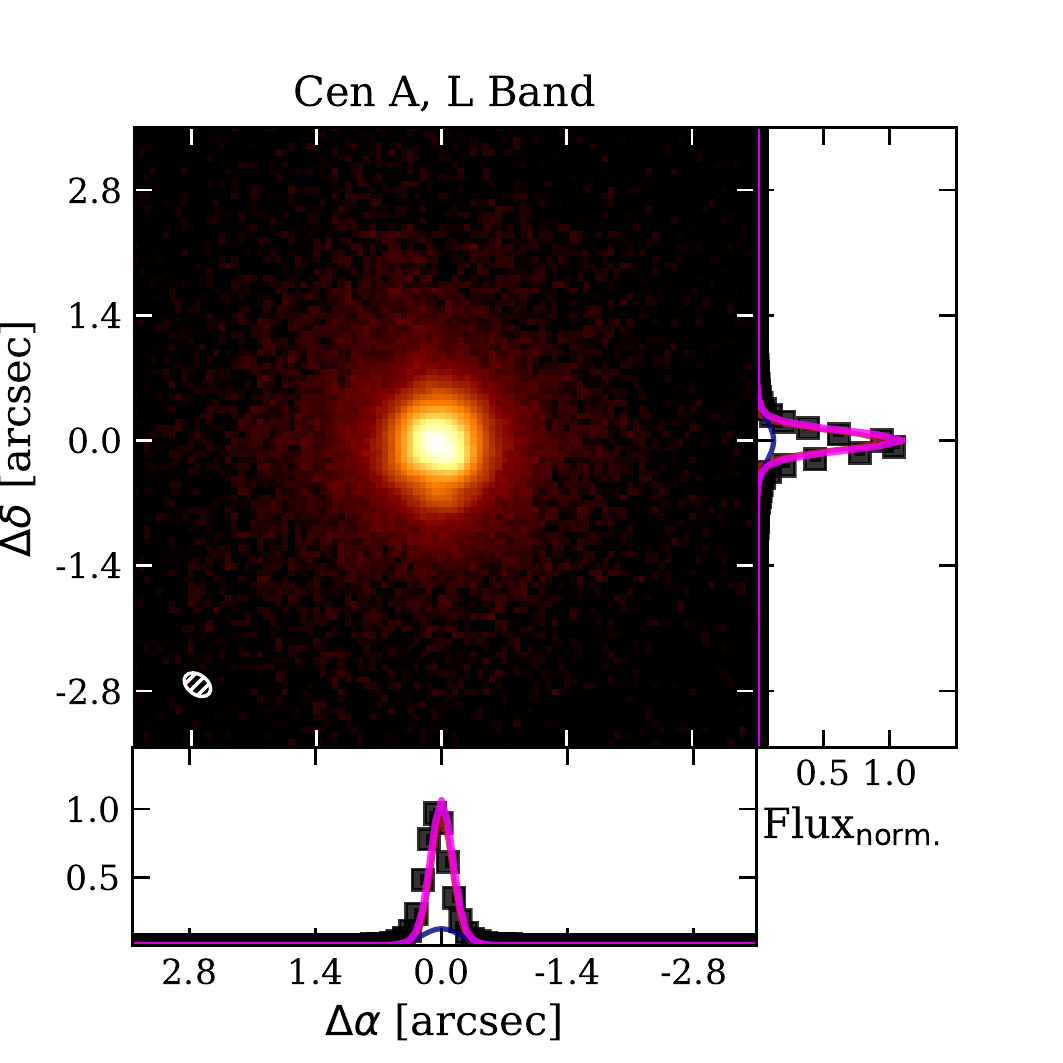}} \\
\subfloat{\includegraphics[width=0.25\hsize]{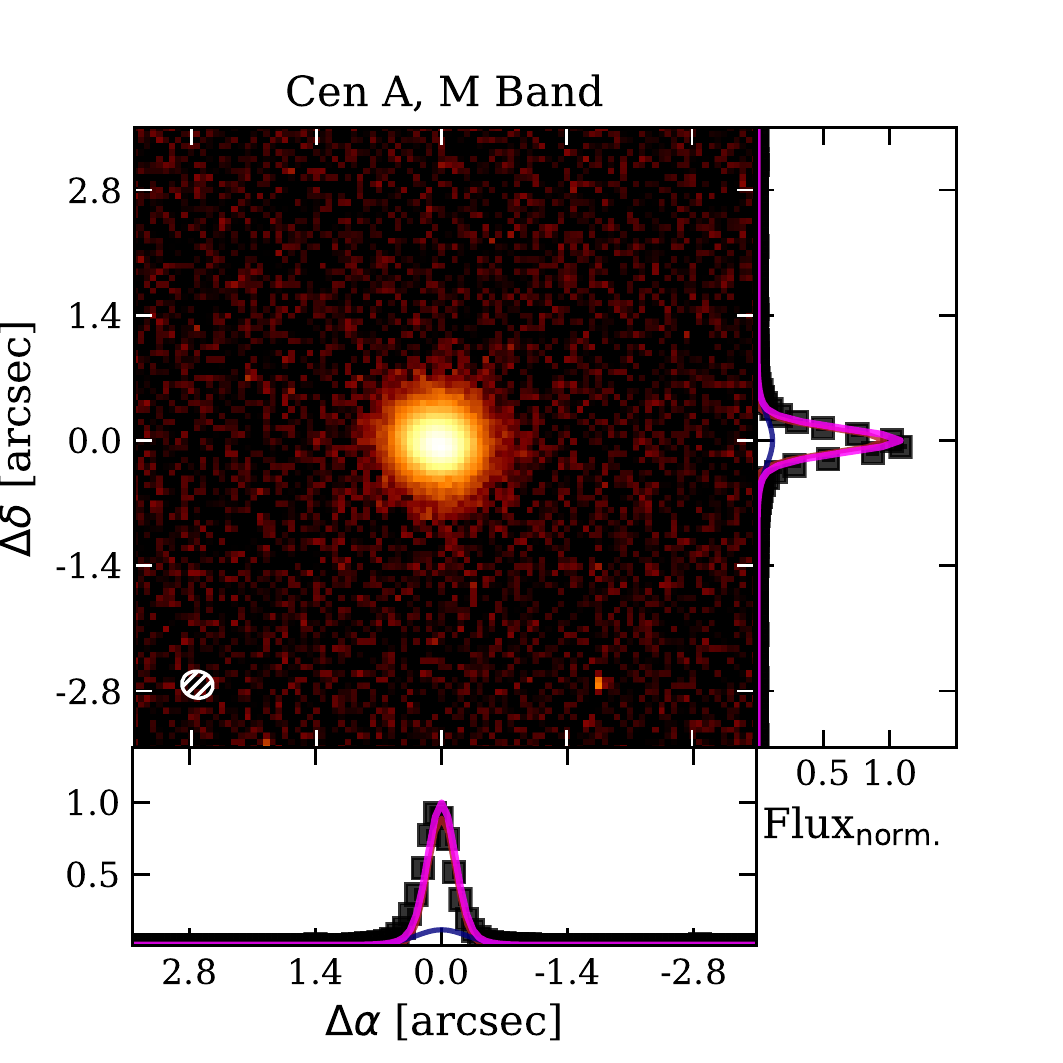}}
\subfloat{\includegraphics[width=0.25\hsize]{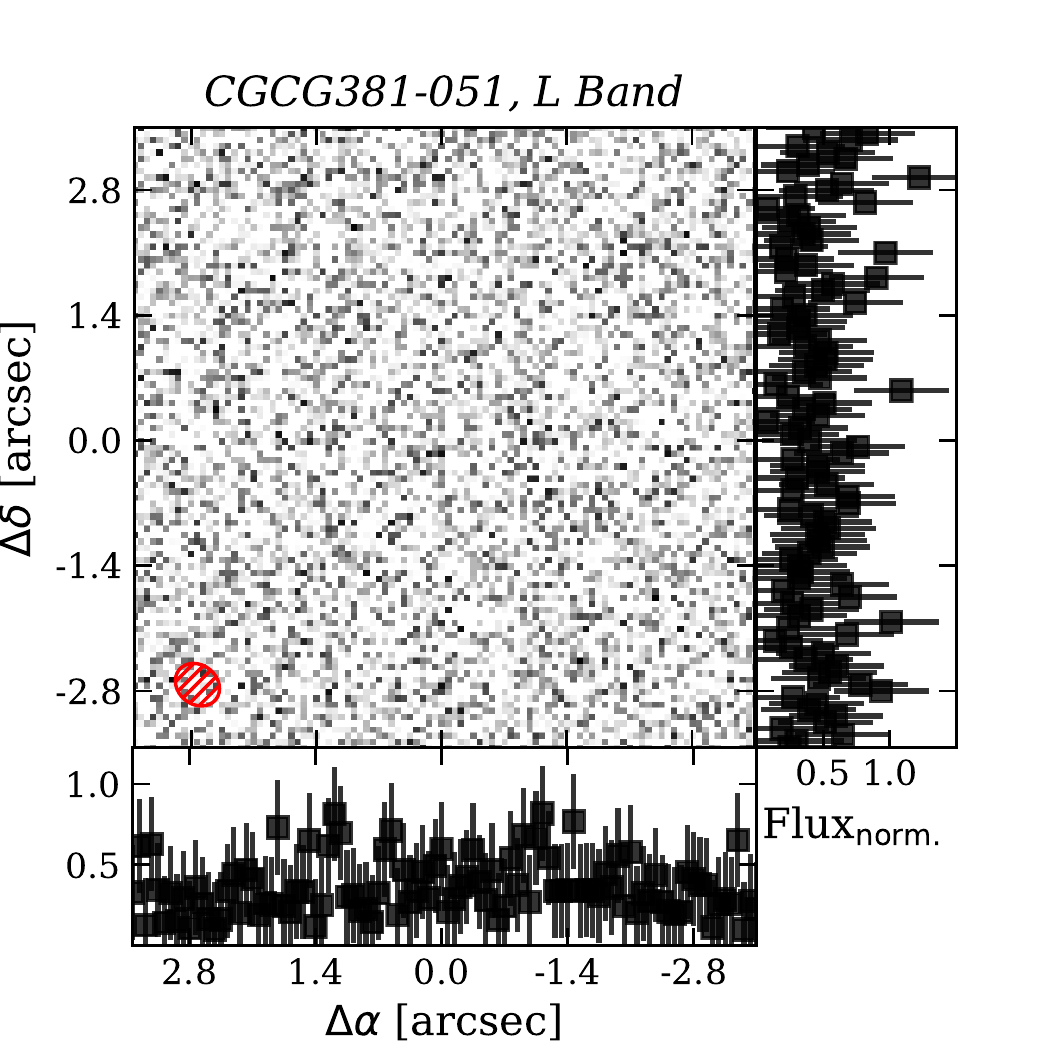}}
\subfloat{\includegraphics[width=0.25\hsize]{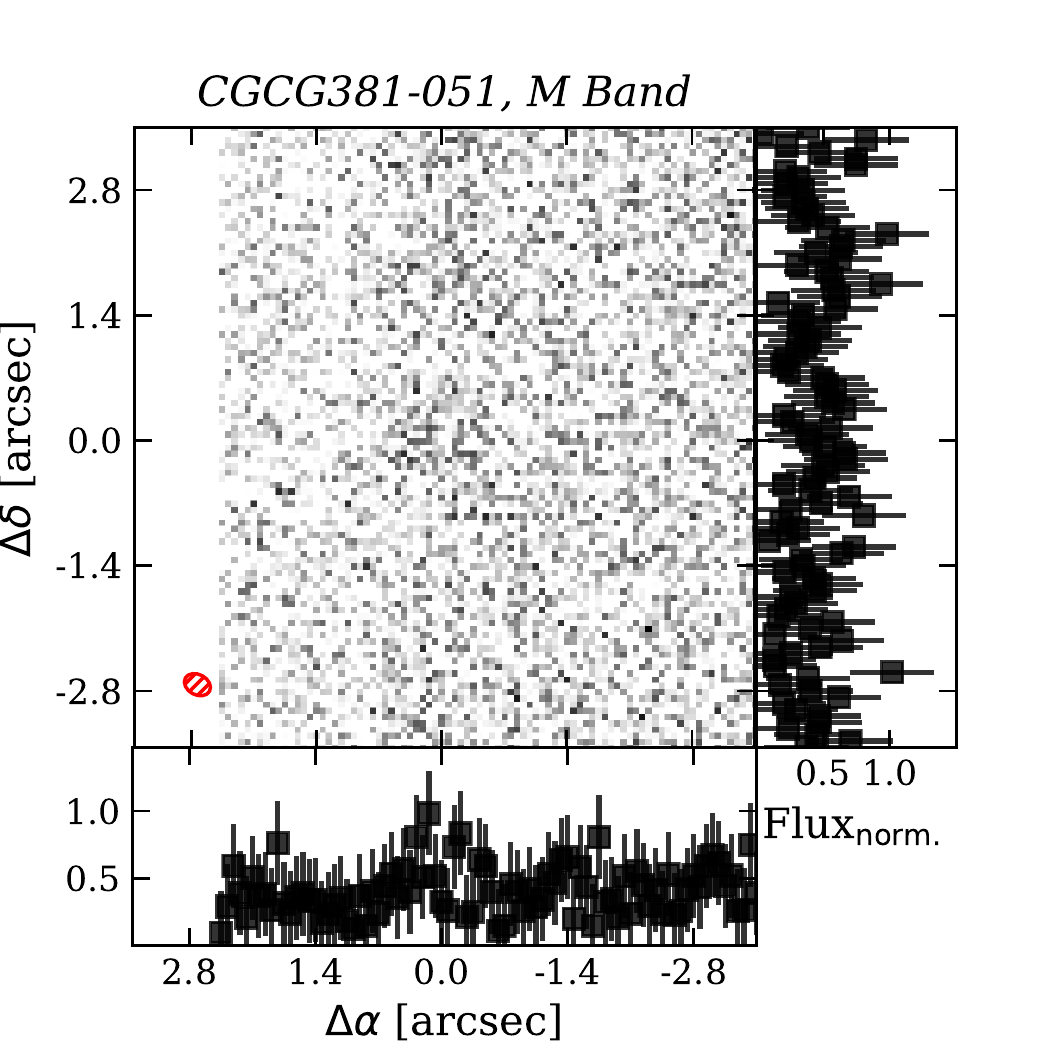}}
\subfloat{\includegraphics[width=0.25\hsize]{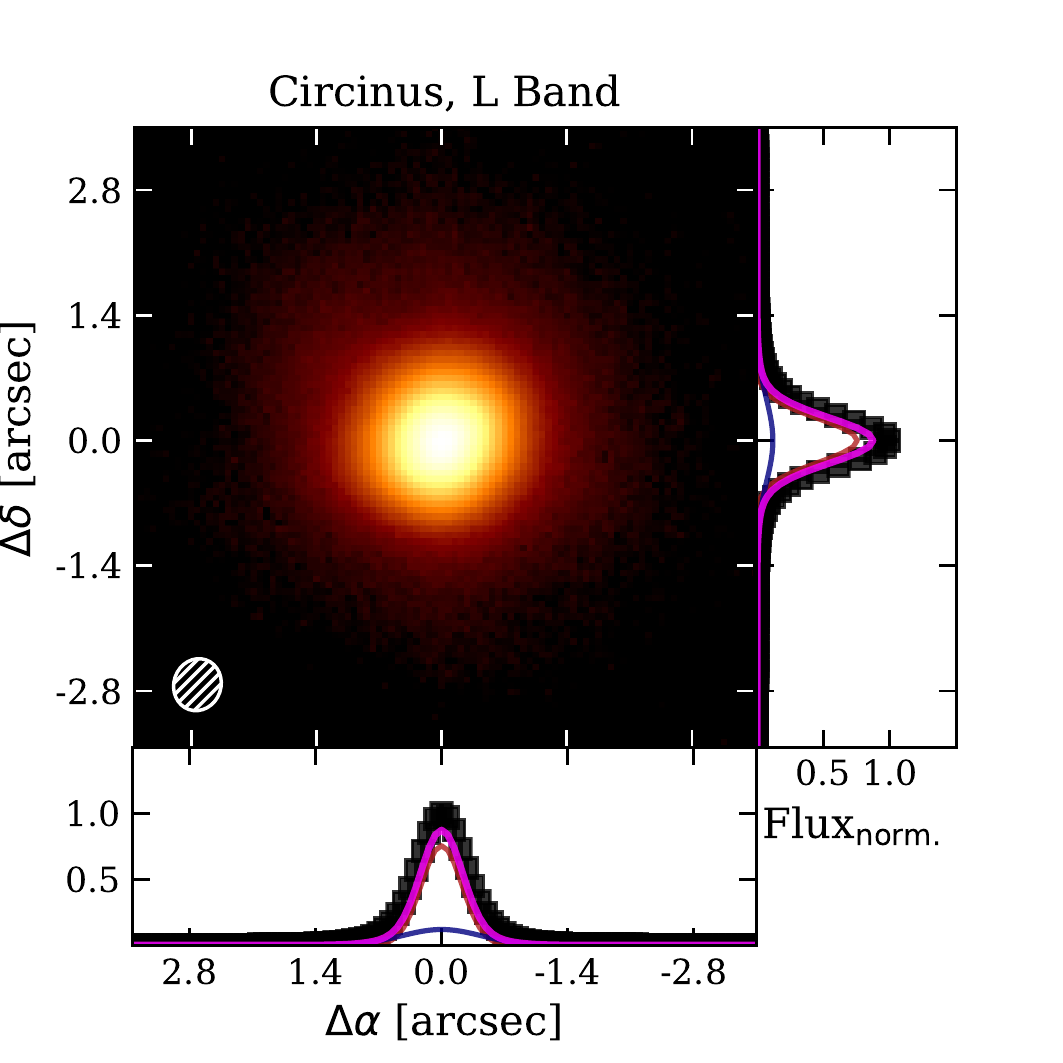}} \\
\label{fig:allcutouts}
\caption{ As Fig \ref{fig:cutouts_one} but for all sources.}
%Cutouts for all sources: Panels in greyscale (with titles in {\it italic}) are classified as non-detections, while those in color have SNR$_{\rm gauss}\geq 3$ in either Gaussian component and are classified as detections. With each cutout we present 1-D slices across the center of the image in both the x- and y-directions. Data in these slices are shown in black, and the profiles of the fitted elliptical Gaussians are plotted in red (nuclear), blue (extended), and magenta (sum). All images are presented with log-scaling. The ellipse in the bottom left of each cutout represents the fitted FWHM of the PSF calibrator. North is up and East is left. }

\end{figure*}
\begin{figure*}
\centering
\subfloat{\includegraphics[width=0.25\hsize]{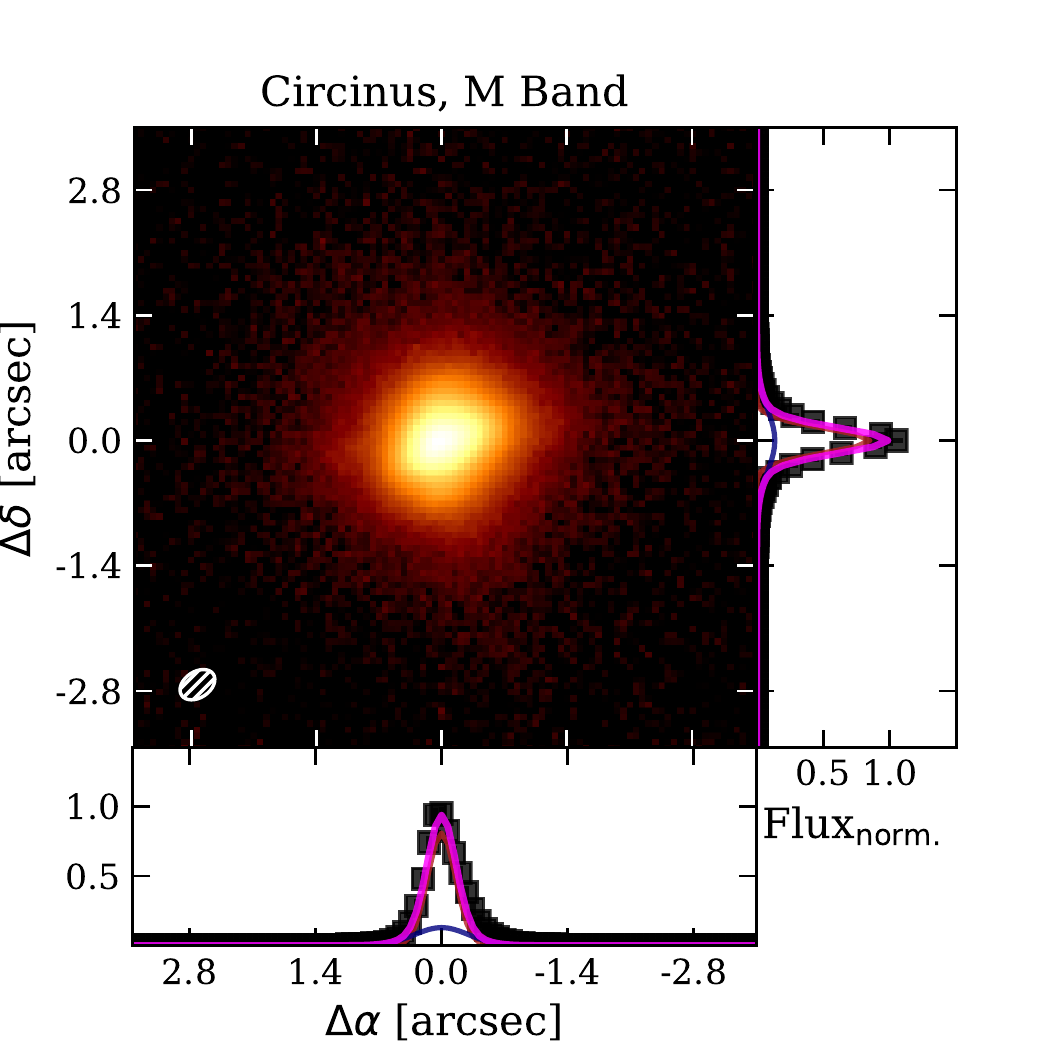}}
\subfloat{\includegraphics[width=0.25\hsize]{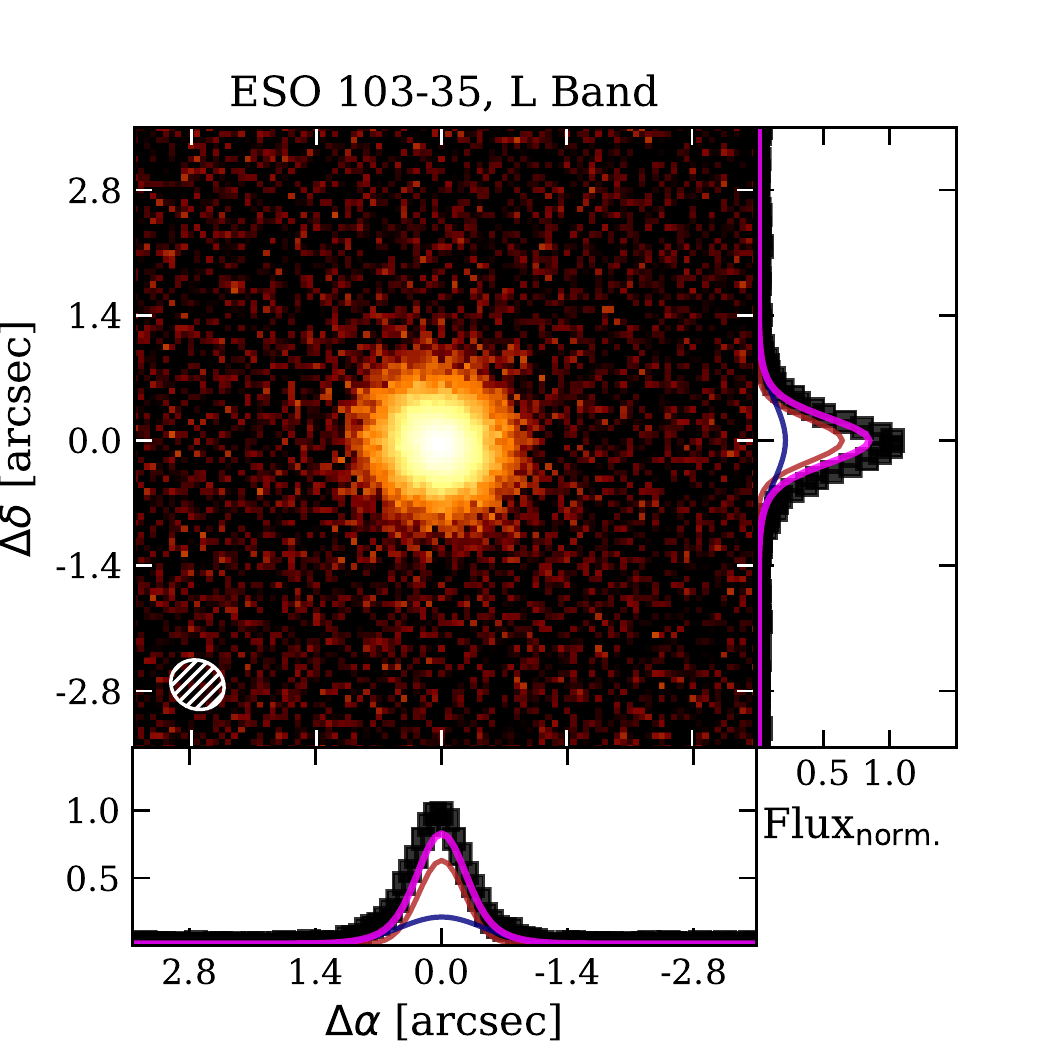}}
\subfloat{\includegraphics[width=0.25\hsize]{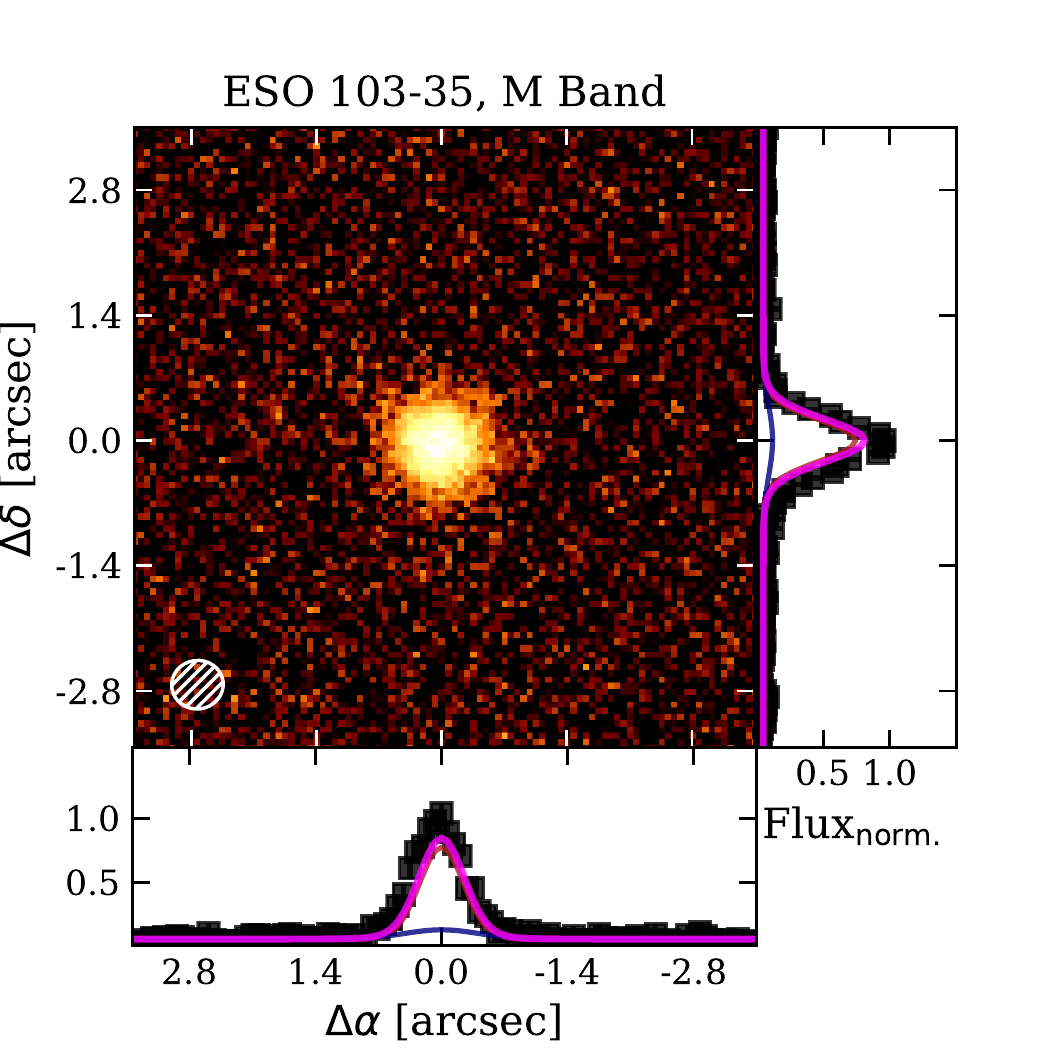}}
\subfloat{\includegraphics[width=0.25\hsize]{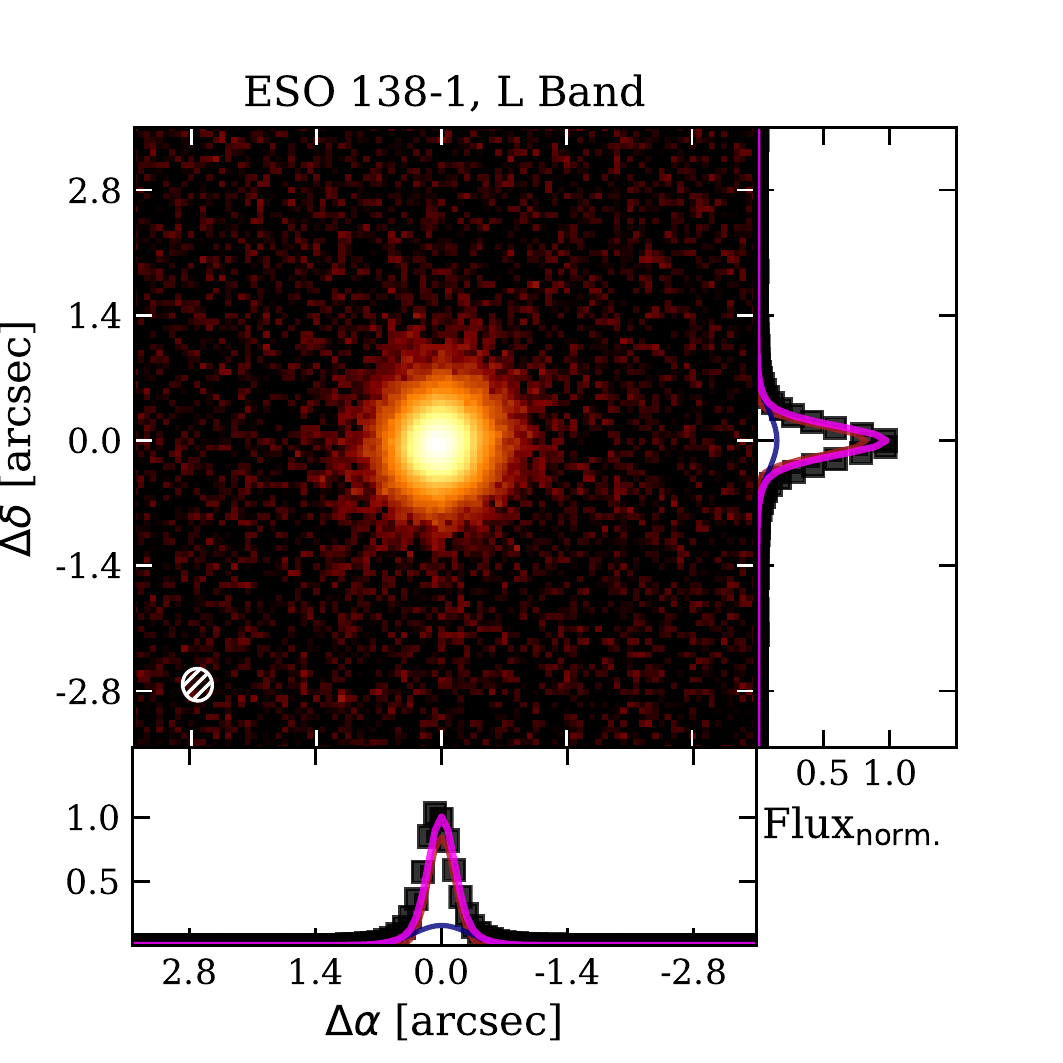}} \\
\subfloat{\includegraphics[width=0.25\hsize]{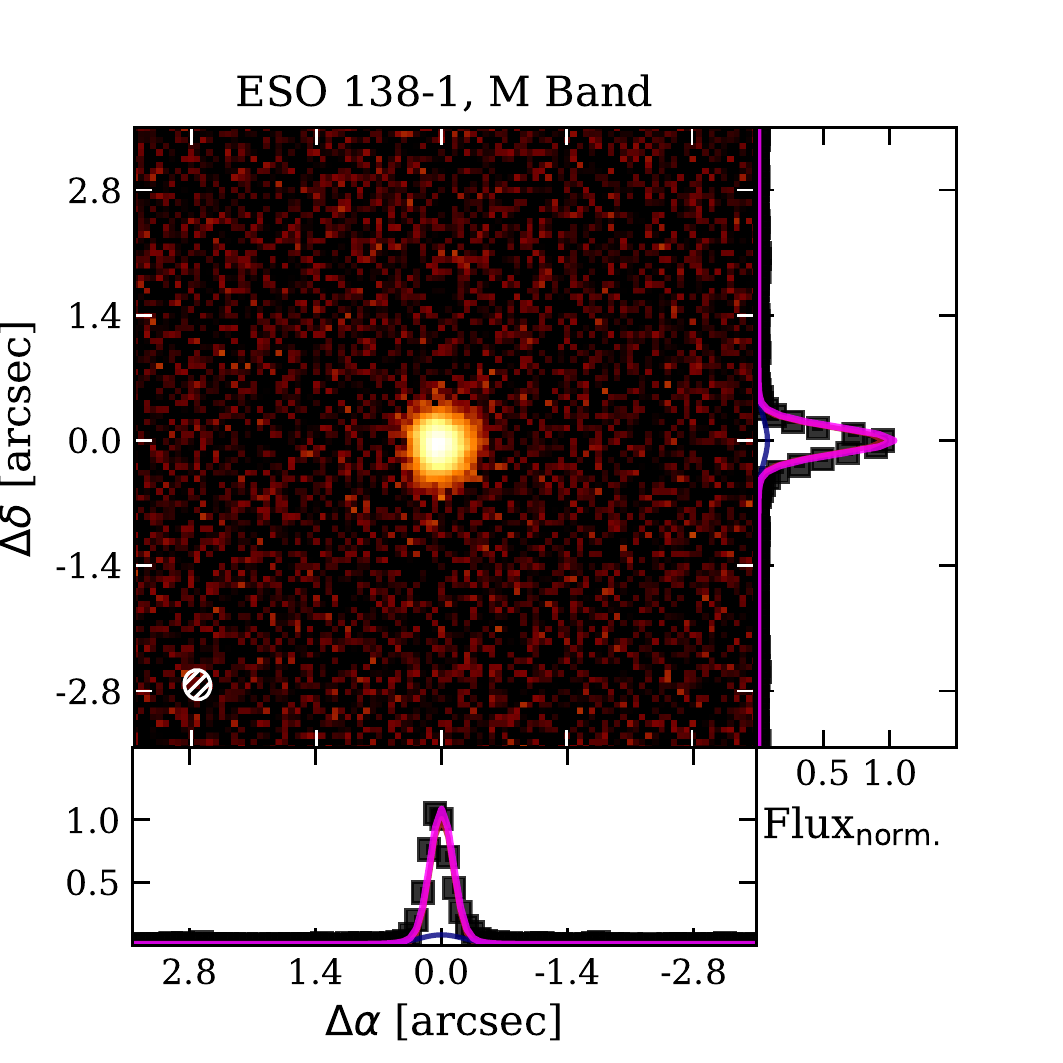}}
\subfloat{\includegraphics[width=0.25\hsize]{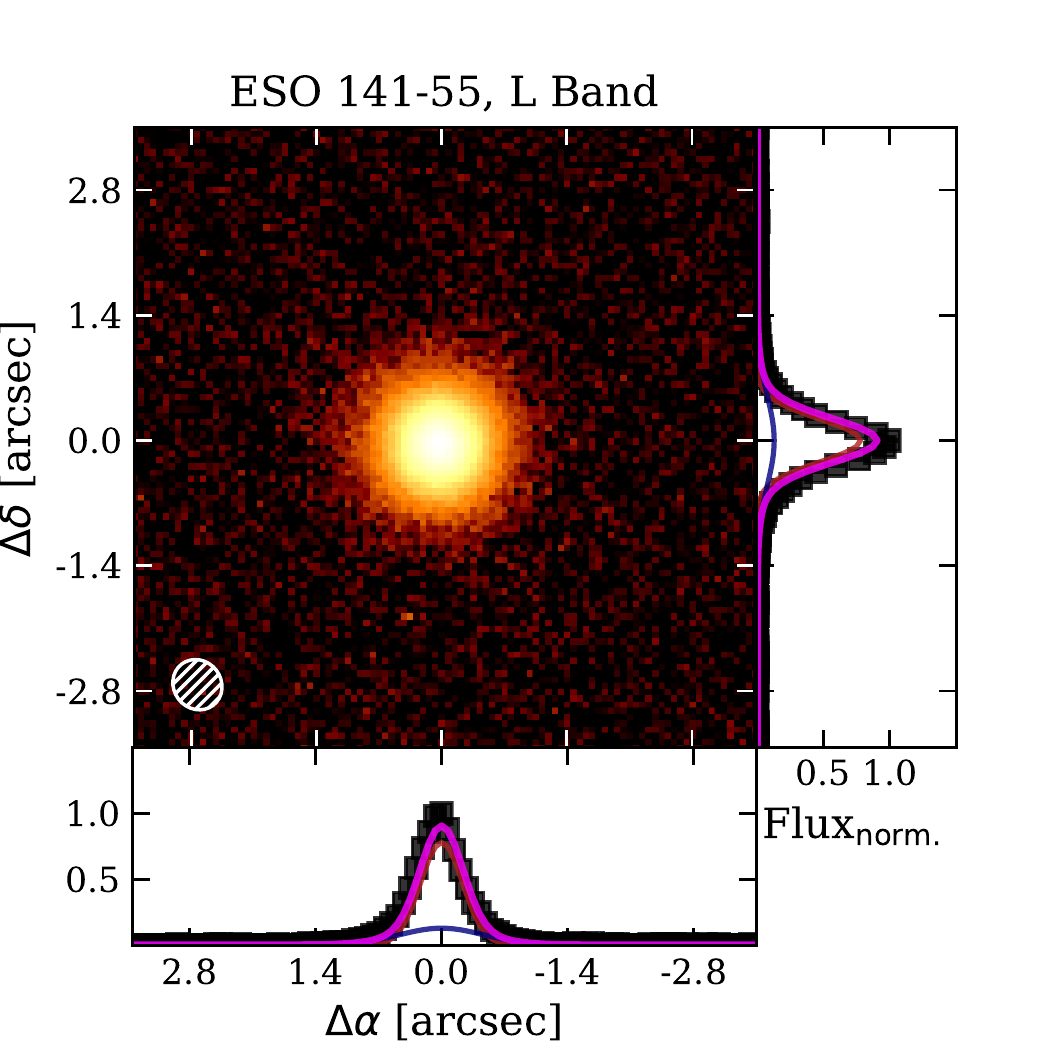}}
\subfloat{\includegraphics[width=0.25\hsize]{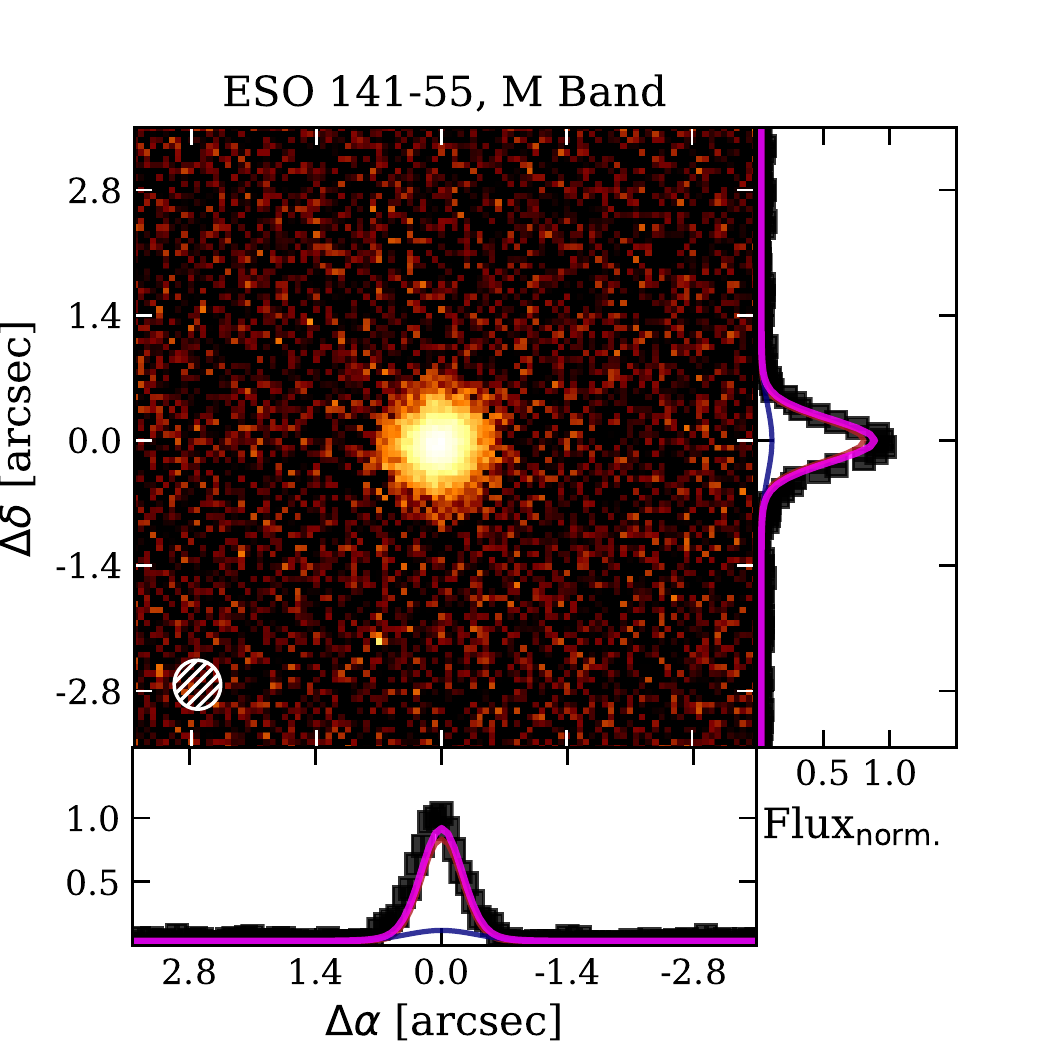}}
\subfloat{\includegraphics[width=0.25\hsize]{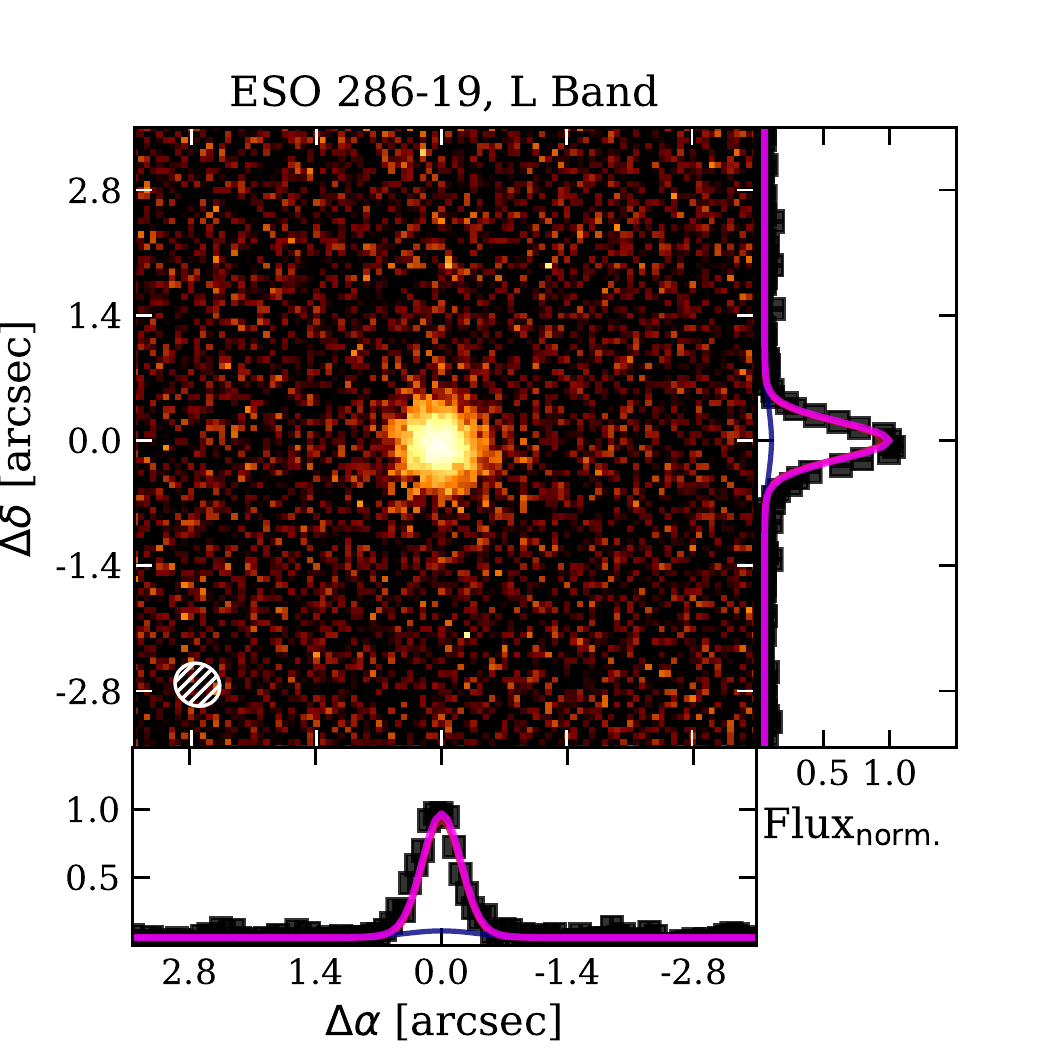}} \\
\subfloat{\includegraphics[width=0.25\hsize]{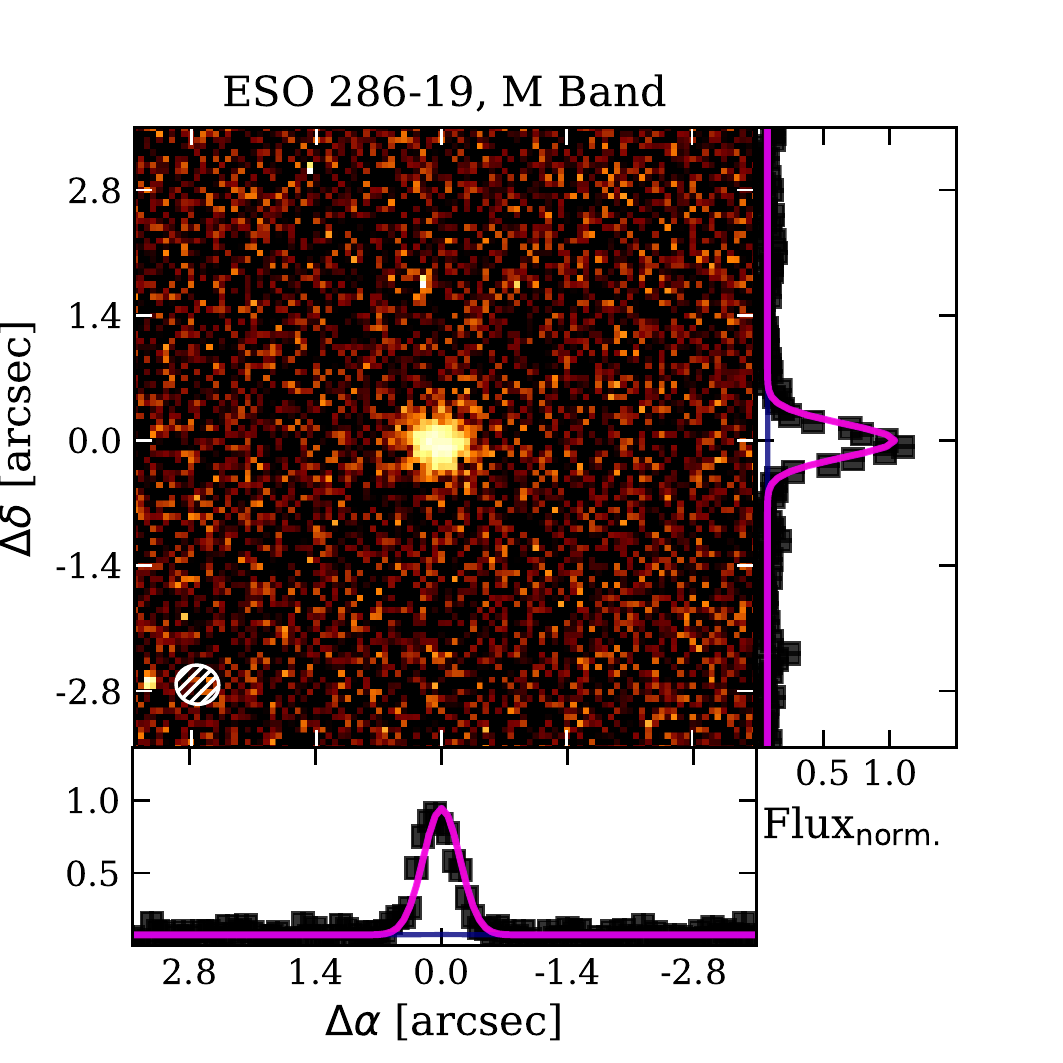}}
\subfloat{\includegraphics[width=0.25\hsize]{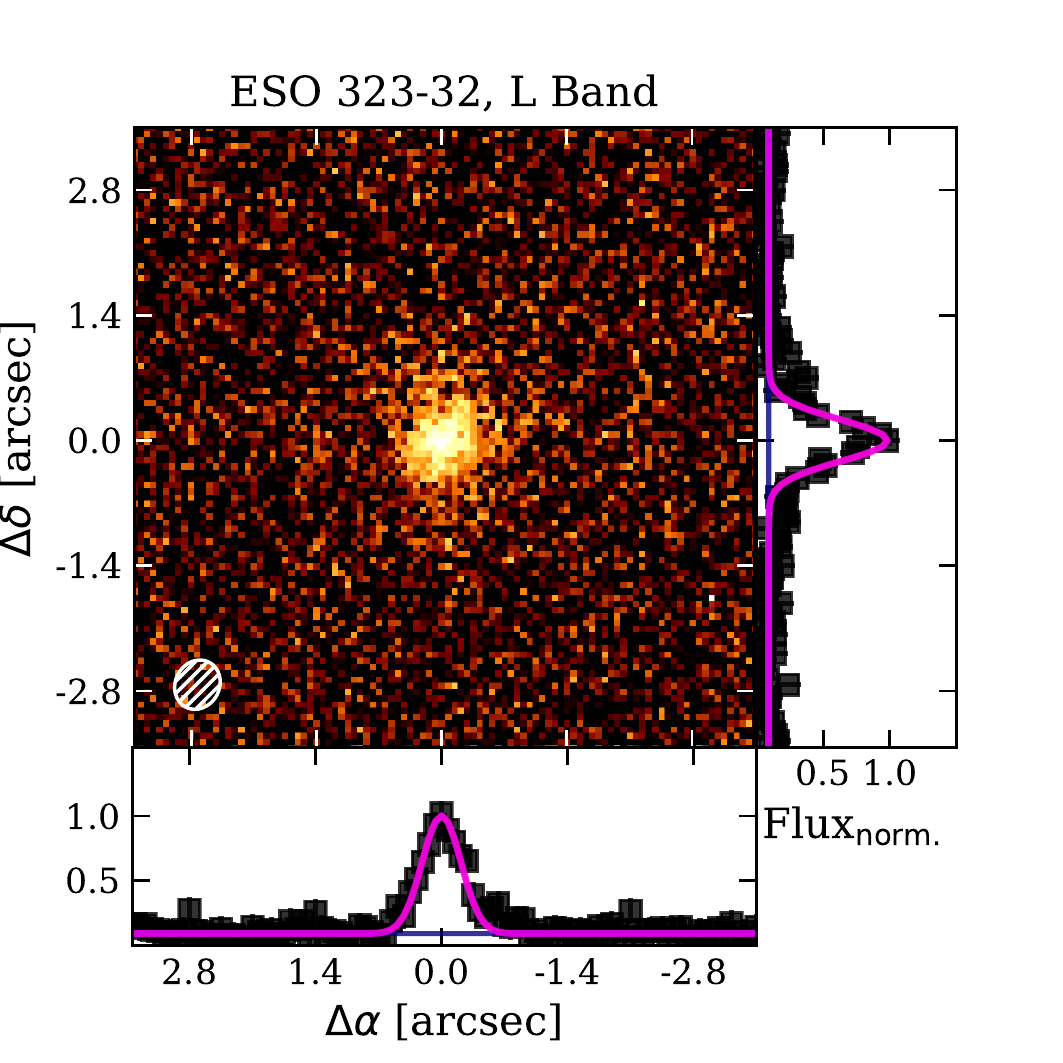}}
\subfloat{\includegraphics[width=0.25\hsize]{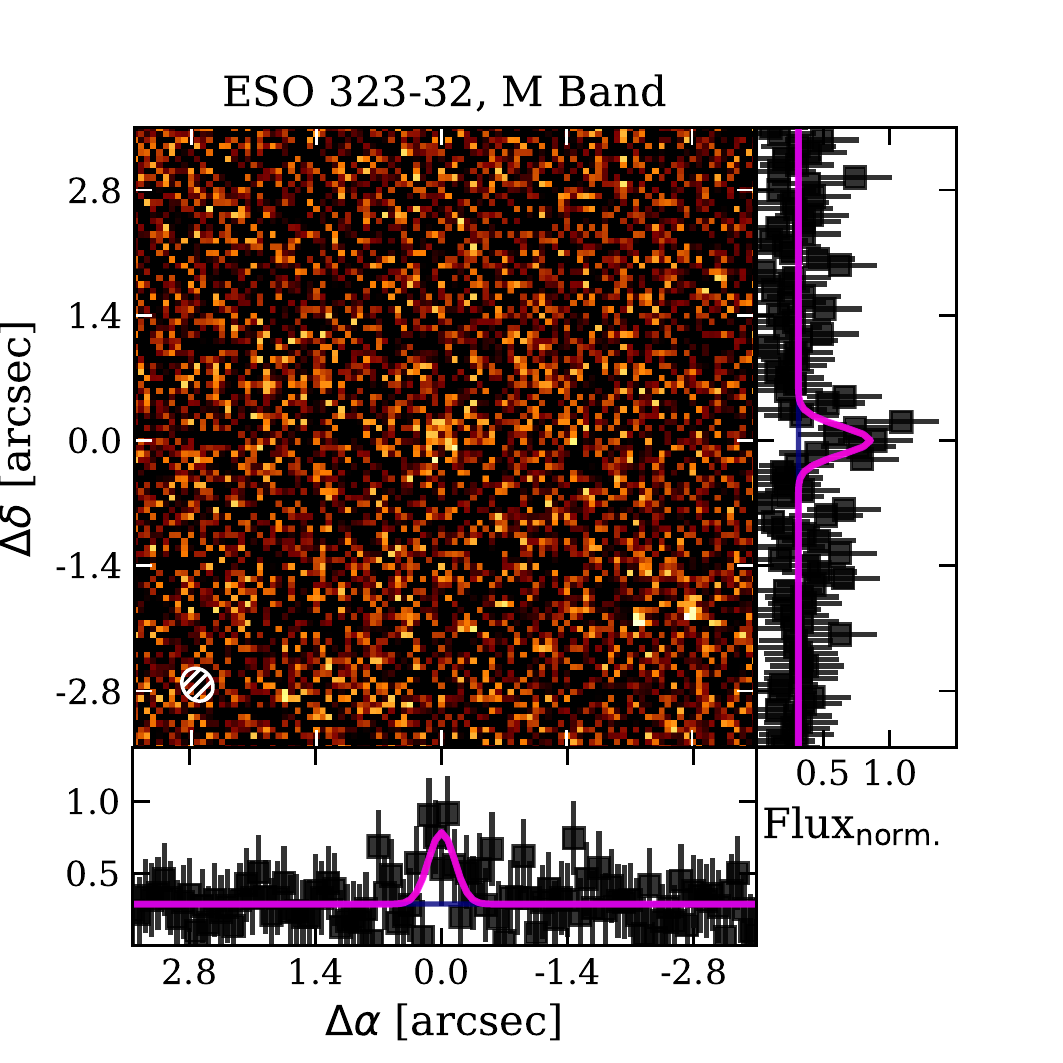}}
\subfloat{\includegraphics[width=0.25\hsize]{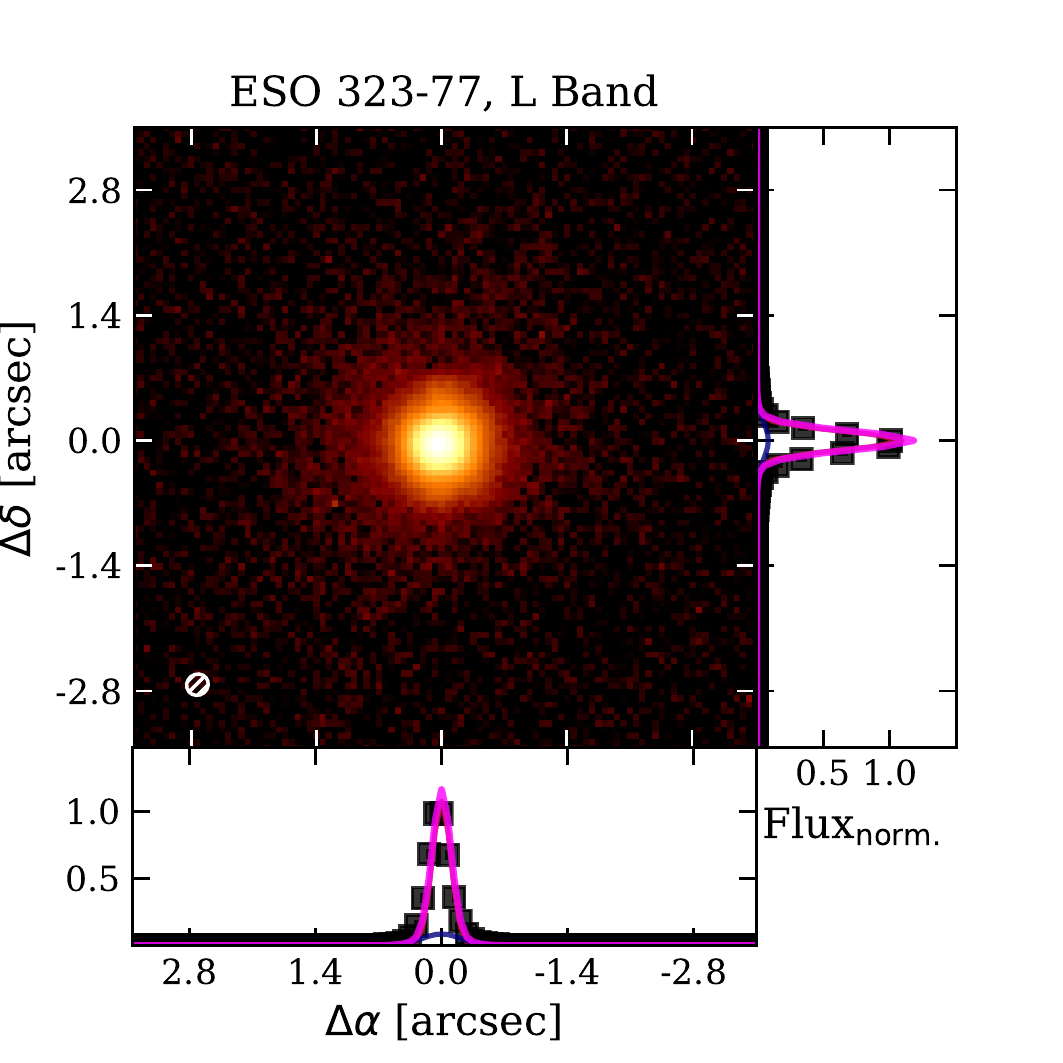}} \\
\subfloat{\includegraphics[width=0.25\hsize]{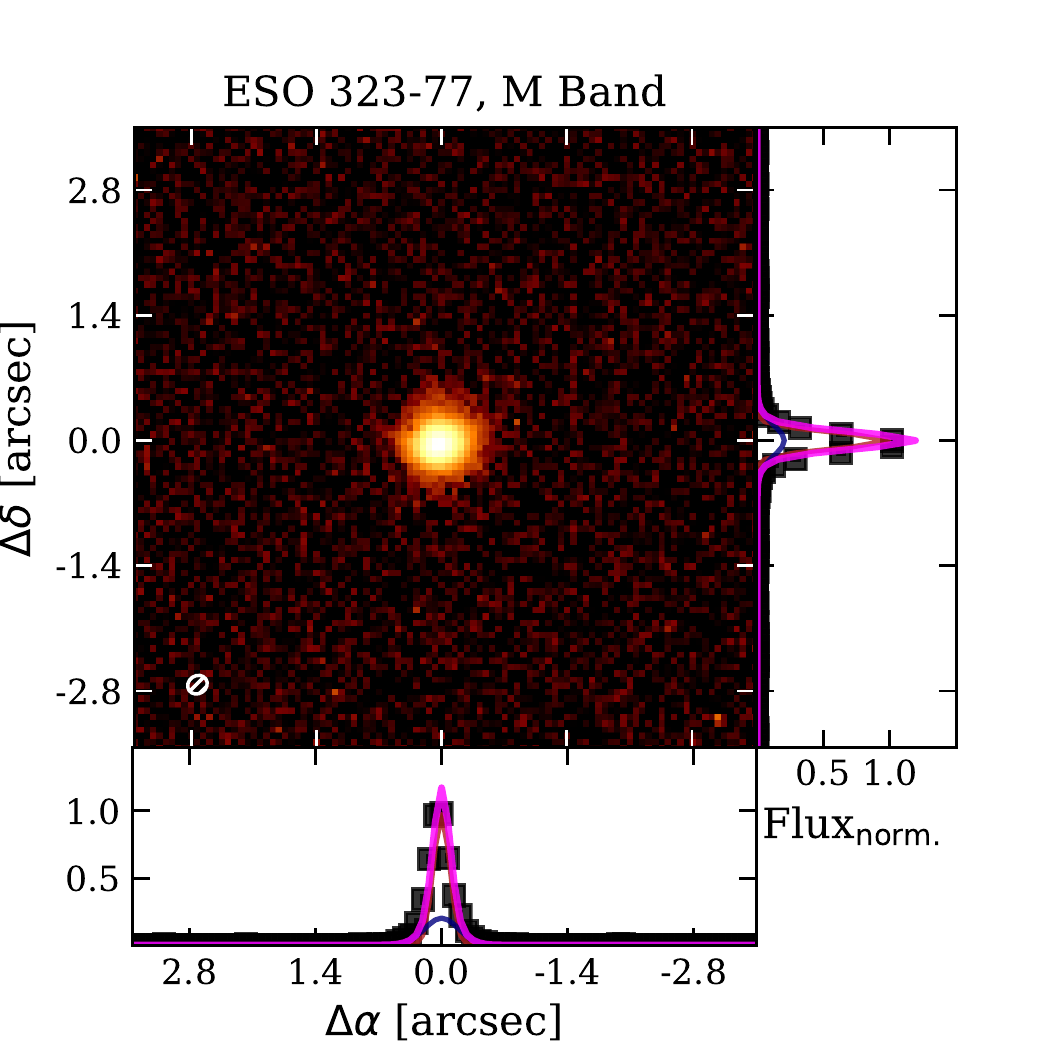}}
\subfloat{\includegraphics[width=0.25\hsize]{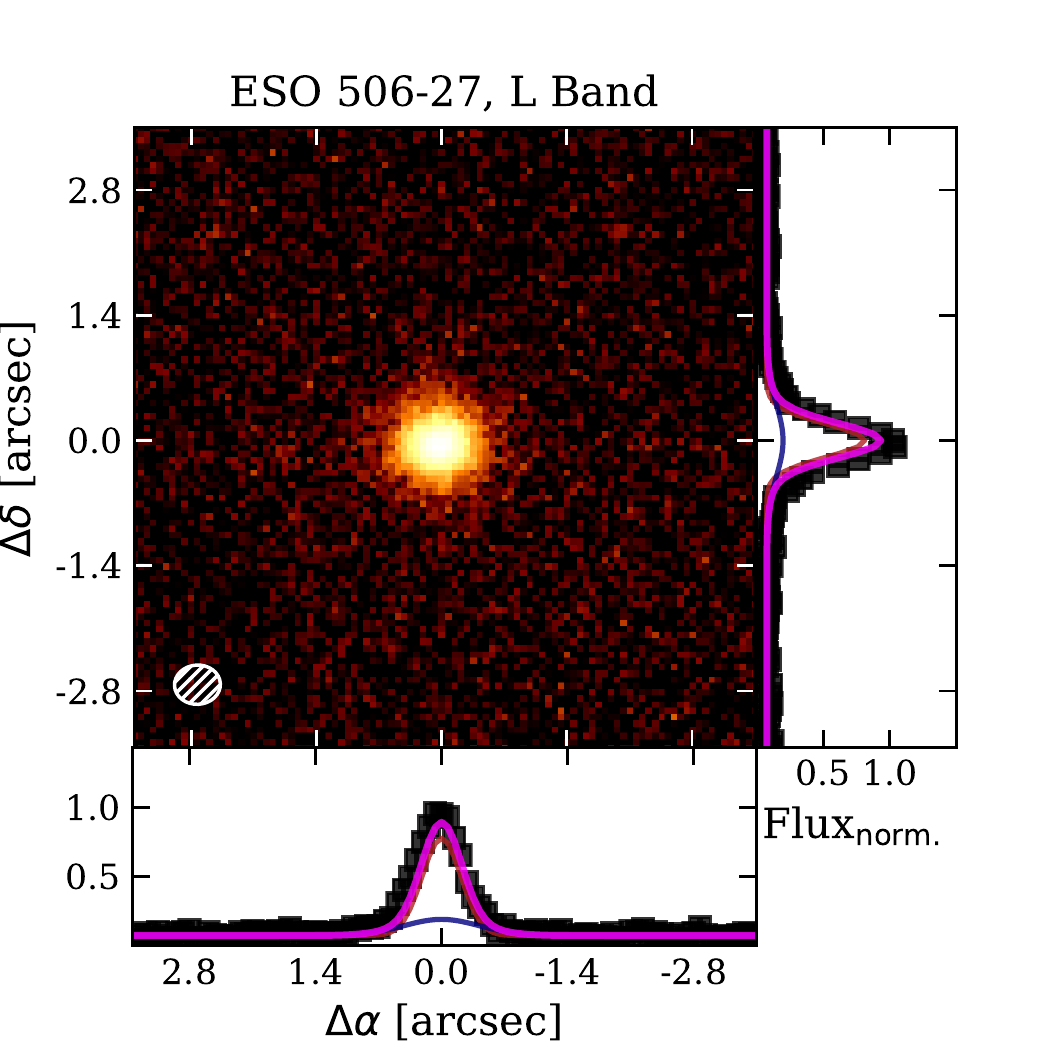}}
\subfloat{\includegraphics[width=0.25\hsize]{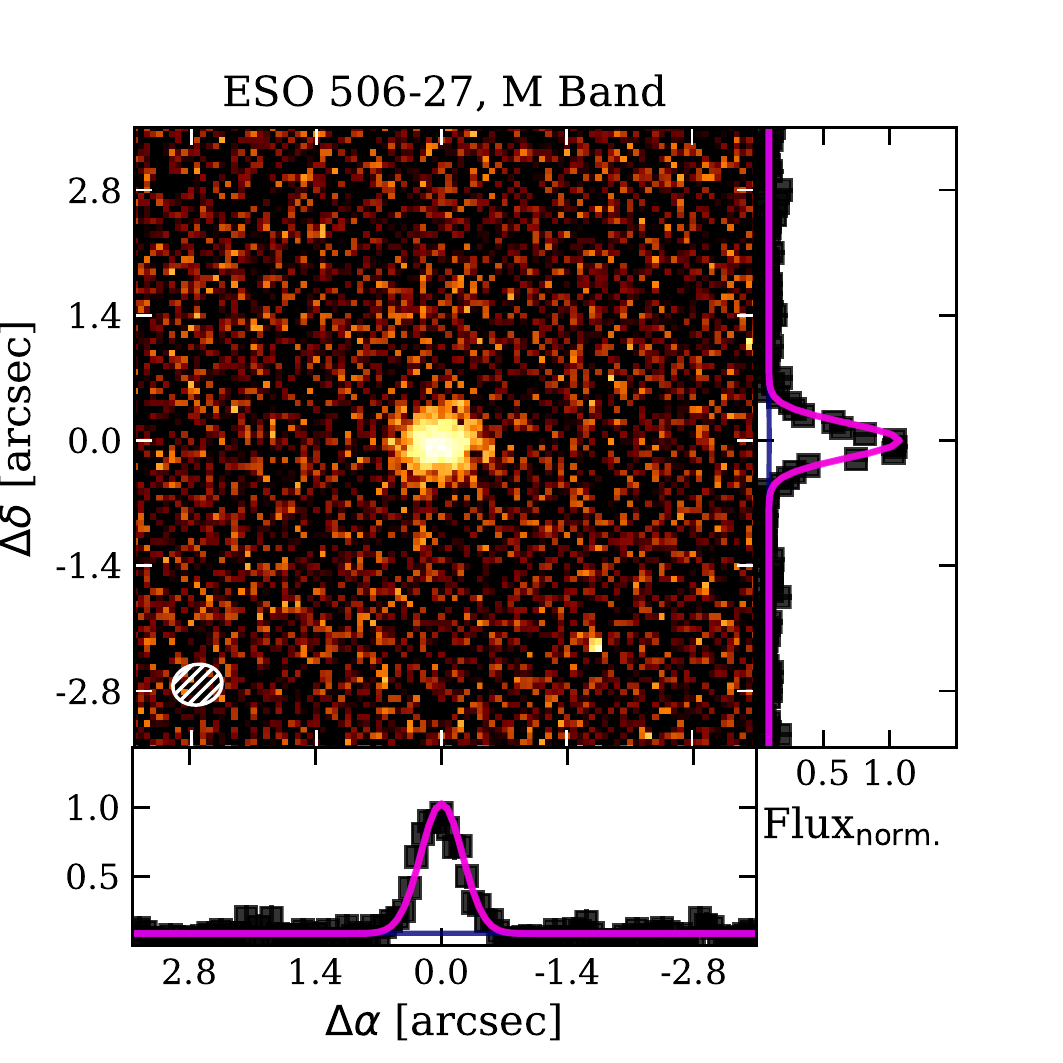}}
\subfloat{\includegraphics[width=0.25\hsize]{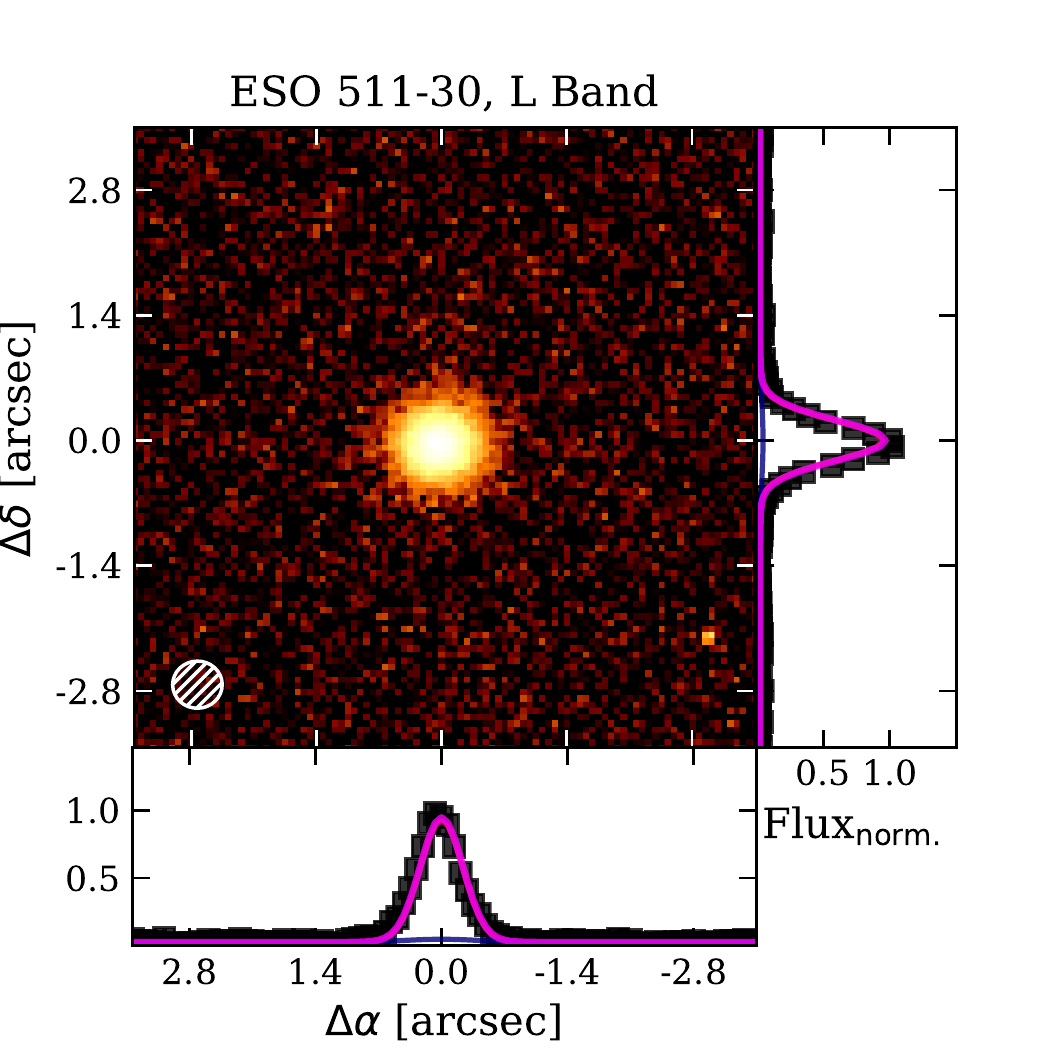}} \\
\subfloat{\includegraphics[width=0.25\hsize]{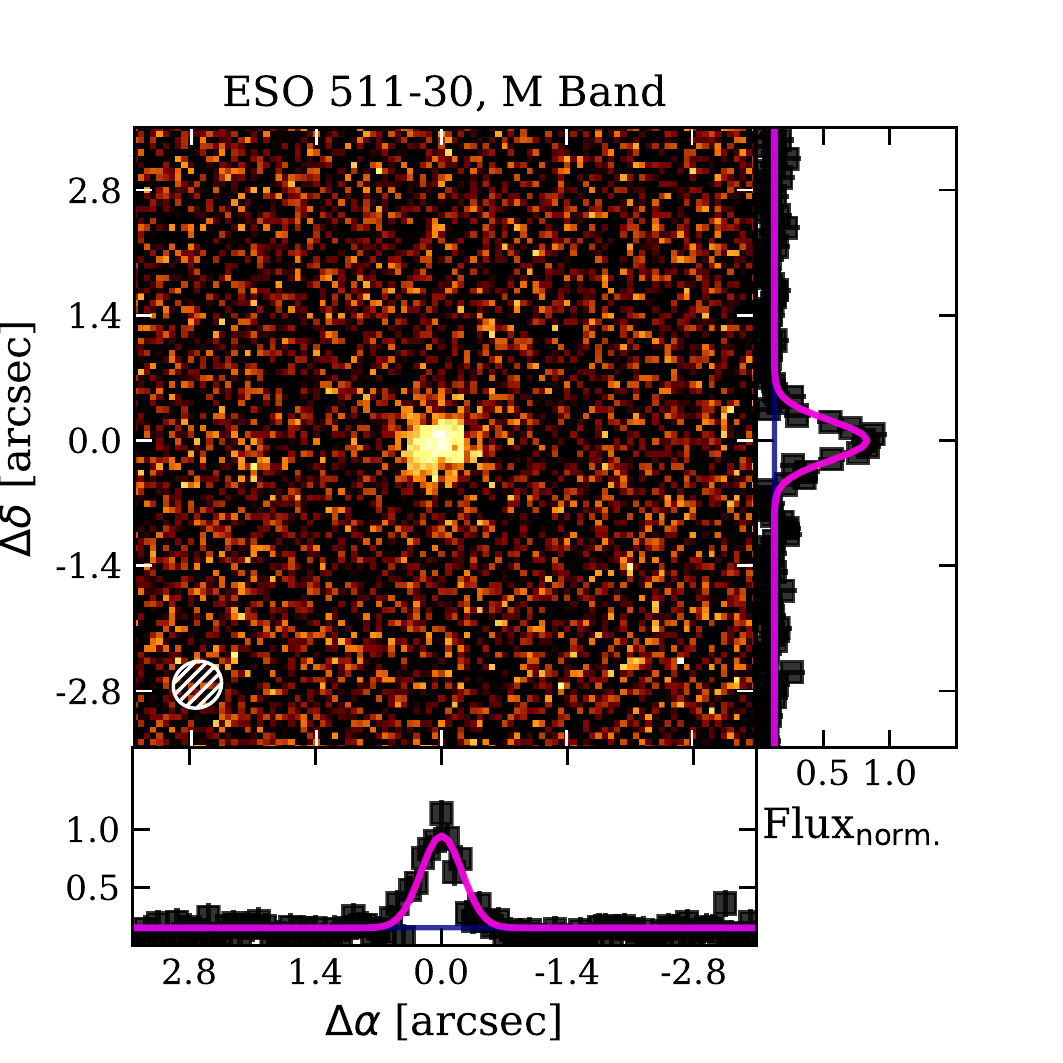}}
\subfloat{\includegraphics[width=0.25\hsize]{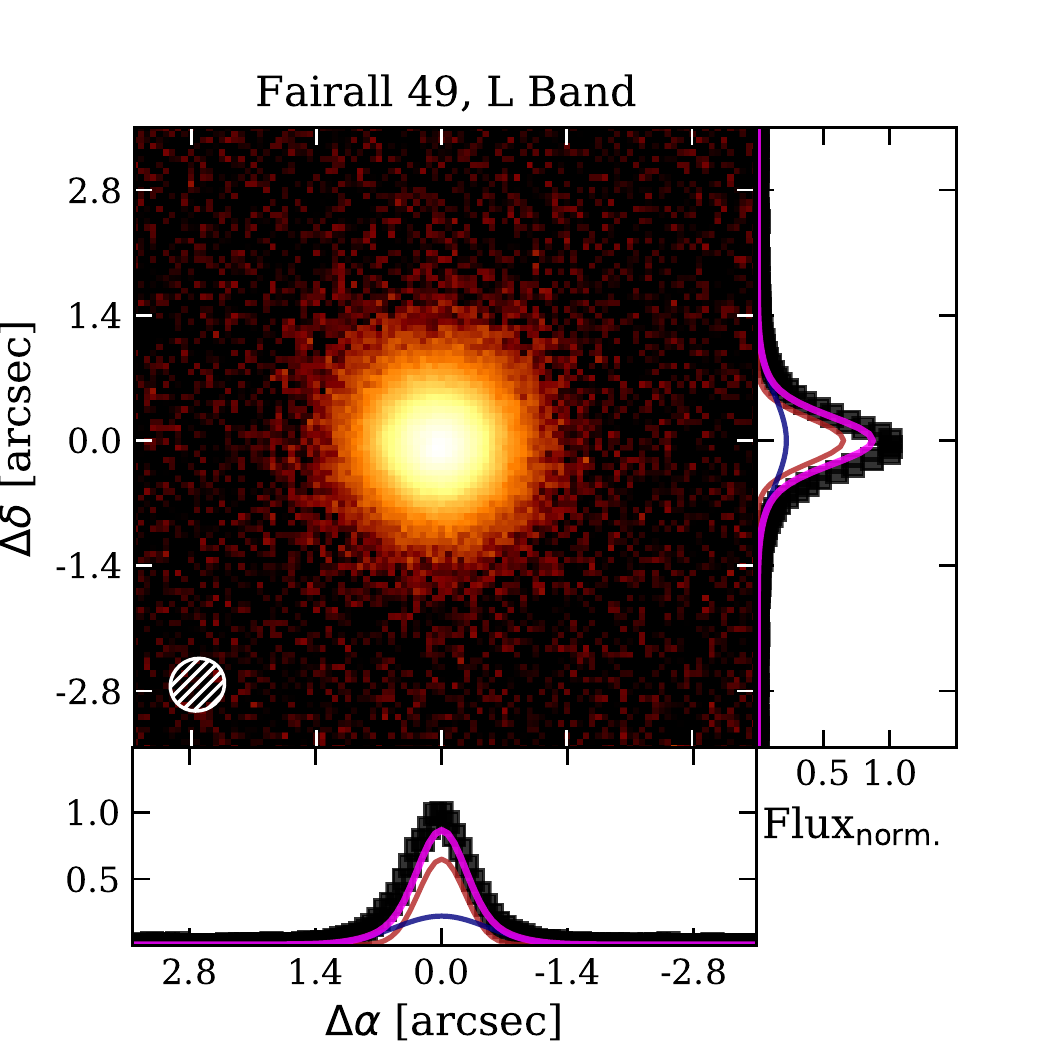}}
\subfloat{\includegraphics[width=0.25\hsize]{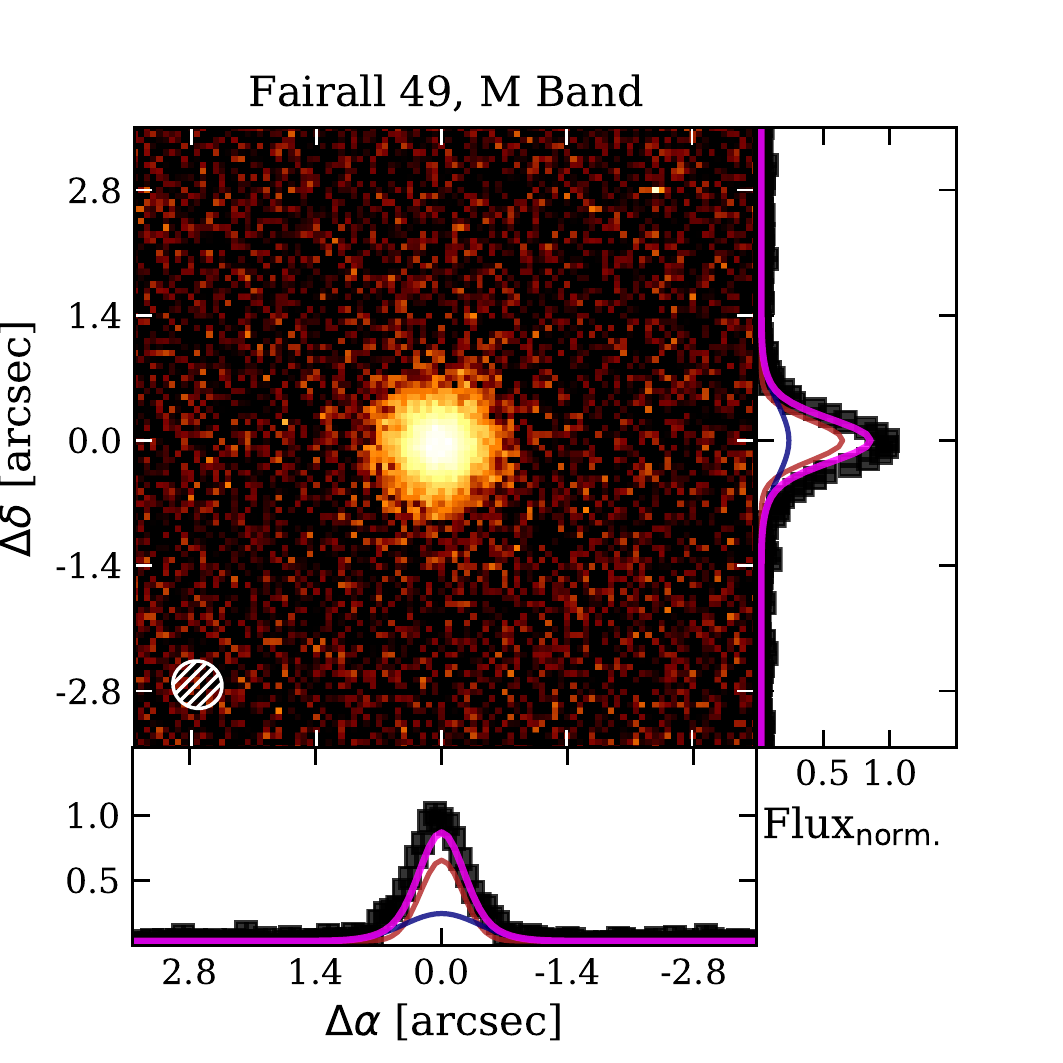}}
\subfloat{\includegraphics[width=0.25\hsize]{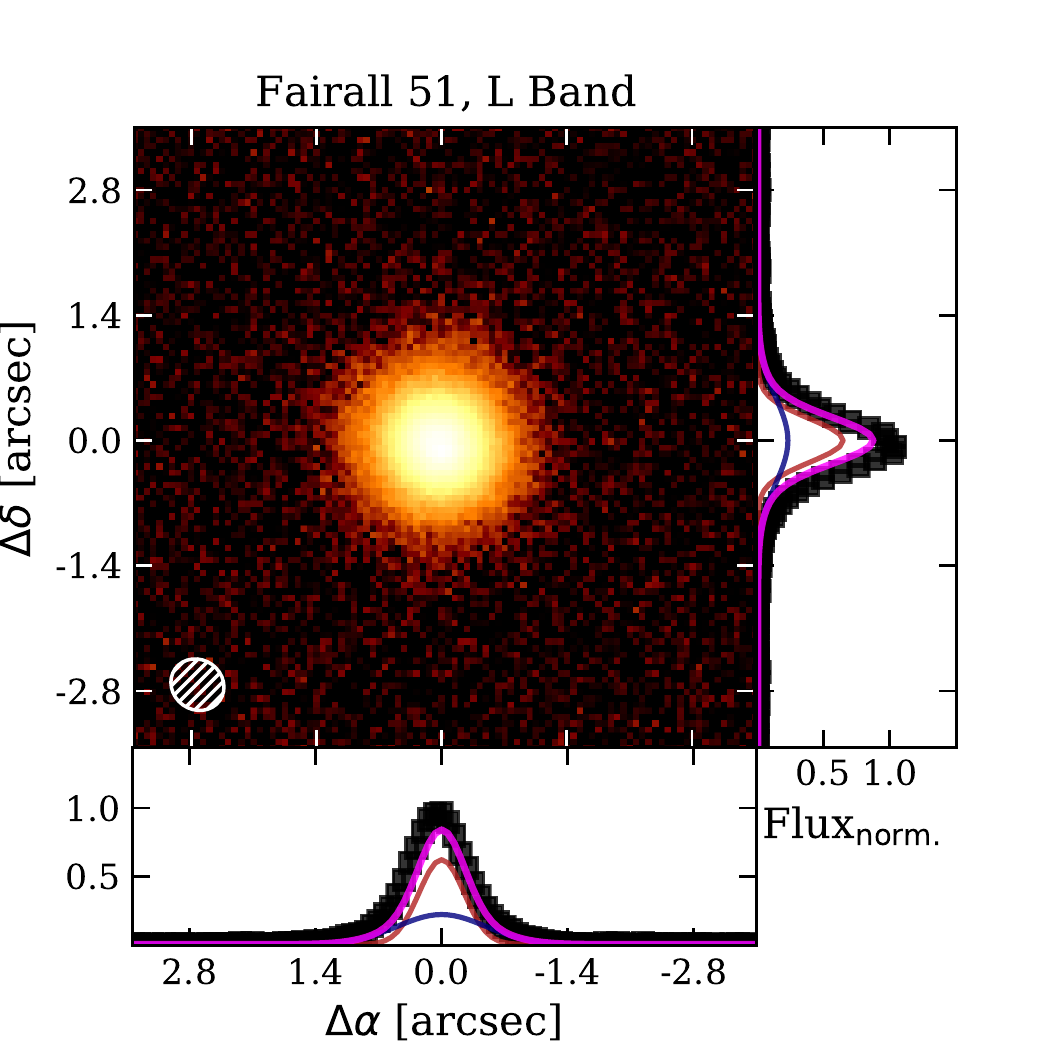}} \\
\caption{As Fig. \ref{fig:cutouts_one}.}
\end{figure*}
\begin{figure*}
\centering
\subfloat{\includegraphics[width=0.25\hsize]{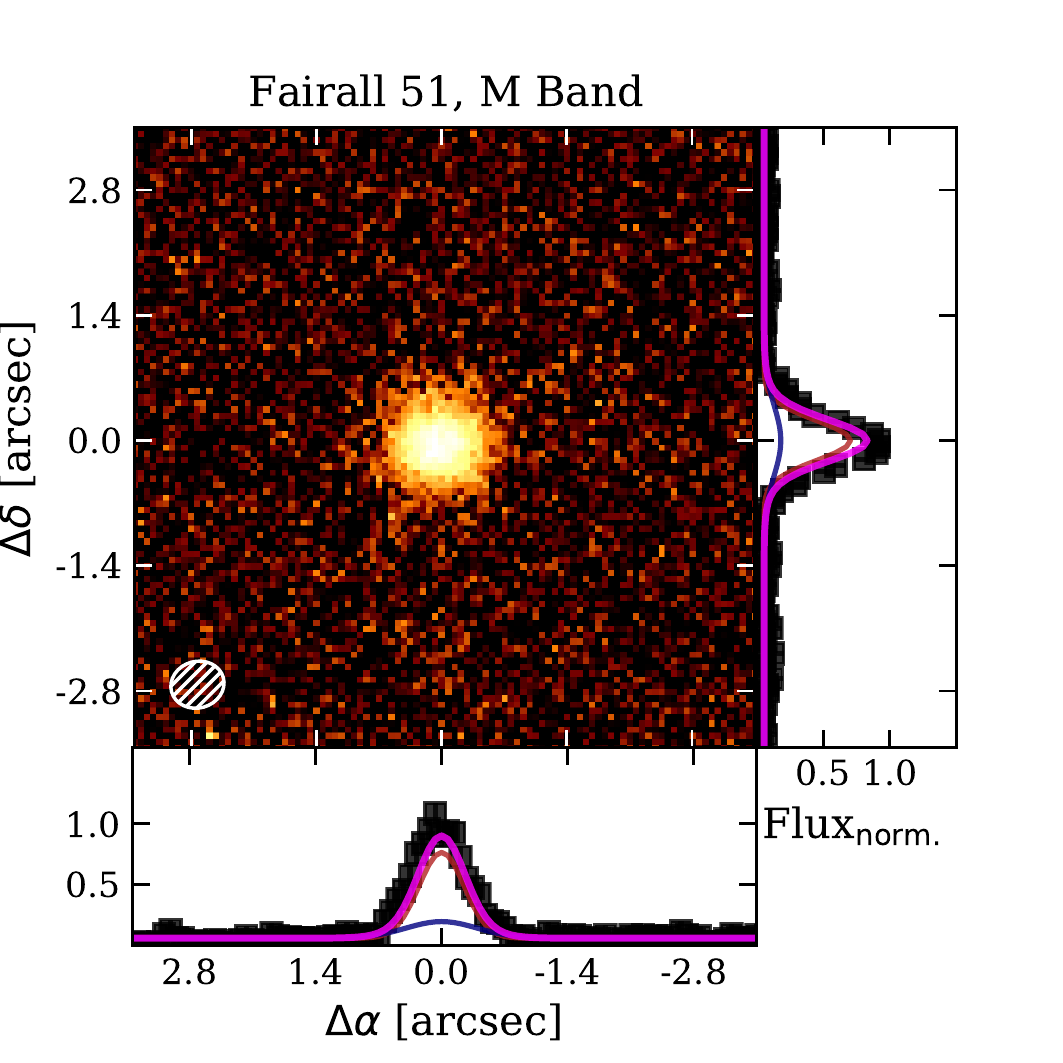}}
\subfloat{\includegraphics[width=0.25\hsize]{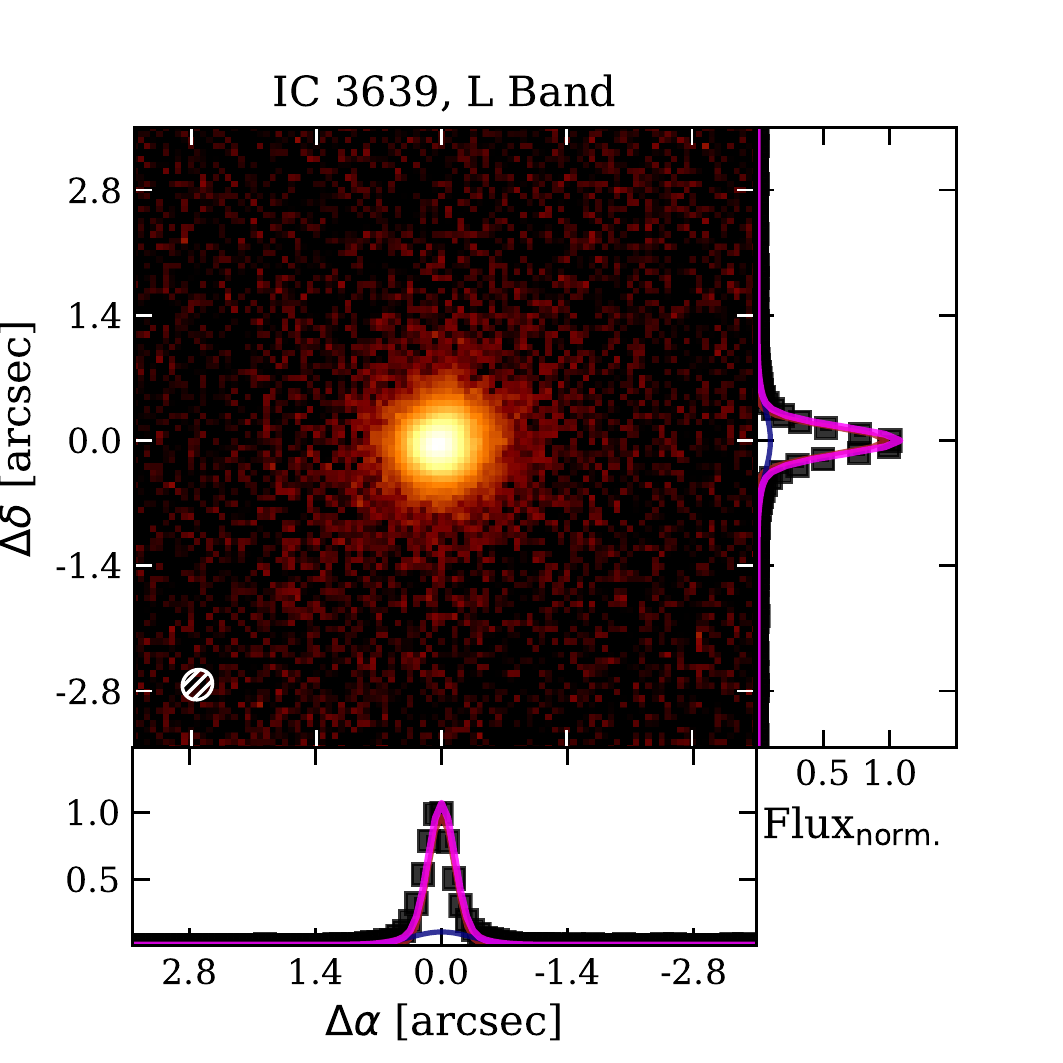}}
\subfloat{\includegraphics[width=0.25\hsize]{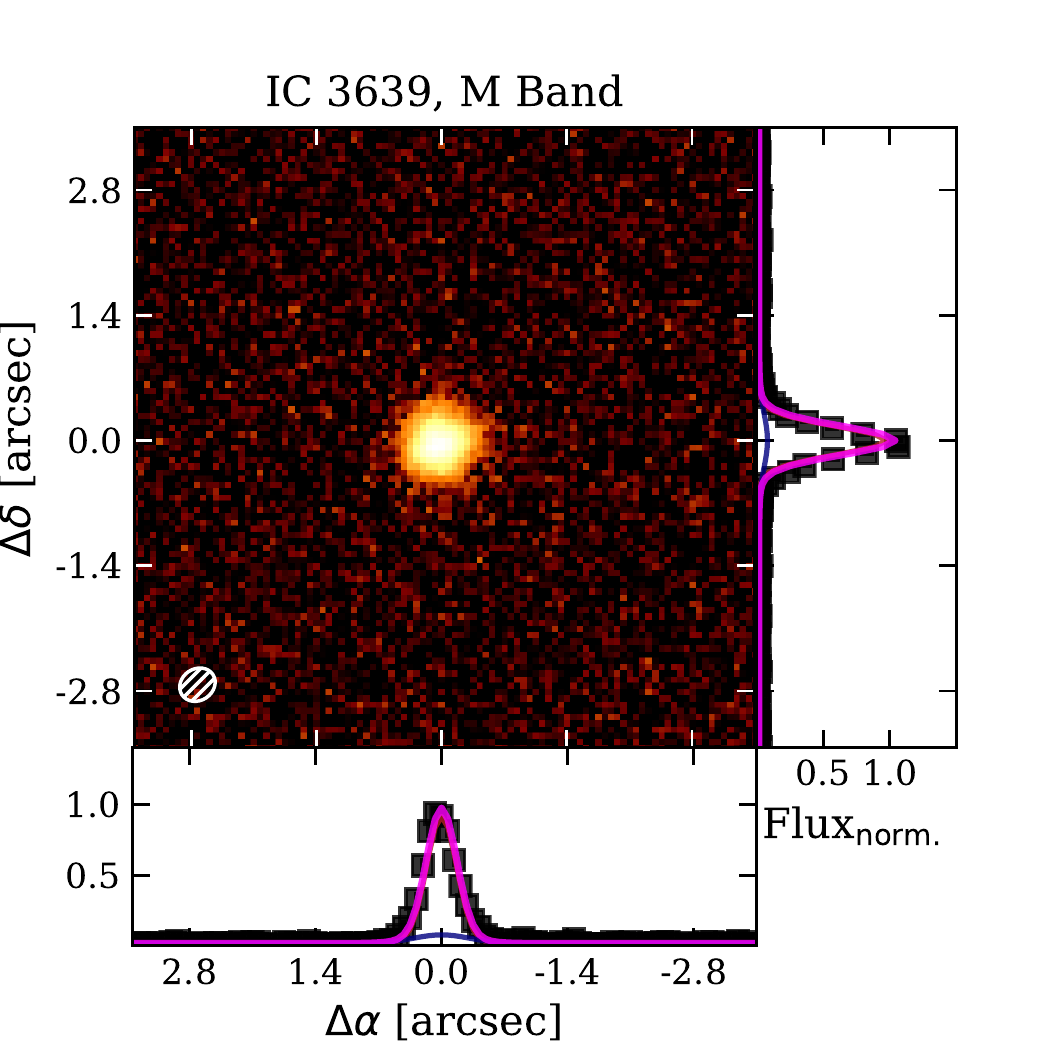}}
\subfloat{\includegraphics[width=0.25\hsize]{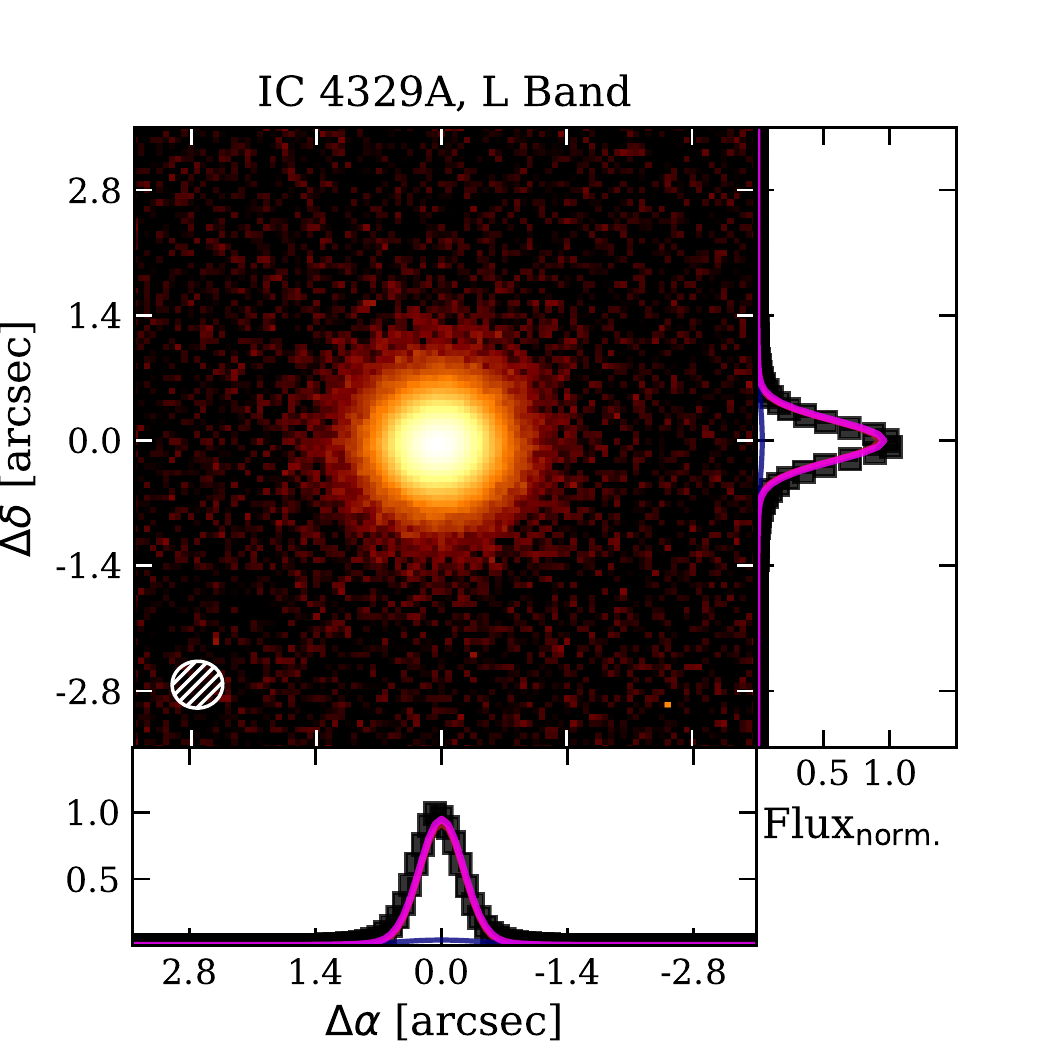}} \\
\subfloat{\includegraphics[width=0.25\hsize]{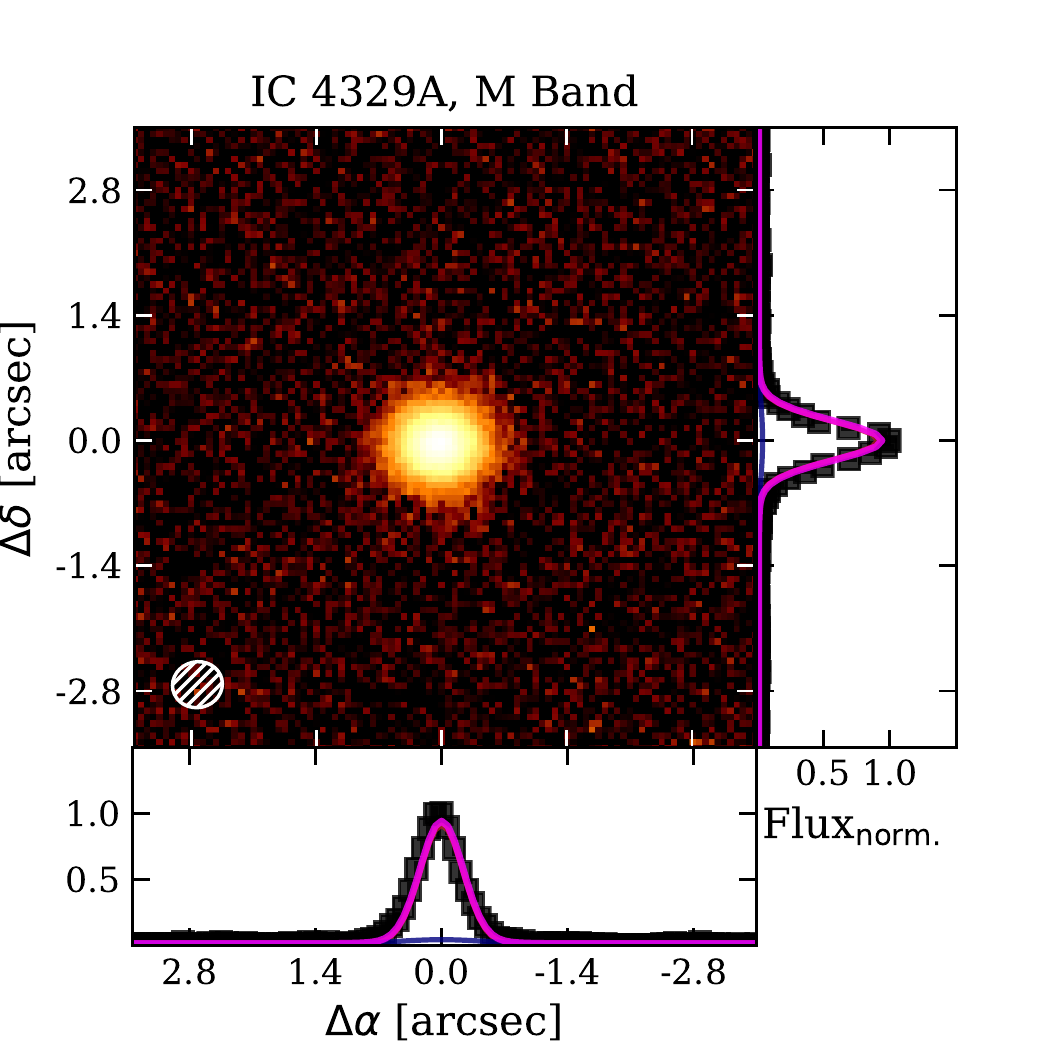}}
\subfloat{\includegraphics[width=0.25\hsize]{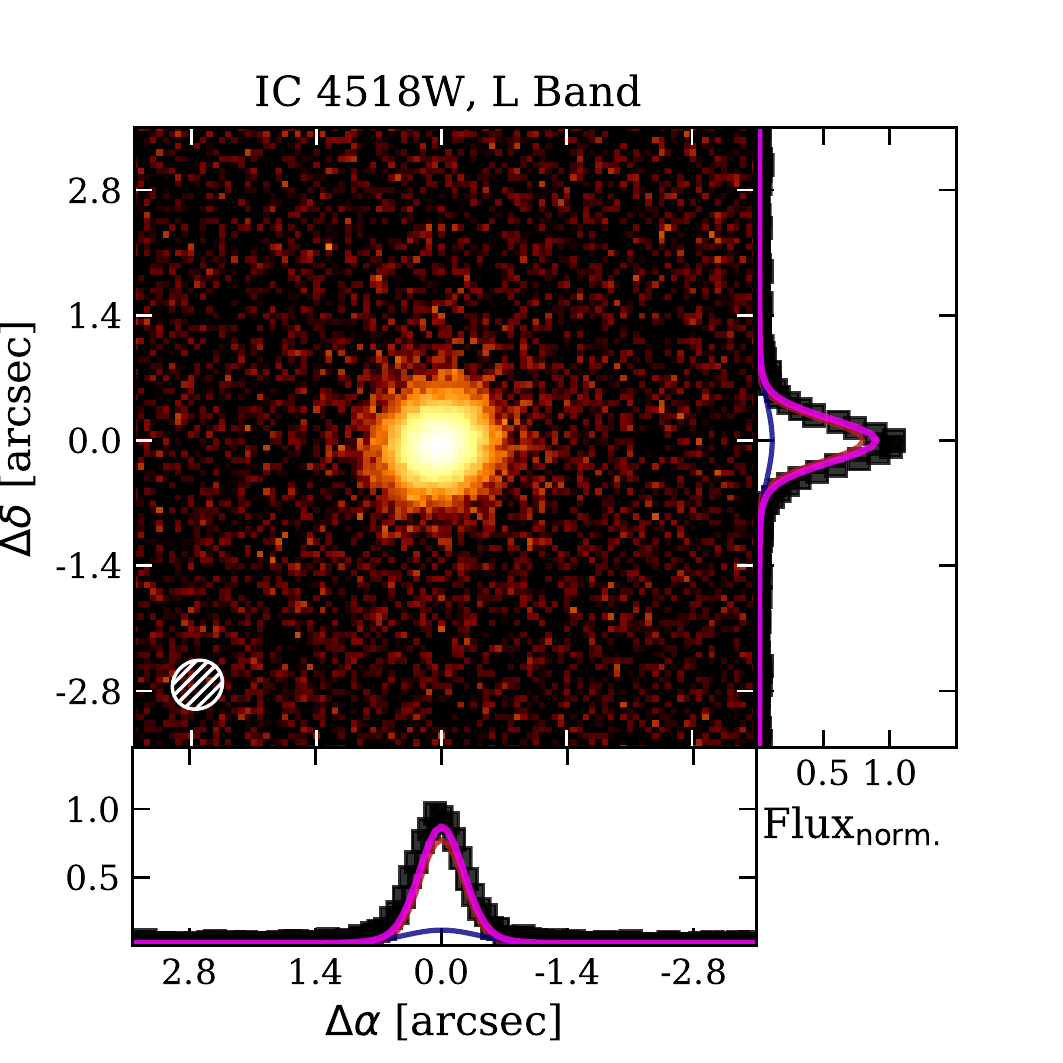}}
\subfloat{\includegraphics[width=0.25\hsize]{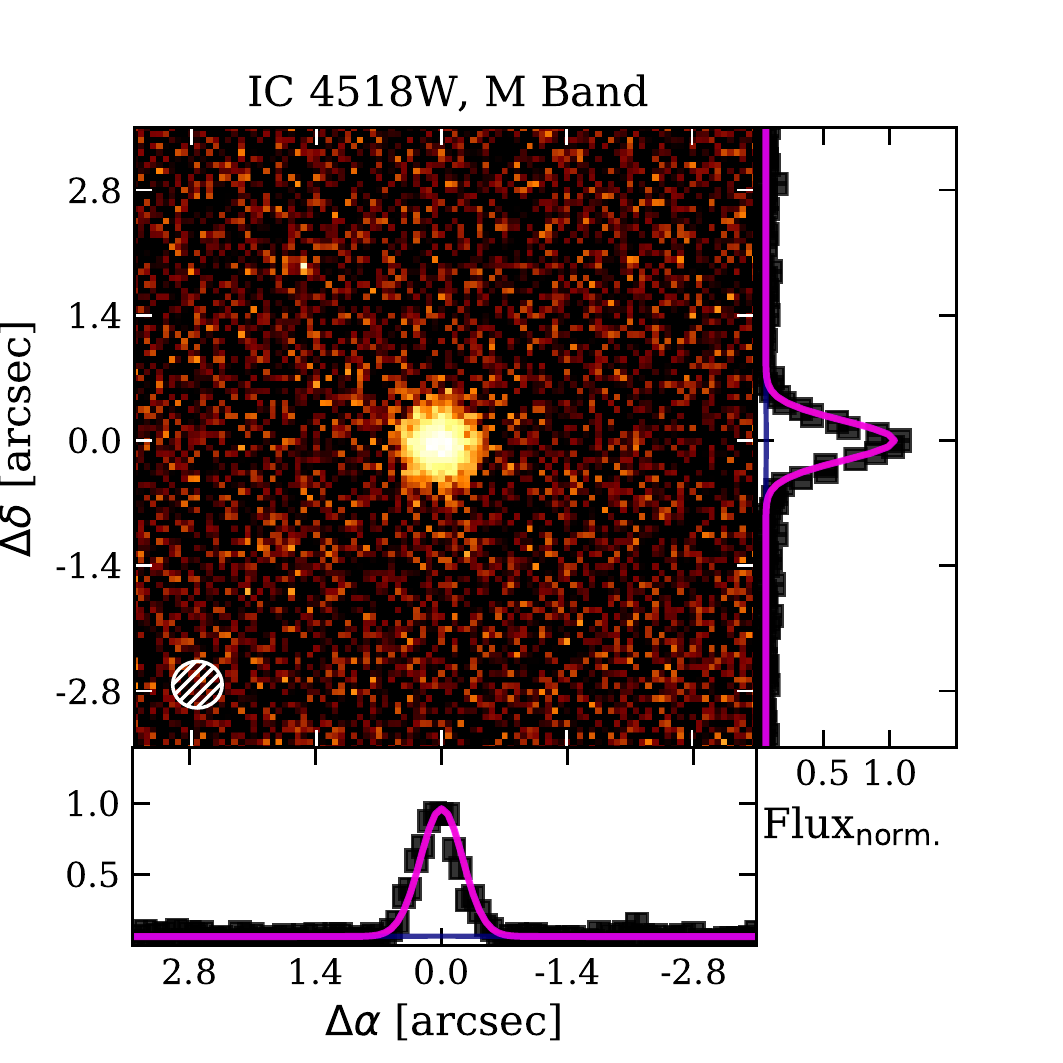}}
\subfloat{\includegraphics[width=0.25\hsize]{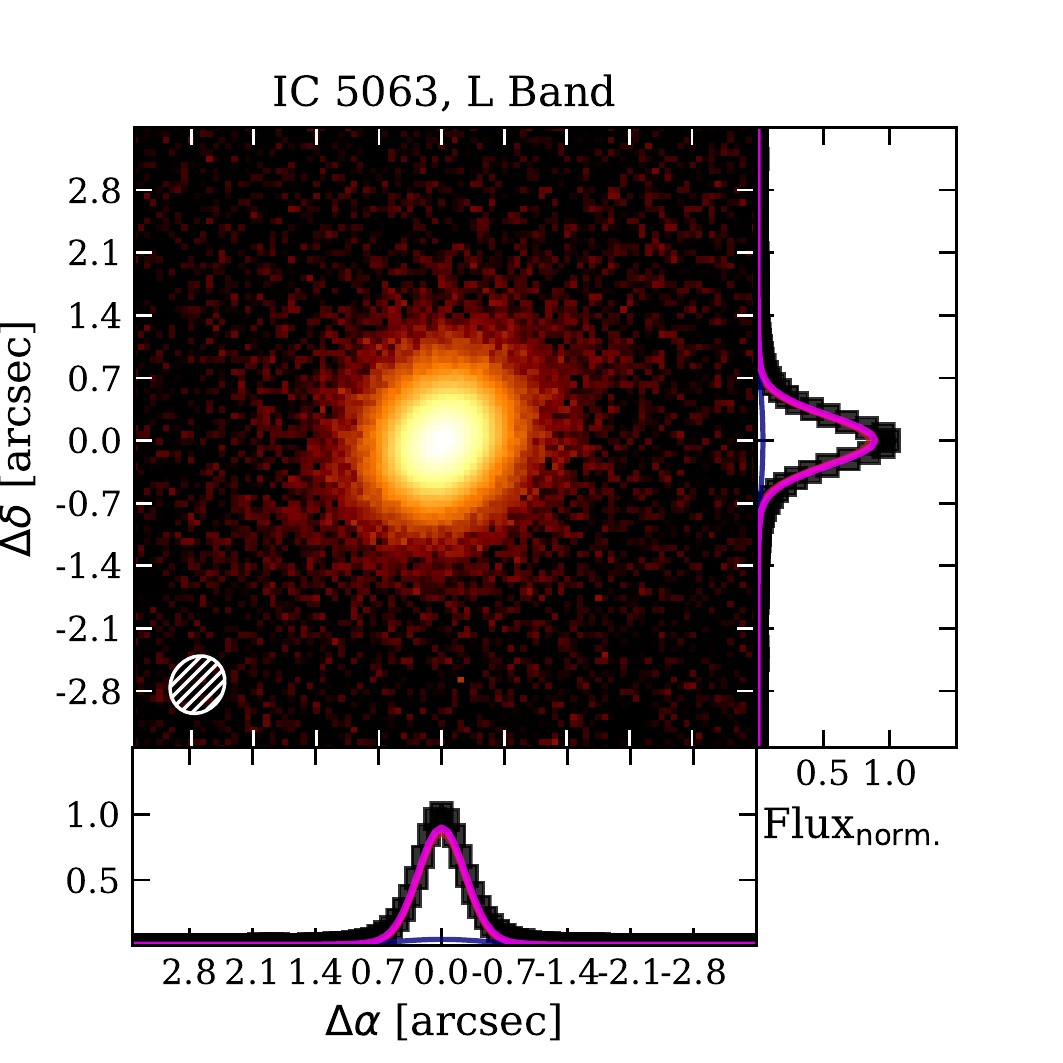}} \\
\subfloat{\includegraphics[width=0.25\hsize]{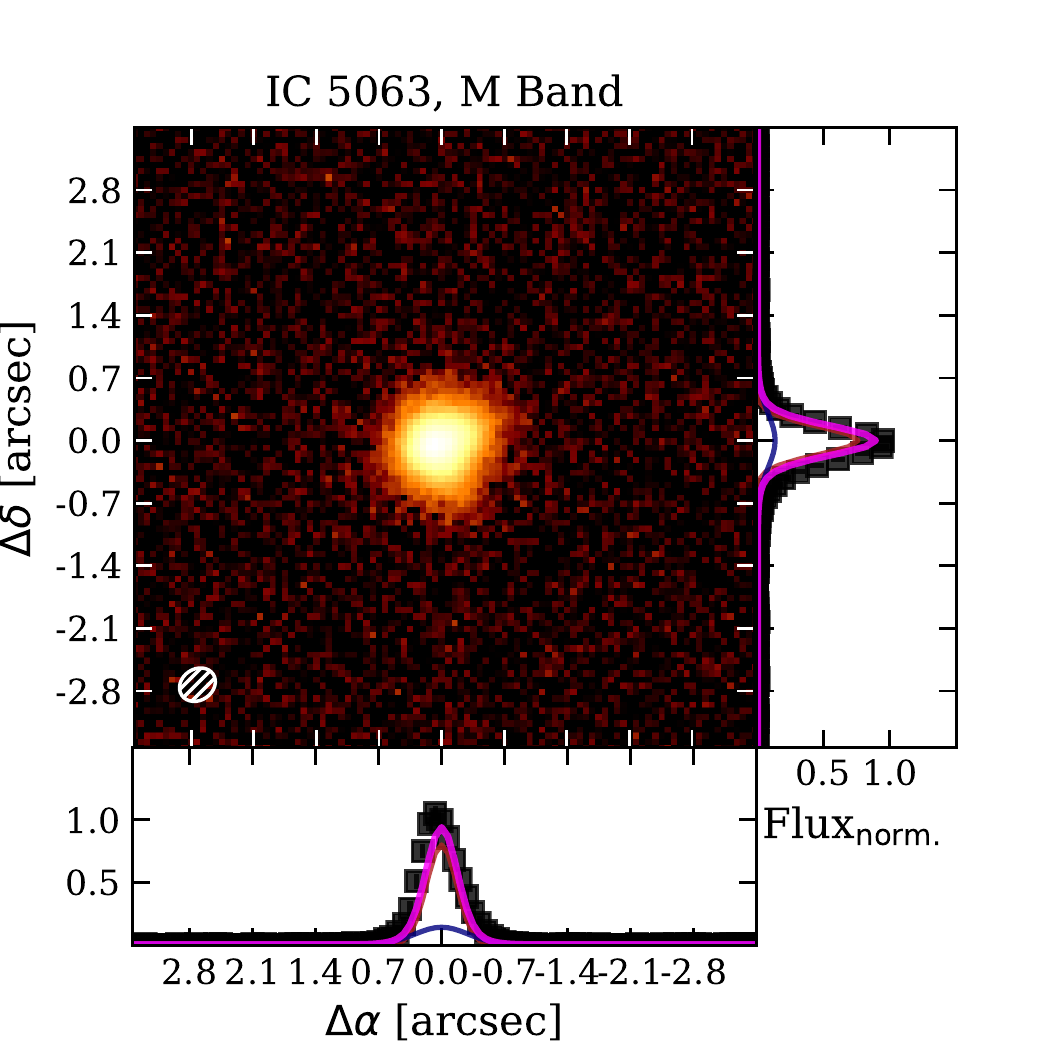}}
\subfloat{\includegraphics[width=0.25\hsize]{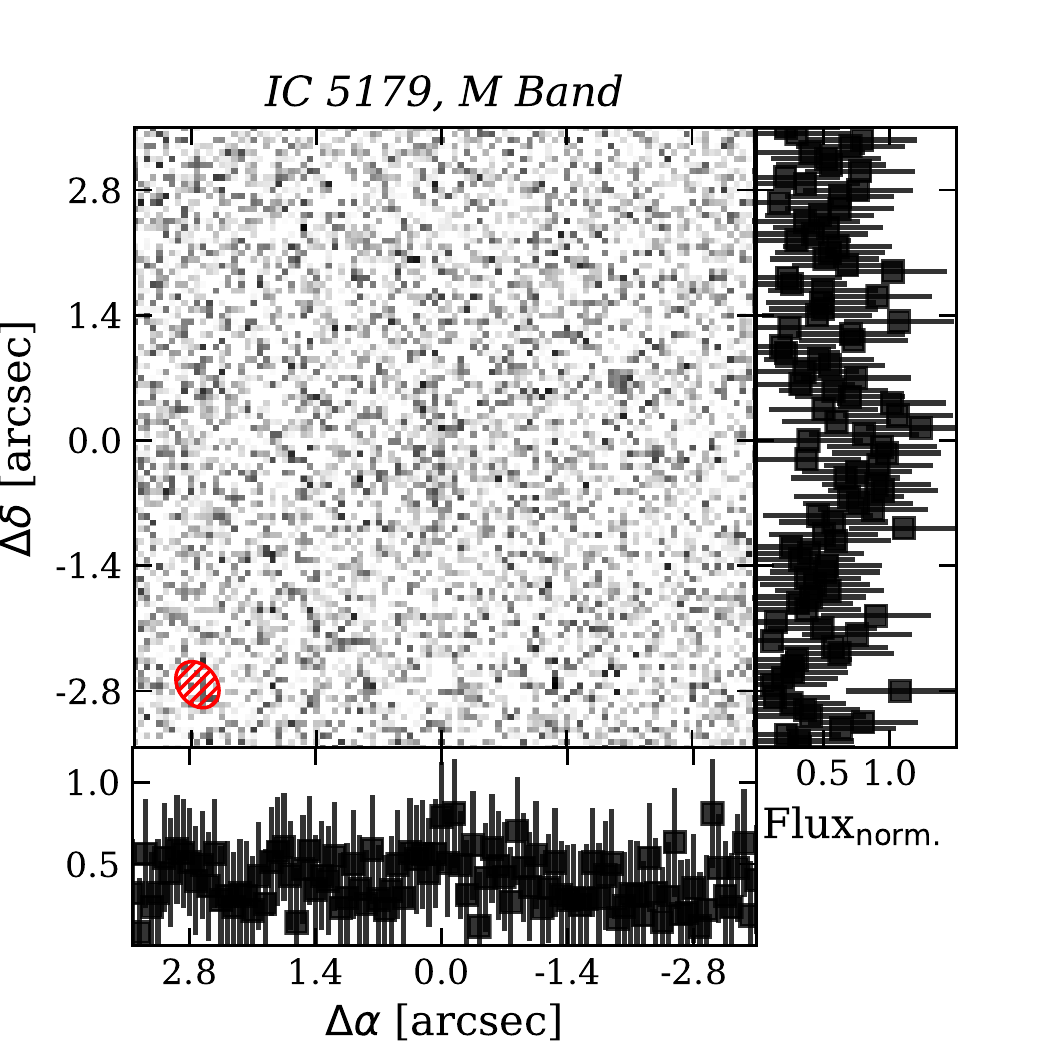}}
\subfloat{\includegraphics[width=0.25\hsize]{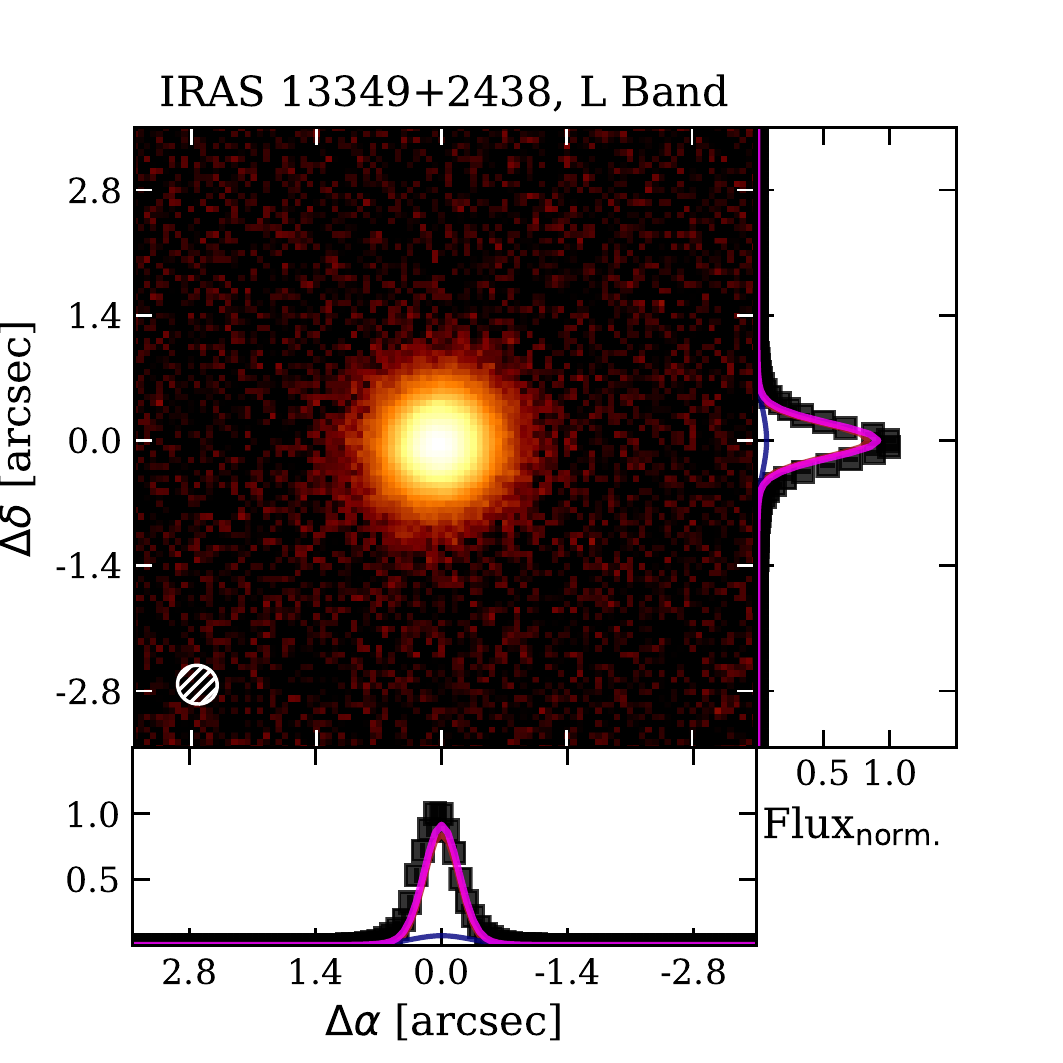}}
\subfloat{\includegraphics[width=0.25\hsize]{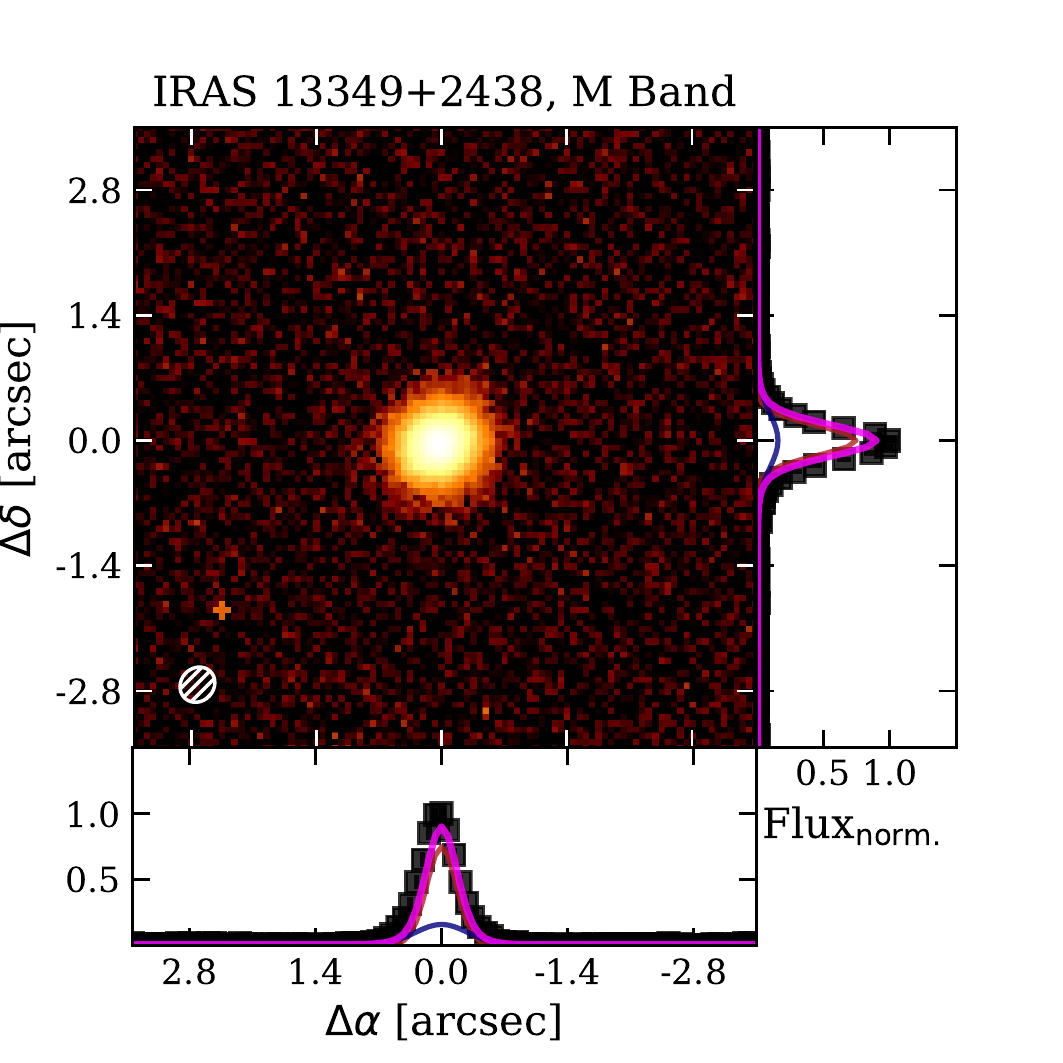}} \\
\subfloat{\includegraphics[width=0.25\hsize]{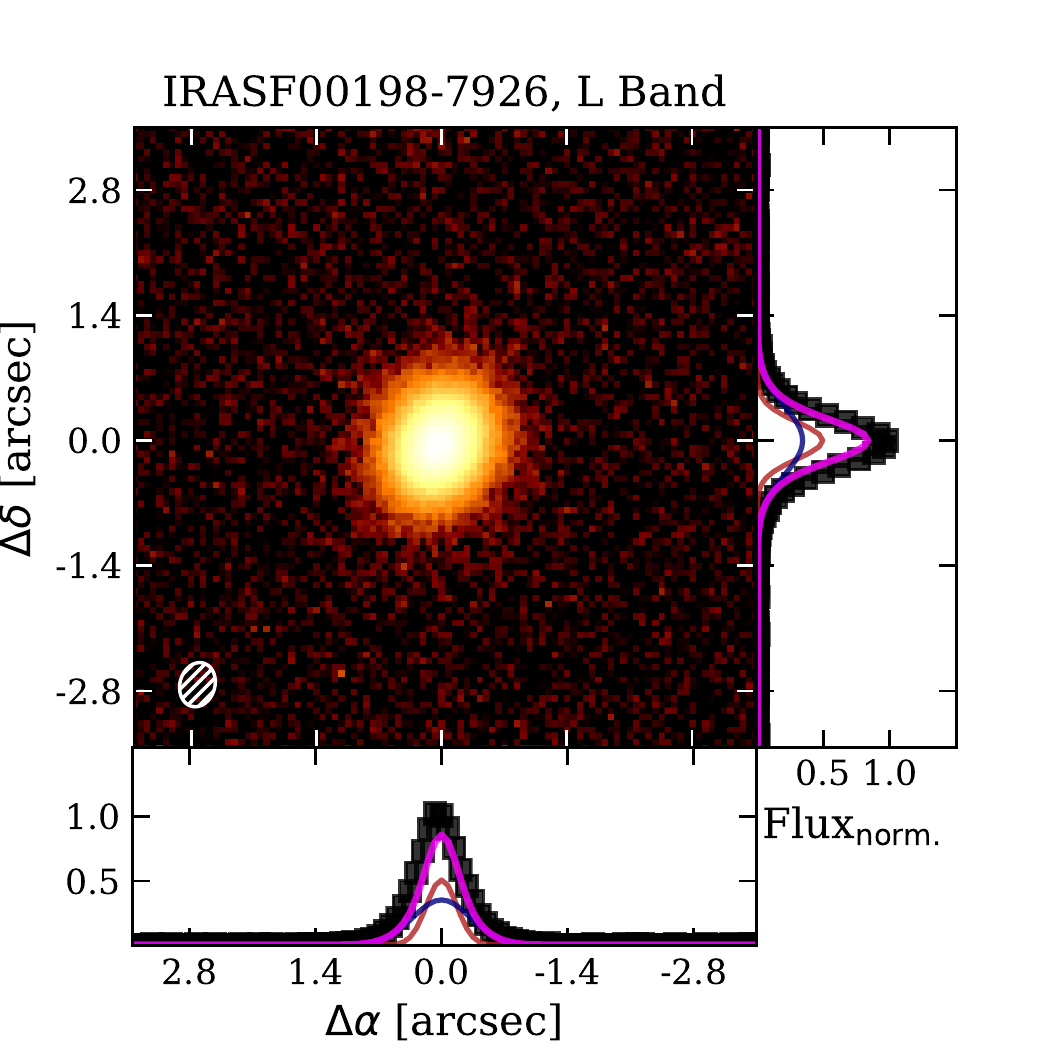}}
\subfloat{\includegraphics[width=0.25\hsize]{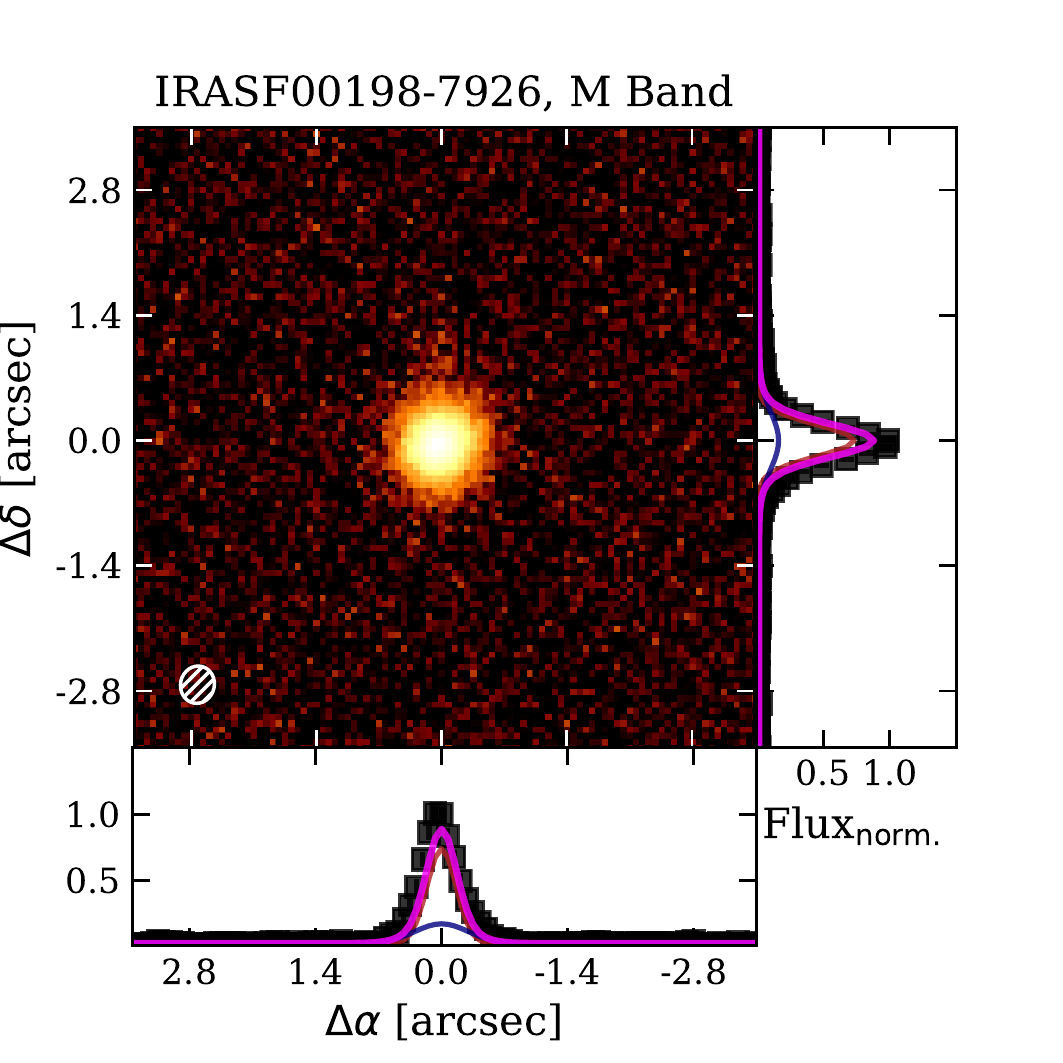}}
\subfloat{\includegraphics[width=0.25\hsize]{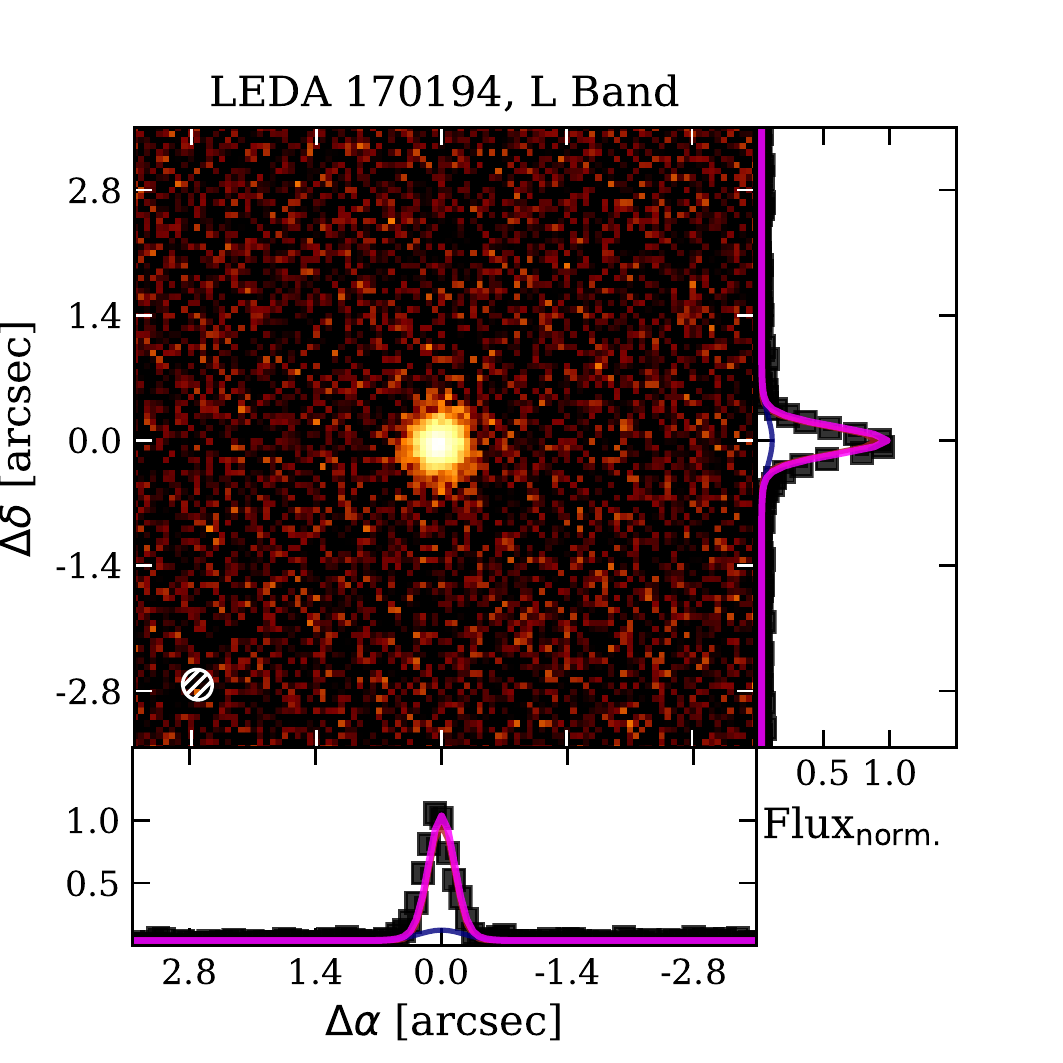}}
\subfloat{\includegraphics[width=0.25\hsize]{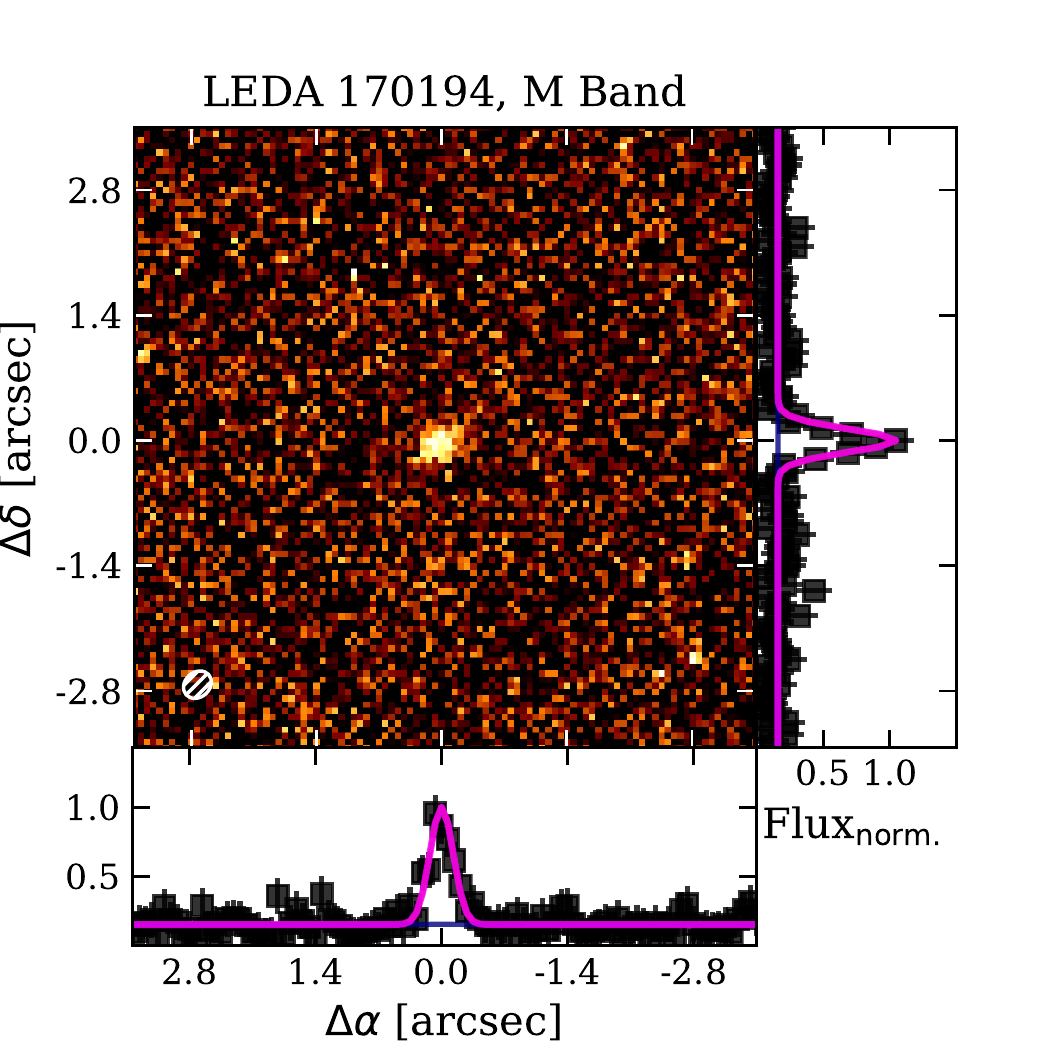}} \\
\subfloat{\includegraphics[width=0.25\hsize]{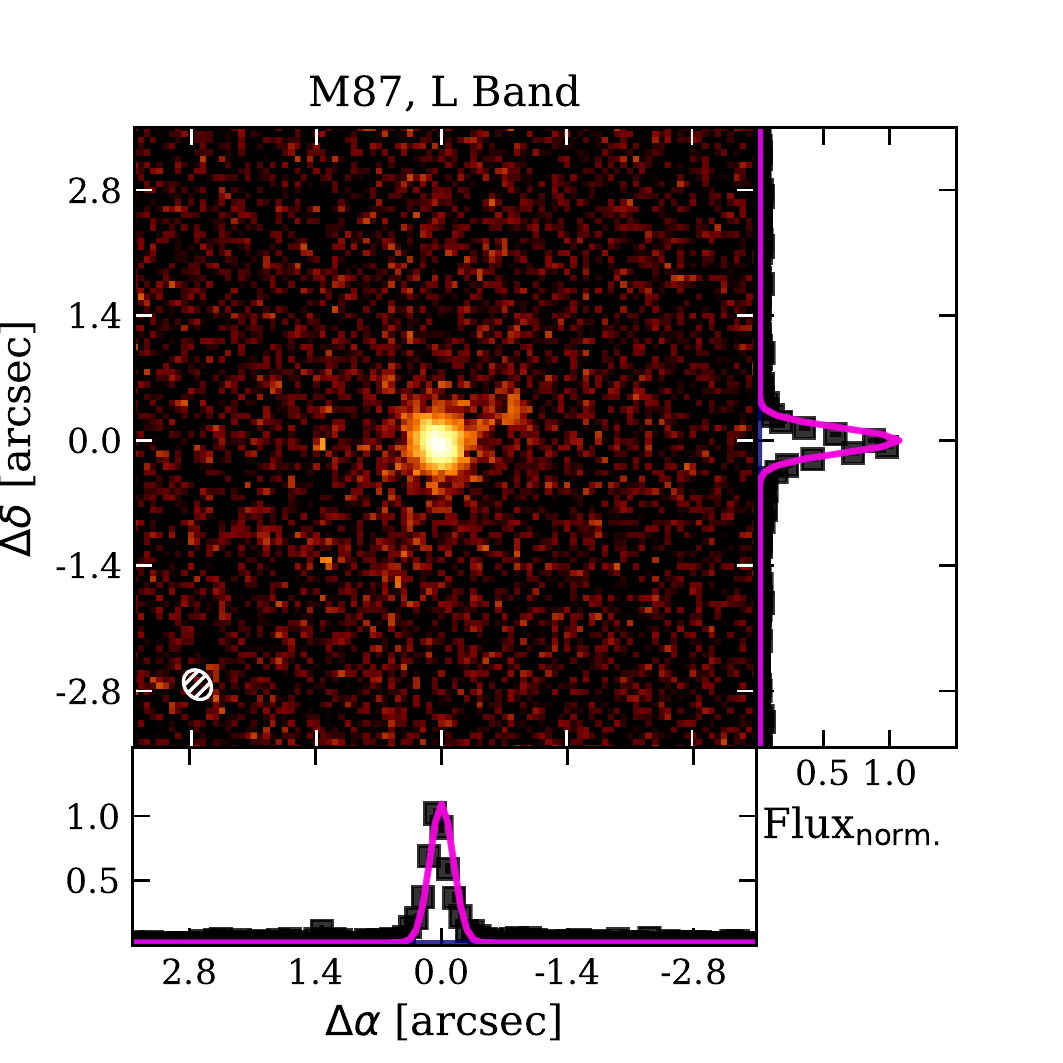}}
\subfloat{\includegraphics[width=0.25\hsize]{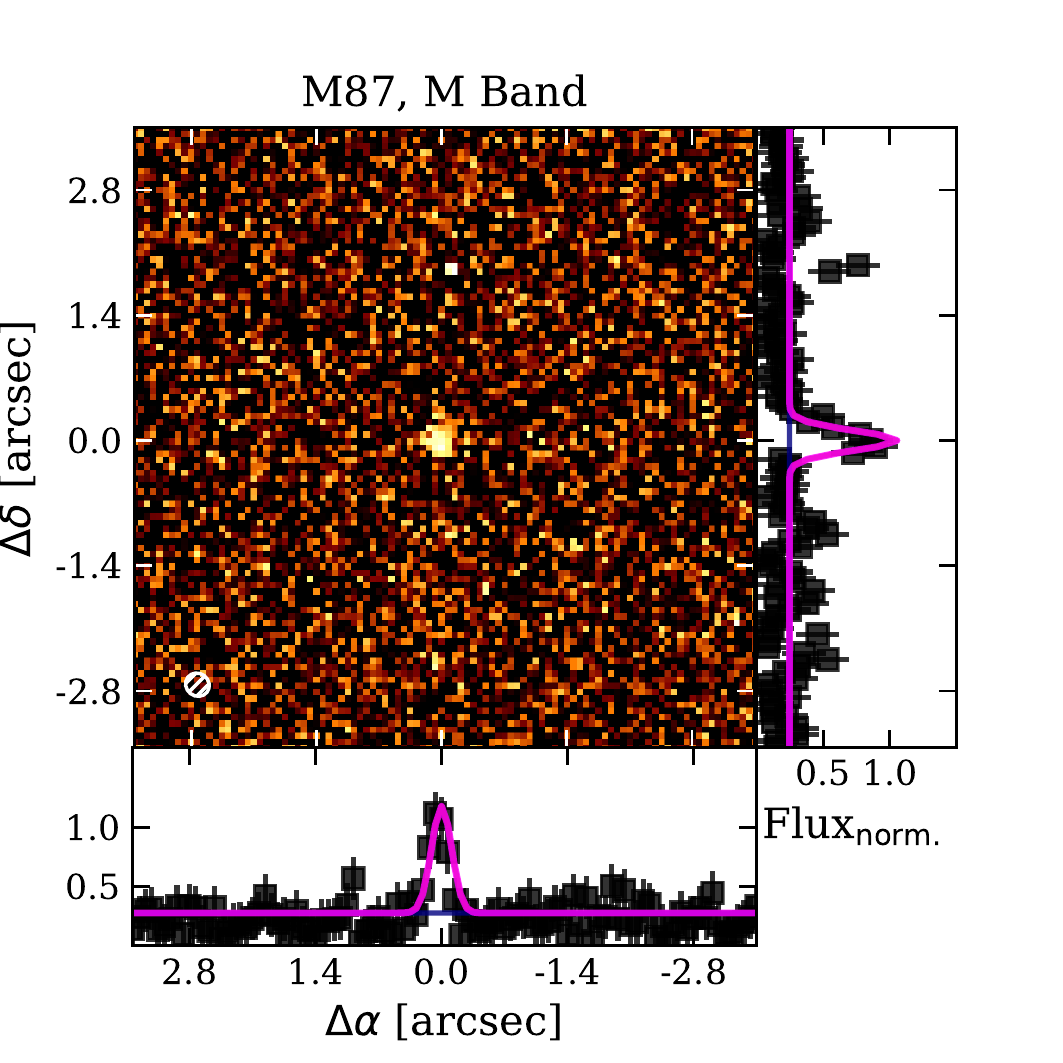}}
\subfloat{\includegraphics[width=0.25\hsize]{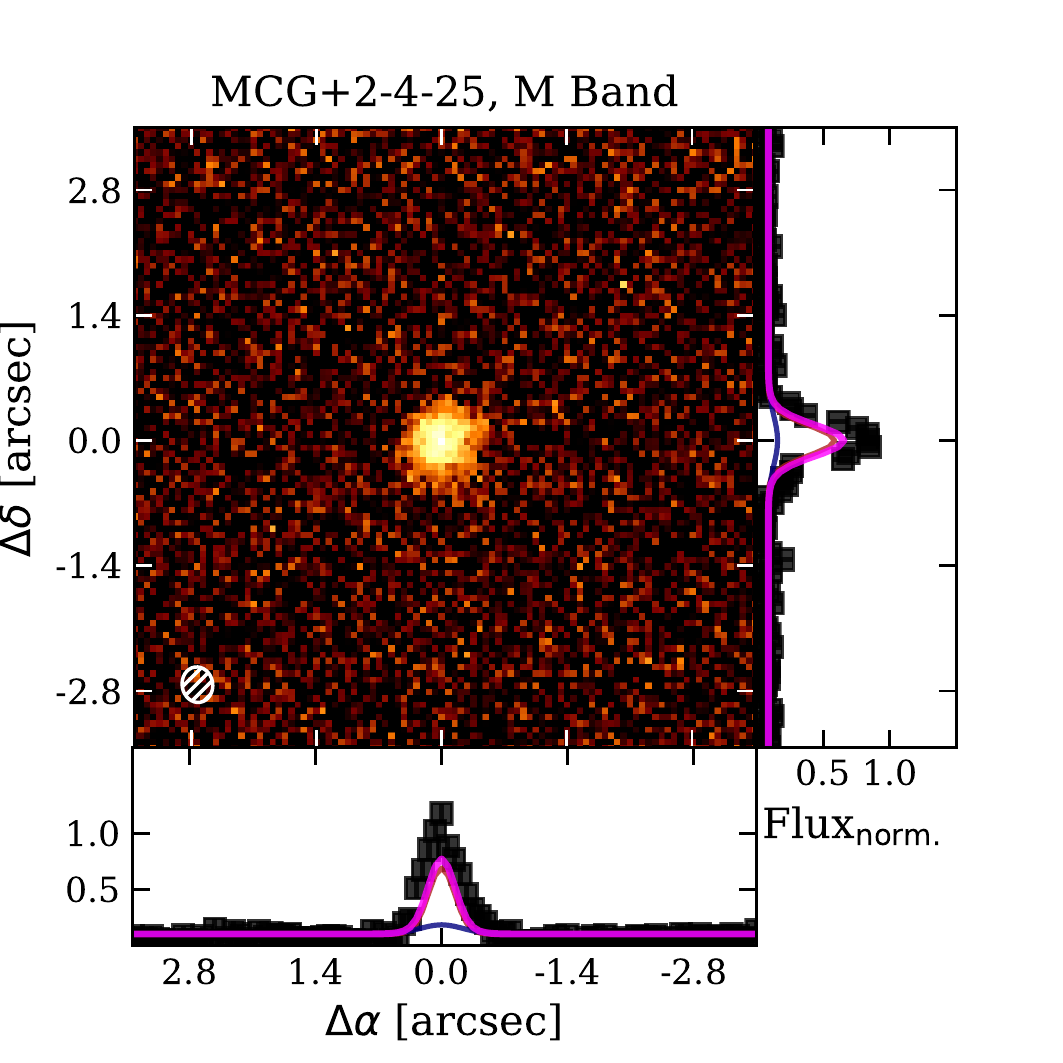}}
\subfloat{\includegraphics[width=0.25\hsize]{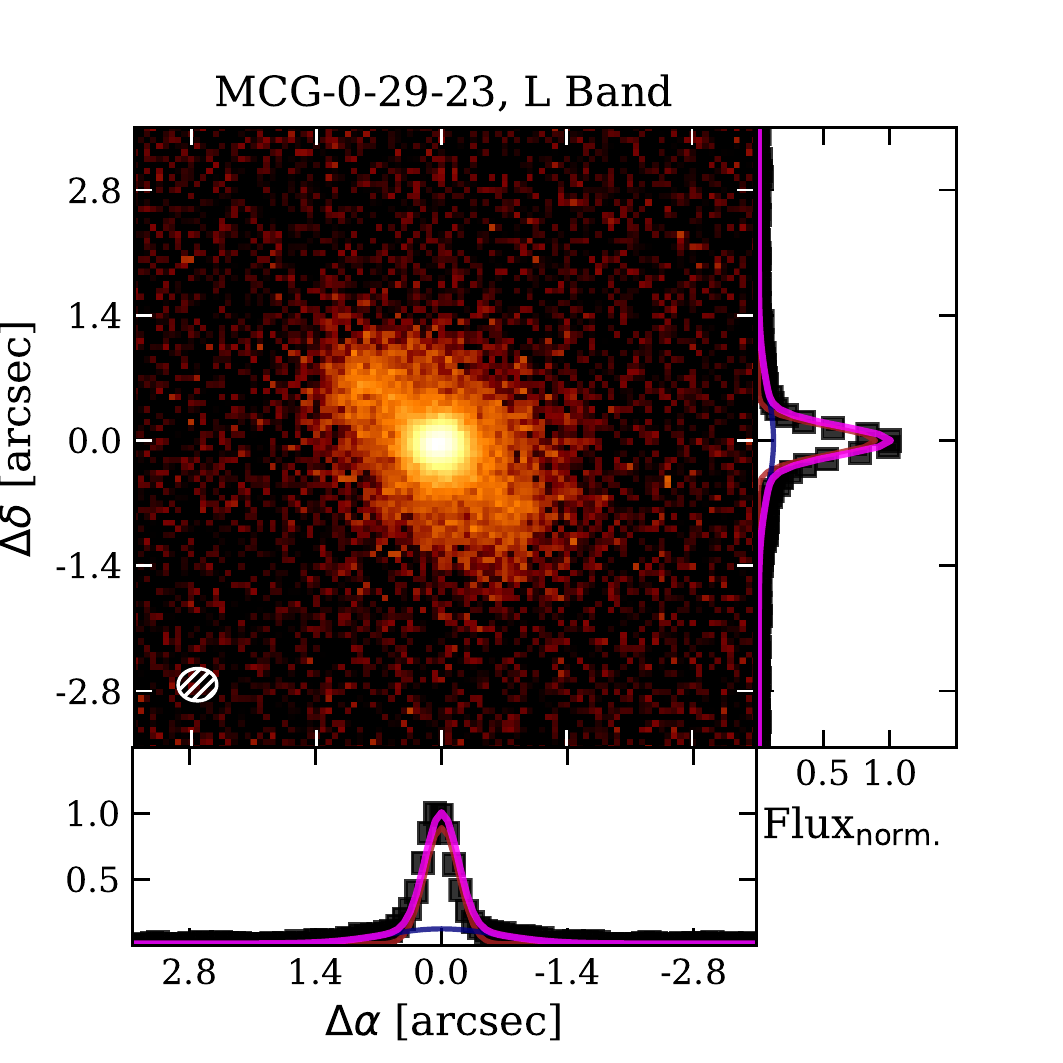}} \\
\caption{As Fig. \ref{fig:cutouts_one}.}
\end{figure*}
\begin{figure*}
\centering
\subfloat{\includegraphics[width=0.25\hsize]{appendix_fig/cutouts/mcg-00-29-23mnb.pdf}}
\subfloat{\includegraphics[width=0.25\hsize]{appendix_fig/cutouts/mcg-03-34-064l.pdf}}
\subfloat{\includegraphics[width=0.25\hsize]{appendix_fig/cutouts/mcg-03-34-064mnb.pdf}}
\subfloat{\includegraphics[width=0.25\hsize]{appendix_fig/cutouts/mcg-06-30-015l.pdf}} \\
\subfloat{\includegraphics[width=0.25\hsize]{appendix_fig/cutouts/mcg-06-30-015mnb.pdf}}
\subfloat{\includegraphics[width=0.25\hsize]{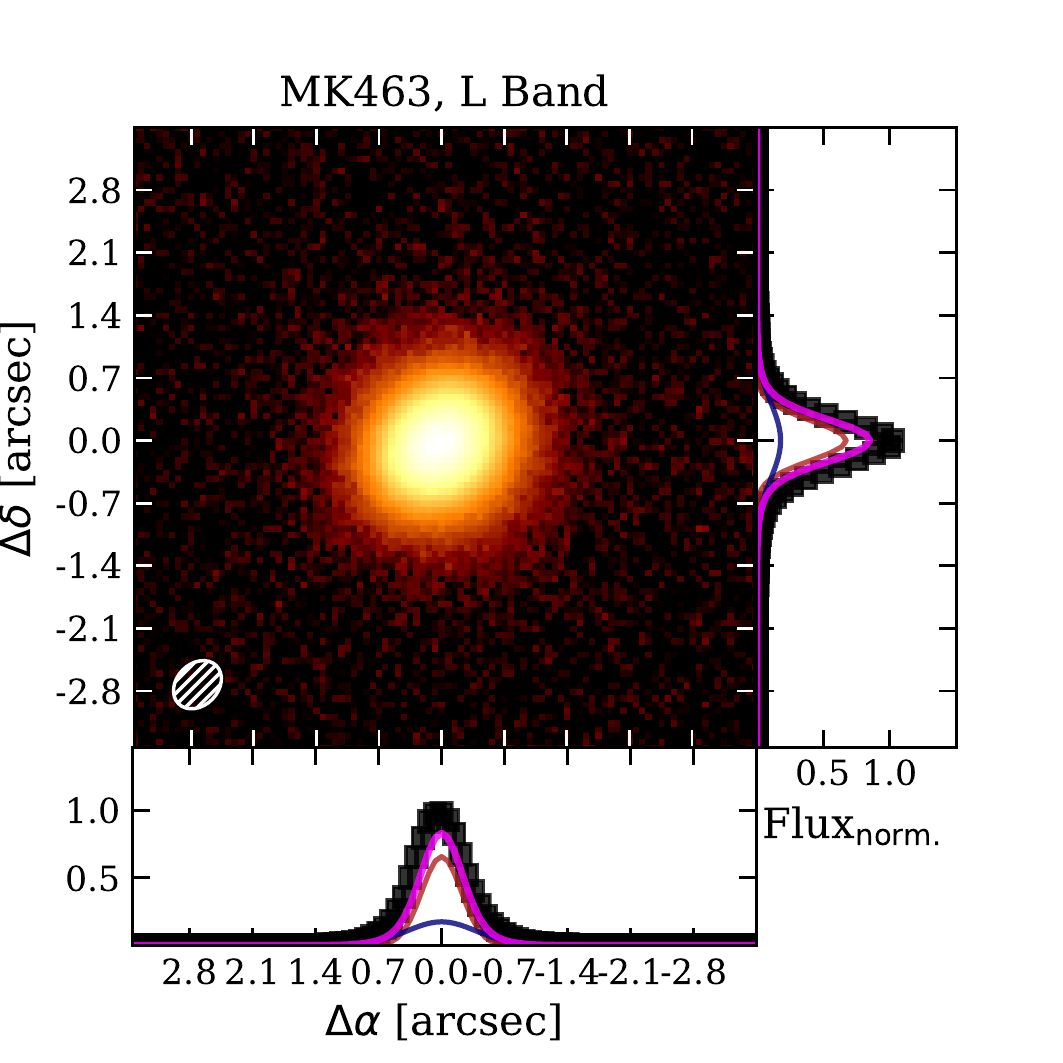}}
\subfloat{\includegraphics[width=0.25\hsize]{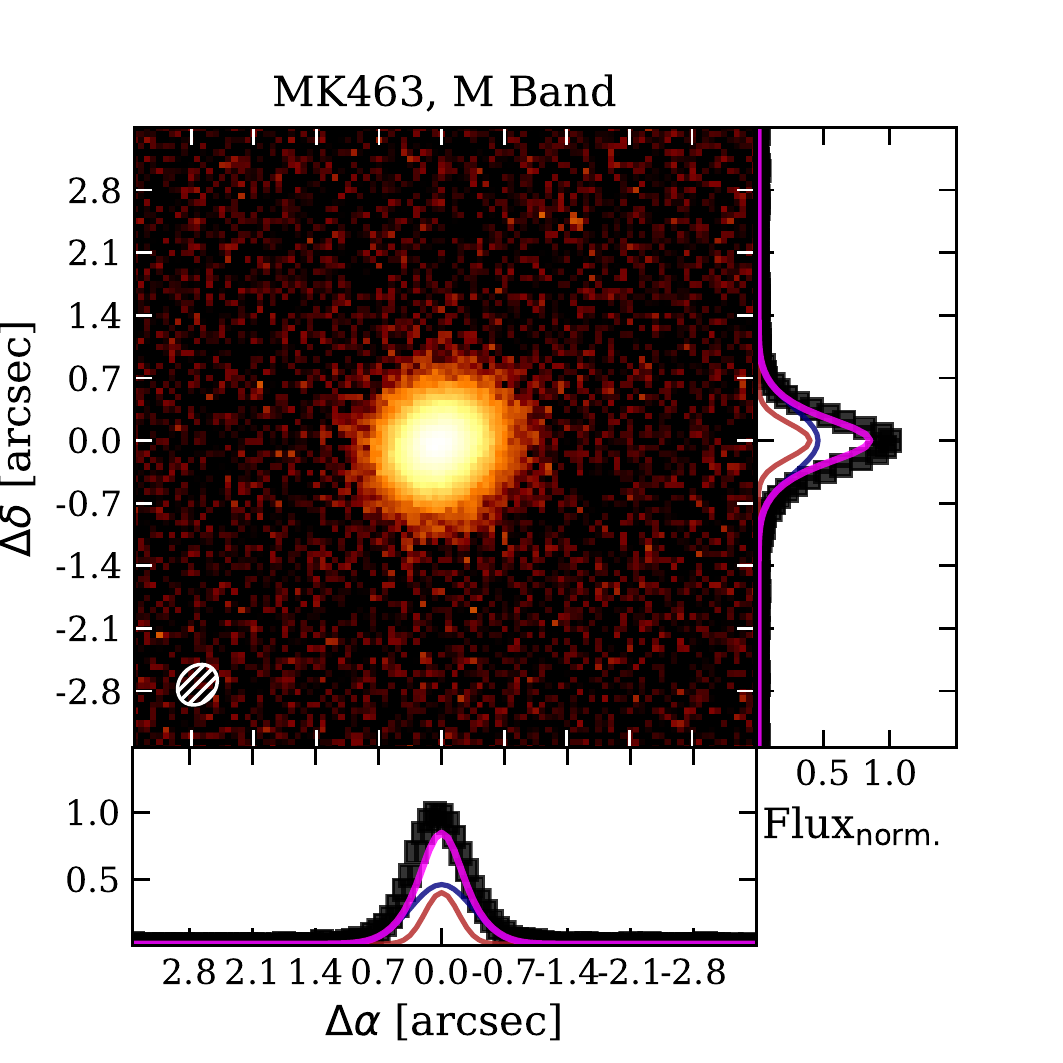}}
\subfloat{\includegraphics[width=0.25\hsize]{appendix_fig/cutouts/mrk0509l.pdf}} \\
\subfloat{\includegraphics[width=0.25\hsize]{appendix_fig/cutouts/mrk0509mnb.pdf}}
\subfloat{\includegraphics[width=0.25\hsize]{appendix_fig/cutouts/mrk0841l.pdf}}
\subfloat{\includegraphics[width=0.25\hsize]{appendix_fig/cutouts/mrk0841mnb.pdf}}
\subfloat{\includegraphics[width=0.25\hsize]{appendix_fig/cutouts/mrk0897l.pdf}} \\
\subfloat{\includegraphics[width=0.25\hsize]{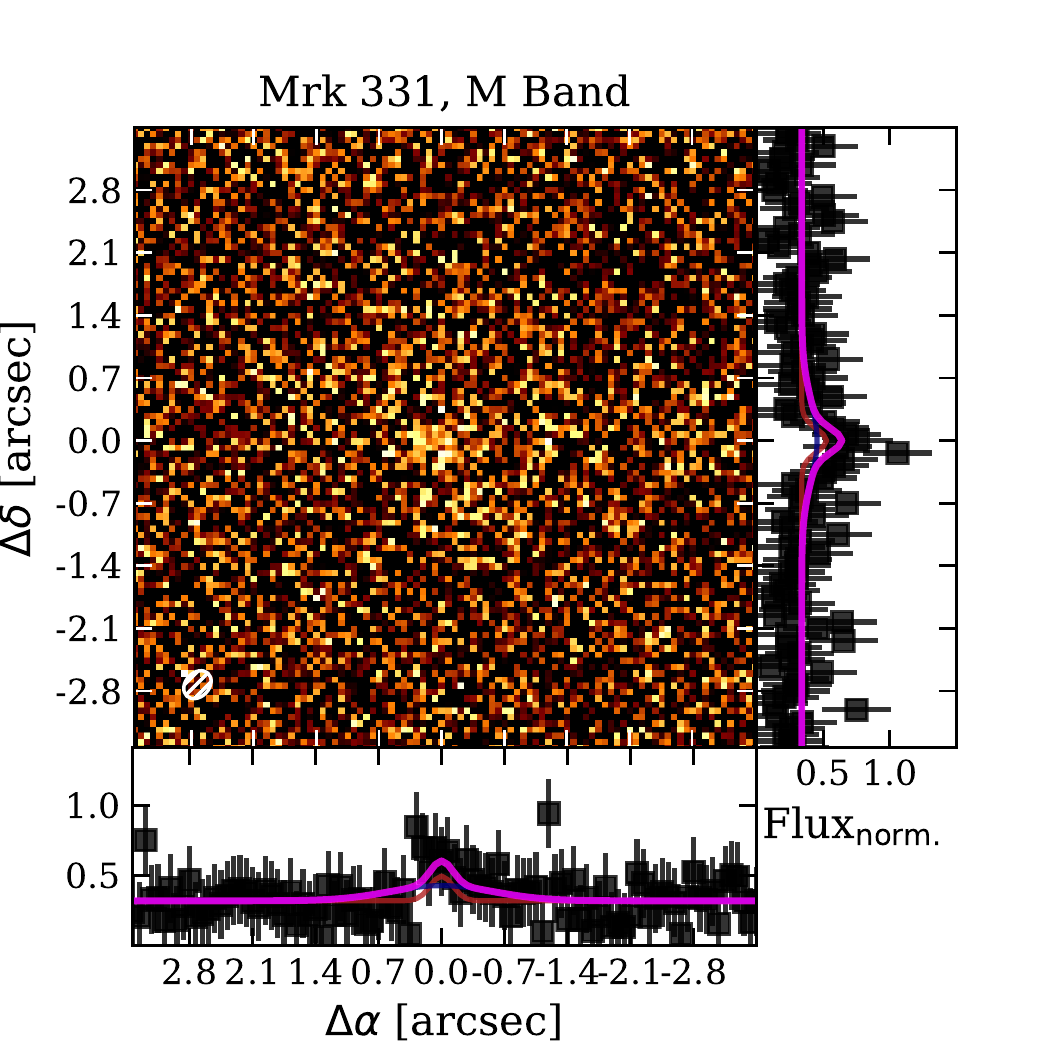}}
\subfloat{\includegraphics[width=0.25\hsize]{appendix_fig/cutouts/ngc0063mnb.pdf}}
\subfloat{\includegraphics[width=0.25\hsize]{appendix_fig/cutouts/ngc0253l.pdf}}
\subfloat{\includegraphics[width=0.25\hsize]{appendix_fig/cutouts/ngc0253mnb.pdf}} \\
\subfloat{\includegraphics[width=0.25\hsize]{appendix_fig/cutouts/ngc0424l.pdf}}
\subfloat{\includegraphics[width=0.25\hsize]{appendix_fig/cutouts/ngc0424mnb.pdf}}
\subfloat{\includegraphics[width=0.25\hsize]{appendix_fig/cutouts/ngc0520mnb.pdf}}
\subfloat{\includegraphics[width=0.25\hsize]{appendix_fig/cutouts/ngc0660mnb.pdf}} \\
\caption{As Fig. \ref{fig:cutouts_one}.}
\end{figure*}
\begin{figure*}
\centering
\subfloat{\includegraphics[width=0.25\hsize]{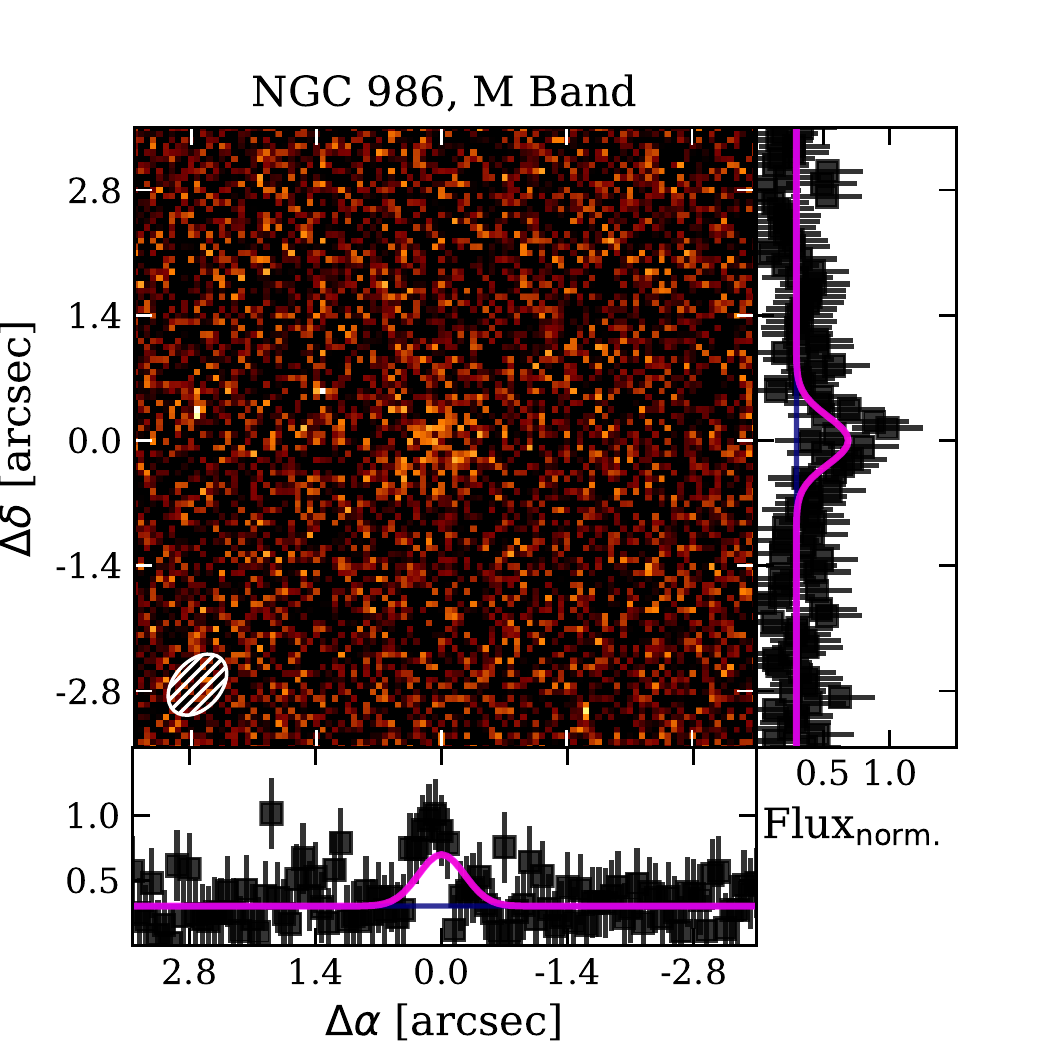}}
\subfloat{\includegraphics[width=0.25\hsize]{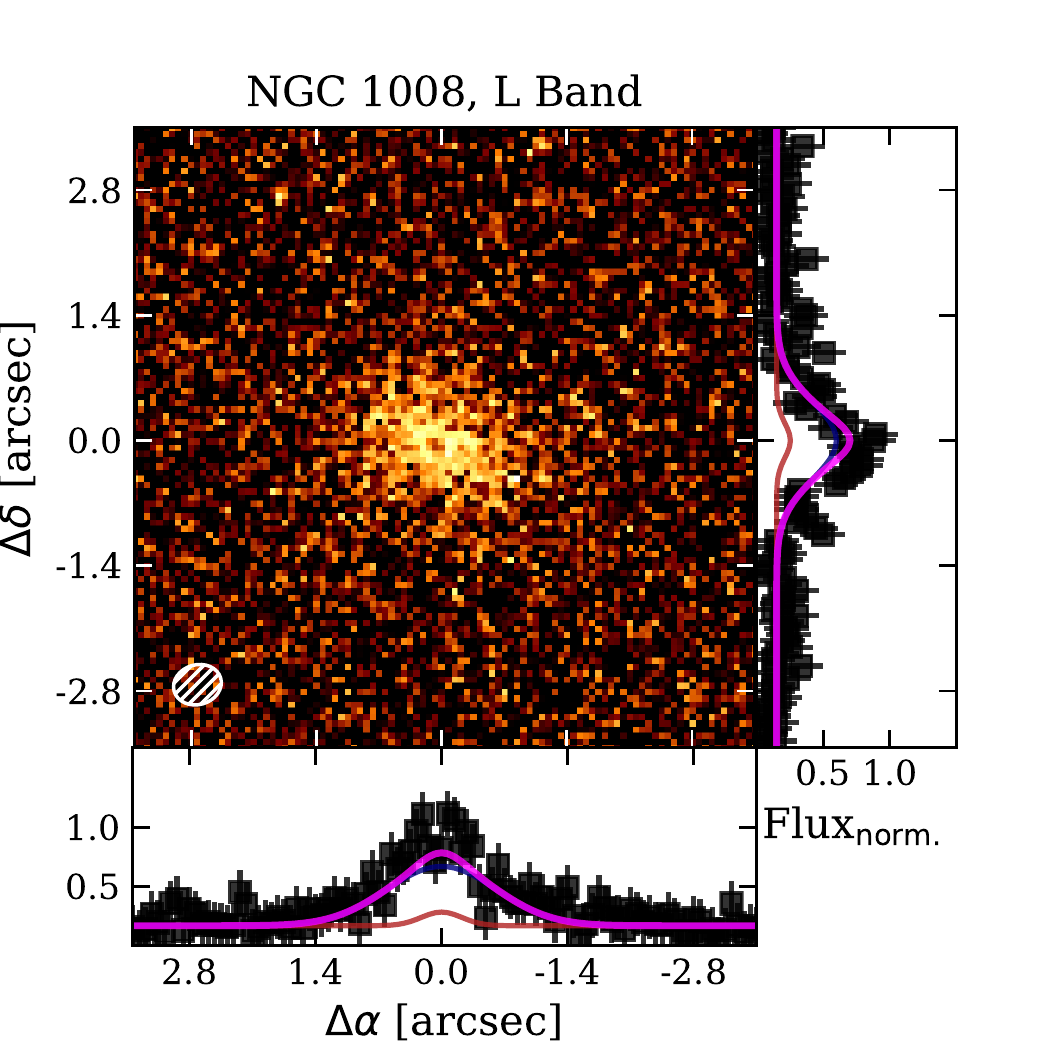}}
\subfloat{\includegraphics[width=0.25\hsize]{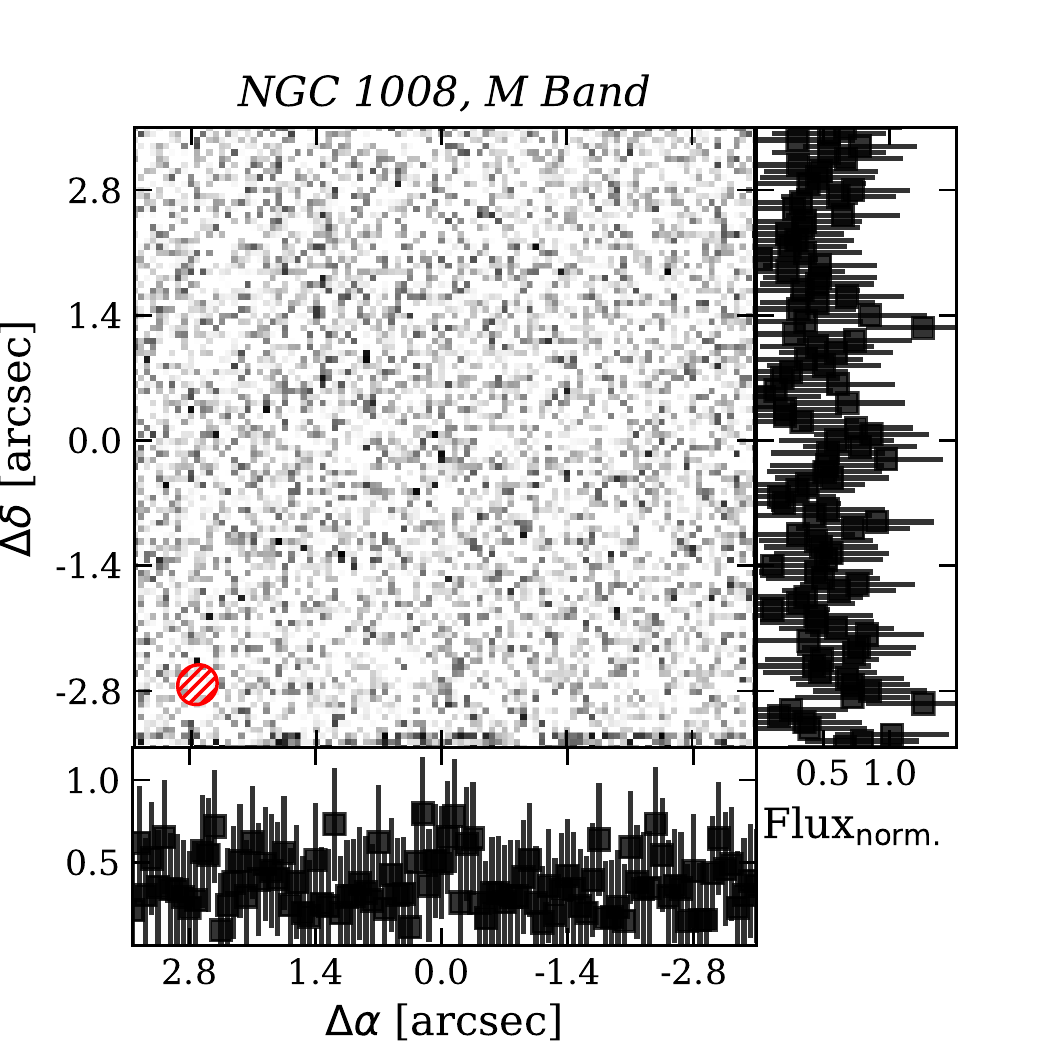}}
\subfloat{\includegraphics[width=0.25\hsize]{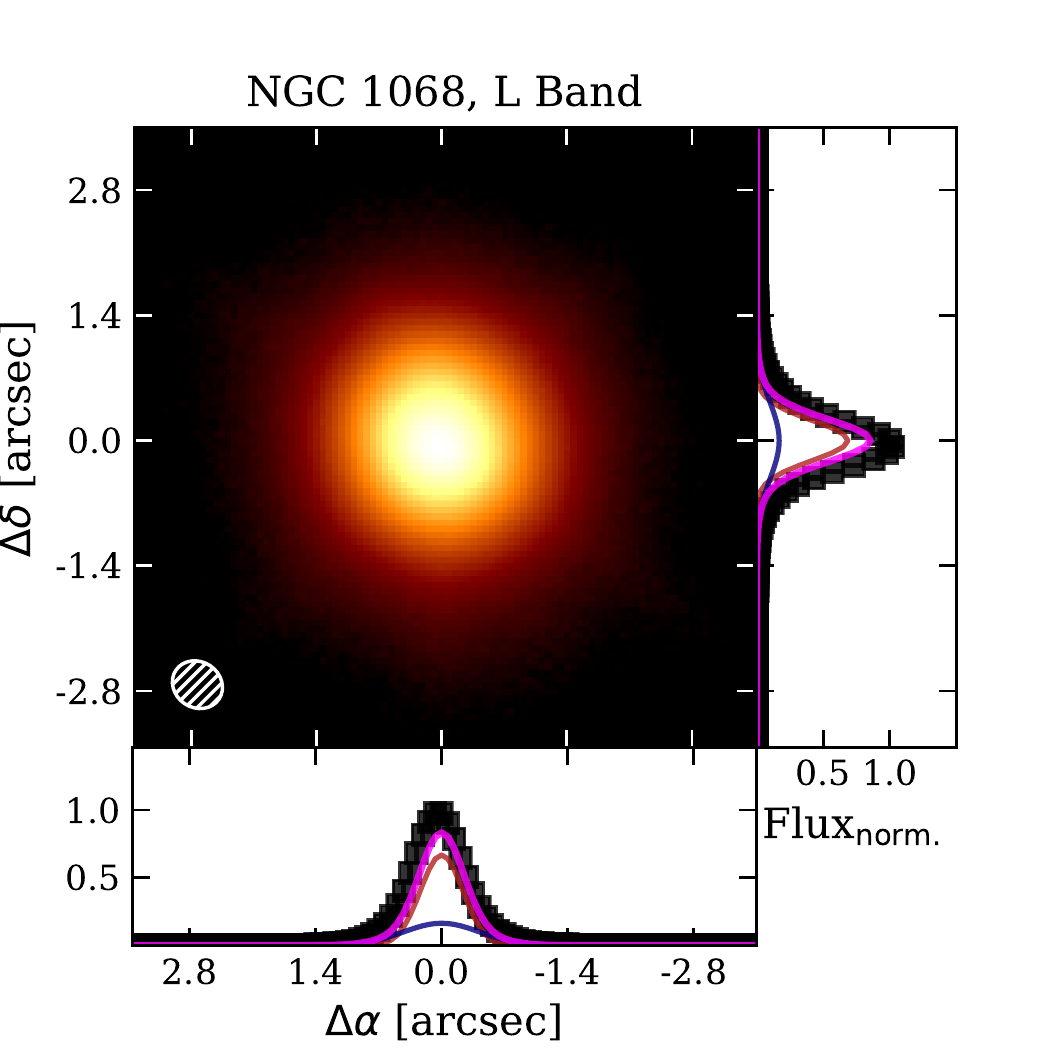}} \\
\subfloat{\includegraphics[width=0.25\hsize]{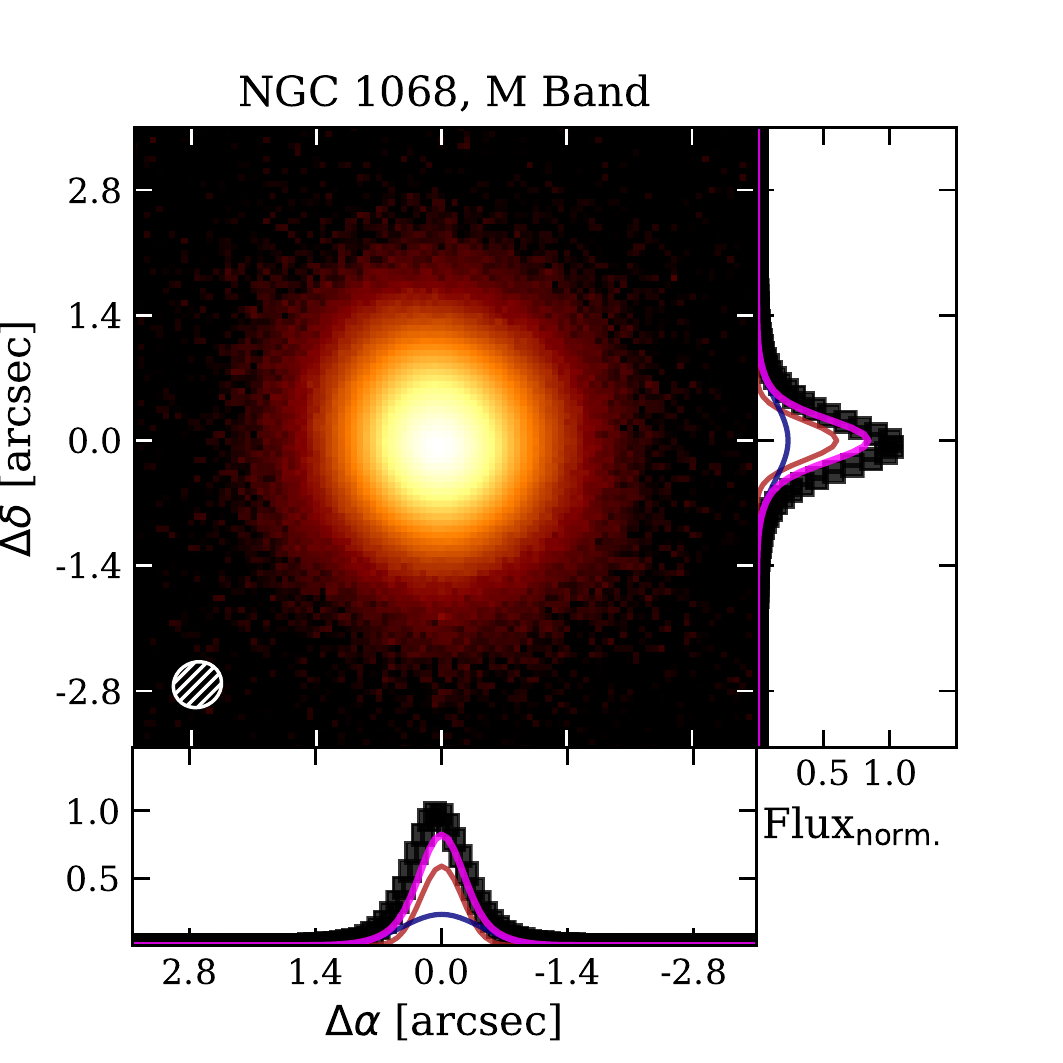}}
\subfloat{\includegraphics[width=0.25\hsize]{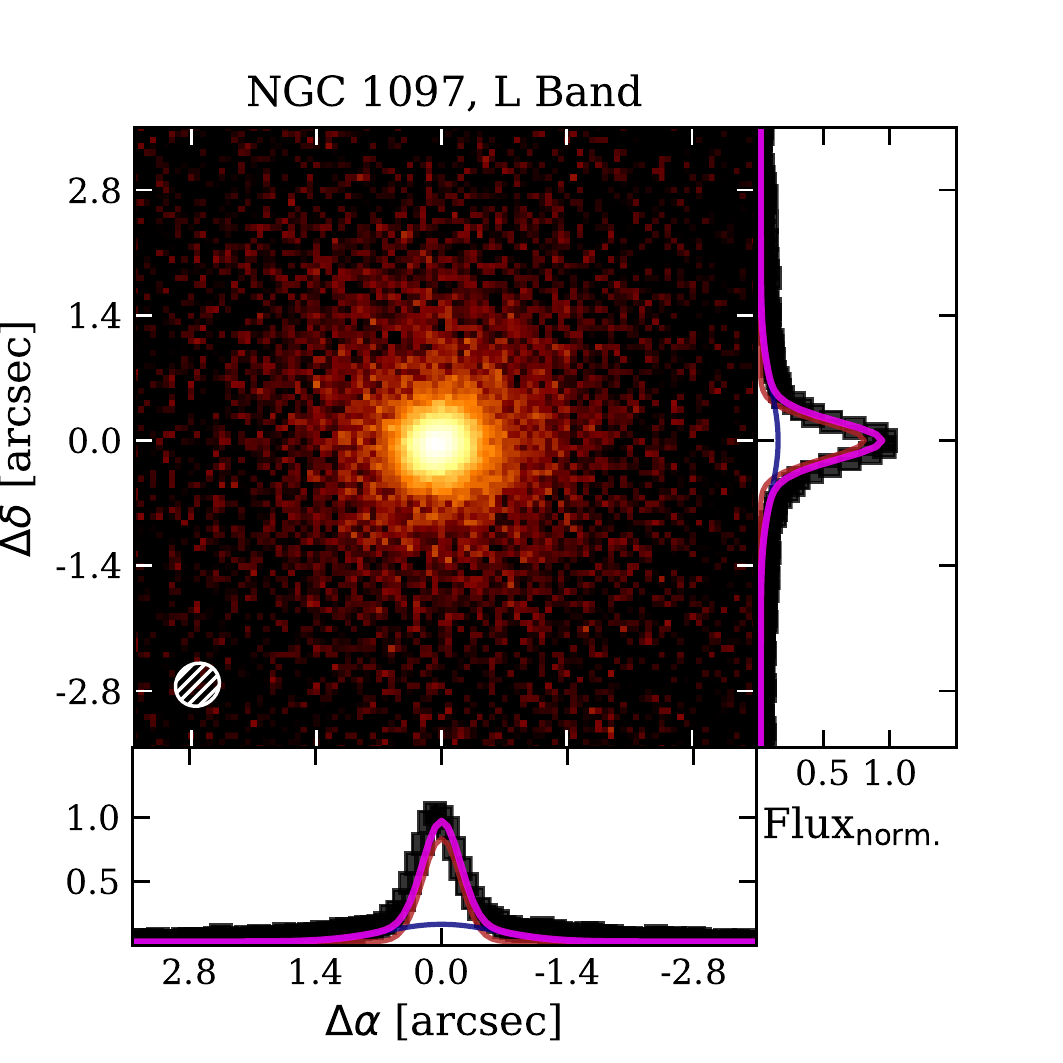}}
\subfloat{\includegraphics[width=0.25\hsize]{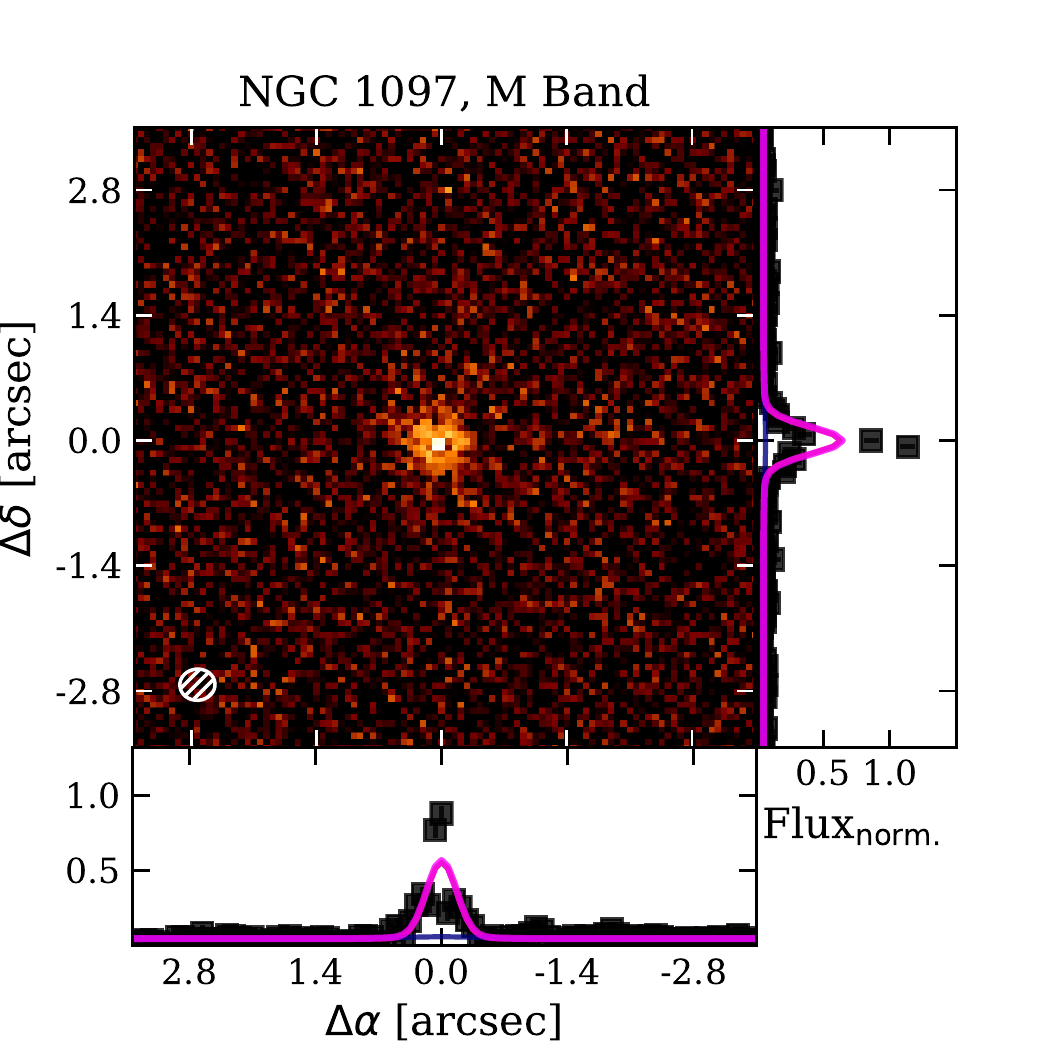}}
\subfloat{\includegraphics[width=0.25\hsize]{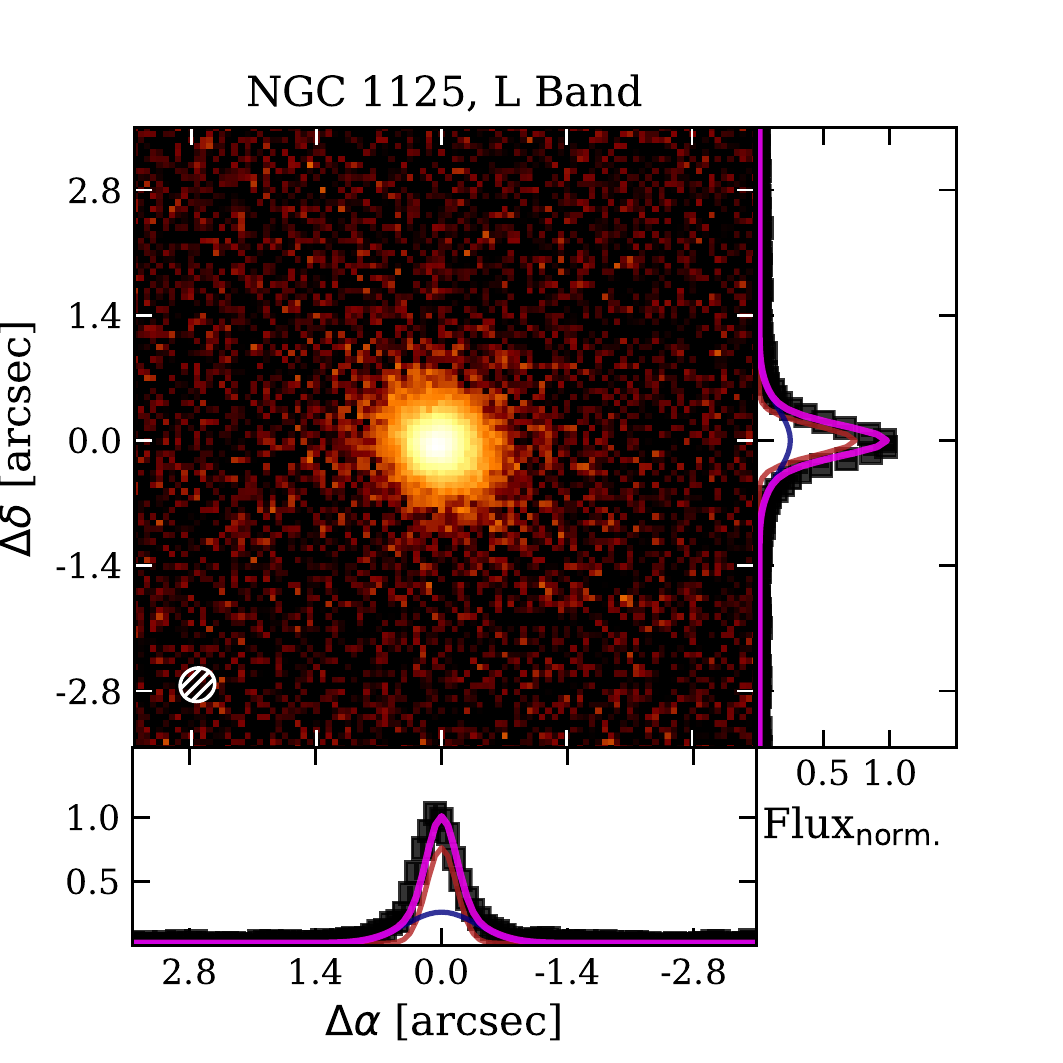}} \\
\subfloat{\includegraphics[width=0.25\hsize]{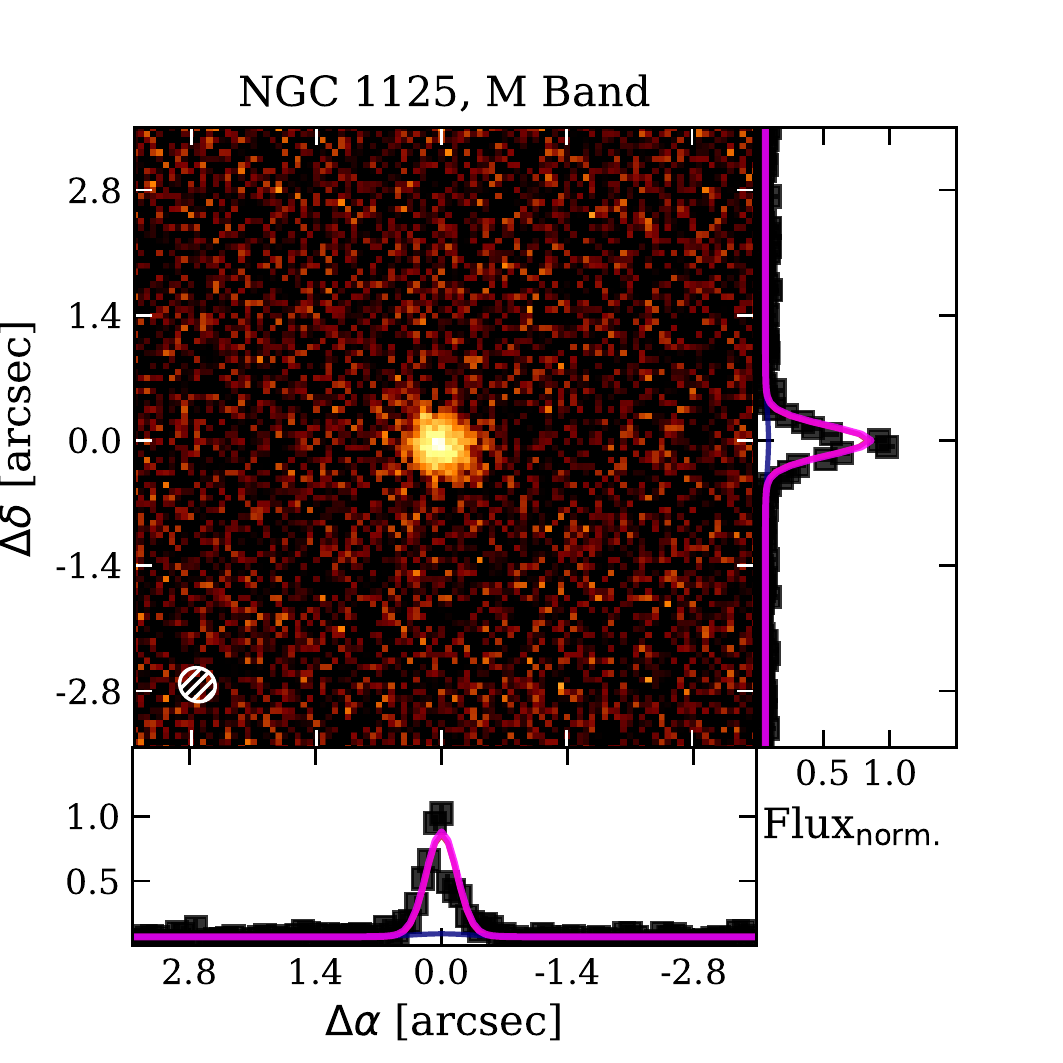}}
\subfloat{\includegraphics[width=0.25\hsize]{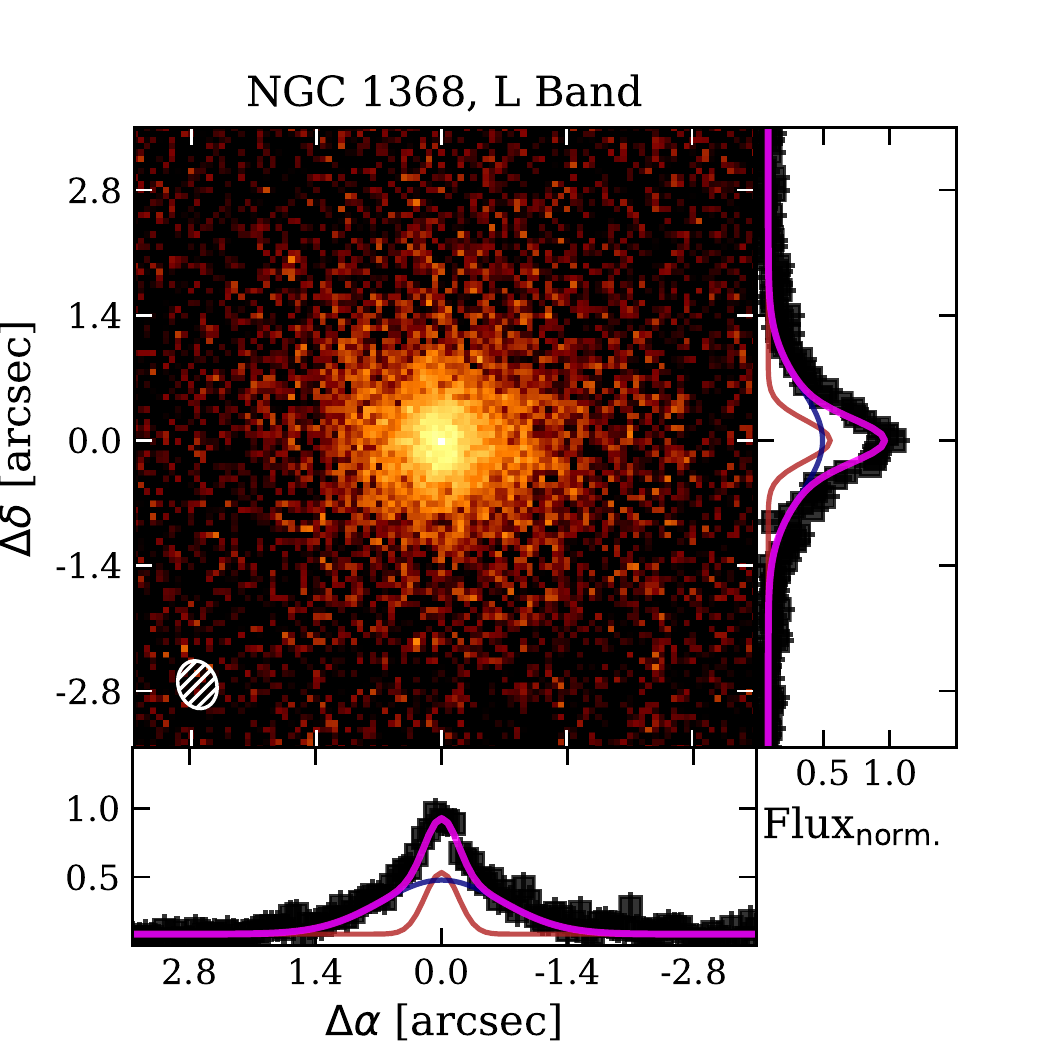}}
\subfloat{\includegraphics[width=0.25\hsize]{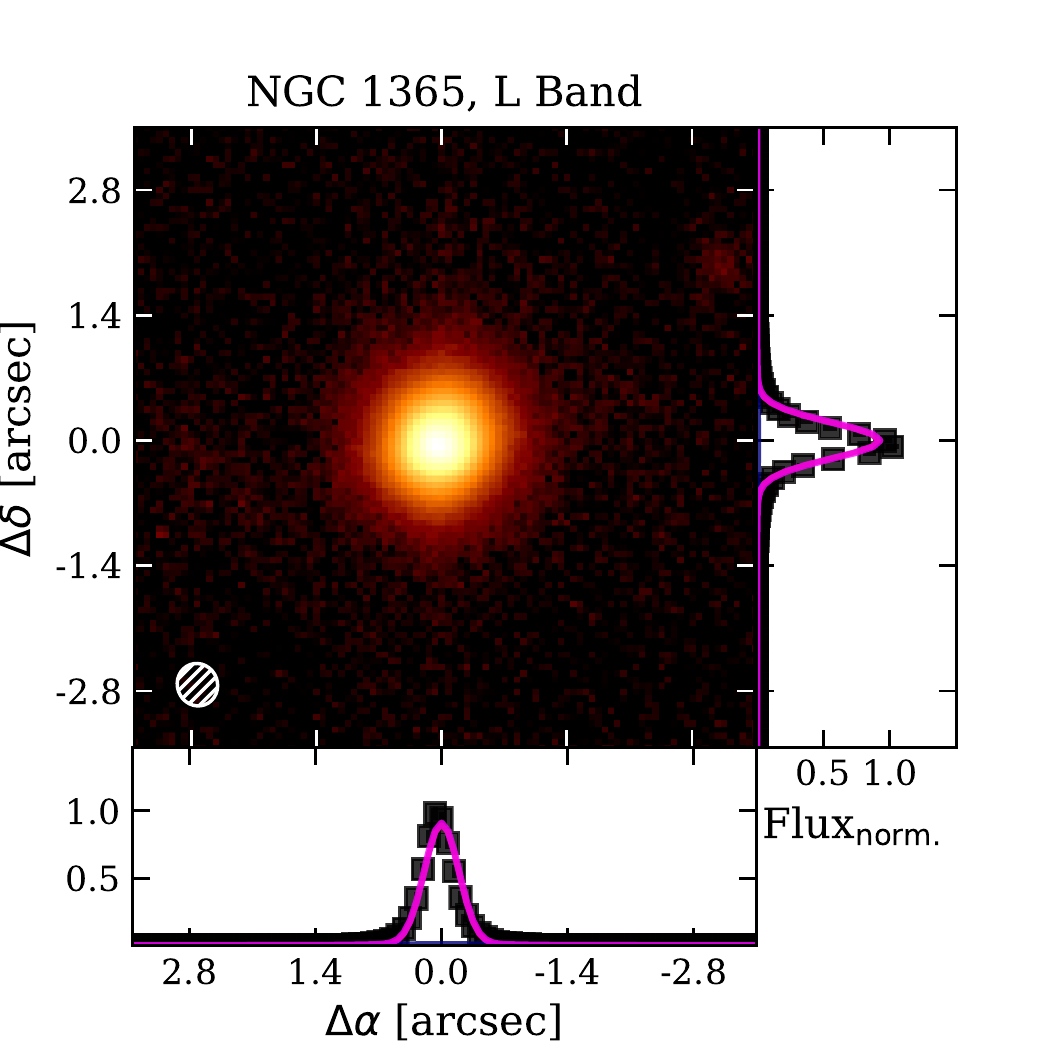}}
\subfloat{\includegraphics[width=0.25\hsize]{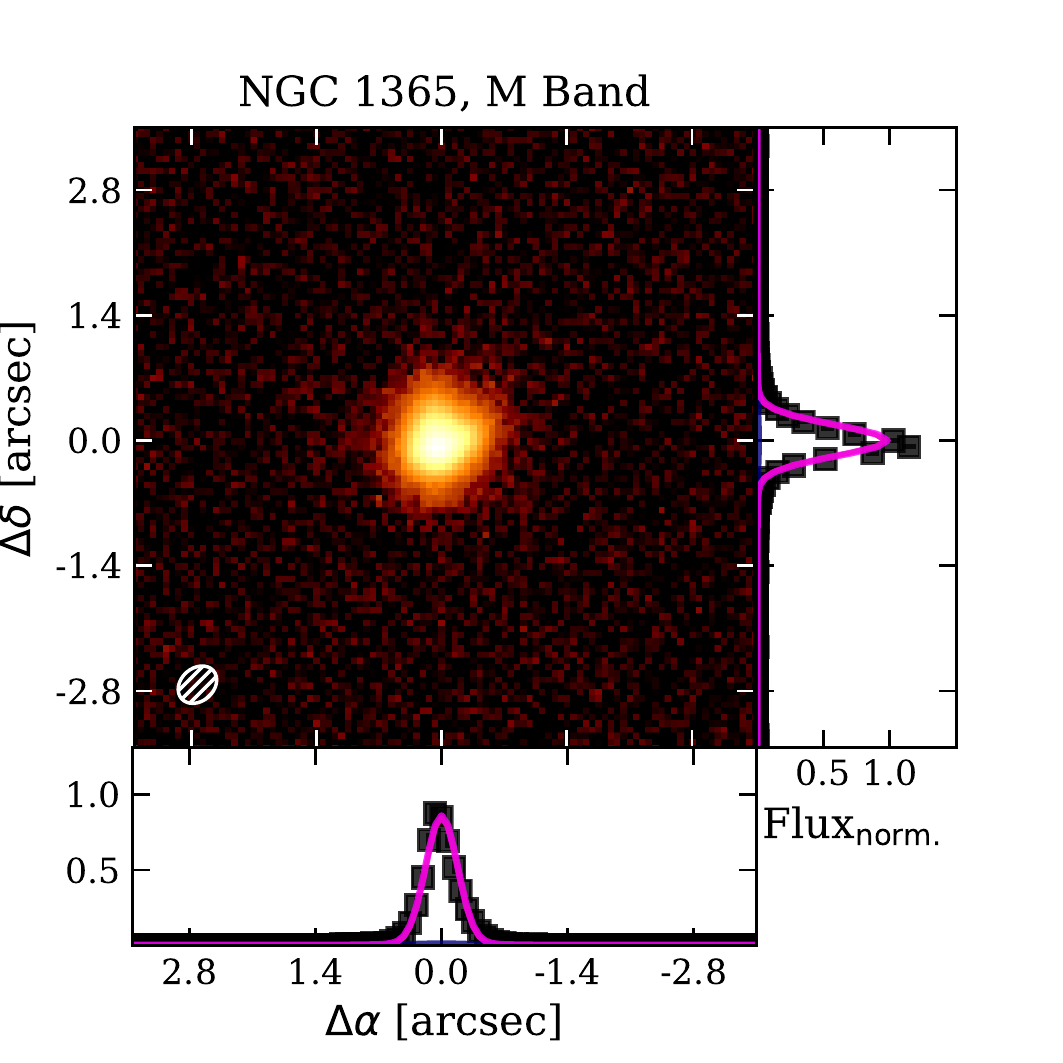}} \\
\subfloat{\includegraphics[width=0.25\hsize]{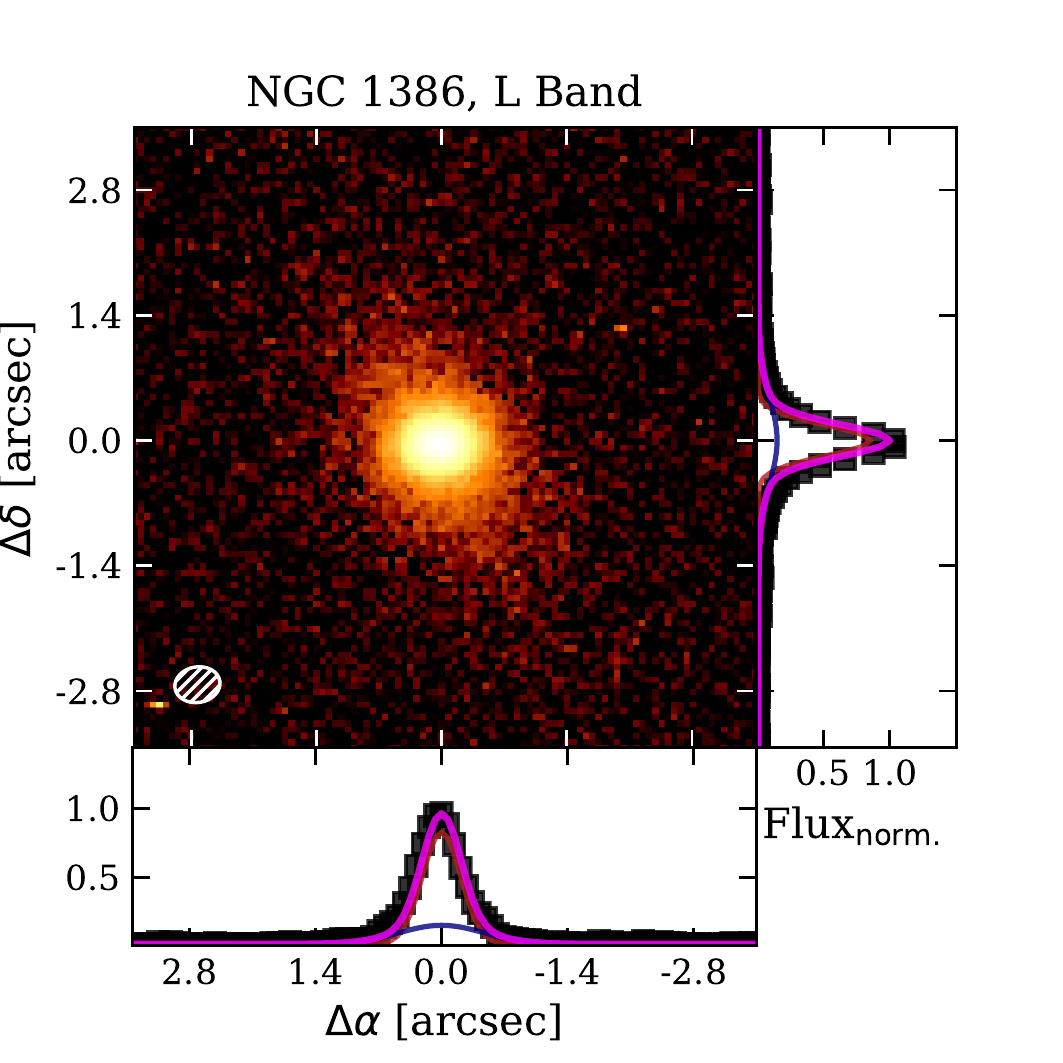}}
\subfloat{\includegraphics[width=0.25\hsize]{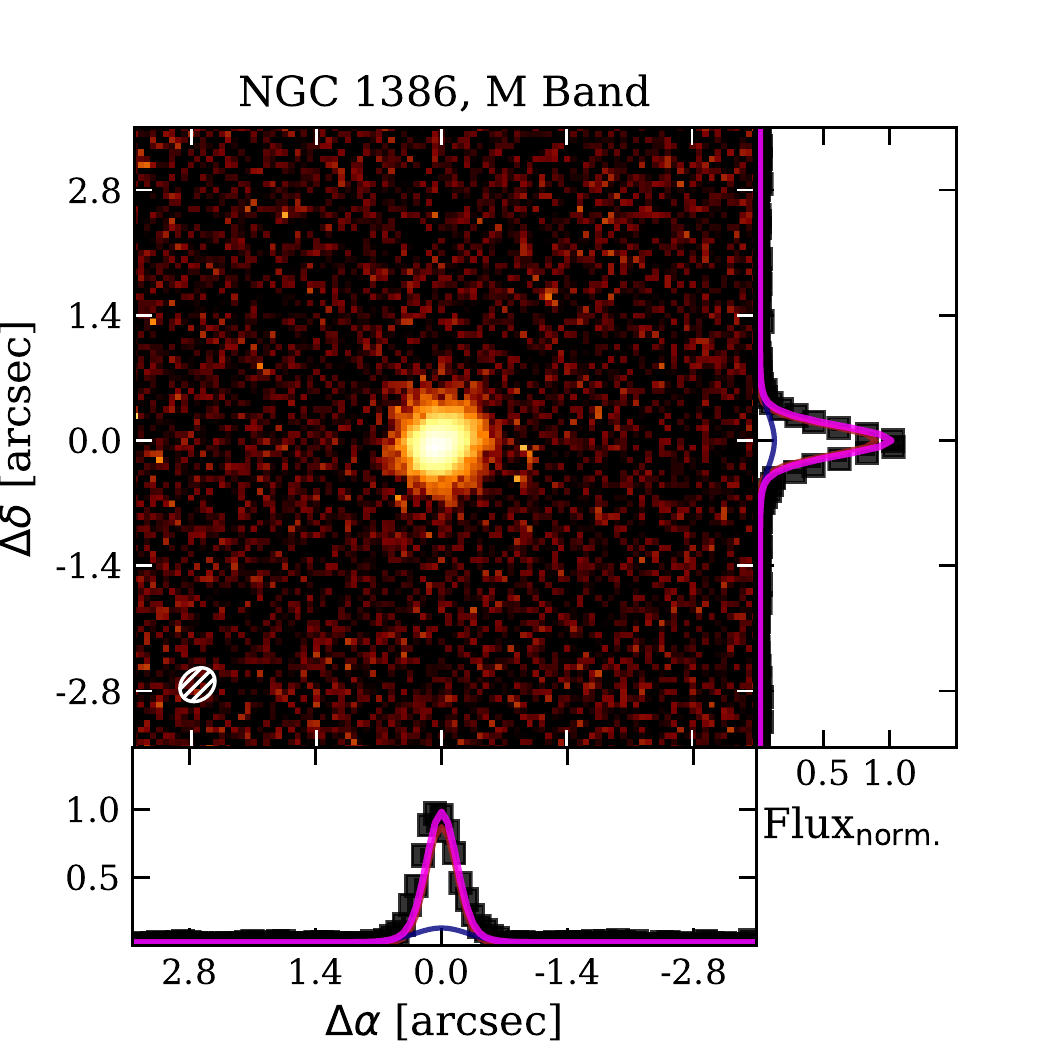}}
\subfloat{\includegraphics[width=0.25\hsize]{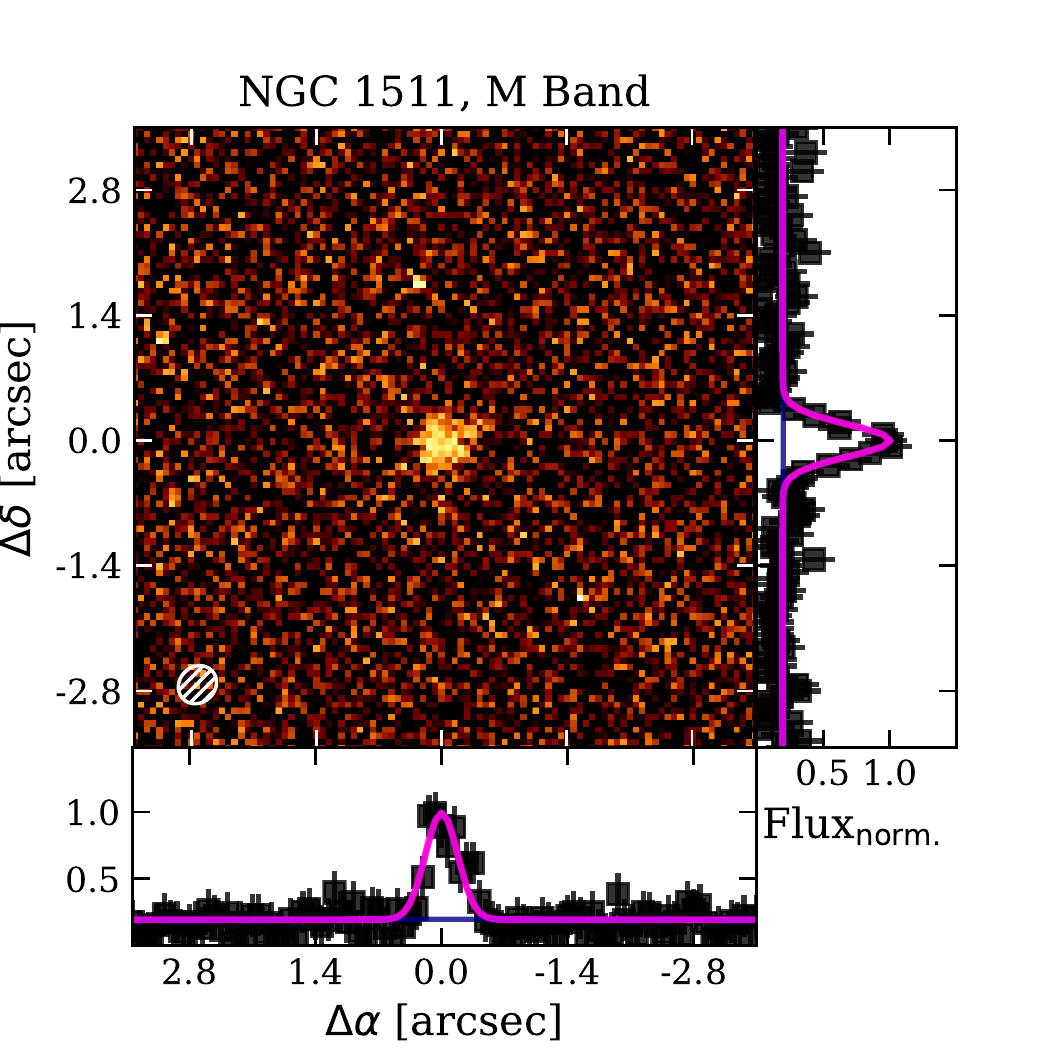}}
\subfloat{\includegraphics[width=0.25\hsize]{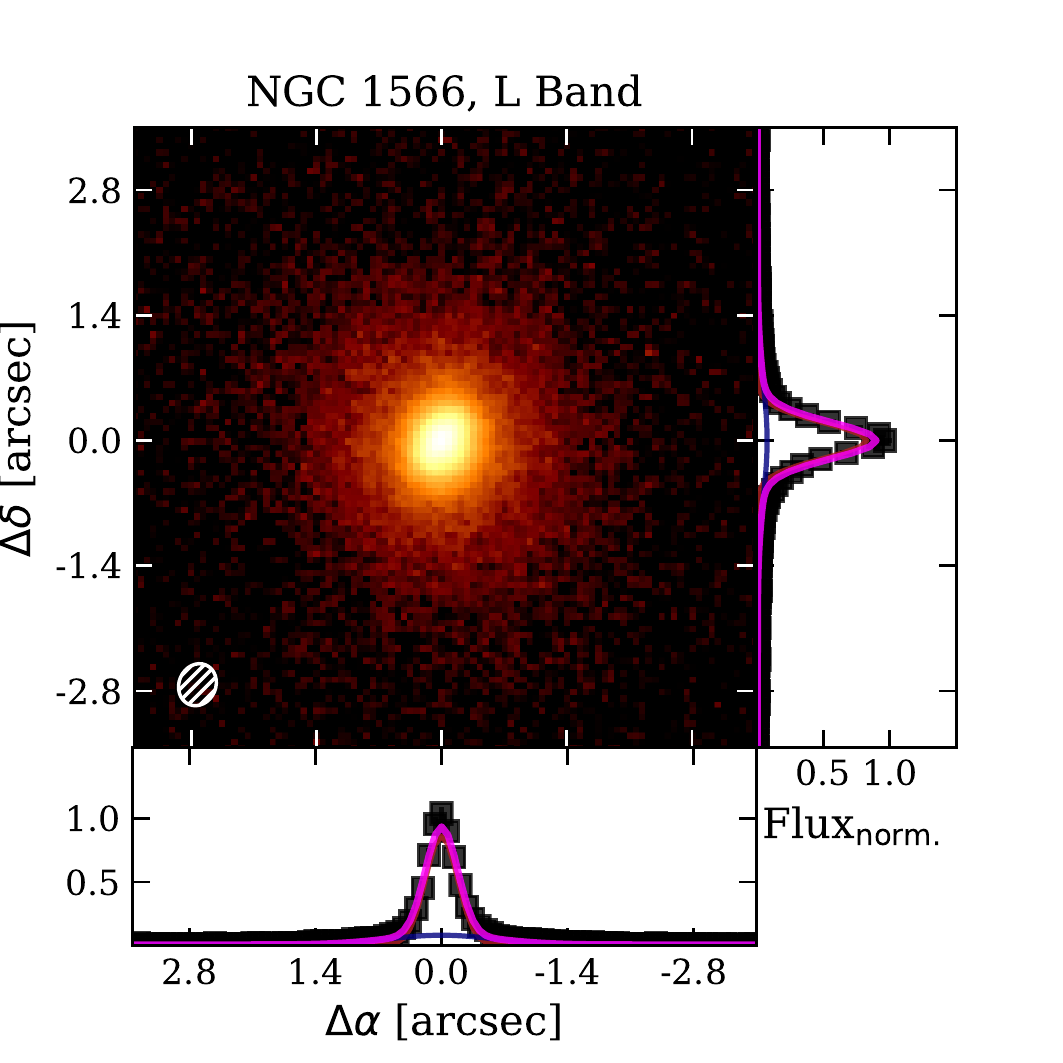}} \\
\subfloat{\includegraphics[width=0.25\hsize]{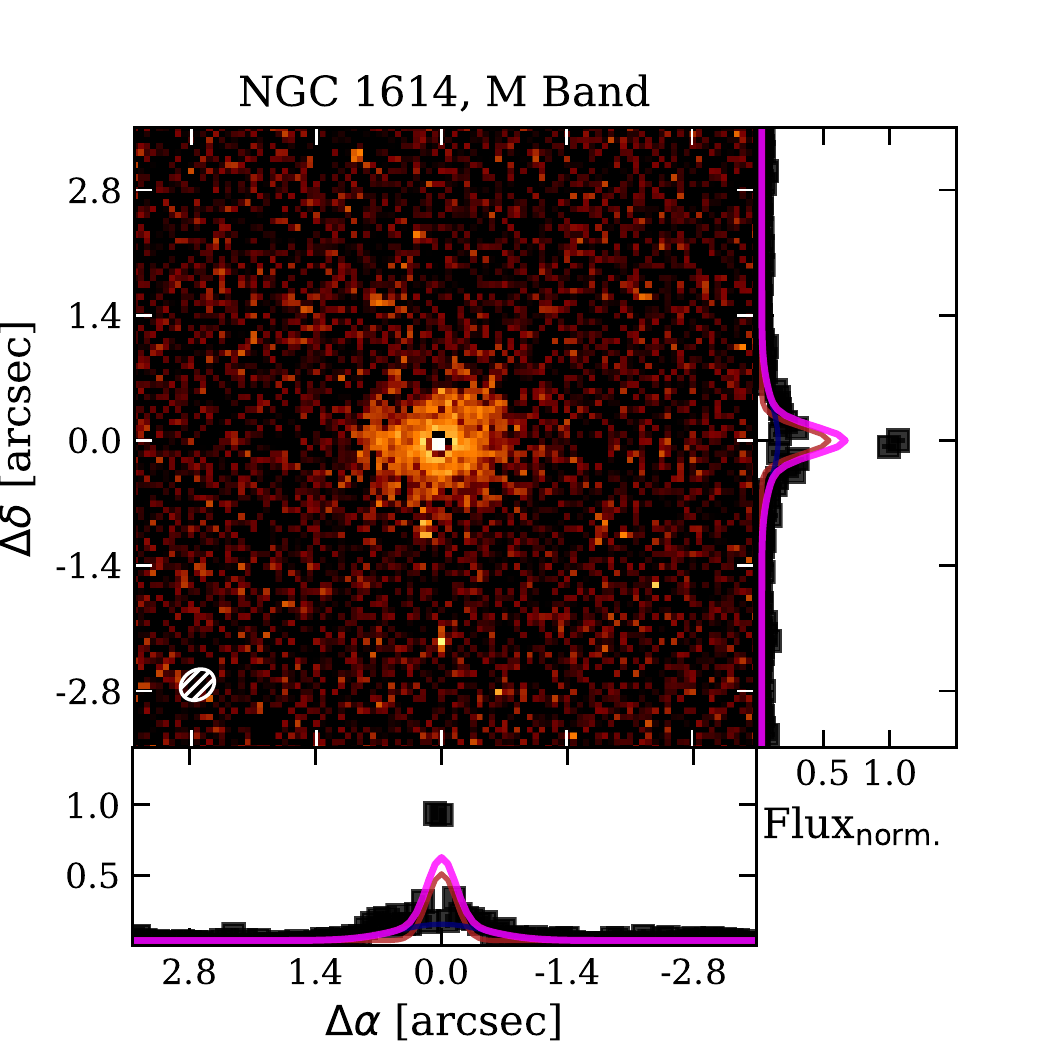}}
\subfloat{\includegraphics[width=0.25\hsize]{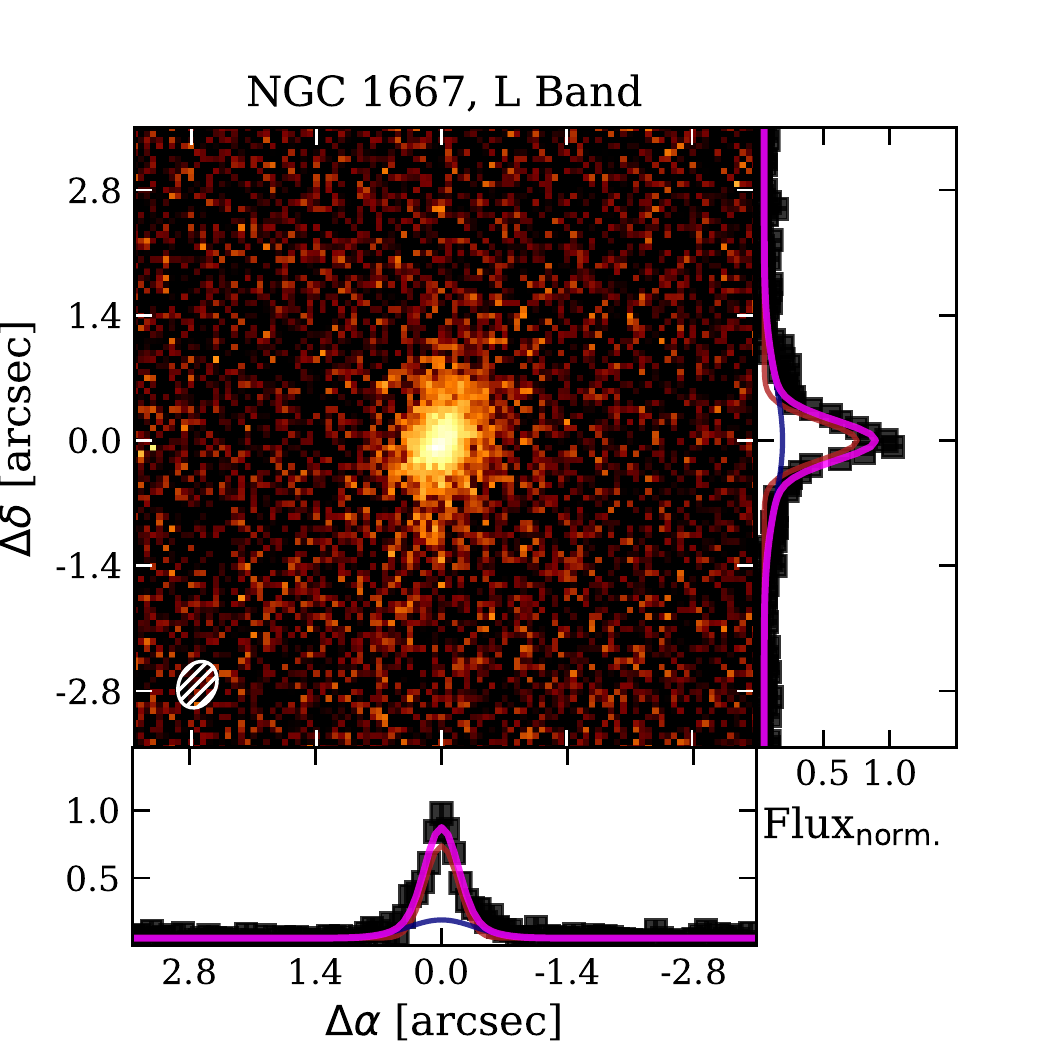}}
\subfloat{\includegraphics[width=0.25\hsize]{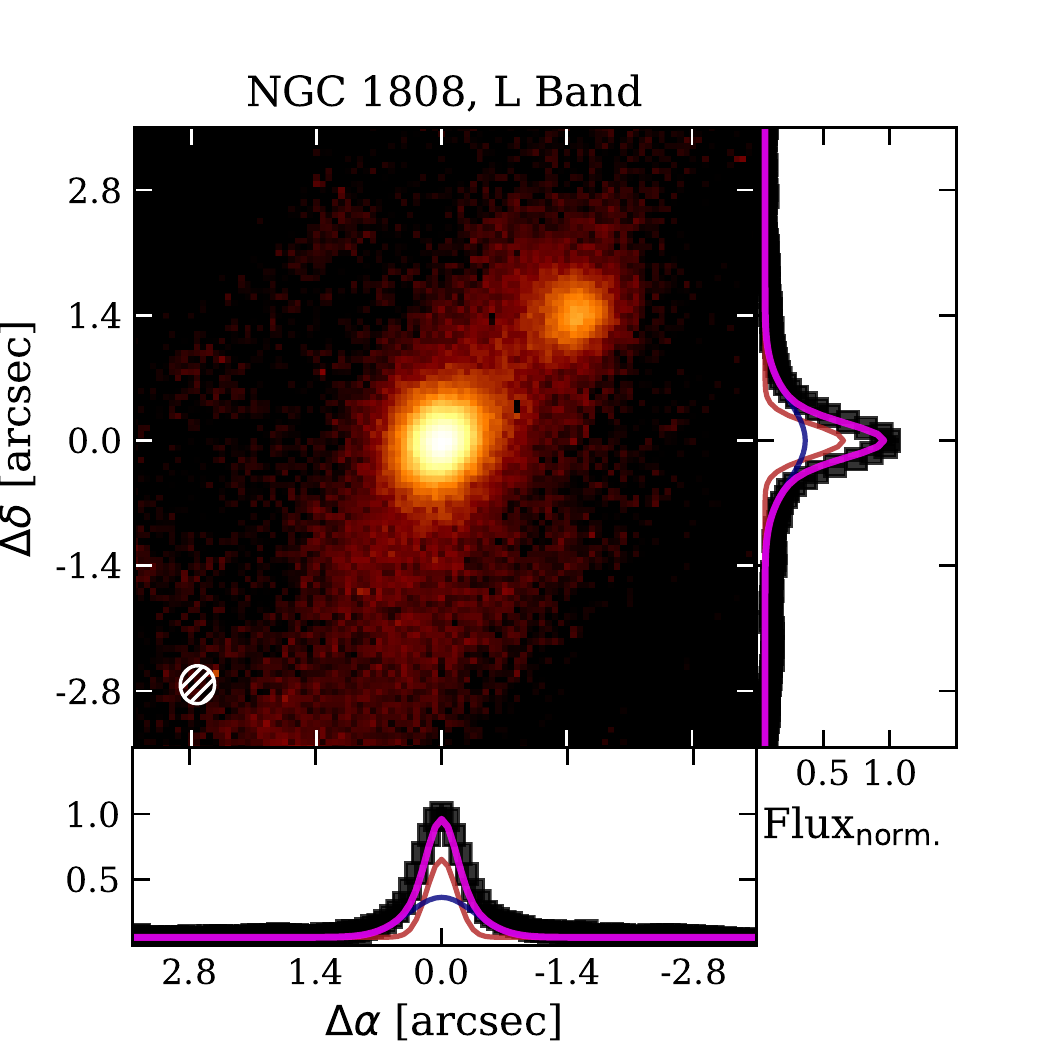}}
\subfloat{\includegraphics[width=0.25\hsize]{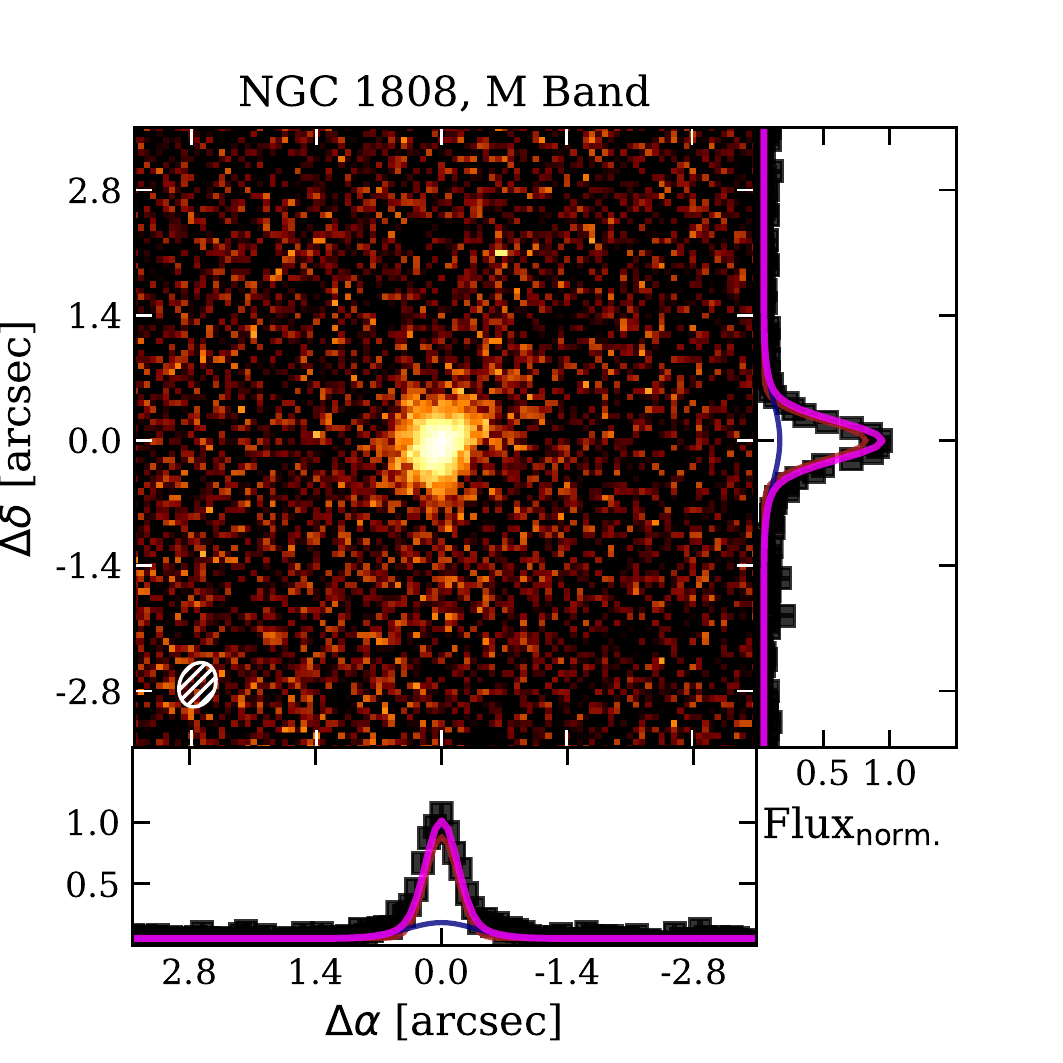}} \\
\caption{As Fig. \ref{fig:cutouts_one}.}
\end{figure*}
\begin{figure*}
\centering
\subfloat{\includegraphics[width=0.25\hsize]{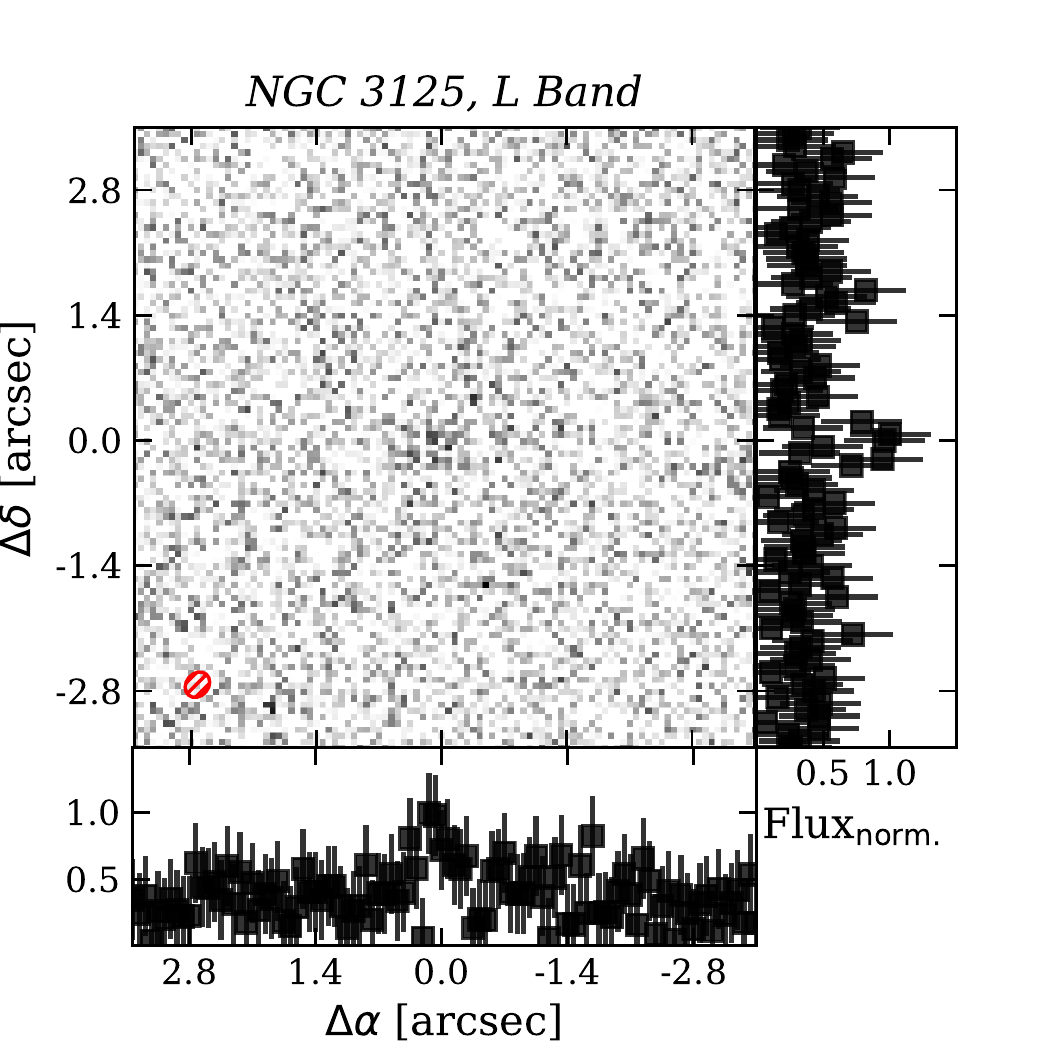}}
\subfloat{\includegraphics[width=0.25\hsize]{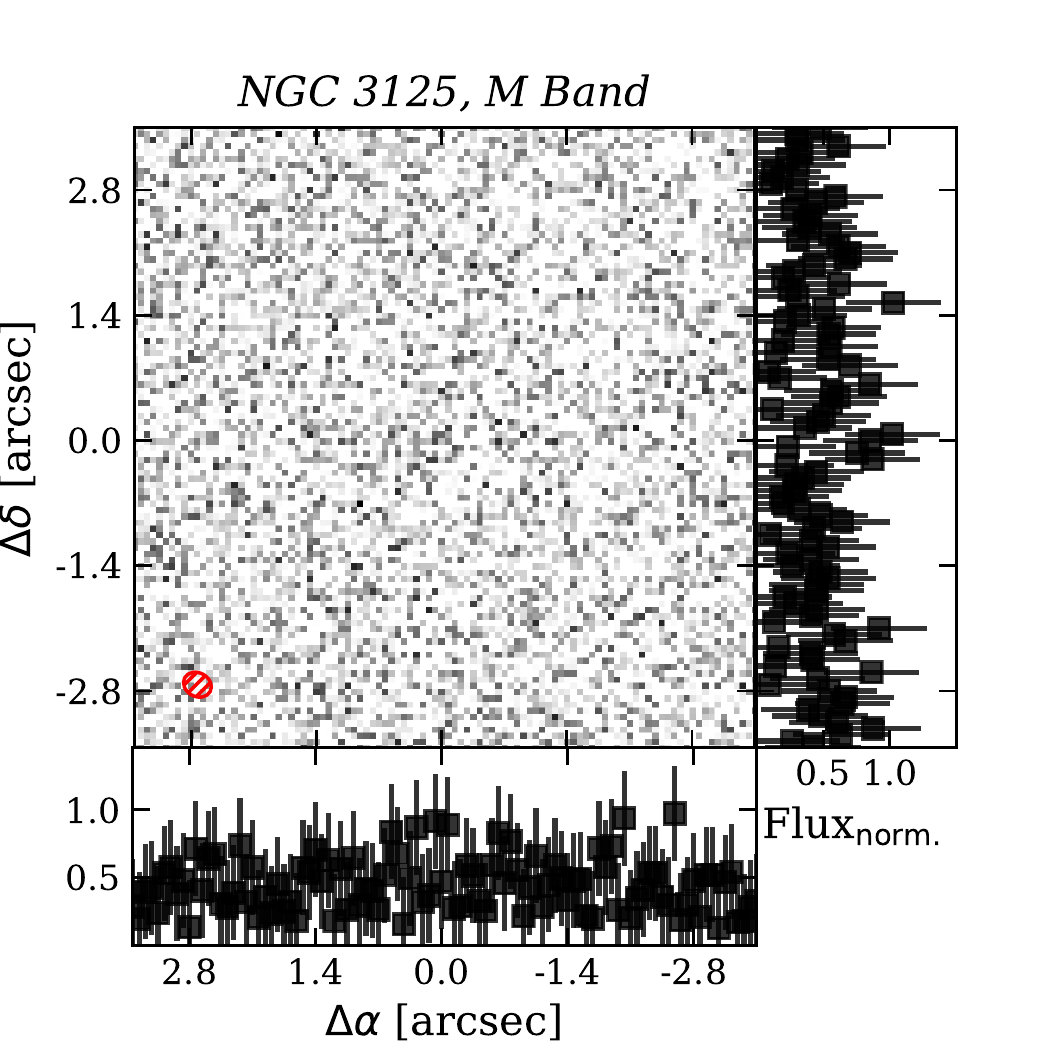}}
\subfloat{\includegraphics[width=0.25\hsize]{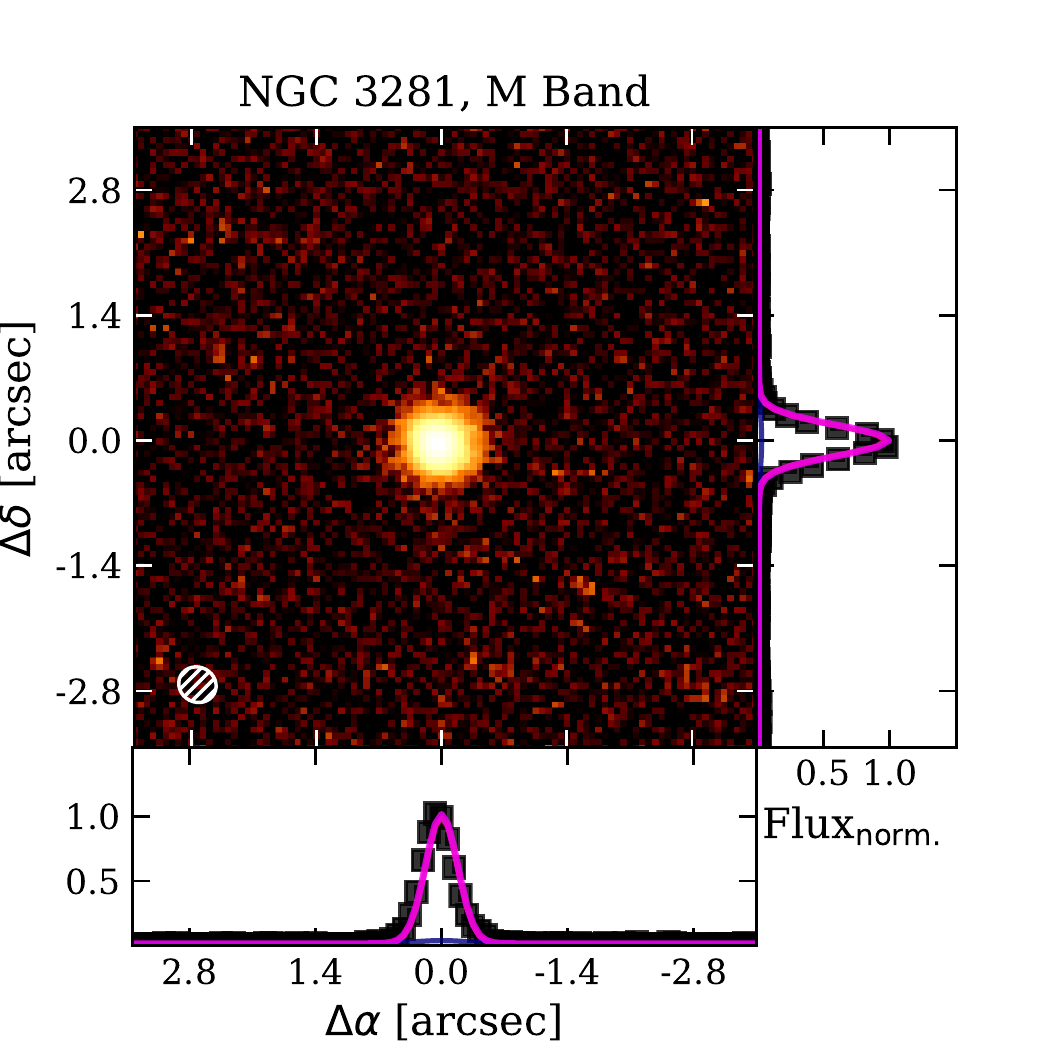}}
\subfloat{\includegraphics[width=0.25\hsize]{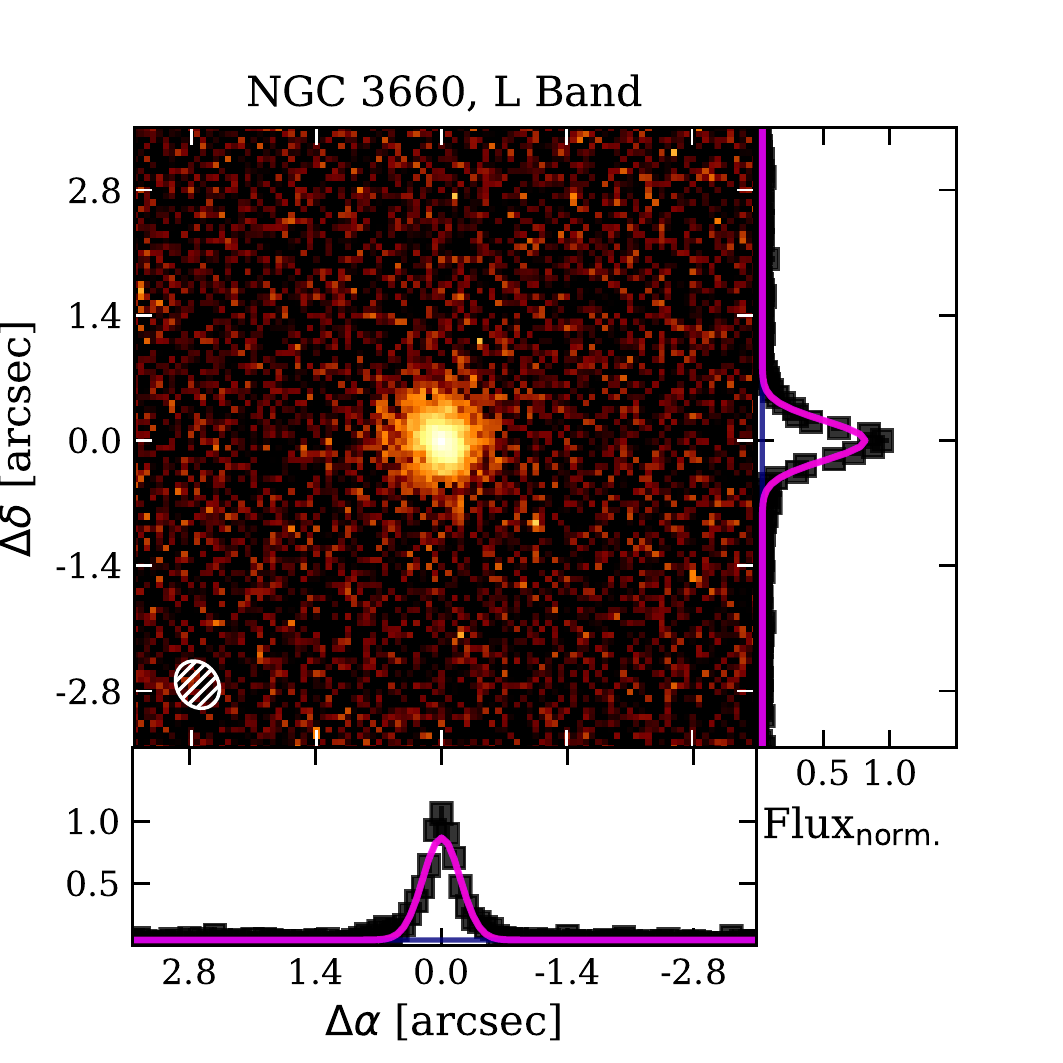}} \\
\subfloat{\includegraphics[width=0.25\hsize]{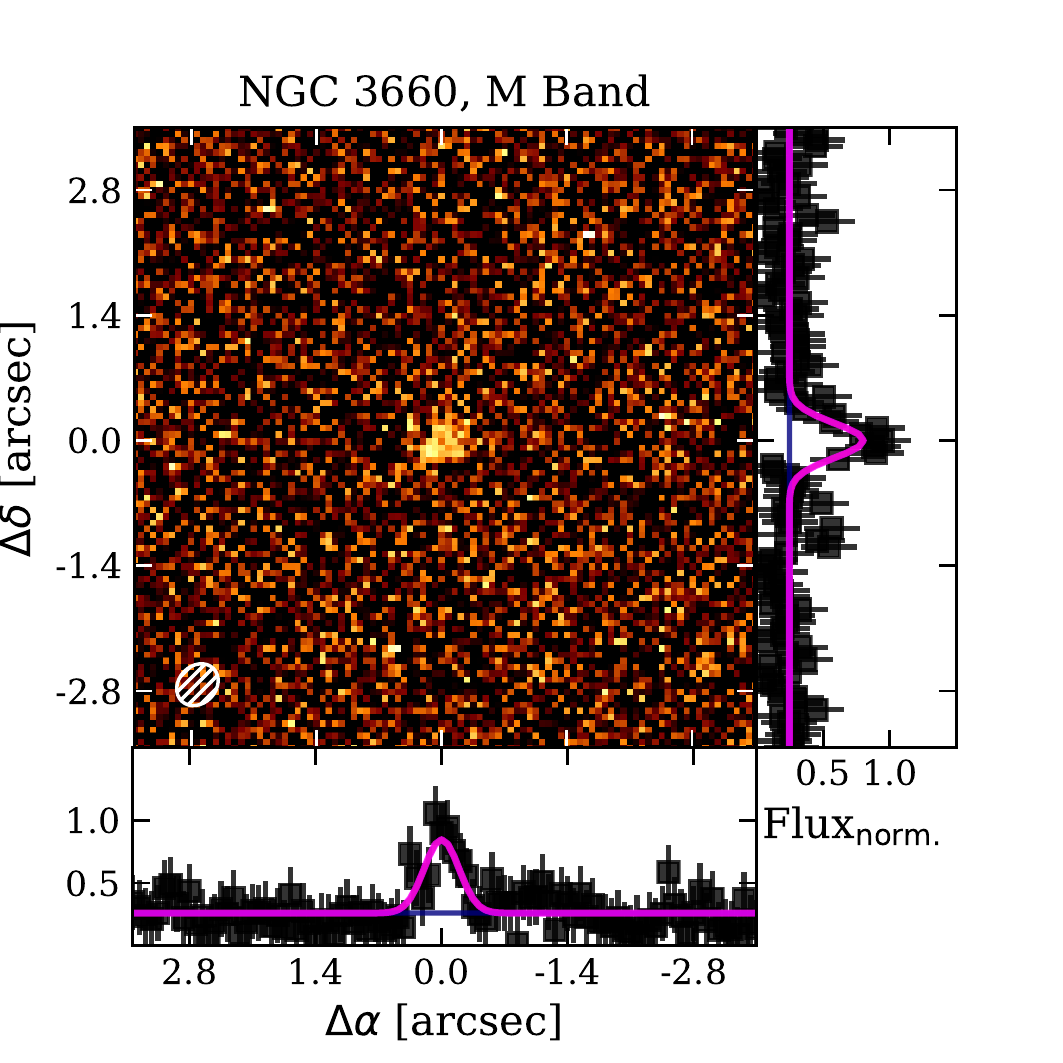}}
\subfloat{\includegraphics[width=0.25\hsize]{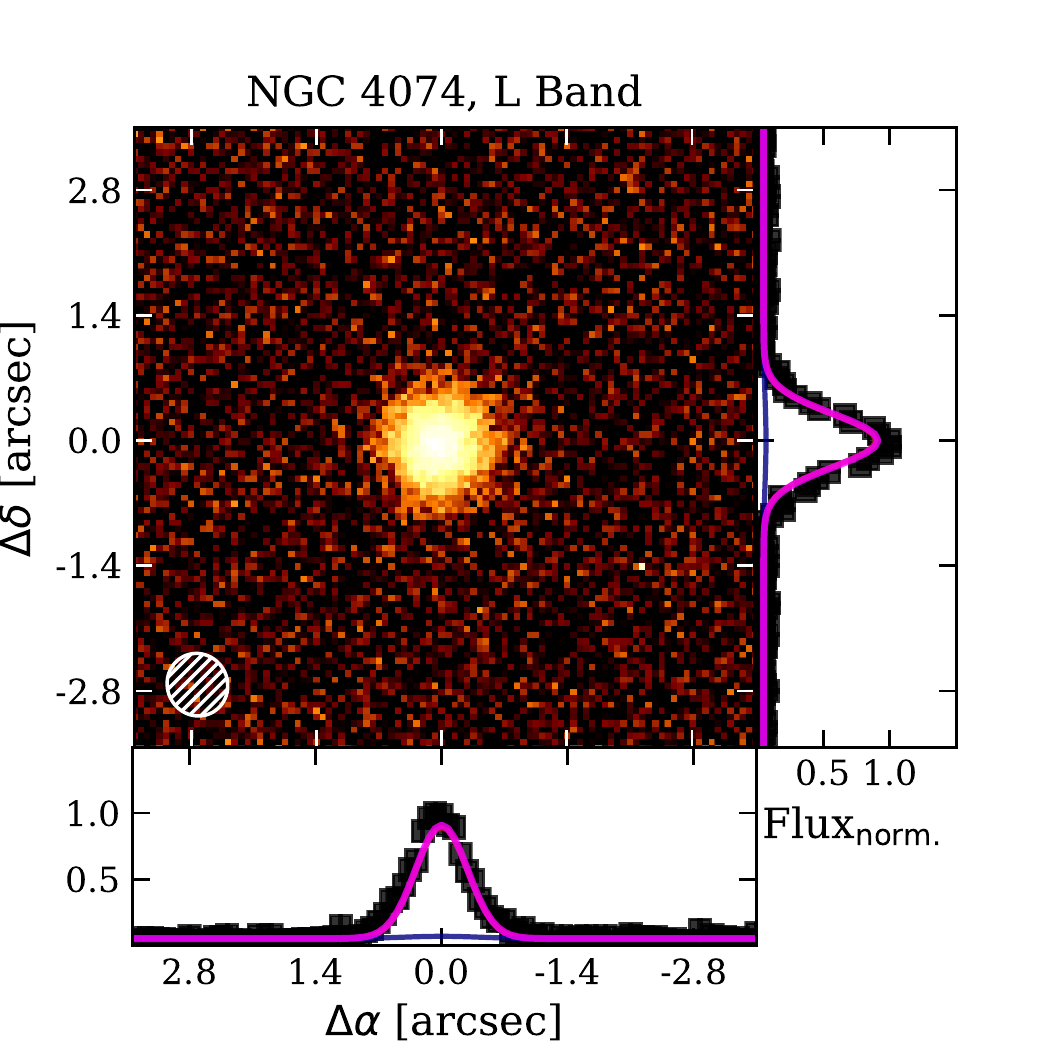}}
\subfloat{\includegraphics[width=0.25\hsize]{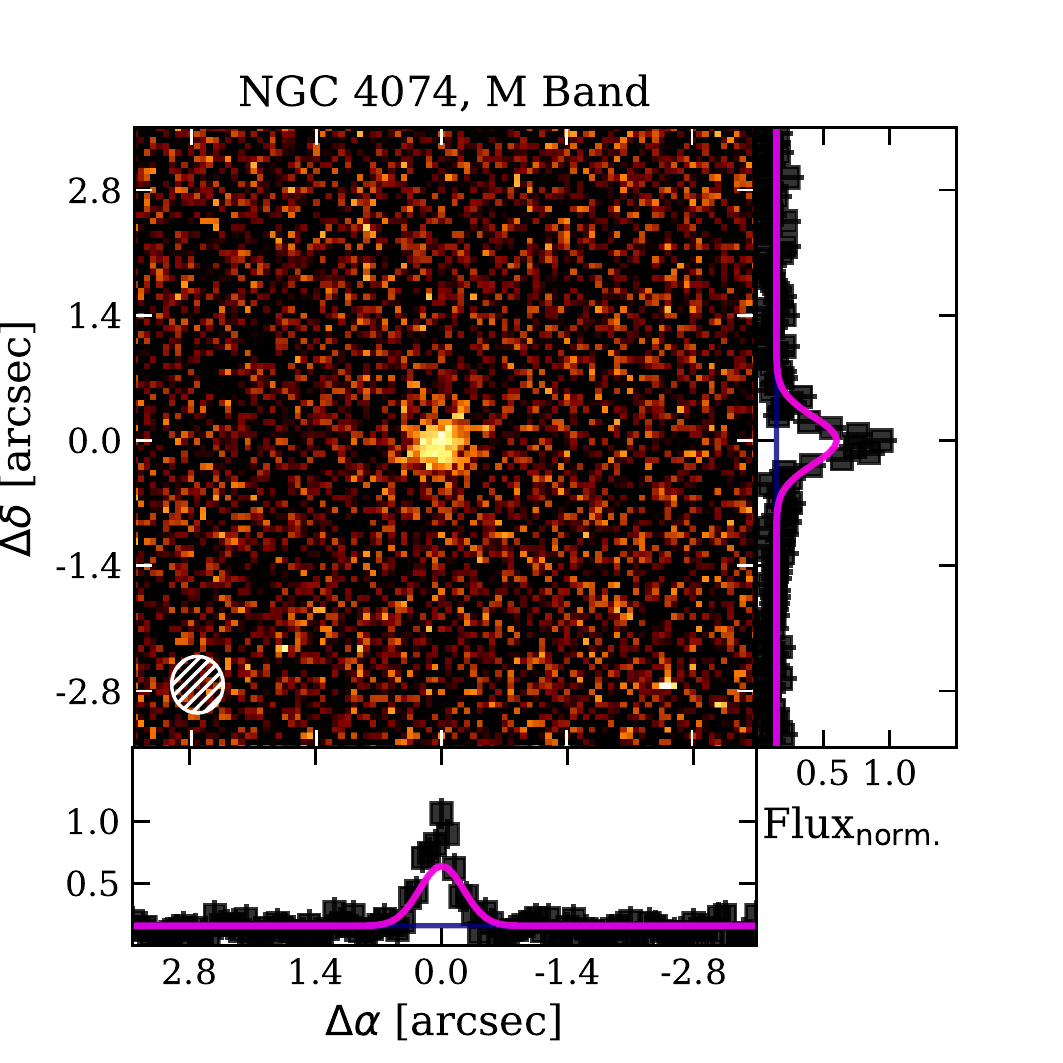}}
\subfloat{\includegraphics[width=0.25\hsize]{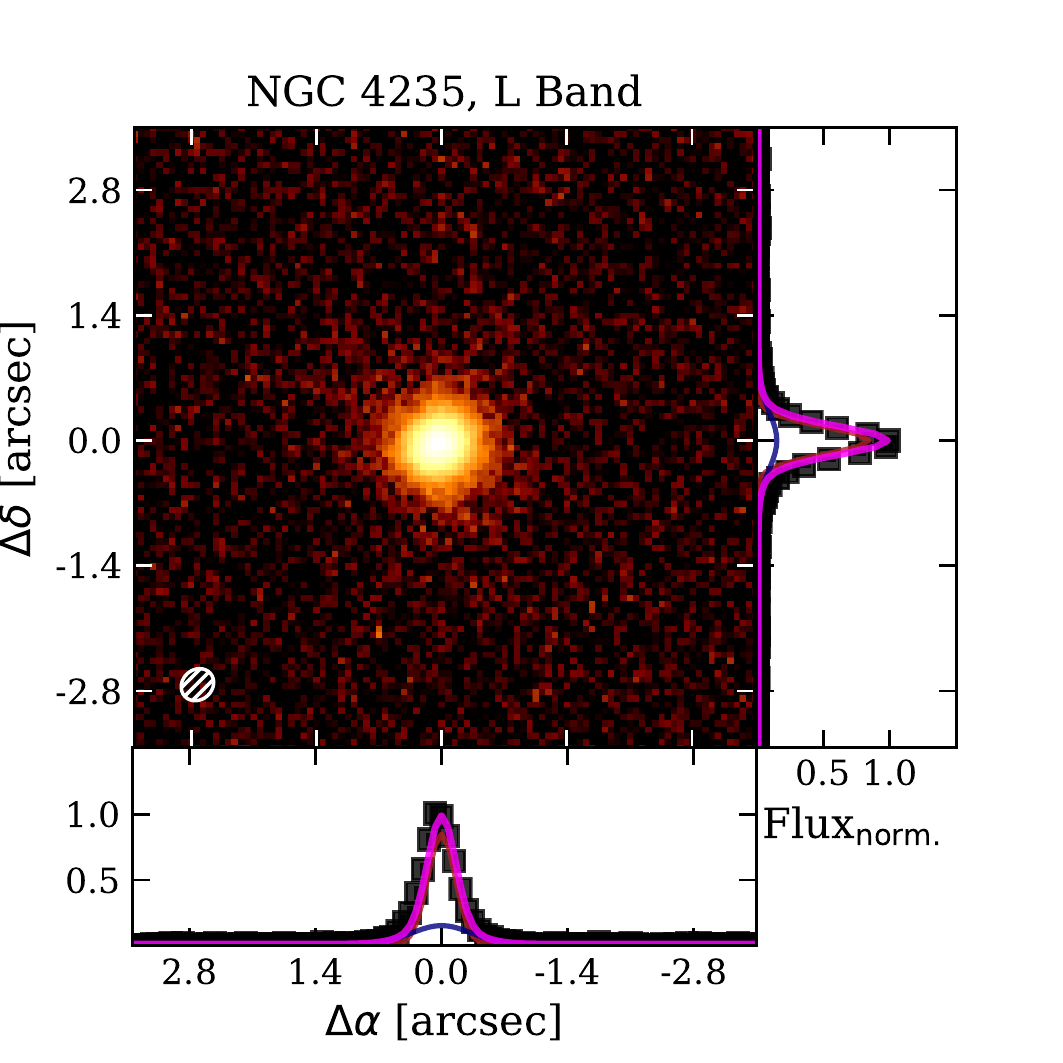}} \\
\subfloat{\includegraphics[width=0.25\hsize]{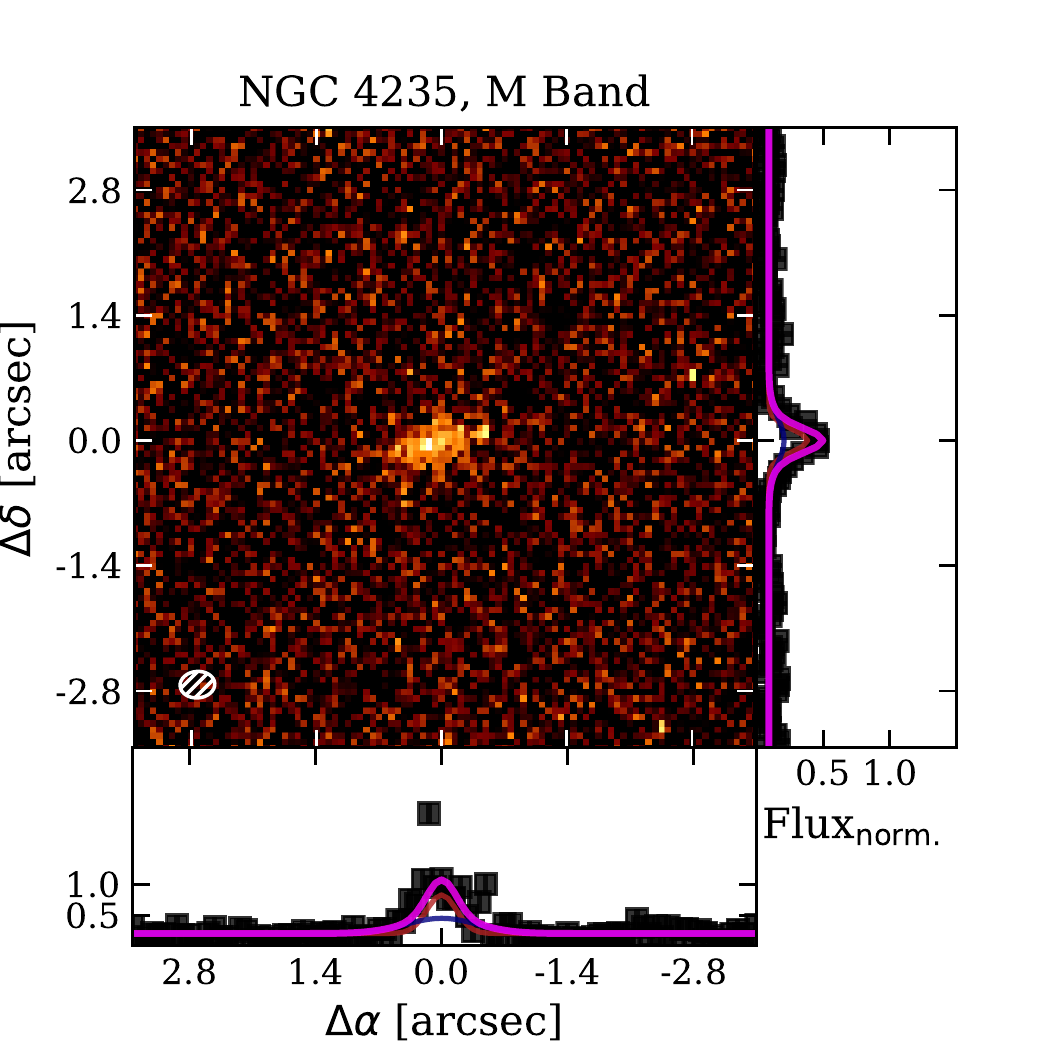}}
\subfloat{\includegraphics[width=0.25\hsize]{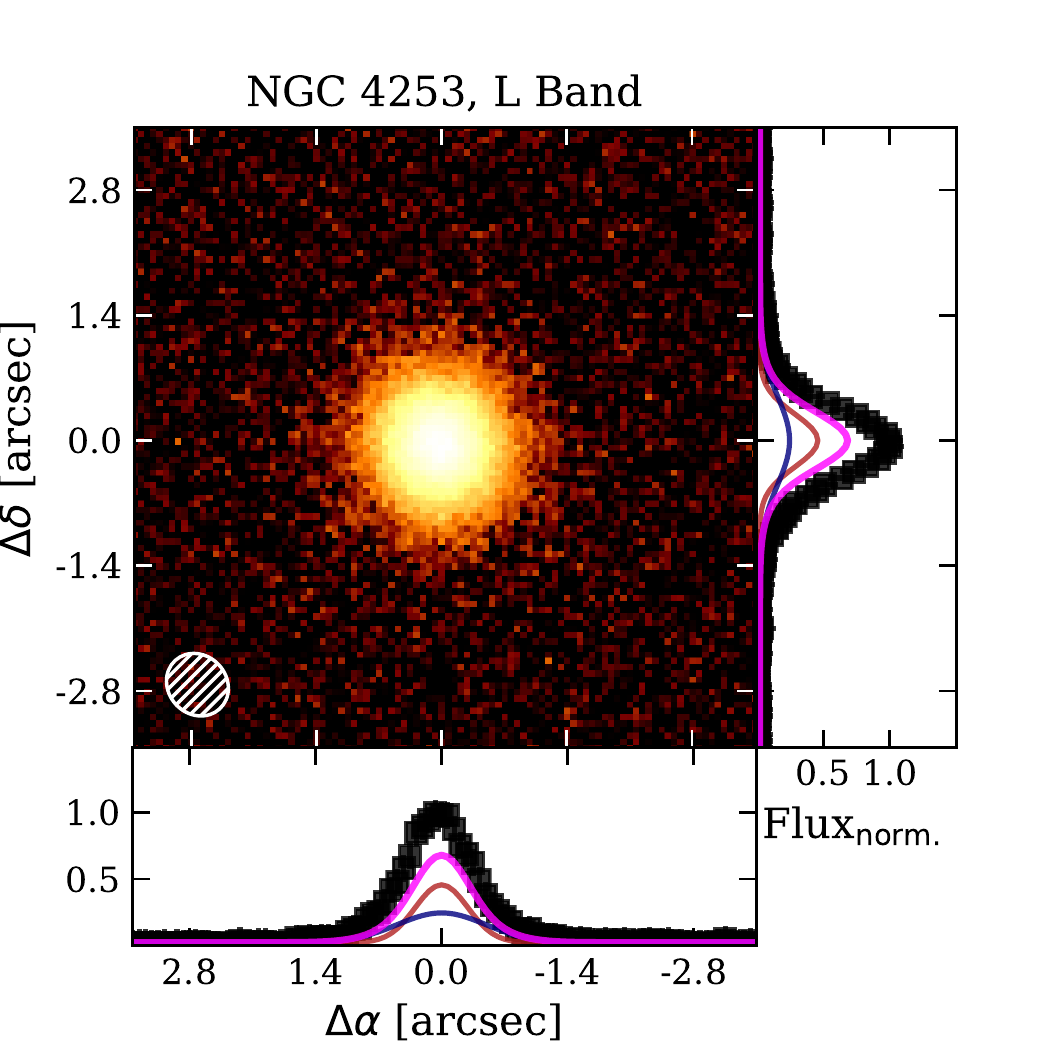}}
\subfloat{\includegraphics[width=0.25\hsize]{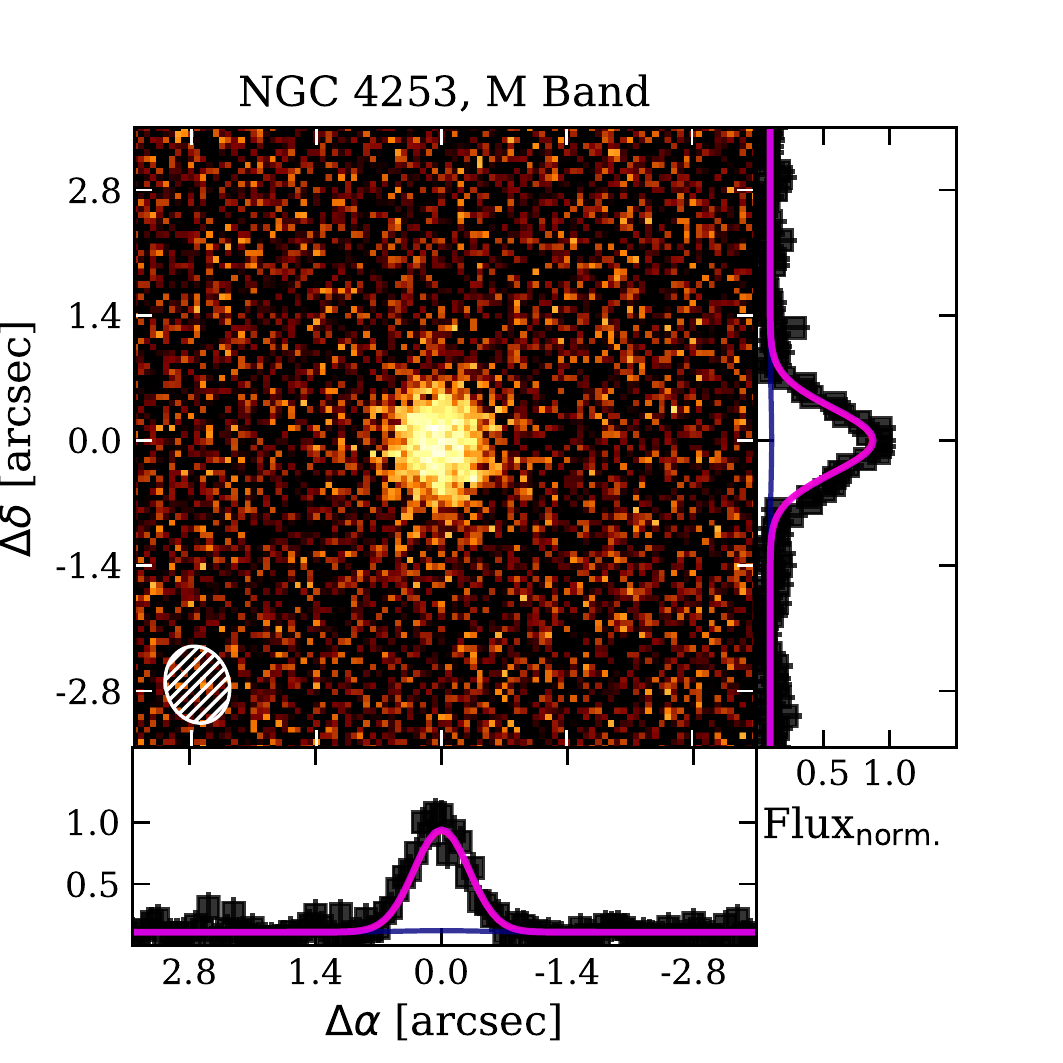}}
\subfloat{\includegraphics[width=0.25\hsize]{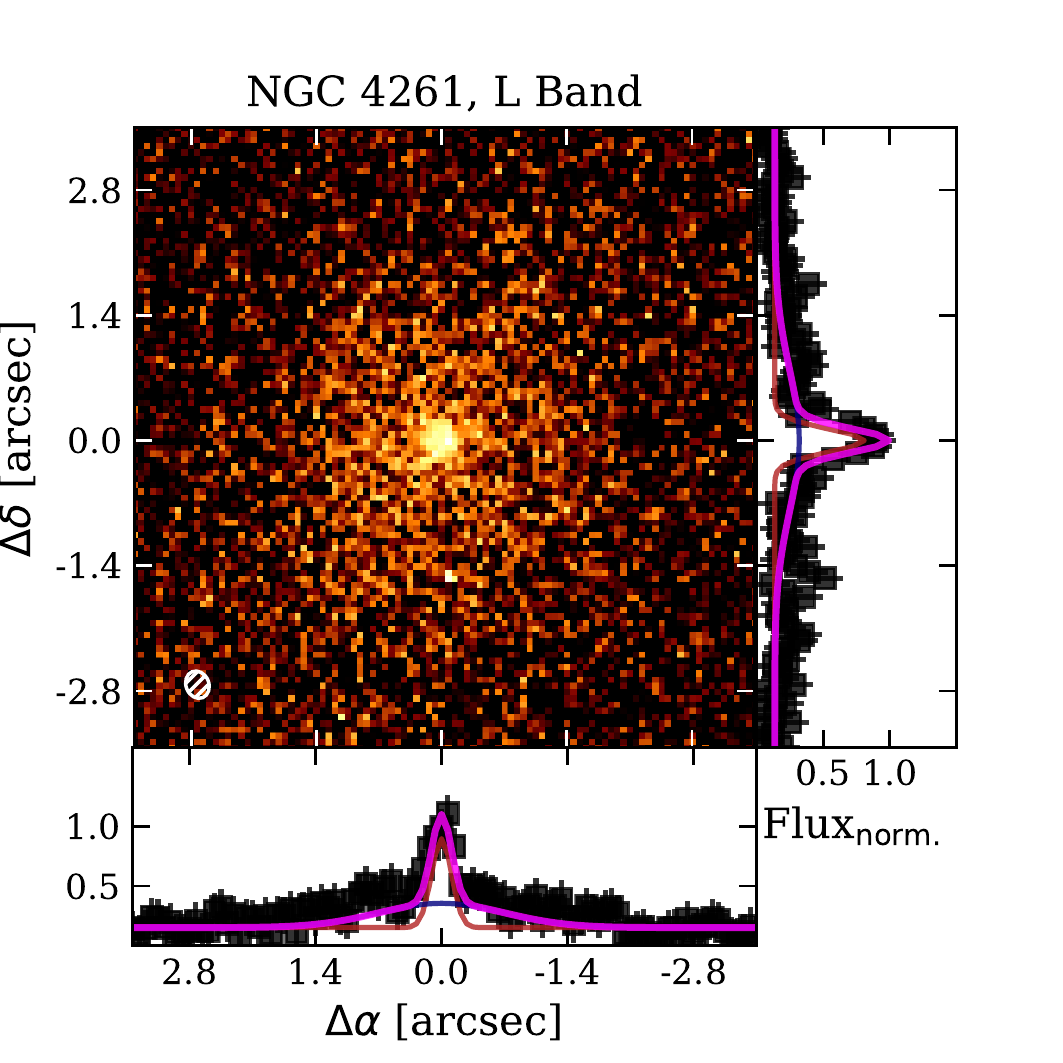}} \\
\subfloat{\includegraphics[width=0.25\hsize]{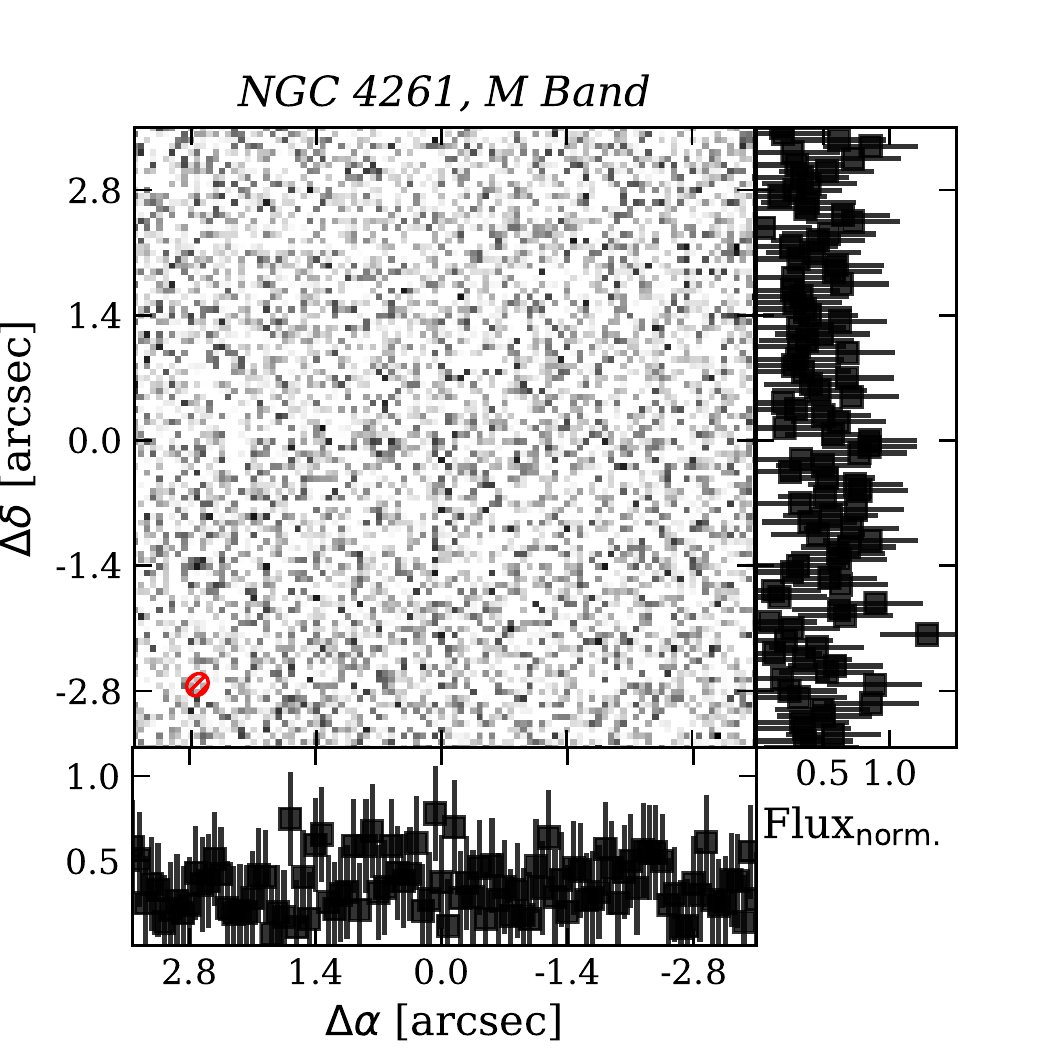}}
\subfloat{\includegraphics[width=0.25\hsize]{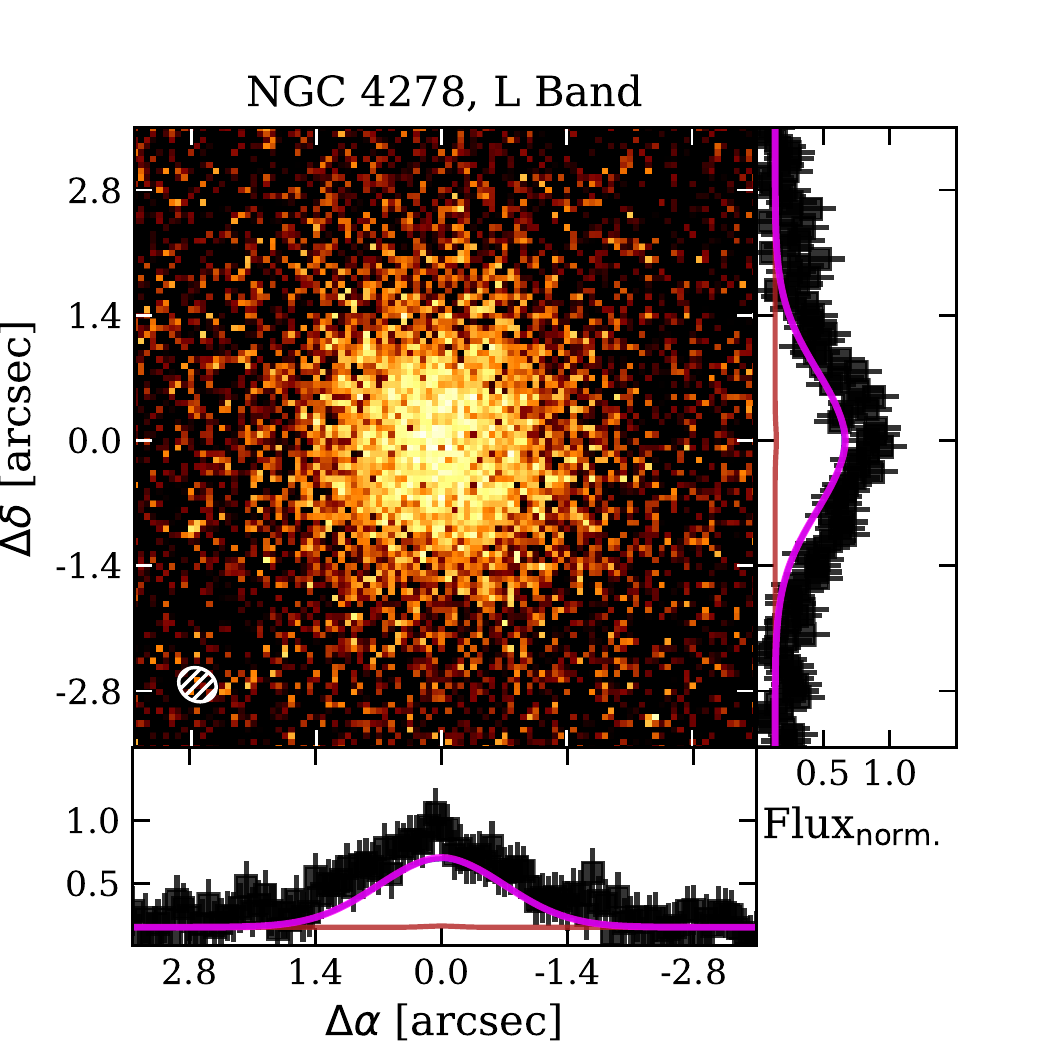}}
\subfloat{\includegraphics[width=0.25\hsize]{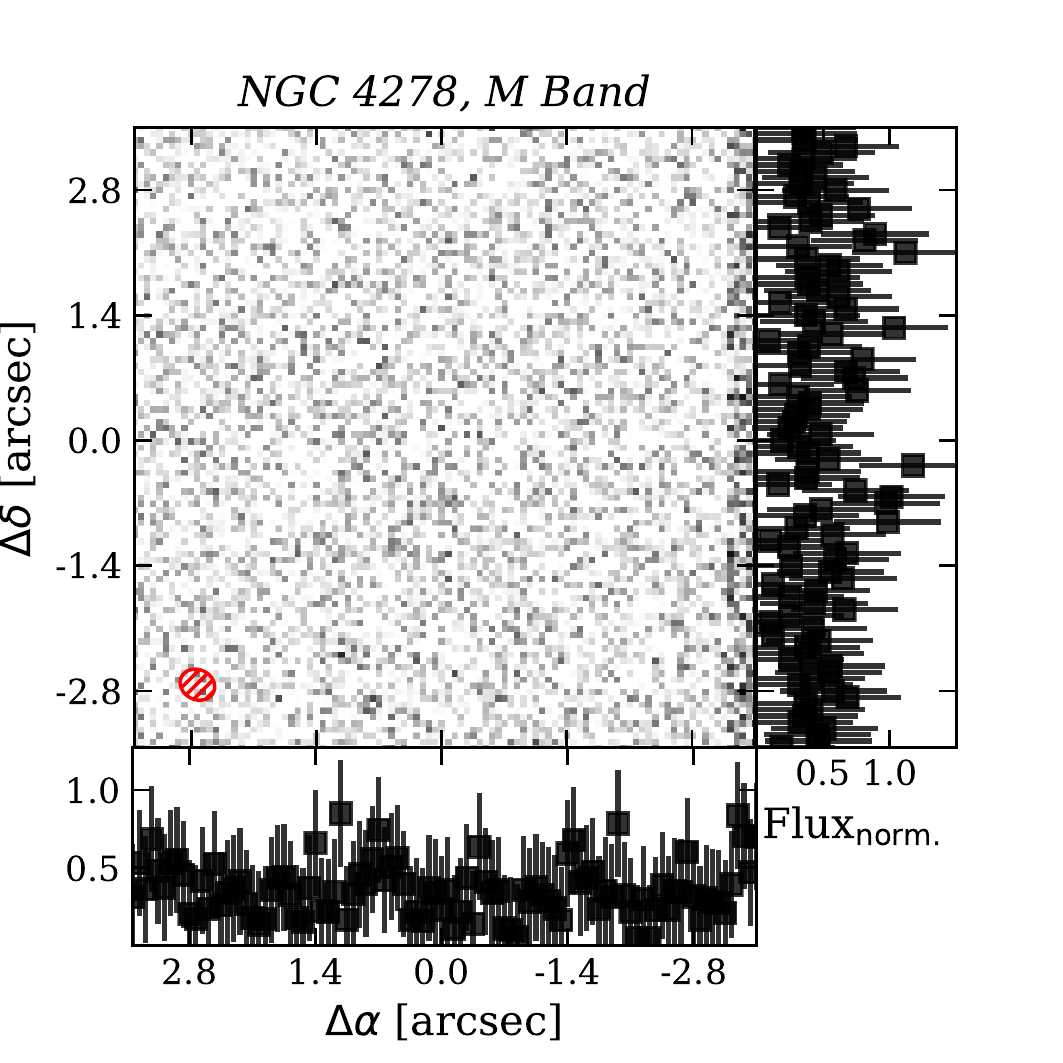}}
\subfloat{\includegraphics[width=0.25\hsize]{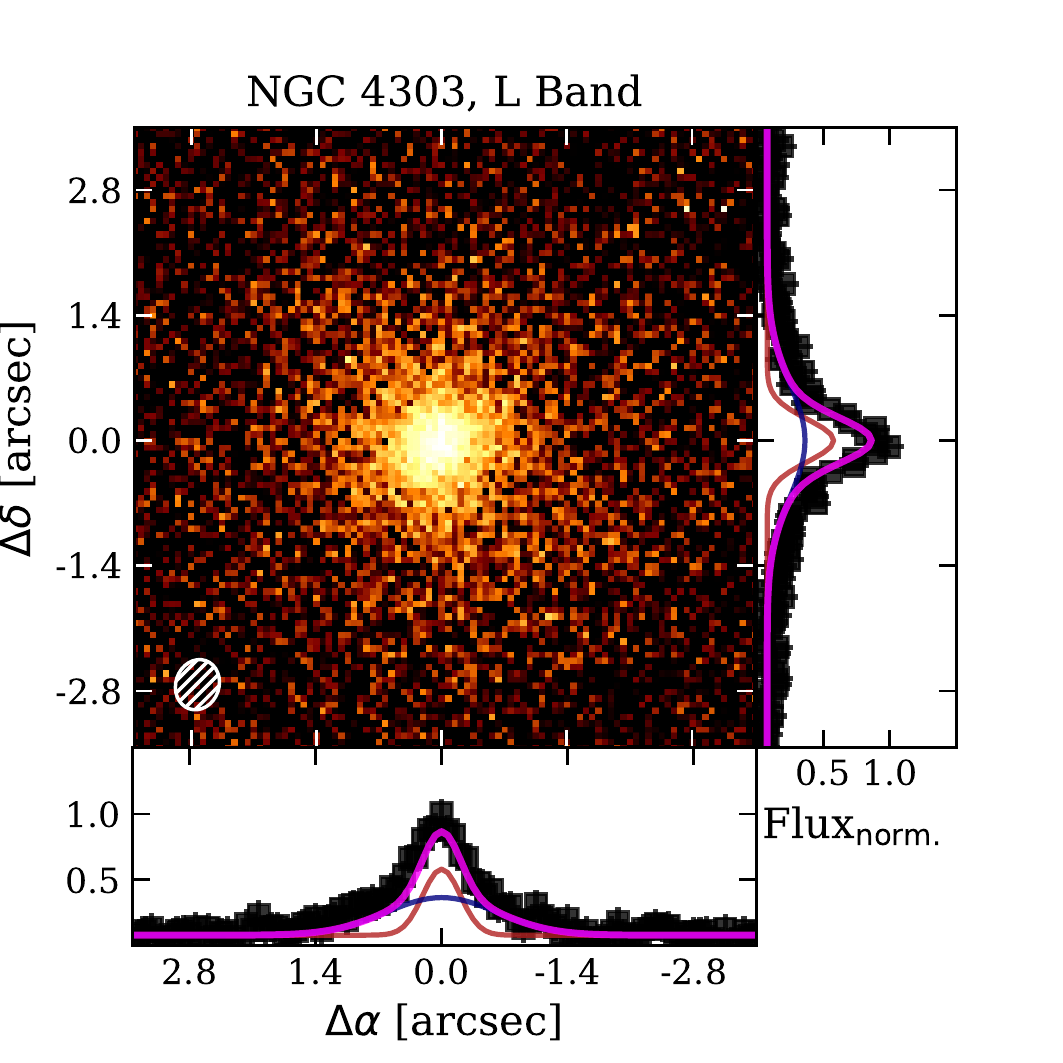}} \\
\subfloat{\includegraphics[width=0.25\hsize]{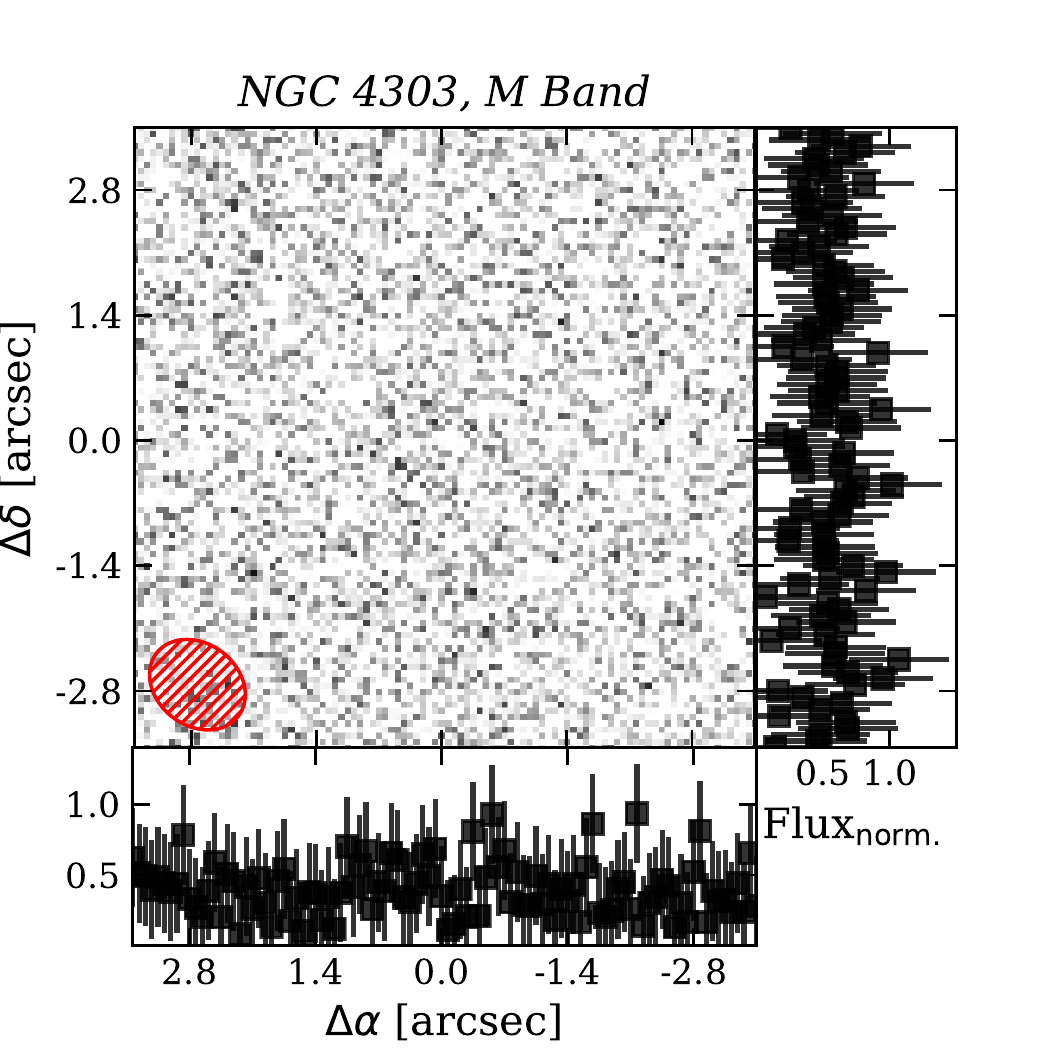}}
\subfloat{\includegraphics[width=0.25\hsize]{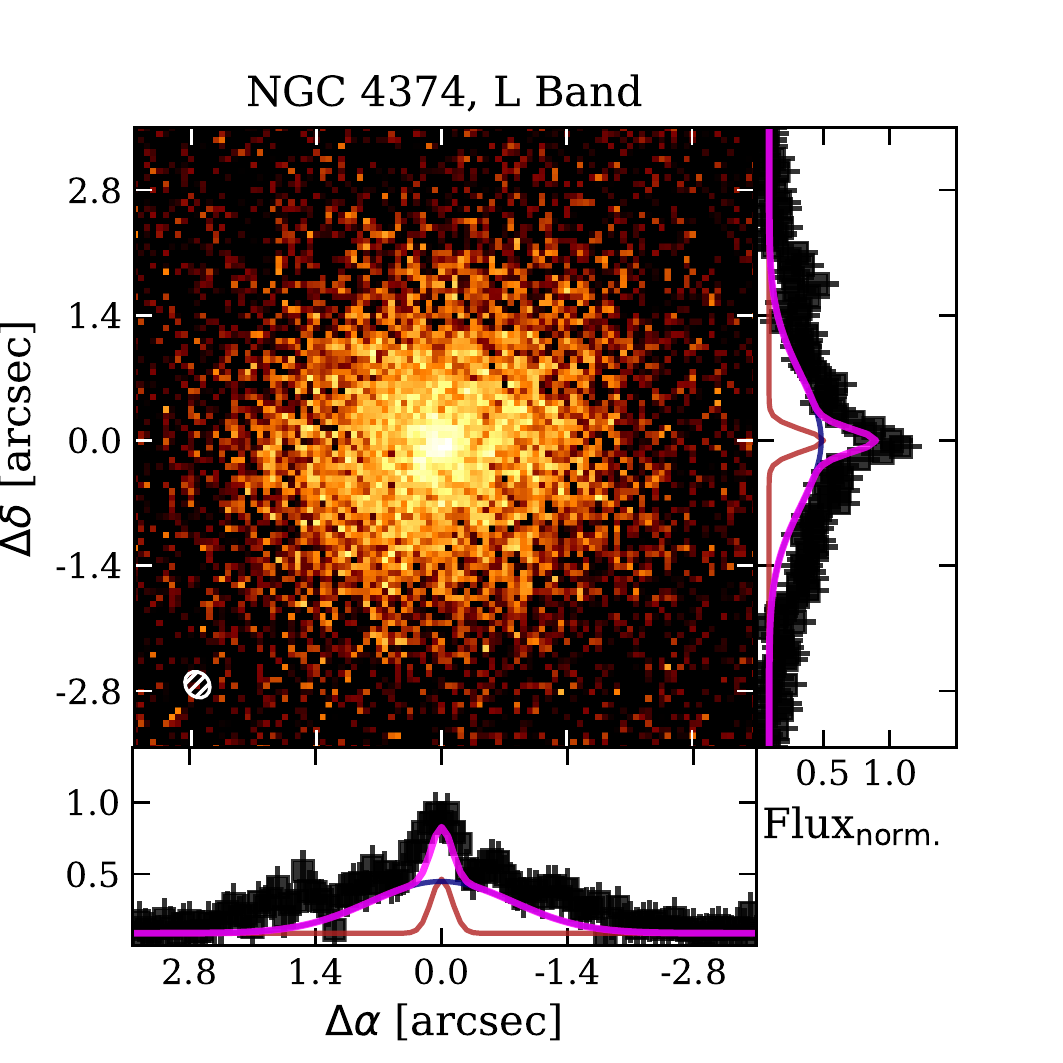}}
\subfloat{\includegraphics[width=0.25\hsize]{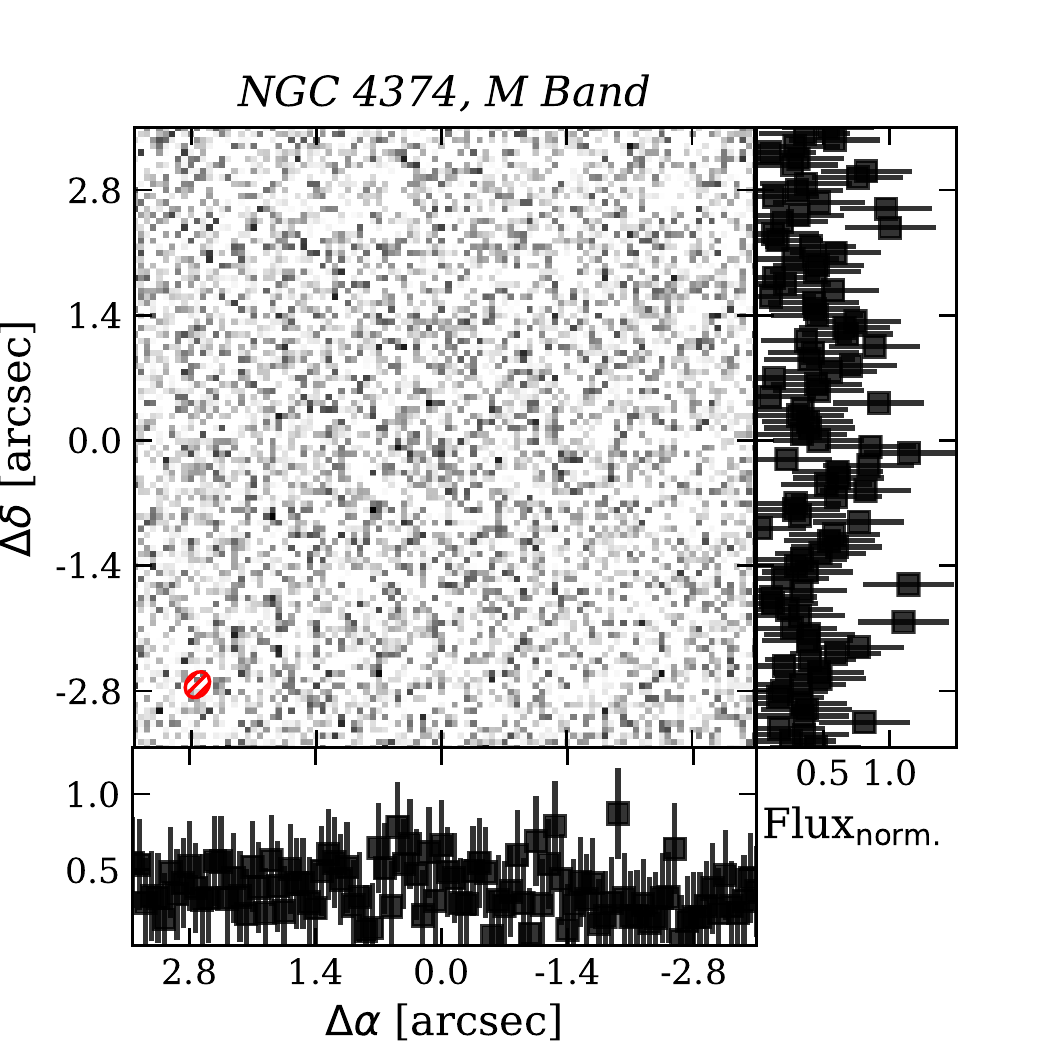}}
\subfloat{\includegraphics[width=0.25\hsize]{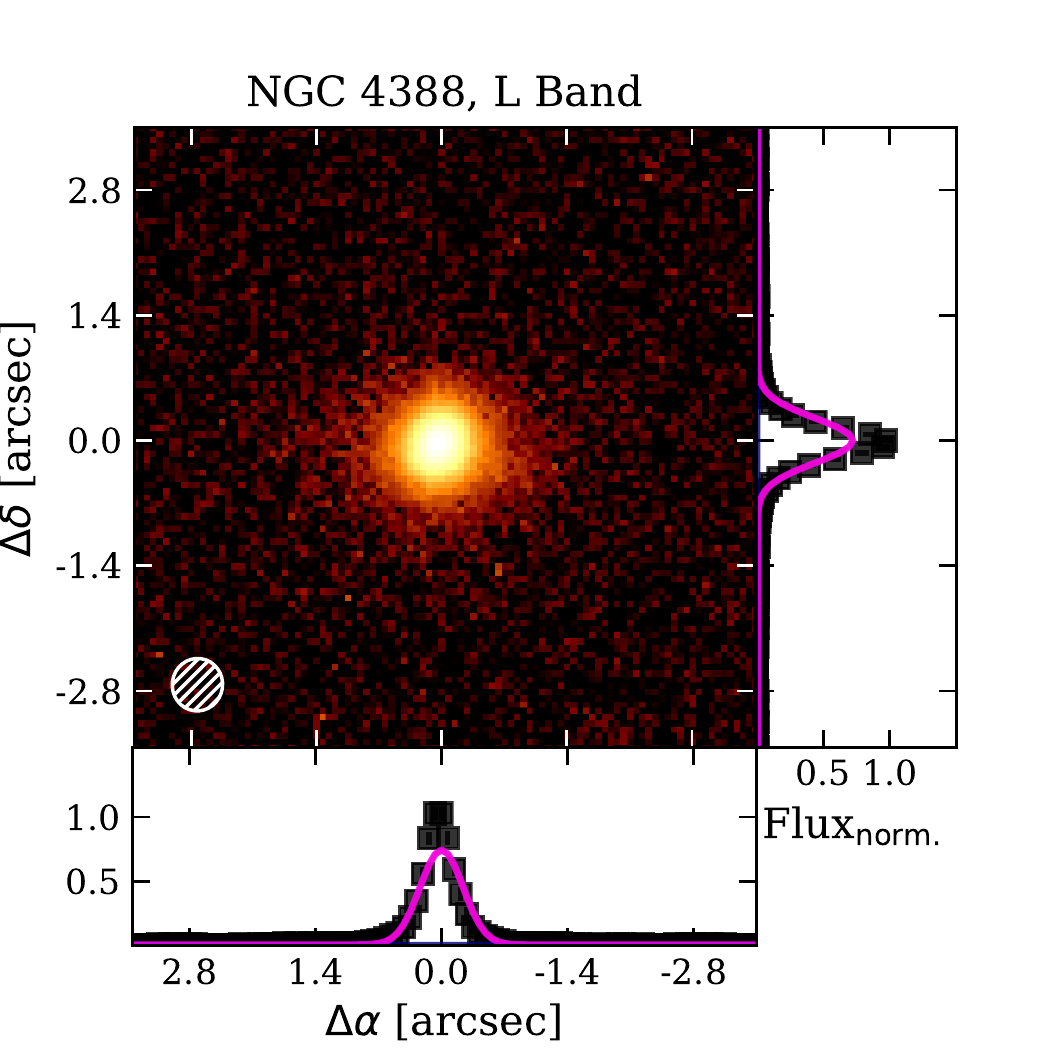}} \\
\caption{As Fig. \ref{fig:cutouts_one}.}
\end{figure*}
\begin{figure*}
\centering
\subfloat{\includegraphics[width=0.25\hsize]{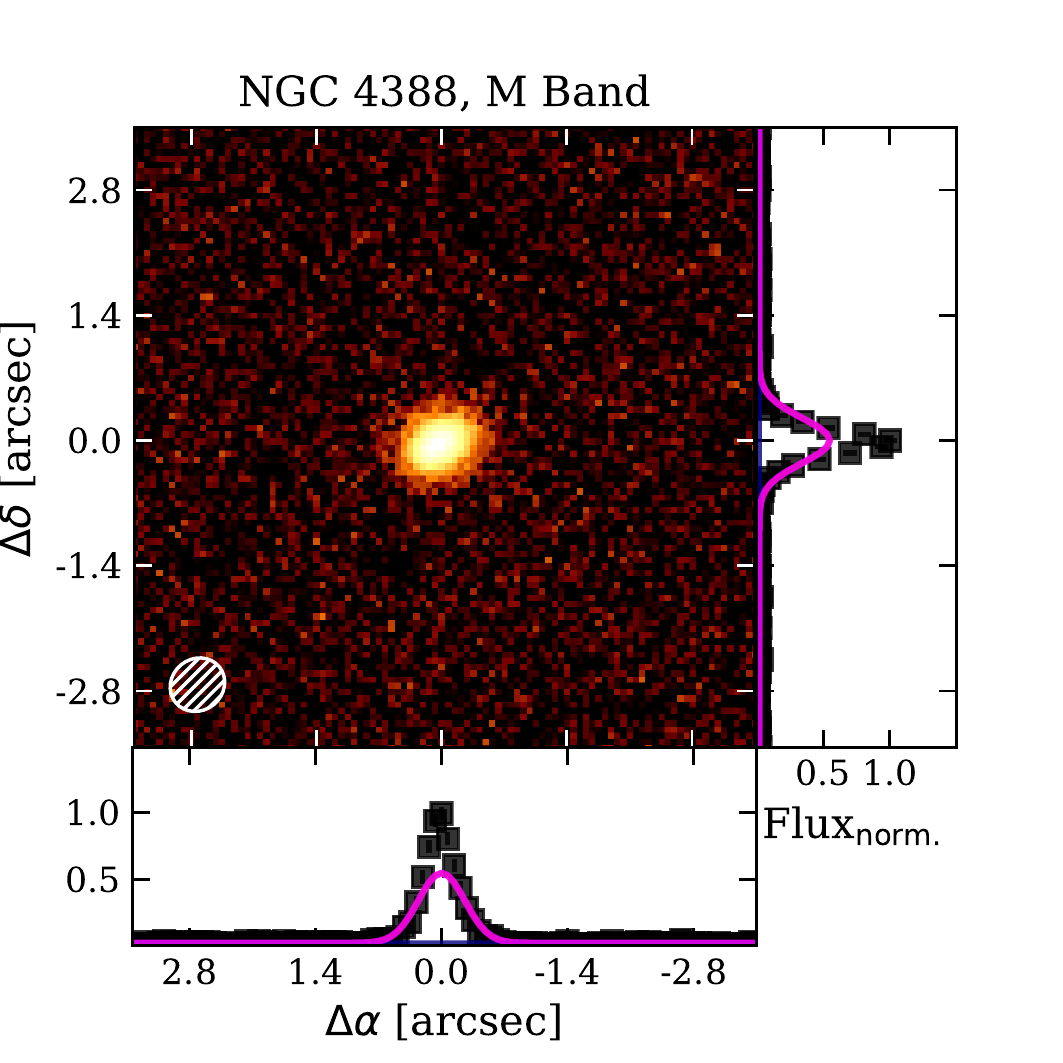}}
\subfloat{\includegraphics[width=0.25\hsize]{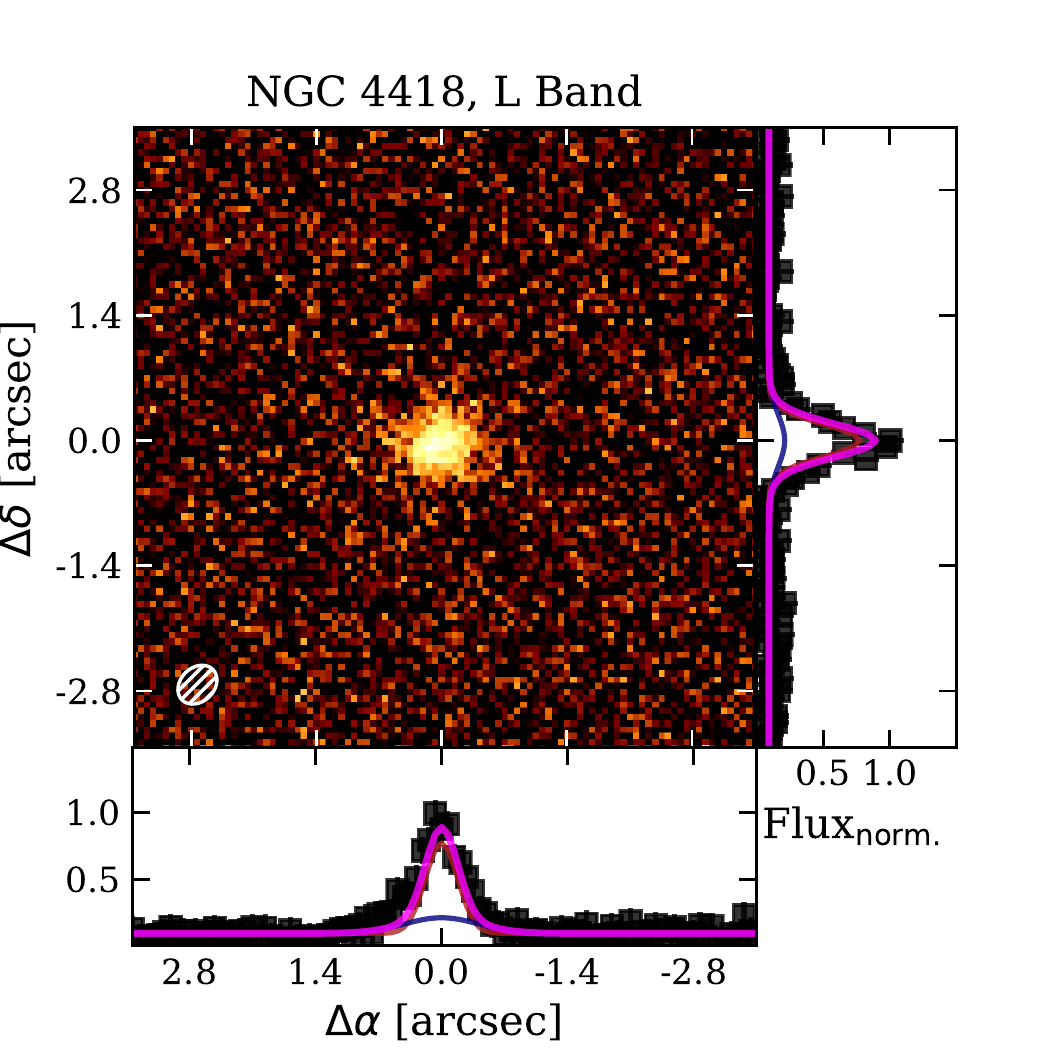}}
\subfloat{\includegraphics[width=0.25\hsize]{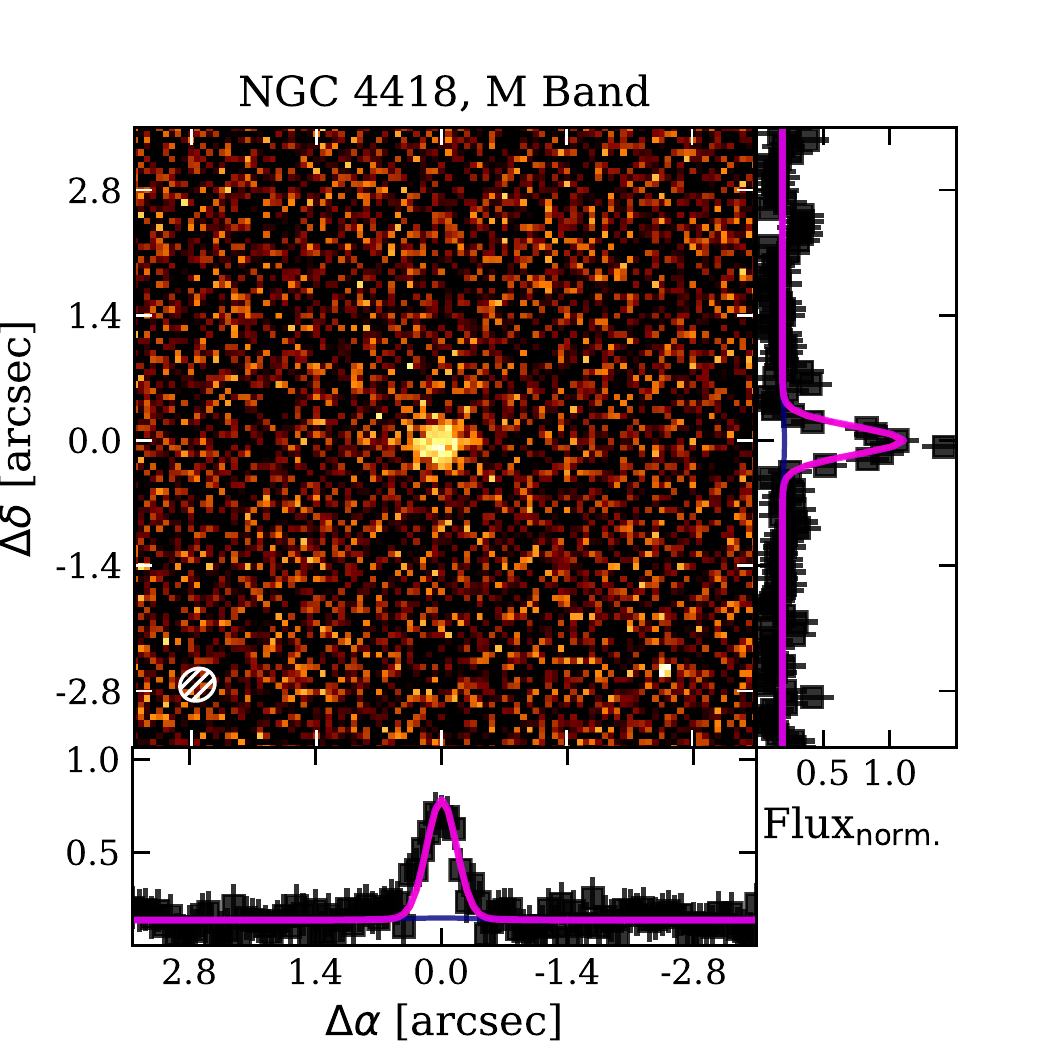}}
\subfloat{\includegraphics[width=0.25\hsize]{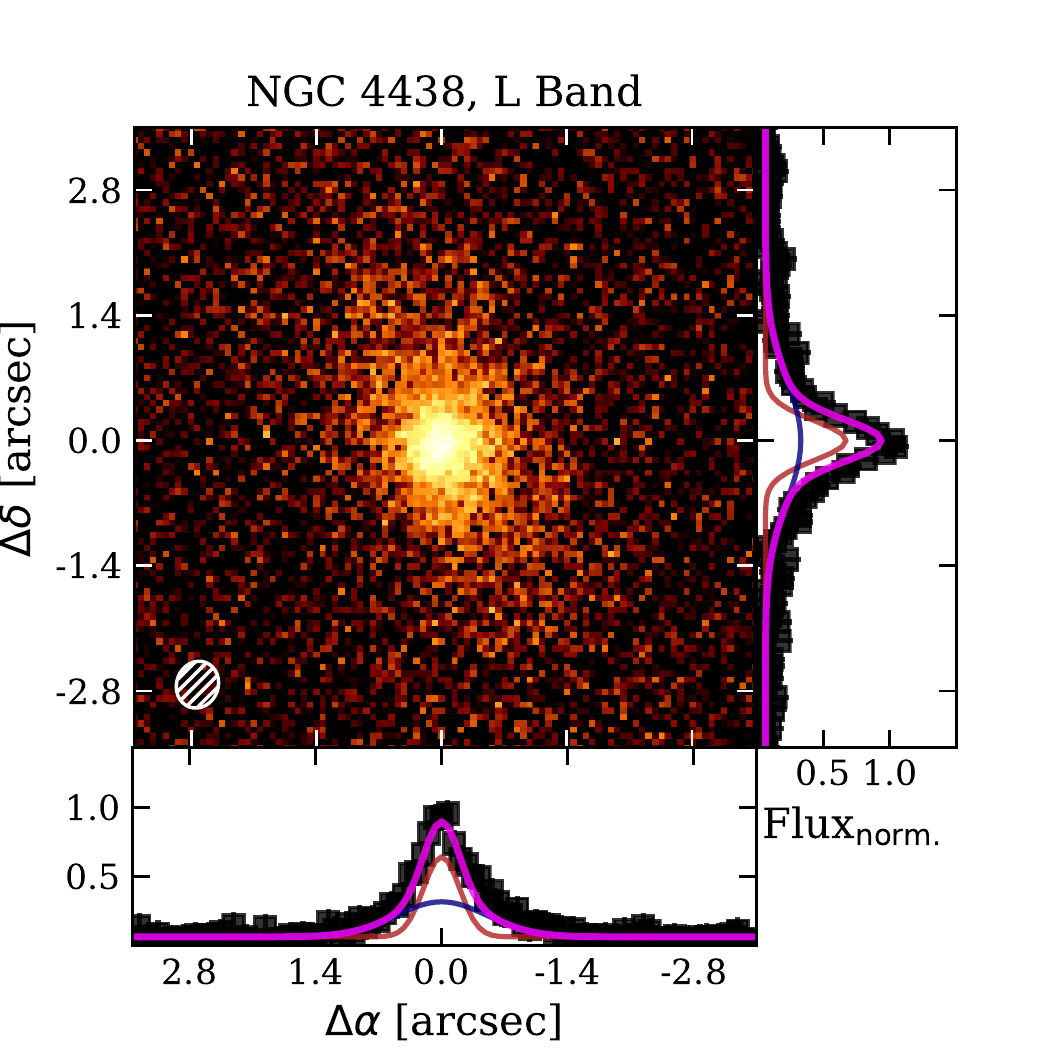}} \\
\subfloat{\includegraphics[width=0.25\hsize]{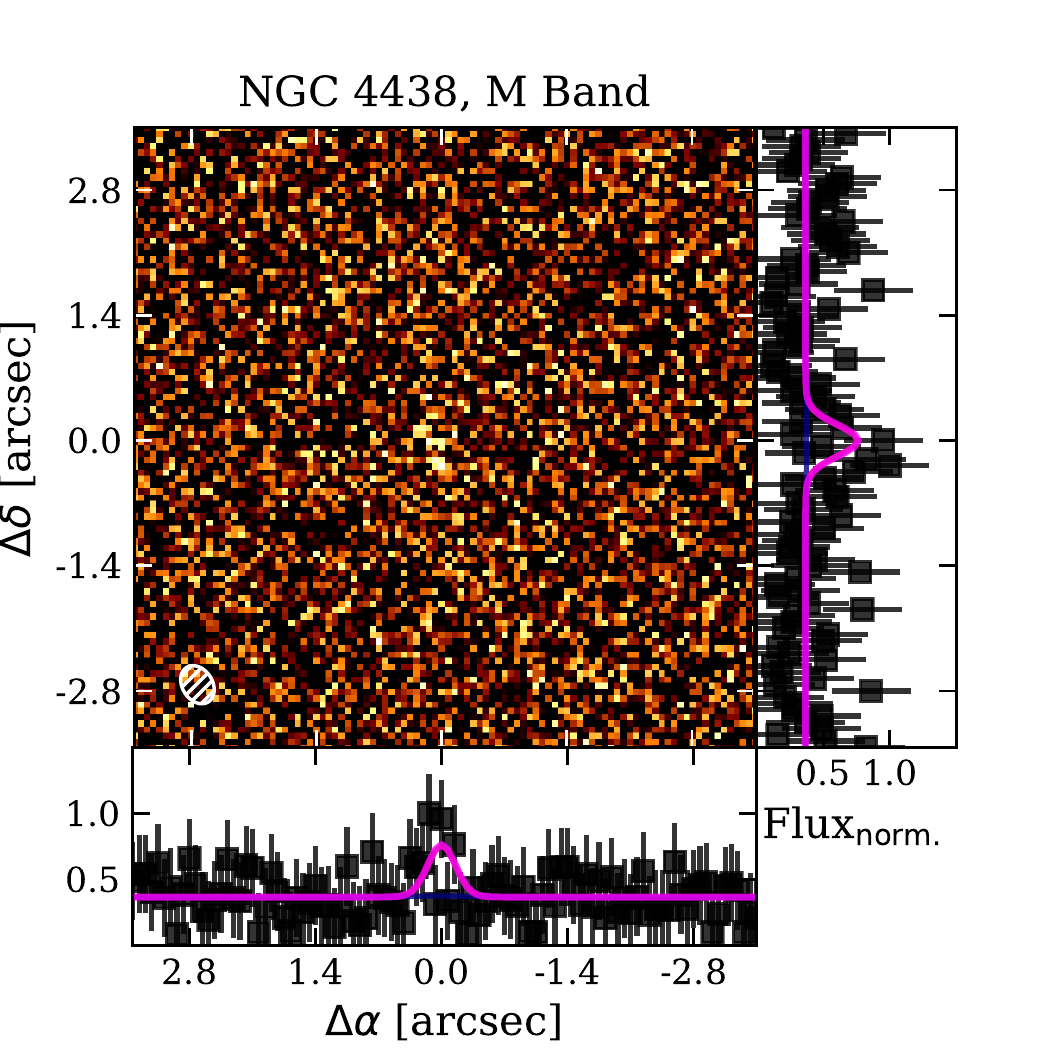}}
\subfloat{\includegraphics[width=0.25\hsize]{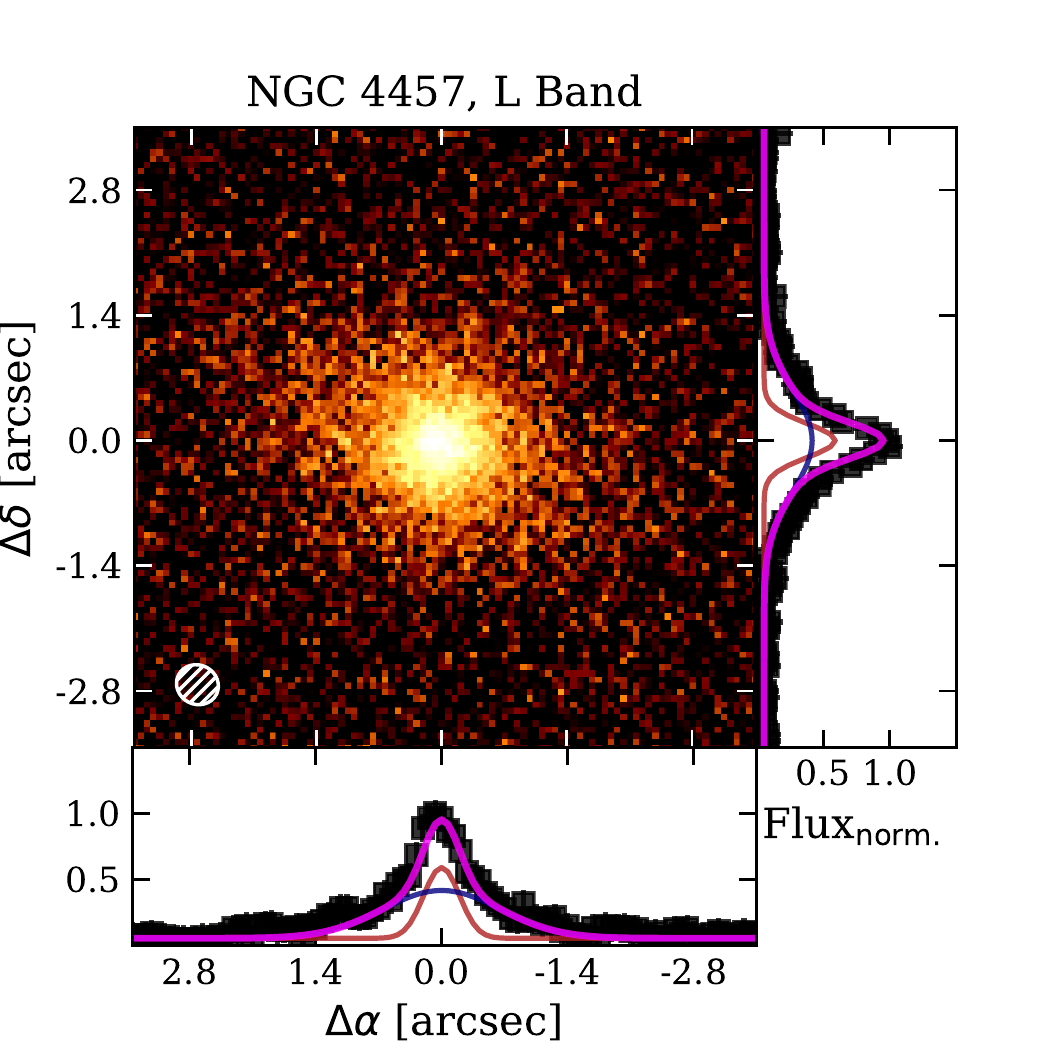}}
\subfloat{\includegraphics[width=0.25\hsize]{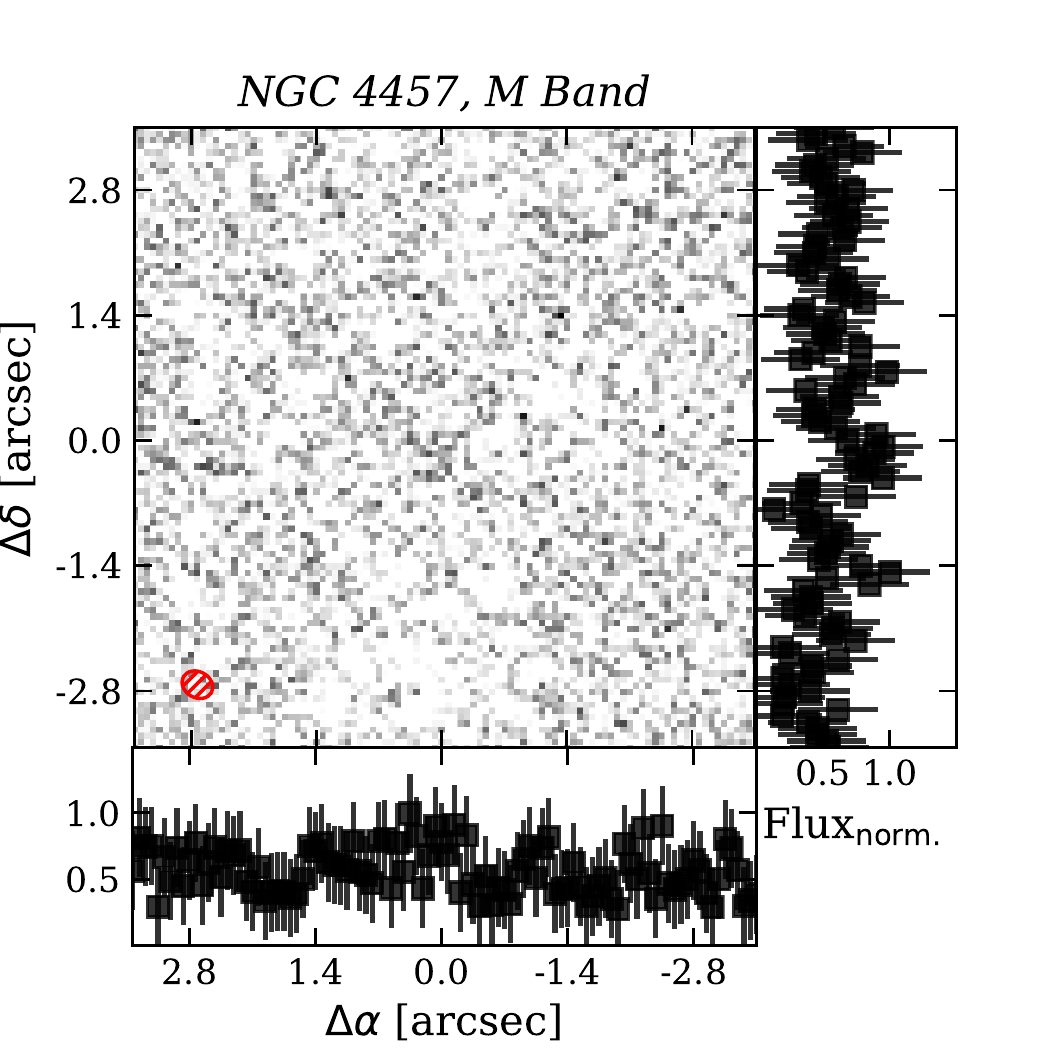}}
\subfloat{\includegraphics[width=0.25\hsize]{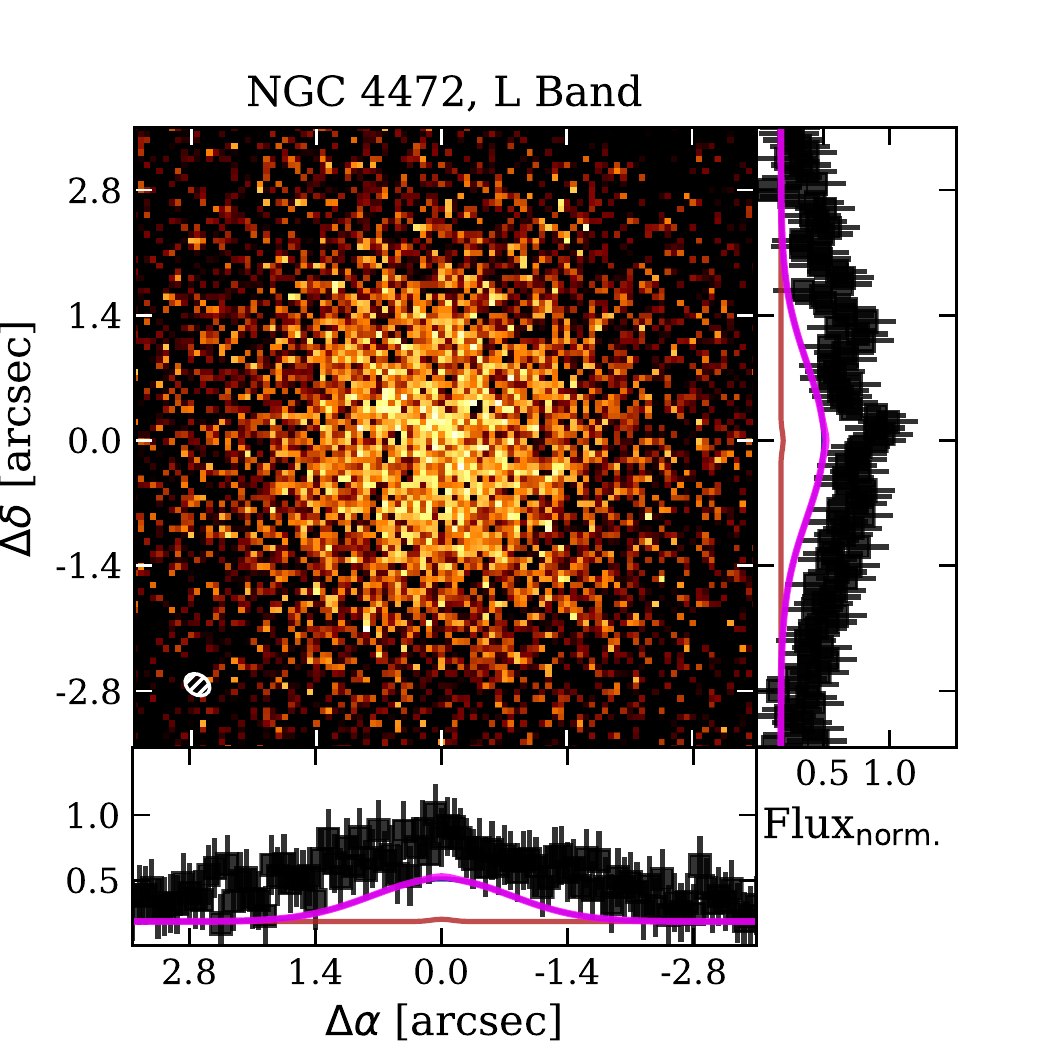}} \\
\subfloat{\includegraphics[width=0.25\hsize]{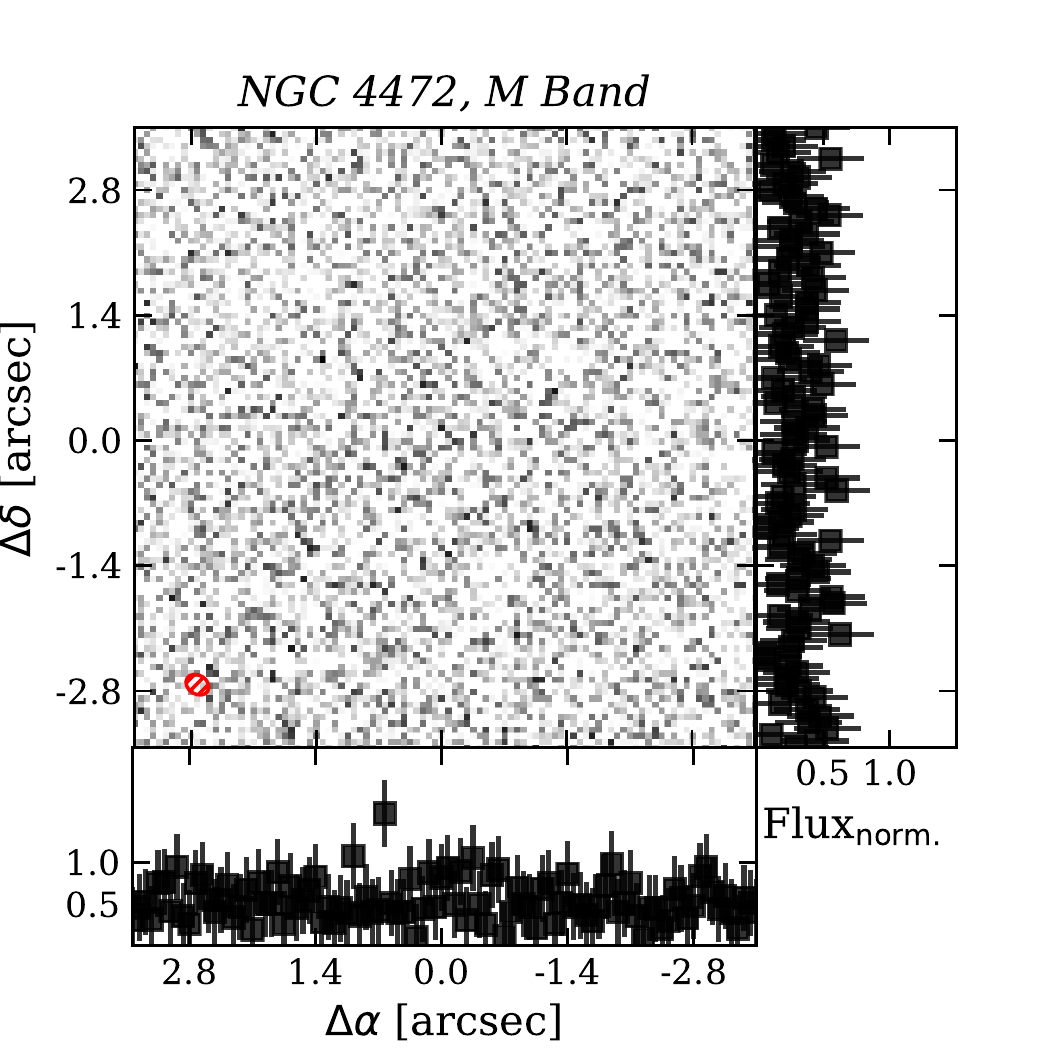}}
\subfloat{\includegraphics[width=0.25\hsize]{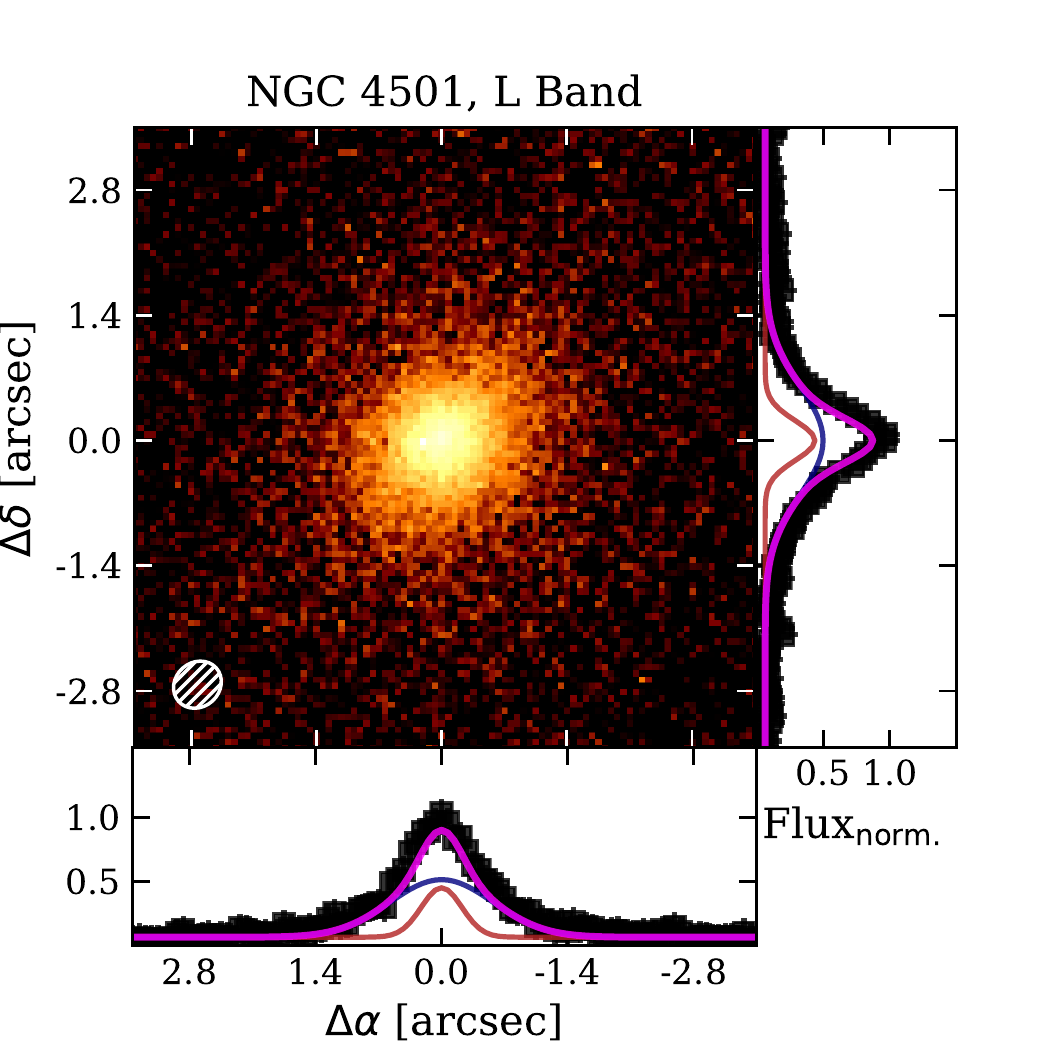}}
\subfloat{\includegraphics[width=0.25\hsize]{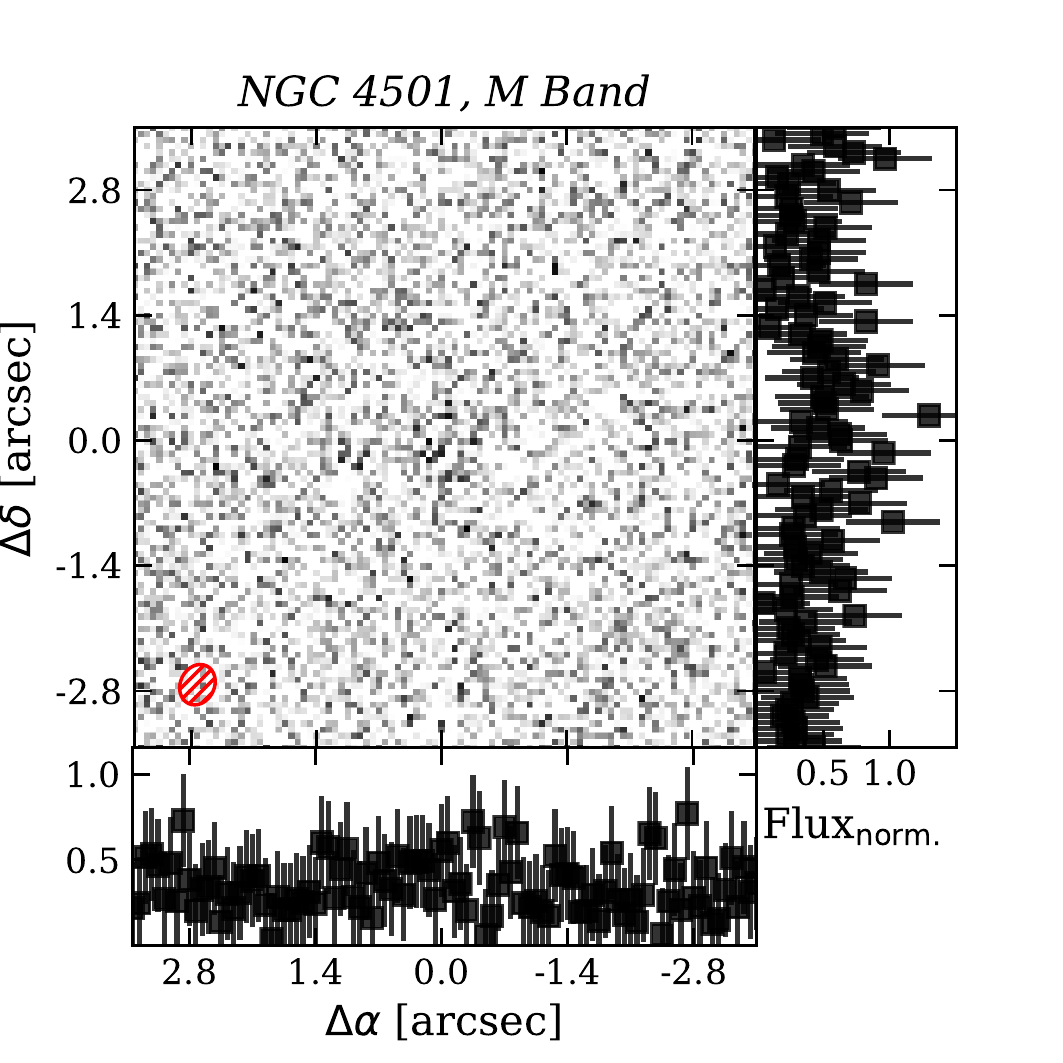}}
\subfloat{\includegraphics[width=0.25\hsize]{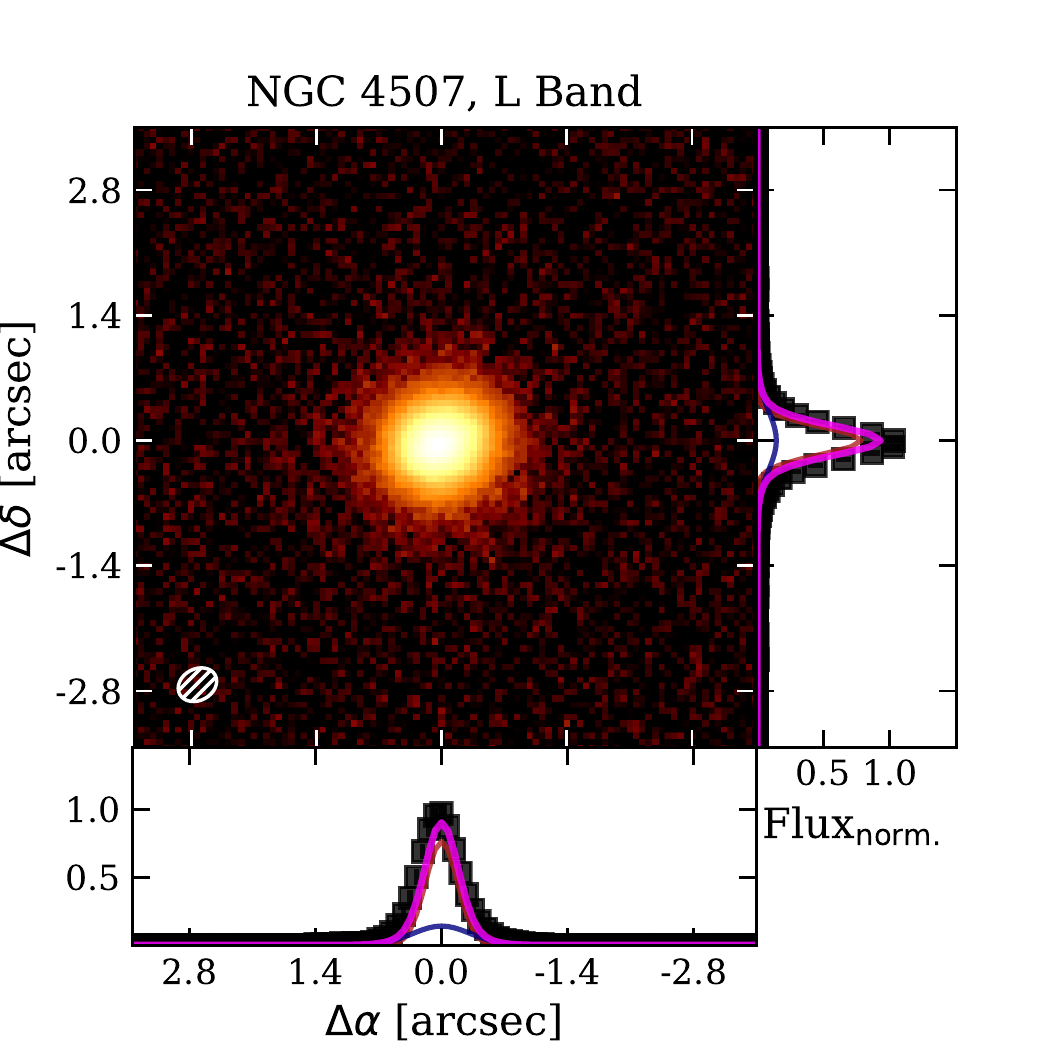}} \\
\subfloat{\includegraphics[width=0.25\hsize]{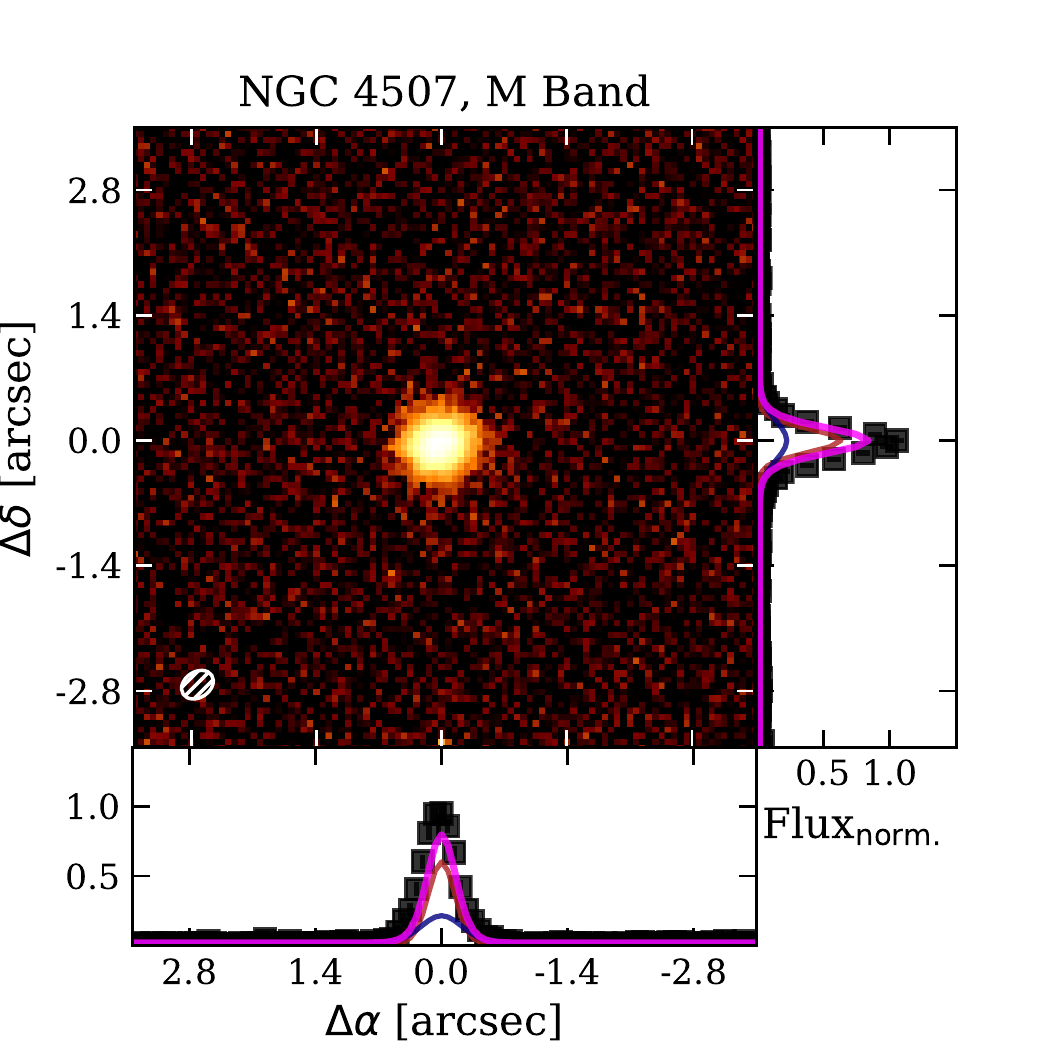}}
\subfloat{\includegraphics[width=0.25\hsize]{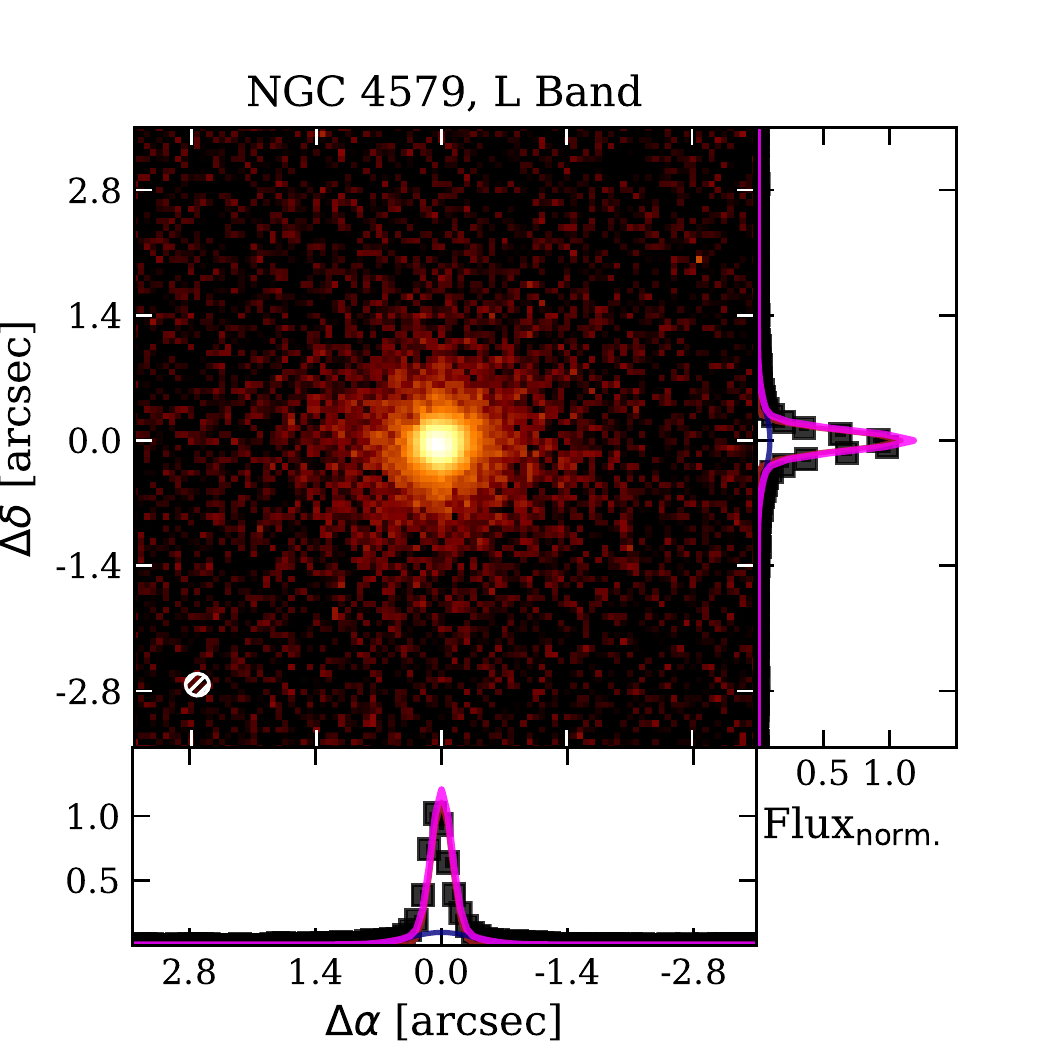}}
\subfloat{\includegraphics[width=0.25\hsize]{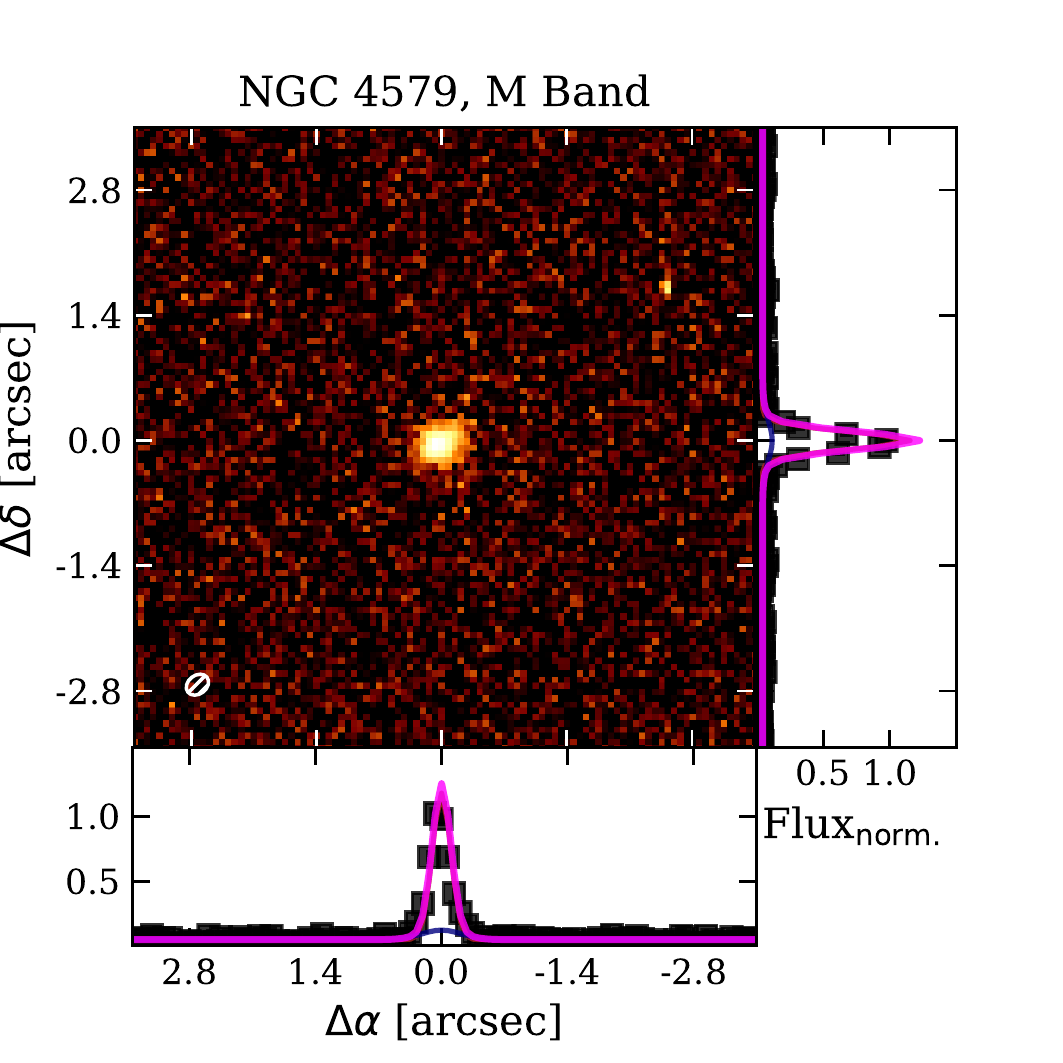}}
\subfloat{\includegraphics[width=0.25\hsize]{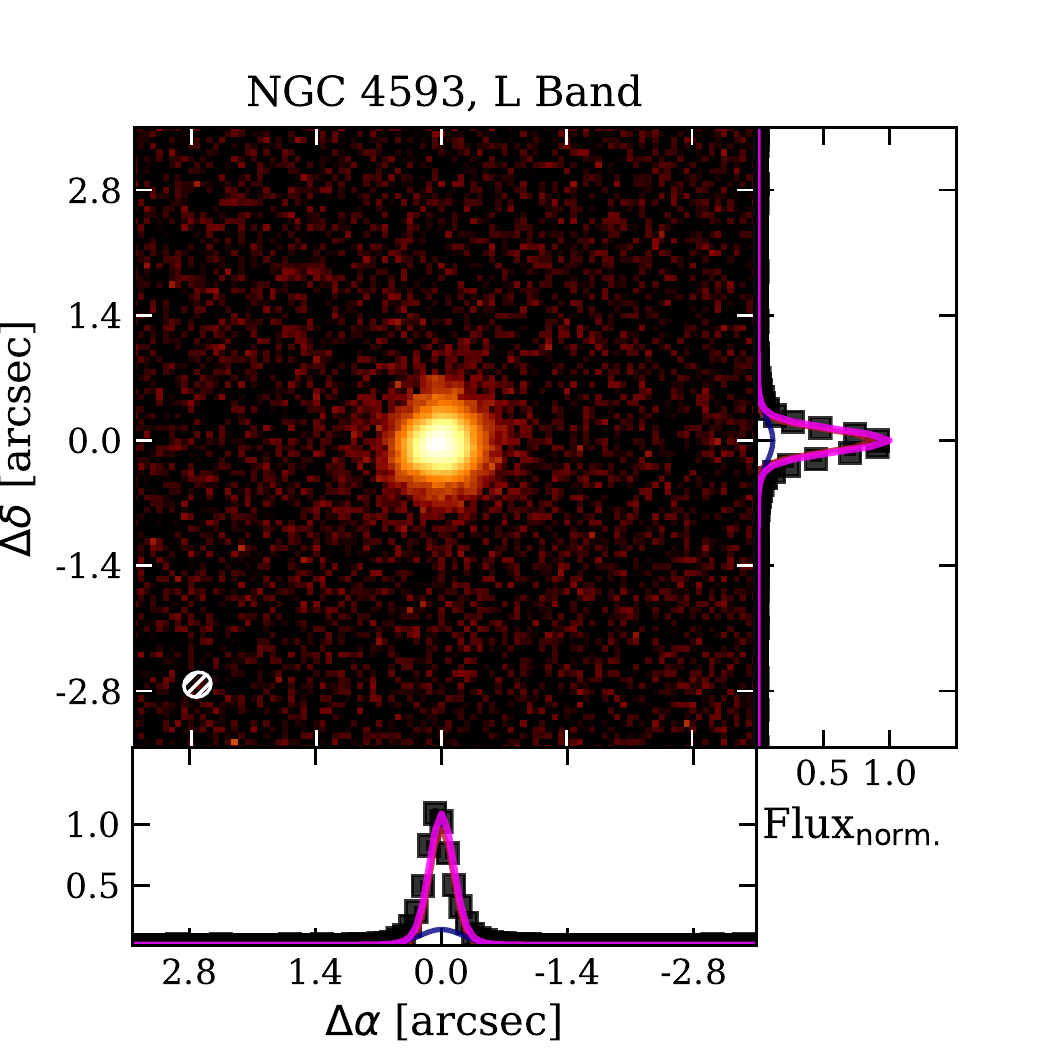}} \\
\subfloat{\includegraphics[width=0.25\hsize]{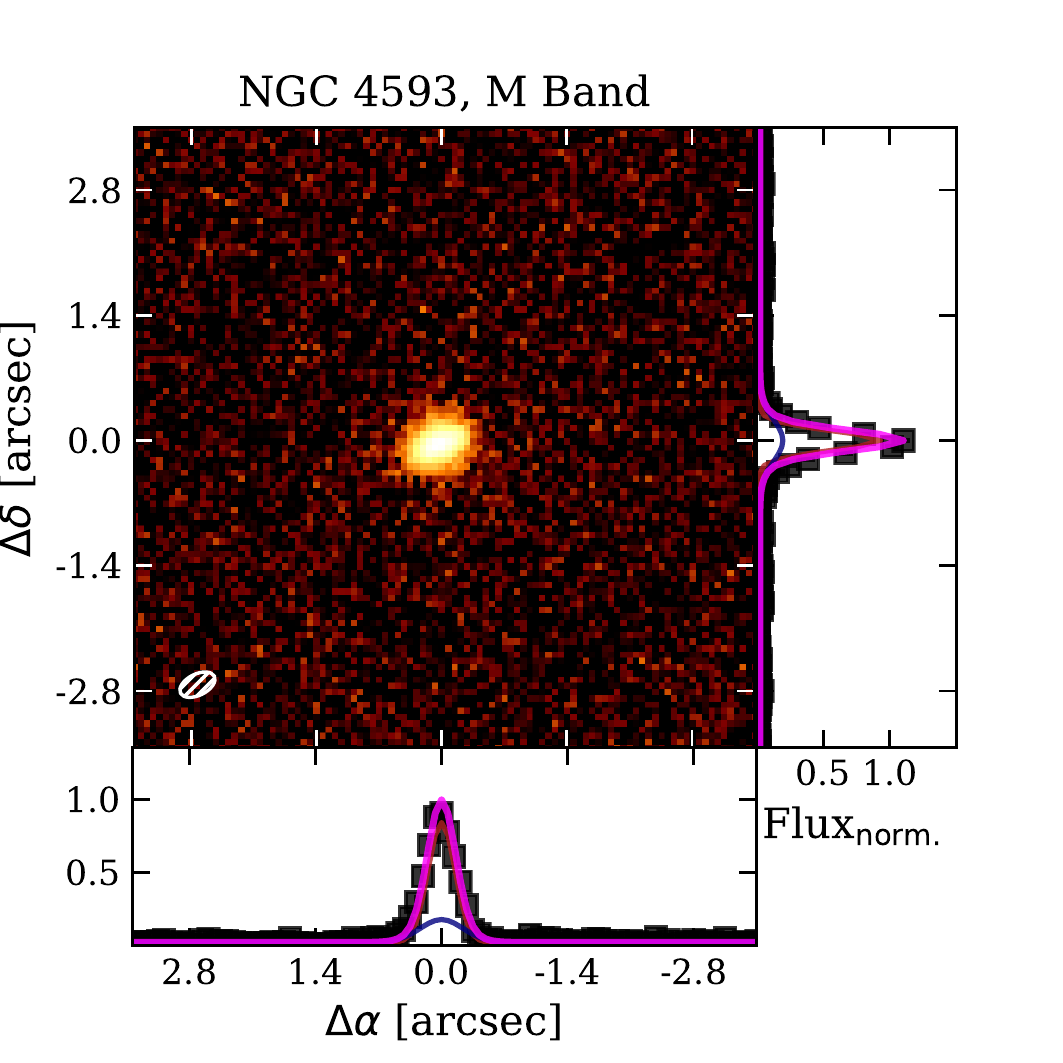}}
\subfloat{\includegraphics[width=0.25\hsize]{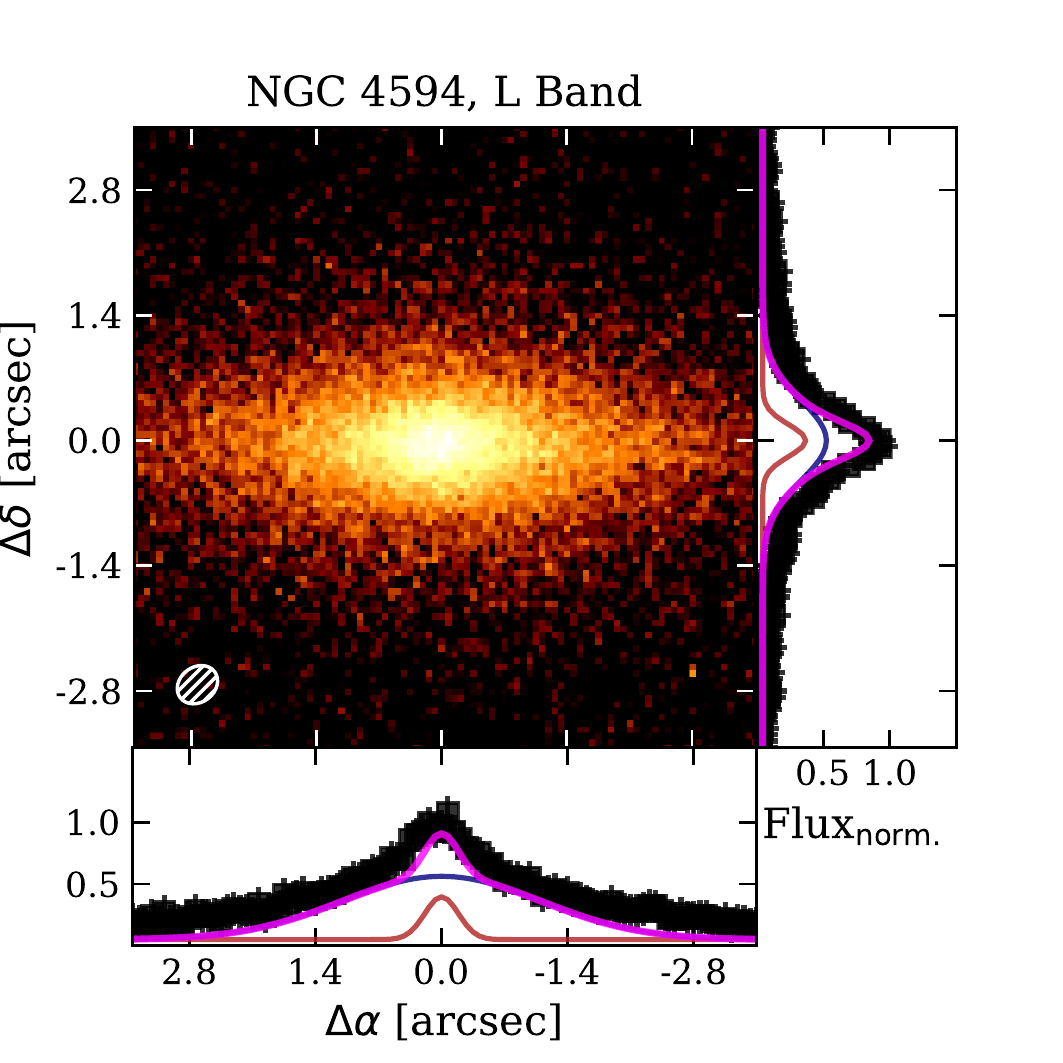}}
\subfloat{\includegraphics[width=0.25\hsize]{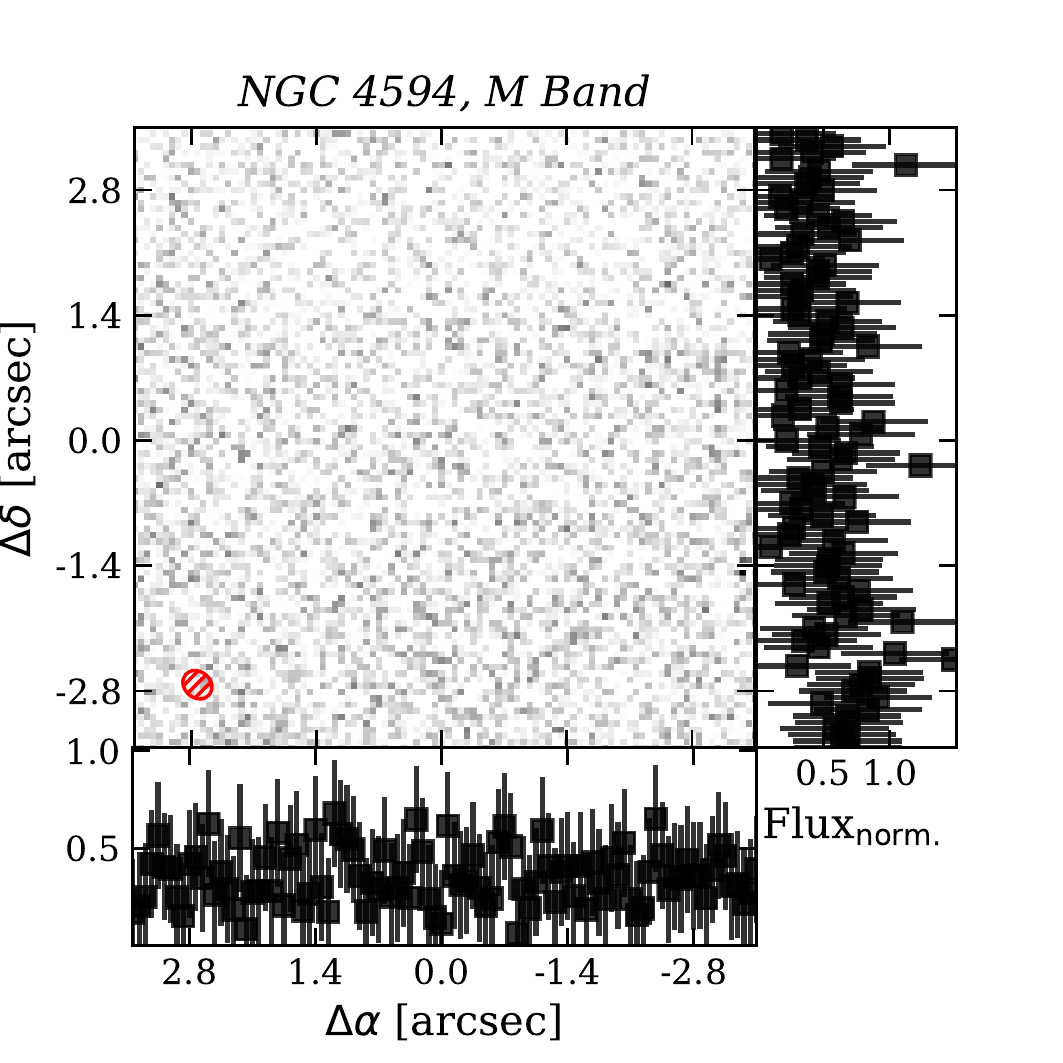}}
\subfloat{\includegraphics[width=0.25\hsize]{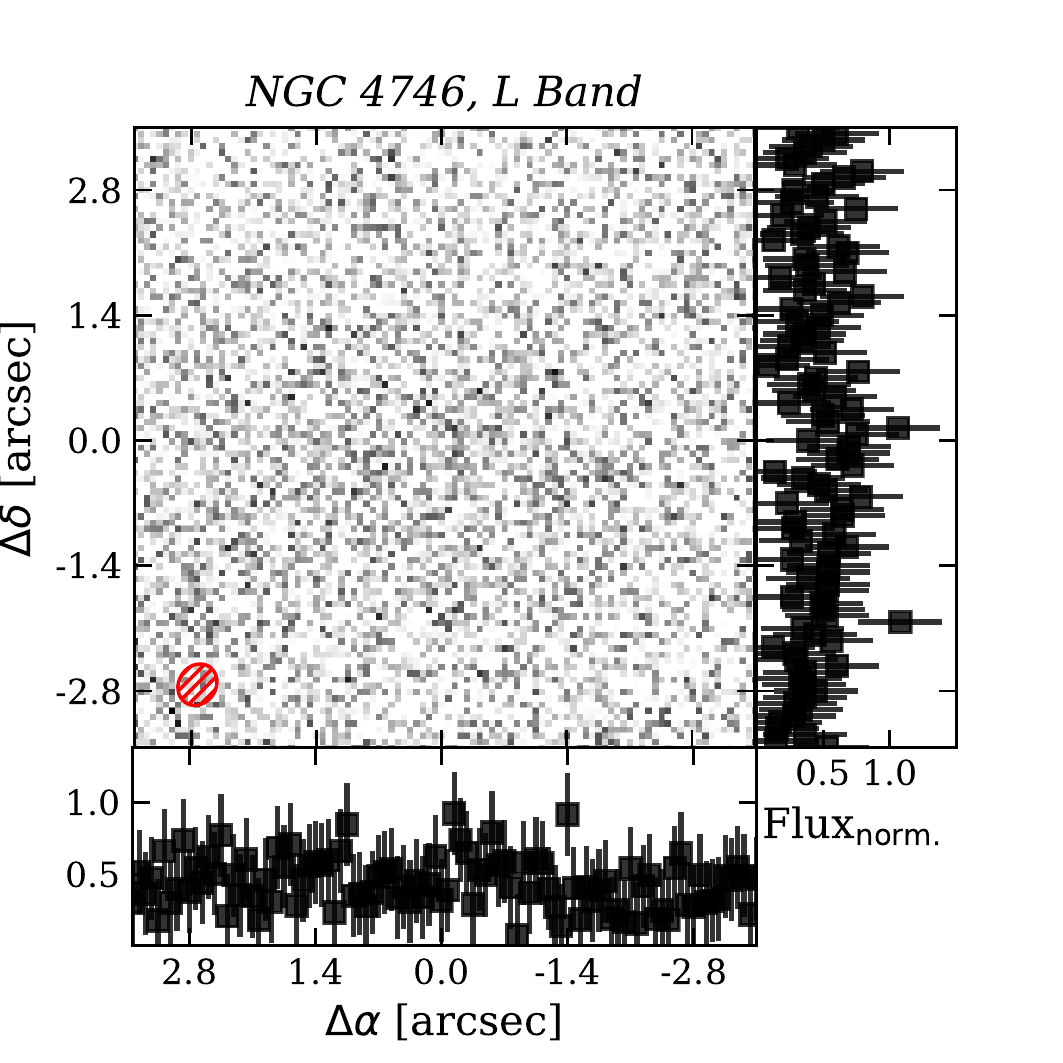}} \\
\caption{As Fig. \ref{fig:cutouts_one}.}
\end{figure*}
\begin{figure*}
\centering
\subfloat{\includegraphics[width=0.25\hsize]{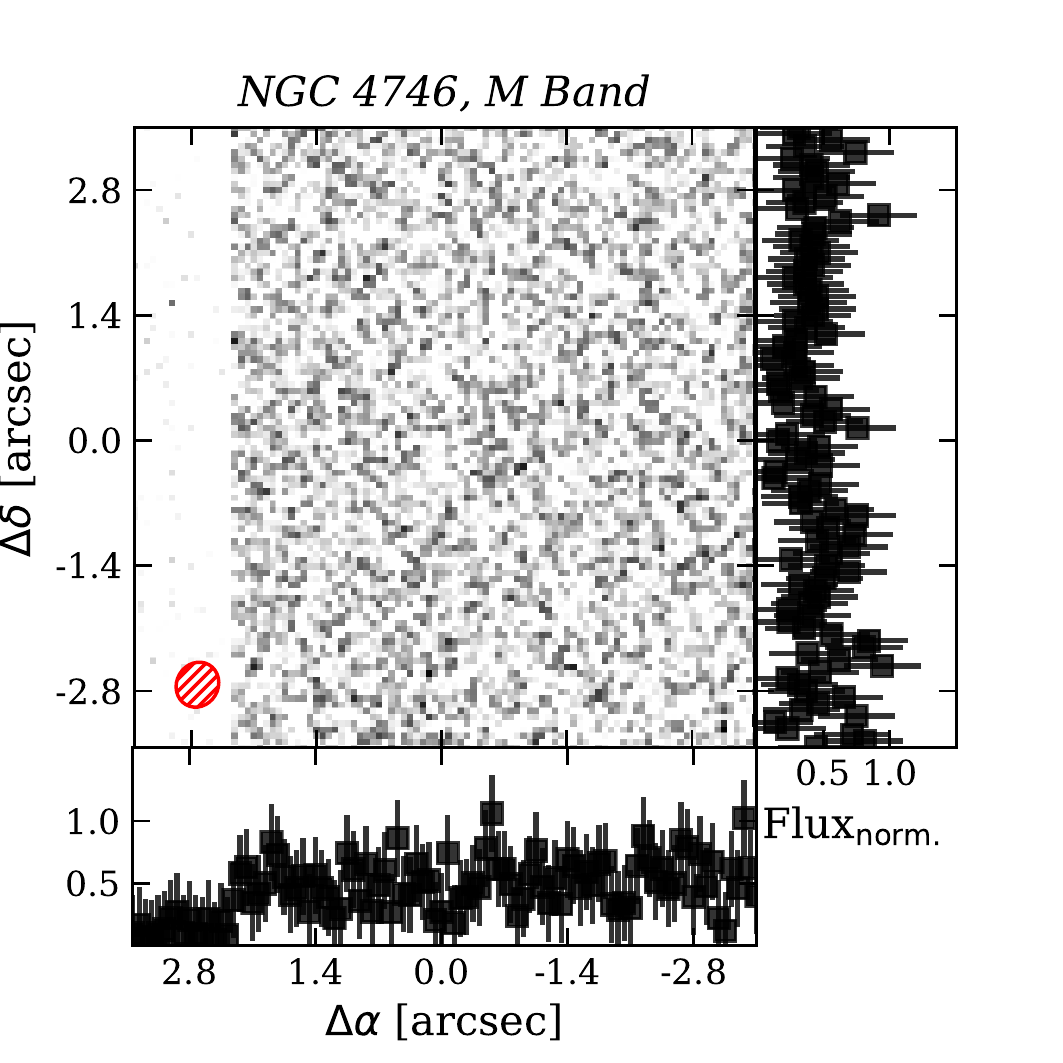}}
\subfloat{\includegraphics[width=0.25\hsize]{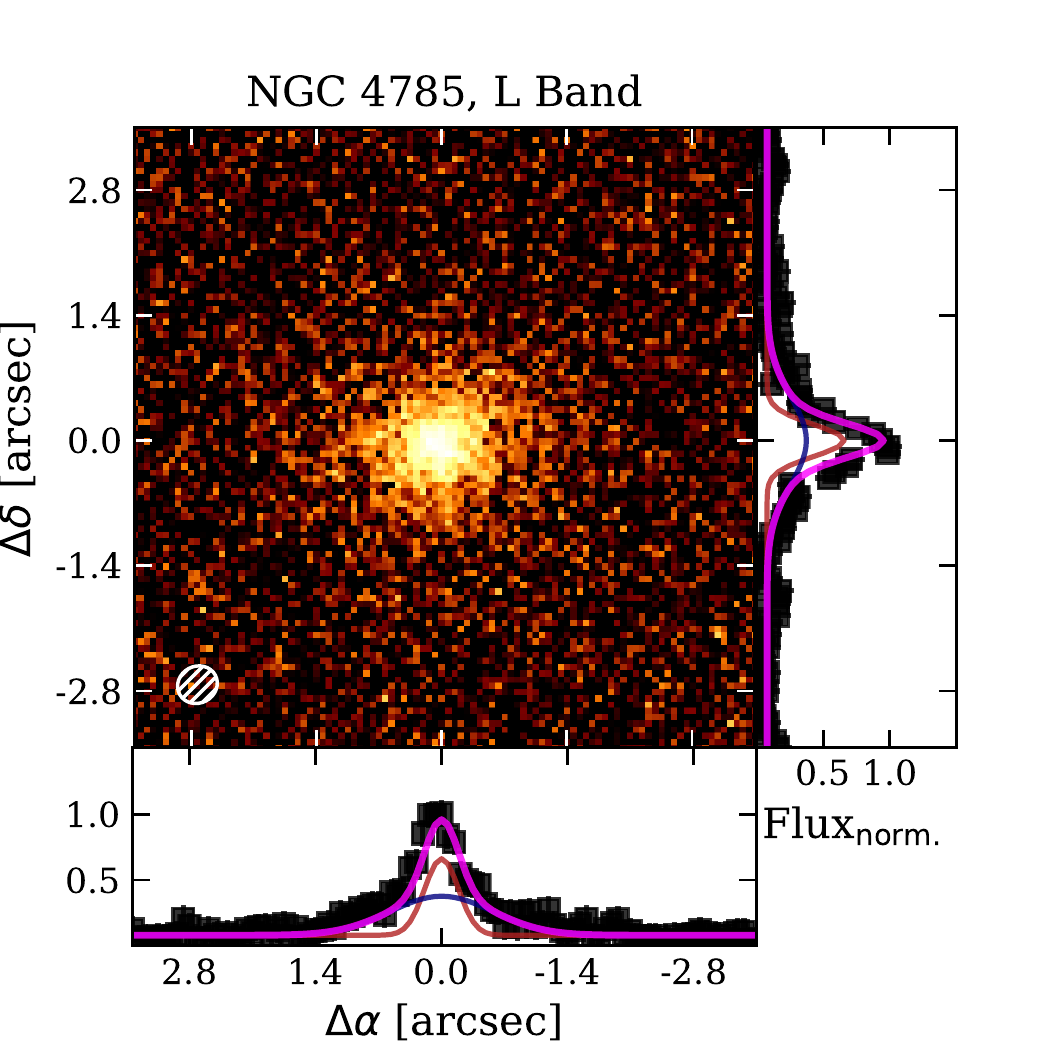}}
\subfloat{\includegraphics[width=0.25\hsize]{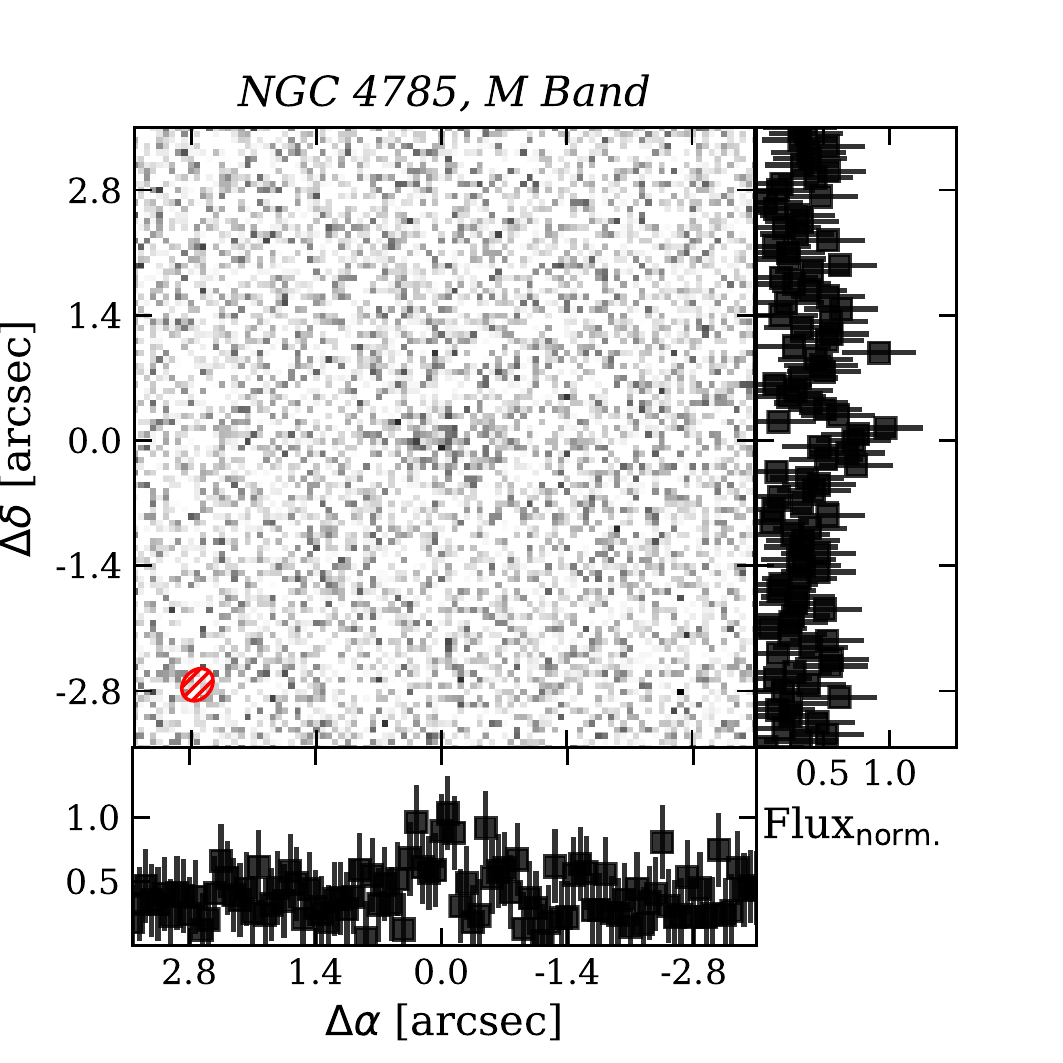}}
\subfloat{\includegraphics[width=0.25\hsize]{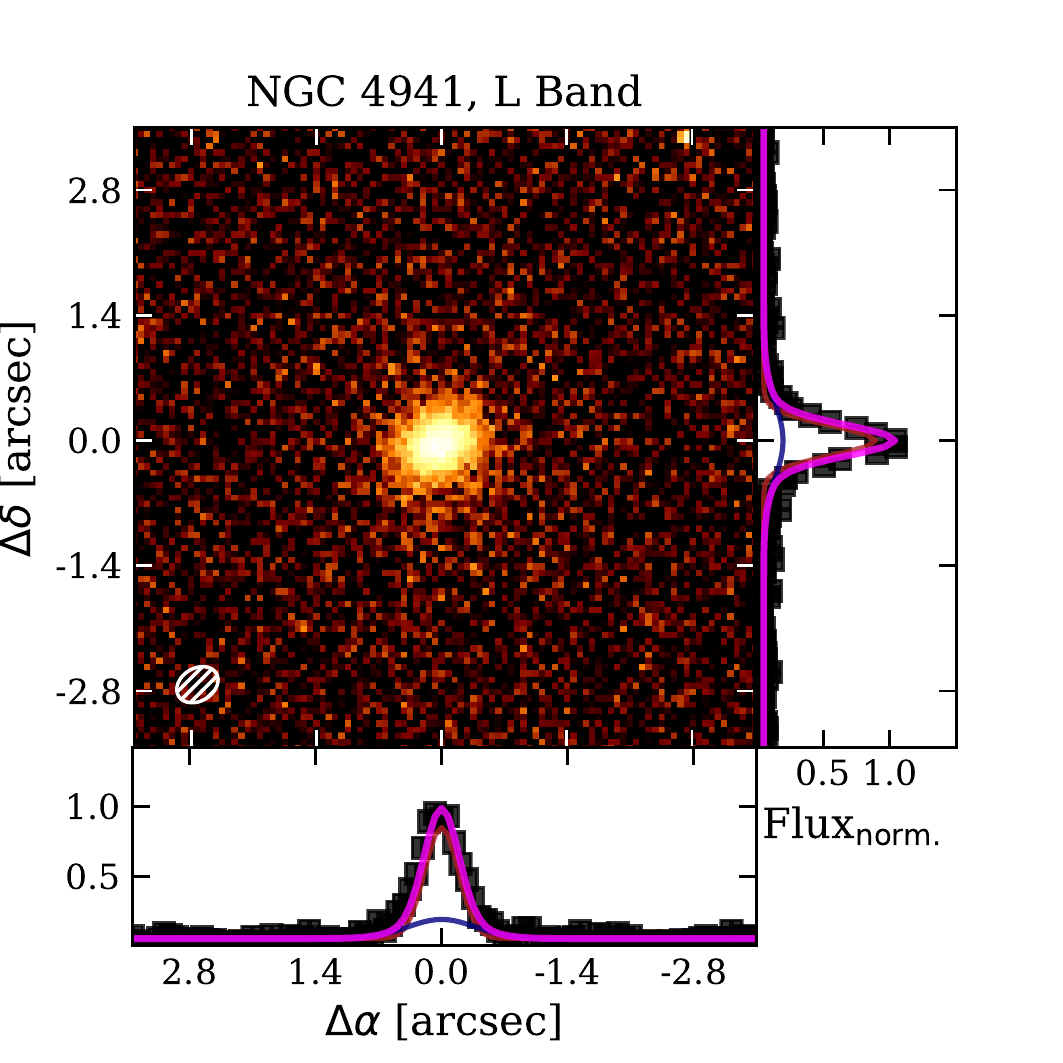}} \\
\subfloat{\includegraphics[width=0.25\hsize]{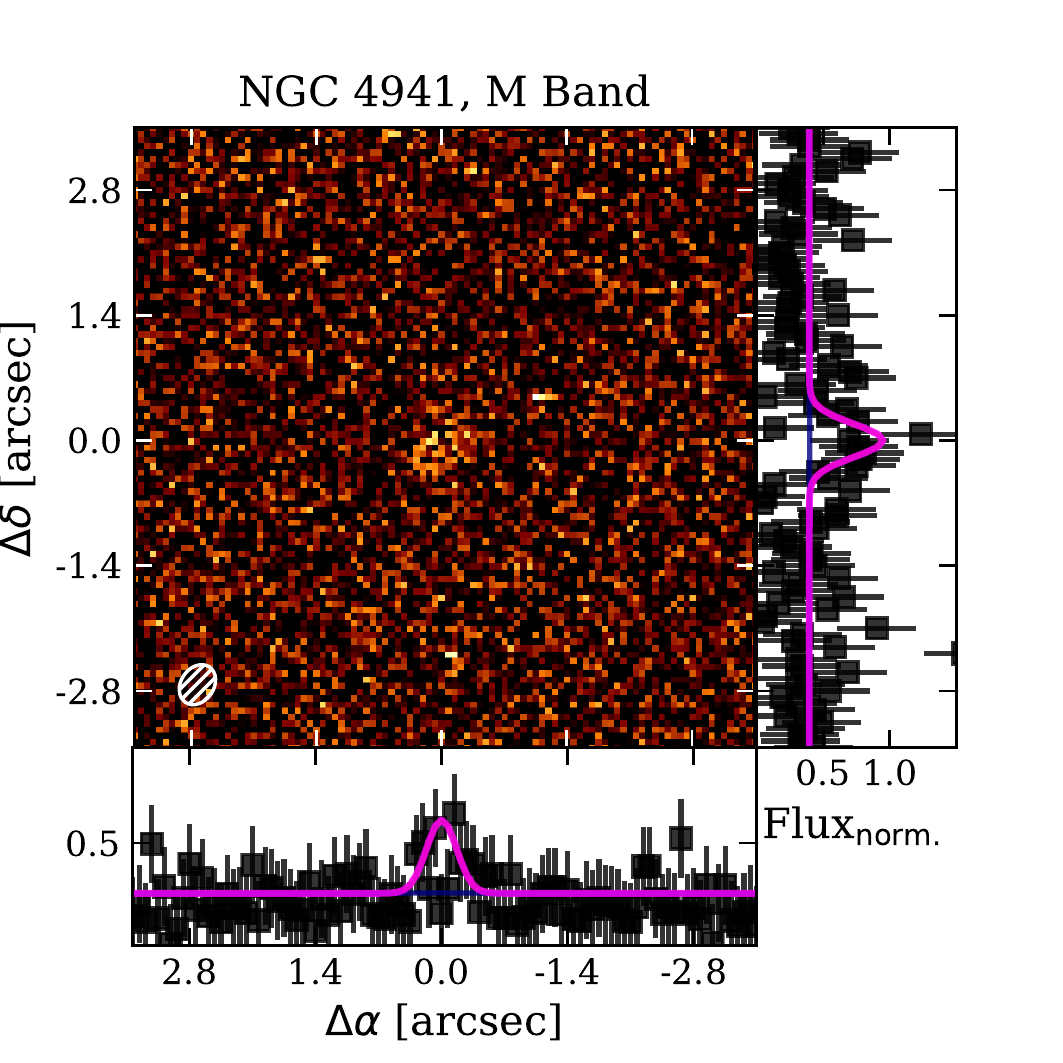}}
\subfloat{\includegraphics[width=0.25\hsize]{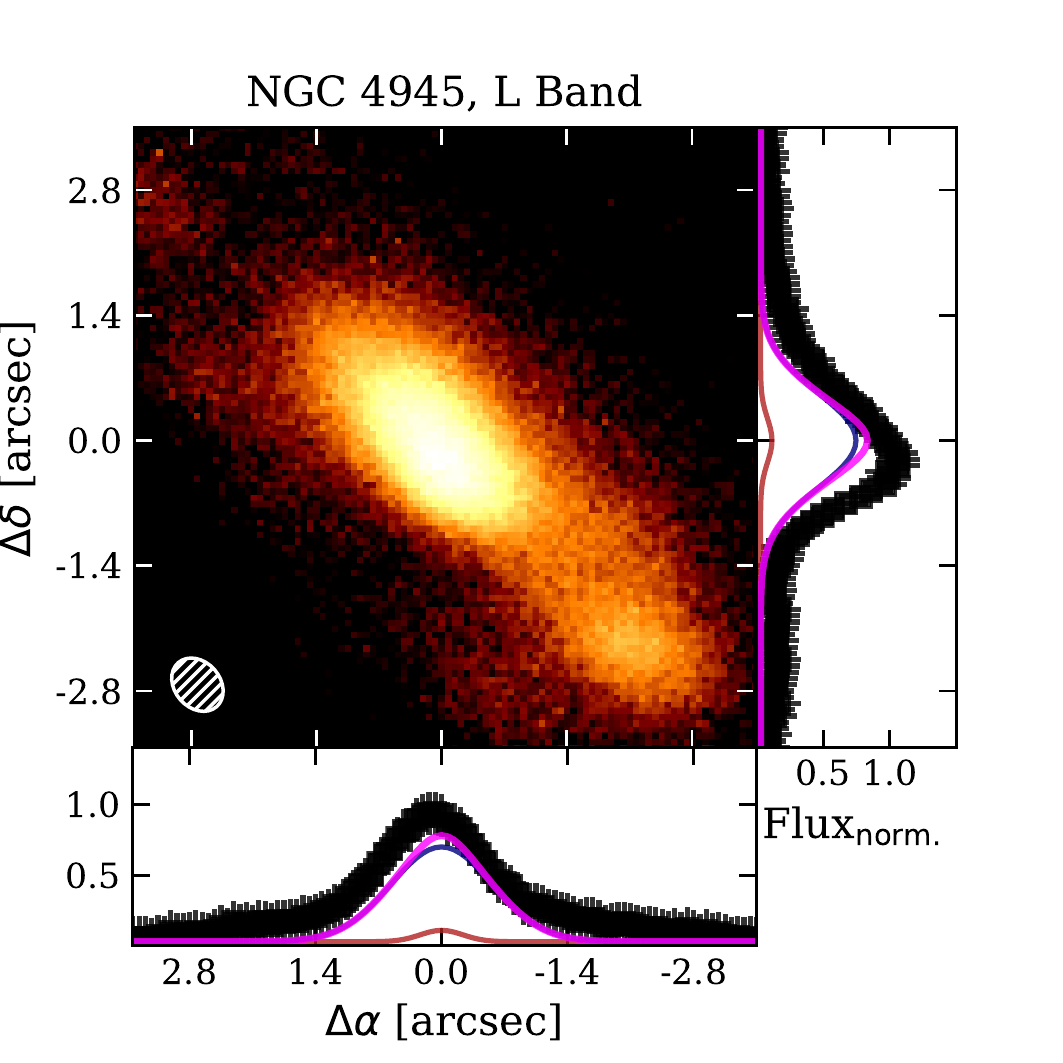}}
\subfloat{\includegraphics[width=0.25\hsize]{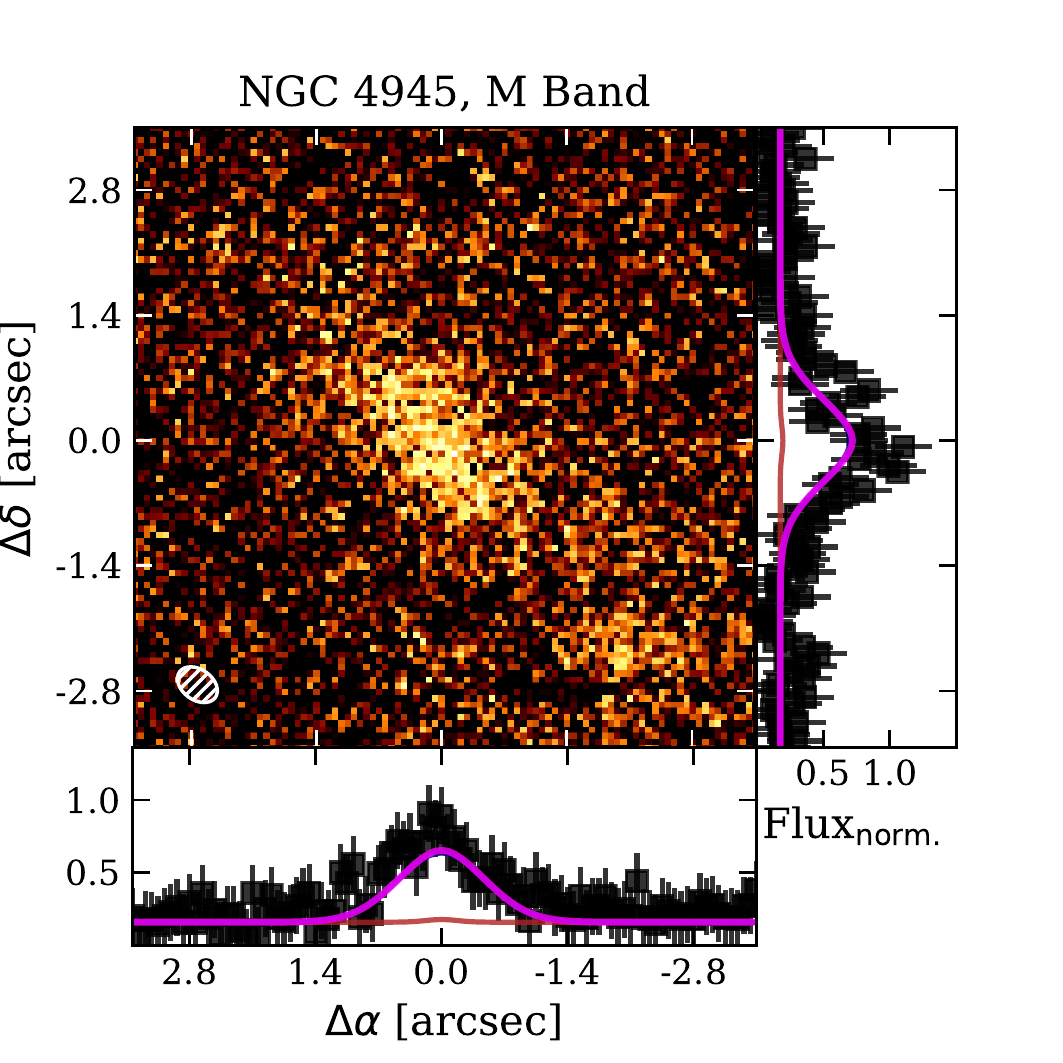}}
\subfloat{\includegraphics[width=0.25\hsize]{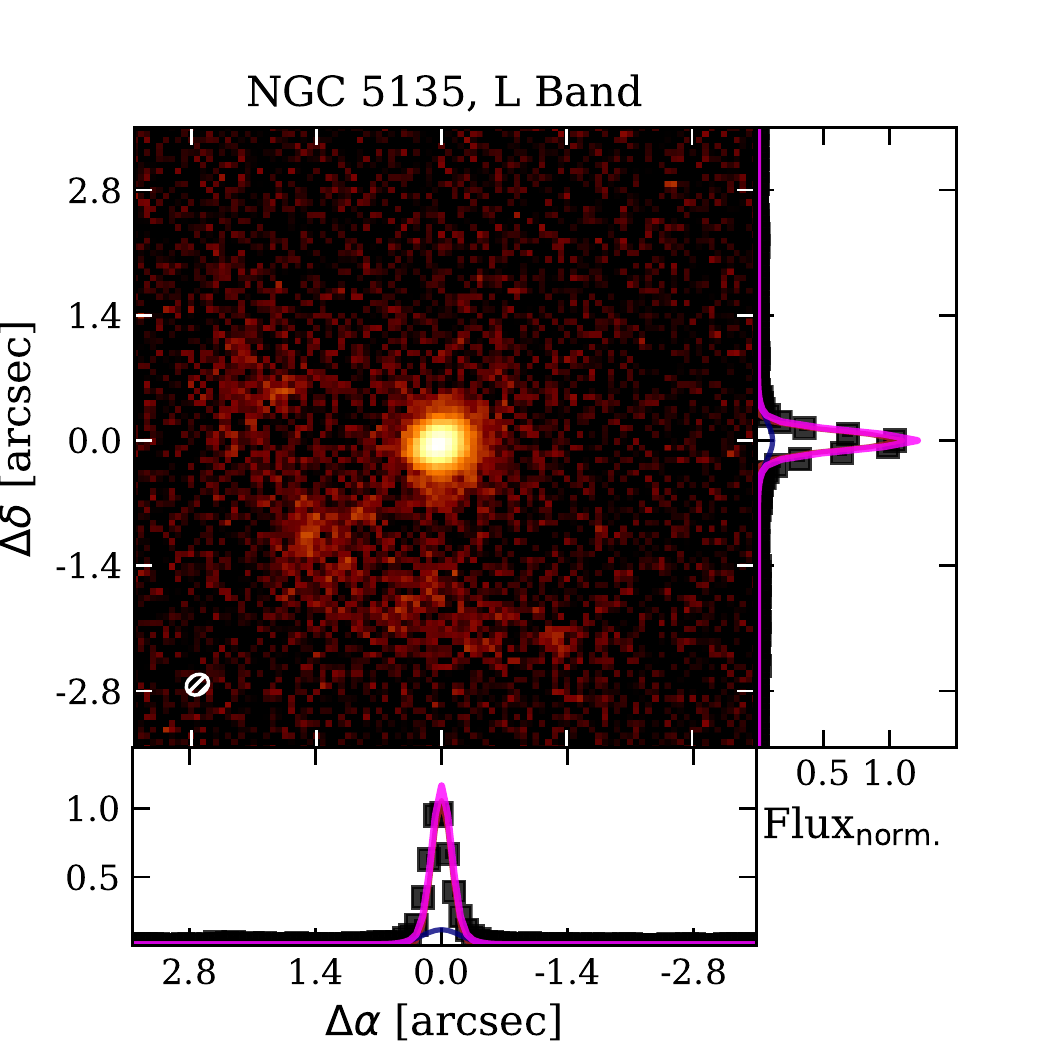}} \\
\subfloat{\includegraphics[width=0.25\hsize]{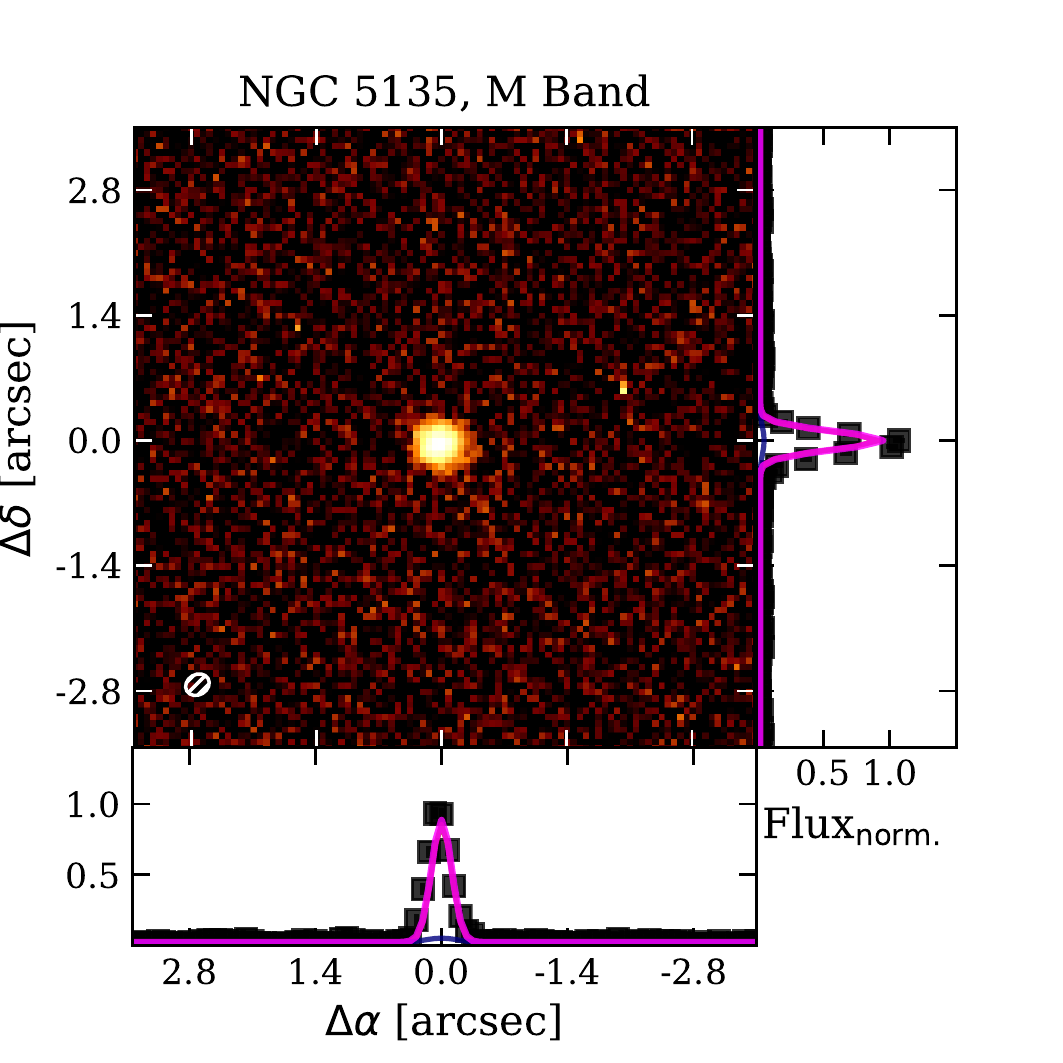}}
\subfloat{\includegraphics[width=0.25\hsize]{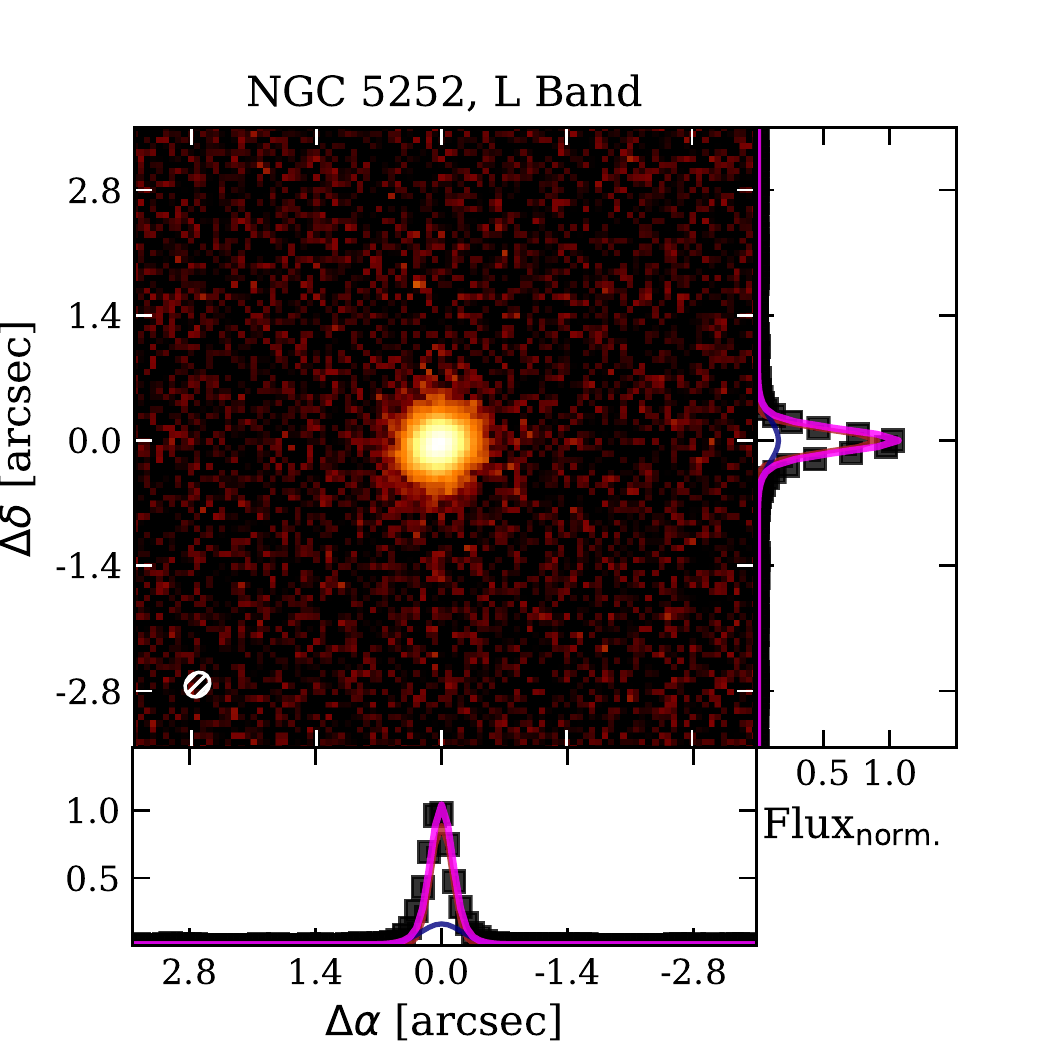}}
\subfloat{\includegraphics[width=0.25\hsize]{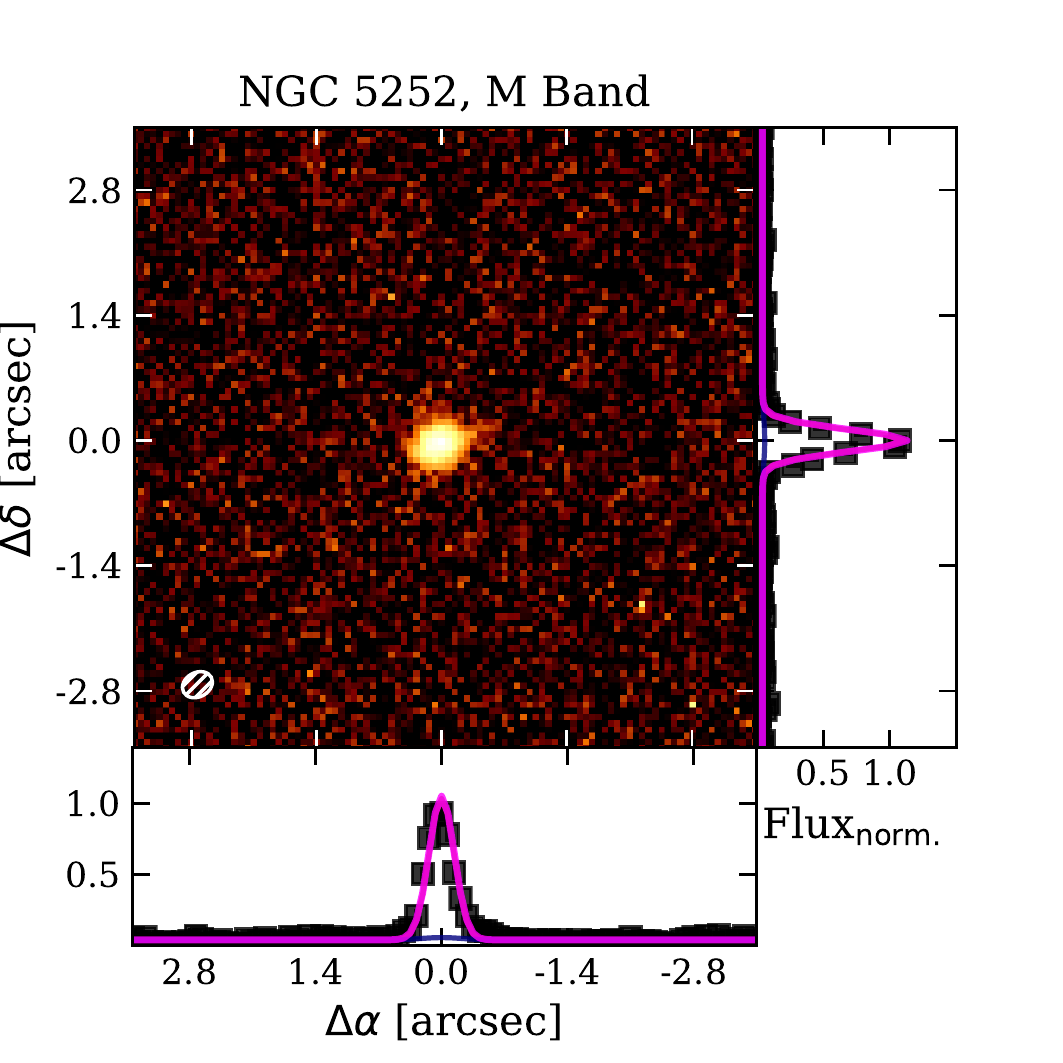}}
\subfloat{\includegraphics[width=0.25\hsize]{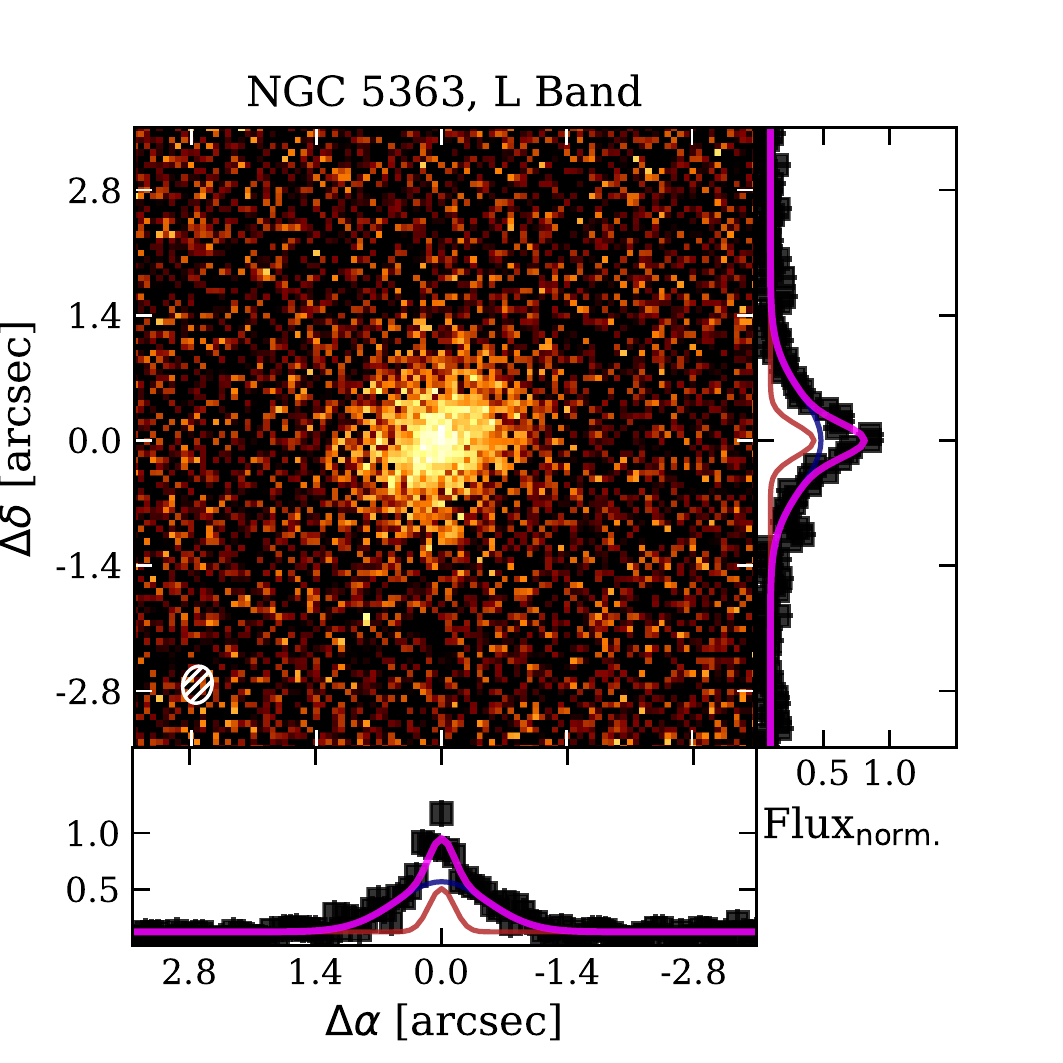}} \\
\subfloat{\includegraphics[width=0.25\hsize]{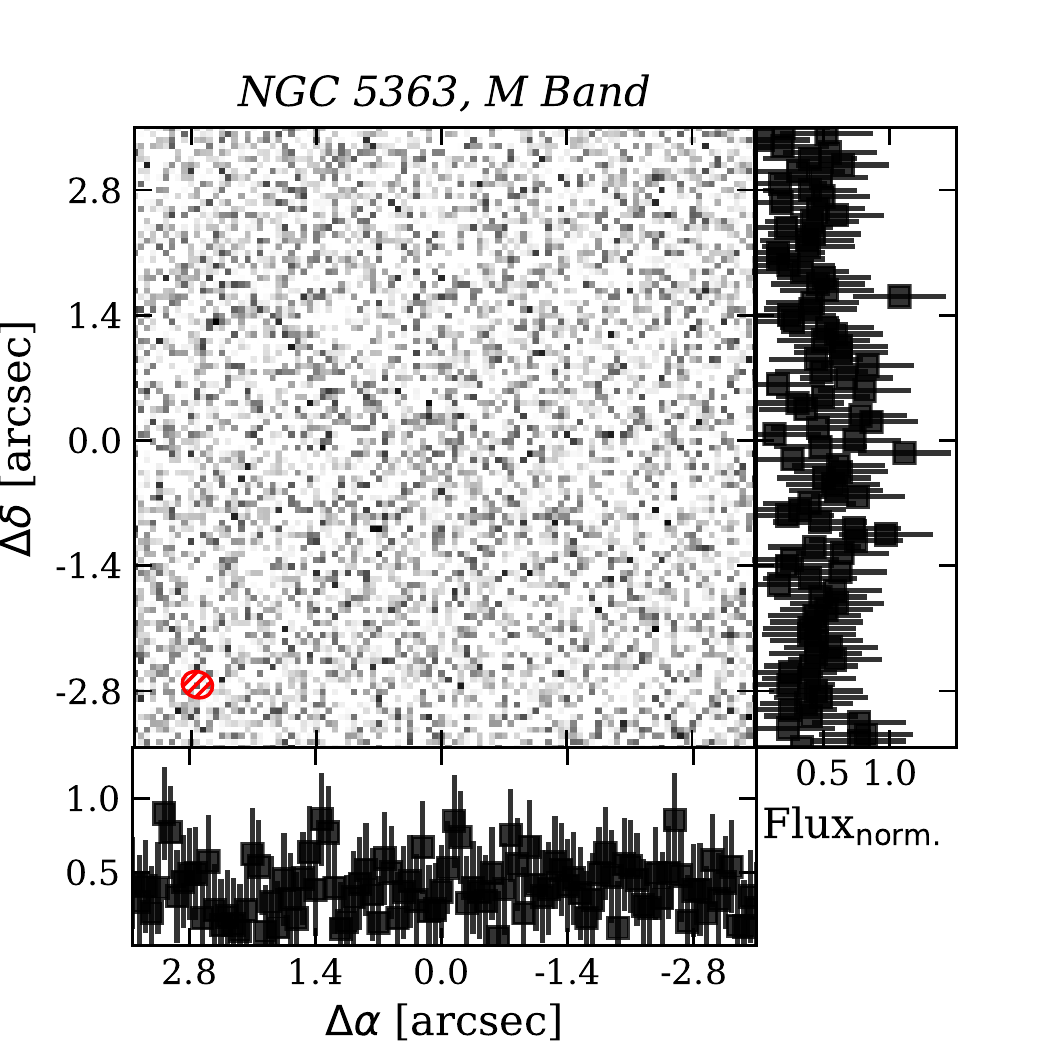}}
\subfloat{\includegraphics[width=0.25\hsize]{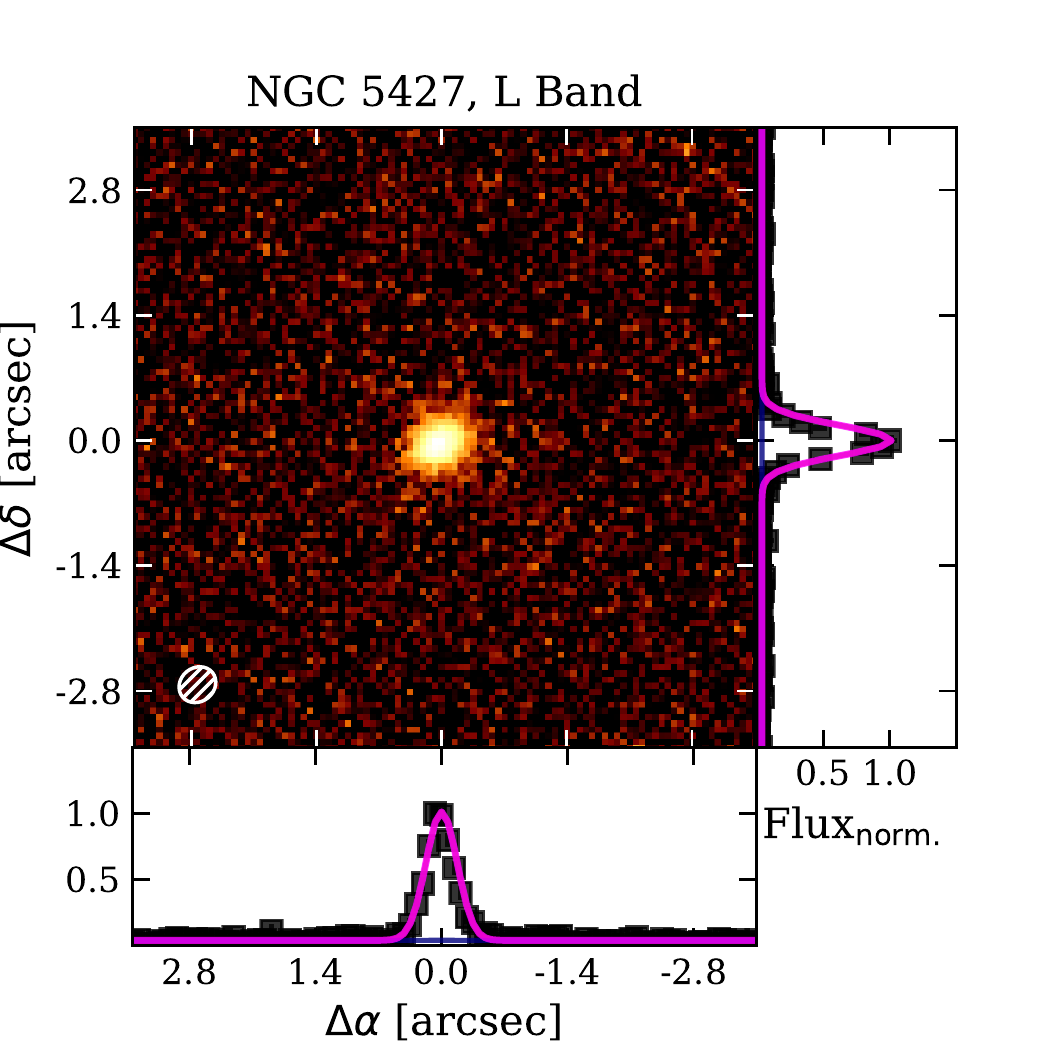}}
\subfloat{\includegraphics[width=0.25\hsize]{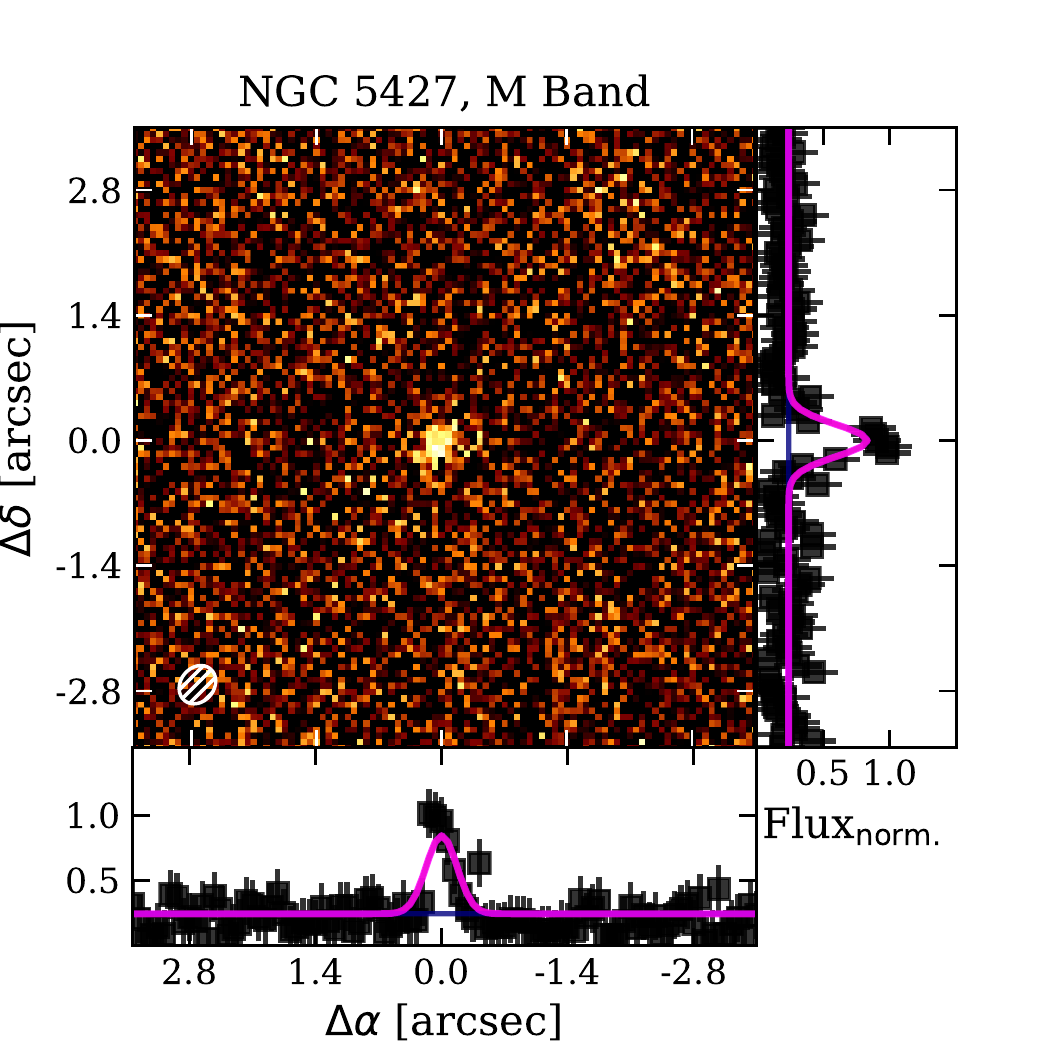}}
\subfloat{\includegraphics[width=0.25\hsize]{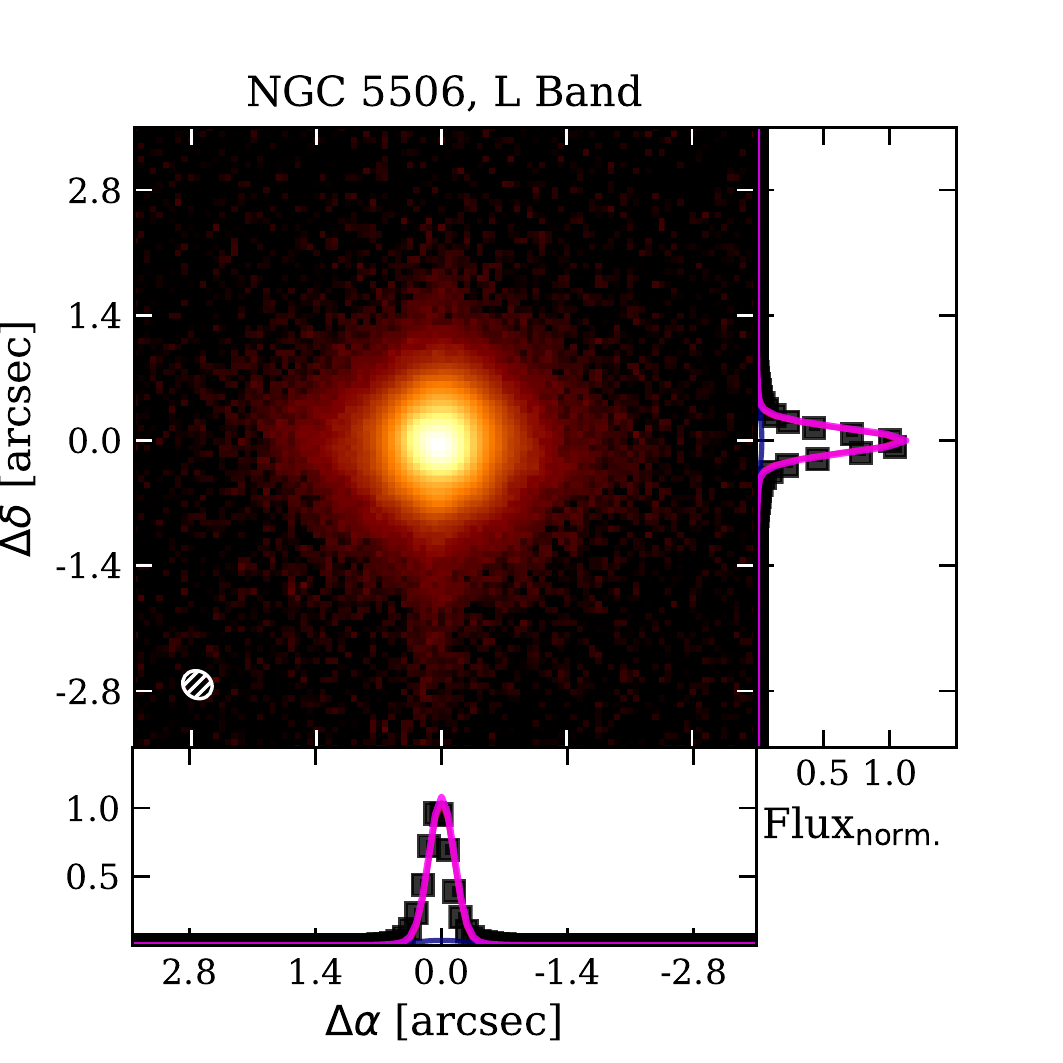}} \\
\subfloat{\includegraphics[width=0.25\hsize]{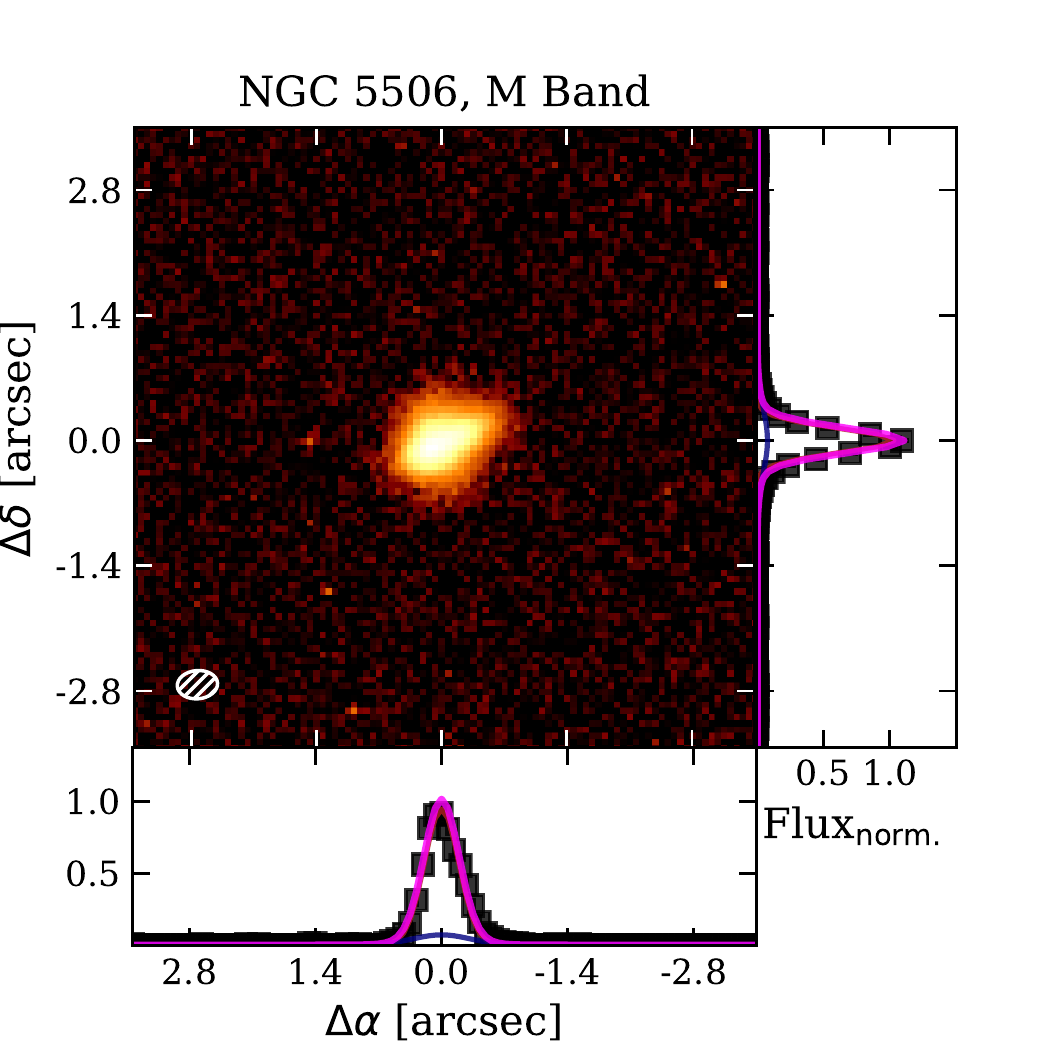}}
\subfloat{\includegraphics[width=0.25\hsize]{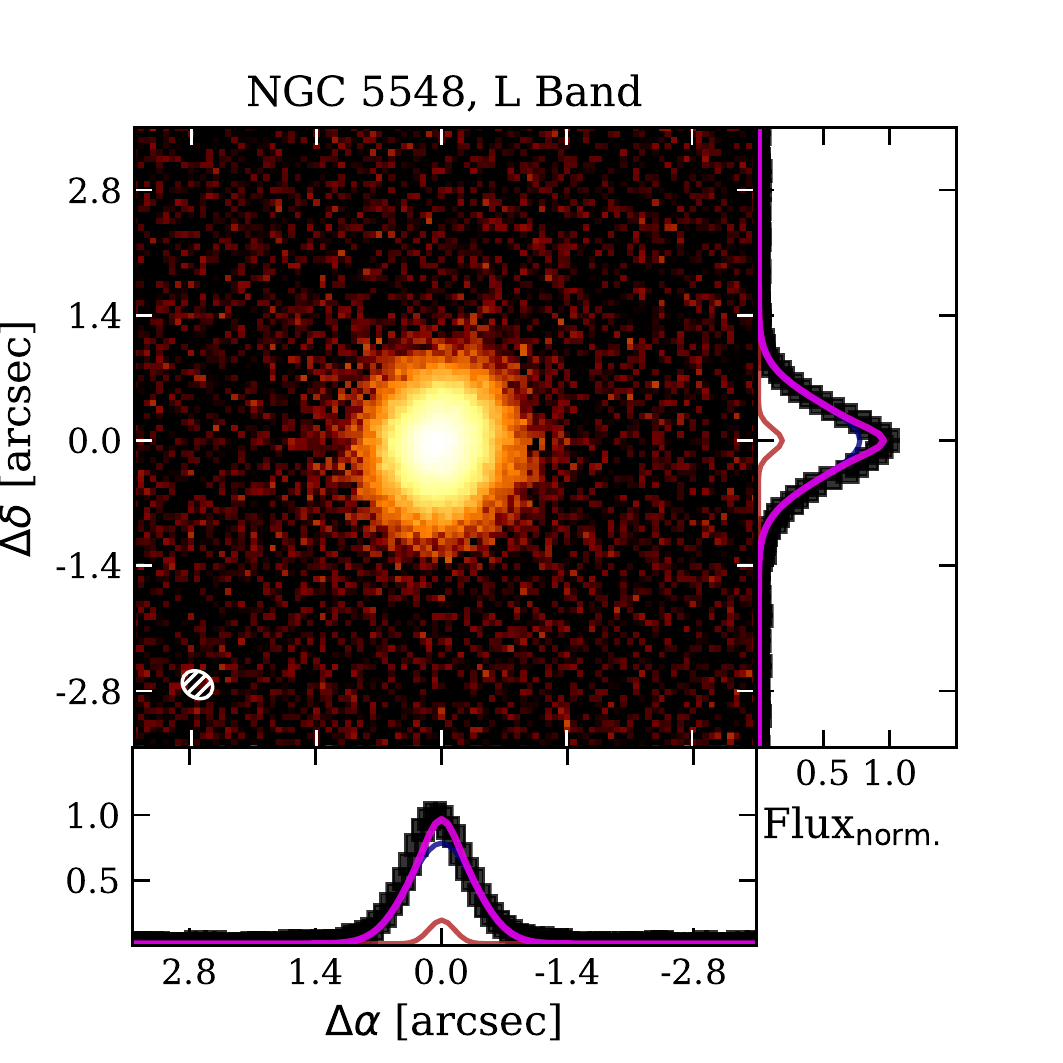}}
\subfloat{\includegraphics[width=0.25\hsize]{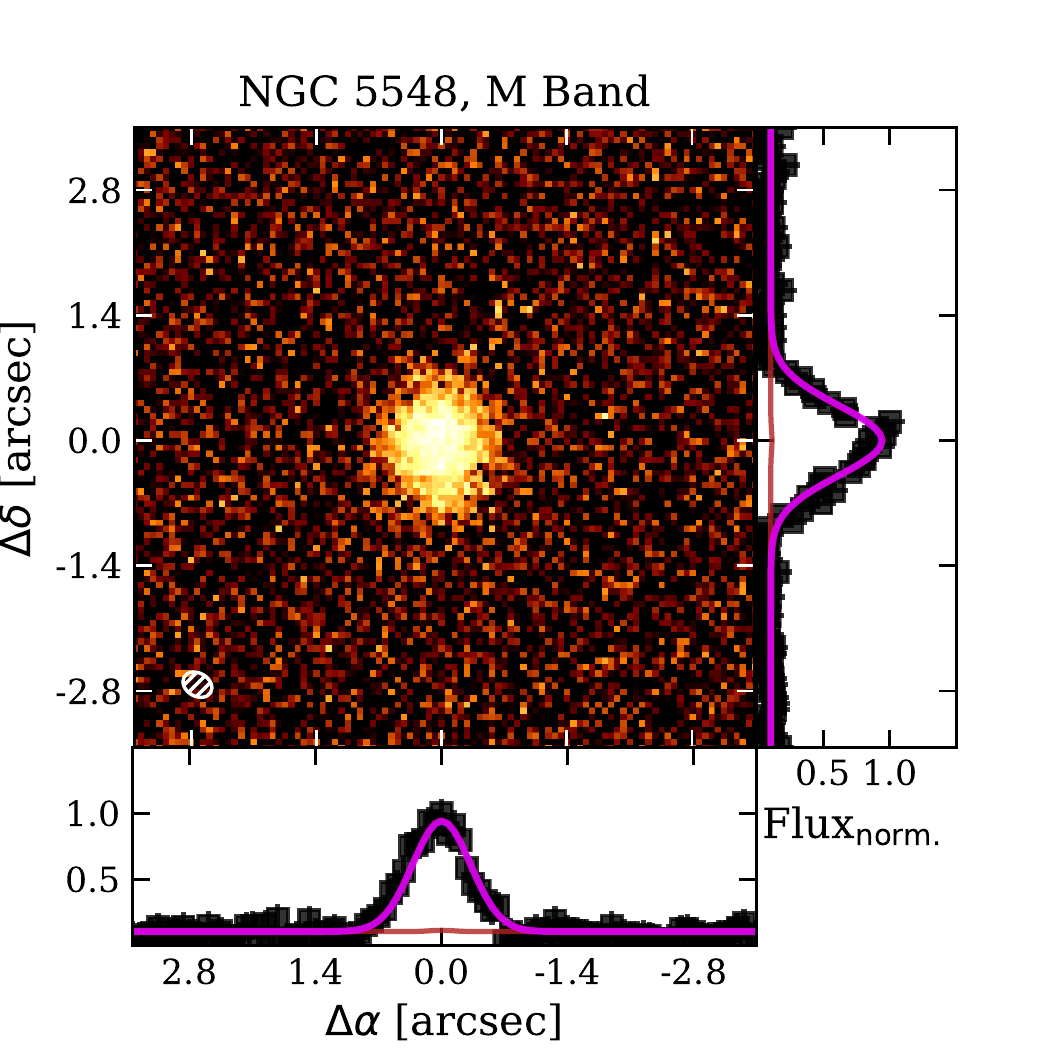}}
\subfloat{\includegraphics[width=0.25\hsize]{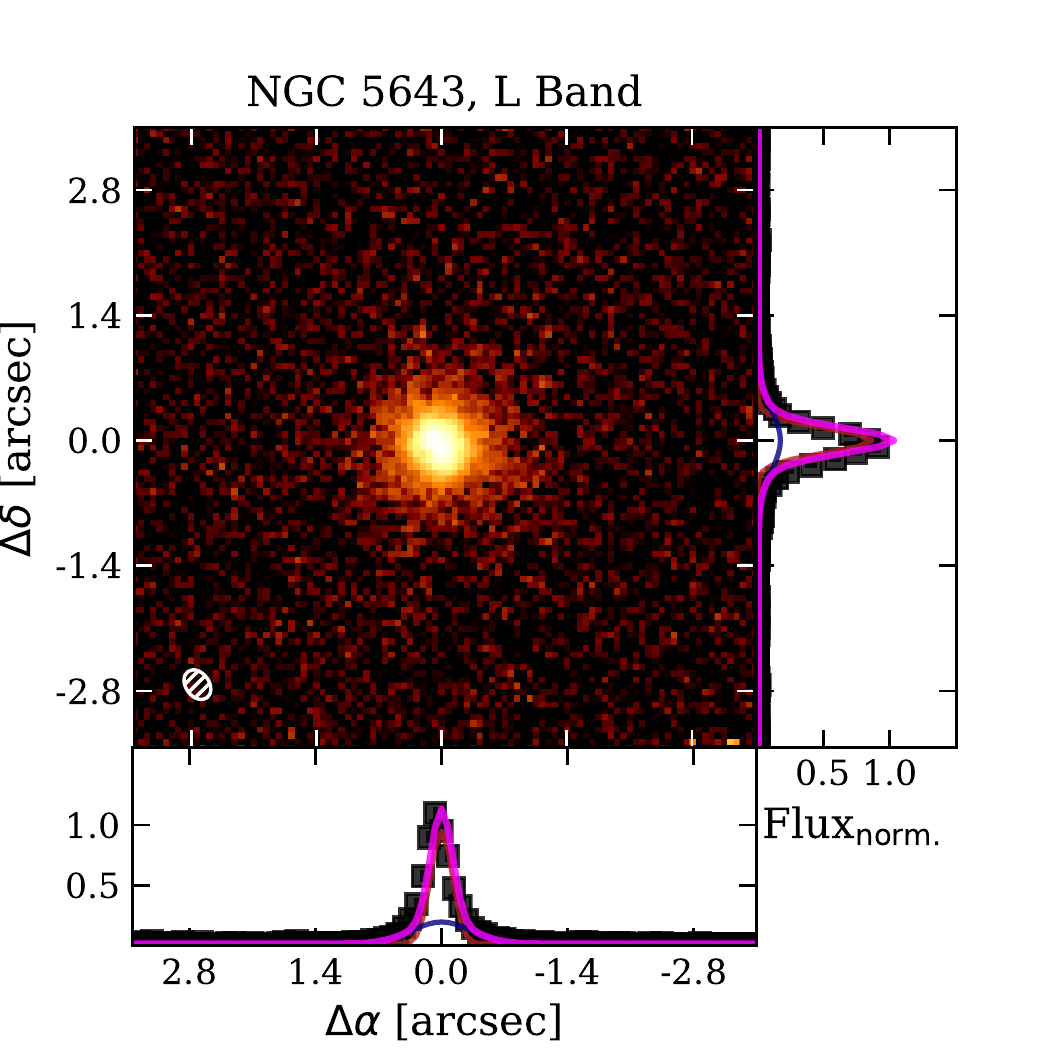}} \\
\caption{As Fig. \ref{fig:cutouts_one}.}
\end{figure*}
\begin{figure*}
\centering
\subfloat{\includegraphics[width=0.25\hsize]{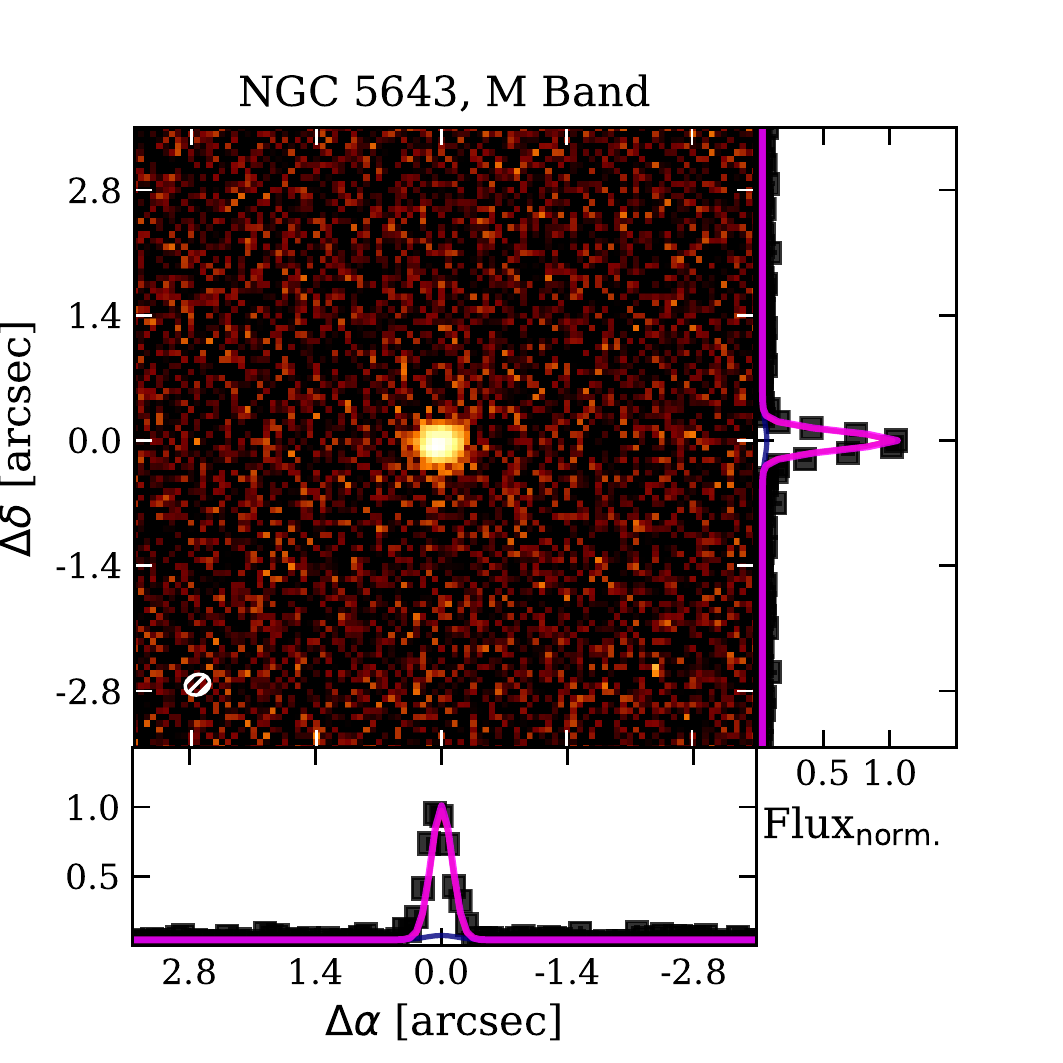}}
\subfloat{\includegraphics[width=0.25\hsize]{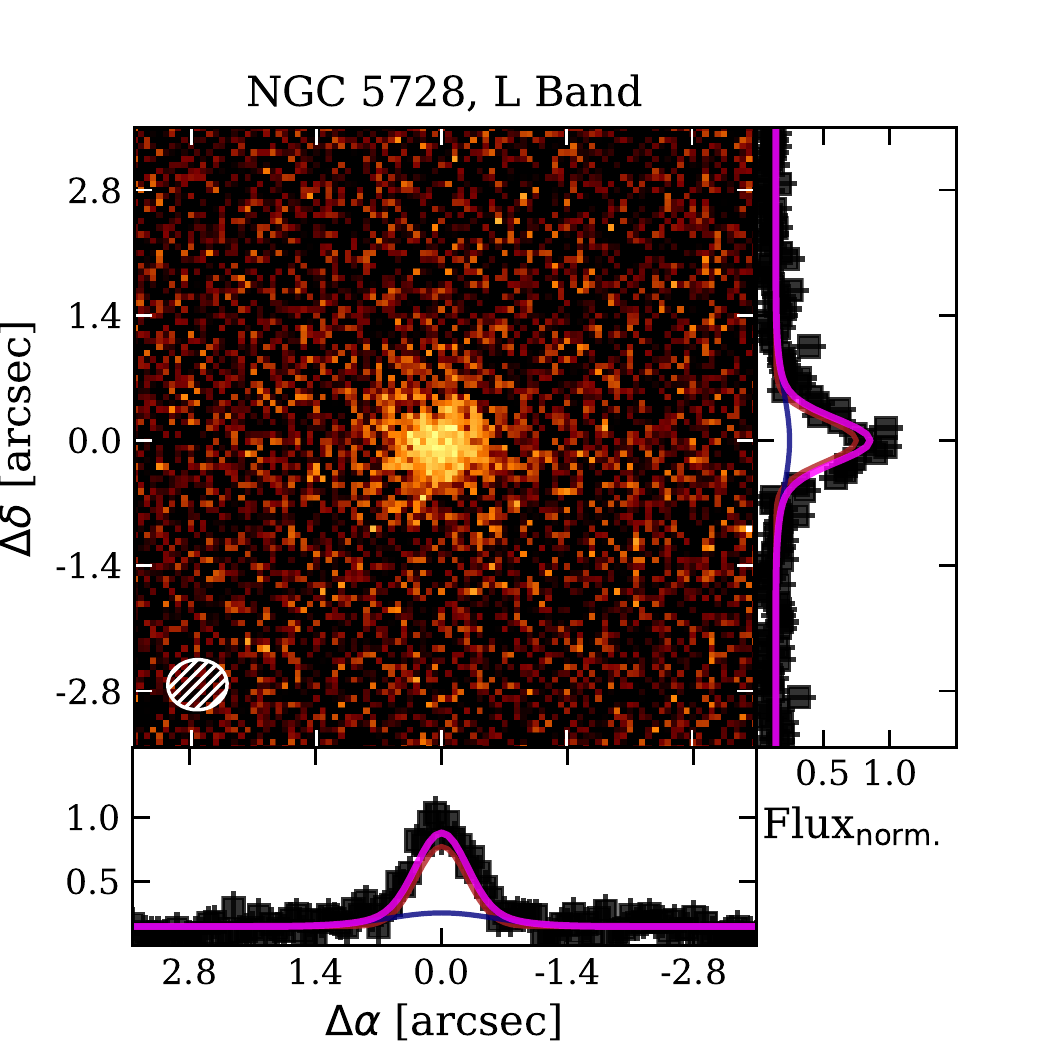}}
\subfloat{\includegraphics[width=0.25\hsize]{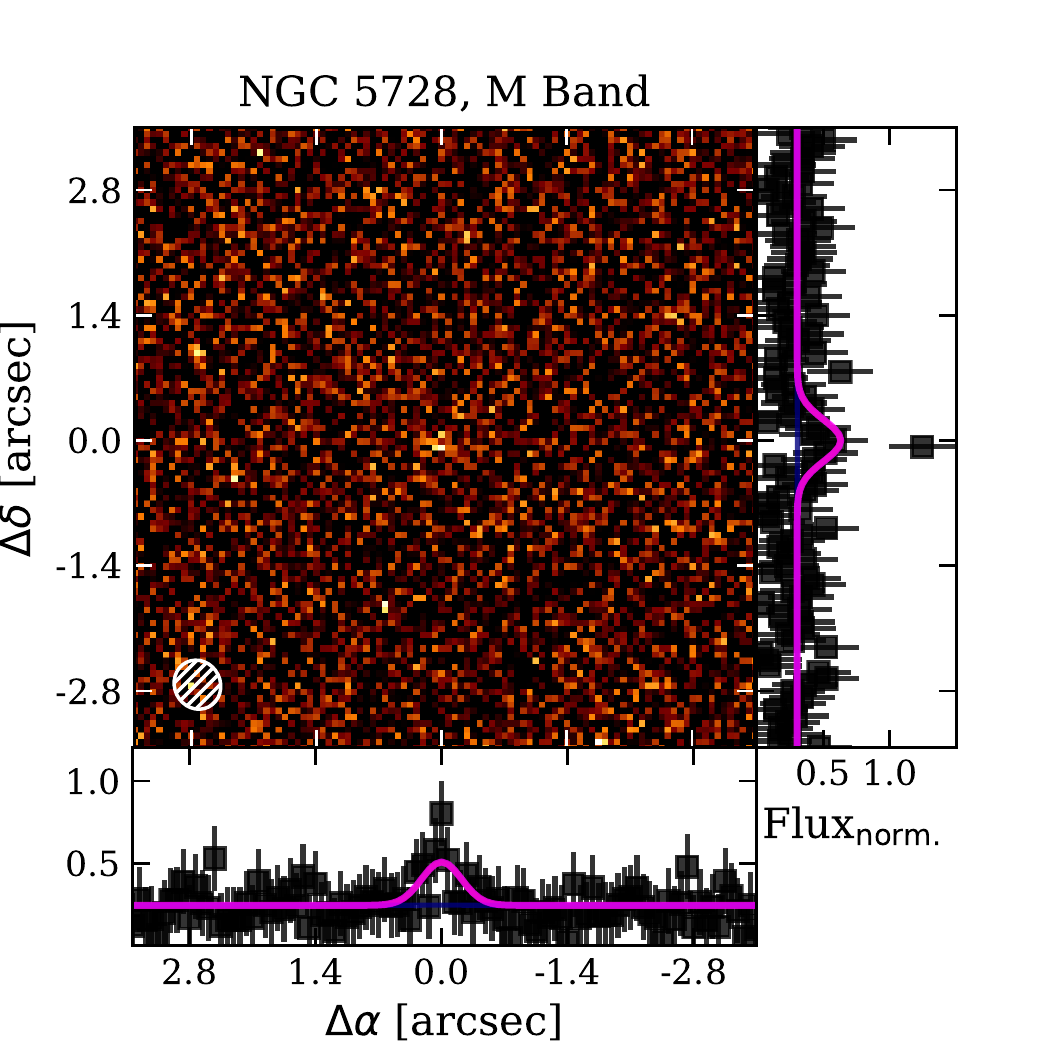}}
\subfloat{\includegraphics[width=0.25\hsize]{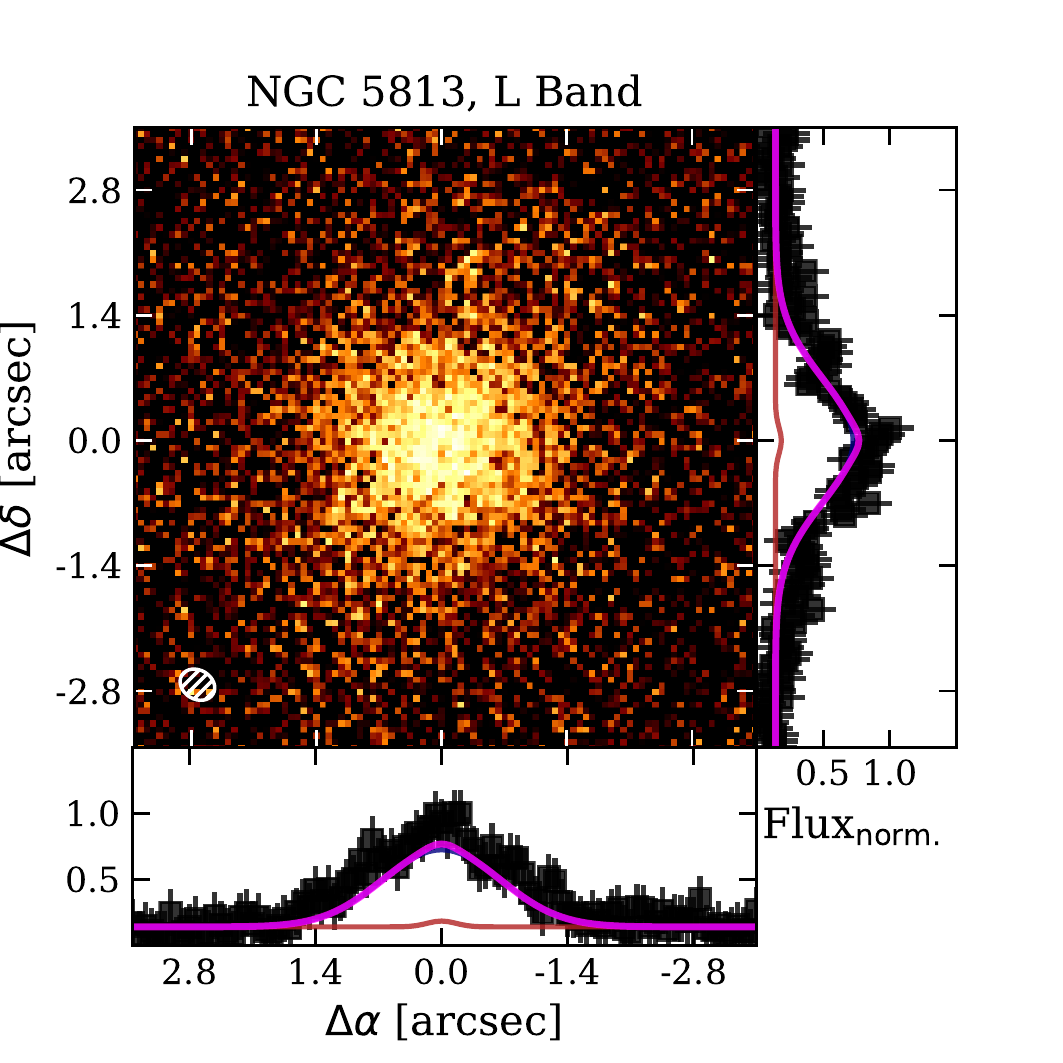}} \\
\subfloat{\includegraphics[width=0.25\hsize]{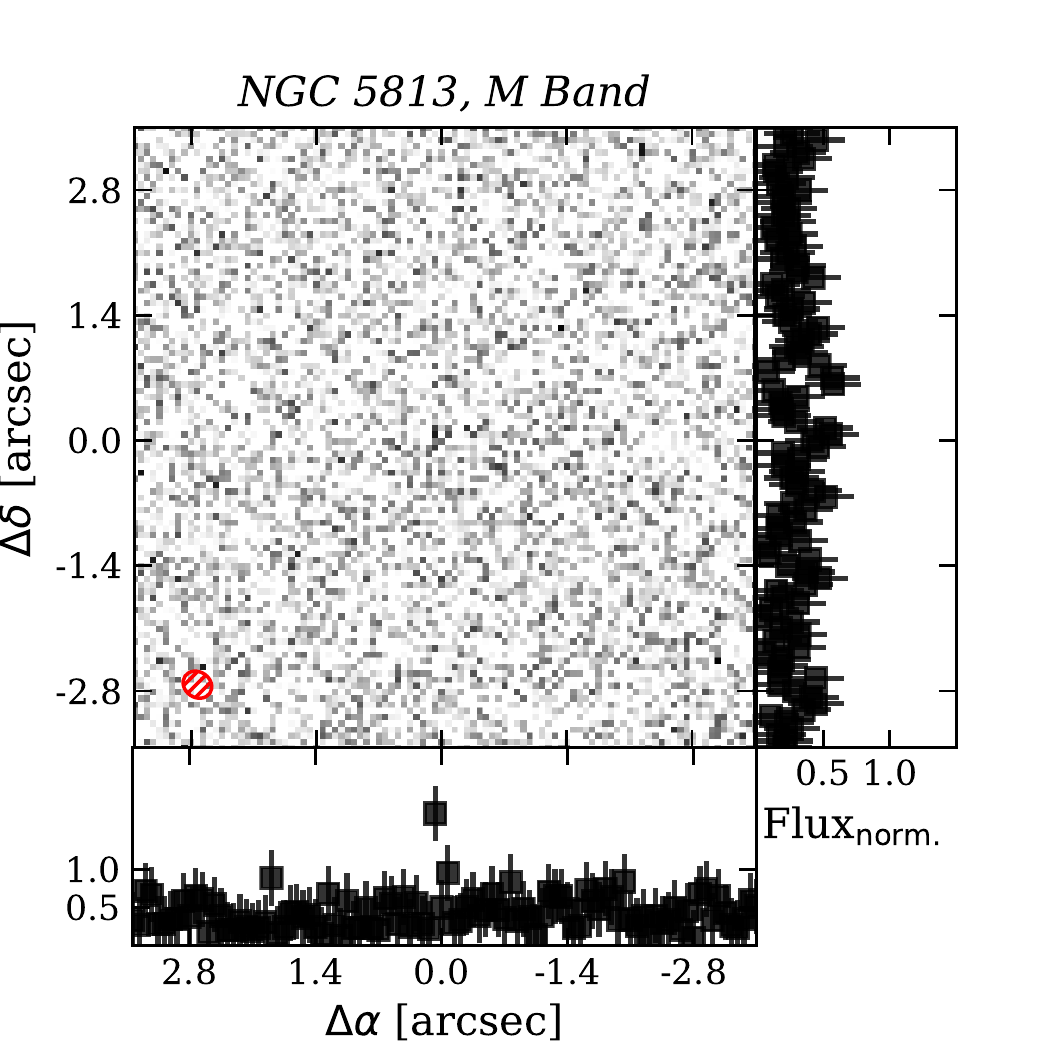}}
\subfloat{\includegraphics[width=0.25\hsize]{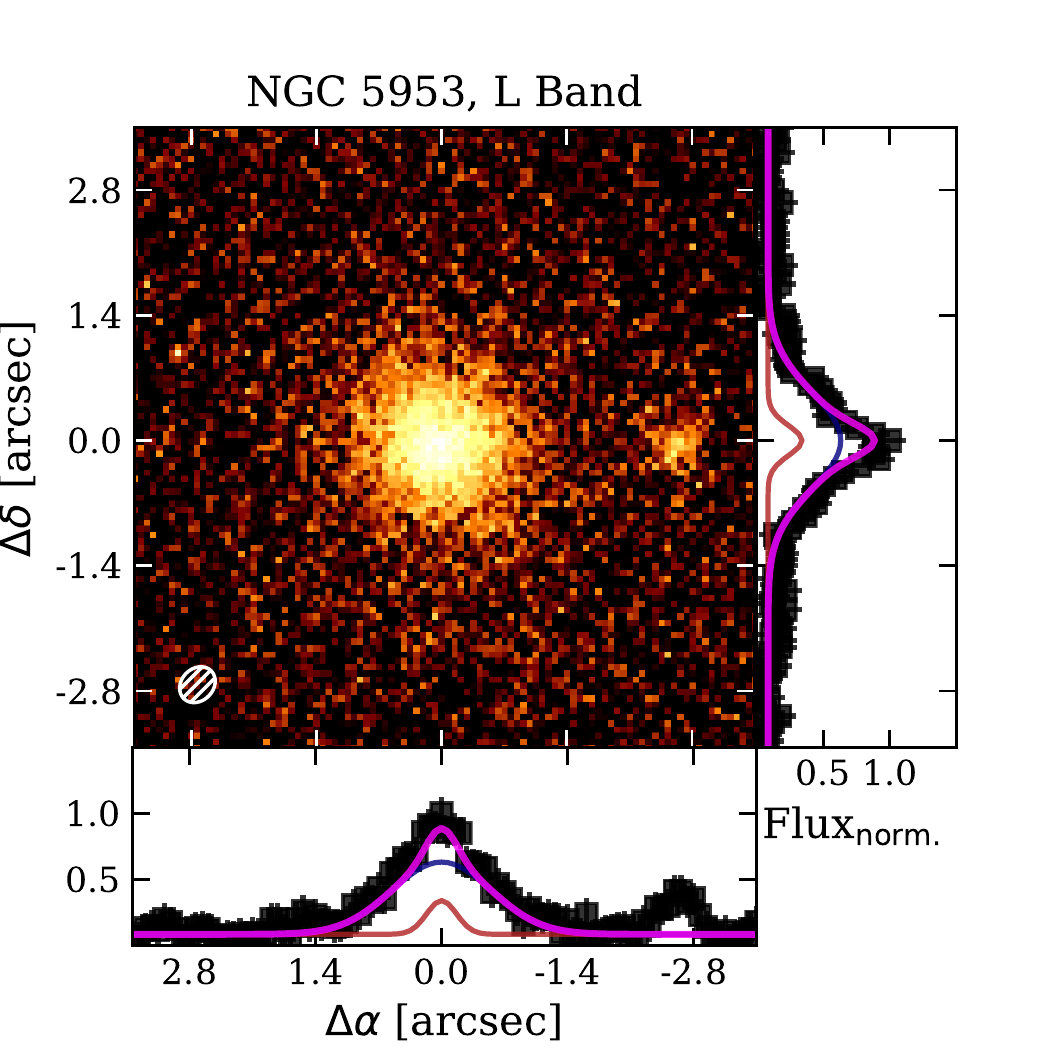}}
\subfloat{\includegraphics[width=0.25\hsize]{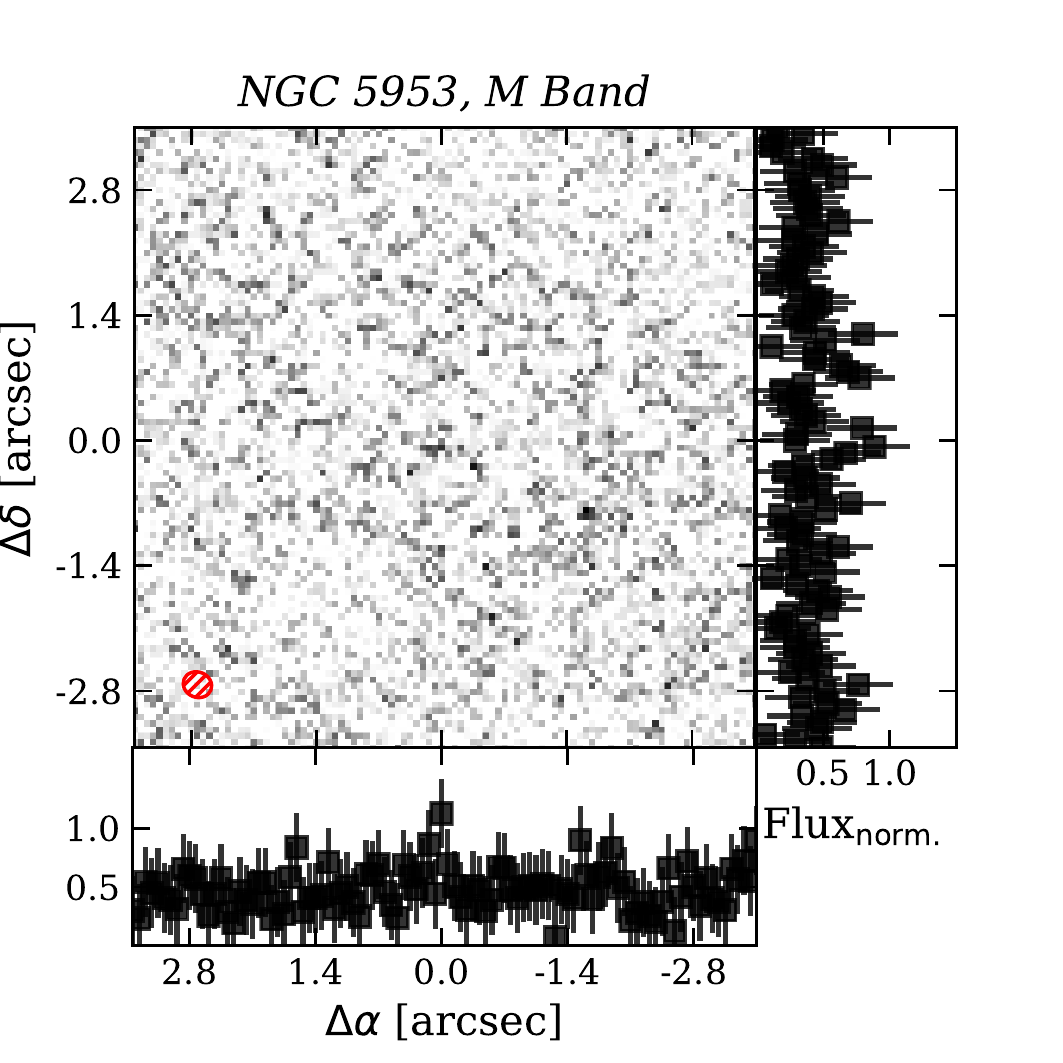}}
\subfloat{\includegraphics[width=0.25\hsize]{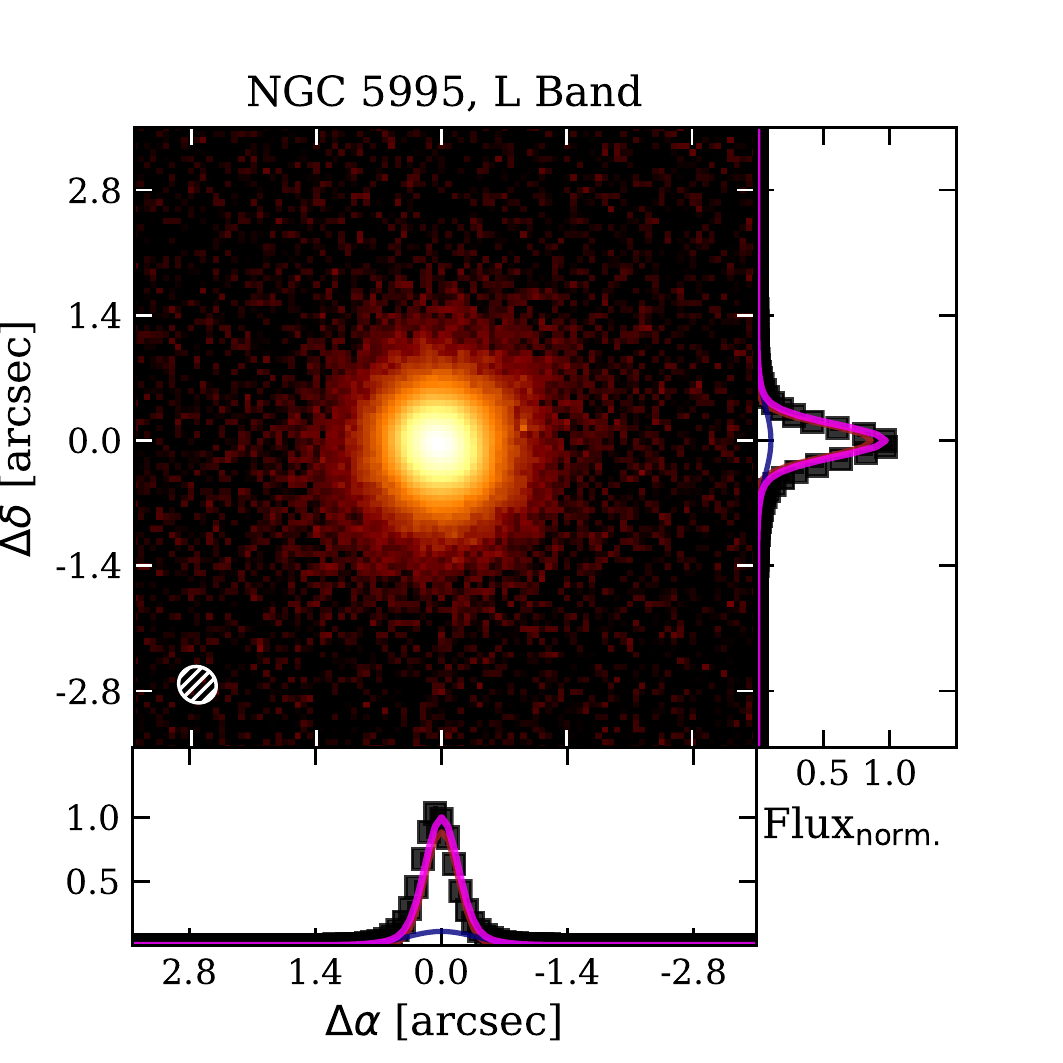}} \\
\subfloat{\includegraphics[width=0.25\hsize]{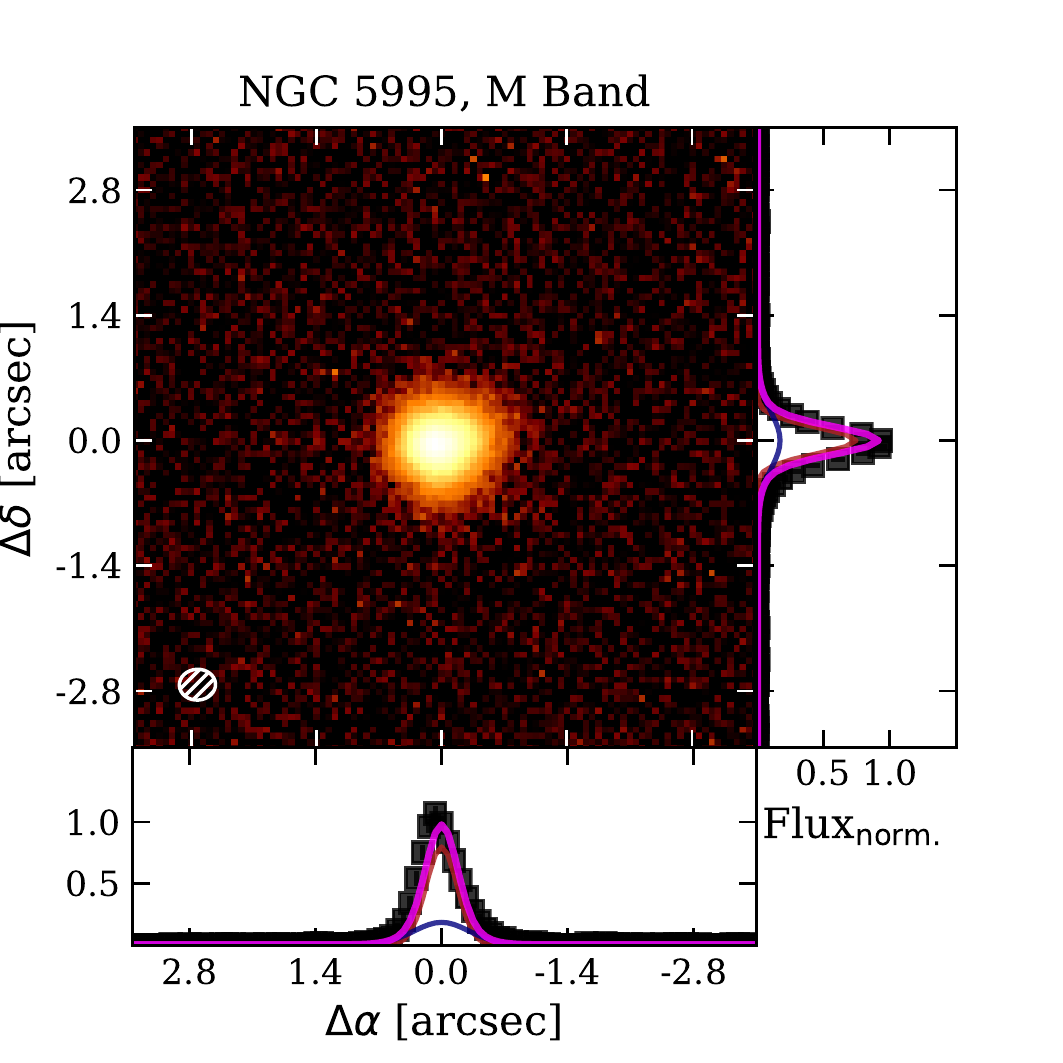}}
\subfloat{\includegraphics[width=0.25\hsize]{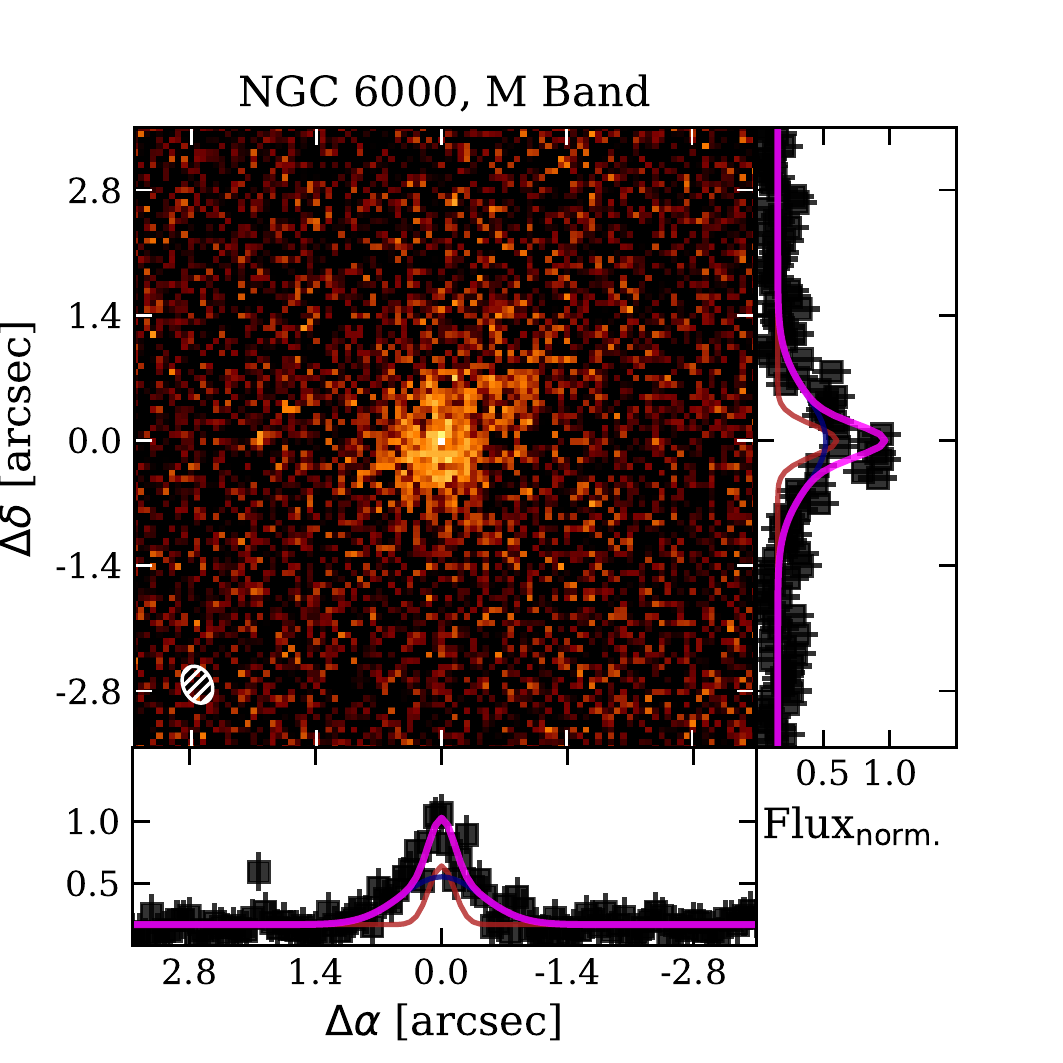}}
\subfloat{\includegraphics[width=0.25\hsize]{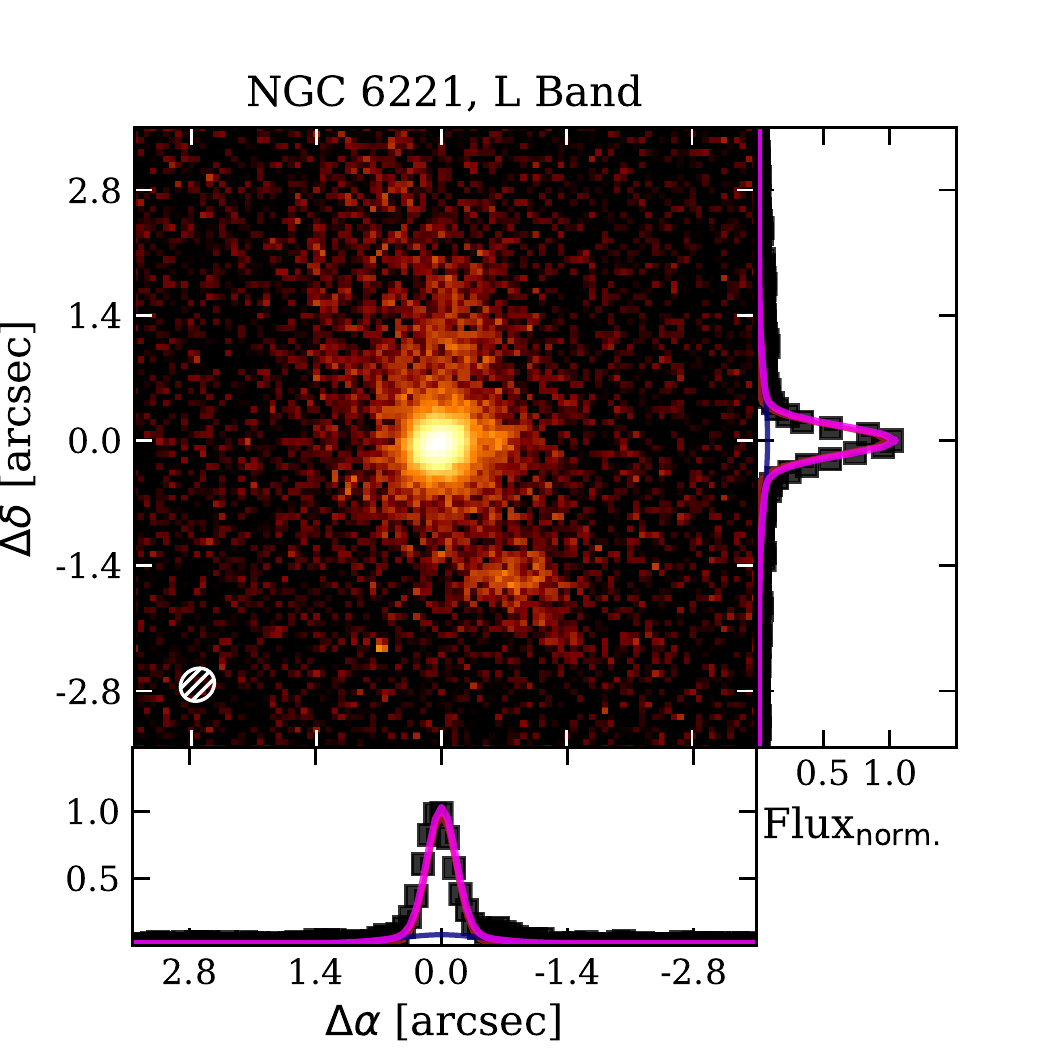}}
\subfloat{\includegraphics[width=0.25\hsize]{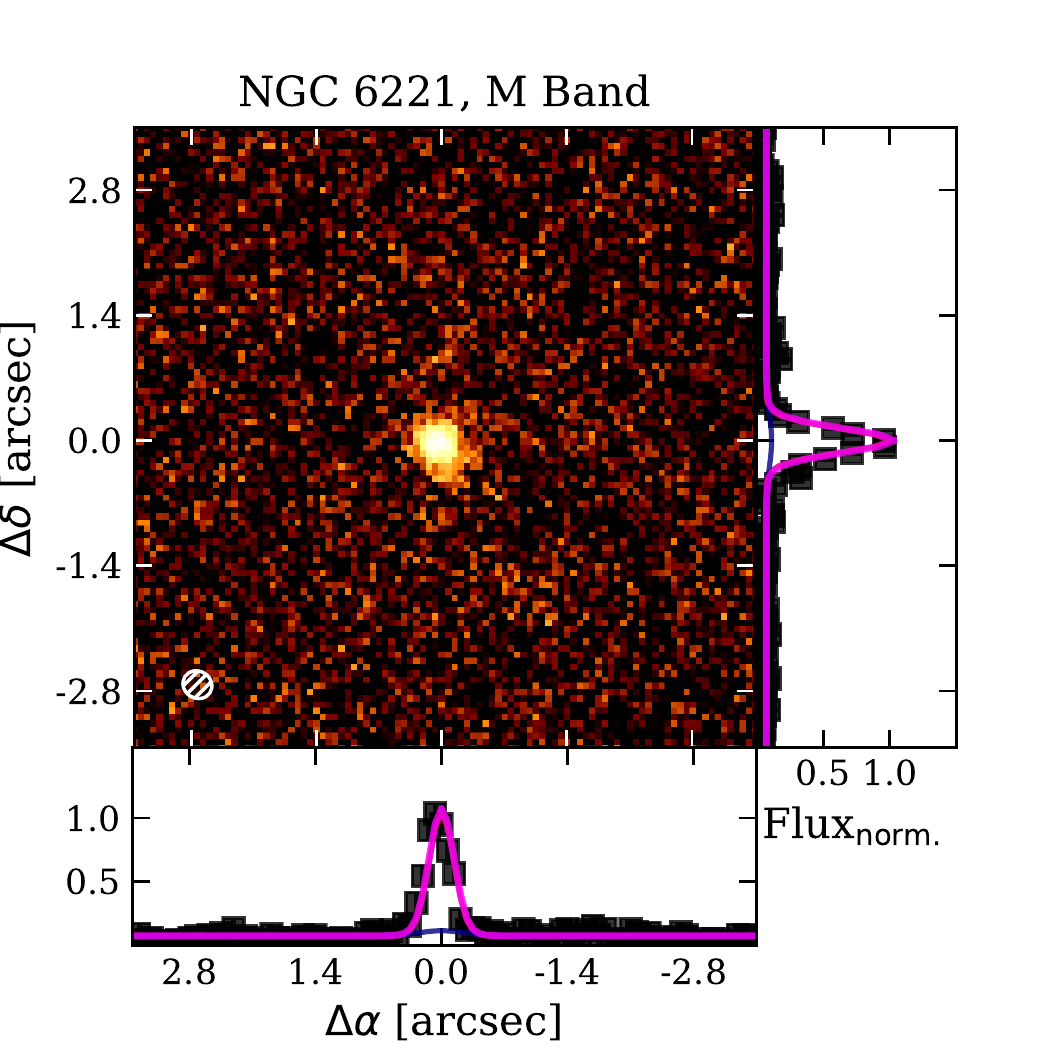}} \\
\subfloat{\includegraphics[width=0.25\hsize]{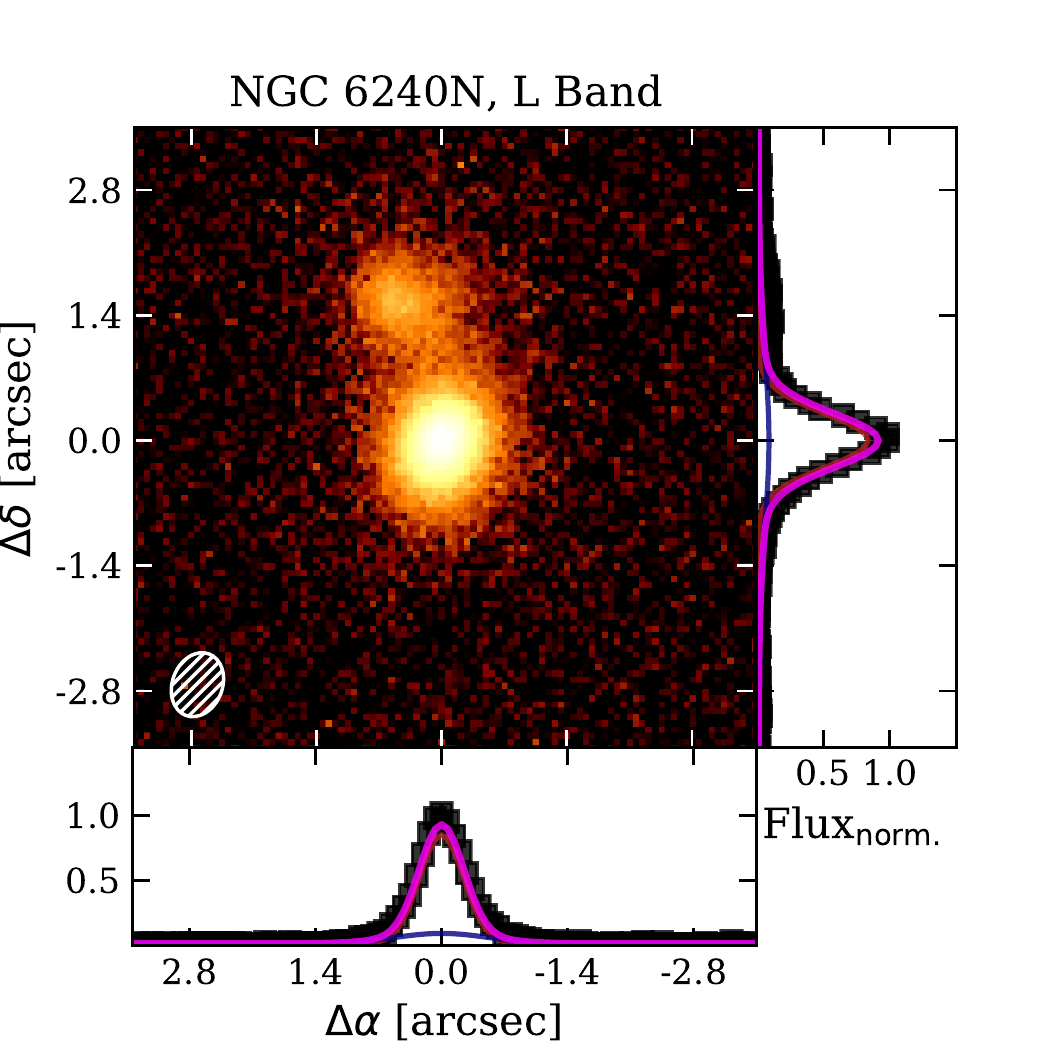}}
\subfloat{\includegraphics[width=0.25\hsize]{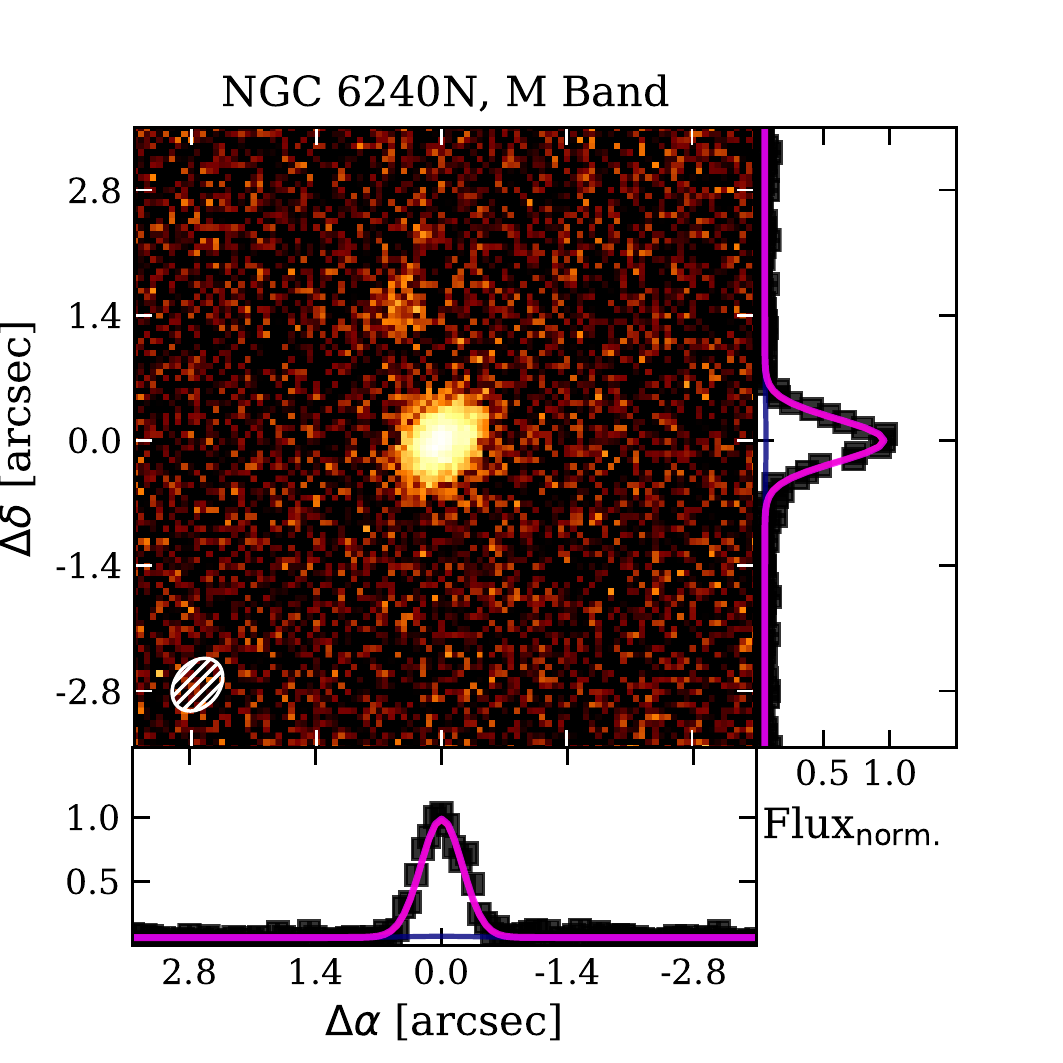}}
\subfloat{\includegraphics[width=0.25\hsize]{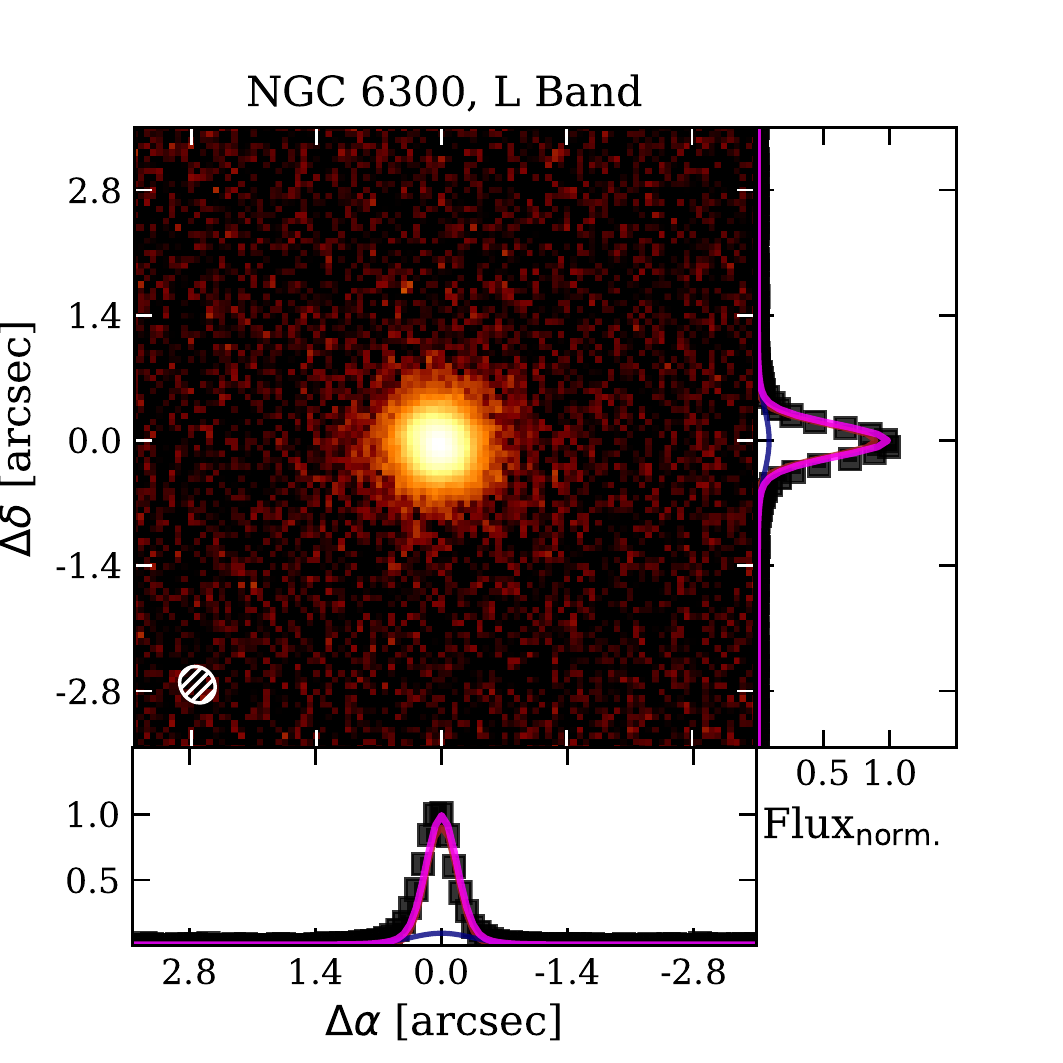}}
\subfloat{\includegraphics[width=0.25\hsize]{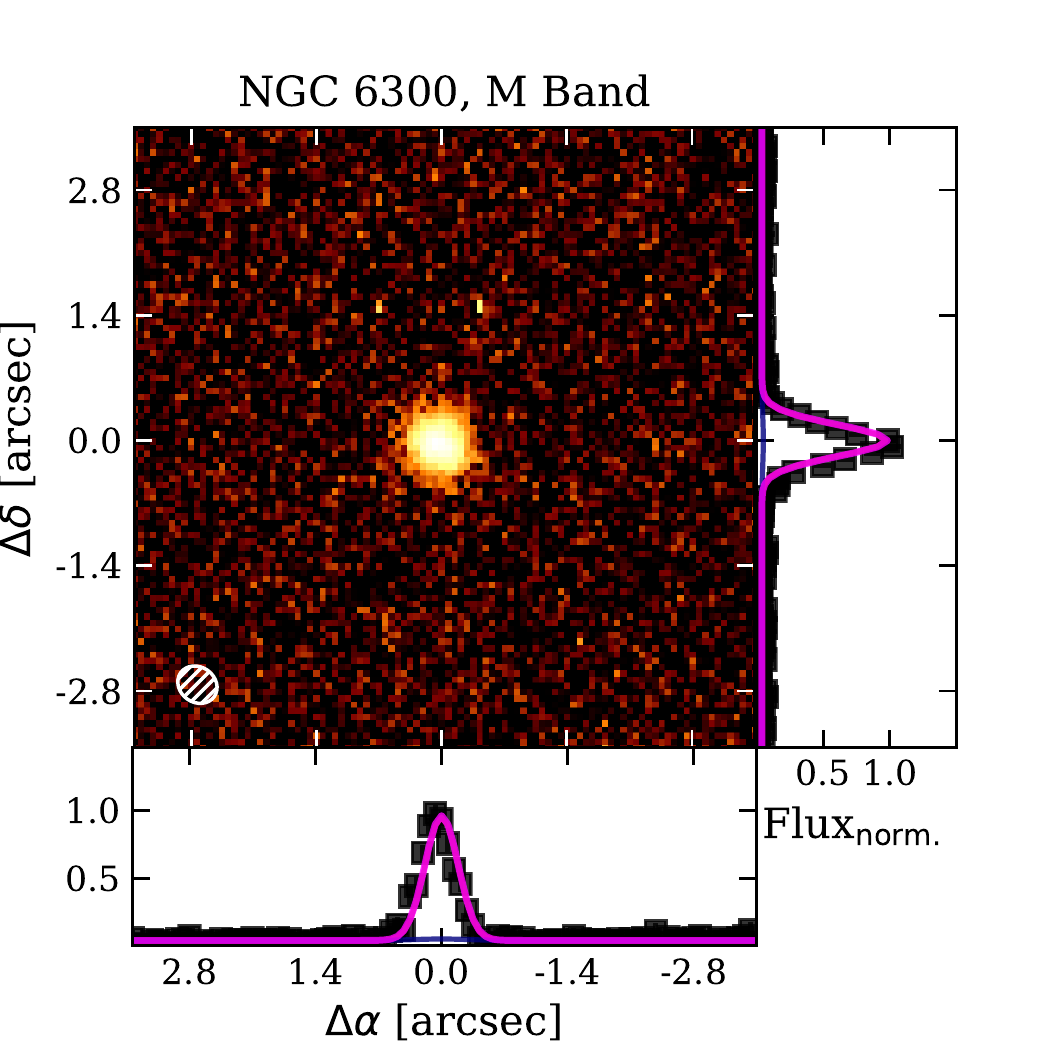}} \\
\subfloat{\includegraphics[width=0.25\hsize]{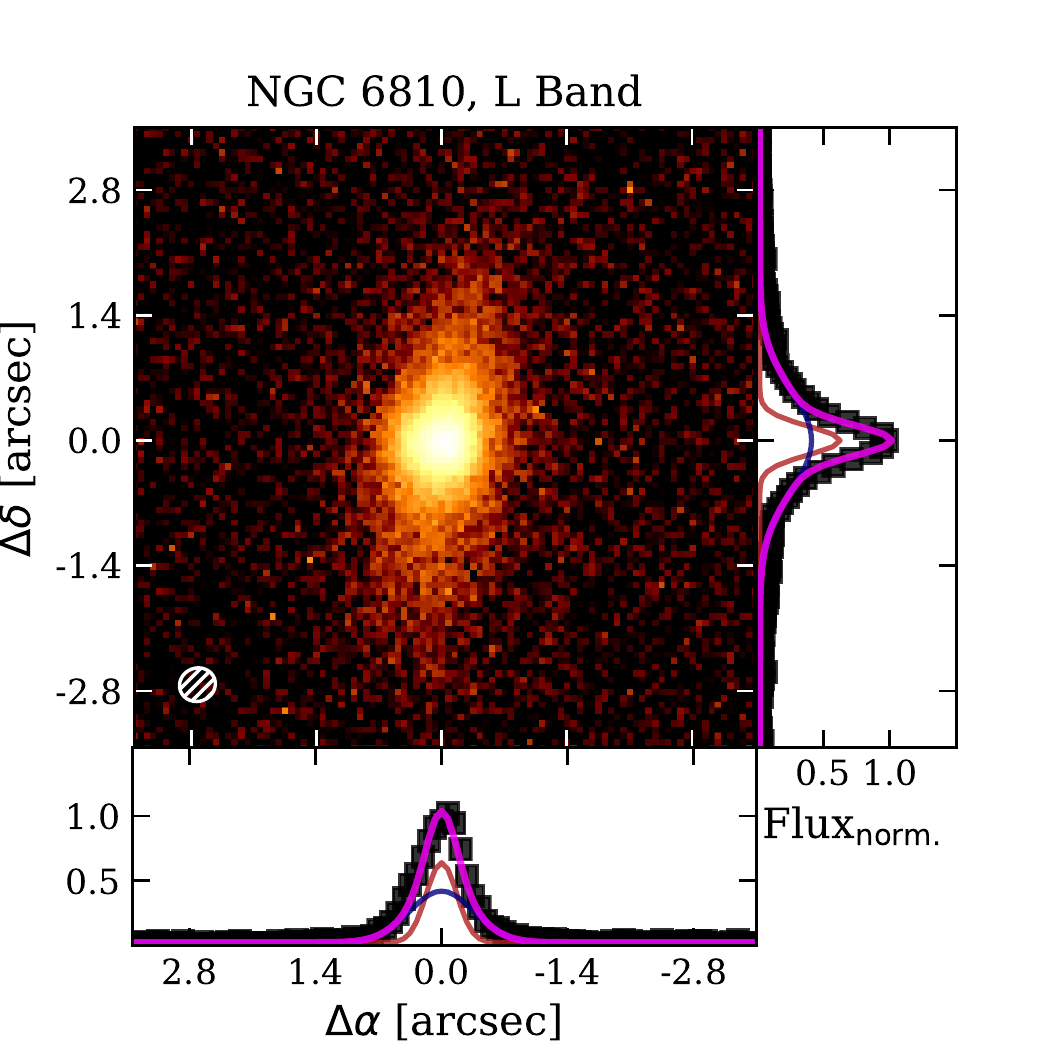}}
\subfloat{\includegraphics[width=0.25\hsize]{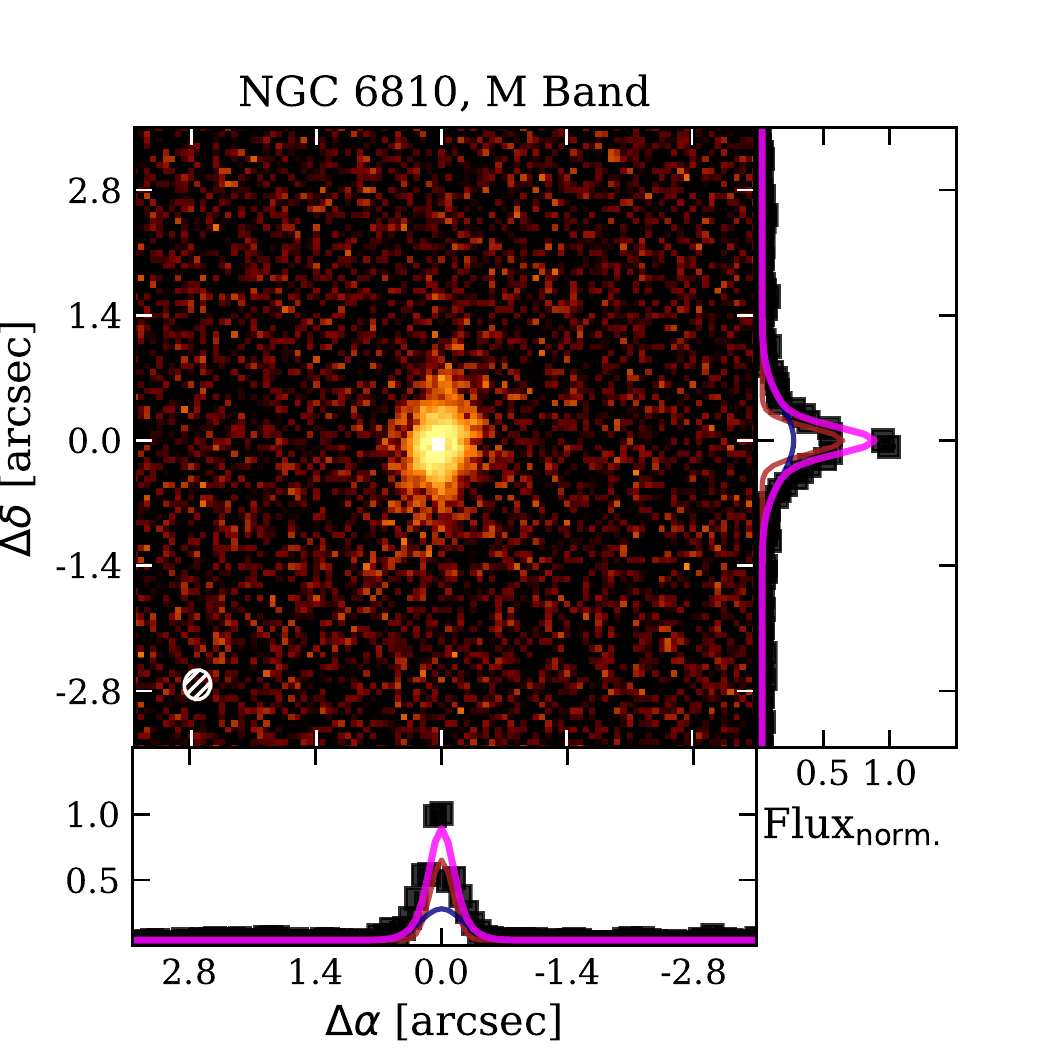}}
\subfloat{\includegraphics[width=0.25\hsize]{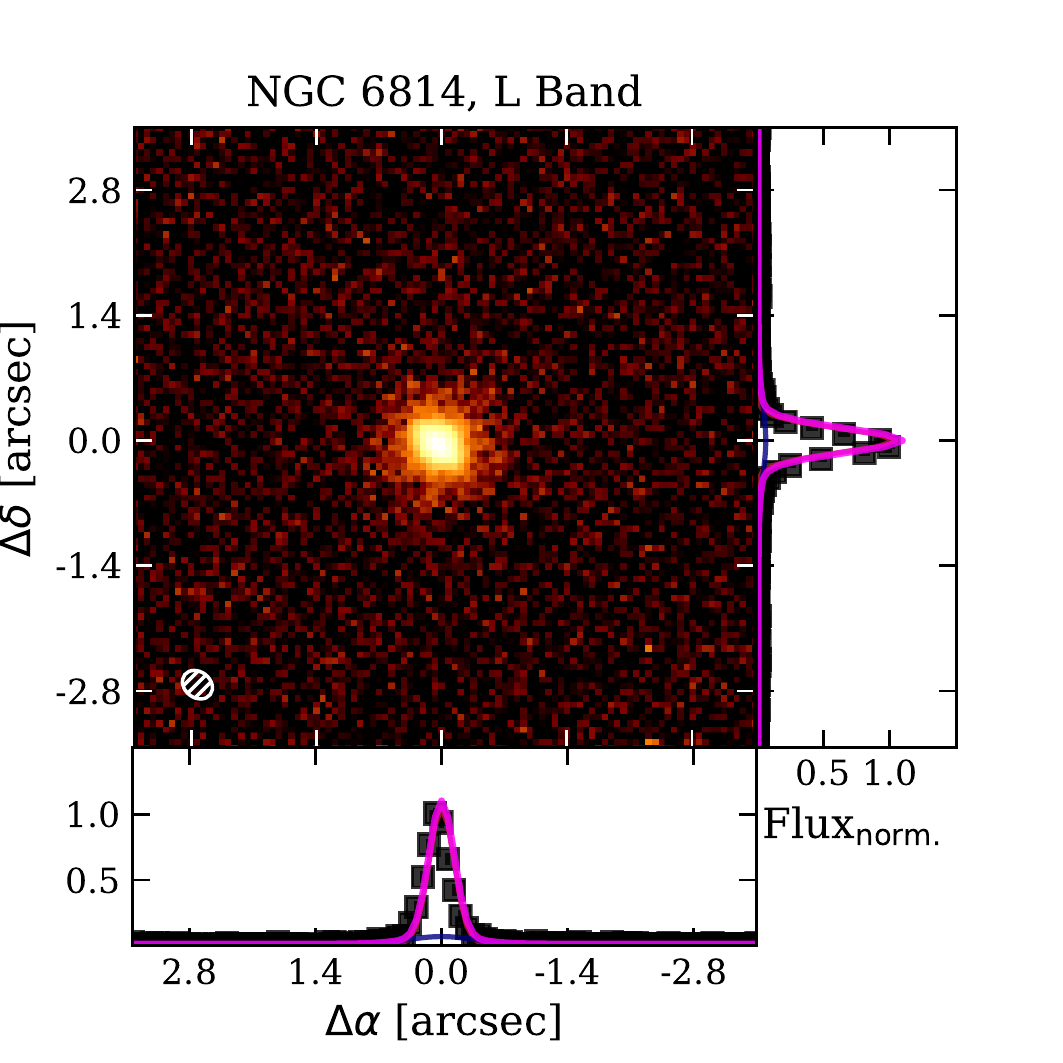}}
\subfloat{\includegraphics[width=0.25\hsize]{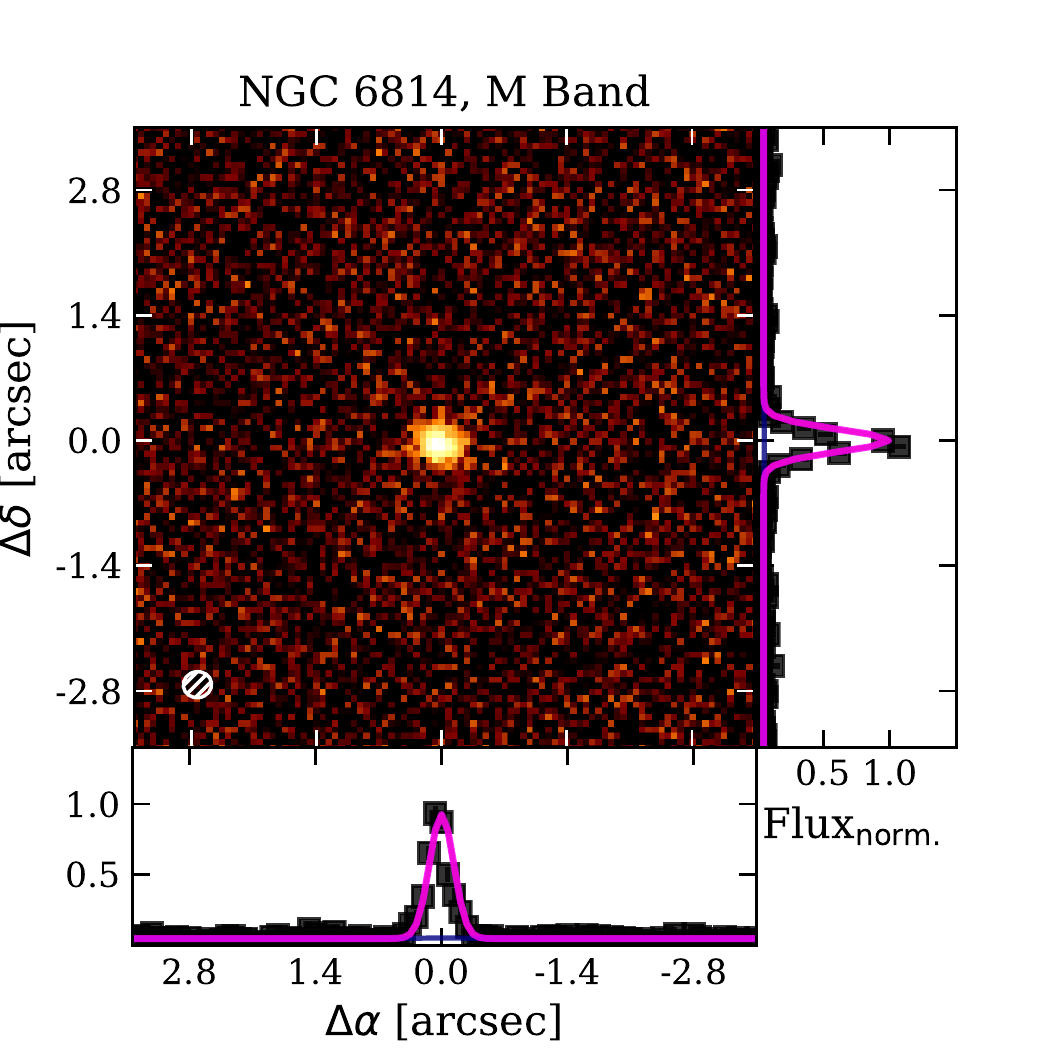}} \\
\caption{As Fig. \ref{fig:cutouts_one}.}
\end{figure*}
\begin{figure*}
\centering
\subfloat{\includegraphics[width=0.25\hsize]{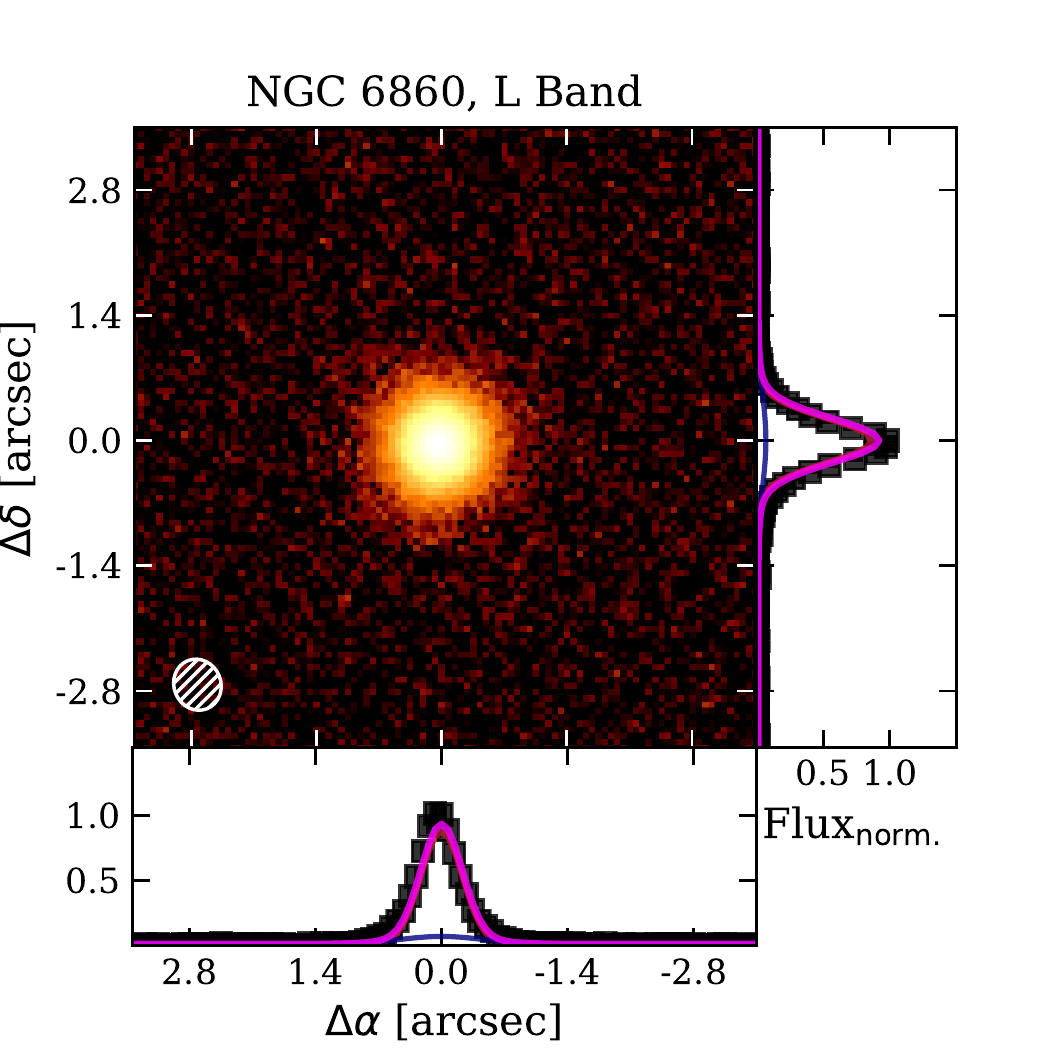}}
\subfloat{\includegraphics[width=0.25\hsize]{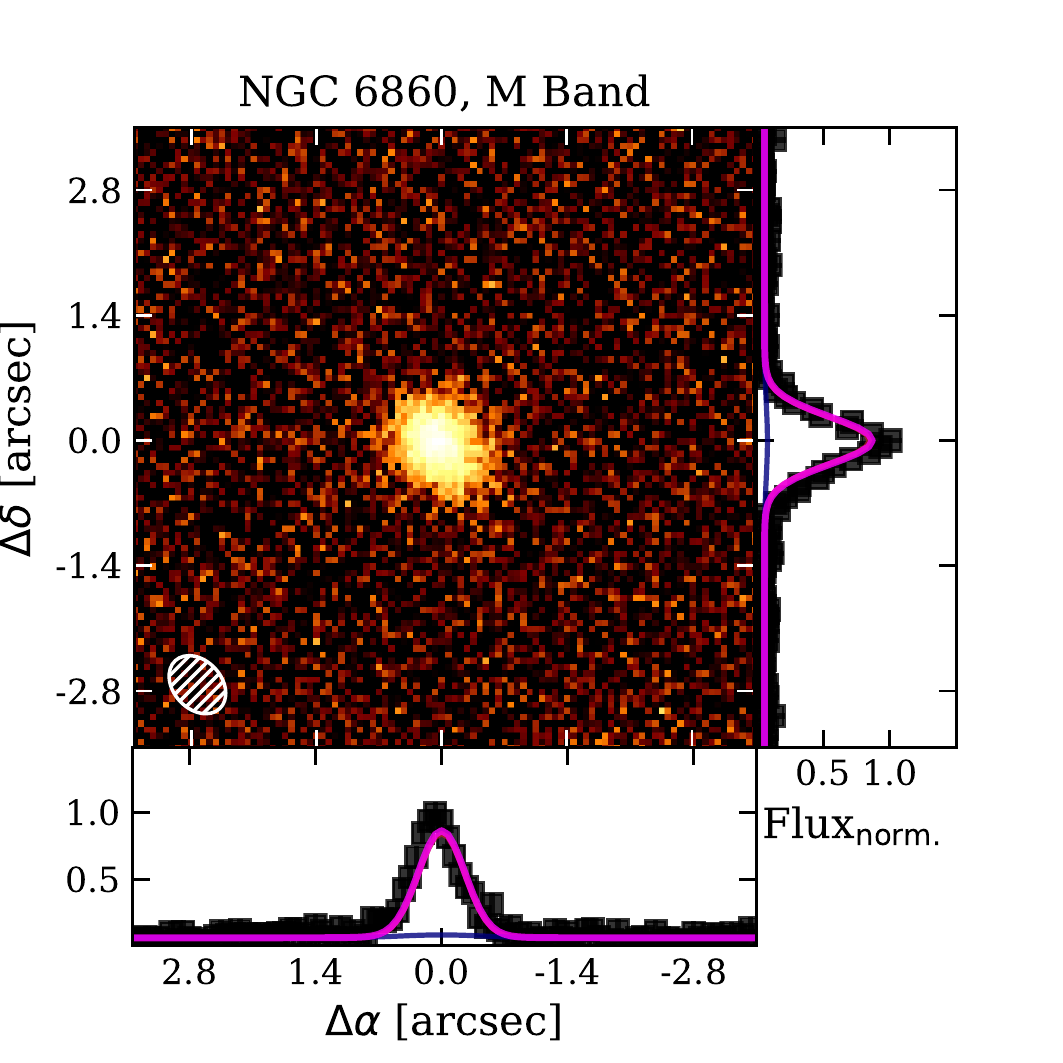}}
\subfloat{\includegraphics[width=0.25\hsize]{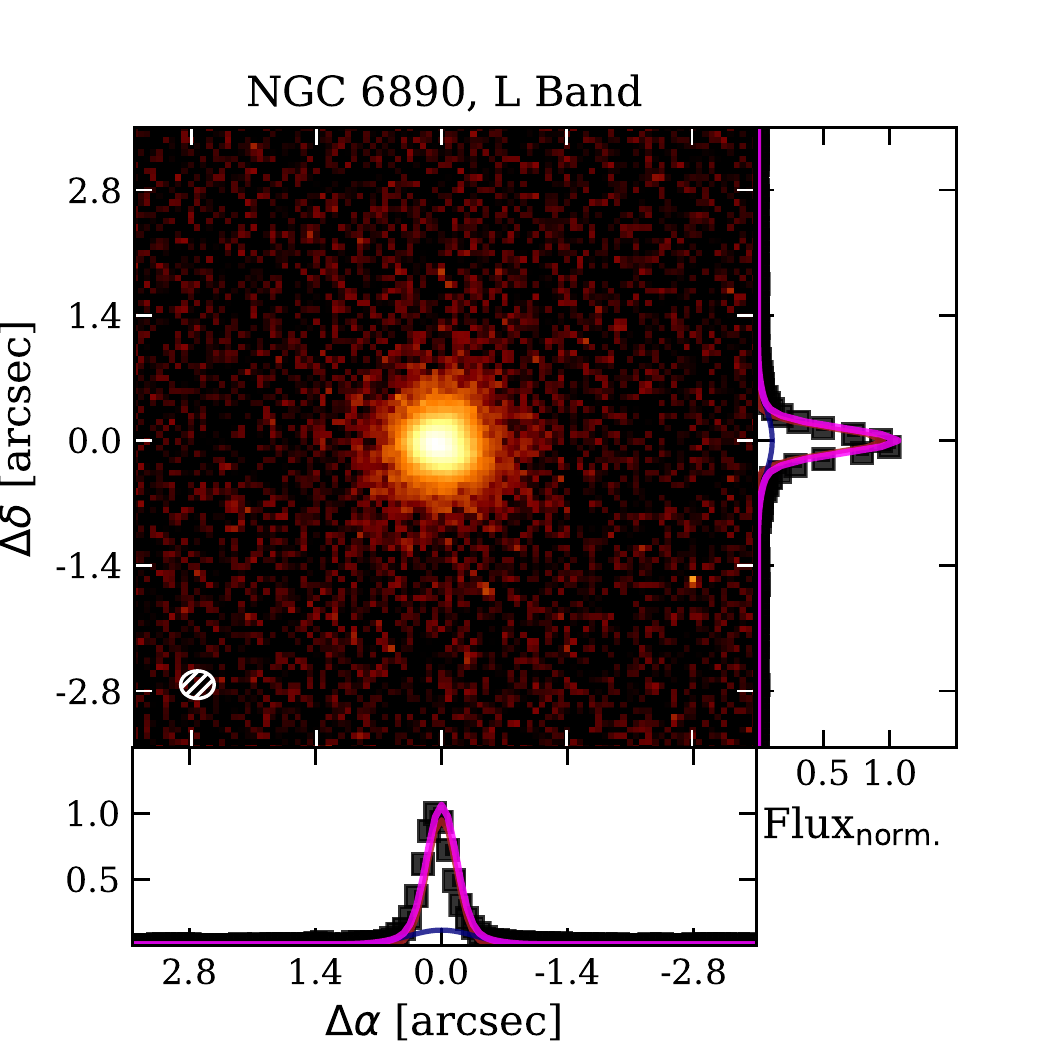}}
\subfloat{\includegraphics[width=0.25\hsize]{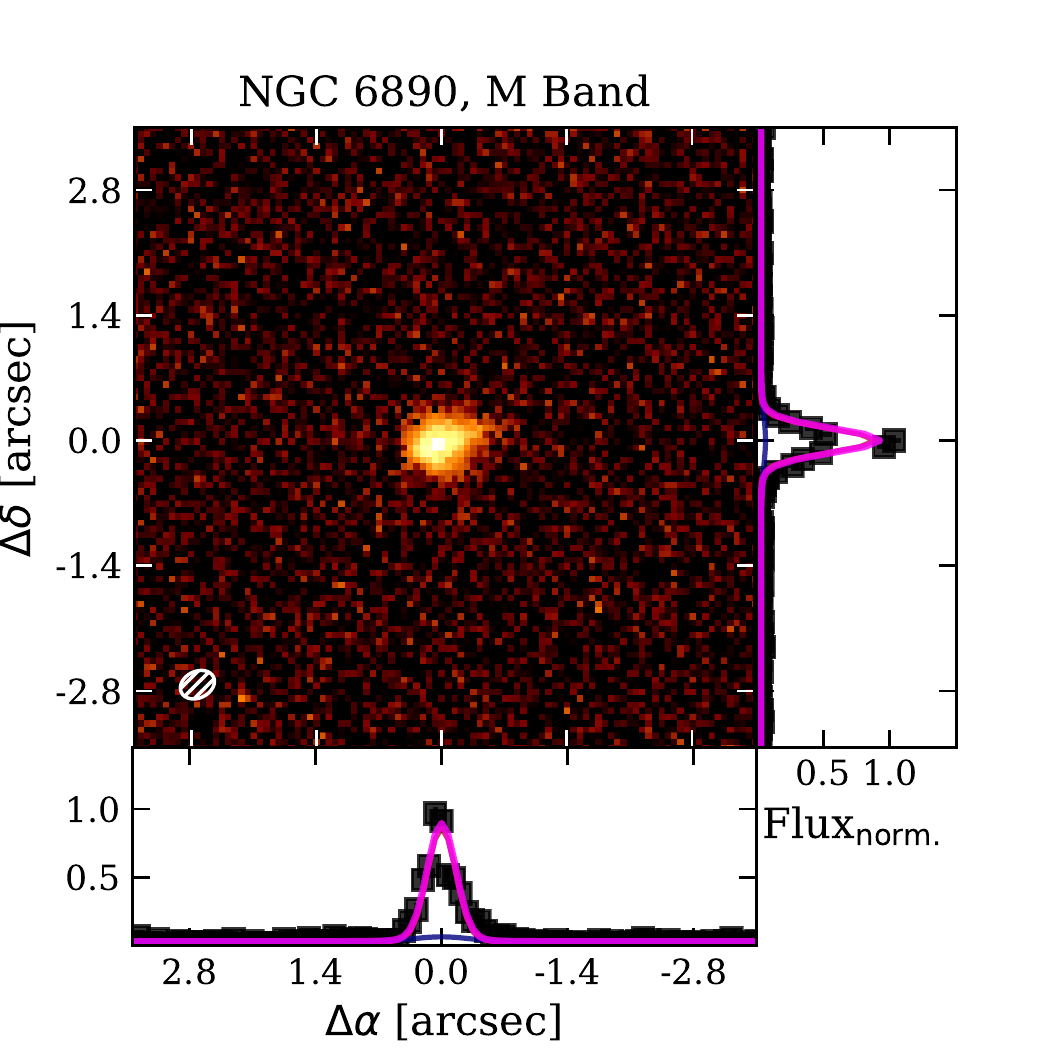}} \\
\subfloat{\includegraphics[width=0.25\hsize]{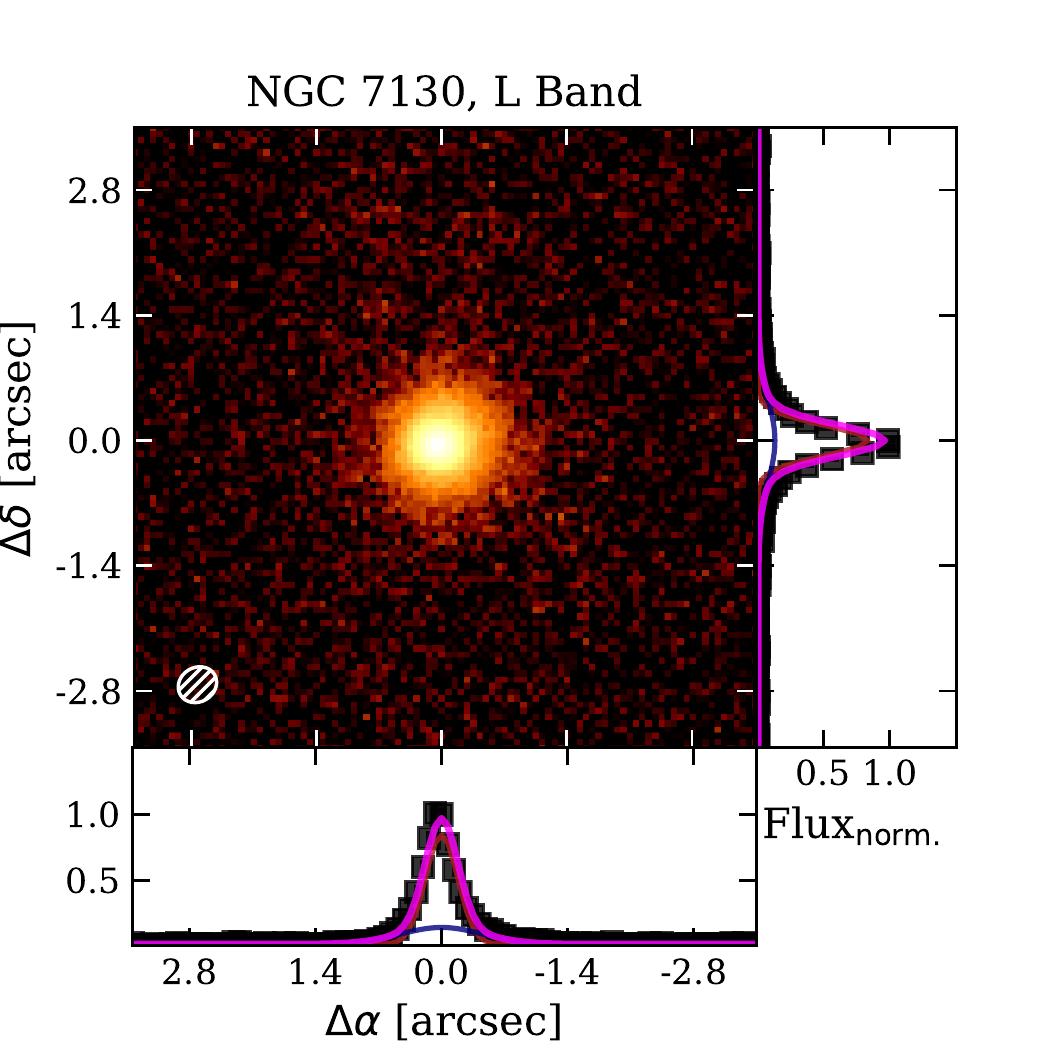}}
\subfloat{\includegraphics[width=0.25\hsize]{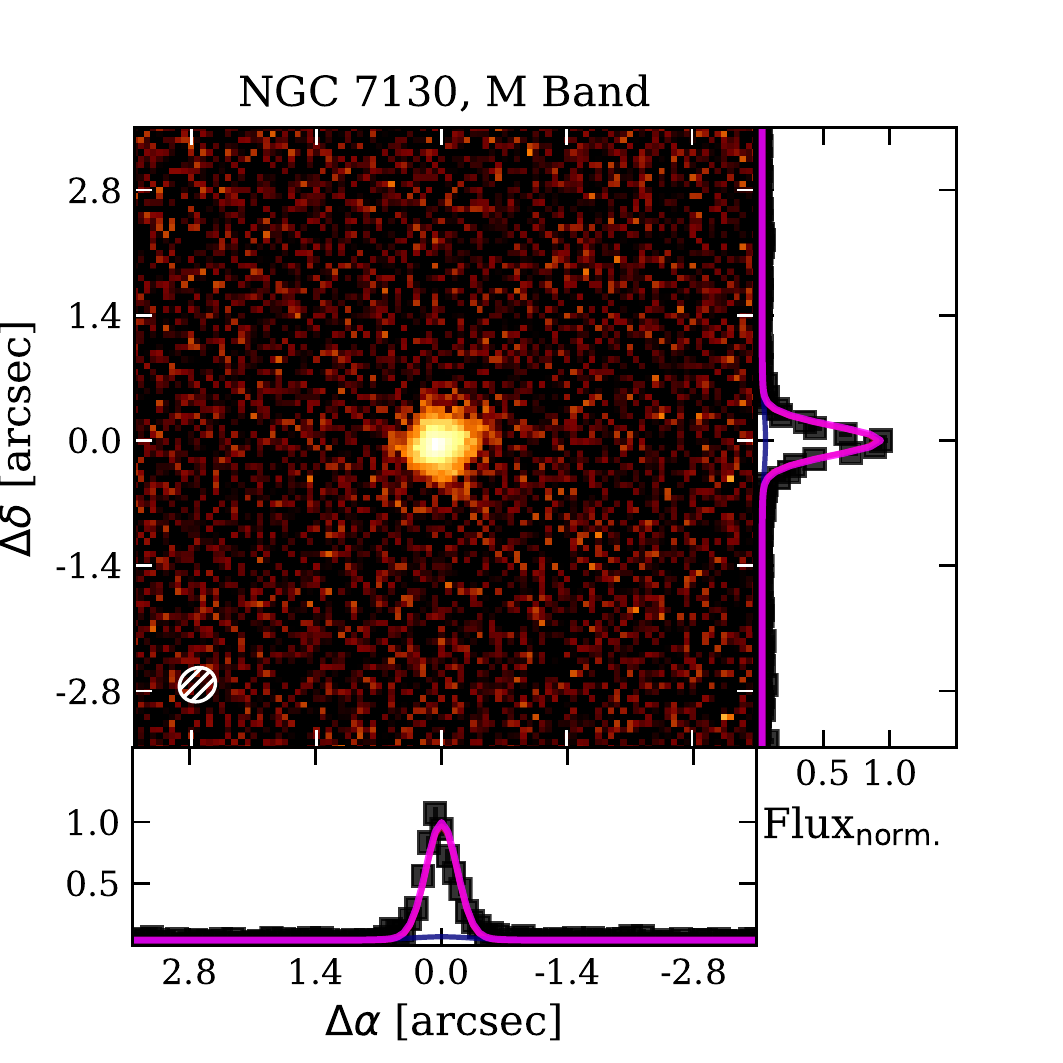}}
\subfloat{\includegraphics[width=0.25\hsize]{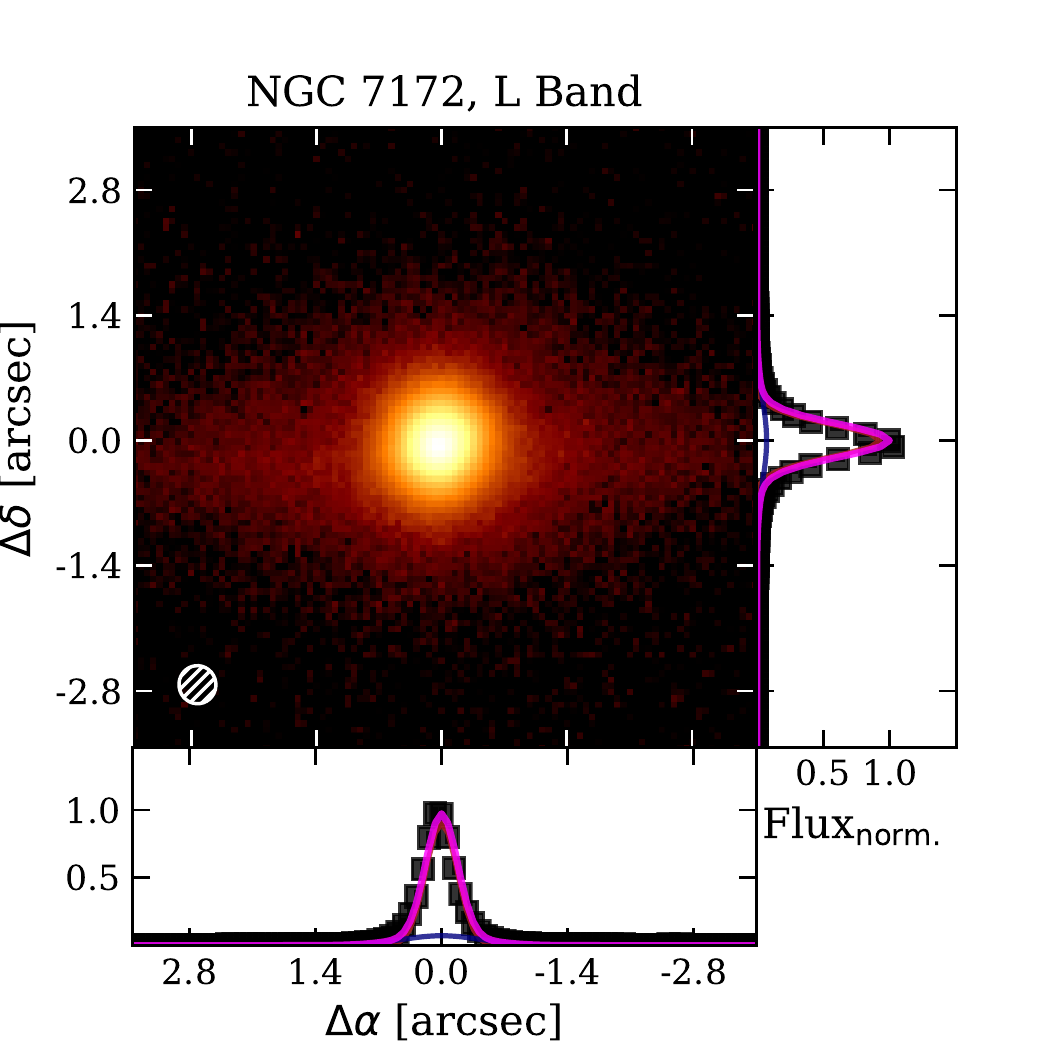}}
\subfloat{\includegraphics[width=0.25\hsize]{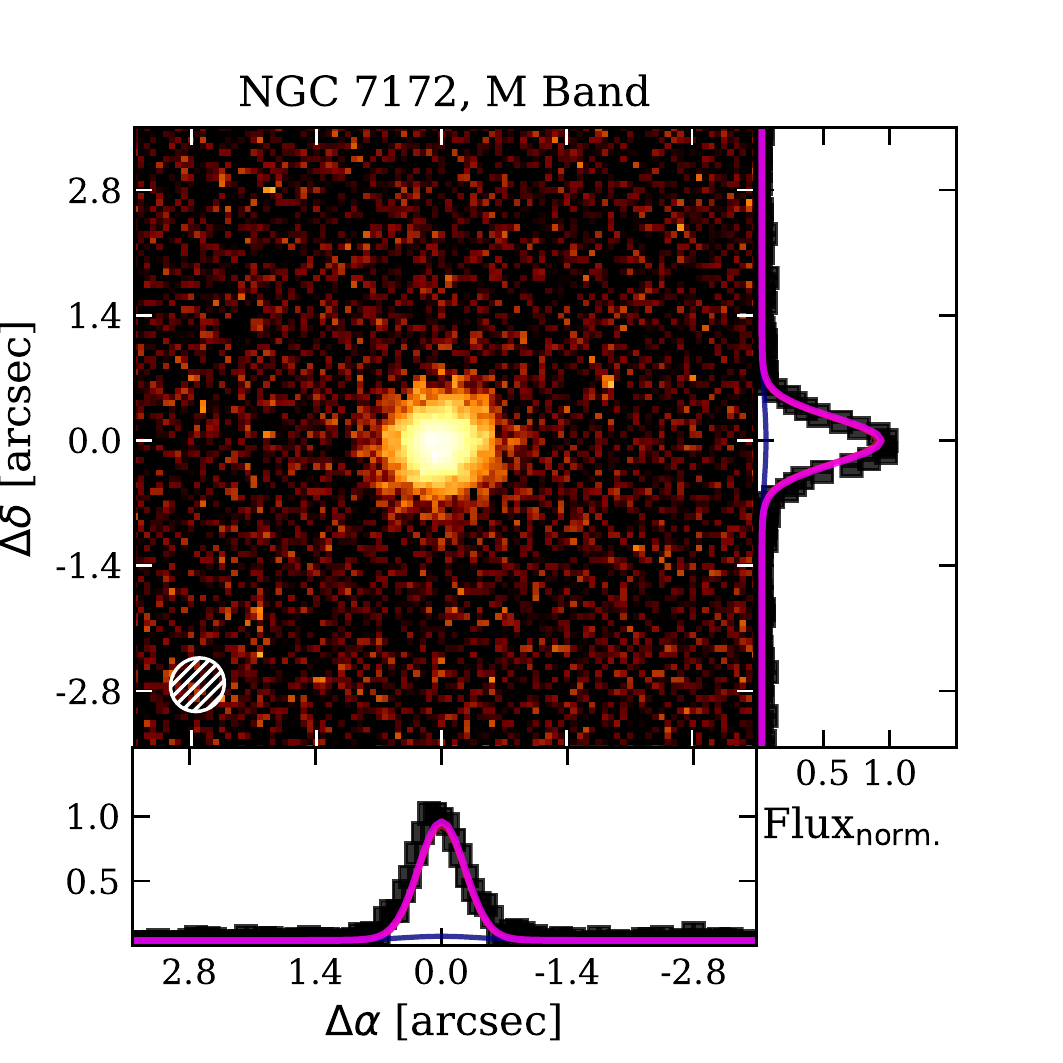}} \\
\subfloat{\includegraphics[width=0.25\hsize]{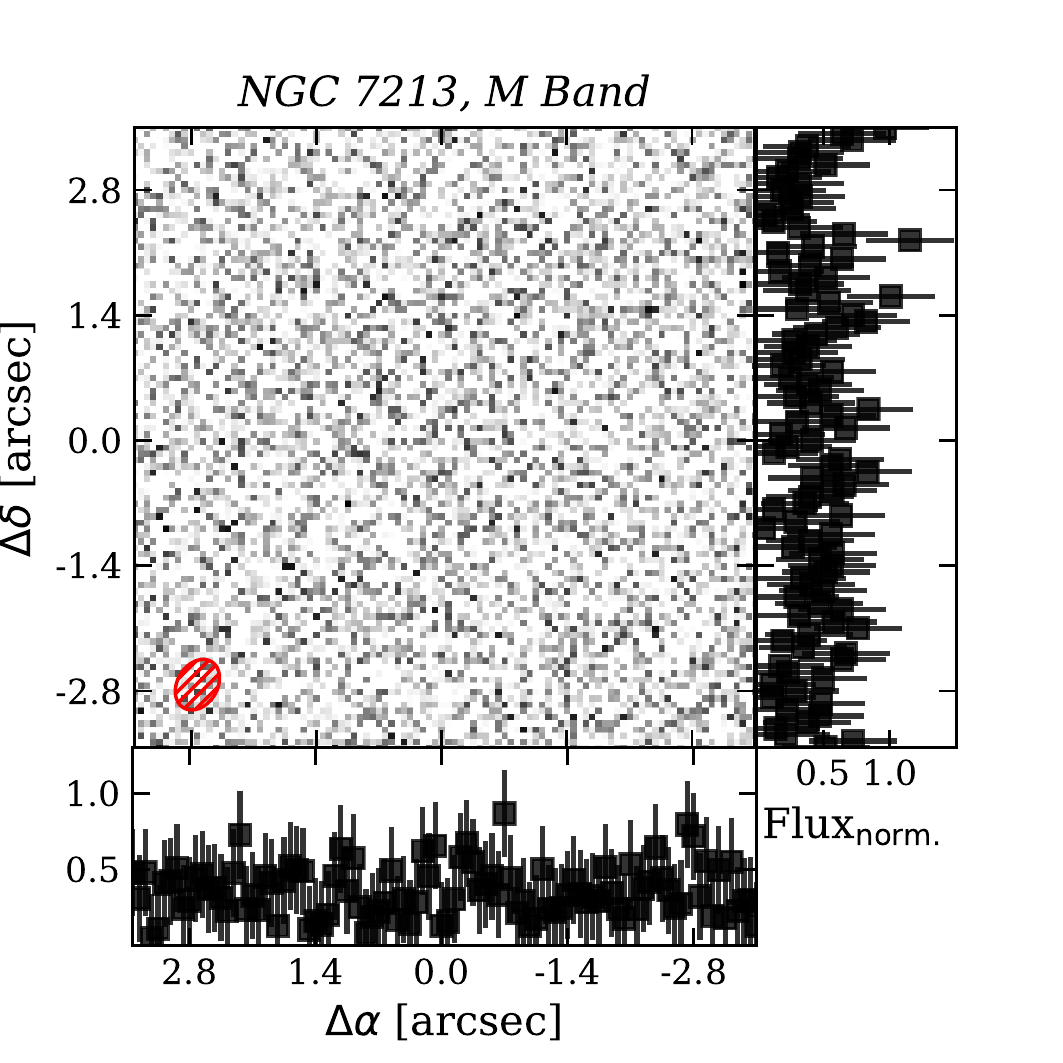}}
\subfloat{\includegraphics[width=0.25\hsize]{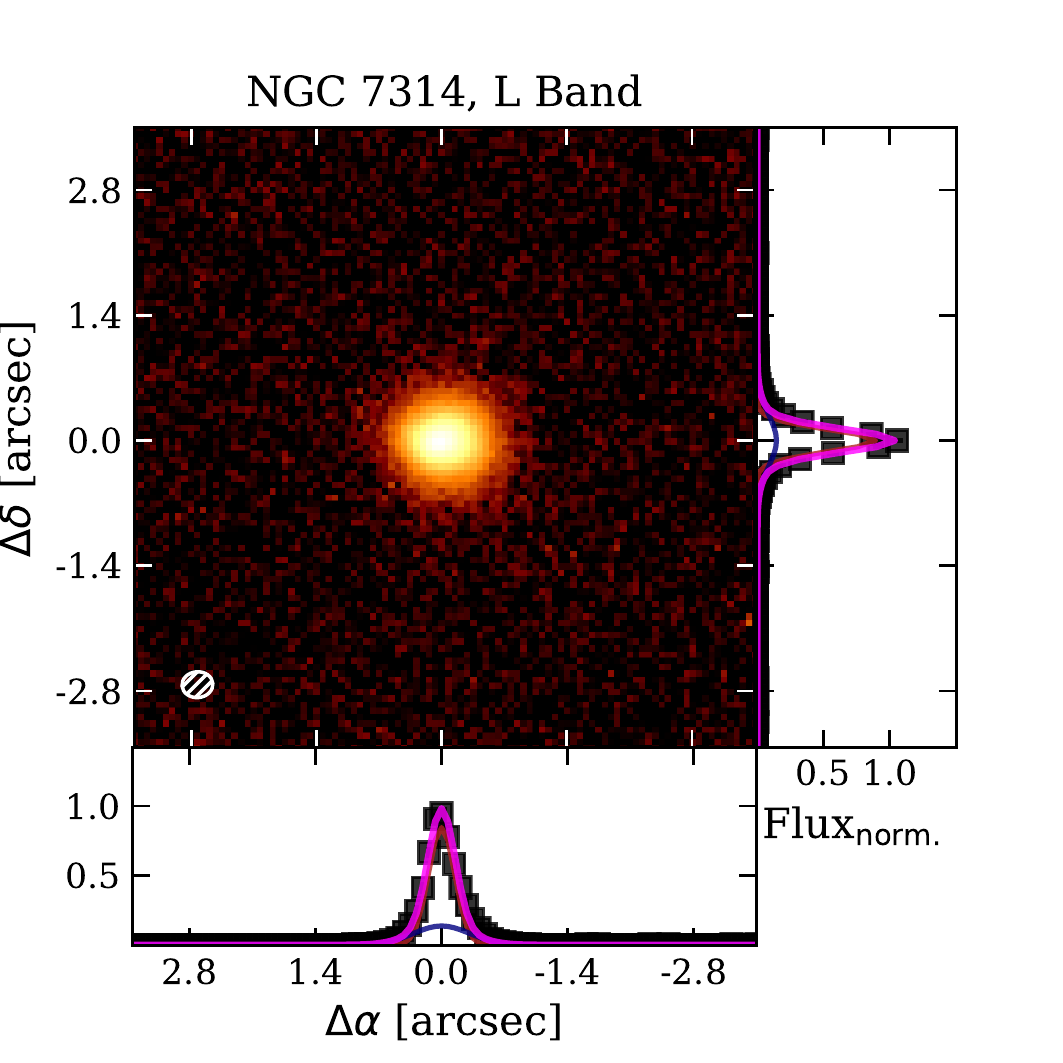}}
\subfloat{\includegraphics[width=0.25\hsize]{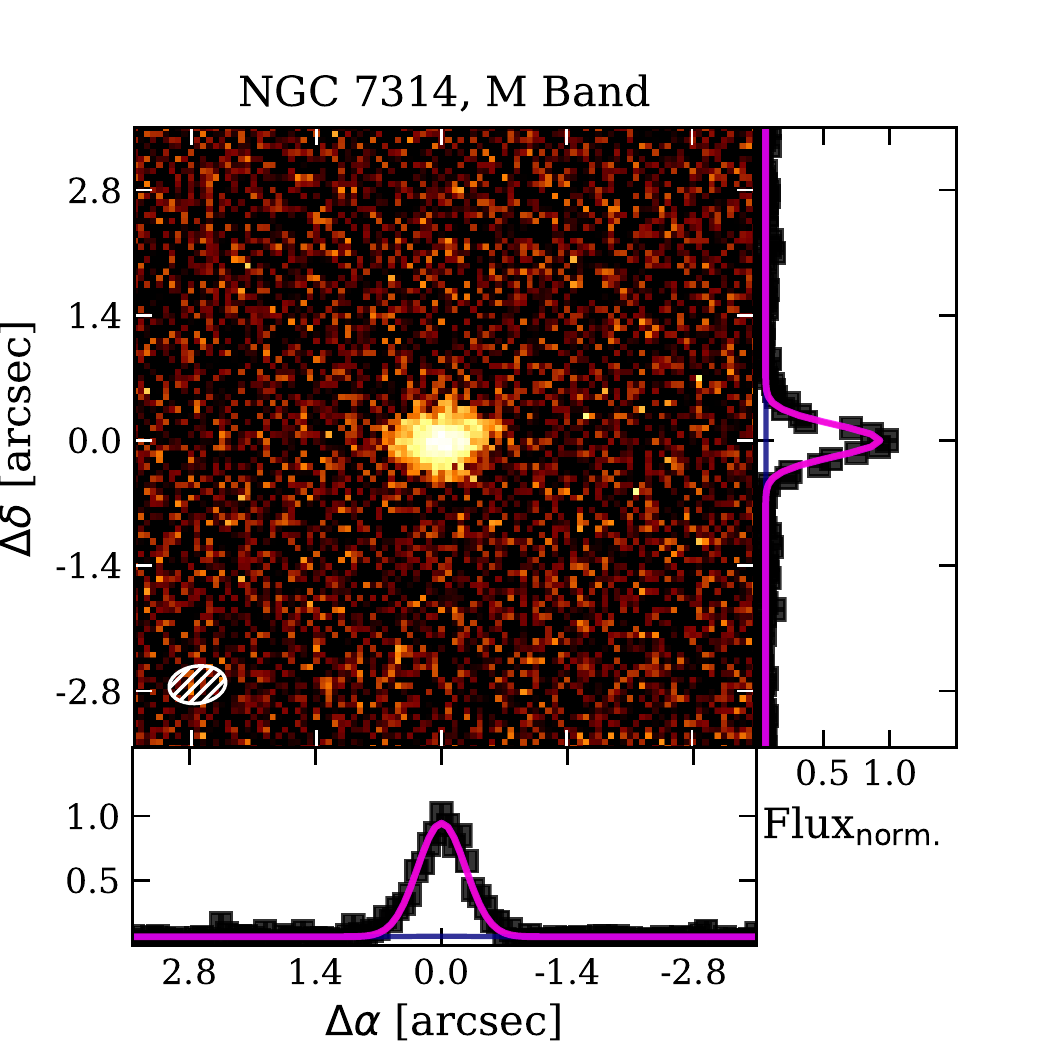}}
\subfloat{\includegraphics[width=0.25\hsize]{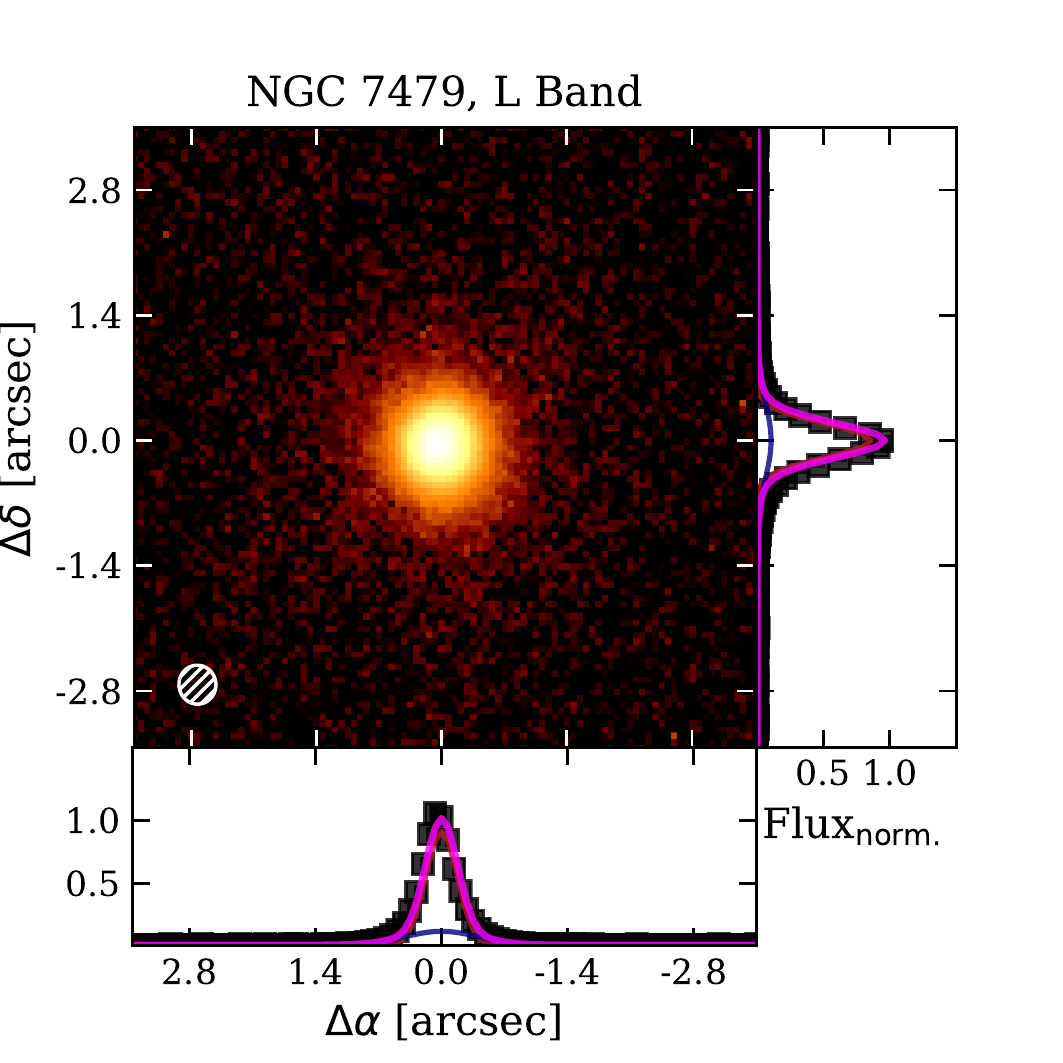}} \\
\subfloat{\includegraphics[width=0.25\hsize]{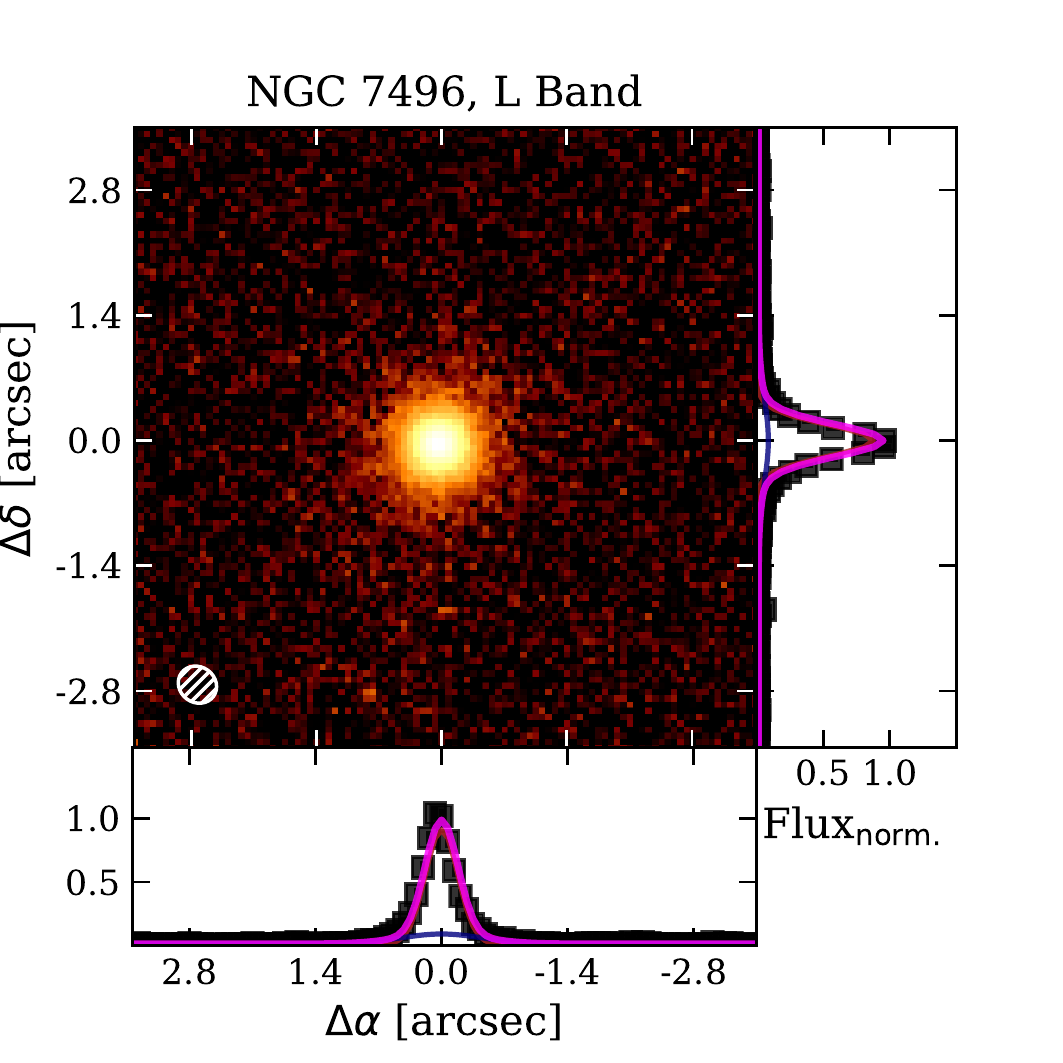}}
\subfloat{\includegraphics[width=0.25\hsize]{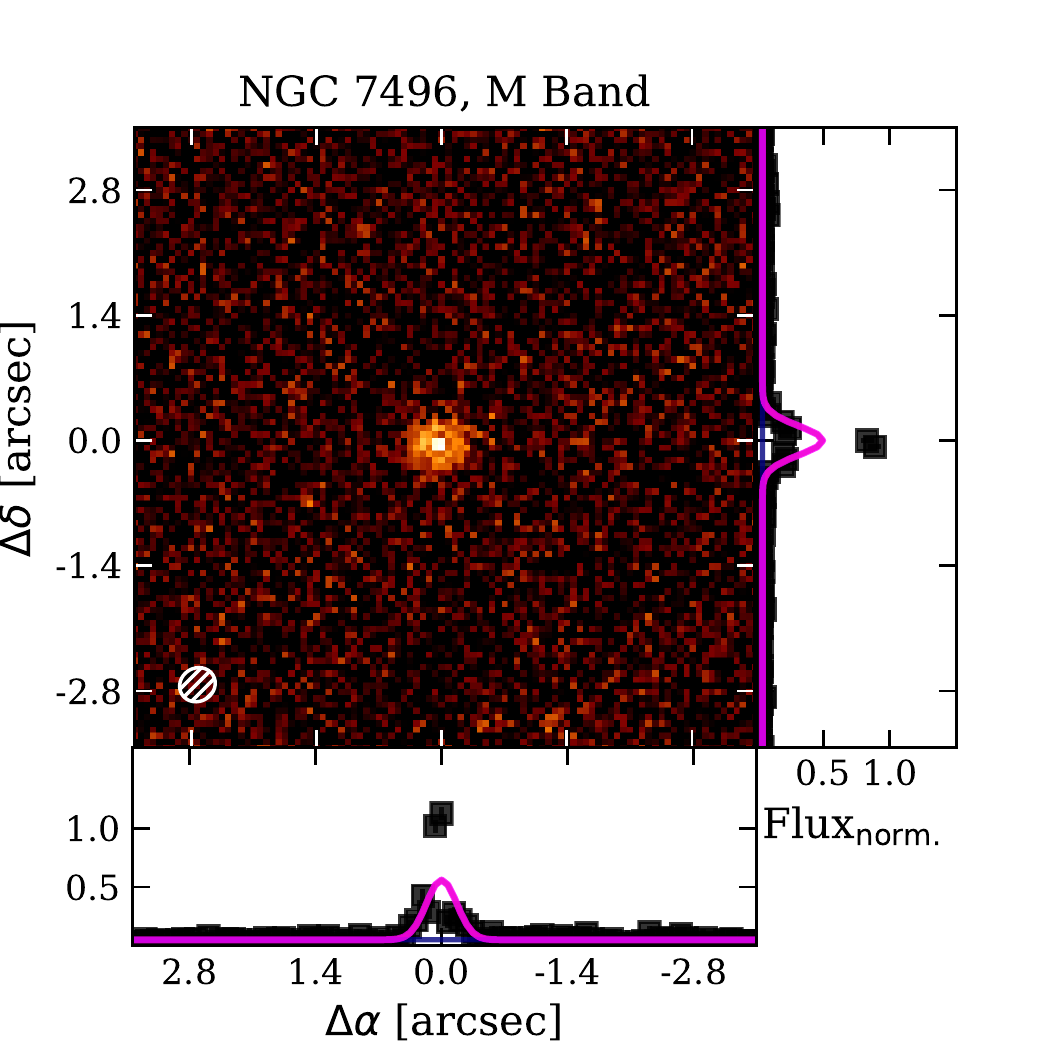}}
\subfloat{\includegraphics[width=0.25\hsize]{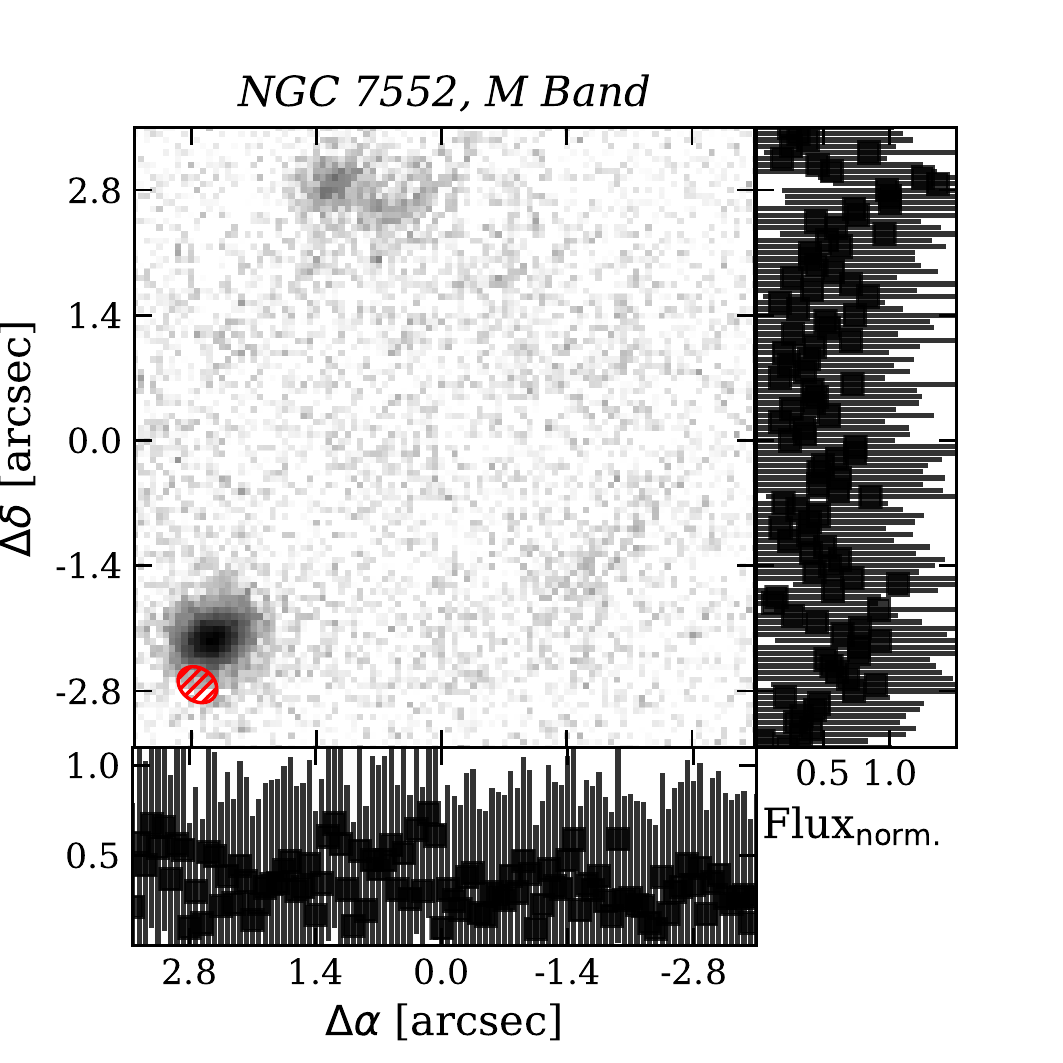}}
\subfloat{\includegraphics[width=0.25\hsize]{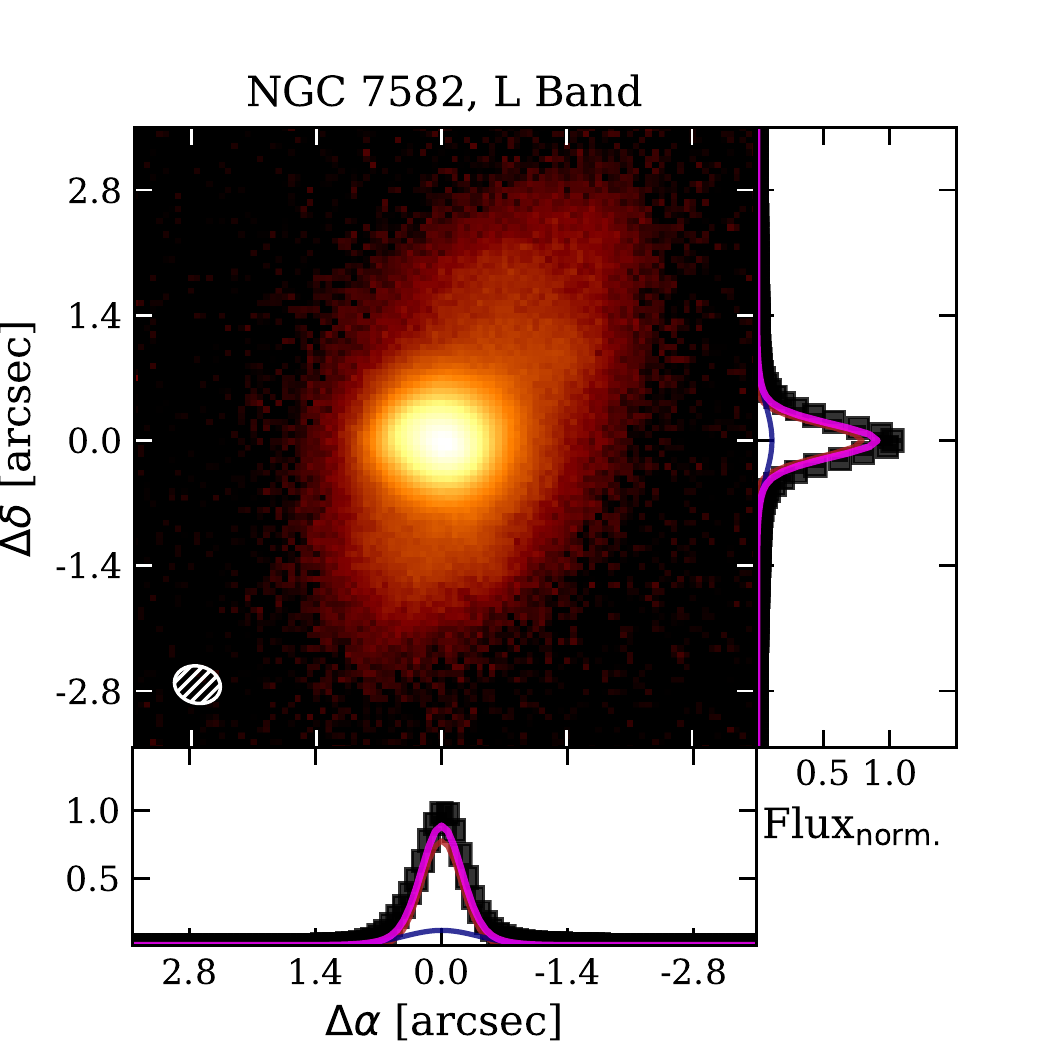}} \\
\subfloat{\includegraphics[width=0.25\hsize]{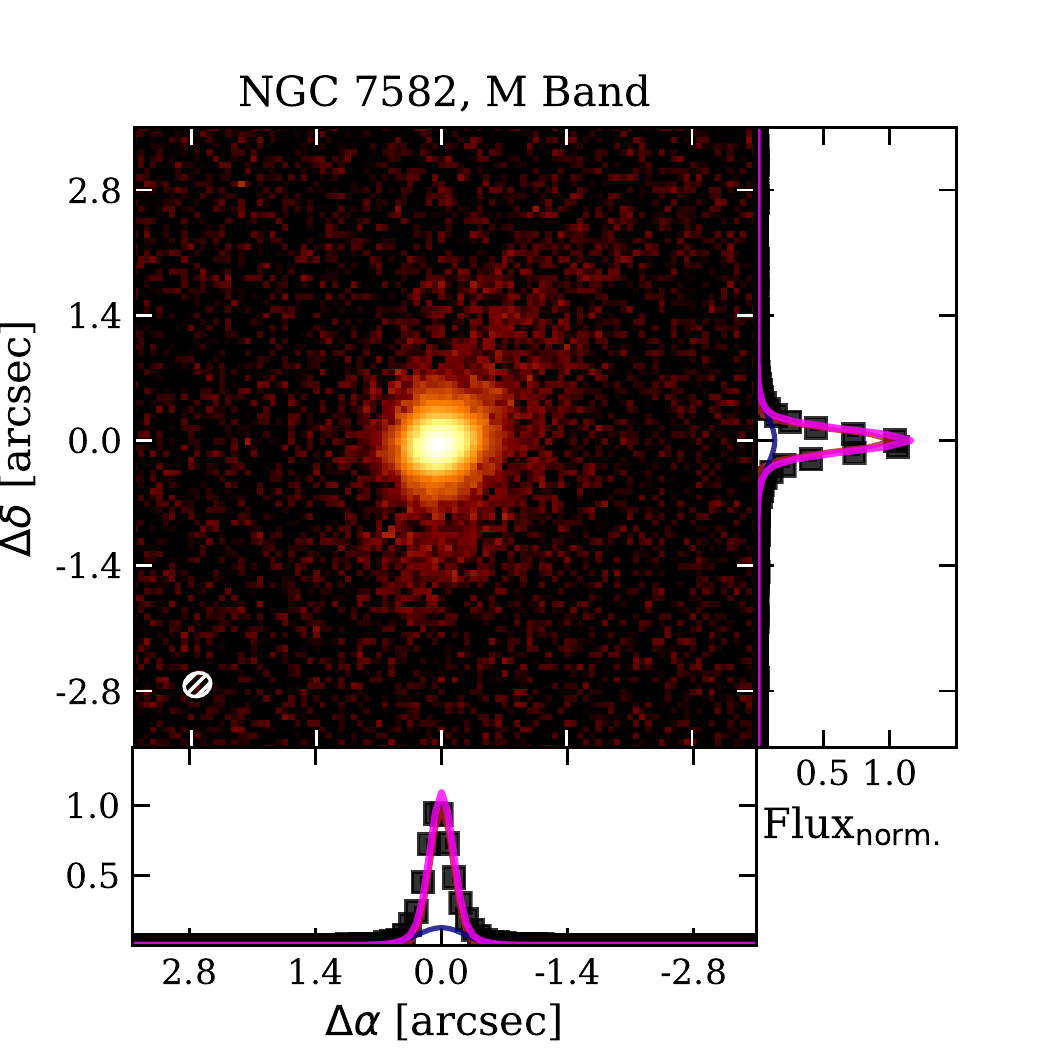}}
\subfloat{\includegraphics[width=0.25\hsize]{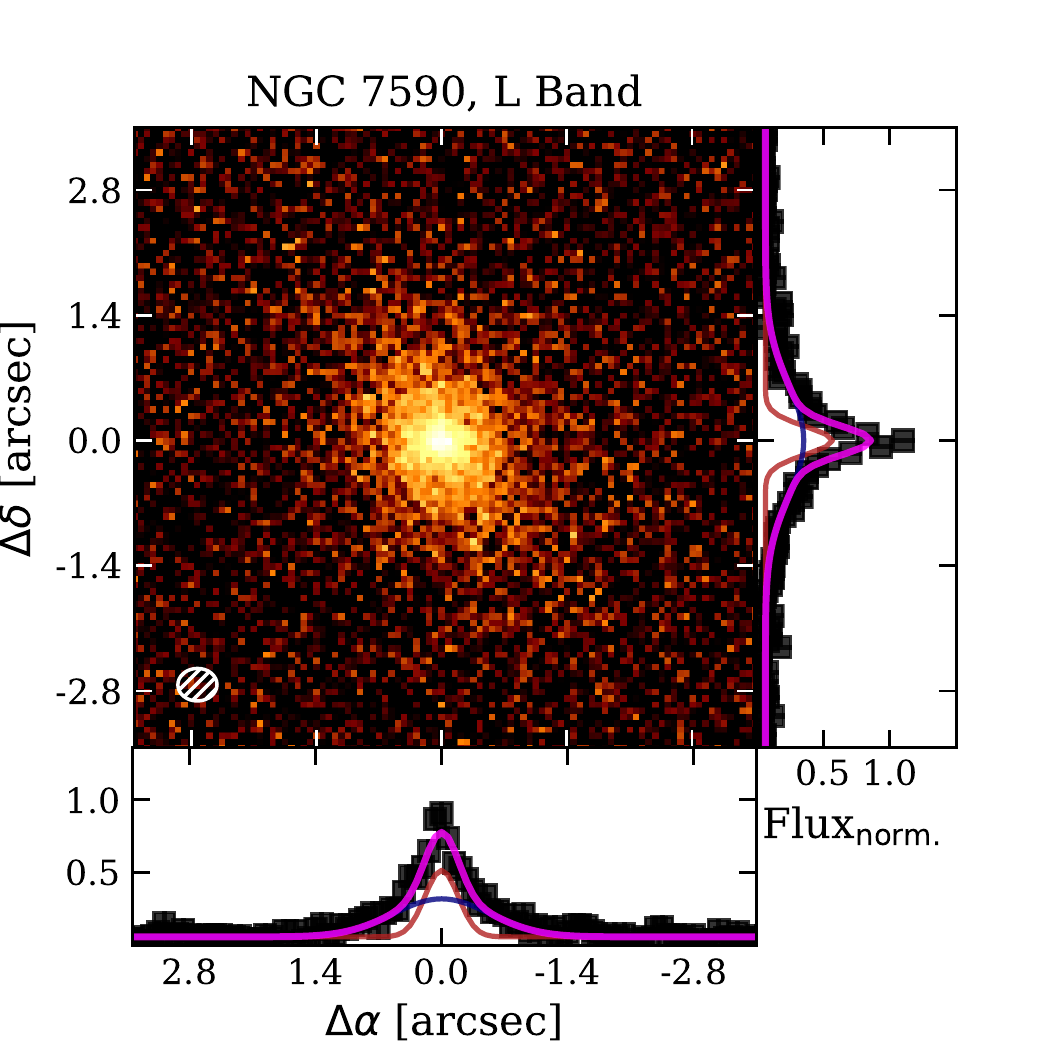}}
\subfloat{\includegraphics[width=0.25\hsize]{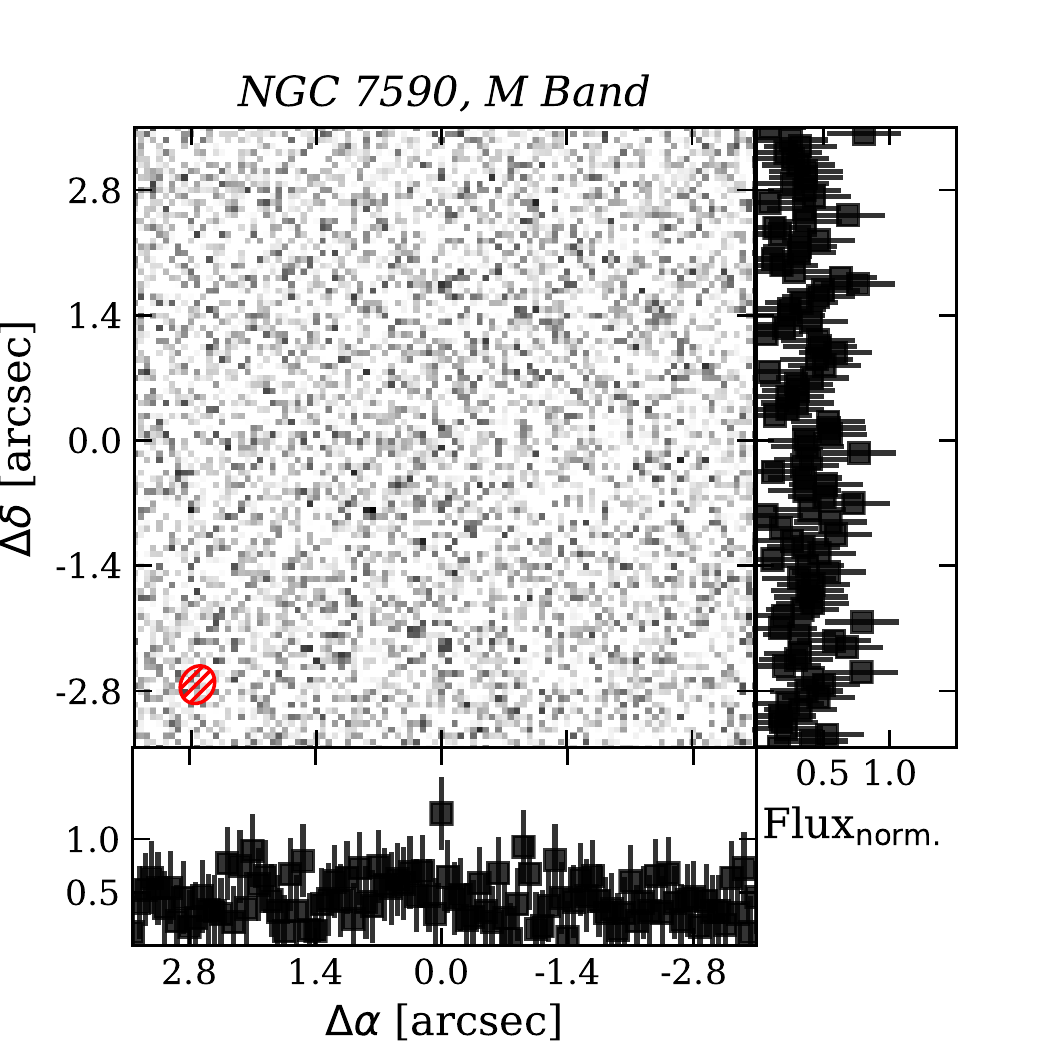}}
\subfloat{\includegraphics[width=0.25\hsize]{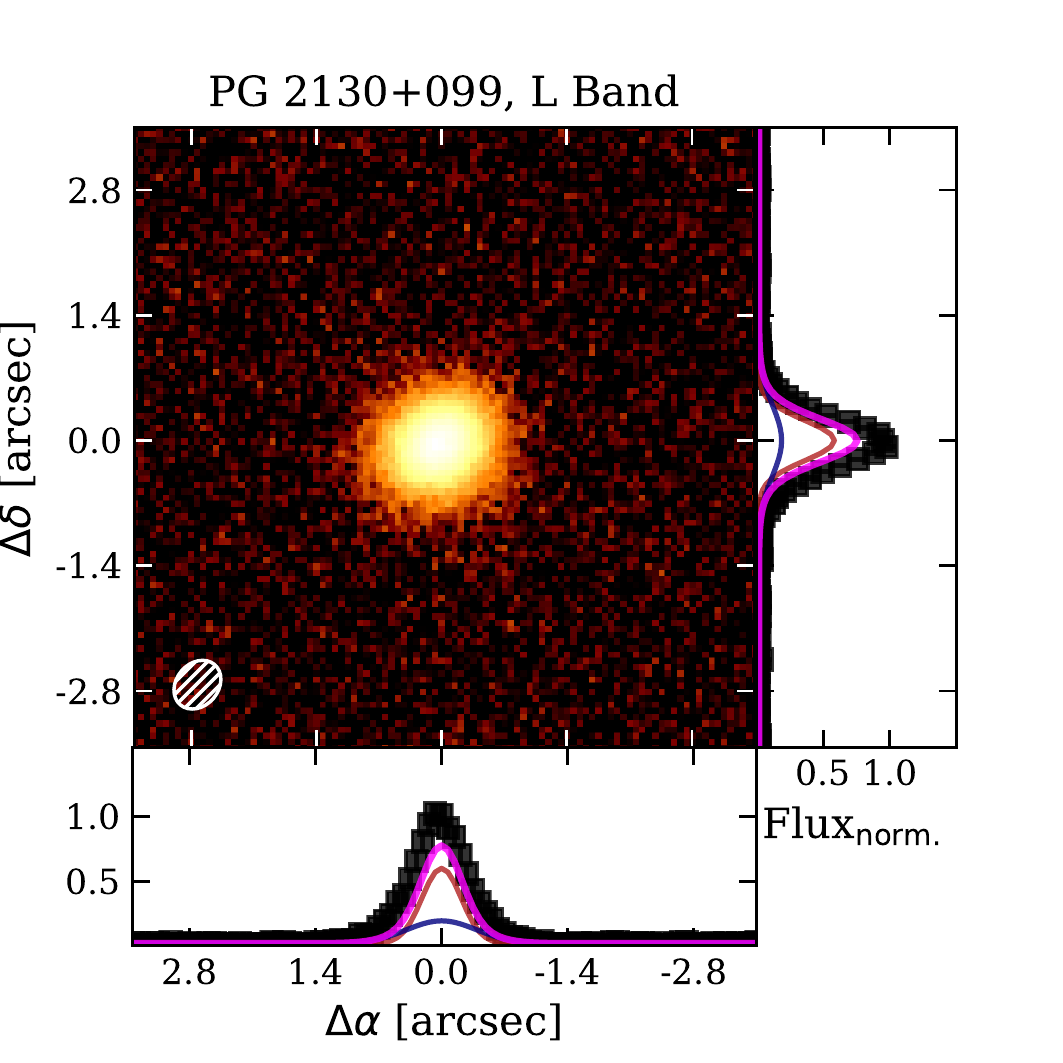}} \\
\caption{As Fig. \ref{fig:cutouts_one}.}
\end{figure*}

\begin{figure*}
\centering
\subfloat{\includegraphics[width=0.25\hsize]{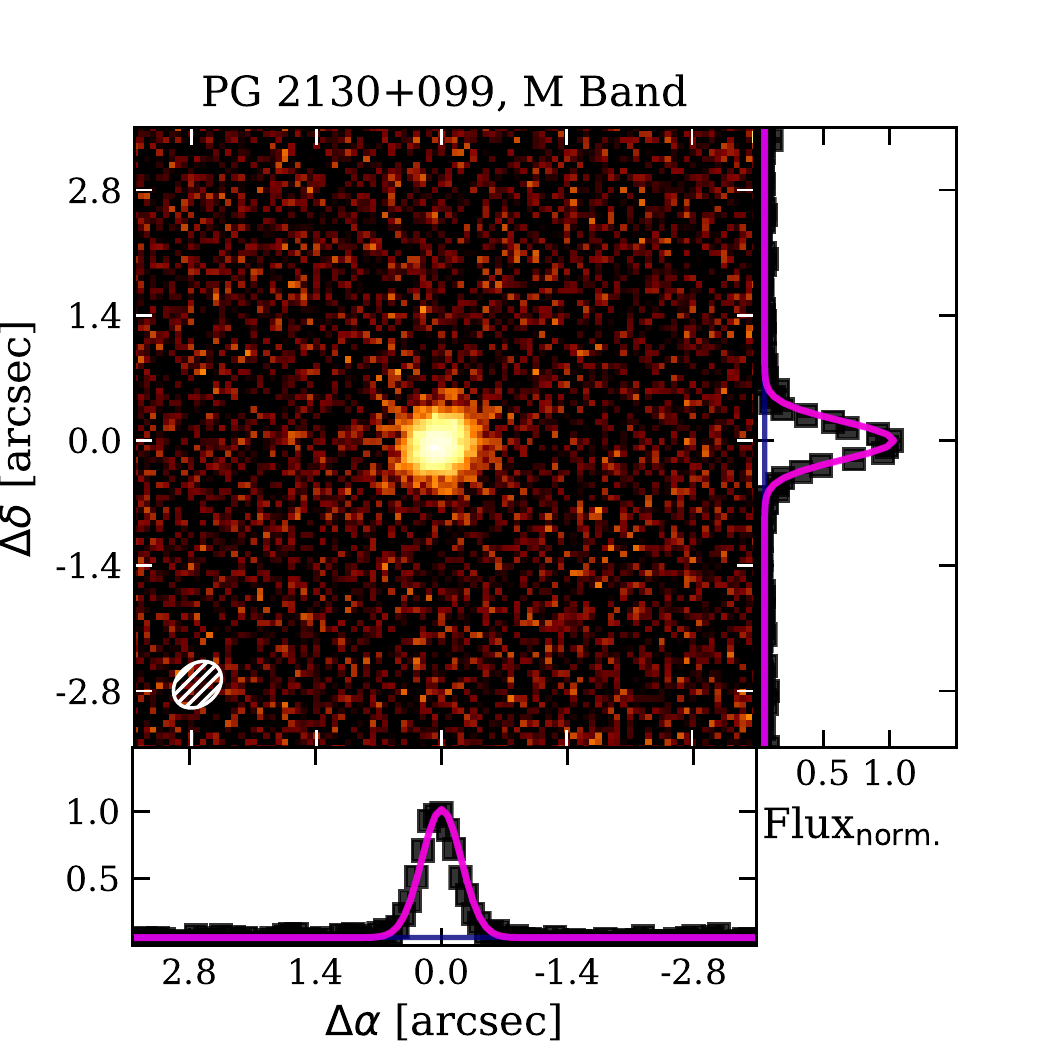}}
\subfloat{\includegraphics[width=0.25\hsize]{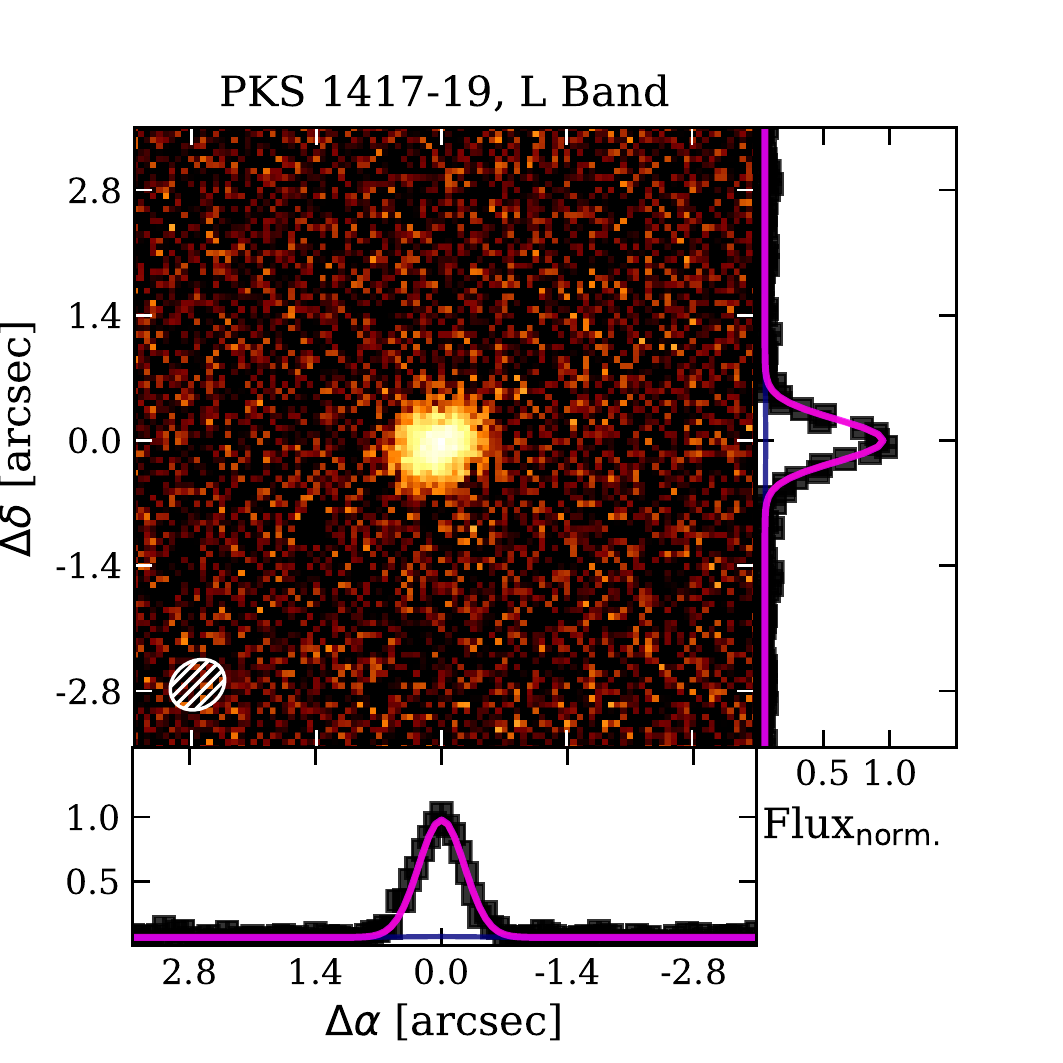}}
\subfloat{\includegraphics[width=0.25\hsize]{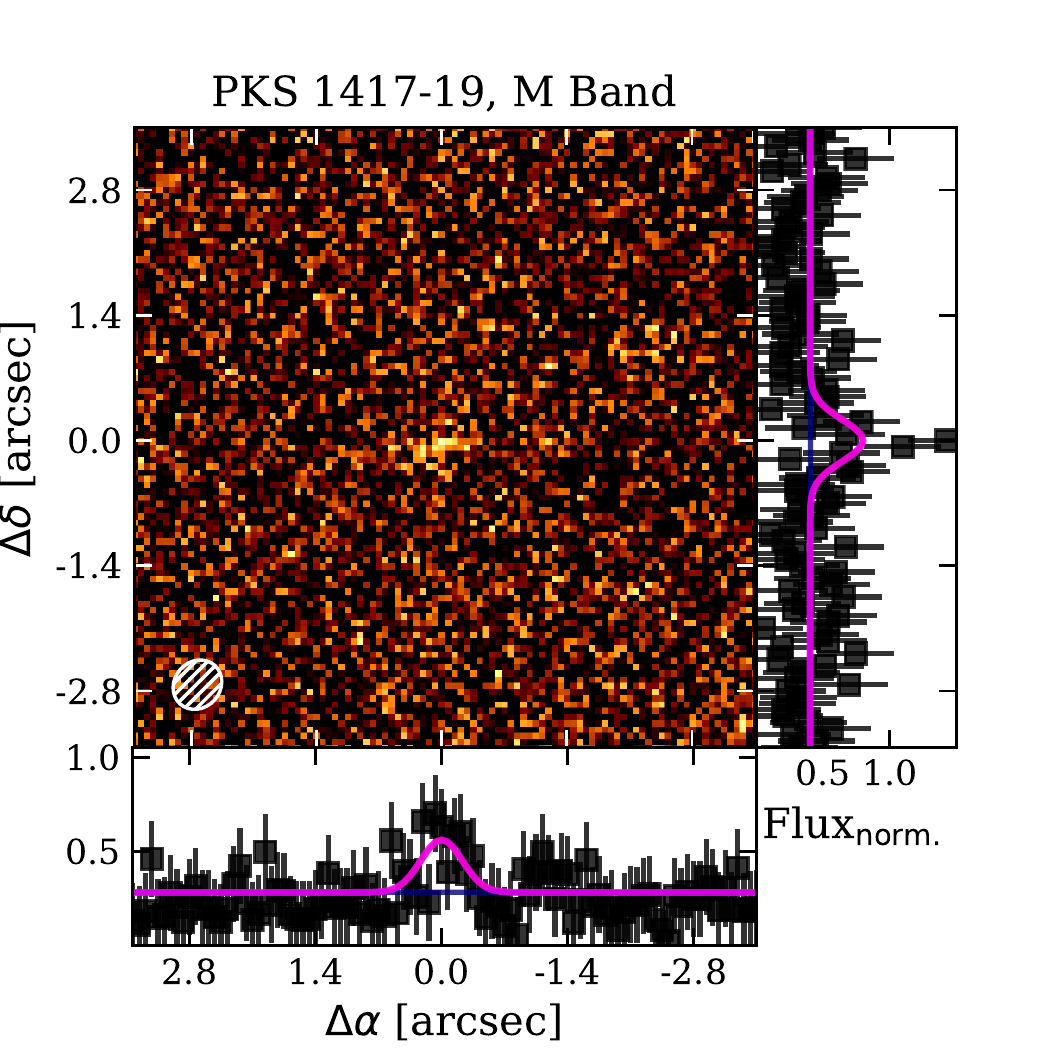}}
\subfloat{\includegraphics[width=0.25\hsize]{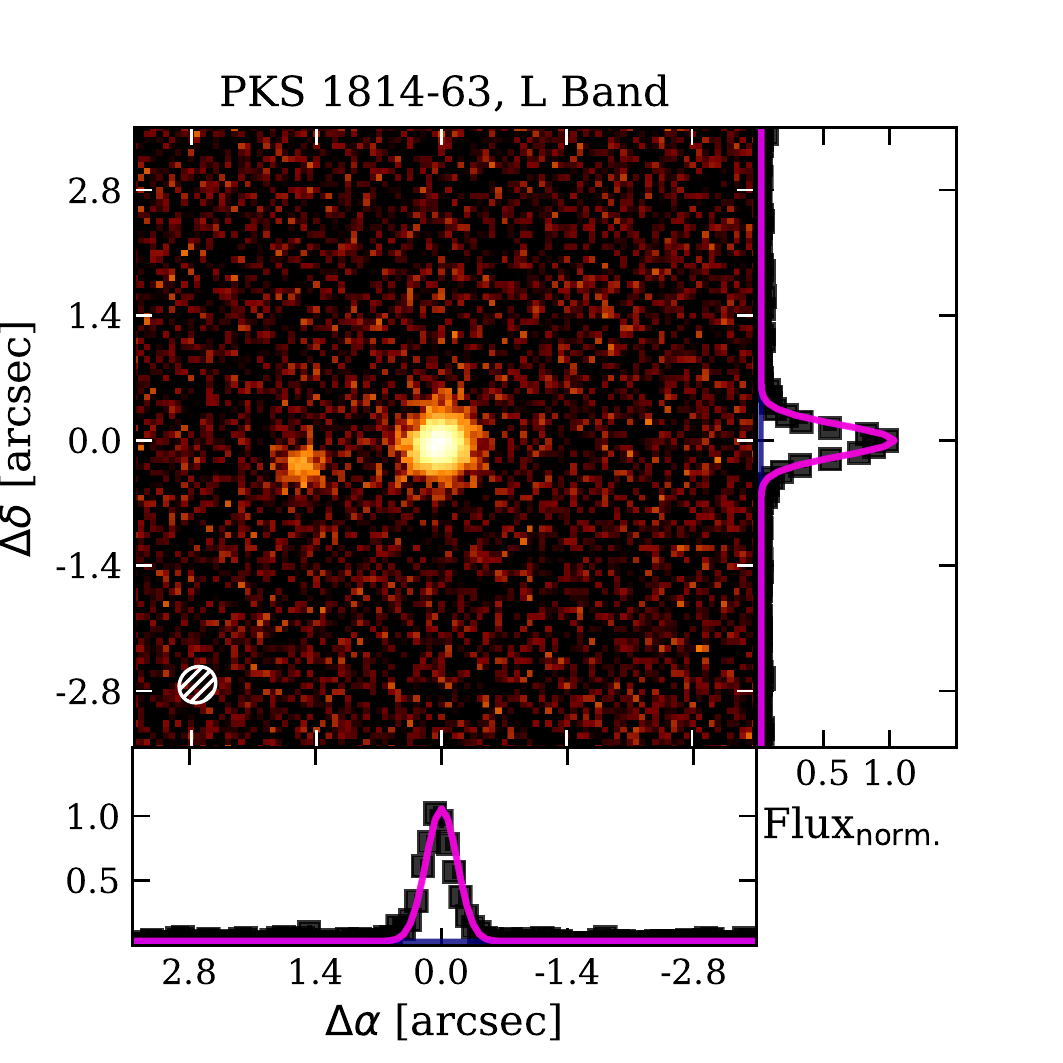}} \\
\subfloat{\includegraphics[width=0.25\hsize]{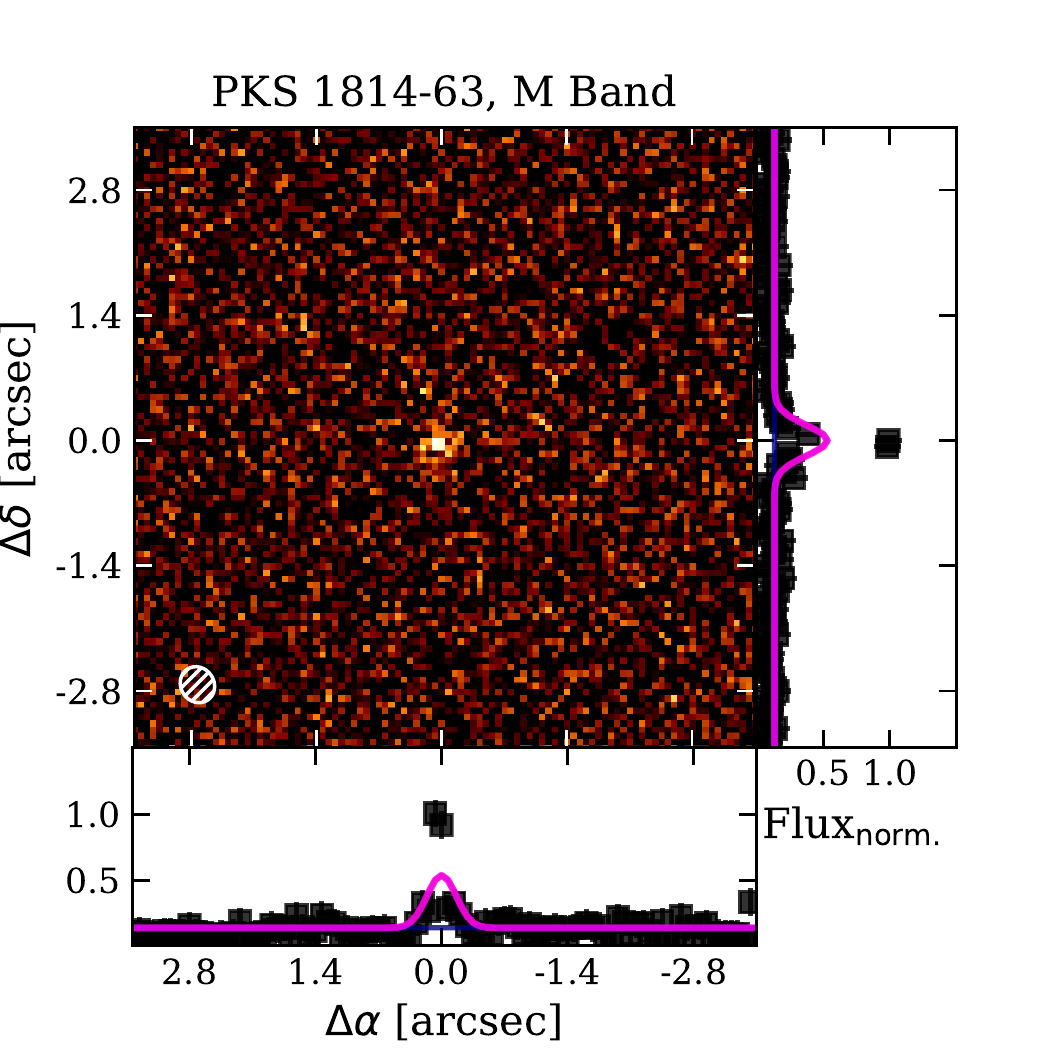}}
\subfloat{\includegraphics[width=0.25\hsize]{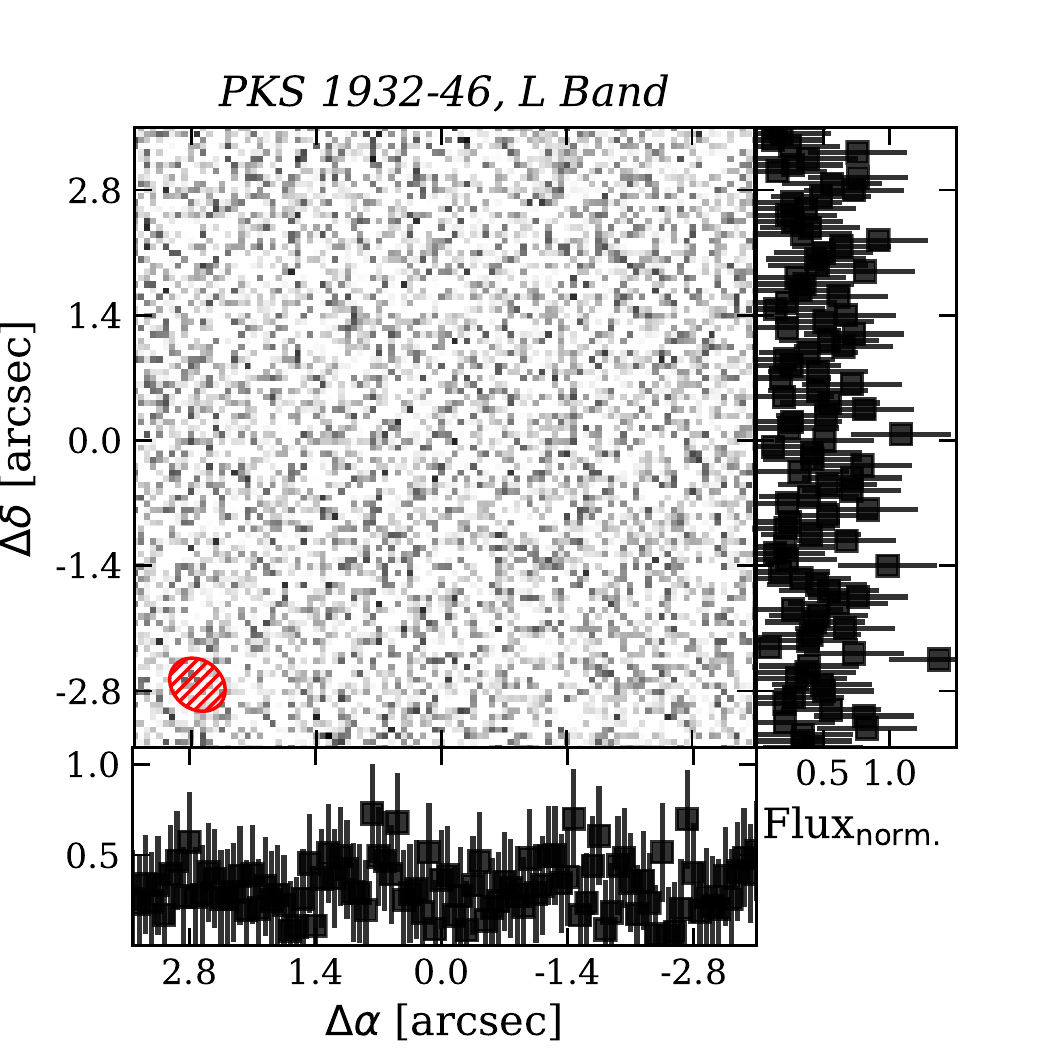}}
\subfloat{\includegraphics[width=0.25\hsize]{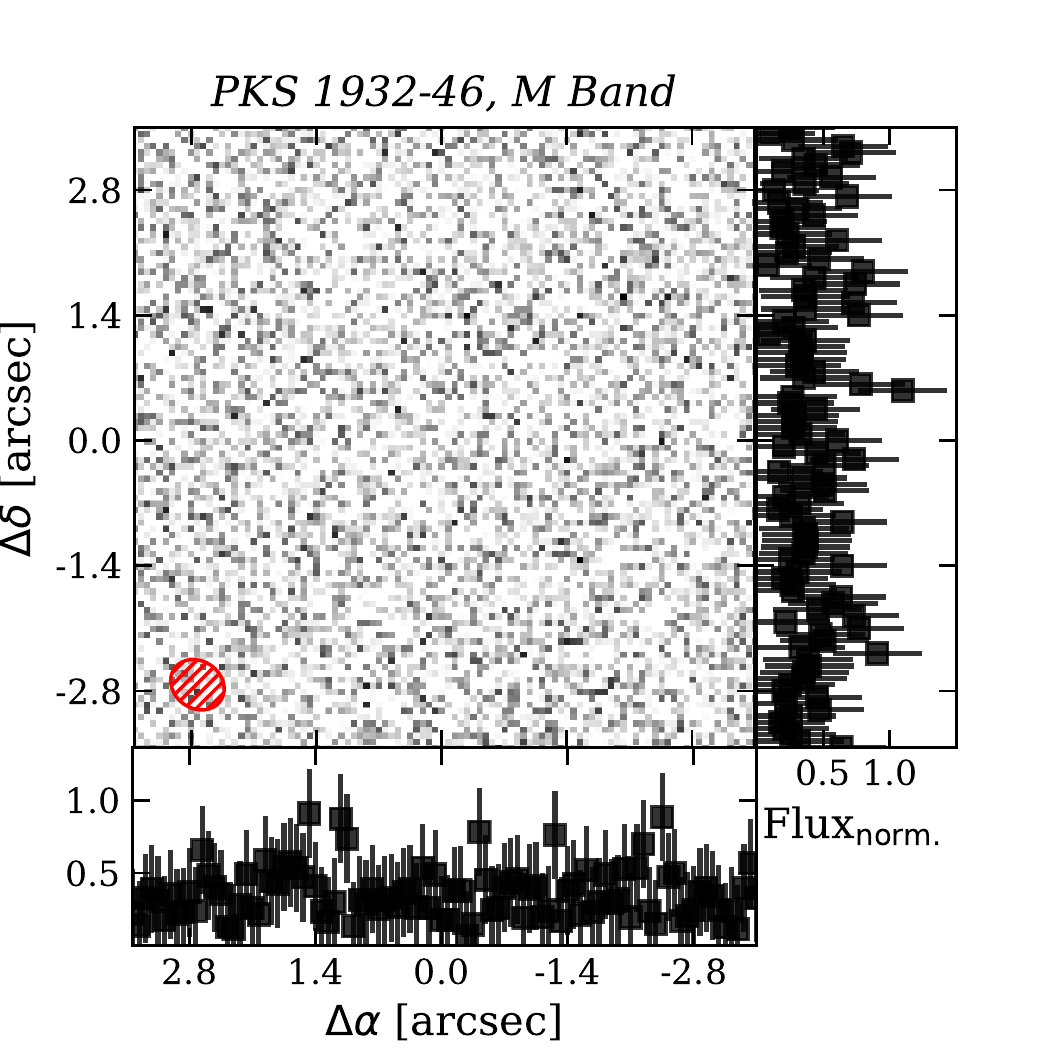}}
\subfloat{\includegraphics[width=0.25\hsize]{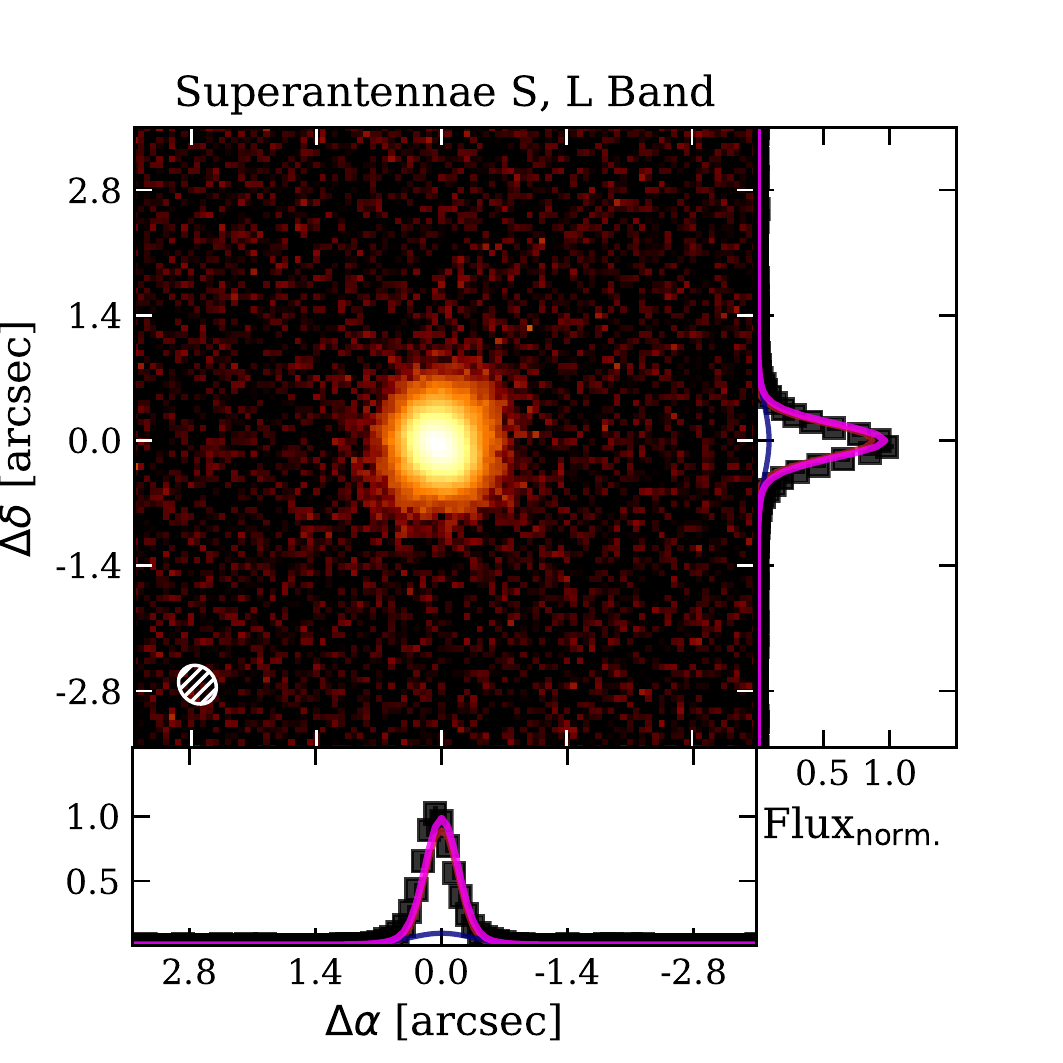}} \\
\subfloat{\includegraphics[width=0.25\hsize]{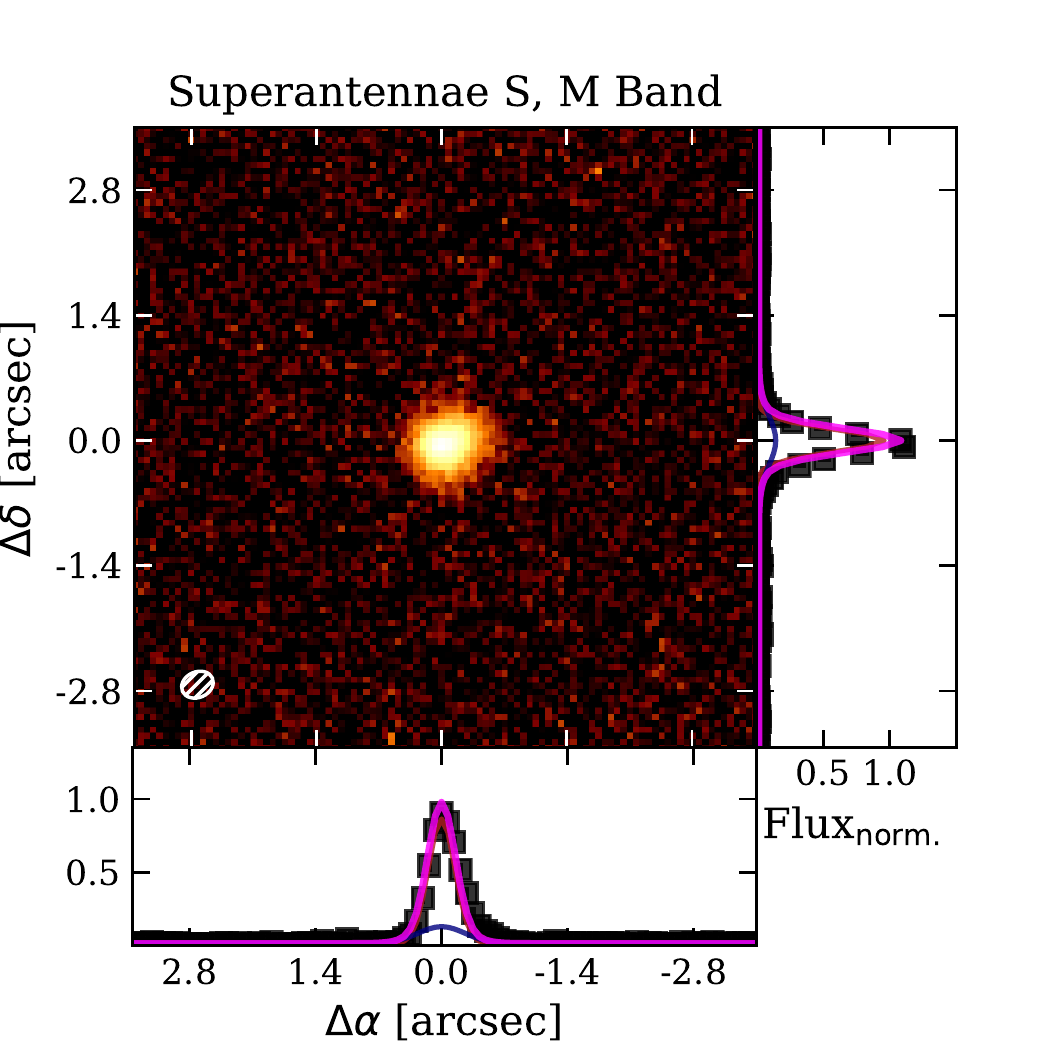}}
\subfloat{\includegraphics[width=0.25\hsize]{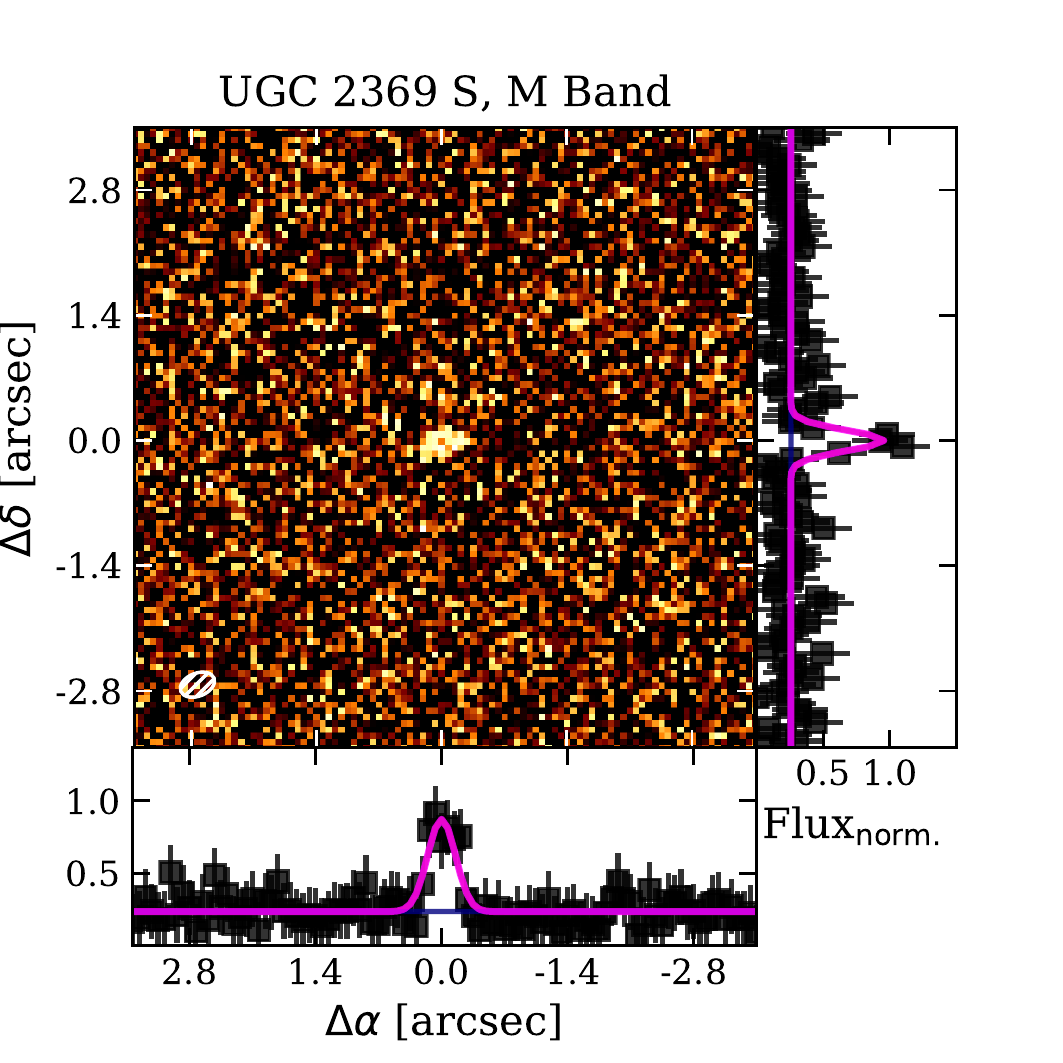}}
\subfloat{\includegraphics[width=0.25\hsize]{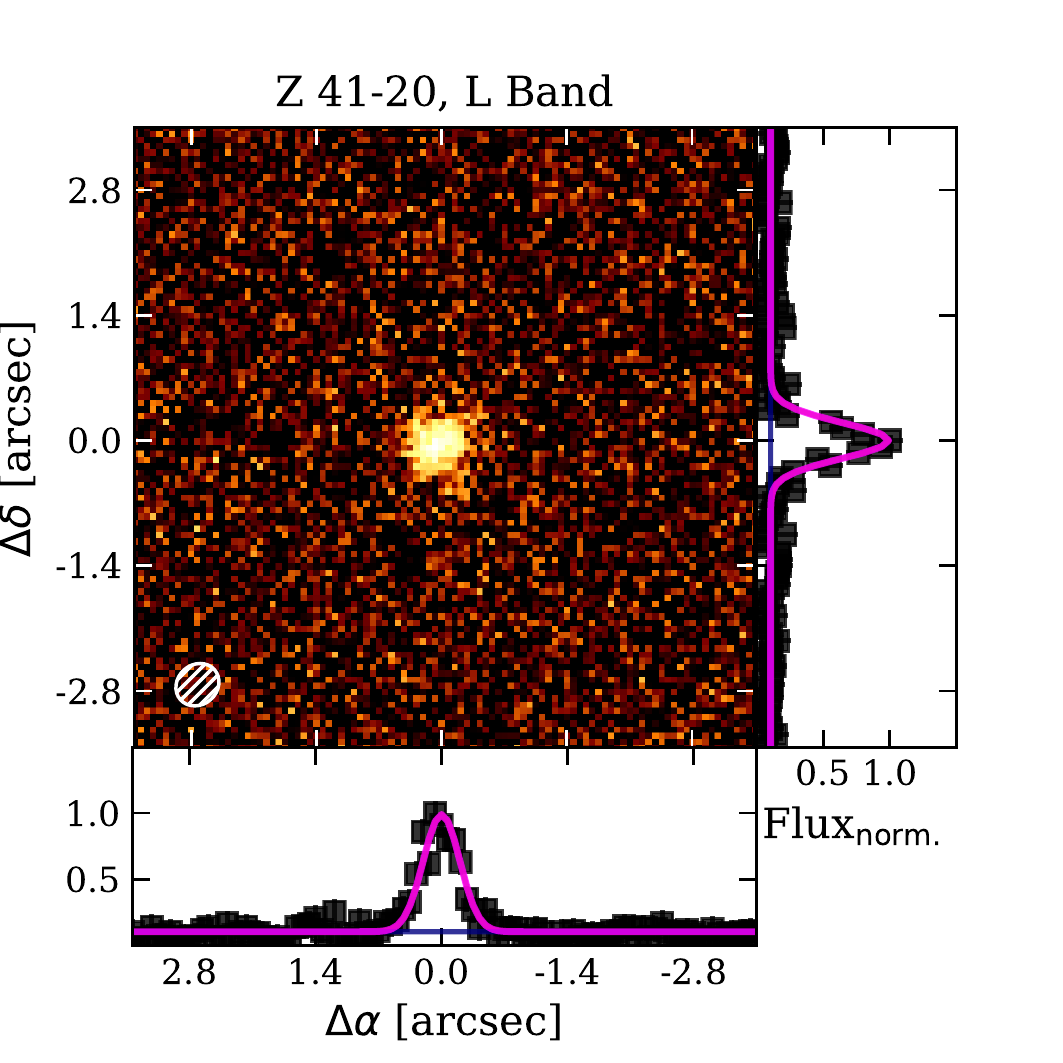}}
\subfloat{\includegraphics[width=0.25\hsize]{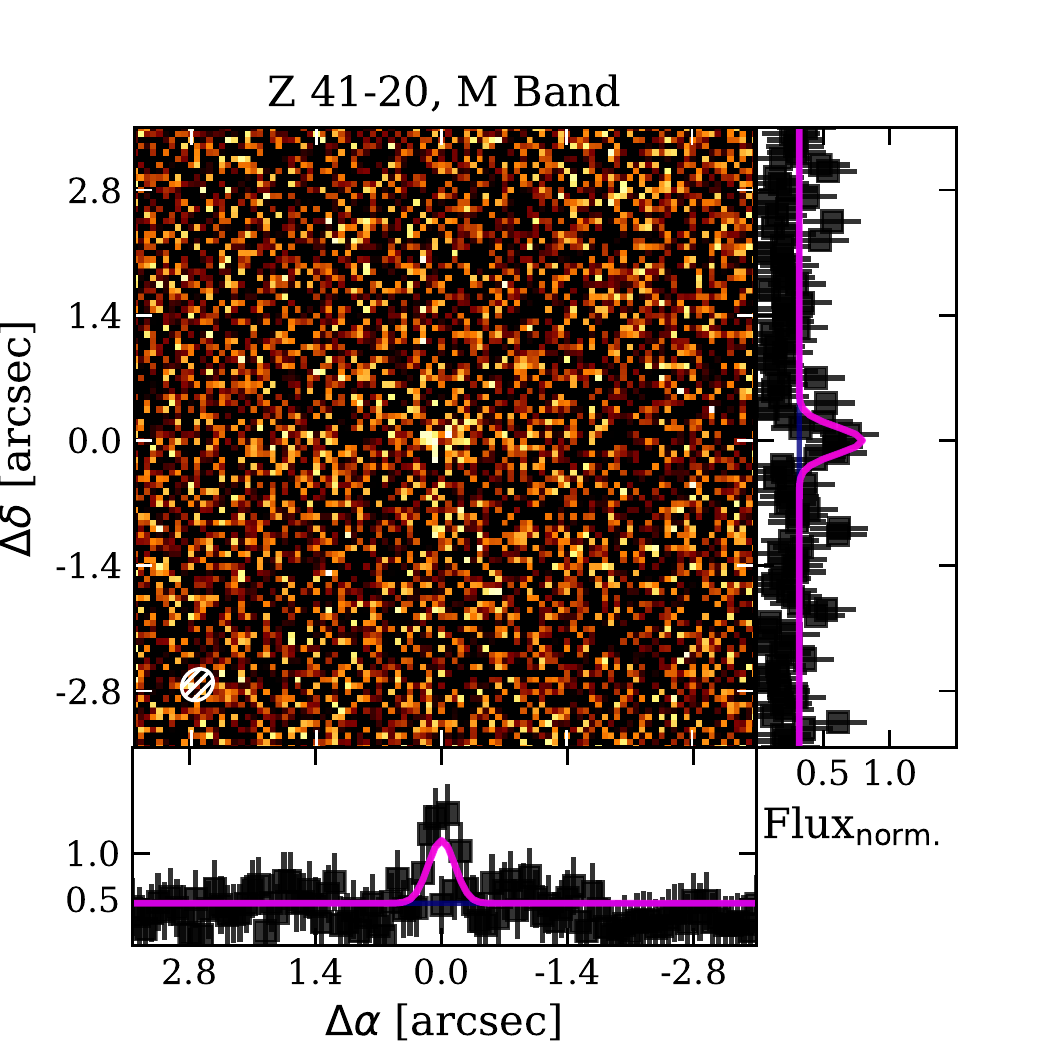}} \\
\caption{As Fig. \ref{fig:cutouts_one}.}
\end{figure*}